\documentclass{aa}  

\usepackage{graphicx}
\usepackage{amsmath,amssymb}
\usepackage{longtable,lscape}
\usepackage{morefloats}

\usepackage{txfonts}
\begin{document}

   \title{A MALT90 study of the chemical properties of massive clumps and 
filaments of infrared dark clouds\thanks{This publication is partly based on data acquired with the Atacama Pathfinder EXperiment (APEX) under programme {\tt 087.F-9315(A)}. APEX is a collaboration between the Max-Planck-Institut f\"{u}r Radioastronomie, the European Southern Observatory, and the Onsala Space Observatory.}}


   \author{O. Miettinen}

   \institute{Department of Physics, University of Helsinki, P.O. Box 64, FI-00014 Helsinki, Finland\\ \email{oskari.miettinen@helsinki.fi}}

   \date{Received ; accepted}

\authorrunning{Miettinen}
\titlerunning{A MALT90 chemical study of clumps within IRDCs}

  \abstract
   {Infrared dark clouds (IRDCs) provide a useful testbed in which to 
investigate the genuine initial conditions and early stages of massive-star 
formation.}
   {We attempt to characterise the chemical properties of a sample of 35 
massive clumps of IRDCs through multi-molecular line observations. We also 
search for possible evolutionary trends among the derived chemical parameters.}
   {The clumps are studied using the MALT90 line survey data obtained with 
the Mopra 22-m telescope. The survey covers 16 different transitions 
near 90 GHz. The spectral-line data are used in concert with 
our previous APEX/LABOCA 870-$\mu$m dust emission data.}
   {Eleven MALT90 transitions are detected towards the clumps at 
least at the $3\sigma$ level. Most of the detected species (SiO, C$_2$H, HNCO, 
HCN, HCO$^+$, HNC, HC$_3$N, and N$_2$H$^+$) show spatially extended emission 
towards many of the sources. The fractional abundances of the molecules with 
respect to H$_2$ are mostly found to be comparable to those determined in 
other recent similar studies of IRDC clumps (Vasyunina et al. 2011, A\&A, 527, 
A88; Sanhueza et al. 2012, ApJ, 756, 60). We found that the 
abundances of SiO, HNCO, and HCO$^+$ are higher in IR-bright clumps than in
IR-dark sources, reflecting a possible evolutionary trend. A hint of such a 
trend is also seen for HNC and HC$_3$N. 
An opposite trend is seen for the C$_2$H and N$_2$H$^+$ 
abundances. Moreover, a positive correlation is found between the abundances 
of HCO$^+$ and HNC, and between those of HNC and HCN. The HCN and HNC 
abundances also appear to increase as a function of the N$_2$H$^+$ abundance. 
The HNC/HCN and N$_2$H$^+$/HNC abundance ratios 
are derived to be near unity on average, while that of HC$_3$N/HCN is 
$\sim10\%$. The N$_2$H$^+$/HNC ratio appears to increase as the clump evolves, 
while the HNC/HCO$^+$ ratio shows the opposite behaviour. }
   {The detected SiO emission is likely caused by shocks driven by outflows in 
most cases, although shocks resulting from the cloud formation process 
could also play a role. Shock-origin for the HNCO, HC$_3$N, and CH$_3$CN 
emission is also plausible. The average HNC/HCN ratio 
is in good agreement with those seen in other IRDCs, but gas temperature 
measurements would be neeeded to study its temperature dependence. Our results 
support the finding that C$_2$H can trace the cold gas, and not just 
the photodissociation regions. The HC$_3$N/HCN ratio appears to be comparable 
to the values seen in other types of objects, such as T Tauri disks and 
comets.} 


   \maketitle

%

\section{Introduction}

The so-called infrared dark clouds (IRDCs) are, by definition, seen 
as dark absorption features against the Galactic mid-IR background 
radiation field (\cite{perault1996}; \cite{egan1998}; \cite{simon2006}; 
\cite{peretto2009}). Infrared dark clouds are a relatively new class of 
molecular clouds and, like molecular clouds in general, IRDCs represent the 
cradles of new stars. While the majority of IRDCs may serve as sites for the 
formation of low- to intermediate-mass stars and stellar clusters 
(\cite{kauffmann2010}), observations have shown that some of them are capable 
of giving birth to high-mass ($M_{\star}\gtrsim8$ M$_{\sun}$; spectral type 
B3 or earlier) stars (e.g., \cite{rathborne2006}; \cite{beuther2007}; 
\cite{chambers2009}; \cite{battersby2010}; \cite{zhang2011}). 
Although IRDCs often show clear signs of star-formation activity 
(such as point sources emitting IR radiation), some of the clumps and 
cores\footnote{Throughout the present paper, we use the term ``clump'' to 
refer to sources whose typical radii, masses, and mean densities are 
$\sim0.2-1$ pc, $\sim10^2-10^3$ M$_{\sun}$, and $\sim10^3-10^4$ cm$^{-3}$, 
respectively (cf.~\cite{bergin2007}). The term ``core'' is used to describe a 
smaller (radius $\sim0.1$ pc) and denser object within a clump.} 
of IRDCs are found to be candidates of high-mass starless objects 
(e.g., \cite{ragan2012}; \cite{tackenberg2012}; \cite{beuther2013}; 
\cite{sanhueza2013}). Such sources are ideal targets to 
examine the pristine initial conditions of high-mass star formation which are 
still rather poorly understood compared to those of the formation of 
solar-type stars.

Besides the initial conditions, IRDCs provide us with the possibility to 
investigate the subsequent early stages of high-mass star formation. These 
include the high-mass young stellar objects (YSOs), hot molecular cores 
(HMCs; e.g., \cite{kurtz2000}), and hyper- and ultracompact (UC) \ion{H}{ii} 
regions (e.g., \cite{churchwell2002}; \cite{hoare2007}). From a chemical point 
of view, dense ($\gtrsim10^4$ cm$^{-3}$) and cold ($\gtrsim10$ K) 
starless IRDCs are expected to be 
characterised by the so-called dark-cloud chemistry which is dominated by 
reactions between electrically charged species (ions) and neutral species 
(e.g., \cite{herbst1973}; \cite{vandishoeck1998}). During this phase, the dust 
grains that are mixed with the gas are expected to accumulate icy mantles 
around them due to freeze-out of some of the gas-phase species onto grain 
surfaces. If the source evolves to the HMC phase 
characterised by the dust temperature of $\gtrsim100$ K, the ice mantles of 
dust grains are evaporated into the gas-phase leading to a rich and complex 
chemistry (e.g., \cite{charnley1995}). Moreover, shocks occuring during the 
course of star formation, e.g., due to outflows, compress and heat the gas, 
and can fracture the grain mantles or even the grain cores leading to shock 
chemistry (e.g., \cite{bachiller1997}). As the chemistry of a star-forming 
region is very sensitive to prevailing physical conditions (temperature, 
density, ionisation degree), understanding the chemical composition is of 
great importance towards unveiling the physics of the early stages of 
high-mass star formation. Clearly, the chemical composition of the source 
changes with time, so the evolutionary timescale of the star-formation 
process can also be constrained through estimating the chemical age.

Establishing the chemical properties of certain types of interstellar clouds 
requires large surveys to be conducted. In the past years, some multi-molecular
line surveys of IRDC sources have already been published [\cite{ragan2006}; 
\cite{beuthersridharan2007}; \cite{sakai2008}; \cite{gibson2009}; 
\cite{sakai2010}; \cite{vasyunina2011} (henceforth called VLH11); 
\cite{miettinen2011}; \cite{sanhueza2012} (hereinafter called SJF12); 
\cite{liu2013}]. However, most of the 
line survey studies of IRDCs performed so far 
are based on single-pointing observations in which case the spatial 
distribution of the studied species cannot be explored. 
To further characterise the chemical properties of IRDCs, the present paper 
presents a multi-line study of a sample of massive clumps within IRDCs 
selected from Miettinen (2012b; hereafter, Paper I). 
Similarly to Liu et al. (2013), the spectral-line data presented here were 
taken from the Millimetre Astronomy Legacy Team 90 GHz (MALT90) survey 
(\cite{foster2011}; \cite{jackson2013}). As these data are based on 
mapping observations, we are able to study the spatial distribution of the 
line emission and the possible correlation between the emission of different 
species. This way we can examine the chemistry of several different species 
on clump scales and how the chemical properties vary among different sources 
or different evolutionary stages. After describing the source sample and data 
in Sect.~2, the observational results and analysis are presented in Sect.~3. 
In Sect.~4, we discuss the results until summarising the paper in Sect.~5.

\section{Data}

\subsection{Source selection}

The source sample of the present paper was selected among the sources 
studied in Paper I where a sample of IRDC regions were investigated 
through mapping observations of the 870-$\mu$m dust continuum emission with 
the APEX/LABOCA bolometer array. From the four fields mapped with LABOCA, 
containing 91 clumps in total, 
altogether 35 clumps are included in the MALT90 survey\footnote{This overlap 
is not a coincidence in the sense that the MALT90 target sources were selected 
from the ATLASGAL (APEX Telescope Large Area Survey of the Galaxy) 870-$\mu$m 
survey (\cite{schuller2009}; \cite{contreras2013}).}. However, three of 
these clumps are only partly covered by 
MALT90 maps. The selected clumps are likely to encompass different 
evolutionary stages, ranging from IR-dark sources (13) to \ion{H}{ii} regions 
with bright IR emission (22 sources are associated with either IR point 
sources and/or extended-like IR emission). In Paper I, the clumps were 
classified into IR-dark and YSO-hosting ones depending on the  
\textit{Spitzer} IRAC-colours of the point sources. Some of the studied clumps 
belong to filamentary IRDCs, most notable in the case of G11.36+0.80 
(hereafter, G11.36 etc.). Moreover, the sample includes clumps associated with 
the mid-IR bubble pair N10/11 (\cite{churchwell2006}), a potential site of 
ongoing triggered high-mass star formation (\cite{watson2008}).

In Fig.~\ref{figure:irac}, we show the \textit{Spitzer} 8-$\mu$m images of our 
sources overlaid with contours showing the LABOCA 870-$\mu$m dust emission. 
The angular sizes of the mid-IR images shown in Fig.~\ref{figure:irac} 
correspond to the MALT90 map sizes. 
The sources with their LABOCA peak positions are listed in Table~\ref{clumps}. 
In this table, we also give the source kinematic distance ($d$), 
effective radius ($R_{\rm eff}$), mass ($M$), H$_2$ column density 
[$N({\rm H_2})$], average H$_2$ number density [$\langle n({\rm H_2}) \rangle$],
and comments on the source appearance at IR wavelengths. The physical 
parameters shown in Table~\ref{clumps} were revised from those presented in 
Paper I, and are briefly described in Appendix~A.

\begin{figure*}
\begin{center}
\includegraphics[width=0.33\textwidth]{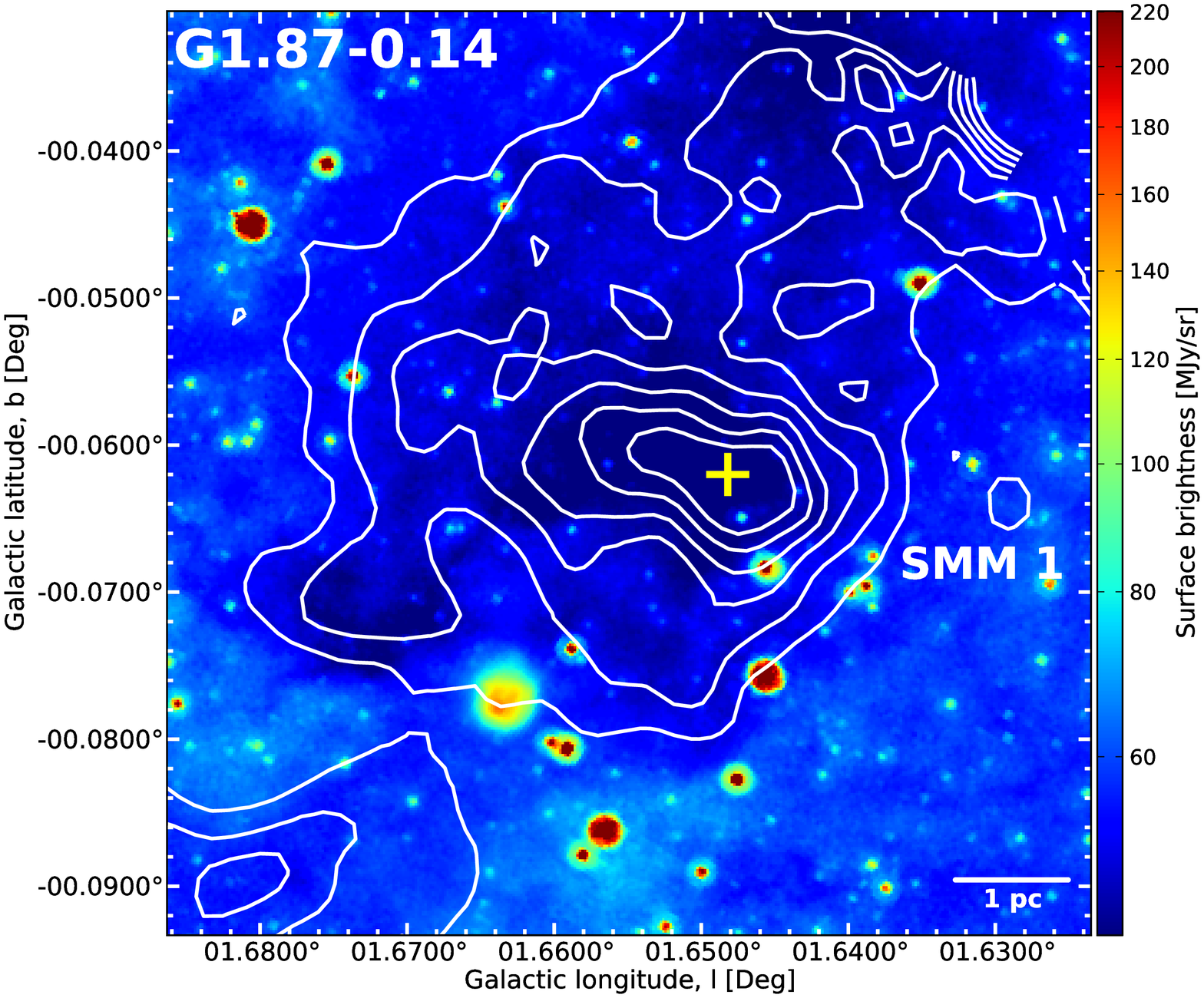}
\includegraphics[width=0.33\textwidth]{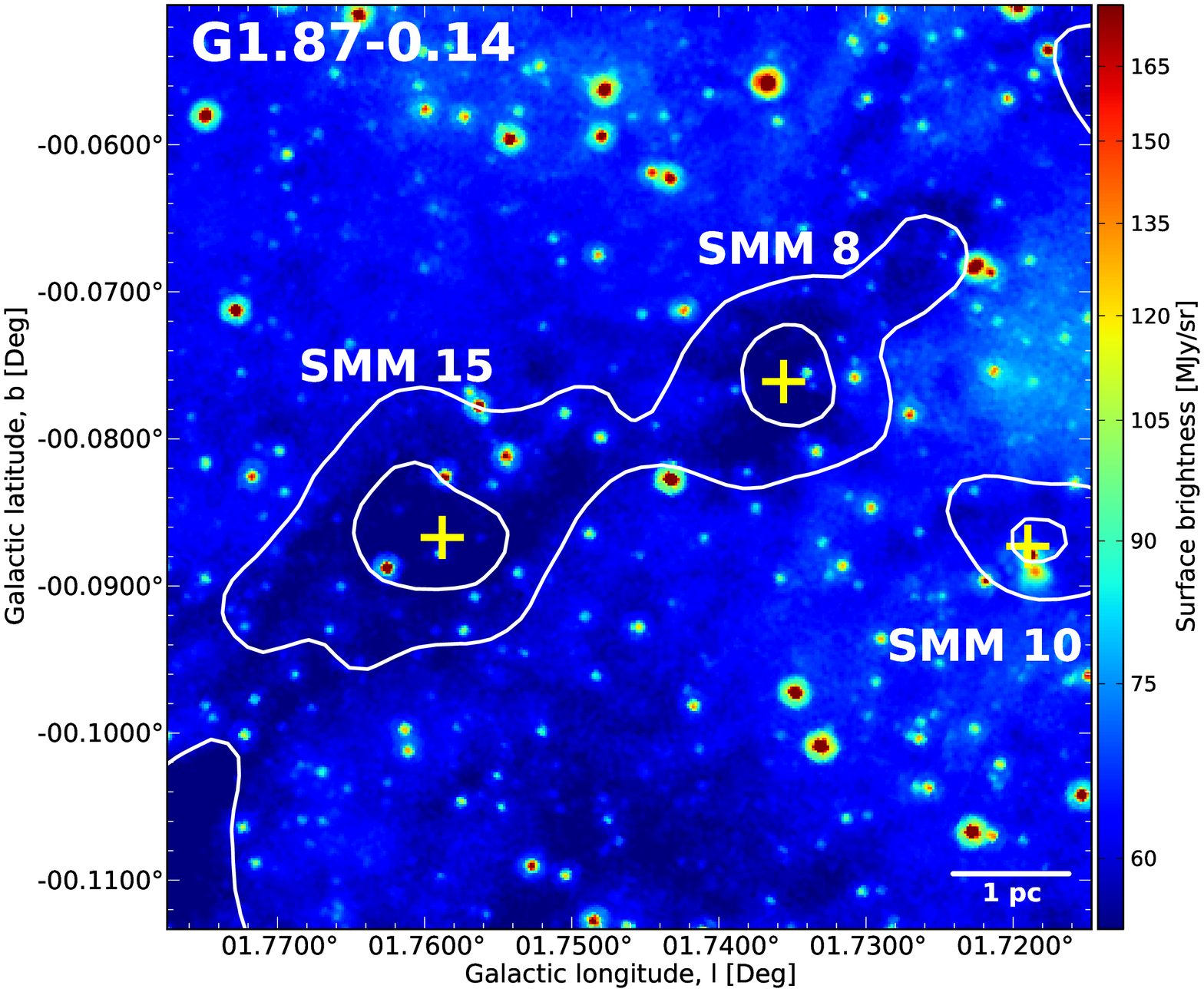}
\includegraphics[width=0.33\textwidth]{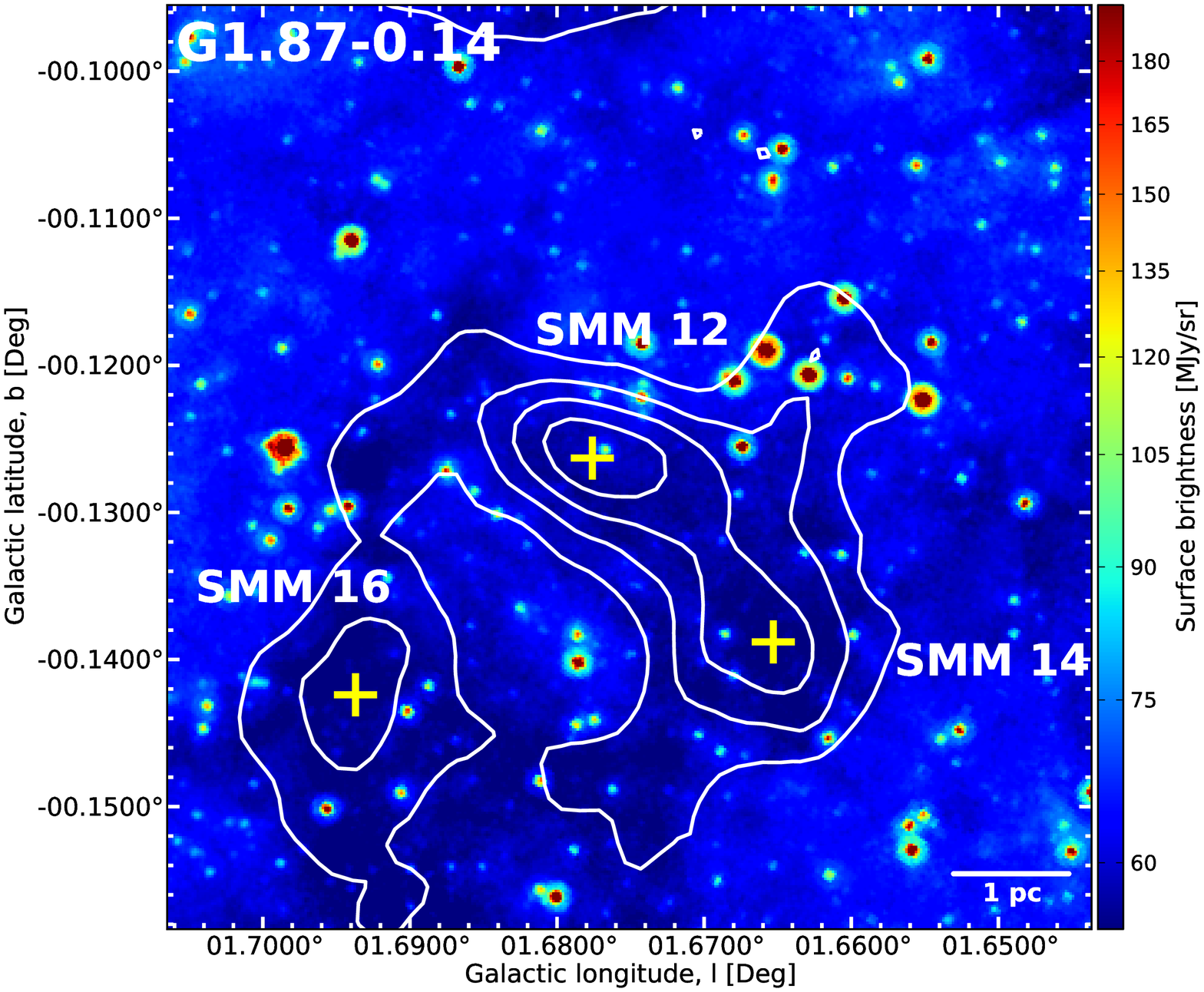}
\includegraphics[width=0.33\textwidth]{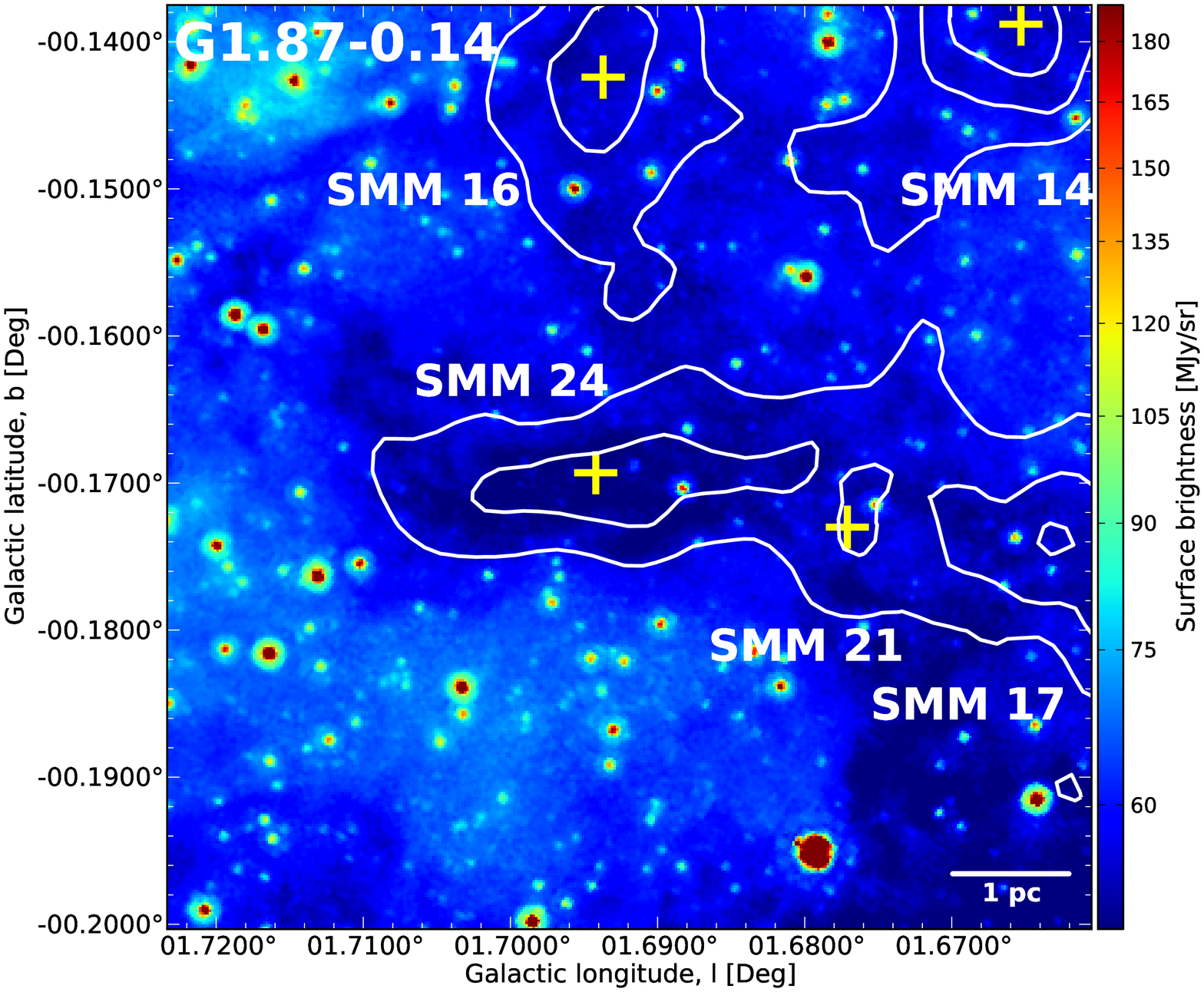}
\includegraphics[width=0.33\textwidth]{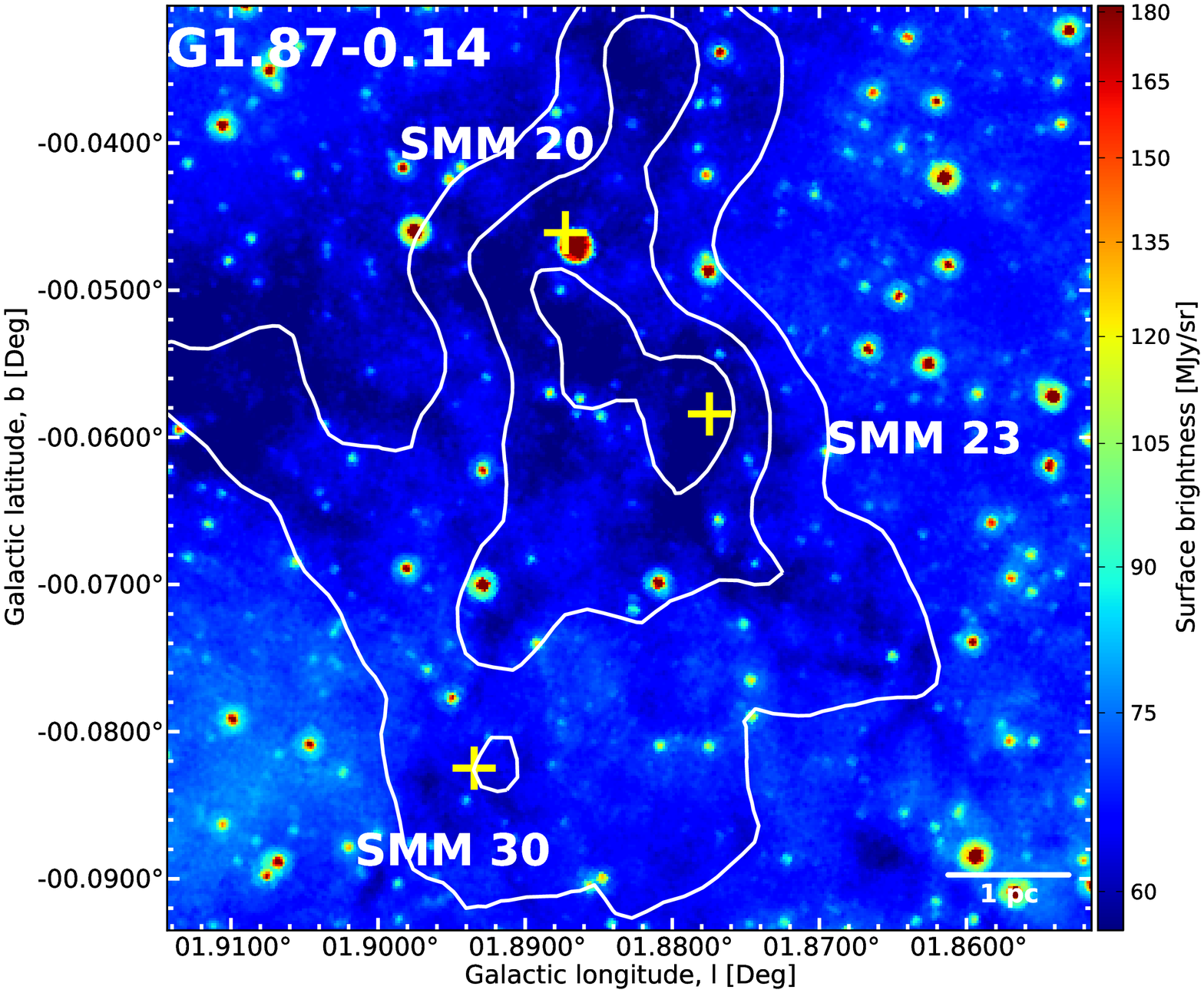}
\includegraphics[width=0.33\textwidth]{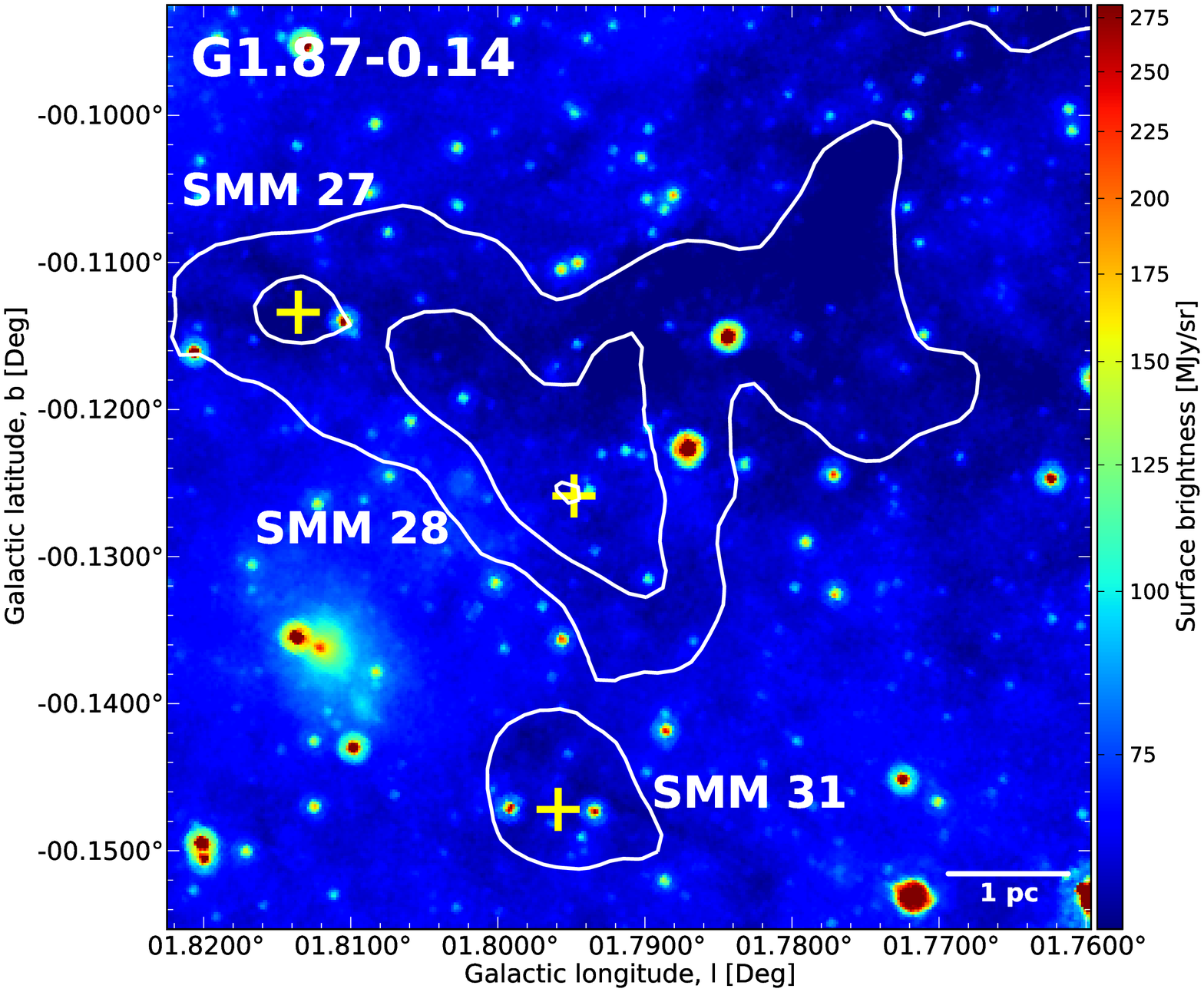}
\includegraphics[width=0.33\textwidth]{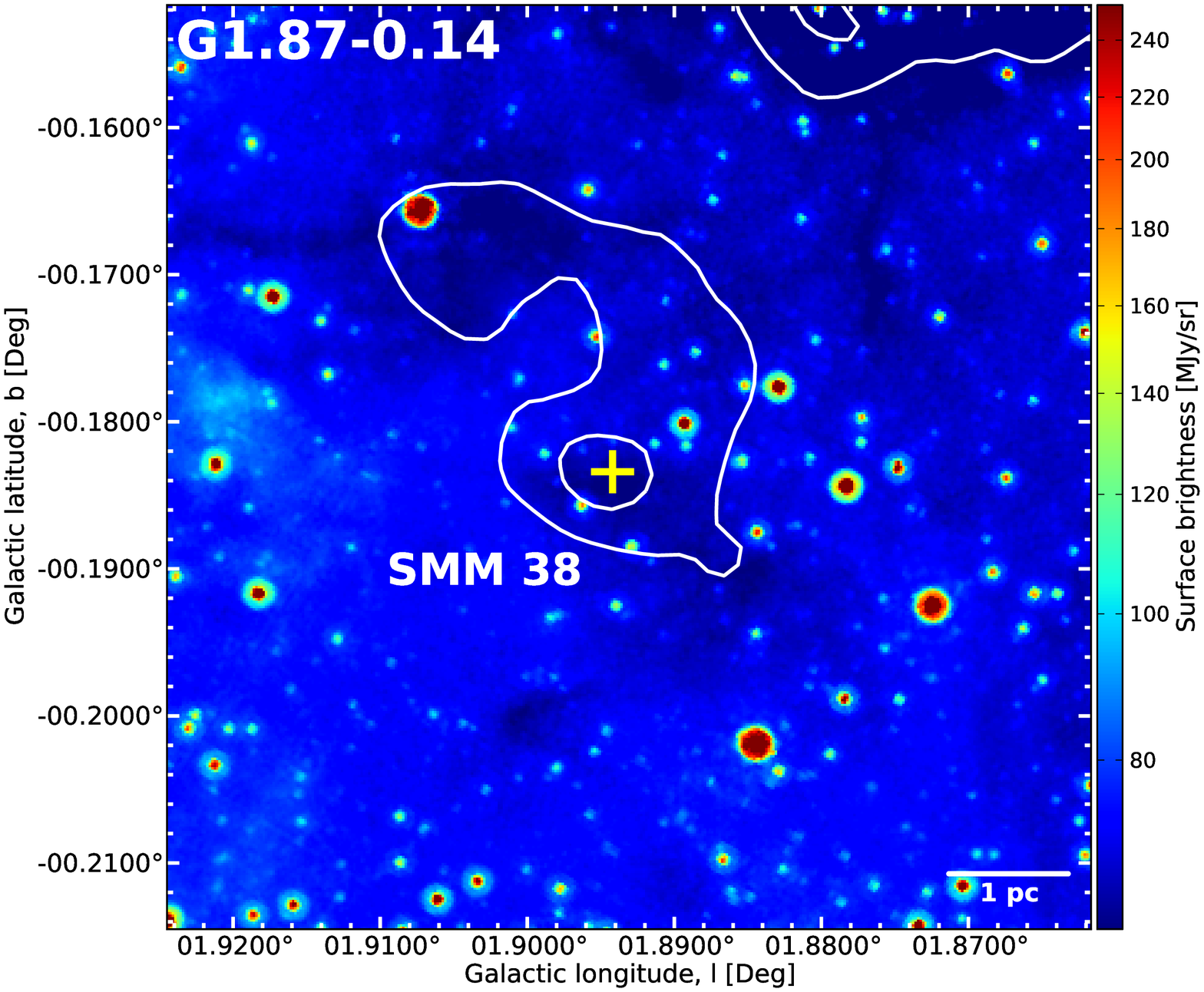}
\includegraphics[width=0.33\textwidth]{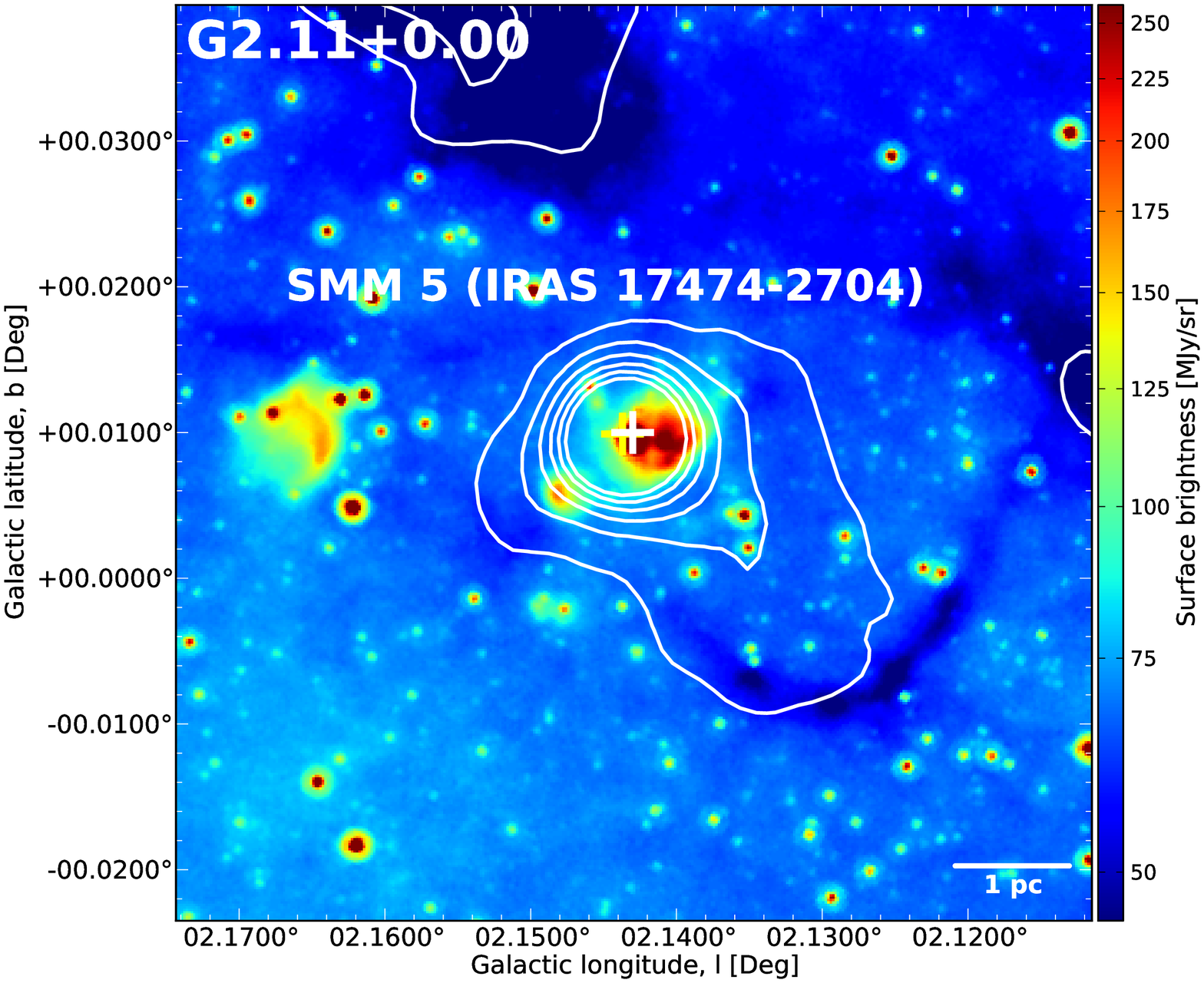}
\includegraphics[width=0.33\textwidth]{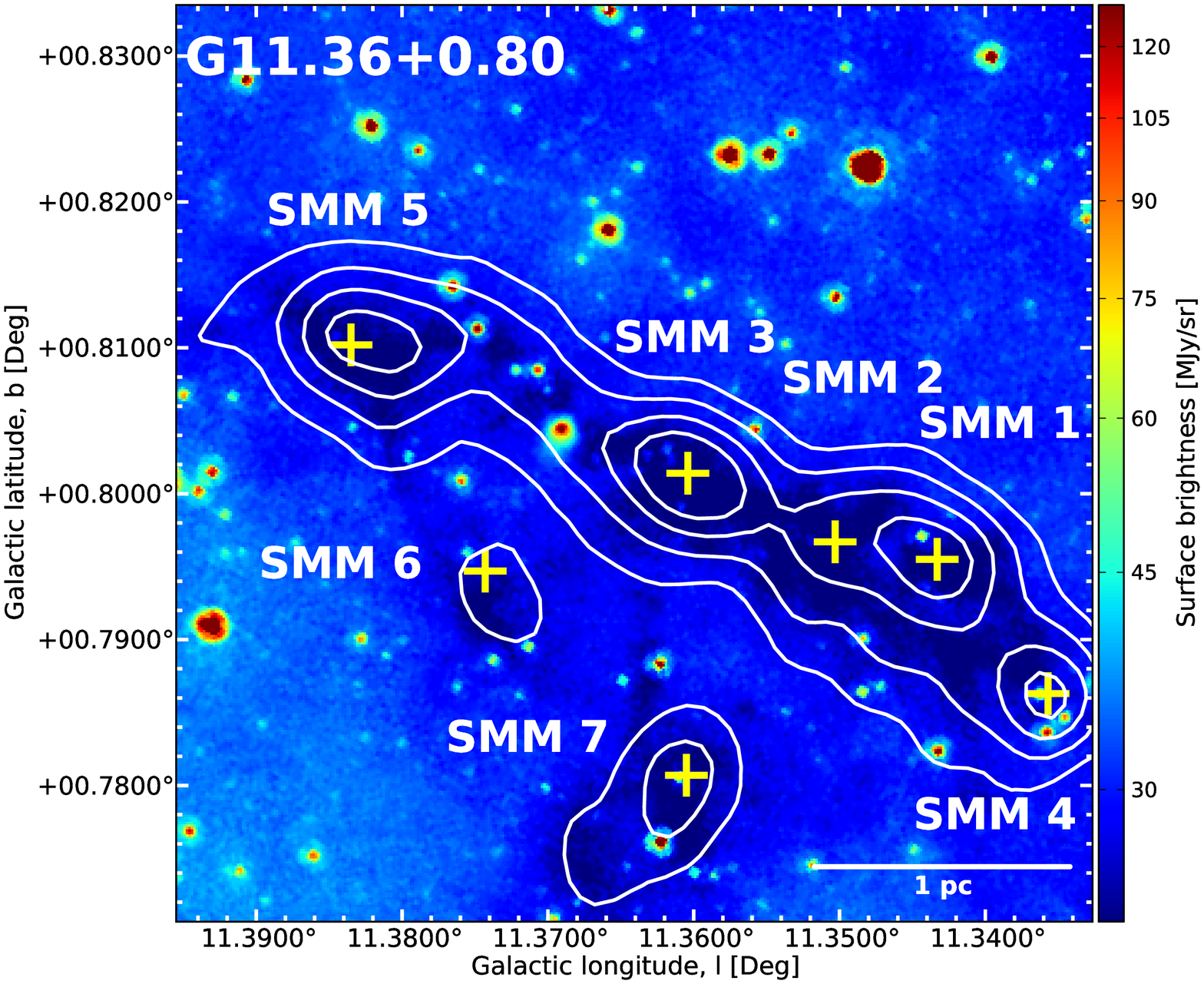}
\caption{\textit{Spitzer}/IRAC 8-$\mu$m images of the clumps and filaments 
studied in the present paper. The images are shown with logarithmic scaling, 
and the colour bars indicate the surface-brightness scale in MJy~sr$^{-1}$. 
The images are overlaid with contours of LABOCA 870-$\mu$m dust continuum 
emission as in Paper I (starting from $3\sigma$ and going in steps of 
$3\sigma$, where $3\sigma$ is 0.27, 0.18, 0.12, and 0.14 Jy~beam$^{-1}$ for 
the fields G1.87, G2.11, G11.36, and G13.22, respectively). 
The 870-$\mu$m peak positions of the clumps are denoted by yellow 
plus signs. In each panel, a scale bar indicating the 1 pc projected length 
is shown, with the assumption of line-of-sight distance given in Col.~(4) of 
Table~\ref{clumps}. The source nomenclature follows that in Paper I. 
The clumps G1.87--SMM 10, 17, and G13.22--SMM 10 are only partly 
covered by the MALT90 maps. The white plus sign towards G2.11--SMM 5 shows the 
position of the UC \ion{H}{ii} region from Becker et al. (1994; as seen at 
5 GHz) and 18-cm OH maser from Argon et al. (2000); the two positions overlap, 
and are very close to the 870-$\mu$m peak position.}
\label{figure:irac}
\end{center}
\end{figure*}

\addtocounter{figure}{-1}
\begin{figure*}
\begin{center}
\includegraphics[width=0.33\textwidth]{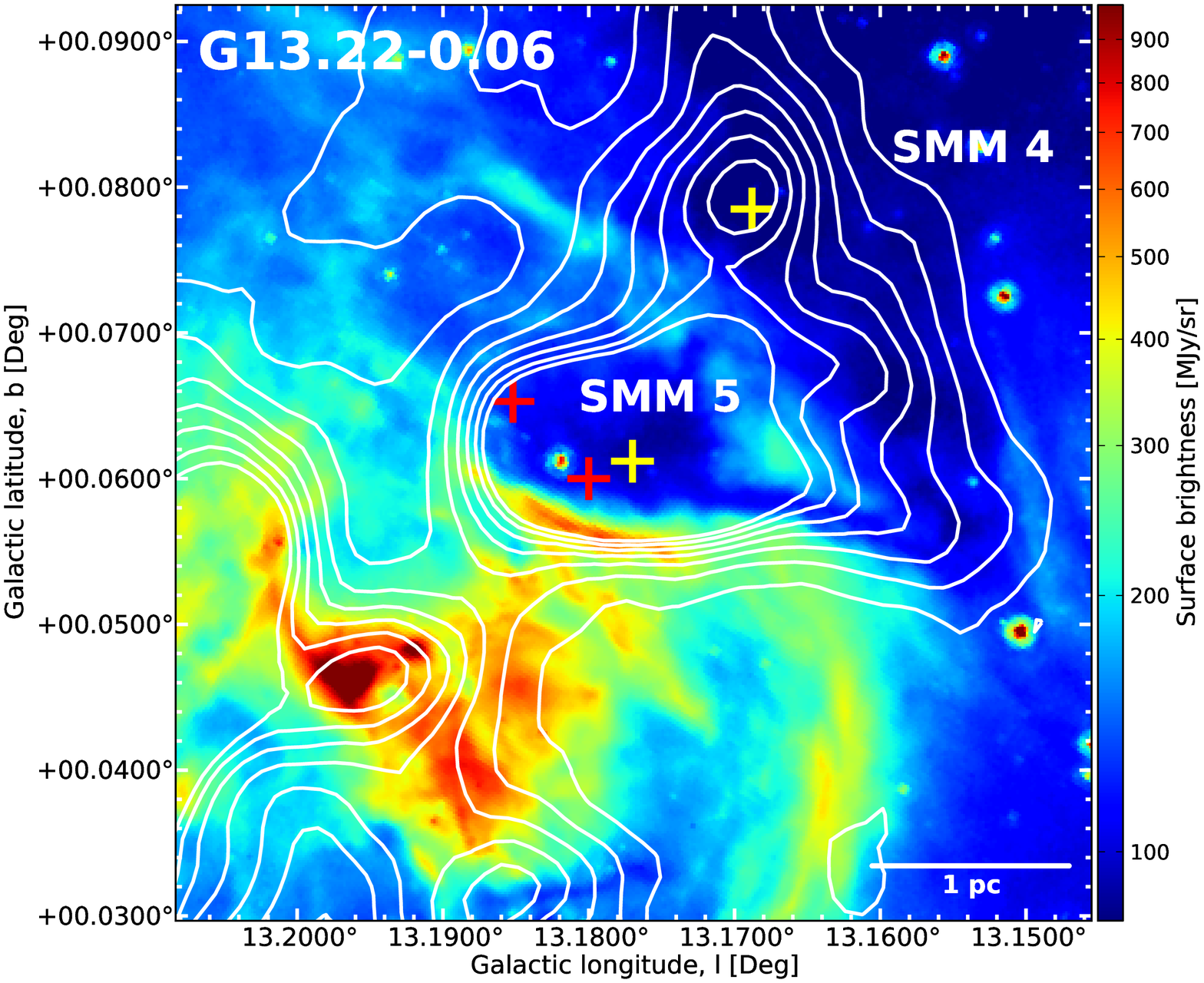}
\includegraphics[width=0.33\textwidth]{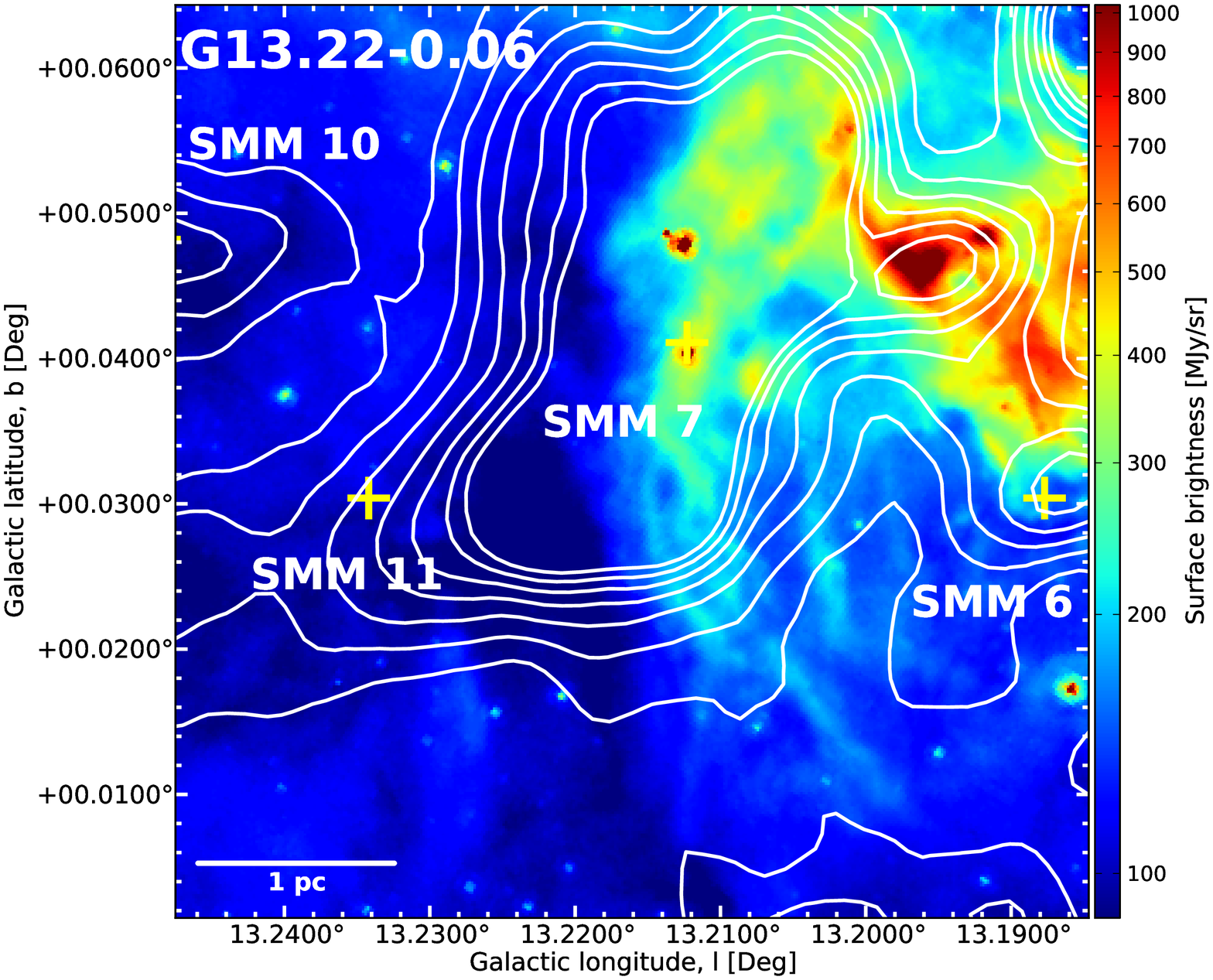}
\includegraphics[width=0.33\textwidth]{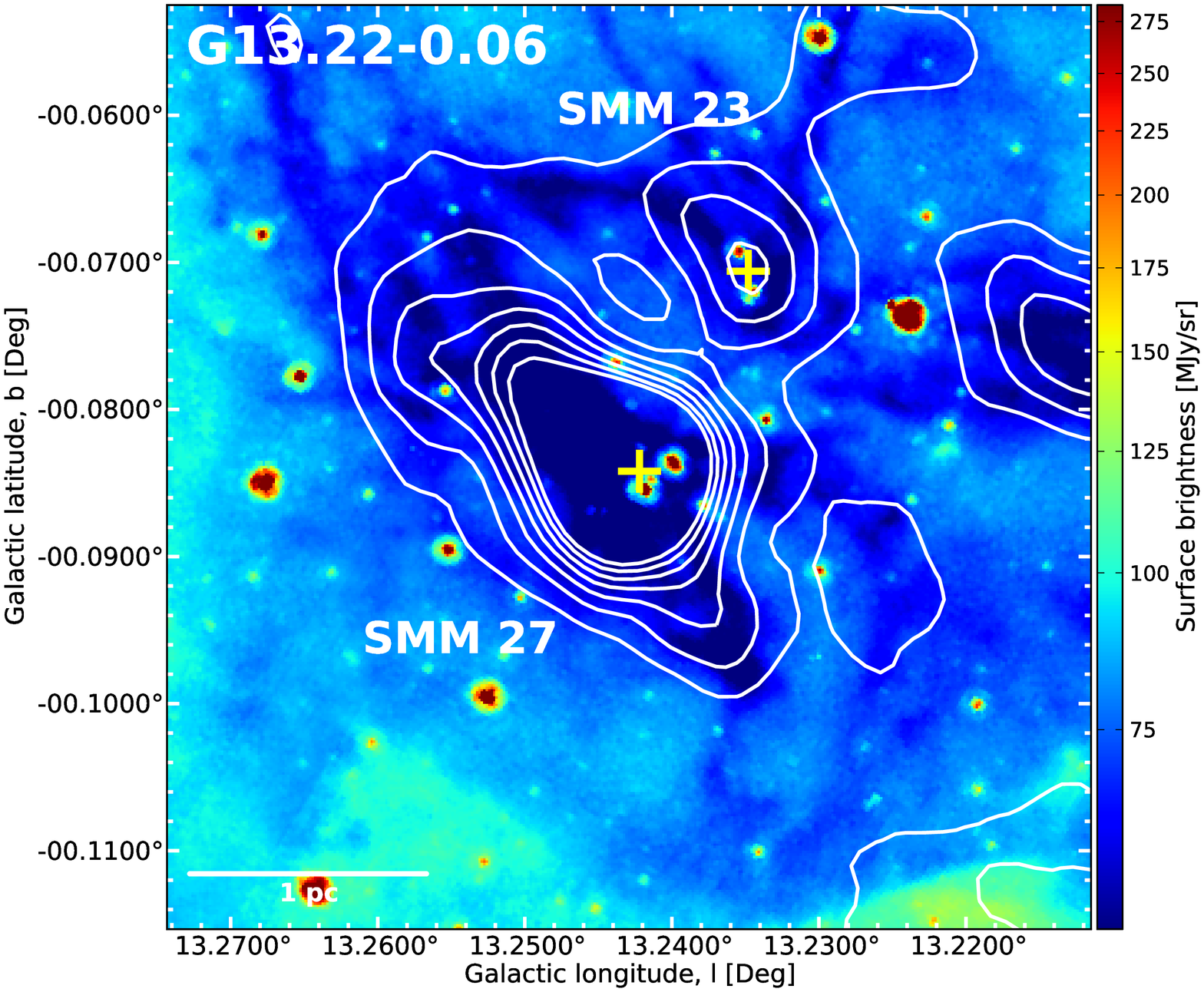}
\includegraphics[width=0.33\textwidth]{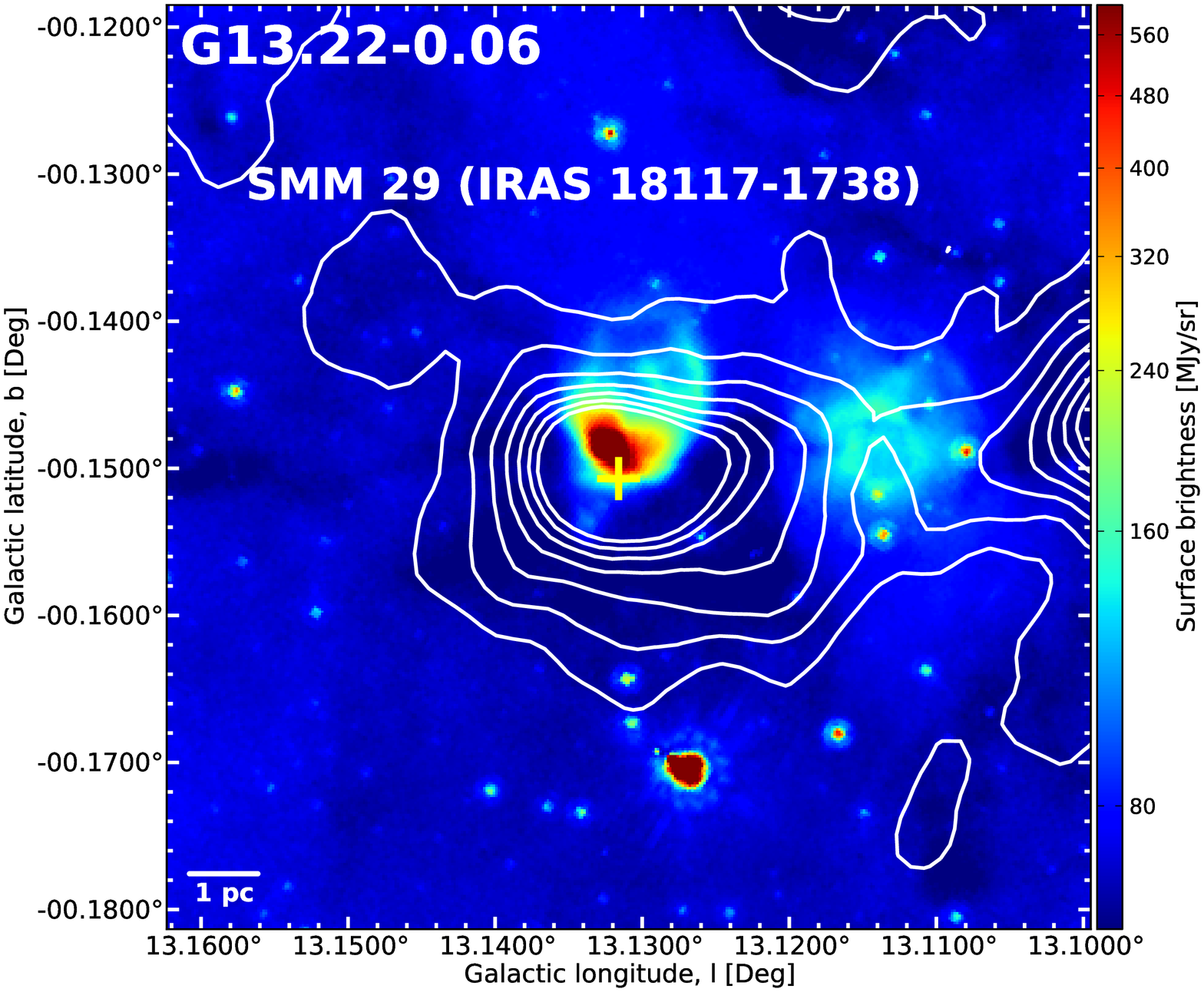}
\includegraphics[width=0.33\textwidth]{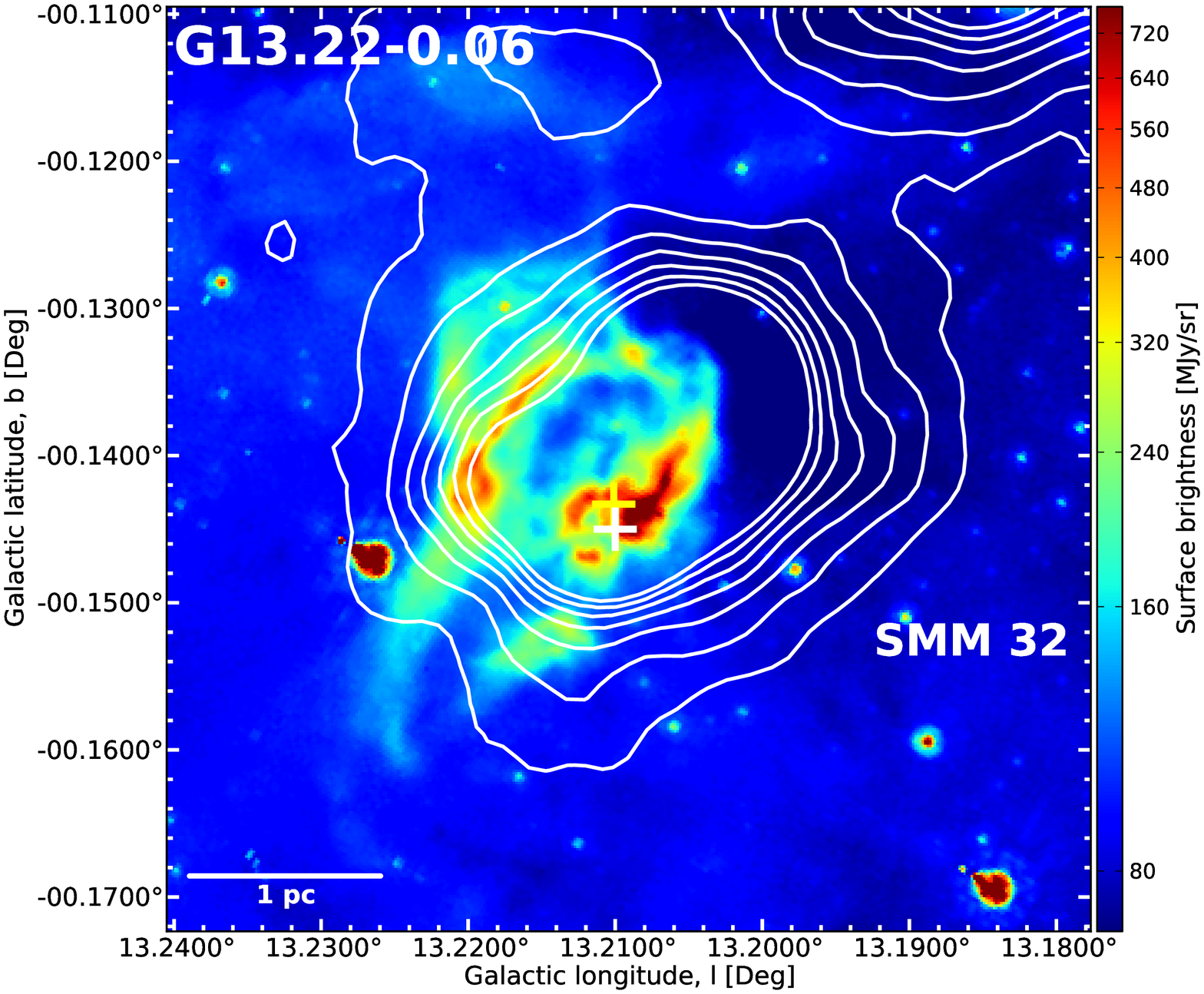}
\caption{continued. The red plus signs near G13.22--SMM 5 indicate the 
positions of the 6.7-GHz Class {\scriptsize II} methanol masers from 
Szymczak et al. (2000; upper; see also \cite{deharveng2010}) and Pandian et 
al. (2008; lower). In the G13.22--SMM 32 panel, the white 
plus sign marks the position of the compact \ion{H}{ii} region from Wink et 
al. (1982), as observed at 4.9 GHz.}
\label{figure:irac}
\end{center}
\end{figure*}

\longtabL{1}{
\begin{landscape}
{\small
\renewcommand{\footnoterule}{}
\begin{longtable}{c c c c c c c c c}
\caption{\label{clumps} Characteristics of the LABOCA 870-$\mu$m clumps.}\\
\hline\hline
Source & $\alpha_{2000.0}$ & $\delta_{2000.0}$ & $d$\tablefootmark{a} & $R_{\rm eff}$ & $M$ & $N({\rm H_2})$ & $\langle n({\rm H_2}) \rangle$ & Comments\\
       & [h:m:s] & [$\degr$:$\arcmin$:$\arcsec$] & [kpc] & [pc] & [M$_{\sun}$] & [$10^{22}$ cm$^{-2}$] & [$10^4$ cm$^{-3}$] &\\
\hline 
\hline
\endfirsthead
\caption{continued.}\\
\hline\hline
Source & $\alpha_{2000.0}$ & $\delta_{2000.0}$ & $d$ & $R_{\rm eff}$ & $M$ & $N({\rm H_2})$ & $\langle n({\rm H_2}) \rangle$ & Comments\\
       & [h:m:s] & [$\degr$:$\arcmin$:$\arcsec$] & [kpc] & [pc] & [M$_{\sun}$] & [$10^{22}$ cm$^{-2}$] & [$10^4$ cm$^{-3}$] &\\
\hline
\endhead
\hline
\endfoot
G1.87-0.14 & \\
SMM 1 \ldots & 17 49 44.0 & -27 33 28 & $7.47_{-0.13}^{+0.10}$ & $1.50\pm0.02$ & $13\,891\pm2\,161$ & $16.0\pm2.0$ & $50.9\pm7.9$ & IR-dark\\
SMM 8 \ldots & 17 49 59.4 & -27 29 24 & $7.29_{-0.17}^{+0.13}$ & $1.09\pm0.02$ & $2\,679\pm291$ & $5.1\pm0.8$ & $0.9\pm0.1$ & IR-dark\\
SMM 10 \ldots & 17 49 59.7 & -27 30 36 & $7.28_{-0.18}^{+0.14}$ & $0.78\pm0.02$ & $865\pm102$ & $3.1\pm0.5$ & $0.8\pm0.1$ & 8 and 24-$\mu$m source\\
SMM 12 \ldots & 17 50 03.0 & -27 33 56 & $7.32_{-0.17}^{+0.13}$ & $1.18\pm0.02$ & $3\,218\pm351$ & $6.5\pm0.7$ & $0.9\pm0.1$ & 8 and 24-$\mu$m source\\
SMM 14 \ldots & 17 50 04.2 & -27 34 57 & $7.40_{-0.15}^{+0.12}$ & $0.93\pm0.02$ & $2\,617\pm285$ & $7.7\pm1.0$ & $1.5\pm0.2$ & IR-dark\\
SMM 15 \ldots & 17 50 05.1 & -27 28 32 & $7.10_{-0.24}^{+0.18}$ & $1.00\pm0.03$ & $1\,635\pm195$ & $3.3\pm0.5$ & $0.7\pm0.1$ & group of 8-$\mu$m sources and a 24-$\mu$m source\\
SMM 16 \ldots & 17 50 09.0 & -27 33 36 & $7.20_{-0.21}^{+0.16}$ & $1.15\pm0.03$ & $3\,139\pm352$ & $5.4\pm0.8$ & $0.9\pm0.1$ & IR-dark\\
SMM 17 \ldots & 17 50 12.3 & -27 36 48 & $7.21_{-0.21}^{+0.16}$ & $0.73\pm0.02$ & $1\,506\pm179$ & $5.6\pm0.8$ & $1.8\pm0.2$ & IR-dark\\
SMM 20 \ldots & 17 50 13.5 & -27 20 40 & $7.00_{-0.26}^{+0.19}$ & $1.60\pm0.05$ & $4\,168\pm499$ & $3.8\pm0.5$ & $0.5\pm0.1$ & 8 and 24-$\mu$m source\\
SMM 21 \ldots & 17 50 13.8 & -27 35 24 & $7.21_{-0.21}^{+0.16}$ & $0.72\pm0.02$ & $816\pm76$ & $2.9\pm0.5$ & $1.0\pm0.1$ & 8 and 24-$\mu$m source\\
SMM 23 \ldots & 17 50 15.0 & -27 21 33 & $6.98_{-0.27}^{+0.20}$ & $1.50\pm0.05$ & $8\,014\pm969$ & $7.4\pm1.0$ & $1.1\pm0.1$ & IR-dark\\
SMM 24 \ldots & 17 50 15.3 & -27 34 24 & $7.27_{-0.19}^{+0.14}$ & $1.15\pm0.03$ & $3\,521\pm389$ & $5.2\pm0.8$ & $1.1\pm0.1$ & IR-dark\\
SMM 27 \ldots & 17 50 18.9 & -27 26 32 & $7.23_{-0.18}^{+0.14}$ & $0.73\pm0.02$ & $1\,266\pm148$ & $4.8\pm0.8$ & $1.5\pm0.2$ & IR-dark\\
SMM 28 \ldots & 17 50 19.2 & -27 27 53 & $7.03_{-0.26}^{+0.19}$ & $1.28\pm0.04$ & $3\,190\pm380$ & $3.8\pm0.6$ & $0.7\pm0.1$ & group of 8-$\mu$m sources and a 
24-$\mu$m source\\
SMM 30 \ldots & 17 50 22.8 & -27 21 28 & $7.02_{-0.25}^{+0.18}$ & $1.29\pm0.04$ & $3\,678\pm434$ & $4.1\pm0.8$ & $0.8\pm0.1$ & IR-dark\\
SMM 31 \ldots & 17 50 24.3 & -27 28 29 & $6.80_{-0.36}^{+0.25}$ & $0.66\pm0.03$ & $527\pm83$ & $1.9\pm0.4$ & $0.8\pm0.1$ & 8 and 24-$\mu$m source\\
SMM 38 \ldots & 17 50 46.3 & -27 24 32 & $7.06_{-0.23}^{+0.18}$ & $1.33\pm0.04$ & $3\,606\pm423$ & $5.2\pm0.8$ & $0.7\pm0.1$ & IR-dark\\
G2.11+0.00  & \\
SMM 5 & 17 50 36.0 & -27 05 44 & $7.40_{-0.12}^{+0.10}$ & $1.23\pm0.02$ & $2\,054\pm215$ & $6.9\pm0.7$ & $0.5\pm0.1$ & IRAS 17474-2704; 
UC \ion{H}{ii} region\tablefootmark{b}; OH maser\tablefootmark{c};\\
 & & & & & & & & extended 8-$\mu$m emission\\
G11.36+0.80 & \\
SMM 1 \ldots & 18 07 35.0 & -18 43 46 & $3.27_{-0.56}^{+0.47}$ & $0.38\pm0.06$ & $177\pm59$ & $2.7\pm0.3$ & $1.5\pm0.5$ & 8 and 24-$\mu$m source\\
SMM 2 \ldots & 18 07 35.6 & -18 43 22 & $3.27_{-0.56}^{+0.47}$ & $0.37\pm0.06$ & $217\pm73$ & $3.3\pm0.4$ & $2.0\pm0.7$ & IR-dark\\
SMM 3 \ldots & 18 07 35.8 & -18 42 42 & $3.27_{-0.56}^{+0.47}$ & $0.47\pm0.06$ & $255\pm85$ & $3.3\pm0.4$ & $1.1\pm0.4$ & 8 and 24-$\mu$m source\\
SMM 4 \ldots & 18 07 36.1 & -18 44 26 & $3.27_{-0.56}^{+0.47}$ & $0.38\pm0.06$ & $140\pm47$ & $2.6\pm0.3$ & $1.2\pm0.4$ & group of 8 and 24-$\mu$m sources\\
SMM 5 \ldots & 18 07 36.7 & -18 41 14 & $3.27_{-0.56}^{+0.47}$ & $0.52\pm0.08$ & $265\pm89$ & $2.7\pm0.3$ & $0.9\pm0.3$ & 8 and 24-$\mu$m source\\
SMM 6 \ldots & 18 07 39.0 & -18 42 10 & $3.27_{-0.56}^{+0.47}$ & $0.18\pm0.03$ & $35\pm13$ & $1.4\pm0.3$ & $2.7\pm1.0$ & IR-dark\\
SMM 7 \ldots & 18 07 40.4 & -18 43 18 & $3.27_{-0.56}^{+0.47}$ & $0.34\pm0.05$ & $81\pm27$ & $1.6\pm0.2$ & $0.9\pm0.3$ & 8 and 24-$\mu$m source\\
G13.22-0.06 & \\
SMM 4 \ldots & 18 13 55.9 & -17 28 34 & $4.24_{-0.35}^{+0.31}$ & $0.93\pm0.07$ & $1\,724\pm319$ & $5.3\pm0.7$ & $1.0\pm0.2$ & 8 and 24-$\mu$m source\\
SMM 5 \ldots & 18 14 00.7 & -17 28 38 & $4.24_{-0.35}^{+0.31}$ & $1.08\pm0.08$ & $4\,876\pm902$ & $24.7\pm2.5$ & $1.8\pm0.3$ & 8-$\mu$m source and extended 24-$\mu$m emission\\
SMM 6 \ldots & 18 14 08.8 & -17 28 57 & $4.24_{-0.35}^{+0.31}$ & $0.74\pm0.05$ & $920\pm171$ & $3.6\pm0.5$ & $1.0\pm0.2$ & extended 8 and 24-$\mu$m emission\\
SMM 7 \ldots & 18 14 09.4 & -17 27 21 & $4.24_{-0.35}^{+0.31}$ & $1.55\pm0.11$ & $9\,070\pm1\,678$ & $13.3\pm1.4$ & $1.1\pm0.2$ & extended 8 and 24-$\mu$m emission\\
SMM 10 \ldots & 18 14 12.2 & -17 25 14 & $4.24_{-0.35}^{+0.31}$ & $0.53\pm0.04$ & $536\pm102$ & $3.7\pm0.7$ & $1.6\pm0.3$ & IR-dark\\
SMM 11 \ldots & 18 14 14.4 & -17 26 30 & $4.24_{-0.35}^{+0.31}$ & $0.81\pm0.06$ & $767\pm143$ & $2.2\pm0.4$ & $0.7\pm0.1$ & diffuse 8 and 24-$\mu$m emission\\
SMM 23 \ldots & 18 14 36.8 & -17 29 22 & $3.56_{-0.44}^{+0.39}$ & $0.43\pm0.05$ & $252\pm66$ & $2.9\pm0.4$ & $1.4\pm0.4$ & 8 and 24-$\mu$m source\\
SMM 27 \ldots & 18 14 40.7 & -17 29 22 & $3.56_{-0.44}^{+0.39}$ & $0.76\pm0.09$ & $1\,757\pm451$ & $13.9\pm1.5$ & $1.8\pm0.5$ & group of 8-$\mu$m sources and a 24-$\mu$m source\\
SMM 29 \ldots & 18 14 42.1 & -17 37 06 & $12.33_{-0.33}^{+0.38}$\tablefootmark{d} & $2.77\pm0.09$ &$17\,927\pm2\,094$ & $10.6\pm1.1$ & $0.4\pm0.04$ & IRAS 18117-1738; extended 8-$\mu$m emission\\
SMM 32 \ldots & 18 14 49.9 & -17 32 45 & $4.41_{-0.32}^{+0.29}$ & $1.33\pm0.09$ & $3\,694\pm638$ & $10.8\pm1.1$ & $0.7\pm0.1$ & \ion{H}{ii} region\tablefootmark{e}; extended 8-$\mu$m emission\\
\hline
\end{longtable}
\tablefoot{\tablefoottext{a}{Near kinematic distance unless otherwise stated.}\tablefoottext{b}{\cite{becker1994}; \cite{forster2000}.}\tablefoottext{c}{\cite{argon2000}.}\tablefoottext{d}{Far distance; the distance ambiguity of the \textit{IRAS} source was resolved by Sewilo et al. (2004).}\tablefoottext{e}{\cite{wink1982}; \cite{chini1987}; \cite{white2005}; \cite{urquhart2009}.}
}
}
\end{landscape}
}

\subsection{MALT90 survey data}

The spectral-line data of the sources employed in the present study 
were observed as part of the MALT90 survey (PI: J.~M. Jackson; 
see \cite{foster2011}, 2013; \cite{jackson2013}). MALT90 observations 
cover the Galactic longitude ranges $3\degr < l < 20\degr$ (1st quadrant) and 
$300\degr < l < 357\degr$ (4th quadrant), and were targeting high-mass 
star-forming clumps in different stages of evolution.  
The survey was conducted with the 22-m Mopra telescope\footnote{The Mopra 
radio telescope is part of the Australia Telescope National Facility which 
is funded by the Commonwealth of Australia for operation as a National 
Facility managed by CSIRO.} in the on-the-fly (OTF) mapping mode during the 
austral winter in 2010--2012, covering the months of May to October. 
The OTF mapping was performed with the beam centre scanning 
in Galactic coordinates $(l,\,b)$ on a $3\farcm4 \times 3\farcm4$ 
grid, where the beam FWHM is $38\arcsec$ at 90 GHz. The scanning speed was 
$3\farcs92$~s$^{-1}$. The step size between adjacent scanning rows was 
$12\arcsec$, i.e., $\sim1/3$ of the beam FWHM, resulting in 17 rows per map. 
Each source was mapped twice by scanning in orthogonal directions 
($b$ versus $l$). One map took about half an hour to complete, and the 
total time spent on each field (two maps) was 1.18 hr. The mapping was carried 
out when the clump elevation was more than $35\degr$ but less than $70\degr$. 
The telescope pointing was checked every 1--1.5 hr on SiO maser sources, and 
was found to be better than $\sim10\arcsec$.

The spectrometer used was the MOPra Spectrometer (MOPS) which is a digital 
filter bank\footnote{The University of New South Wales Digital Filter Bank 
used for the observations with the Mopra Telescope was provided with support 
from the Australian Research Council.}. The MOPS spectro\-meter was 
tuned to a central frequency of 89.690 GHz, and the 8 GHz wide frequency band 
of MOPS was split into 16 subbands of 137.5 MHz each (4\,096 
channels), resulting in a velo\-city resolution of $\sim0.11$ km~s$^{-1}$ in 
each band (the so-called zoom mode). The typical system 
temperatures during the observations were in the range $T_{\rm sys}\sim180-300$ 
K, and the typical rms noise level is $\sim250$ mK per 0.11 km~s$^{-1}$ channel.
The output intensity scale given by the Mopra/MOPS system is $T_{\rm A}^{\star}$, 
i.e., the antenna temperature corrected for the atmospheric attenuation. 
The observed intensities were converted to the main-beam brightness 
temperature scale by $T_{\rm MB}=T_{\rm A}^{\star}/\eta_{\rm MB}$, where 
$\eta_{\rm MB}$ is the main-beam efficiency. The value of $\eta_{\rm MB}$ is 
0.49 at 86 GHz and 0.44 at 110 GHz (\cite{ladd2005}). Extrapolation using the 
Ruze formula gives the $\eta_{\rm MB}$ values in the range 0.49--0.46 for the 
86.75--93.17 GHz frequency range of MALT90.

The 16 spectral-line transitions mapped simultaneously in the MALT90 survey 
are listed in Table~\ref{table:lines}. In this table, we give some 
spectroscopic parameters of the spectral lines, and in the last column we also 
provide comments on each transition and information provided by the lines. 
The MALT90 datafiles are publicly available and can be downloaded through the 
Australia Telescope Online Archive 
(ATOA)\footnote{{\tt http://atoa.atnf.csiro.au/MALT90}}.

\begin{table*}
\caption{MALT90 spectral-line transitions.}
\begin{minipage}{2\columnwidth}
\centering
\renewcommand{\footnoterule}{}
\label{table:lines}
\begin{tabular}{c c c c c}
\hline\hline 
Transition\tablefootmark{a} & $\nu$\tablefootmark{b} & $E_{\rm u}/k_{\rm B}$\tablefootmark{c} & $n_{\rm crit}$\tablefootmark{d} & Comments\tablefootmark{e} \\
      & [MHz] & [K] & [cm$^{-3}$] & \\
\hline        
H$^{13}$CO$^+(J=1-0)$ & 86\,754.330 & 4.16 & $1.6\times10^5$ & high-density and 
ionisation tracer; \\
                     &             &      &                 & $J=1-0$ is split into six hyperfine (hf) components\tablefootmark{f}\\
SiO$(J=2-1)$ & 86\,847.010 & 6.25 & $2\times10^6$ & shocked-gas/outflow tracer\\
HN$^{13}$C$(J=1-0)$ & 87\,090.859 & 4.18 & $2\times10^5$\tablefootmark{g} & high-density tracer; \\
                     &             &      &                 & $J=1-0$ is split into 11 hf components \\
                     &             &      &                 & with four having a different frequency\tablefootmark{h}\\
C$_2$H$(N_{J,\,F}=1_{3/2,\,2}-0_{1/2,\,1})$ & 87\,316.925 & 4.19 & $2\times10^5$\tablefootmark{i} & a tracer of photodissociation regions (PDRs); \\
                     &             &      &                 & $N_J=1_{3/2}-0_{1/2}$ is split into three hf components\tablefootmark{j}\\
HNCO$(J_{K_a,\,K_b}=4_{0,\,4}-3_{0,\,3})$ & 87\,925.238 & 10.55 & $4.5\times10^6$\tablefootmark{k} & hot core and shock-chemistry tracer; \\
                     &             &      &                 & six hf components\tablefootmark{l}; $a$-type transition ($\Delta K_a=0$)\\
HNCO$(J_{K_a,\,K_b}=4_{1,\,3}-3_{1,\,2})$ & 88\,239.027 & 53.86 & $2.3\times10^6$\tablefootmark{k} & six hf components\tablefootmark{l}; $a$-type transition ($\Delta K_a=0$)\\
HCN$(J=1-0)$ & 88\,631.847 & 4.25 & $2.4\times10^6$ & high-density and 
infall tracer;\\
                     &             &      &                 & $J=1-0$ is split into three hf components\tablefootmark{m}\\
HCO$^+(J=1-0)$ & 89\,188.526 & 4.28 & $1.7\times10^5$ & high-density, infall, and ionisation tracer; \\
                     &             &      &                 & enhanced in outflows due to shock-induced UV radiation\tablefootmark{n}\\
HC$^{13}$CCN$(J=10-9,\,F=9-8)$ & 90\,593.059 & 23.91 & $1.7\times10^5$\tablefootmark{o} & hot-core tracer; five hf components\\
HNC$(J=1-0)$ & 90\,663.572 & 4.35 & $2.9\times10^5$ & high-density tracer; three hf components\\
$^{13}$C$^{34}$S$(J=2-1)$ & 90\,926.036 & 6.54 & $4.3\times10^5$\tablefootmark{p} & high-density tracer\\
HC$_3$N$(J=10-9)$ & 91\,199.796 & 24.01 & $5.3\times10^5$ & high-density/hot-core tracer;\\
                     &             &      &                 & six hf components\\
CH$_3$CN$(J_K=5_1-4_1)$ & 91\,985.316 & 20.39 & $4\times10^5$\tablefootmark{q} & hot-core tracer; seven hf components\\
H$41\alpha$ & 92\,034.475 & 89.5\tablefootmark{r} & $\sim5$\tablefootmark{s} & ionised gas tracer; the principal quantum number \\
                     &             &      &                 & changes from $n=42$ to 41 $\rightarrow$ $\alpha$-type radio recombination line\\
$^{13}$CS$(J=2-1)$ & 92\,494.303 & 6.66 & $5\times10^5$\tablefootmark{k} & high-density tracer; three hf components\\
N$_2$H$^+(J=1-0)$ & 93\,173.480 & 4.47 & $1.5\times10^5$ & high-density/CO-depleted gas tracer;\\
                     &             &      &                 & $J=1-0$ line has 15 hf components out of which\\
                     &             &      &                 & seven have a different frequency\tablefootmark{t}\\
\hline 
\end{tabular} 
\tablefoot{\tablefoottext{a}{The rotational transitions here occur in the 
vibrational ground state (${\rm v}=0$).}\tablefoottext{b}{Rest frequencies 
adopted from the MALT90 webpage ({\tt http://malt90.bu.edu/parameters.html}).}\tablefoottext{c}{Upper-state energy divided by the Boltzmann constant.}\tablefoottext{d}{Critical density at 15 K unless otherwise stated. Unless otherwise stated, the Einstein $A$ coefficients and collision rates were adopted from the Leiden Atomic and Molecular Database [LAMDA (\cite{schoier2005}); {\tt http://home.strw.leidenuniv.nl/$\sim$moldata/}].}\tablefoottext{e}{Comments on the species and transition in question.}\tablefoottext{f}{See, e.g., \cite{schmid2004}.}\tablefoottext{g}{Collision rate for HNC from LAMDA was used to estimate $n_{\rm crit}$.}\tablefoottext{h}{\cite{vandertak2009}; \cite{padovani2011}.}\tablefoottext{i}{From \cite{lo2009}.}\tablefoottext{j}{\cite{reitblat1980}; \cite{padovani2009}; \cite{spielfiedel2012}.}\tablefoottext{k}{$n_{\rm crit}$ at 20 K.}\tablefoottext{l}{\cite{lapinov2007}.}\tablefoottext{m}{See, e.g., \cite{cao1993}.}\tablefoottext{n}{\cite{rawlings2000}, 2004.}\tablefootmark{o}{$n_{\rm crit}$ was estimated using the Einstein $A$ coefficient from the Cologne Database for Molecular Spectroscopy [CDMS (\cite{muller2005}); {\tt http://www.astro.uni-koeln.de/cdms}] and the HC$_3$N collision rate at 15 K from LAMDA.}\tablefootmark{p}{$n_{\rm crit}$ was estimated using the Einstein $A$ coefficient from CDMS and the $^{12}$C$^{34}$S collision rate at 20 K from LAMDA.}\tablefoottext{q}{From SJF12.}\tablefoottext{r}{The energy of the $n=42$ level ($E=-E_0/n^2$, where $E_0=13.6$ eV).}\tablefoottext{s}{Critical electron density at $10^4$ K [see Appendix~D.1 in Gordon \& Sorochenko (2009)].}\tablefoottext{t}{See Table~2 in Pagani et al. (2009) and Table~1 in Keto \& Rybicki (2010).}}
\end{minipage}
\end{table*}

\section{Results and analysis}

\subsection{Spatial distributions of the spectral-line emission}

In this subsection, we present the integrated intensity maps of the spectral 
lines detected towards the clumps. Besides the maps of integrated intensity, 
or the 0th moment maps, the MALT90 data archive contains the uncertainty maps 
of the 0th moment images. The typical $1\sigma$ error, in units of integrated 
$T_{\rm MB}$, was found to be $\sim0.6-0.7$ K~km~s$^{-1}$. 
The HNCO$(4_{1,\,3}-3_{1,\,2})$ maps were an exception, however, as  
in many cases they were found to be very noisy 
with $1\sigma$ values of $\sim3-8$ K~km~s$^{-1}$. Moreover, the HCN and HCO$^+$ 
maps towards G1.87--SMM 38 were corrupted and could not be used. 

The 0th moment maps are presented in 
Figs.~\ref{figure:G187SMM1lines}--\ref{figure:G1322SMM32lines} of Appendix~B, 
where the white contours showing the spectral-line emission are overlaid on the 
\textit{Spitzer} 8-$\mu$m images. The red plus signs mark the LABOCA peak 
positions to guide the eye. The line emission was deemed to be real if 
the integrated intensity was detected at least at the $3\sigma$ level. 
Only maps of detected line emission are presented. In some cases, the contours 
are plotted to start at a stronger emission level than $3\sigma$ for 
illustrative purposes.

As can be seen from Figs.~\ref{figure:G187SMM24lines} and 
\ref{figure:G1322SMM7lines} (cf.~Fig.~\ref{figure:irac}), 
the LABOCA 870-$\mu$m emission peaks of G1.87--SMM 17 and G13.22--SMM 10 are 
not covered by the MALT90 maps. For most of the lines the detection rate is 
generally high. In parti\-cular, HNC$(1-0)$ and N$_2$H$^+(1-0)$ were detected 
towards all fields. Also, SiO$(2-1)$, C$_2$H$(1_{3/2,\,2}-0_{1/2,\,1})$, 
HCN$(1-0)$, and HCO$^+(1-0)$ were seen towards $\sim86-93$\% of the fields. 
As mentioned above, the HNCO$(4_{1,\,3}-3_{1,\,2})$ maps were often very noisy, 
and emission was not detected even in the cases where the noise level was at 
the normal level of $0.6-0.7$ K~km~s$^{-1}$. The HC$^{13}$CCN$(10-9)$, 
H$41\alpha$, and $^{13}$C$^{34}$S$(2-1)$ lines were not detected in any of the 
sources. Moreover, only two fields show weak traces of $^{13}$C$^{32}$S$(2-1)$ 
emission, which explains the non-detections of the rarer isotopologue 
$^{13}$C$^{34}$S.


\subsection{Spectra and line parameters}

The beam-averaged ($38\arcsec$) spectra were extracted from the data cubes 
towards the LABOCA peak positions of the clumps and towards selected line 
emission peaks. The spectra were ana\-lysed using the CLASS90 programme
of the GILDAS software package\footnote{Grenoble Image and Line Data Analysis 
Software is provided and actively developed by IRAM, and is available at
{\tt http://www.iram.fr/IRAMFR/GILDAS}}. Linear (first-order) to third-order 
baselines were determined from velocity ranges without line-emission features,
and then subtracted from the spectra. The resulting $1\sigma$ rms noise 
le\-vels were in the range $\sim0.2-0.4$ K on the $T_{\rm MB}$ scale. 
The spectra are presented in 
Figs.~\ref{figure:G187SMM1_spectra}--\ref{figure:G1322SMM32_spectra} of 
Appendix~C. The spectra are overlaid with single Gaussian/hf-structure fits 
(see below). The coordinates of the line emission peaks in the clump regions 
are shown in the upper left corners of the corresponding panels (e.g., the 
HCO$^+$ and HNC spectra in Fig.~\ref{figure:G187SMM1_spectra}). When only weak 
trace of emission was seen in the 0th moment map, there was no detectable line 
in the extracted spectrum. This was particularly the case for the 
$^{13}$CS$(2-1)$ transition. 

As described in Table~\ref{table:lines}, of the detected lines only 
SiO$(2-1)$ and HCO$^+(1-0)$ have no hf structure. The SiO and HCO$^+$ lines 
were therefore fitted with a single Gaussian profile using CLASS90. 
For the rest of the detected lines, we used the {\tt hfs} method of CLASS90 to 
fit the hf structure, although the hf components were not fully resolved in 
any of the sources due to large linewidths typical of massive clumps. 
The relative positions (in frequency or velocity) and re\-lative strengths of 
the hf components were searched from the literature (see the refe\-rences in 
Table~\ref{table:lines}) or via the Splatalogue spectral line 
database\footnote{{\tt http://splatalogue.net/}}. The derived spectral-line 
parameters are listed in Table~\ref{table:lineparameters} at the end of the 
paper. In Cols.~(3) and (4) we give the LSR velocity of the emission 
(${\rm v}_{\rm LSR}$) and FWHM linewidth ($\Delta {\rm v}$), respectively. 
Columns~(5) and (6) list the peak intensities 
($T_{\rm MB}$) and integrated line intensities ($\int T_{\rm MB} {\rm dv}$). 
The quoted uncertainties in these parameters represent the formal fitting 
errors (i.e., calibration uncertainties are not taken into account). 
The values of $T_{\rm MB}$ and $\int T_{\rm MB} {\rm dv}$ were either determined 
for a blended groups of hf components or, when resolved, for the strongest 
line which itself could be a multiplet of indivi\-dual hf lines.

Some of the HCN, HCO$^+$, HNC, and N$_2$H$^+$ lines were found to show 
double-peaked profiles caused by gas kinematics -- not by hf splitting. 
The blue-skewed profiles, i.e., those with blue-shifted peaks stronger than 
red peaks (e.g., the HCN spectrum towards G1.87--SMM 31; 
Fig.~\ref{figure:G187SMM31_spectra}) could be the manifestation of large-scale 
collapse motions (e.g., \cite{zhou1993}; \cite{myers1996}; \cite{lee1999}; 
\cite{gao2009}). In contrast, the red-skewed profiles with 
stronger red peaks and weaker blue peaks suggest that the envelope is 
expanding (e.g., \cite{thompson2004}; \cite{velusamy2008}; \cite{gao2010}); 
see, for example, the HCN and HNC lines towards G1.87--SMM 1 
(Fig.~\ref{figure:G187SMM1_spectra}). 
In Table~\ref{table:lines}, for double-peaked lines we also give the line 
parameters of both the blue and red peaks separately derived through fitting 
a single Gaussian to each peak. Finally, in a few cases (e.g., the HCO$^+$ 
line towards G1.87--SMM 1 and G2.11--SMM5) we observe more than 
one velocity component along the line of sight. These additional velocity 
components are typically much weaker than the main component and do not 
significantly contribute to the integrated intensity maps that were 
constructed by integrating over the whole velocity range.

\subsection{Line optical thicknesses and excitation temperatures}

The optical thickness of the line emission ($\tau$) and the excitation 
temperature ($T_{\rm ex}$) could be derived through fitting the hf structure in 
only some cases. The main reasons for this are the blending of the hf 
components and limited signal-to-noise (S/N) ratio of the spectra. The average 
of the $T_{\rm ex}$ values that could be directly derive via the {\tt hfs} 
method was adopted for the rest of the given lines. 
In a few cases we were able to derive the optical thickness by comparing the 
intensities of two different isotopologues of the same species, namely 
HCO$^+$/H$^{13}$CO$^+$ and HNC/HN$^{13}$C [cf.~SJF12; their Eq.~(6)]. 
For this analysis, we adopted the galactocentric distance-dependent 
$[^{12}{\rm C}]/[^{13}{\rm C}]$ ratio from Wilson \& Rood (1994):

\begin{equation}
\label{eq:carbon}
\frac{[^{12}{\rm C}]}{[^{13}{\rm C}]}=7.5\times R_{\rm GC}[{\rm kpc}]+7.6\,.
\end{equation}
The optical thickness ratio between the two isotopologues was assumed to be 
equal to that given by Eq.~(\ref{eq:carbon}). Using the derived value of 
$\tau$, the value of $T_{\rm ex}$ was calculated using the familiar antenna 
equation [see, e.g., Eq.~(A.1) of Miettinen (2012a)].

When $T_{\rm ex}$ could not be derived/assumed as described above, we assumed 
that it is equal to $E_{\rm u}/k_{\rm B}$ in the case of linear molecules (SiO 
and HC$_3$N in our case), and that $T_{\rm ex}=2/3\times E_{\rm u}/k_{\rm B}$ 
in the case of HNCO, which is a nearly prolate asymmetric top molecule, 
and CH$_3$CN, which is a prolate symmetric top. These $T_{\rm ex}$ values, 
used to estimate the line optical thickness, lead to the lower limit to the 
molecular column density (e.g., \cite{hatchell1998}; see also 
\cite{miettinen2012a}). The values of $\tau$ and $T_{\rm ex}$ are listed in 
Cols.~(7) and (8) of Table~\ref{table:lineparameters}. In case the line has a 
hf structure, the $\tau$ value refers to the sum of the peak optical 
thicknesses of individual hf components. For lines with blended hf multiplets, 
this total optical thickness was derived by dividing the optical thickness of 
the strongest hf component by its statistical weight.

\subsection{Column densities and fractional abundances}

The beam-averaged column densities of the molecules, $N({\rm mol})$, were 
calculated by using the standard local thermodynamic equilibrium (LTE) 
formulation:

\begin{equation}
\label{eq:N}
N({\rm mol})=\frac{3h\epsilon_0}{2\pi^2}\frac{1}{\mu^2S}\frac{Z_{\rm rot}(T_{\rm ex})}{g_Kg_I}e^{E_u/k_{\rm B}T_{\rm ex}}F(T_{\rm ex})\int \tau({\rm v}){\rm dv} \, , 
\end{equation}
where $h$ is the Planck constant, $\epsilon_0$ is the vacuum permittivity, 
$\mu$ is the permanent electric dipole moment, $S$ is the line strength, 
$Z_{\rm rot}$ is the rotational partition function, $g_K$ is the $K$-level 
degeneracy, $g_I$ is the reduced nuclear spin degeneracy (see, e.g., 
\cite{turner1991}), and 
$F(T_{\rm ex})\equiv \left(e^{h\nu/k_{\rm B}T_{\rm ex}}-1\right)^{-1}$. 
Here, the electric dipole moment matrix element is defined as 
$\left|\mu_{\rm ul} \right|\equiv \mu^2S/g_{\rm u}$, where $g_{\rm u}=2J+1$ 
is the rotational degeneracy of the upper state (\cite{townes1975}).     
The values of the pro\-duct $\mu^2S$ were taken from the Splatalogue database. 
For linear molecules, $g_K=g_I=1$ for all levels 
(\cite{turner1991}). As an asymmetric top, HNCO has $g_K=1$ (no $K$-level 
degeneracy), and due to absence of identical interchangeable nuclei $g_I$ is 
also equal to unity. For the detected CH$_3$CN line, $g_K=2$ because 
$K \neq 0$ (degeneration among the $K$-type doublets), and $g_I=1/4$ because 
$K \neq 3n$, where $n$ is an integer (\cite{turner1991}).

The partition function of the linear molecules was approximated as 

\begin{equation}
\label{eq:Z1}
Z_{\rm rot}(T_{\rm ex}) \simeq \frac{k_{\rm B}T_{\rm ex}}{hB}+\frac{1}{3}\,,
\end{equation}
where $B$ is the rotational constant. The above expression is appropriate
for heteropolar molecules at the high temperature limit of 
$hB/k_{\rm B}T_{\rm ex} \ll 1$. For HNCO, the partition function was 
calculated as

\begin{equation}
\label{eq:Z1}
Z_{\rm rot}(T_{\rm ex})=\sqrt{\frac{\pi(k_{\rm B}T_{\rm ex})^3}{h^3ABC}}\,,
\end{equation}
where $A$, $B$, and $C$ are the three rotational constants. For CH$_3$CN, 
the partition function is given by Eq.~(\ref{eq:Z1}) multiplied by $1/3$ due to 
the three interchangeable H-nuclei (\cite{turner1991}). We note that for 
the prolate symmetric top molecule CH$_3$CN, $B=C$ in Eq.~(\ref{eq:Z1}).

In case the line profile has a Gaussian shape, the last integral term in 
Eq.~(\ref{eq:N}) can be expressed as a function of the FWHM linewidth and peak 
optical thickness of the line as

\begin{equation}
\label{eq:tau}
\int \tau({\rm v}){\rm dv}=\frac{\sqrt{\pi}}{2\sqrt{\ln 2}}\Delta {\rm v} \tau \simeq1.064\Delta {\rm v} \tau \,.
\end{equation}
Moreover, if the line emission is optically thin ($\tau \ll 1$), 
$T_{\rm MB}\propto \tau$, and $N({\rm mol})$ can be computed from the integrated 
line intensity [see, e.g., Eq.~(A.4) of Miettinen (2012a)]. The values of 
$\tau$ listed in Table~\ref{table:lines} were used to decide by which method 
(from the linewidth or integrated intensity) the column density was computed. 
Our analysis assumed that the line emission fills the telescope beam, i.e., 
that the beam filling factor is unity. As can be seen in the 0th moment maps, 
the line emission is often extended with respect to the $38\arcsec$ 
($0\fdg011$) beam size. However, this does not necessarily mean that 
the assumption of unity filling factor is correct. If the gas has clumpy 
structure within the beam area, the true filling factor is $<1$. 
In this case, the derived beam-averaged column density is only a lower limit 
to the source-averaged value.

The fractional abundances of the molecules were calculated by dividing the 
molecular column density by the H$_2$ column density, 
$x({\rm mol})=N({\rm mol})/N({\rm H_2})$. To be directly comparable with the 
line observations, the $N({\rm H_2})$ values were derived from the LABOCA dust 
continuum maps smoothed to the MALT90 re\-solution of $38\arcsec$. 

The beam-averaged column densities and abundances with respect to H$_2$ are 
given in the last two columns of Table~\ref{table:lineparameters}. 
Statistics of these parameters are given in Table~\ref{table:stat}, where we 
provide the mean, median, standard deviation (std), and minimum and maximum 
values of the sample (the values for additional velocity components have been 
neglected). This table provides an easier way to compare the derived molecular 
column densities and abundances with those found in other studies.

\setcounter{table}{3}
\begin{table}
\renewcommand{\footnoterule}{}
\caption{Statistics of column densities and fractional abundances. 
The notation $a(b)$ means $a\times10^b$.}
{\tiny
\begin{minipage}{1\columnwidth}
\centering
\label{table:stat}
\begin{tabular}{c c c c c c}
\hline\hline 
\multicolumn{6}{c}{All clumps} \\
\hline
Quantity & Mean & Median & Std\tablefootmark{a} & Min. & Max. \\
\hline
$N({\rm H^{13}CO^+})$\tablefootmark{b} & $7.4(12)$ & $6.4(12)$ & $2.3(12)$ & $5.8(12)$ & $1.0(13)$\\ 
$x({\rm H^{13}CO^+})$\tablefootmark{b} & $9.7(-11)$ & $1.1(-10)$ & $5.1(-11)$ & $4.0(-11)$ & $1.4(-10)$ \\ 
$N({\rm SiO})$ & $1.9(13)$ & $1.5(13)$ & $1.3(13)$ & $5.5(12)$ & $4.8(13)$\\ 
$x({\rm SiO})$ & $5.7(-10)$ & $3.6(-10)$ & $5.1(-10)$ & $4.0(-11)$ & $1.8(-9)$\\ 
$N({\rm HN^{13}C})$ & $1.1(13)$ & $1.1(13)$ & $2.5(12)$ & $9.6(12)$ & $1.5(13)$\\ 
$x({\rm HN^{13}C})$ & $1.2(-10)$ & $1.2(-10)$ & $1.6(-11)$ & $1.0(-10)$ & $1.4(-10)$\\
$N({\rm C_2H})$ & $4.3(14)$ & $3.4(14)$ & $3.3(14)$ & $7.8(13)$ & $1.2(15)$\\
$x({\rm C_2H})$ & $8.1(-9)$ & $6.3(-9)$ & $6.1(-9)$ & $1.8(-9)$ & $2.4(-8)$\\
$N({\rm HNCO})$ & $2.2(14)$ & $1.4(14)$ & $2.2(14)$ & $5.4(12)$ & $7.9(14)$\\
$x({\rm HNCO})$ & $6.6(-9)$ & $5.0(-9)$ & $5.8(-9)$ & $1.3(-10)$ & $1.9(-8)$\\
$N({\rm HCN})$ & $5.9(14)$ & $8.7(13)$ & $1.4(15)$ & $1.6(13)$ & $5.5(15)$\\
$x({\rm HCN})$ & $2.1(-8)$ & $2.0(-9)$ & $5.2(-8)$ & $2.7(-10)$ & $2.1(-7)$\\
$N({\rm HCO^+})$ & $3.0(13)$ & $9.7(12)$ & $5.6(13)$ & $1.1(12)$ & $2.6(14)$\\
$x({\rm HCO^+})$ & $5.6(-10)$ & $3.3(-10)$ & $6.9(-10)$ & $6.0(-11)$ & $3.6(-9)$\\
$N({\rm HNC})$ & $7.1(13)$ & $4.8(13)$ & $9.4(13)$ & $3.7(12)$ & $4.4(14)$\\
$x({\rm HNC})$ & $2.2(-9)$ & $1.6(-9)$ & $2.1(-9)$ & $2.3(-10)$ & $8.4(-9)$\\
$N({\rm HC_3N})$ & $1.6(14)$ & $2.8(13)$ & $2.8(14)$ & $7.0(12)$ & $7.9(14)$\\
$x({\rm HC_3N})$ & $5.0(-9)$ & $6.1(-10)$ & $8.5(-9)$ & $1.0(-10)$ & $2.7(-8)$\\
$N({\rm CH_3CN})$\tablefootmark{b} & $5.5(11)$ & $5.4(11)$ & $1.4(11)$ & $4.2(11)$ & $7.0(11)$ \\
$x({\rm CH_3CN})$\tablefootmark{b} & $1.1(-11)$ & $1.0(-11)$ & $8.1(-12)$ & $4.0(-12)$ & $2.0(-11)$ \\
$N({\rm N_2H^+})$ & $5.4(13)$ & $2.8(13)$ & $8.8(13)$ & $6.6(12)$ & $5.4(14)$ \\
$x({\rm N_2H^+})$ & $1.6(-9)$ & $9.9(-10)$ & $1.8(-9)$ & $2.8(-10)$ & $9.8(-9)$\\
\hline 
\multicolumn{6}{c}{IR-dark clumps} \\
\hline
Quantity & Mean & Median & Std\tablefootmark{a} & Min. & Max. \\
\hline
$N({\rm SiO})$ & $1.5(13)$ & $1.1(13)$ & $7.9(12)$ & $8.2(12)$ & $2.8(13)$\\
$x({\rm SiO})$ & $3.6(-10)$ & $3.5(-10)$ & $1.4(-10)$ & $1.6(-10)$ & $5.7(-10)$\\
$N({\rm HN^{13}C})$ & $9.6(12)$\tablefootmark{c} \\
$x({\rm HN^{13}C})$ & $1.0(-10)$\tablefootmark{c} \\
$N({\rm C_2H})$ & $4.5(14)$ & $1.7(14)$ & $4.7(14)$ & $7.8(13)$ & $1.2(15)$\\ 
$x({\rm C_2H})$ & $8.6(-9)$ & $6.3(-9)$ & $4.0(-9)$ & $5.1(-9)$ & $1.3(-8)$\\
$N({\rm HNCO})$ & $2.4(14)$ & $1.3(14)$ & $2.7(14)$ & $5.4(12)$ & $7.9(14)$\\
$x({\rm HNCO})$ & $5.5(-9)$ & $4.1(-9)$ & $5.7(-9)$ & $1.3(-10)$ & $1.9(-8)$\\
$N({\rm HCN})$ & $6.5(14)$ & $2.0(14)$ & $7.5(14)$ & $3.4(13)$ & $1.7(15)$\\  
$x({\rm HCN})$ & $2.0(-8)$ & $4.4(-9)$ & $2.5(-8)$ & $9.3(-10)$ & $5.6(-8)$\\
$N({\rm HCO^+})$ & $2.2(13)$ & $1.3(13)$ & $2.1(13)$ & $1.4(12)$ & $6.3(13)$\\
$x({\rm HCO^+})$ & $4.3(-10)$ & $3.5(-10)$ & $3.3(-10)$ & $6.0(-11)$ & $1.1(-9)$\\
$N({\rm HNC})$ & $9.6(13)$ & $6.3(13)$ & $1.4(14)$ & $3.7(12)$ & $5.3(14)$\\
$x({\rm HNC})$ & $2.0(-9)$ & $1.7(-9)$ & $1.5(-9)$ & $4.2(-10)$ & $5.5(-9)$\\
$N({\rm HC_3N})$ & $1.6(14)$ & $3.3(13)$ & $3.1(14)$ & $1.4(13)$ & $7.9(14)$\\
$x({\rm HC_3N})$ & $5.1(-9)$ & $5.7(-10)$ & $1.1(-8)$ & $2.8(-10)$ & $2.7(-8)$\\
$N({\rm CH_3CN})$ & $5.6(11)$ & $5.6(11)$ & $2.0(11)$ & $4.2(11)$ & $7.0(11)$\\
$x({\rm CH_3CN})$ & $1.2(-11)$ & $1.2(-11)$ & $1.1(-11)$ & $4.0(-12)$ & $2.0(-11)$\\
$N({\rm N_2H^+})$ & $9.0(13)$ & $4.0(13)$ & $1.6(14)$ & $1.6(13)$ & $5.4(14)$\\
$x({\rm N_2H^+})$ & $1.8(-9)$ & $1.3(-9)$ & $1.5(-9)$ & $7.3(-10)$ & $5.6(-9)$\\
\hline 
\multicolumn{6}{c}{IR-bright clumps} \\
\hline
Quantity & Mean & Median & Std\tablefootmark{a} & Min. & Max. \\
\hline
$N({\rm H^{13}CO^+})$ & $7.4(12)$ & $6.4(12)$ & $2.3(12)$ & $5.8(12)$ & $1.0(13)$\\
$x({\rm H^{13}CO^+})$ & $9.7(-11)$ & $1.1(-10)$ & $5.1(-11)$ & $4.0(-11)$ & $1.4(-10)$\\
$N({\rm SiO})$ & $2.4(13)$ & $2.2(13)$ & $1.8(13)$ & $5.5(12)$ & $4.8(13)$\\ 
$x({\rm SiO})$ & $9.4(-10)$ & $9.7(-10)$ & $7.3(-10)$ & $4.0(-11)$ & $1.8(-9)$\\
$N({\rm HN^{13}C})$ & $1.2(13)$ & $1.1(13)$ & $2.6(12)$ & $1.0(13)$ & $1.5(13)$\\
$x({\rm HN^{13}C})$ & $1.3(-10)$ & $1.2(-10)$ & $1.2(-11)$ & $1.2(-10)$ & $1.4(-10)$\\
$N({\rm C_2H})$ & $4.2(14)$ & $4.0(14)$ & $2.3(14)$ & $1.1(14)$ & $8.8(14)$\\ 
$x({\rm C_2H})$ & $7.9(-9)$ & $5.2(-9)$ & $7.3(-9)$ & $1.8(-9)$ & $2.4(-8)$\\
$N({\rm HNCO})$ & $1.8(14)$ & $1.5(14)$ & $1.3(14)$ & $1.7(13)$ & $4.3(14)$\\
$x({\rm HNCO})$ & $7.9(-9)$ & $6.2(-9)$ & $5.9(-9)$ & $6.1(-10)$ & $1.9(-8)$\\
$N({\rm HCN})$ & $5.7(14)$ & $6.1(13)$ & $1.6(15)$ & $1.6(13)$ & $5.5(15)$\\
$x({\rm HCN})$ & $2.1(-8)$ & $1.7(-9)$ & $6.1(-8)$ & $2.7(-10)$ & $2.1(-7)$\\
$N({\rm HCO^+})$ & $3.3(13)$ & $9.2(12)$ & $6.5(13)$ & $1.1(12)$ & $2.6(14)$\\
$x({\rm HCO^+})$ & $6.2(-10)$ & $3.2(-10)$ & $8.0(-10)$ & $7.0(-11)$ & $3.6(-9)$\\
$N({\rm HNC})$ & $7.7(13)$ & $3.3(13)$ & $1.1(14)$ & $4.0(12)$ & $4.4(14)$\\
$x({\rm HNC})$ & $2.4(-9)$ & $1.4(-9)$ & $2.4(-9)$ & $2.3(-10)$ & $8.4(-9)$\\
$N({\rm HC_3N})$ & $1.6(14)$ & $2.8(13)$ & $2.7(14)$ & $7.0(12)$ & $7.6(14)$\\
$x({\rm HC_3N})$ & $4.7(-9)$ & $7.8(-10)$ & $7.2(-9)$ & $1.0(-10)$ & $1.8(-8)$\\
$N({\rm CH_3CN})$ & $5.4(11)$\tablefootmark{c} \\
$x({\rm CH_3CN})$ & $1.0(-11)$\tablefootmark{c} \\
$N({\rm N_2H^+})$ & $4.1(13)$ & $2.5(13)$ & $3.7(13)$ & $6.6(12)$ & $1.4(14)$\\
$x({\rm N_2H^+})$ & $1.5(-9)$ & $9.0(-10)$ & $1.9(-9)$ & $2.8(-10)$ & $9.8(-9)$\\
\hline 
\end{tabular} 
\tablefoot{\tablefoottext{a}{Standard deviation.}\tablefoottext{b}{Only three 
detections.}\tablefoottext{c}{Only one detection.}}
\end{minipage} 
}
\end{table}

\subsection{Abundance ratios and correlations}

As the purpose of the present study is to examine the chemistry of the 
sources, we computed the abundance ratios between selected molecules. In 
Table~\ref{table:ratios}, we list the HNC/HCN, HNC/HCO$^+$, 
N$_2$H$^+$/HCO$^+$, N$_2$H$^+$/HNC, and HC$_3$N/HCN column density ratios for 
the clumps. The quoted uncertainties were propagated from those of the column 
densities. 

\begin{figure}[!h]
\centering
\resizebox{0.48\hsize}{!}{\includegraphics{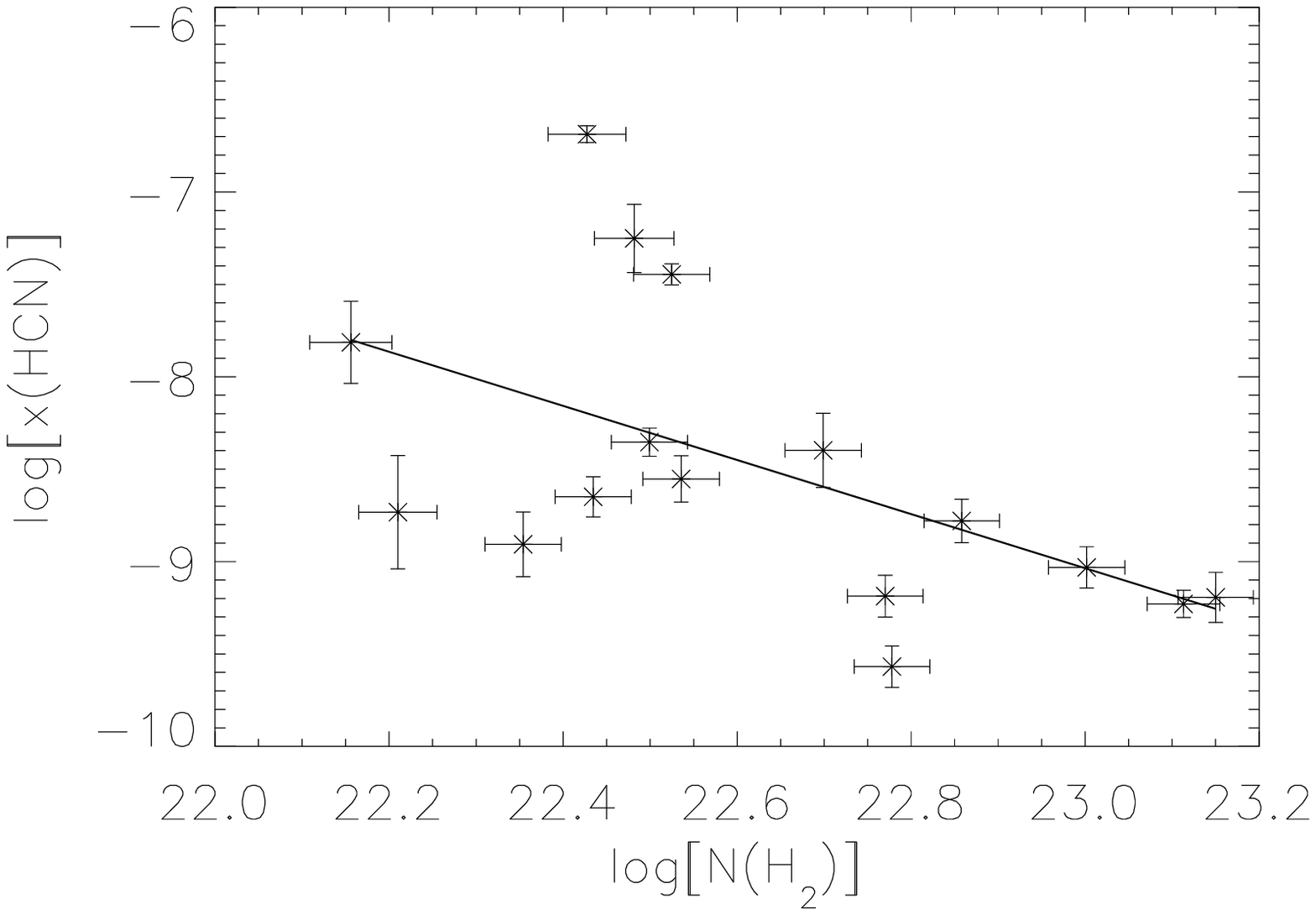}}
\resizebox{0.48\hsize}{!}{\includegraphics{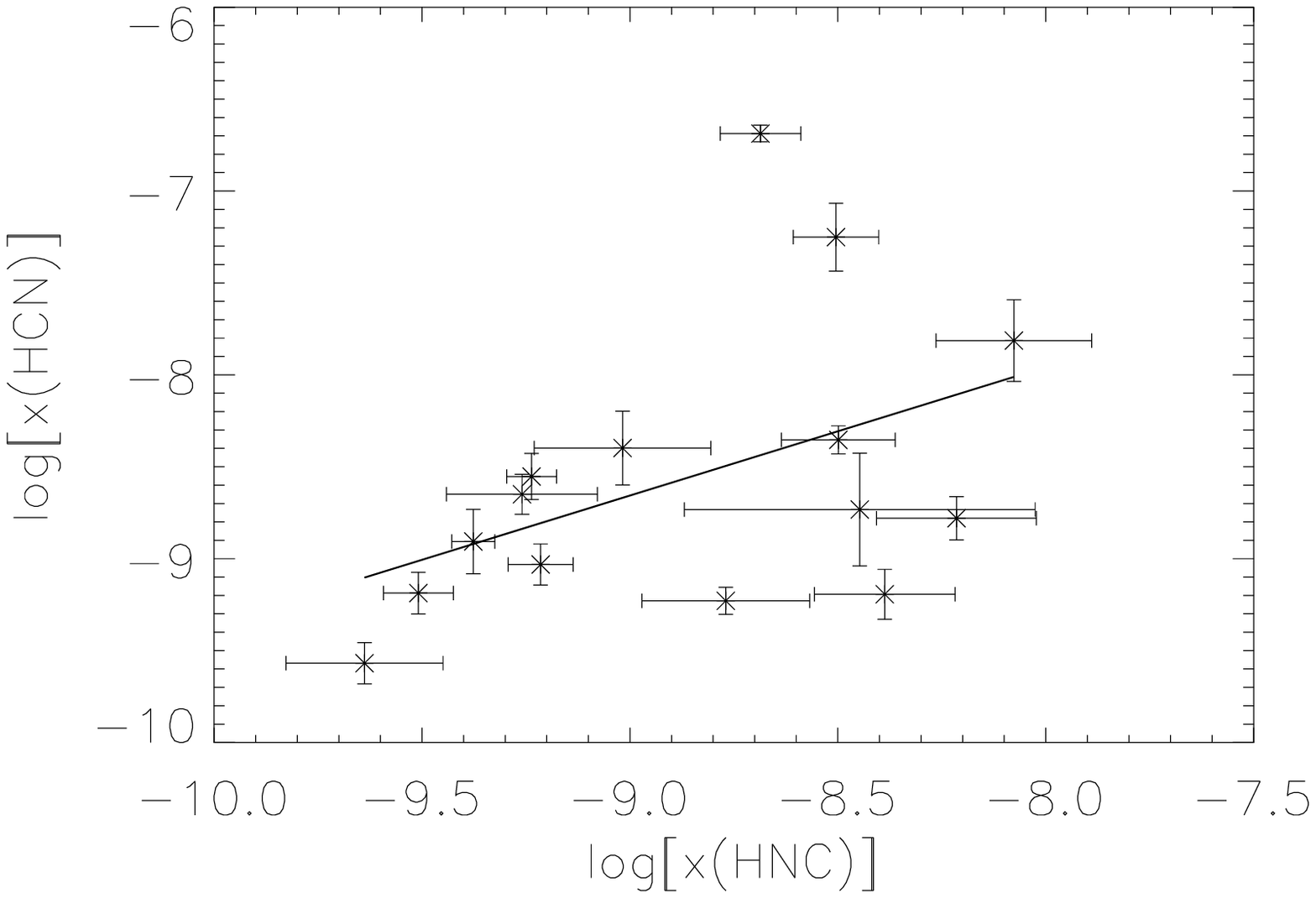}}
\resizebox{0.48\hsize}{!}{\includegraphics{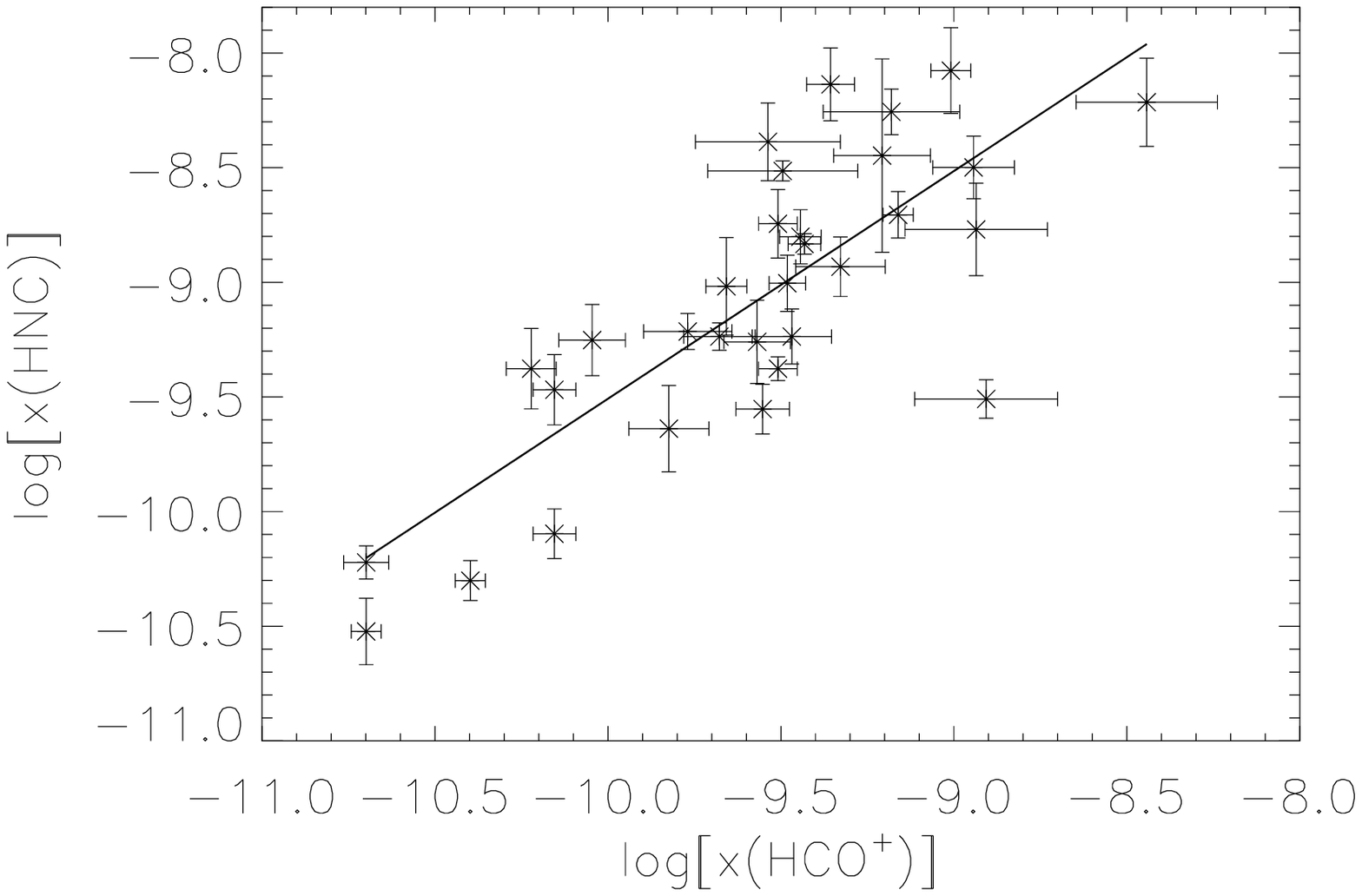}}
\resizebox{0.48\hsize}{!}{\includegraphics{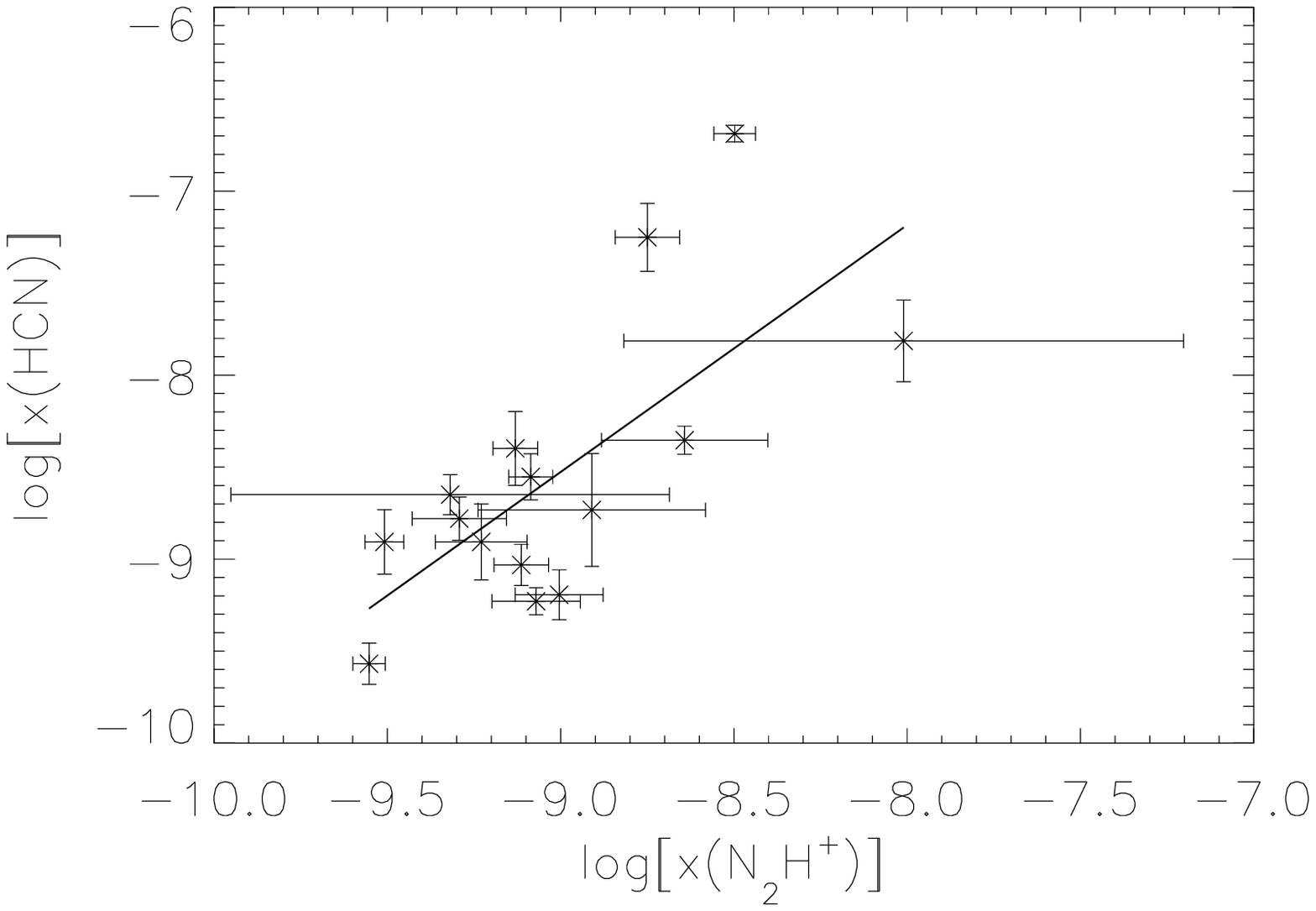}}
\resizebox{0.48\hsize}{!}{\includegraphics{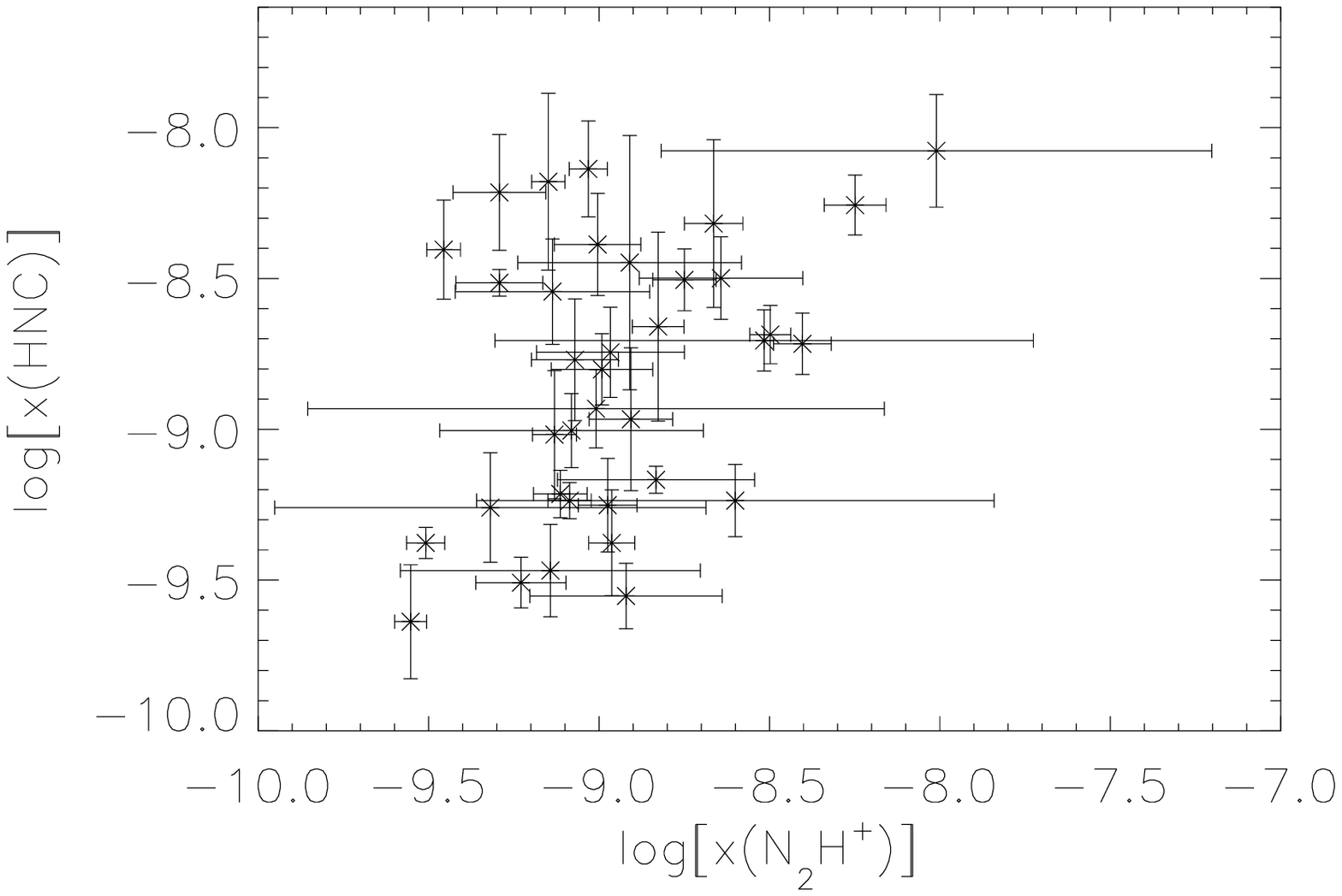}}
\caption{The upper left panel shows the HCN fractional abundance plotted as a 
function of H$_2$ column density in logarithmic scales. The rest of the panels 
show the found correlations between the derived fractional abundances of 
the molecules. Form top right to bottom panel, the panels plot $x({\rm HCN})$ 
versus $x({\rm HNC})$, $x({\rm HNC})$ versus $x({\rm HCO^+})$, $x({\rm HCN})$ 
versus $x({\rm N_2H^+})$, and $x({\rm HNC})$ versus $x({\rm N_2H^+})$ in 
logarithmic scales. The solid lines show the least squares fit to the data 
(see text for details).}
\label{figure:correlations}
\end{figure}

We also searched for possible correlations between different parameter pairs. 
As shown in the upper left panel of Fig.~\ref{figure:correlations}, there is 
a hint that the fractional abundance of HCN decreases as a function of the 
H$_2$ column density. A least squares fit to the data points yields 
$\log\left[x({\rm HCN})\right]=(24.64\pm13.53)-(1.46\pm0.60)\log \left[N({\rm H_2})\right]$, with the linear Pearson correlation coefficient of $r=-0.55$. 
For this plot, the H$_2$ column densities were derived from the LABOCA maps 
smoothed to the resolution of the MALT90 data. In the rest of the 
Fig.~\ref{figure:correlations} panels, we show the correlations found between 
different molecular fractional abundances. The top right panel plots the 
HCN abundance as a function of $x({\rm HNC})$. 
The overplotted linear regression model is of the form $\log\left[x({\rm HCN})\right]=(-2.36\pm3.46)+(0.70\pm0.39)\log \left[x({\rm HNC})\right]$, with the Pearson's $r$ of 0.45. 
The middle left panel plots the HNC abundance as a function of the HCO$^+$ 
abundance. A positive corrrelation is found, and the fitted linear relationship is of the form 
$\log\left[x({\rm HNC})\right]=(0.42\pm1.32)+(0.99\pm0.14)\log \left[x({\rm HCO^+})\right]$ ($r=0.80$). The middle right panel shows the HCN abundance plotted 
as a function of $x({\rm N_2H^+})$. Again, the data suggest a positive 
correlation, and the functional form of the linear fit is 
$\log\left[x({\rm HCN})\right]=(3.56\pm3.38)+(1.34\pm0.37)\log \left[x({\rm N_2H^+})\right]$ ($r=0.71$). The bottom panel shows the HNC abundance as a 
function of the N$_2$H$^+$ abundance. Here, the correlation coefficient is only 
0.39, and no linear fit is shown. 

\begin{table*}
\caption{Column density ratios.}
\begin{minipage}{2\columnwidth}
\centering
\renewcommand{\footnoterule}{}
\label{table:ratios}
\begin{tabular}{c c c c c c}
\hline\hline 
Source & $\frac{N({\rm HNC})}{N({\rm HCN})}$ & $\frac{N({\rm HNC})}{N({\rm HCO^+})}$ & $\frac{N({\rm N_2H^+})}{N({\rm HCO^+})}$ & $\frac{N({\rm N_2H^+})}{N({\rm HNC})}$ & $\frac{N({\rm HC_3N})}{N({\rm HCN})}$\\
\hline
G1.87-0.14 &\\
SMM 1 & \ldots & $8.41\pm4.13$ & $8.57\pm4.13$ & $1.02\pm0.28$ & \ldots\\
SMM 1\tablefootmark{a} & \ldots & $1.83\pm0.06$ & \ldots & \ldots & \ldots\\
SMM 8 & $0.71\pm0.24$ & $2.78\pm1.08$ & $2.00\pm1.19$ & $0.72\pm0.45$ & \ldots\\
SMM 10 & $0.55\pm0.36$ & $8.57\pm3.62$ & $10.00\pm18.59$ & $1.17\pm2.22$ & \ldots\\
SMM 12 & \ldots & $2.84\pm0.61$ & $4.40\pm8.00$ & $1.55\pm2.84$ & \ldots\\
SMM 14 & \ldots & $5.87\pm1.97$ & $3.53\pm2.41$ & $0.60\pm0.45$& \ldots \\
SMM 15 & \ldots & \ldots & \ldots & $0.11\pm0.07$ & \ldots\\
SMM 20 & \ldots & \ldots & \ldots & $0.09\pm0.03$ & \ldots\\
SMM 21 & \ldots & \ldots & \ldots & $2.05\pm0.55$ & \ldots\\
SMM 23 & $0.24\pm0.16$ & $4.36\pm2.13$ & $3.36\pm0.48$ & $0.77\pm0.38$ & $0.07\pm0.03$\\
SMM 23\tablefootmark{a} & \ldots & \ldots & \ldots & $1.74\pm1.15$ & \ldots\\
SMM 24 & \ldots & \ldots & \ldots & $0.26\pm0.19$ & \ldots \\
SMM 27 & $0.06\pm0.03$ & \ldots & \ldots & $0.57\pm0.16$ & $0.02\pm0.01$\\
SMM 28 & $0.01\pm0.002$ & \ldots & \ldots & $1.55\pm0.34$ & $0.02\pm0.01$\\
SMM 30 & \ldots & $4.36\pm1.16$ & $2.82\pm0.94$ & $0.65\pm0.26$ & \ldots\\
SMM 31 & \ldots & $16.49\pm6.13$ & $2.11\pm0.31$ & $0.13\pm0.05$ & \ldots\\
SMM 38 & \ldots & \ldots & \ldots & $0.69\pm0.51$ & \ldots\\
SMM 38\tablefootmark{a} & \ldots & \ldots & \ldots & $0.48\pm0.31$ & \ldots\\
G2.11+0.00 &\\
SMM 5 & $0.25\pm0.11$ & $2.03\pm0.90$ & $1.76\pm2.59$ & $0.87\pm1.31$ & \ldots\\
G11.36+0.80 &\\
SMM 1  & \ldots & $5.94\pm2.19$ & $11.25\pm2.34$ & $1.89\pm0.73$ & \ldots\\
SMM 2 & \ldots & $7.14\pm3.03$ & $18.57\pm3.41$  & $2.60\pm1.08$ & \ldots\\
SMM 3 & \ldots & $5.18\pm1.97$ & $10.91\pm11.09$ & $2.11\pm2.22$ & \ldots\\
SMM 4 & \ldots & \ldots & \ldots & $1.15\pm0.69$& \ldots \\
SMM 5 & \ldots & $1.00\pm0.26$ & $4.25\pm2.80$ & $4.25\pm2.91$ & \ldots\\
SMM 6 & \ldots & $1.68\pm0.56$ & $7.27\pm12.83$ & $4.32\pm7.64$ & \ldots\\
SMM 7 & \ldots & $2.47\pm0.98$ & $2.06\pm4.07$ & $0.84\pm1.65$ & \ldots\\
G13.22-0.06 &\\
SMM 4 & $0.21\pm0.06$ & $2.78\pm0.71$ & $3.89\pm1.01$ & $1.40\pm0.21$ & \ldots\\
SMM 5 & $2.86\pm1.35$ & $1.47\pm0.96$ & $0.73\pm0.40$ & $0.50\pm0.27$ & $0.17\pm0.03$\\
SMM 6 & $0.34\pm0.13$ & $1.36\pm0.13$ & $1.00\pm0.09$ & $0.73\pm0.36$ & \ldots\\
SMM 7 & $6.38\pm3.05$ & $14.23\pm8.49$ & $3.42\pm1.85$ & $0.24\pm0.11$ & \ldots\\
SMM 7(37 km~s$^{-1}$)\tablefootmark{b} & \ldots & $3.67\pm0.83$ & \ldots & \ldots & \ldots\\
SMM 11 & \ldots & $3.02\pm0.83$ & $2.54\pm2.23$ & $0.84\pm0.77$ & \ldots\\
SMM 11(37 km~s$^{-1}$)\tablefootmark{b} & \ldots & $1.07\pm0.26$ & \ldots & \ldots & \ldots\\
SMM 23 & $1.93\pm2.31$ & $5.80\pm5.86$ & $2.00\pm1.62$ & $0.34\pm0.42$ & \ldots\\
SMM 23(53 km~s$^{-1}$)\tablefootmark{b} & \ldots & $1.68\pm0.34$ & \ldots & \ldots & \ldots\\
SMM 27 & $3.67\pm1.83$ & $1.69\pm1.07$ & $0.14\pm0.08$ & $0.08\pm0.04$ & $0.06\pm0.02$\\
SMM 29 & $0.88\pm0.43$ & $1.52\pm0.76$ & $1.85\pm0.47$ & $1.21\pm0.52$ & \ldots\\
SMM 29(13.6 km~s$^{-1}$)\tablefootmark{b} & \ldots & $1.68\pm0.43$ & \ldots & \ldots & \ldots \\
SMM 29(37 km~s$^{-1}$)\tablefootmark{b} & \ldots & $1.25\pm0.17$ & \ldots & \ldots & \ldots\\
SMM 29\tablefootmark{a} & \ldots & $1.55\pm1.11$ & $2.58\pm1.44$ & $1.67\pm1.00$& \ldots\\
SMM 32 & $0.47\pm0.14$ & $0.25\pm0.12$ & $0.48\pm0.26$  & $1.94\pm0.64$ & $0.24\pm0.06$\\
SMM 32\tablefootmark{a} & $0.38\pm0.11$ & $1.40\pm0.43$ & $4.95\pm1.53$ & $3.54\pm0.77$ & \ldots\\
SMM 32(14 km~s$^{-1}$)\tablefootmark{a, b} & \ldots & $1.83\pm0.51$ & \ldots & \ldots & \ldots \\
\hline 
\end{tabular} 
\tablefoot{\tablefoottext{a}{Towards the line emission peak.}\tablefoottext{b}{For the additional velocity component.} } 
\end{minipage}
\end{table*}


\section{Discussion}

In this section, we discuss the obtained results for each indivi\-dual species 
separately. We mostly compare our results with those obtained by VLH11 and 
SJF12, because they also employed the Mopra telescope observations for their 
studies.

\subsection{HCO$^+$ and H$^{13}$CO$^+$ (Formylium)}

In dense molecular clouds, HCO$^+$ is mainly formed through the gas-phase 
ion-neutral reaction ${\rm H}_3^+ + {\rm CO}\rightarrow {\rm HCO^+}+{\rm H}_2$ 
(e.g., \cite{herbst1973}). The HCO$^+$ abundance can be increased in regions 
where shocks are generated, e.g., due to outflows from embedded YSOs. 
When the shock heats the gas and produces UV radiation through Ly-$\alpha$ 
emission ($\lambda=121.6$ nm), the icy grain mantles 
are evaporated and the HCO$^+$ abundance gets enhanced (\cite{rawlings2000}, 
2004). This is due to evaporated CO and H$_2$O, where the latter species can 
form HCO$^+$ in the reaction with photoionised carbon 
(${\rm C^+}+{\rm H_2O}\rightarrow {\rm HCO^+}+{\rm H}$).
The destruction of the HCO$^+$ molecules is, in turn, 
mainly caused by the dissociative recombination with electrons, 
${\rm HCO^+}+{\rm e}^-\rightarrow {\rm CO}+{\rm H}$.  

Extended HCO$^+$ emission is seen particularly around the submm peaks 
G1.87--SMM 1, 8, 10, 12, 14--16, and G13.22--SMM 4, 5, 6, 7, 10, 11, 23, 27, 
and 32. Moreover, the submm clump/UC \ion{H}{ii} region G2.11--SMM 5 is 
associated with an elongated HCO$^+$ clump (Fig.~\ref{figure:G211SMM5lines}). 
In the case of G2.11--SMM 5, G11.36--SMM 5, G13.22--SMM 5, and 
G13.22--SMM 32, the HCO$^+$ emission peak is close to the LABOCA 
870-$\mu$m dust emission peak. 
The clump G1.87--SMM 1 is classified as IR-dark, but the HCO$^+$ line towards 
the submm peak position shows non-Gaussian wing emission, indicative of 
outflows/shocks. The clumps G13.22--SMM 4--7 and 11 are associated with the 
\textit{Spitzer} IR-bubble system N10/11 (Figs.~\ref{figure:G1322SMM5lines} and 
\ref{figure:G1322SMM7lines}; \cite{churchwell2006}), which is suggested to 
represent a site of triggered massive-star formation (\cite{watson2008}). 
The strong HCO$^+$ emission seen towards the bubble surroundings could 
originate in the swept-up bubble shells where shock fronts are 
expanding into the surrounding medium. Moreover, the high-mass stars in the 
system produce a strong radiation field of UV photons. 
Some of the IRDCs studied by Liu et al. 
(2013) show similar extended HCO$^+$ emission as the sources studied here. 
The HCO$^+$ column densities derived by Liu et al. (2013), 
$2.24\times10^{12}-1.31\times10^{13}$ cm$^{-2}$ ($\sim5.1\times10^{12}$ cm$^{-2}$ 
on average), are lower by a factor of about six on average than those we 
derived. On the other hand, the values derived by SJF12, 
$5.8\times10^{13}-1\times10^{15}$ cm$^{-2}$ with the median of 
$1.88\times10^{14}$ cm$^{-2}$, exceed our values.

The HCO$^+$ abundances we derive lie in the range 
$6\times10^{-11}-3.6\times10^{-9}$, with the mean (median) value of 
$5.6\times10^{-10}$ ($3.3\times10^{-10}$). For their sample of the 4th 
quadrant IRDC sources, VLH11 derived the abundances of 
$3.5\times10^{-9}-5.1\times10^{-8}$ with an average value of $1.7\times10^{-8}$. 
Also SJF12 determined higher values of $x({\rm HCO^+})$ for 
their sample of IRDC clumps, ranging from $3.9\times10^{-9}$ to 
$2.8\times10^{-7}$ (with the median of $2.51\times10^{-8}$). 
The latter authors found that both the HCO$^+$ column density and abundance 
increase as the clump 
evolves from the quiescent state with no IR emission (as seen by 
\textit{Spitzer}) to ``red'' state with bright 8-$\mu$m emission and when the 
central source has likely formed an \ion{H}{ii} region. More recently, 
the MALT90 study of 333 massive clumps by Hoq et al. (2013) revealed a similar 
evolutionary trend in $x({\rm HCO^+})$ (their Fig.~5). We also derive  
higher HCO$^+$ column densities and abundances on average for IR-bright clumps 
as compared to IR-dark ones, although the median values are quite similar 
between the two classes. The lowest value of $x({\rm HCO^+})$ in our sample is 
derived towards the IR-dark clump SMM 2 in the G11.36 filament, while the 
highest abundance is seen towards the IR-bright clump G13.22--SMM 27. 
The G13.22--SMM 32 clump, associated with an \ion{H}{ii} region, also shows a 
relatively high value of $x({\rm HCO^+})$ compared to the rest of our sources. 
These findings are in agreement with the evolutionary trend found by SJF12.
The fact that we derive lower values of $x({\rm HCO^+})$ than in VLH11 and 
SJF12 could mean that our sources are, on average, less evolved. The depletion 
of CO molecules would, at least partly, explain the meagre amount of HCO$^+$ 
found in the present study. In Paper I, the CO depletion factor was derived 
towards some of our clumps. For example, towards 
G11.36--SMM 1 and G13.22--SMM 27 the values $f_{\rm D}({\rm CO})=3.9\pm0.6$ 
and $9.9\pm1.5$ were determined.  

The $^{13}$C isotopologue H$^{13}$CO$^+$ is formed in a similar way as the main 
$^{12}$C-form except from $^{13}$CO. The isotope transfer via 
${\rm HCO^+}+{\rm ^{13}CO}\leftrightarrow {\rm H^{13}CO^+}+{\rm ^{12}CO}$ can 
also play a role in the formation of H$^{13}$CO$^+$ (\cite{langer1984}). 
Only weak emission of H$^{13}$CO$^+$, if any, is detected towards our clumps 
(only three detections). Towards G13.22--SMM 27, the line emission is quite 
well correlated with the 8-$\mu$m absorption 
(Fig.~\ref{figure:G1322SMM23lines}). The column densities and fractional 
abundances are derived to be $5.8\times10^{12}-1\times10^{13}$ cm$^{-2}$ 
($7.4\times10^{12}$ cm$^{-2}$ on average) and $4\times10^{-11}-1.4\times10^{-10}$ 
($9.7\times10^{-11}$ on average). The former values are comparable to those 
found by Sakai et al. (2010) for their sample of clumps within IRDCs 
($1.3\times10^{12}-1.4\times10^{13}$ cm$^{-2}$). Vasyunina et al. (2011) 
derived H$^{13}$CO$^+$ abundances of $7.6\times10^{-11}-7.4\times10^{-10}$ 
with an average of $3.4\times10^{-10}$, which is 3.5 times higher than our 
average abundance.

\subsection{SiO (Silicon Monoxide)}

In star-forming regions, SiO emission is usually believed to be linked to the 
action of high-velocity ($\sim20-50$ km~s$^{-1}$) shocks 
(\cite{martinpintado1992}; \cite{schilke1997}; \cite{gusdorf2008a}, b). 
SiO emission can also trace irradiated medium-velocity ($\sim10-20$ 
km~s$^{-1}$) shocks in PDRs (e.g., \cite{schilke2001}).

SiO can form via sputtering of Si atoms from the grain 
cores, which then undergo oxidation through the neutral-neutral gas-phase
reactions ${\rm Si}+{\rm O}_2 \rightarrow {\rm SiO}+{\rm O}$ and 
${\rm Si}+{\rm OH} \rightarrow {\rm SiO}+{\rm H}$. If Si is present in the 
icy grain mantles, a clearly lower shock velocity is sufficient to release 
it into the gas phase (\cite{gusdorf2008b}). Alternatively, SiO can be 
directly formed through dust destruction by vaporisation in grain-grain
collisions (e.g., \cite{guillet2009}). In the hot post-shock gas, OH molecules 
are abundant due to the reaction 
${\rm O}+{\rm H}_2 \rightarrow {\rm OH}+{\rm O}$. When SiO reacts with OH, 
a conversion to SiO$_2$ takes place 
(${\rm SiO}+{\rm OH} \rightarrow {\rm SiO}_2+{\rm H}$; \cite{schilke1997}). 
This limits the SiO abundance in the shocked gas. 

The SiO emission appears to be quite widespread/extended particularly towards 
G1.87--SMM 27, 28, 31 (Fig.~\ref{figure:G187SMM28lines}). Moreover, 
G1.87--SMM 20, 23, 30 (Fig.~\ref{figure:G187SMM23lines}) and G1.87--SMM 38 
(Fig.~\ref{figure:G187SMM38lines}) are associated with a few parsec-scale SiO 
clump. It is also worth noting that the submm peaks G1.87--SMM 28, 30, and 
38 are coincident with the local SiO peak positions. 
Jim\'enez-Serra et al. (2010) proposed that 
the extended SiO emission they observed along the filamentary IRDC 
G035.39-00.33 is the result of a low-velocity shock produced by colliding 
flows (see also \cite{henshaw2013}). The widespread SiO emission could 
therefore originate in the cloud formation process instead of star formation. 
Sanhueza et al. (2013) recently detected SiO emission from the candidate 
\textit{starless} IRDC G028.23-00.19. The authors suggested that the SiO 
emission with narrow linewidths, coincident with the subclouds' interface 
within the source, could be caused by vaporisation of icy grain mantles in 
grain-grain collisions. Some of our clumps around which extended SiO emission 
is detected are, however, associated with IR sources, and are likely hosting 
embedded YSOs. Outflows from these forming stars are likely to be responsible 
for the detected SiO emission. This is supported by the fact that the SiO and 
LABOCA emission peaks are coincident in G1.87--SMM 28, 30, and 38, and that 
some of the line profiles show wing emission. We also note that the 
filamentary IRDC G11.36 does not show extended SiO emission along its long 
axis (and neither do the other filaments of this study). The clumps SMM 4 and 
5 around the bubble system N10/11 are neither associated with extended SiO 
emission, although expanding shock fronts are expected to be present there.

Sakai et al. (2010) found that towards their IRDC sources, the SiO column 
densities are $\sim4.6\times10^{12}-3.8\times10^{13}$ cm$^{-2}$ with an average 
value of $1.5\times10^{13}$ cm$^{-2}$. This is very similar to our clumps, for 
which $N({\rm SiO})=5.5\times10^{12}-4.8\times10^{13}$ cm$^{-2}$ 
($1.9\times10^{13}$ cm$^{-2}$ on average). However, given the assumption made 
for $T_{\rm ex}$ (Sect.~3.3), our $N({\rm SiO})$ values should be taken as 
lower limits. Vasyunina et al. (2011) derived SiO 
abundances in the range of $1.6\times10^{-10}-1.6\times10^{-9}$ (average value 
$9.5\times10^{-10}$) towards their IRDCs. Our values, 
$4\times10^{-11}-1.8\times10^{-9}$ (average $5.7\times10^{-10}$), are mostly 
comparable to them. Sanhueza et al. (2012) found that the SiO column densities 
and abundances for their whole sample are $1.36\times10^{12}-3.47\times10^{13}$ 
cm$^{-2}$ and $2.78\times10^{-10}-9.20\times10^{-10}$ (median values are 
$7.72\times10^{12}$ cm$^{-2}$ and $5.79\times10^{-10}$). In particular, their 
median abundance is close to what we found ($3.6\times10^{-10}$). 

The mean and median SiO abundances are found to be almost three times higher 
towards IR-bright clumps as compared to IR-dark sources 
(Table~\ref{table:stat}), which is in accordance with the results by 
SJF12. However, there are distinct exceptions from the average SiO trend 
among our sources, calling its statistical significance into question. The 
lowest SiO abundance is derived towards the IR-\textit{bright} clump 
G13.22--SMM 5, and the second highest SiO abundance is observed towards the 
IR-\textit{dark} clump G1.87--SMM 31. Indeed, Sakai et al. (2010) found a trend 
opposite of ours, and they proposed that the SiO emission from the mid-IR 
dark sources originates in newly formed shocks, while the SiO emission from 
more evolved, mid-IR bright sources could originate in gas shocked earlier in 
time. This could be related to the discovery by Miettinen et al. (2006), 
namely that the SiO abundance in massive clumps appears to decrease as a 
function of gas kinetic temperature, possibly reflecting an evolutionary trend. 

\subsection{HCN, HNC, and HN$^{13}$C [Hydrogen (Iso-)cyanide]}

Gas-phase chemical models suggest that HCN and its metastable geometrical 
isomer, HNC (hydrogen isocyanide), are primarily produced via the dissociative 
recombination reaction ${\rm HCNH^+}+{\rm e^-}\rightarrow {\rm HCN}+{\rm H}$ or 
${\rm HNC}+{\rm H}$ (e.g., \cite{herbst1978}). The resulting HNC/HCN abudance 
ratio is predicted to be 0.9 in this case, i.e., close to unity. The 
formation of HNC (and only HNC, not HCN) can also take place via the reactions 
${\rm H_2CN^+}+{\rm e^-}$ or 
${\rm H_2NC^+}+{\rm e^-}\rightarrow {\rm HNC}+{\rm H}$ (\cite{pearson1974}; 
\cite{allen1980}). As a result of this additional HNC production channel, 
the HNC/HCN ratio can rise above unity. Another ways to form HCN and HNC are 
the neutral-neutral reactions 
${\rm CH_2}+{\rm N}\rightarrow {\rm HCN}+{\rm H}$ and 
${\rm NH_2}+{\rm C}\rightarrow {\rm HNC}+{\rm H}$ (\cite{herbst2000}). After 
these reactions, the species are able to undergo rapid isomerisation 
reactions, again leading to the near unity HNC/HCN ratio. 

In dense clouds and PDRs, HCN can be photodissociated into CN, either directly 
or via cosmic-ray induced photodissociation (\cite{boger2005}). Additional 
destruction processes of HCN in dense clouds are 
${\rm HCN}+{\rm H^+}\rightarrow {\rm HCN^+}+{\rm H}$ and 
${\rm HCN}+{\rm HCO^+}\rightarrow {\rm H_2CN^+}+{\rm CO}$ (\cite{boger2005}).
The HNC molecule can be destroyed via the reactions with hydrogen 
and oxygen atoms, ${\rm HNC}+{\rm H}\rightarrow {\rm HCN}+{\rm H}$ and 
${\rm HNC}+{\rm O}\rightarrow {\rm NH}+{\rm CO}$ (\cite{schilke1992}; 
\cite{talbi1996}). HNC also converts to HCN through the reaction 
${\rm HNC}+{\rm H^+}\rightarrow {\rm HCN}+{\rm H^+}$ (\cite{herbst2000}).

As can be seen in the 0th moment maps in Appendix~B, the spatial 
distributions of HCN and HNC emissions are generally extended. Moreover, for 
example in the case of G2.11--SMM 5 and G13.22--SMM 5, the submm peak position 
is close to the peak HCN and HNC emissions. The HN$^{13}$C emission is 
extended, although weak, in G1.87--SMM 20, 23, 30 
(Fig.~\ref{figure:G187SMM23lines}). Similarly, Jackson et al. (2010) 
found that HNC$(1-0)$ emission traces well the filamentary IRDC G338.4-0.4 or 
the Nessie Nebula. Sanhueza et al. (2012) found that HNC is ubiquitous in the 
clumps of IRDCs, in agreement with our results. Also Liu et al. (2013) found 
a good correlation between the 8-$\mu$m absorption and HNC and HCN 
emissions towards many of their IRDCs. 

The HCN and HNC abundances we derived are in the ranges of 
$2.7\times10^{-10}-2.1\times10^{-7}$ and $2.3\times10^{-10}-8.4\times10^{-9}$, 
respectively. The HN$^{13}$C abundances are found to be $\sim10^{-10}$ on 
average. Sakai et al. (2010) found HN$^{13}$C column densities of 
$\sim2.6\times10^{12}-1.4\times10^{13}$ cm$^{-2}$ towards IRDCs, quite similar 
to our values of $9.6\times10^{12}-1.5\times10^{13}$ cm$^{-2}$. 
Vasyunina et al. (2011) derived $x({\rm HCN})$ and 
$x({\rm HNC})$ values of $3.3\times10^{-10}-6.8\times10^{-9}$ (average 
$2\times10^{-9}$) and $2.4\times10^{-10}-6.3\times10^{-9}$ (average 
$1.6\times10^{-9}$), respectively. In general, these are comparable to our 
values, although the average HCN abundance we found is an order magnitude 
higher than derived by VLH11. The column densities and fractional 
abundances of HNC found by SJF12 are $5.1\times10^{13}-1.36\times10^{15}$ 
cm$^{-2}$ and $5.1\times10^{-9}-1.85\times10^{-7}$. Their median values of  
$2.42\times10^{14}$ cm$^{-2}$ and $3.73\times10^{-8}$ are both clearly higher 
than our values. Among the sample of SJF12, $N({\rm HNC})$ was found to 
slightly increase with the clump evolution up to the active phase (when the 
clump shows an extended 4.5-$\mu$m emission and hosts an embedded 24-$\mu$m 
source), possibly as a result of accretion of the ambient material as 
suggested by the authors. However, the authors did not found evidence 
of increasing $x({\rm HNC})$ as a function of source evolution. 
In contrast, we found higher $N({\rm HNC})$ values 
towards IR-dark clumps on average than towards IR-bright clumps (the same 
holds for the median values also). However, the average HNC abundance 
(but not the median value) appears to be slightly higher in IR-bright clumps.

As shown in the top left panel of Fig.~\ref{figure:correlations}, there is a 
slight hint that $x({\rm HCN})$ decreases when the H$_2$ column density 
increases. Although the correlation coefficient is quite low ($r=-0.55$), 
this could be related to the enhanced abundance of HCO$^+$ in more 
evolved (i.e., denser) sources\footnote{A higher H$_2$ column 
density may not necessarily indicate a more advanced evolutionary stage. 
For example, Hoq et al. (2013) found no clear tendency for more evolved clumps 
to have a higher H$_2$ column density (their Fig.~3).}; HCO$^+$ destroys HCN 
producing H$_2$CN$^+$ ions (see above). The middle left panel of 
Fig.~\ref{figure:correlations} shows that the HNC abundance increases when 
the HCO$^+$ abundance is enhanced. This can be understood as an increased
production of H$_2$CN$^+$ from HCO$^+$. The dissociative 
recombination of H$_2$CN$^+$ then leads to the production of HNC (see above). 
The top right panel plot of Fig.~\ref{figure:correlations} suggests that the 
HCN abundance increases when that of HNC increases. This is reminiscent of the 
positive correlation between the integrated intensities of HCN and HNC found 
by Liu et al. (2013; their Fig.~17). They also found a similar correlation 
between HCN and HCO$^+$, in agreement with our relationship shown in the 
middle left panel of Fig.~\ref{figure:correlations}. Our positive 
$x({\rm HCO^+})-x({\rm HNC})$ correlation supports the scenario where the HNC 
abundance increases as the clump evolves. However, if $x({\rm HCN})$ also 
increases as our Fig.~\ref{figure:correlations} (top right panel) suggests, 
the possible negative correlation we found between $N({\rm H_2})$ and 
$x({\rm HCN})$ becomes questionable.

The values of the HNC/HCN ratio found towards the clumps lie in the range 
of $0.01\pm0.002-6.38\pm3.05$, with the mean$\pm$std of $1.26\pm1.78$ (median 
is 0.47). The average value resembles the result by VLH11 who found 
that for their IRDCs the HNC/HCN ratio is $\sim1$. Liu et al. (2013) determined 
values in the range $0.71\pm0.11-2.26\pm0.44$ (average $1.47\pm0.50$) towards 
IRDCs, also comparable to our values on average. In the dark cloud cores 
studied by Hirota et al. (1998), the HNC/HCN ratio was found to be 
$0.54\pm0.33-4.5\pm1.2$ ($2.1\pm1.2$ on average), mostly comparable to our 
values within the errors. The above results are consistent with the near unity 
HNC/HCN ratio theoretically expected in cold molecular clouds 
(\cite{sarrasin2010}). However, the abundance ratio between HNC and HCN is 
strongly dependent on the temperature. A good example is the Orion mole\-cular 
cloud, where the HNC/HCN ratio was found to strongly decrease (by more than 
an order of magnitude) when going from the colder quiescent parts of the cloud 
to the warm plateau and hot-core regions where the ratio is much
smaller than unity (\cite{goldsmith1986}; cf.~\cite{schilke1992}). 
Hirota et al. (1998) found that the HNC/HCN ratio starts to rapidly decrease 
when the gas kinetic temperature rises above 24 K, because of HNC conversion
into HCN. The mean$\pm$std (median) of the HNC/HCN ratio towards our IR-dark 
clumps is $0.34\pm0.34$ (0.24), while that for IR-bright clumps is 
$1.49\pm1.93$ (0.55). Due to the large scatter of these values, it is difficult 
to say how well (or poorly) they agree with the earlier studies of 
the temperature dependence of the HNC/HCN ratio. Moreover, the clump 
temperatures should be determined in order to study such behaviour more 
quantitatively. Hoq et al. (2013) examined the integrated intensity 
ratios between HNC and HCN, and found the median values of 0.9, 0.8, and 0.6 
for quiescent IR-dark clumps, clumps containing YSOs, and clumps associated 
with \ion{H}{ii} regions/PDRs. This suggests a possible, although weak, 
evolutionary trend (see Fig.~4 in \cite{hoq2013}).

The column density ratio between HNC and HCO$^+$ is found to lie in the 
range $0.25\pm0.12-16.49\pm6.13$ with a mean$\pm$std of $4.29\pm3.94$ (median 
is 2.78) (the values for the additional velocity components are excluded 
here). Within the errors, this is in agreement with the gas-phase chemical 
models of cold dark clouds which suggest comparable abundances for 
these species (see \cite{roberts2012} and references therein). Moreover, 
the average HN$^{13}$C and H$^{13}$CO$^+$ abundances, $1.2\times10^{-10}$ and 
$9.7\times10^{-11}$, are very similar. This could be indicative of weak 
$^{13}$C fractionation effects. For IR-dark clumps, the mean$\pm$std and median 
values of the HNC/HCO$^+$ ratio are $4.55\pm2.46$ and 4.36. 
For IR-bright sources the corresponding values are $4.19\pm4.47$ and $2.47$. 
It is therefore possible that the HNC/HCO$^+$ ratio decreases slightly as the 
source evolves. This trend can be seen from the data by SJF12 when the clump 
evolves from the so-called intermediate state (either extended 4.5-$\mu$m 
emission or an associated 24-$\mu$m source, but not both) to red state. 
However, the quiescent sources of SJF12 do not follow this trend.

\subsection{C$_2$H (Ethynyl)}

The origin of C$_2$H in the PDR regions of the interstellar medium, 
i.e., at the boundary layers between ionised and molecular gas, 
is believed to be in the photodissociation of acetylene (C$_2$H$_2$): 
${\rm C_2H_2}+h\nu \rightarrow {\rm C_2H}+{\rm H}$ (e.g., \cite{fuente1993}).
The neutral-neutral reaction ${\rm CH}_2+{\rm C}\rightarrow {\rm C_2H}+{\rm H}$
can also produce C$_2$H, where the precursor carbon atom is formed through 
the photodissociation of CO (e.g., \cite{turner2000}).
The dissociative recombination of the ions CH$^+$ (\cite{rimmer2012}), 
C$_2$H$_2^+$, and C$_2$H$_3^+$ also yield ethynyl molecules in dense mole\-cular 
clouds (\cite{mul1980}). The C$_2$H molecules can themselves be 
photodissociated back to form C$_2$ and C$_2$H$^+$ (\cite{fuente1993}). 
C$_2$H is also destroyed through the reactions 
${\rm C_2H}+{\rm O}\rightarrow {\rm CO}+{\rm CH}$ and ${\rm C_2H}+{\rm C^+}\rightarrow {\rm C_3^+}+{\rm H}$ (\cite{watt1988}).

The spatial distribution of the C$_2$H emission is found to be quite extended 
(e.g., the clumps  G1.87--SMM 1, G1.87--SMM 20, 23, 30, and G13.22--SMM 32). 
It is also worth noting that the emission is ridge-like around the IR-bubble 
pair N10/11 on the side of G13.22--SMM 4 and 5, and extended on the other side 
containing G13.22--SMM 6, 7, 10, and 11. In N10/11, 
and towards G13.22--SMM 32, we are probably probing the PDR parts 
of the sources, where the origin of C$_2$H can be understood in terms of 
UV photodissociation. Only weak emission of C$_2$H is seen towards 
G1.87--SMM 38, the G11.36 fi\-lament, and G13.22--SMM 23, 27, and 29. 
In the massive star-forming clumps NGC 6334 E and I, both associated with 
\ion{H}{ii} regions, the absence of C$_2$H emission is suggested to be caused 
by the destruction of the molecules in such harsh environments 
(\cite{walsh2010}). Sanhueza et al. (2013) mapped 
the IRDC G028.23-00.19 in C$_2$H, and found the emission to come from the cold 
central part of the cloud, instead from the outer layers. 
The authors suggested that C$_2$H is tracing the dense and cold gas in their 
IRDC, which seems to be the case also in some of our clumps. 
Beuther et al. (2008) also found that C$_2$H is prevalent during the early 
stages of star formation (and not only in PDRs). They suggested that C$_2$H 
starts to decrease in abundance at the hot-core phase due to transformation to 
other species (e.g., CO from O; see above). Only in the outer layers of the 
source, where UV photons produce elemental C from CO, the abundance of C$_2$H 
can remain high. Furthermore, it was proposed by the authors that C$_2$H could 
be a useful tracer of the initial conditions of high-mass star formation.

The C$_2$H column densities we found, $7.8\times10^{13}-1.2\times10^{15}$ 
cm$^{-2}$ ($4.3\times10^{14}$ cm$^{-2}$ on average), are quite similar to those 
derived by Sakai et al. (2008) towards their IRDC sources 
($5.4\times10^{13}-2.9\times10^{14}$ cm$^{-2}$, the average being 
$1.5\times10^{14}$ cm$^{-2}$). The fractional abundances we found are
$1.8\times10^{-9}-2.4\times10^{-8}$ ($8.1\times10^{-9}$ on average). 
Similarly to us, VLH11 found abundances in the range 
$2.5\times10^{-9}-5.3\times10^{-8}$ ($1.4\times10^{-8}$ on average). 
Sanhueza et al. (2012) derived the values of 
$6.1\times10^{13}-7.75\times10^{14}$ cm$^{-2}$ and 
$7.5\times10^{-9}-1.62\times10^{-7}$ for the column densities and abundances 
of C$_2$H. Their median column density of $2.41\times10^{14}$ cm$^{-2}$ is 
similar to our value, but their median abundance of $3.72\times10^{-8}$ is 
almost six times higher. We found that both the mean and median values of 
the C$_2$H abundance are higher towards IR-dark clumps than towards IR-bright 
clumps, and the lowest C$_2$H abundance ($1.8\times10^{-9}$) is seen towards 
G13.22--SMM 29, the clump associated with IRAS 18117-1738 and probably 
(at least) in the hot-core phase of evolution. This is in accordance with the 
results by Beuther et al. (2008). On the other hand, SJF12 found no clear 
evolutionary trends for $N({\rm C_2H})$ or $x({\rm C_2H})$.

\subsection{HNCO (Isocyanic Acid)}

HNCO is a ubiquitous molecule in the interstellar medium. It was 
detected for the first time over 40 yr ago in Sgr B2 by Snyder \& Buhl (1971).
Since then, HNCO has been detected in, e.g., dark clouds 
(TMC-1; \cite{brown1981}), low-mass protostellar cores (IRAS 16293-2422; 
\cite{vandishoeck1995}), hot cores/UC \ion{H}{ii} regions (e.g., G34.3+0.15; 
\cite{macdonald1996}), translucent clouds (\cite{turner1999}), and outflows 
emanating from massive YSOs (IRAS 17233-3606; \cite{leurini2011}).
Zinchenko et al. (2000) searched for HNCO emission towards a sample of 
81 dense molecular cloud cores, and detected it in 57 (70\%) of them.
Jackson et al. (1984) proposed that HNCO is tracing the densest 
($\gtrsim10^6$ cm$^{-3}$) parts of molecular clouds. Other authors have 
suggested that HNCO could be tracing shocks as, for example, it is found to 
correlate with SiO emission (e.g., \cite{zinchenko2000}; \cite{minh2006}; 
\cite{rodriguez2010}). Association of HNCO clumps with embedded YSOs 
was recently suggested by the Purple Mountain Observatory 13.7-m telescope
observations by Li et al. (2013).

Gas-phase reactions that could be responsible for the formation of HNCO 
include the dissociative recombinations of H$_2$NCO$^+$ 
(\cite{iglesias1977}), H$_3$NCO$^+$, CNH$_3$O$^+$, and CNH$_2$O$^+$, and the 
neutral-neutral reaction between NCO and H$_2$ (\cite{turner1999}). 
HNCO is dominantly destroyed by reactions with H$_3^+$ and He$^+$ 
(\cite{turner1999}). HNCO could also form through grain-surface 
chemistry, namely hydrogenation of accreted OCN (\cite{garrod2008}).
In grain ices containing H$_2$O and NH$_3$, the HNCO molecules can react 
with these species to form OCN$^-$ anions (e.g., \cite{vanbroekhuizen2004}; 
Table~2 therein). More recently, Tideswell et al. (2010) found that gas-phase 
reactions are unlikely to be able to produce HNCO in its observed abundances 
(even at hot-core temperatures). Instead, grain-surface pathways are required, 
although direct evaporation from the icy grain mantles is not sufficent itself. 
More complex molecules formed from HNCO on grain surfaces are expected 
to be evaporated into the gas phase; their dissociation can then lead to the 
formation of HNCO (\cite{tideswell2010}).

The HNCO$(4_{0,\,4}-3_{0,\,3})$ emission is found to be extended in many of 
the clumps (G1.87--SMM 1, 12, 14, 16, 20, 23, 30). Particularly 
in G1.87--SMM 14, the HNCO emission peaks at the position of the strongest dust
emission as traced by LABOCA (Fig.~\ref{figure:G187SMM12lines}). On the other 
hand, towards G1.87--SMM 20, 23, and 30, where the HNCO emission is also 
extended, the emission peak is not coincident with any of the submm peaks 
(Fig.~\ref{figure:G187SMM23lines}). Instead, the HNCO emission has its maximum 
at the position of the SiO, HNC, HC$_3$N, CH$_3$CN, and N$_2$H$^+$ emission 
maxima. We note that there is no \textit{Spitzer} 24-$\mu$m source within 
the Mopra beam in this molecular-line emission peak. Towards G1.87--SMM 1, the 
HNCO emission morphology is quite similar to that of HC$_3$N 
(Fig.~\ref{figure:G187SMM1lines}). Towards the clumps G13.22--SMM 23, 27, and 
32 the HNCO emission is very weak (Figs.~\ref{figure:G1322SMM23lines} and 
\ref{figure:G1322SMM32lines}). 

Based on the $4_{0,\,4}-3_{0,\,3}$ transition, the HNCO column densities 
and fractional abundances are determined to be 
$5.4\times10^{12}-7.9\times10^{14}$ cm$^{-2}$ ($2.2\times10^{14}$ cm$^{-2}$ on 
average) and $1.3\times10^{-10}-1.9\times10^{-8}$ ($6.6\times10^{-9}$ on 
average). The above values should be taken as lower limits only because of 
the assumption made for $T_{\rm ex}$ (Sect.~3.3). Vasyunina et al. (2011) 
derived HNCO abundances of $2.2\times10^{-10}-3.7\times10^{-9}$. 
Their average value, $1.2\times10^{-9}$, 
is 5.5 times lower than ours. Sanhueza et al. (2012) found HNCO column 
densities in the range $\sim9.8\times10^{12}-9.9\times10^{13}$ cm$^{-2}$, 
with the median value of $3.36\times10^{13}$ cm$^{-2}$, i.e., about four times 
lower than our value ($1.4\times10^{14}$ cm$^{-2}$). The authors 
found that the column density increases when the source evolves. Our median 
HNCO column densities agree with this trend, as do the mean and median 
fractional abundances. It is unclear whether our high $N({\rm HNCO})$ values 
could point towards more evolved sources on average compared to those studied 
by SJF12, because the HCO$^+$ data suggest the opposite (Sect.~4.1).
The fractional HNCO abundances found by SJF12 are
$\sim2.7\times10^{-10}-7.6\times10^{-9}$. Their median value, 
$2.8\times10^{-9}$, is only $\sim1.8$ times lower than the value 
$5\times10^{-9}$ we found. In most of the clumps studied by SJF12, the HNCO 
line profiles showed no signatures of shocks. In contrast, we found that 
eleven of our clumps (G1.87--SMM 1, 8, 10, 14, 15, 21, 24, 27, 28, 31, and 38) 
show HNCO wing-emission indicative of outflows (and therefore shocks).

\subsection{HC$_3$N and HC$^{13}$CCN (Cyanoacetylene)}

Cyanopolyynes are organic chemical species that contain a chain of at least one 
C-C triple bonds, alternating with single bonds and ending with a cyanide (CN) 
group. The most simple example of cyanopolyynes is the cyanoacetylene HC$_3$N. 
Its first interstellar detection was made towards Sgr B2 (\cite{turner1971}).
This molecule can trace both the cold molecular clouds (see below) and 
hot cores where it is formed through the gas-phase 
reaction ${\rm C_2H_2}+{\rm CN}\rightarrow {\rm HC_3N}+{\rm H}$ after 
C$_2$H$_2$ is released from the grain mantles due to heating 
(e.g., \cite{chapman2009}). 

The HC$_3$N emission is found to be extended towards many of the observed 
fields. G1.87--SMM 1 and G1.87--SMM 28 are examples where the line emission 
peaks towards the dust emission peak. Similarly, Walsh et al. (2010) found 
that the HC$_3$N emission closely resembles that of the dust continuum emission
in NGC 6334, and suggested that HC$_3$N is a good tracer of quiescent dense 
gas (see also \cite{pratap1997} for the HC$_3$N emission along the 
quiescent TMC-1 ridge). The sources G1.87--SMM 20, 23, 30 are associated with 
a few parsec-scale HC$_3$N clump, but the line emission peak is not coincident 
with any of the submm peaks. Instead, the emission morphology resembles those 
of SiO, HNCO, HNC, and N$_2$H$^+$. Also, the CH$_3$CN emission towards the 
field peaks at the HC$_3$N peak. There is also a small HC$_3$N clump in 
G1.87--SMM 38, and its maximum is coincident with the dust peak.
The clump G13.22--SMM 5, belonging to the N10/11 bubble system, is associated 
with a small HC$_3$N clump. On the other hand, towards G1.87--SMM 8, 10, 
14--17, 21, 24, and G13.22--SMM 23, 27 the HC$_3$N emission is very weak or 
absent. The HC$_3$N line wings detected towards some of our clumps suggest 
the shock-origin for the emission (e.g., the case of the IR-dark clump 
G1.87--SMM 27)

The HC$_3$N column densities and abundances we derive are 
$7\times10^{12}-7.9\times10^{14}$ cm$^{-2}$ and 
$1\times10^{-10}-2.7\times10^{-8}$. The mean values are $1.6\times10^{14}$ 
cm$^{-2}$ and $5\times10^{-9}$. Similarly to the linear SiO molecule, we 
assumed that $T_{\rm ex}=E_{\rm u}/k_{\rm B}$ (Sect.~3.3), so the above values 
should be interpreted as lower limits. Sakai et al. (2008) determined 
$N({\rm HC_3N})$ values in the range $<2.2\times10^{12}-5.4\times10^{13}$ 
cm$^{-2}$ for their sources. Our mean column density exceeds the highest 
value found by Sakai et al. (2008) by a factor of three. Vasyunina et al. 
(2011) derived HC$_3$N abundances of $1\times10^{-10}-1.5\times10^{-9}$ with 
an average value of $5.4\times10^{-10}$. Our average value is almost an order of 
magnitude higher. The $N({\rm HC_3N})$ values derived by SFJ12 are 
$1.4\times10^{12}-3.33\times10^{13}$ cm$^{-2}$ (median $4.77\times10^{12}$ 
cm$^{-2}$). Their median $N({\rm HC_3N})$ is about six times lower than ours 
but the median abundance for their sample, $4.23\times10^{-10}$, is comparable 
to our value of $6.1\times10^{-10}$.

In agreement with the chemical model by Nomura \& Millar (2004), SJF12 found 
that $N({\rm HC_3N})$ increases as a function of clump evolutionary stage, 
but their median abundances did not show such a trend. 
Among the quiescent clumps they studied, HC$_3$N was detected in only 
one source. The trend between $N({\rm HC_3N})$ and evolutionary stage found by 
SJF12 is not seen among our sources. On the other hand, we found that the 
average HC$_3$N abundance is very similar between the IR-dark and 
-bright clumps ($5.1\times10^{-9}$ and $4.7\times10^{-9}$), while the median 
value is about 1.4 times higher towards the latter sources.

The HC$_3$N/HCN ratios are found to lie in the range 
$0.02\pm0.01-0.24\pm0.06$, where the mean$\pm$std is $0.10\pm0.09$ (median 
is 0.07). For comparison, Dickens et al. (2000) found that the HC$_3$N/HCN 
ratio is about 0.05--0.07 in the L134N (L183) prestellar core. In 
comet Hale-Bopp, the above ratio is found to be 8.2\% (\cite{bockelee2000}). 
More recently, Chapillon et al. (2012), who made the first detection of 
HC$_3$N in protoplanetary disks, derived the HC$_3$N/HCN ratios of $\leq0.05$, 
0.075, and 0.55 towards DM Tau, LkCa 15, and MWC 480, respectively. 
Interestingly, the abundance ratio between HC$_3$N and HCN appears to be 
quite similar in IRDC clumps, low-mass prestellar cores, disks around T Tauri 
stars, and comets\footnote{The high HC$_3$N/HCN ratio in the MWC 480 disk is 
an exception due to the warmer dust present, which 
enhances the diffusion of radicals on grain surfaces and leads to a higher 
abundance of solid HC$_3$N (\cite{chapillon2012}).}.

The $^{13}$C isotopologue of HC$_3$N, HC$^{13}$CCN, forms via the isotope 
exchange reaction 
${\rm ^{13}C^+}+{\rm HC_3N}\rightarrow {\rm C^+}+{\rm HC^{13}CCN}$ 
(\cite{takano1998}). No HC$^{13}$CCN emission was detected in any of 
our sources. The $E_{\rm u}/k_{\rm B}$ value of the observed HC$^{13}$CCN 
transition is very similar to that of HC$_3$N ($\sim24$ K), so the 
non-detection is not likely to be caused by excitation conditions. 
The probable reasons are the relative rareness of HC$^{13}$CCN, and the limited 
S/N ratio of mapping observations. A positive detection is expected with a 
longer integration time/single-pointing observation, at least towards the 
strongest HC$_3$N sources. HC$^{13}$CCN was first detected in Sgr B2 by 
Gardner \& Winnewisser (1975), and Gibb et al. (2000) make a detection towards 
the hot core G327.3-0.6. 

\subsection{$^{13}$CS and $^{13}$C$^{34}$S (Carbon Monosulfide)}

Only traces of $^{13}$CS$(2-1)$ emission are seen towards the clump 
G1.87--SMM 1 (Fig.~\ref{figure:G187SMM1lines}), and towards the field 
containing G1.87--SMM 27, 28, and 31 (Fig.~\ref{figure:G187SMM28lines}). 
$^{13}$C$^{34}$S is not detected at all. Beuther \& Henning (2009) found only 
weak $^{13}$CS$(2-1)$ emission towards the IRDCs 19175-4 and -5, and derived 
very high CS depletion factors of $\sim100$. Although CS can suffer from 
depletion onto dust grain surfaces in cold sources, its non-detection in 
the present study is likely caused by limited S/N ratio. Also VLH11 detected 
only very weak $^{13}$CS$(2-1)$ emission in only three sources of their IRDC 
sample, with an average $x({\rm ^{13}CS})$ of $3\times10^{-10}$.

\subsection{CH$_3$CN (Methyl Cyanide)}

Methyl cyanide is a good tracer of warm/hot and dense parts of molecular 
clouds (e.g., \cite{araya2005}; \cite{purcell2006}). 
The formation of CH$_3$CN is likely taking place on the grain surfaces 
during the earliest stages of YSO evolution. The most efficient route for this 
is the reaction between CH$_3$ and CN (e.g., \cite{garrod2008}). 
Later, when the central star starts to heat its surrounding medium, 
CH$_3$CN gets evaporated into the gas phase.
In the gas phase, CH$_3$CN can form when CH$_3^+$ and HCN first form 
H$_4$C$_2$N$^+$ ions through radiative association, and which then 
dissociatively recombine with electrons to form CH$_3$CN molecules and H-atoms 
(e.g., \cite{mackay1999}). Shocks associated with YSO outflows can also be 
responsible for the production of CH$_3$CN (\cite{codella2009}).
The CH$_3$CN molecules can be photodissociated into CH$_3$ and CN 
molecules (\cite{mackay1999}).

Inspection of the CH$_3$CN maps shows that the emission is quite weak in most 
cases (e.g., G1.87--SMM 12, 14, 16, 27, 28, and 31). Towards G1.87--SMM 1, 
the CH$_3$CN emission is extended (although weak) around the submm peak 
(Fig.~\ref{figure:G187SMM1lines}). Interestingly, the clump 
appears to be IR-dark, so perhaps the emission has its origin in 
shocks. Between the submm peaks 
G1.87--SMM 20, 23, and 30, there is a pc-scale CH$_3$CN clump, with the 
emission peak being coincident with the SiO, HNCO, HNC, HC$_3$N, and 
N$_2$H$^+$ maxima (Fig.~\ref{figure:G187SMM23lines}). Also here, the 
shock-origin of CH$_3$CN is supported by the coincidence with the SiO 
and HNCO emission peaks.

The CH$_3$CN column densities and fractional abundances could be derived 
only towards three positions (where the extracted spectra showed visible lines).
These are $5.5\times10^{11}$ cm$^{-2}$ and $\sim10^{-11}$ on average. These are 
only lower limits because we assumed that $T_{\rm ex}=2E_{\rm u}/3k_{\rm B}$ 
(Sect.~3.3). Beuther \& Sridharan (2007) found an order of magnitude higher 
average values for both the column density and abundance towards their IRDC 
sources. However, they used single-pointing observations with the IRAM 30-m 
telescope, while we have used OTF-data with lower S/N ratios. It is also 
possible that our sources are in an earlier stage of evolution and therefore 
show only weak CH$_3$CN emission. For example, Nomura \& Millar (2004) 
modelled the chemistry of the hot core G34.3+0.15, and found that the CH$_3$CN 
column density is at the level we have derived when the source's age is only 
$\lesssim10^3$ yr. Moreover, Vasyunina et al. (2011) detected no CH$_3$CN 
emission towards their IRDCs, and SJF12 found CH$_3$CN emission in only one 
active (extended 4.5-$\mu$m emission in addition to a YSO seen at 24 $\mu$m) 
clump (G034.43 MM1) of their large sample of IRDC sources. The hot cores 
studied by Bisschop et al. (2007) show CH$_3$CN abundances of 
$1.5\times10^{-8}-1.5\times10^{-7}$, much higher than seen towards IRDCs.

\subsection{H41$\alpha$}

The H41$\alpha$ recombination line emission is not detected in this study. 
It is known that some of our clumps do contain \ion{H}{ii} regions, and 
in these partially or fully ionised gas regions protons can capture electrons. 
Therefore, these \ion{H}{ii} regions are expected to emit recombination lines.
The non-detection towards the clumps with \ion{H}{ii} regions is 
likely caused by our low S/N ratio (similarly to HC$^{13}$CCN and $^{13}$CS).
For example, Araya et al. (2005), who used single-pointing 15-m SEST telescope 
observations, detected H41$\alpha$ line emission in nine sites of high-mass 
star formation. More recently, Klaassen et al. (2013) detected H41$\alpha$ 
towards the high-mass star-forming region K3-50A using the CARMA 
interferometer, and suggested that the line is tracing the outflow entrained 
by photoionised gas.

\subsection{N$_2$H$^+$ (Diazenylium)}

Due to its resistance of depletion at low temperatures and high densities, 
N$_2$H$^+$ is an excellent tracer of cold and dense molecular clouds (e.g., 
\cite{caselli2002}). Diazenylium molecules are primarily formed through the 
gas-phase reaction ${\rm H_3^+}+{\rm N_2}\rightarrow {\rm N_2H^+}+{\rm H_2}$.
When there are CO molecules in the gas phase, they can destroy N$_2$H$^+$ 
producing HCO$^+$ (${\rm N_2H^+}+{\rm CO}\rightarrow {\rm HCO^+}+{\rm N_2}$).
When CO has depleted via freeze-out onto dust grains, N$_2$H$^+$ is mainly 
destroyed in the electron recombination 
(${\rm N_2H^+}+{\rm e^-}\rightarrow {\rm N_2}+{\rm H}$ or 
${\rm NH}+{\rm N}$).

The N$_2$H$^+$ emission is found to be extended towards many of our fields 
(e.g., the fields covering G1.87--SMM 1; G1.87--SMM 12, 14, 16; G1.87--SMM 
20, 23, 30; G1.87--SMM 27, 28, 31; G1.87--SMM 38; G13.22--SMM 23, 27; 
G13.22--SMM 32). The N$_2$H$^+$ emission is also tracing well the G11.36 
filament as seen by LABOCA at 870 $\mu$m and in absorption by \textit{Spitzer} 
at 8 $\mu$m (Fig.~\ref{figure:G1136lines}). Very strong and extended 
N$_2$H$^+$ emission is seen around the N10/11 double bubble, and the emission 
peaks are coincident with the submm peaks G13.22--SMM 5 and 7 
(Figs.~\ref{figure:G1322SMM5lines} and \ref{figure:G1322SMM7lines}).
Also in G1.87--SMM 38 the line emission peaks at the dust peak 
(Fig.~\ref{figure:G187SMM38lines}). 
Interestingly, towards G1.87--SMM 20, 23, 30, the N$_2$H$^+$ emission peaks 
in between the submm dust emission maxima, where also line emission 
from several other species has its maximum as discussed earlier. Also in 
G13.22--SMM 23, 27, and 32 the N$_2$H$^+$ emission peaks are offset from the 
submm peaks by $17\arcsec$, $15\farcs9$, and $30\farcs5$, respectively.
The pc-scale N$_2$H$^+$ clump in G2.11--SMM 5 has its 
maximum about $11\farcs6$ (0.42 pc) in projection from the strongest dust 
emission. Ragan et al. (2006) and Liu et al. (2013), 
who mapped IRDCs in the $J=1-0$ line of N$_2$H$^+$, also found good 
correspondence between the line emission and 8-$\mu$m absorption.

We found N$_2$H$^+$ column densities in the range of 
$6.6\times10^{12}-5.4\times10^{14}$ cm$^{-2}$, and abundances of 
$2.8\times10^{-10}-9.8\times10^{-9}$. The average values are $5.4\times10^{13}$ 
cm$^{-2}$ and $1.6\times10^{-9}$, respectively. Ragan et al. (2006) derived 
comparable abundances of $5.7\times10^{-11}-2\times10^{-9}$ for their sample 
of IRDCs. The column density range found by Sakai et al. (2008) for their IRDCs,
$\sim4.3\times10^{12}-1.3\times10^{14}$ cm$^{-2}$, is also very similar to our 
range of values. Vasyunina et al. (2011) derived fractional N$_2$H$^+$ 
abundances of $1.9\times10^{-10}-8.5\times10^{-9}$ with an average of 
$2.8\times10^{-9}$, remarkably similar to our values. More recently, SJF12 
derived column densities of $3.6\times10^{12}-1.37\times10^{14}$ cm$^{-2}$ 
(median is $1.6\times10^{13}$ cm$^{-2}$) and abundances in the range 
$1.9\times10^{-10}-1.68\times10^{-8}$ (the median being $2.4\times10^{-9}$), 
again similar to our results. They also found that both the values of 
$N({\rm N_2H^+})$ and $x({\rm N_2H^+})$ increase as the clump evolves from 
quiescent (IR-dark) to star-forming stage. Similarly, Hoq et al. 
(2013) found that $x({\rm N_2H^+})$ rises as a function of 
evolutionary stage (their Fig.~5). The reason for such a correlation 
is somewhat unclear because of the destroying effect of CO (see above), and 
in contrast, our clumps show the opposite trend. Sanhueza et al. (2012) 
suggested that the rate coefficients of the relevant chemical reactions 
might be inaccurate, and/or the large beam size of Mopra observations is seeing 
the cold N$_2$H$^+$ gas around the warmer central YSOs, thus leading to their 
observed correlation. On the other hand, higher temperature could lead to an 
enhanced evaporation of N$_2$ from the dust grains, thus allowing N$_2$H$^+$ 
to increase in abundance (\cite{chen2013}). Finally, Liu et al. (2013) derived 
N$_2$H$^+$ column densities of $\sim3.3\times10^{12}-2.7\times10^{13}$ cm$^{-2}$ 
with the mean value being $\sim9.4\times10^{12}$ cm$^{-2}$ -- a factor of 
about six lower than ours.

The N$_2$H$^+$/HCO$^+$ ratios we derived lie in range 
$0.14\pm0.08-18.57\pm3.41$, i.e., the ratio varies a lot. The mean$\pm$std 
is $4.48\pm4.26$, and the median value is 3.36. Separately for IR-dark and 
-bright clumps, these values are $6.59\pm5.82$ and 3.53, and $8.49\pm22.61$ 
and 2.54, respectively. Due to the large dispersion in these values, there is 
no clear correlation between the clump evolutionary stage and the value 
of the N$_2$H$^+$/HCO$^+$ ratio among our source sample. However, one 
would expect to observe a higher ratio of N$_2$H$^+$/HCO$^+$ towards clumps in 
the earliest stages of evolution, which is actually mildly suggested 
by our median values. This is because CO is expected to be depleted in 
starless clumps, so that HCO$^+$ cannot form via the reaction between 
${\rm H_3^+}$ and ${\rm CO}$. When the source evolves, it gets warmer and CO 
should be evaporated from the dust grains when the dust temperature exceeds 
about $\sim20$ K (see \cite{tobin2013}). Two effects can then follow: 
the CO molecules can start to form HCO$^+$, and destroy N$_2$H$^+$. 
As a consequence, the N$_2$H$^+$/HCO$^+$ ratio should decrease. 
Sanhueza et al. (2012) indeed found that the N$_2$H$^+$/HCO$^+$ 
abundance ratio decreases from intermediate to active and red clumps, 
and therefore acts as a chemical clock in accordance with the scenario 
described above. However, the quiescent clumps studied by SJF12 do not follow 
this trend. As an explanation for this, the authors suggested, for example, 
that some of them could contain YSOs not detected by \textit{Spitzer}, or that 
CO has not had enough time to deplete from the gas phase. 
During the starless phase of evolution, however, the 
N$_2$H$^+$/HCO$^+$ ratio is expected to increase (CO gets more and more 
depleted and cannot destroy N$_2$H$^+$). Hoq et al. (2013) also inspected 
whether there is an evolutionary trend in the N$_2$H$^+$/HCO$^+$ abundance 
ratio, and the median value appeared to be slightly \textit{higher} in more 
evolved clumps, although, as the authors stated, the result is not 
statistically significant (their Fig.~6). Hoq et al. (2013) 
employed the chemical evolution model by Vasyunina et al. 
(2012) to further investigate the behaviour of the N$_2$H$^+$/HCO$^+$ ratio. 
As shown in their Fig.~10, the abundance ratio appears to increase as a 
function of time until the peak value is reached at $\sim10^4$ yr.

We also derived the N$_2$H$^+$/HNC ratios, ranging from $0.08\pm0.04$ to 
$4.32\pm7.64$. The mean$\pm$std is $1.26\pm1.07$, and the median value is 
0.87. For IR-dark clumps these values are $1.20\pm1.17$ and 0.72, while 
for IR-bright clumps they are $1.26\pm1.05$ and 1.17.
The median N$_2$H$^+$/HNC ratio found by SJF12 for all their clumps 
(0.07) is comparable to the lowest values we found, while our median 
is over 12 times higher. For the IRDC sample 
examined by Liu et al. (2013), the ratio ranges from $\sim0.54-3.86$ with 
the mean$\pm$std of $1.20\pm0.79$ and median of 1.07 (computed from the 
values in their Table~4). These are comparable with our values.
Sanhueza et al. (2012) suggested that there is a trend of increasing 
N$_2$H$^+$/HNC ratio with the clump evolution, but the change in their median 
value is very small, only from 0.06 for quiescent clumps to 0.08 for active 
and red clumps. Our clumps show a similar, except stronger, trend in median 
values (also the average ratios suggest this trend).
Sanhueza et al. (2012) proposed that HNC could be preferentially formed during 
the cold phase, while their median N$_2$H$^+$ abundance appeared to increase 
when the source evolves. We found that there is a positive N$_2$H$^+$ -- HCN 
abundance correlation, as shown in the middle right panel of 
Fig.~\ref{figure:correlations}. A weaker, although possible positive 
correlation, is also found between N$_2$H$^+$ and HNC 
(Fig.~\ref{figure:correlations}, bottom panel). Accoring to our and 
Liu et al. (2013) results, the N$_2$H$^+$/HNC ratio is near unity on average, 
although the std is very large. If the increasing trend in $x({\rm HNC})$ as a 
function of $x({\rm N_2H^+})$ is real, the abundance ratio is not expected to 
change significantly, which conforms to the results by SJF12 and of the 
present study.

\section{Summary and conclusions}
   
Altogether 14 subfields from the LABOCA 870-$\mu$m survey of IRDCs 
by Miettinen (2012b) are covered by the MALT90 molecular-line survey. 
These IRDC fields contain 35 clumps in total, ranging from quiescent (IR-dark) 
sources to clumps associated with \ion{H}{ii} regions. In the present study, 
the MALT90 observations were used to investigate the chemical properties of the 
clumps. Our main results and conclusions are summarised as follows:

\begin{enumerate}

\item Of the 16 transitions at $\lambda \approx3$ mm included in the MALT90 
survey, all except five [HNCO$(4_{1,\,3}-3_{1,\,2})$, HC$^{13}$CCN$(10-9)$, 
$^{13}$C$^{34}$S$(2-1)$, H$41\alpha$, $^{13}$CS$(2-1)$] are detected towards 
our sources. 

\item The HCO$^+(1-0)$ emission is extended in many of the clumps, resembling 
the MALT90 mapping results of IRDCs by Liu et al. (2013). The fractional 
HCO$^+$ abundances appear to be lower than in the IRDC clumps studied by 
Vasyunina et al. (2011) and Sanhueza et al. (2012), who also used the Mopra 
telescope observations. We found that the average HCO$^+$ abundance increases 
when the clump evolves (the median HCO$^+$ abundance is very similar between 
the IR-dark and -bright clumps), resembling the trend discovered by Sanhueza 
et al. (2012) and Hoq et al. (2013). The H$^{13}$CO$^+(1-0)$ emission 
is generally weak in our clumps, and its average abundance is a factor of 3.5 
lower than in the sources of Vasyunina et al. (2011).

\item Extendend or clump-like SiO$(2-1)$ emission is seen towards several 
clumps. In three cases, the maximum of the integrated intensity of SiO 
is coincident with the 
LABOCA 870-$\mu$m peak position. As supported by the observed line wings, 
SiO emission is likely caused by outflow activity which generates shocks 
releasing SiO into the gas phase. However, some of the widespread SiO 
emission could result from shocks associated with cloud formation as suggested 
in the IRDC G035.39-00.33 (\cite{jimenez2010}). The SiO abundances derived for 
our clumps are mostly comparable to those seen in other IRDCs 
(\cite{vasyunina2011}; \cite{sanhueza2012}). No trend of decreasing SiO 
abundance with the clump evolution is seen, as suggested by some earlier 
studies of massive clumps (\cite{miettinen2006}; \cite{sakai2010}). 
Instead, our data suggest the opposite trend, as seen in the data by Sanhueza 
et al. (2012).

\item The $J=1-0$ line emission of HCN and HNC is generally found to be 
spatially extended. The fractional abundances of these species are mostly 
comparable to those found by Vasyunina et al. (2011), but Sanhueza et al. 
(2012) derived clearly higher HNC abundances. We found no evidence of 
increasing HNC column density as the source evolves as the latter authors did.
However, our average HNC abundance (but not the median value) appears to be 
slightly higher in more evolved sources. This is perhaps related to the 
accumulation of gas from the surrounding mass reservoir (\cite{sanhueza2012}). 
We found a hint that the HCN abundance gets lower when the column density of 
molecular hydrogen becomes higher. This might be related to the enhanced 
HCO$^+$ abundance in more evolved sources, where the species can destroy HCN 
molecules to form H$_2$CN$^+$ and CO. The HNC abundance is found to increase 
as a function of the HCO$^+$ abundance. This is likely related to the 
increased production of H$_2$CN$^+$ -- a molecular ion that forms HNC in the 
dissociative recombination with an electron. The HCN fractional abundance 
appears to increase when that of HNC increases, in agreement with the result 
by Liu et al. (2013). In this case, however, the decrease in $x({\rm HCN})$ 
as a function of time is questionable, if HNC gets more abundant as the clump 
evolves. The HNC/HCN ratio is found to lie in the range 
$\sim0.01-6.38$ with an average value near unity, as seen in other IRDCs 
(\cite{vasyunina2011}; \cite{liu2013}). It is also in agreement with the 
theore\-tical prediction (\cite{sarrasin2010}). The gas kinetic temperature 
measurements would be needed to study how the HNC/HCN ratio depends on the 
temperature. Moreover, the average 
HNC/HCO$^+$ and HN$^{13}$C/H$^{13}$CO$^+$ ratios, $\sim4.3\pm3.9$ and 
$\sim1.8\pm1.1$, are in reasonable agreement with the gas-phase chemical 
models which suggest similar abundances for HNC and HCO$^+$ and their $^{13}$C 
isotopologues in the case of weak $^{13}$C fractionation (see, e.g., 
\cite{roberts2012} and references therein). The median HNC/HCO$^+$ ratio is 
found to decrease as the clump evolves, in agreement with data from Sanhueza 
et al. (2012).

\item The C$_2$H$(N=1-0)$ emission is also found to be extended. This molecule 
is believed to be a good tracer of the PDR regions. Towards the IR-binary 
bubble system N10/11 and the clump G13.22--SMM 32, the detected 
extended C$_2$H emission is likely related to the UV photodissociation 
process. Recently, C$_2$H is found to trace cold gas in the IRDC G028.23-00.19 
by Sanhueza et al. (2013), and this also appears to be the case in some of our 
sources. The fractional C$_2$H abundances are found to be comparable with 
the values obtained by Vasyunina et al. (2011), but the median abundance 
derived by Sanhueza et al. (2012) exceeds our value by a factor of $\sim6$. 
We found that the average and median C$_2$H abundances are lower towards more 
evolved clumps, some of which are likely to be in the hot-core phase or 
even more evolved. This agrees with the suggestion that C$_2$H starts to 
decrease in abundance in hot cores, and could therefore be used to probe the 
initial conditions of massive-star formation (\cite{beuther2008}). 

\item The HNCO$(4_{0,\,4}-3_{0,\,3})$ line emission shows extended morphology 
towards many of our sources. The average HNCO abundance we derive is a factor 
of 5.5 times higher than the value obtained by Vasyunina et al. (2011). 
However, the median value we found is comparable to the one derived by 
Sanhueza et al. (2012) (within a factor of $\sim1.8$). 
According to models, grain-surface chemistry appears to be required for the 
origin of gas-phase HNCO (\cite{tideswell2010}). 
The extended-like emission of HNCO could have its origin in shocks, as 
suggested in some other molecular clouds (e.g., \cite{zinchenko2000}). In some 
cases, this is supported by the similarity to the SiO emission, and also by 
non-Gaussian line wing emission. Our HNCO data support the discovery 
that the molecule's column density increases as the clump evolves 
(\cite{sanhueza2012}); also the fractional abundance shows such a trend in the 
present study.

\item The $J=10-9$ emission of cyanoacetylene (HC$_3$N) shows extended 
morphology towards many of our clumps. In some cases the emission maxima 
correlate well with the dust emission peaks. However, in some sources the 
HC$_3$N emission is found to be weak or completely absent. The deteced line 
wings suggest that the HC$_3$N emision has its origin in shocks. On average, 
we derived an order of magnitude higher abundance of HC$_3$N than what 
Vasyunina et al. (2011) determined for their sources. Our median HC$_3$N 
abundance is very close to that derived by Sanhueza et al. (2012; within a 
factor of $\sim1.4$). The latter authors found that the column density of 
HC$_3$N increases as the clump evolves, but our column density data do not 
show such a trend. Only the median abundance shows a hint of positive 
correlation with the clump evolution. Our results support the finding that, 
besides tracing hot cores, HC$_3$N can also exist in cold molecular clouds 
(e.g., in the prestellar core L183; \cite{dickens2000}). 
The HC$_3$N/HCN ratios are derived to be 
$0.02\pm0.01-0.24\pm0.06$ with an average of $\sim0.1$. Interestingly, 
this is similar to what has been detected in low-mass starless cores 
(\cite{dickens2000}), T Tauri disks (\cite{chapillon2012}), and comets 
(\cite{bockelee2000}).

\item The CH$_3$CN$(5_1-4_1)$ emission is found to be weak in the three 
sources where it was detected. The fractional abundance, estimated 
in these three targets, was found to be $\sim10^{-11}$ on average. 
As CH$_3$CN is believed to be a hot-core tracer, it is possible that our 
clumps are mostly too cold (i.e., young) to produce any significant CH$_3$CN 
emission. This agrees with the very low detection rates of CH$_3$CN by 
Vasyunina et al. (2011; no sources) and Sanhueza et al. (2012; one source 
with active star formation manifested in 4.5- and 24-$\mu$m emission). 
Shock-origin is possible for G1.87--SMM 1 (line-wing emission) and G1.87--SMM 
20, 23, 30 (peaks at the SiO maximum).

\item The $J=1-0$ emission of N$_2$H$^+$ is also extended in the clumps we have 
studied. For example, the line emission traces well the filamentary IRDC G11.36.
The N$_2$H$^+$ abundances we derived are comparable to those obtained by Ragan 
et al. (2006), Vasyunina et al. (2011), and Sanhueza et al. (2012). The 
correlation found by the latter authors and by Hoq et al. (2013), 
i.e., that the N$_2$H$^+$ abundance 
increases as the source evolves, is not recognised in our sample. Instead, the 
opposite trend is manifested in both the average and median abundances. The 
correlation found by Sanhueza et al. (2012) is also difficult to explain 
because gas-phase CO should destroy the N$_2$H$^+$ molecules as discussed by 
the authors. It could be related to an enhanced N$_2$ abundance, 
however. The derived N$_2$H$^+$/HCO$^+$ ratios are in the 
range $0.14\pm0.08-18.57\pm3.41$, but no evolutionary trend in this parameter 
is found. From a theoretical point of view, the N$_2$H$^+$/HCO$^+$ ratio 
should decrease when CO is evaporated off the grain mantles, starting to 
destroy N$_2$H$^+$ and produce more HCO$^+$. The result by Sanhueza et al. 
(2012), i.e., that the median N$_2$H$^+$/HCO$^+$ ratio decreases slightly as 
the clump evolves further from the so-called intermediate stage (associated 
with an extended 4.5-$\mu$m emission or a 24-$\mu$m point source), is in 
agreement with the above scheme, but in contrast to that of Hoq et al. 
(2013). The N$_2$H$^+$/HNC is found to be near unity 
on average in both IR-dark and -bright clumps of our sample. 
Sanhueza et al. (2012) found generally lower values, 
while similar values can be calculated from the Liu et al. (2013) data of 
IRDCs. We found that the N$_2$H$^+$/HNC abundance ratio icreases slightly as 
the clump evolves (our median value changes from 0.72 for IR-dark sources to 
1.17 for IR-brigth clumps). This is in agreement with the discovery by 
Sanhueza et al. (2012). Our data suggest that as the N$_2$H$^+$ abundance 
increases, the abundances of both HCN and HNC also increase. This conforms 
with the small observed change in the N$_2$H$^+$/HNC ratio as the clump 
evolves further. 

\end{enumerate}

\begin{acknowledgements}
    
This paper is dedicated to the memory of S.~Miettinen. 
I would like to thank the anonymous referee for his/her comments
and suggestions. The author acknowledges the Academy of Finland for the 
financial support through grant 132291. This research has made use of data 
products from the Millimetre Astronomy Legacy Team 90 GHz (MALT90) survey, 
NASA's Astrophysics Data System, and the NASA/IPAC Infrared Science Archive, 
which is operated by the JPL, California Institute of Technology, under 
contract with the NASA. This research made use of APLpy, an open-source 
plotting package for Python hosted at {\tt http://aplpy.github.com}.

\end{acknowledgements}

\begin{appendix} 

\section{Kinematic distances and physical properties of the clumps}

The kinematic distances of the clumps listed in Col.~(4) of Table~\ref{clumps} 
are based on the Galactic rotation curve model by Reid et al. (2009). As 
the sources are associated with IRDCs, it is assumed that they lie at
the near distance in which case there is more IR background radiation against 
which to see the source in absorption. 
The clump distances were adopted 
from Paper I with the following exception. In Paper I, for the first quadrant 
field G1.87 we had radial velocity data only towards a filamentary cloud near 
the field centre (see Figs.~1 and 5 in Paper I). As discussed in Paper I, the 
negative radial velocity of the cloud ($\sim -41$ km~s$^{-1}$) suggests that it 
lies at the far distance. The far distance solution ($\sim10.6$ kpc) was 
therefore adopted for all the clumps in the field. In the present paper, we 
used the MALT90 N$_2$H$^+(1-0)$ radial velocity data to determine the G1.87 
clump distances. As a high-density gas tracer, N$_2$H$^+$ is expected to be 
well-suited for this purpose. Moreover, N$_2$H$^+(1-0)$ was detected towards 
all sources in the present study. The radial velocities of the clumps were 
typically found to be $\sim40-50$ km~s$^{-1}$ (i.e., \textit{positive} rather 
than negative), and the corresponding near kinematic distances were derived 
to be $\sim7$ kpc as shown in Table~\ref{clumps}. We note that the near-far 
kinematic distance ambiguity towards G13.22--SMM 29 (IRAS 18117-1738) was 
resolved by Sewilo et al. (2004). This source was placed at the far distance 
because H$_2$CO absorption was seen between the source velocity and the 
velocity at the tangent point. 

The clump effective radii listed in Table~\ref{clumps} correspond to the 
kinematic distances explained above. The rest of the clump physical properties 
listed in Table~\ref{clumps} were revised from those presented in Paper I by 
making the following modifications. The masses and densities of the clumps 
were previously calculated by assuming that the dust-to-gas mass ratio is 1/100.
However, this value refers to the cano\-nical dust-to-hydrogen mass ratio,
$M_{\rm dust}/M_{\rm H}$ (e.g., \cite{draine2011}; Table~23.1 therein). Assuming 
that the clumps' chemical composition is similar to the solar mixture, i.e., 
the mass fractions for hydrogen, helium, and heavier elements are $X=0.71$, 
$Y=0.27$, and $Z=0.02$, respectively, the ratio of total mass (H+He+metals) 
to hydrogen mass is $1/X\simeq1.41$. The total dust-to-gas mass ratio is 
therefore $M_{\rm dust}/M_{\rm gas} = M_{\rm dust}/(1.41M_{\rm H}) = 1/141$. 
For the assumed gas composition, the mean molecular weight per H$_2$ 
molecule, needed in the calculation of the column and number densities, is 
$\mu_{\rm H_2}\simeq2.82$ (\cite{kauffmann2008}; Appendix~A.1 therein). As 
explained in Paper I, the dust temperature was assumed to be $T_{\rm dust}=15$ K 
for IR-dark clumps, and 20 K for clumps associated with \textit{Spitzer} IR 
emission. For G2.11--SMM 5 and G13.22--SMM 29, which are associated with 
\textit{IRAS} sources, the dust colour temperatures were derived to be 30 and 
18.9 K, respectively (Paper I). For G13.22--SMM 32, which is 
associated with an \ion{H}{ii} region, we adopted the value 
$T_{\rm dust}=30$ K.\footnote{Recently, Hoq et al. (2013) showed that 
the median dust temperature of the MALT90 clumps increases as a function of the 
stage of evolution (their Fig.~2). For quiescent (IR-dark) clumps, clumps 
containing YSOs, and \ion{H}{ii}/PDR sources the median dust temperatures were 
determined to be 13.9, 17.9, and 26.0 K, respectively. These are comparable to 
the values we have adopted.}
Finally, the dust opa\-city per unit dust mass at 870 $\mu$m was taken to be 
$\kappa_{870}=1.38$ cm$^2$~g$^{-1}$ -- the value interpolated
from the widely used Ossenkopf \& Henning (1994) model describing 
graphite-silicate dust grains that have coagulated and accreted thin ice 
mantles over a period of $10^5$ yr at a gas density of $10^5$ cm$^{-3}$.

\section{Maps of spectral-line emission}

\begin{figure*}
\begin{center}
\includegraphics[width=0.245\textwidth]{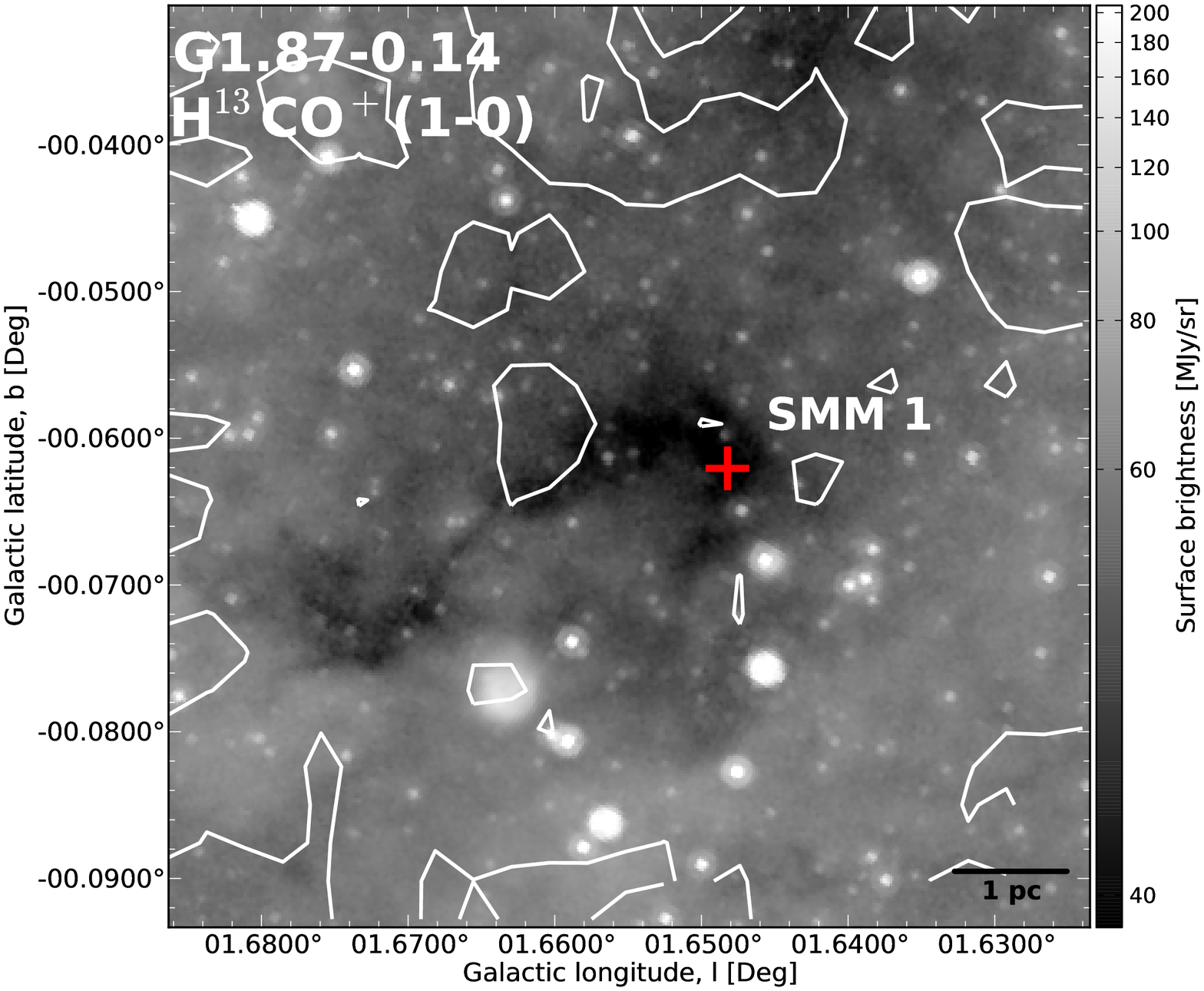}
\includegraphics[width=0.245\textwidth]{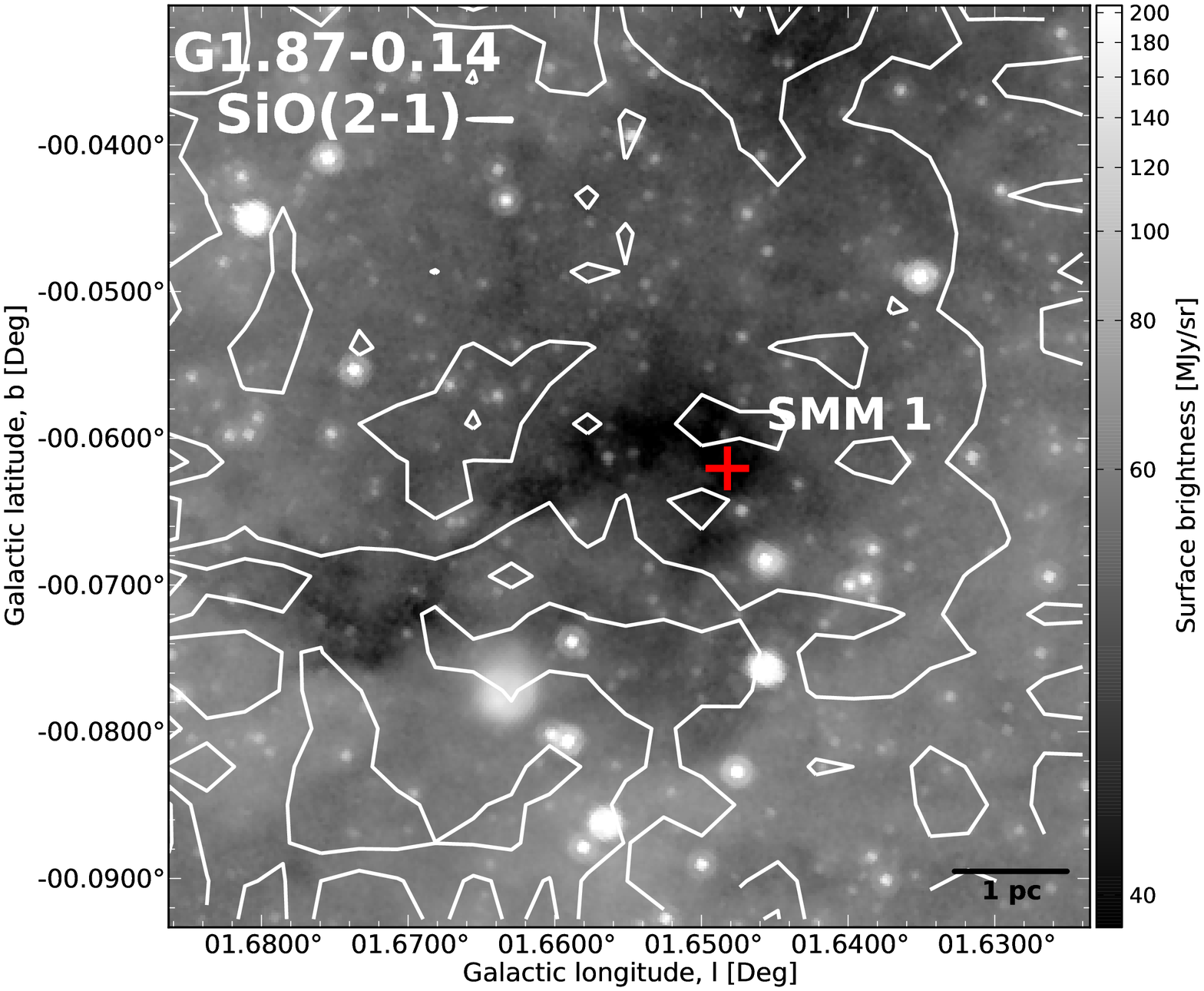}
\includegraphics[width=0.245\textwidth]{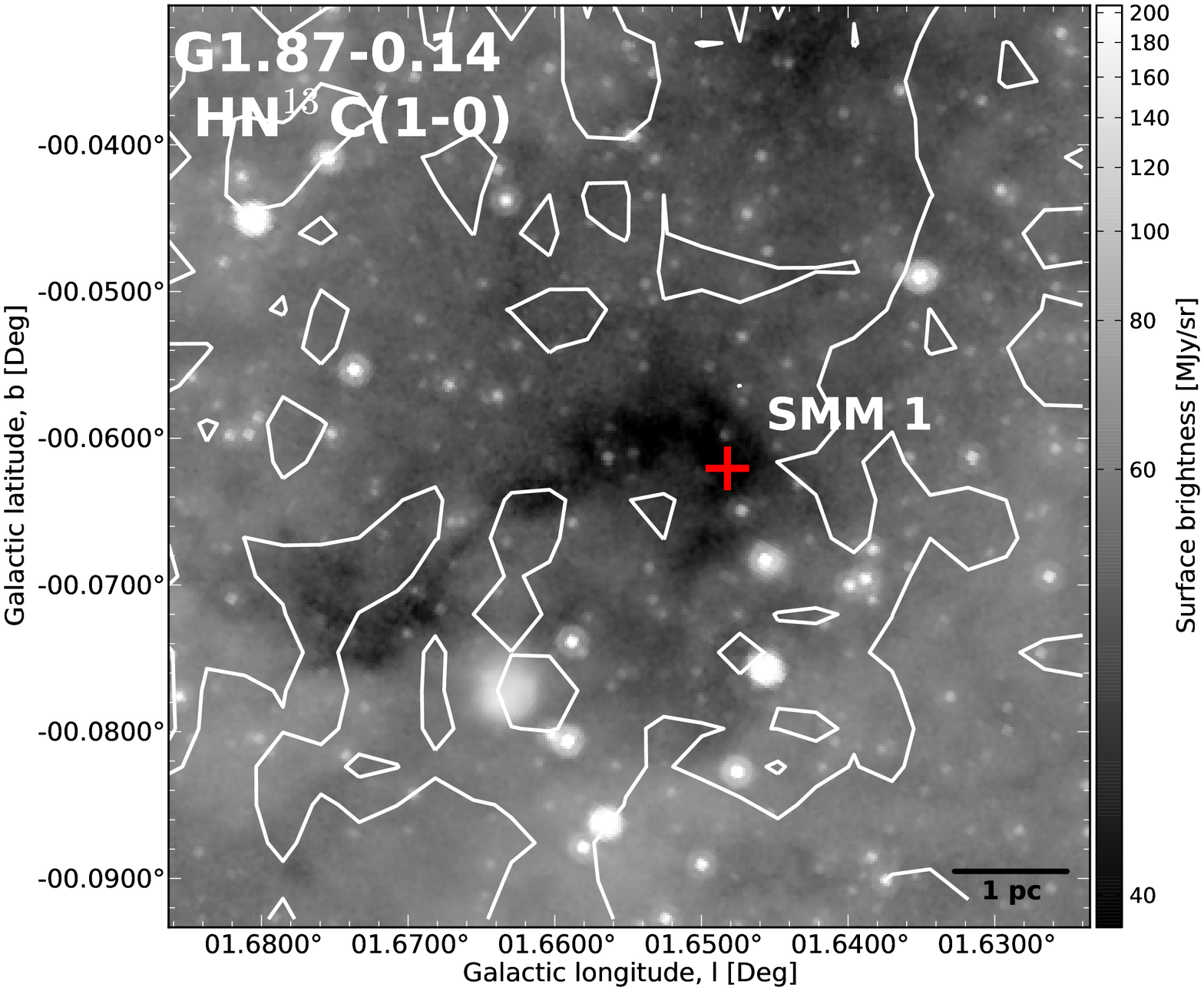}
\includegraphics[width=0.245\textwidth]{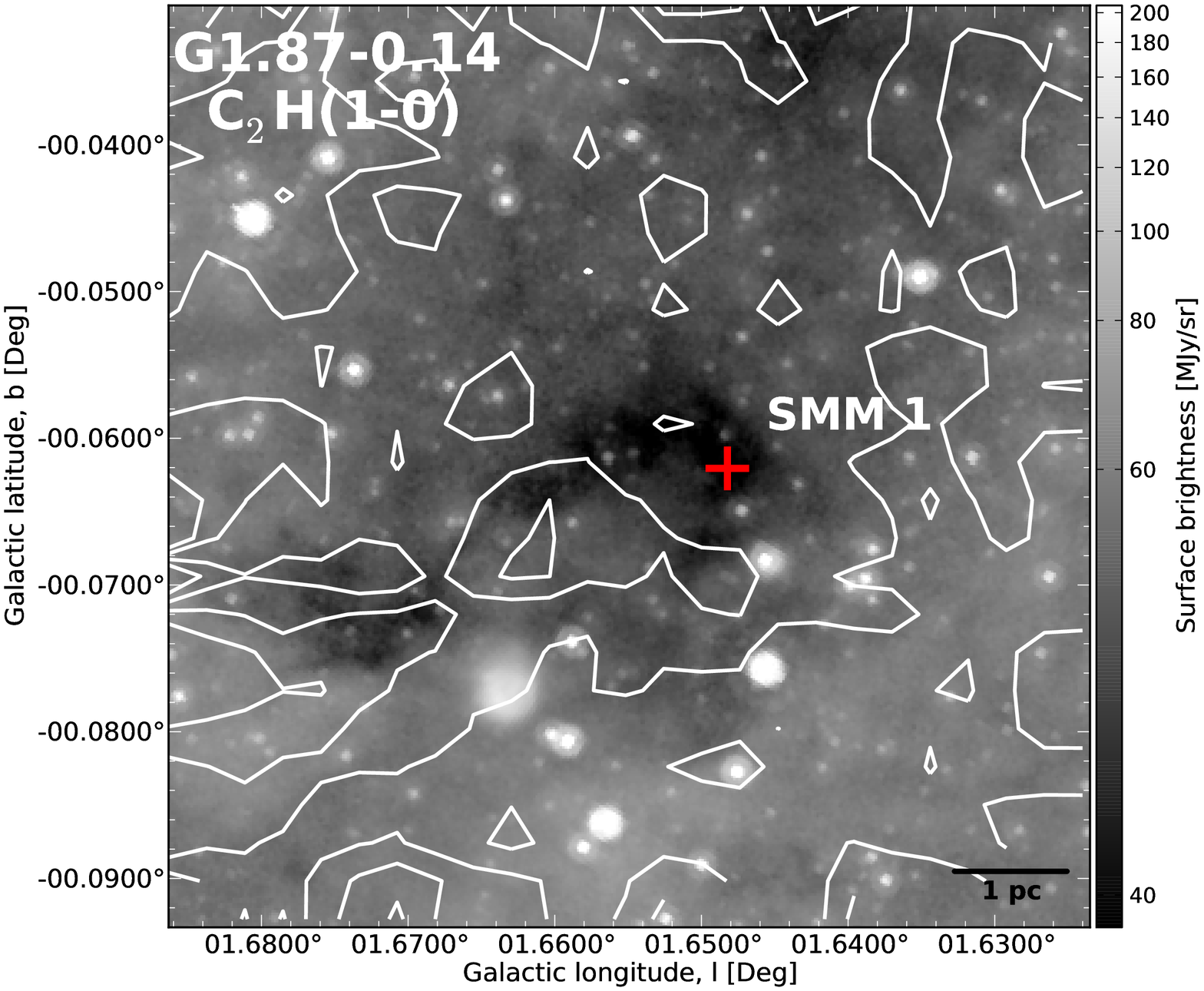}
\includegraphics[width=0.245\textwidth]{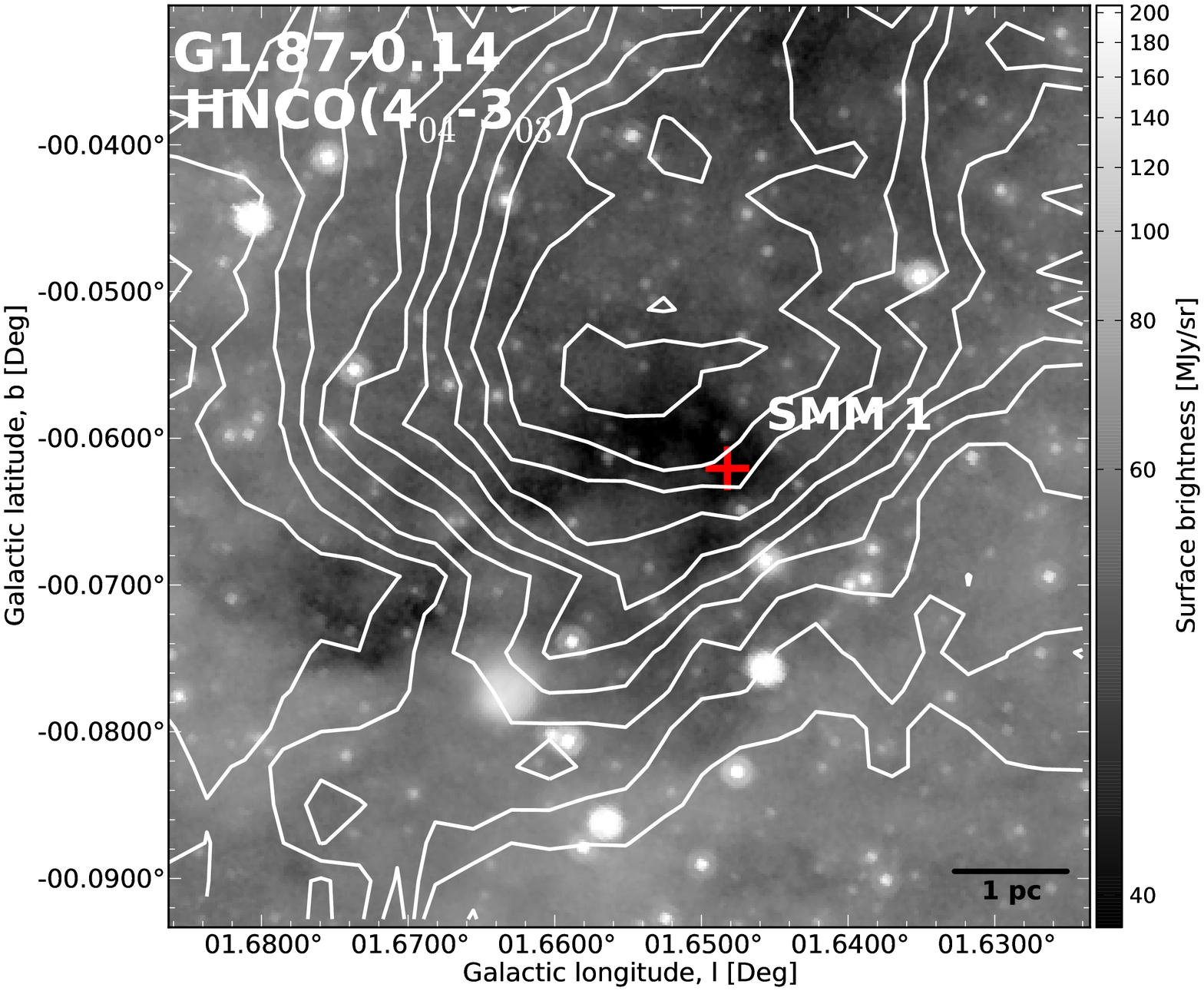}
\includegraphics[width=0.245\textwidth]{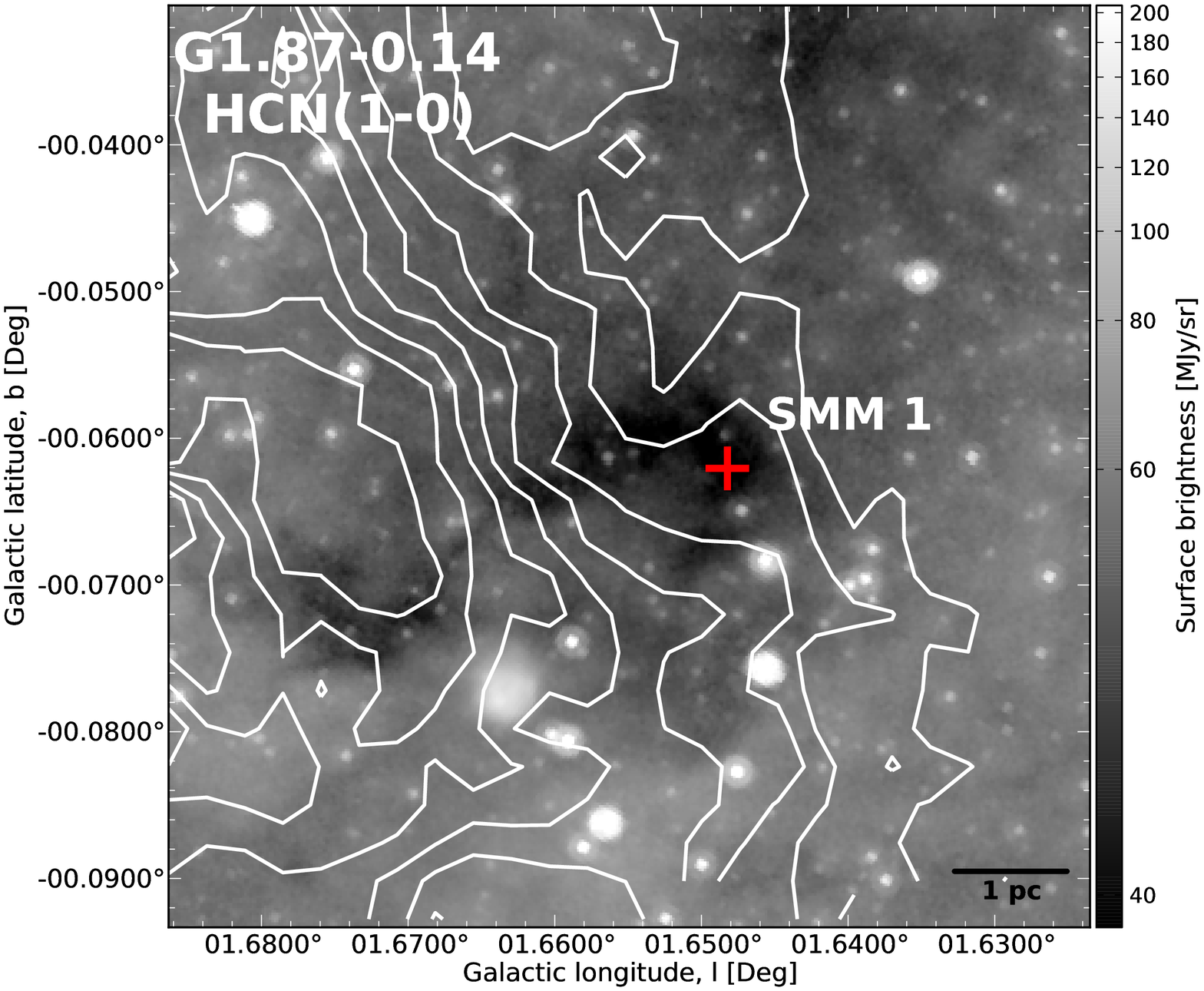}
\includegraphics[width=0.245\textwidth]{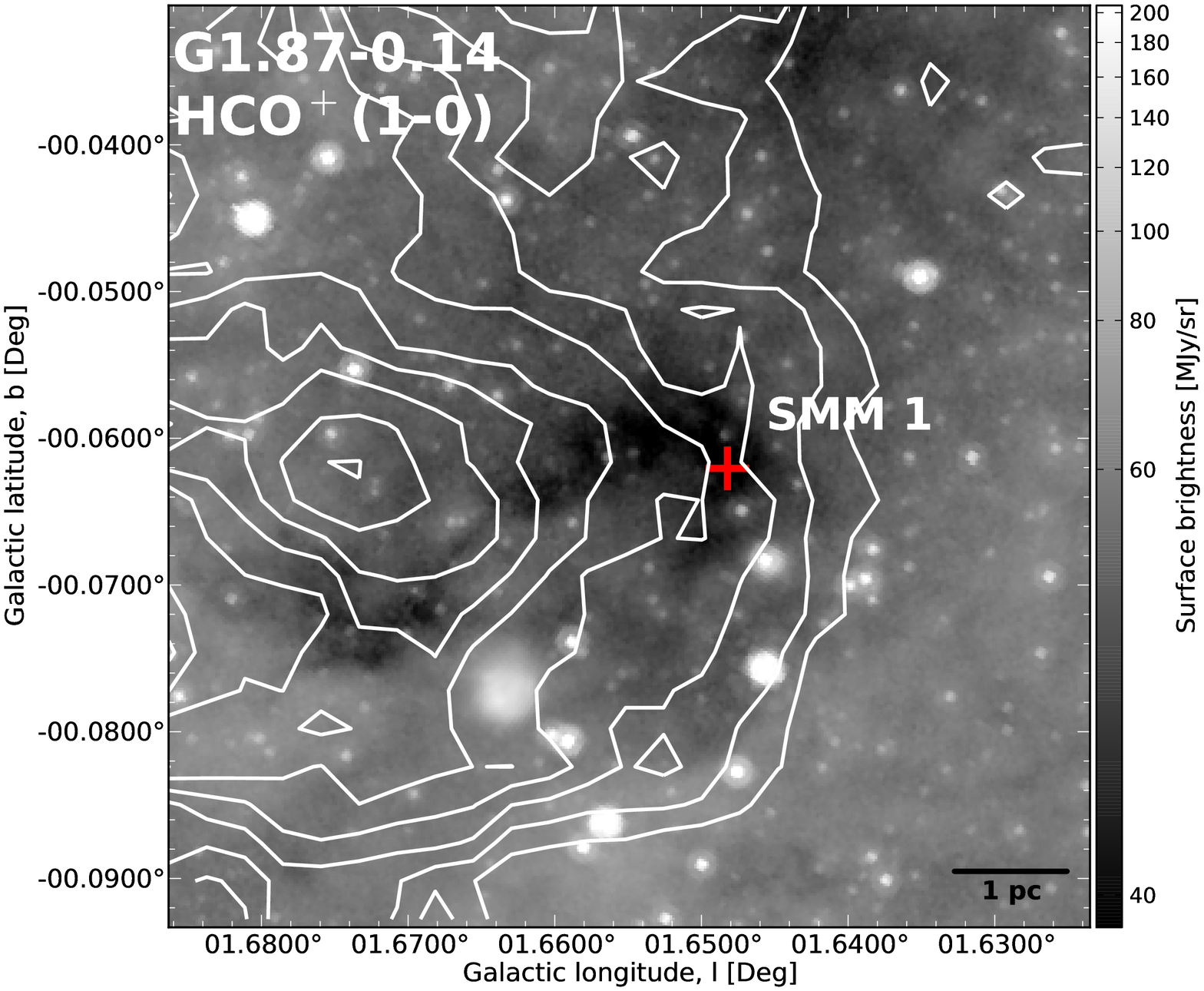}
\includegraphics[width=0.245\textwidth]{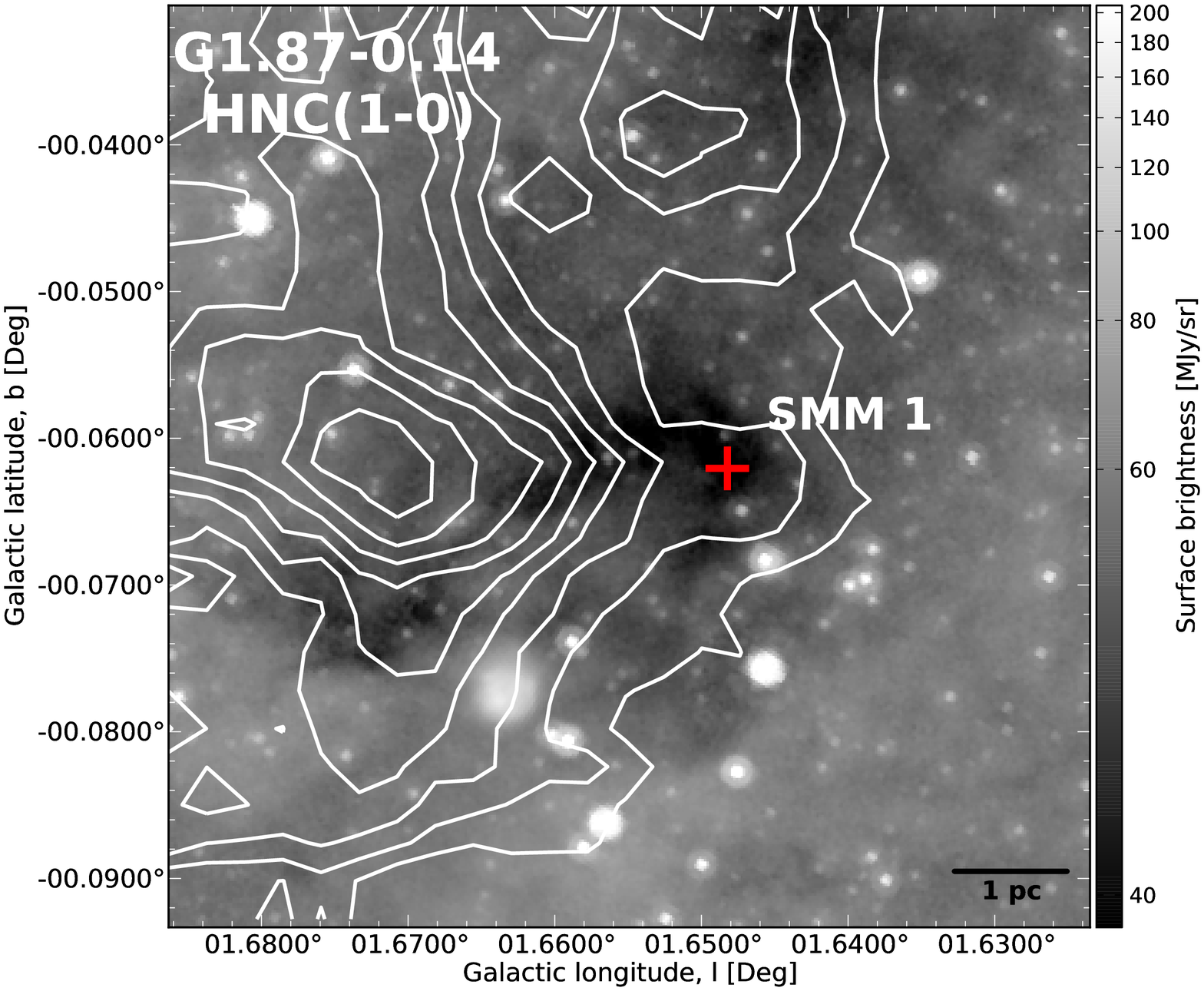}
\includegraphics[width=0.245\textwidth]{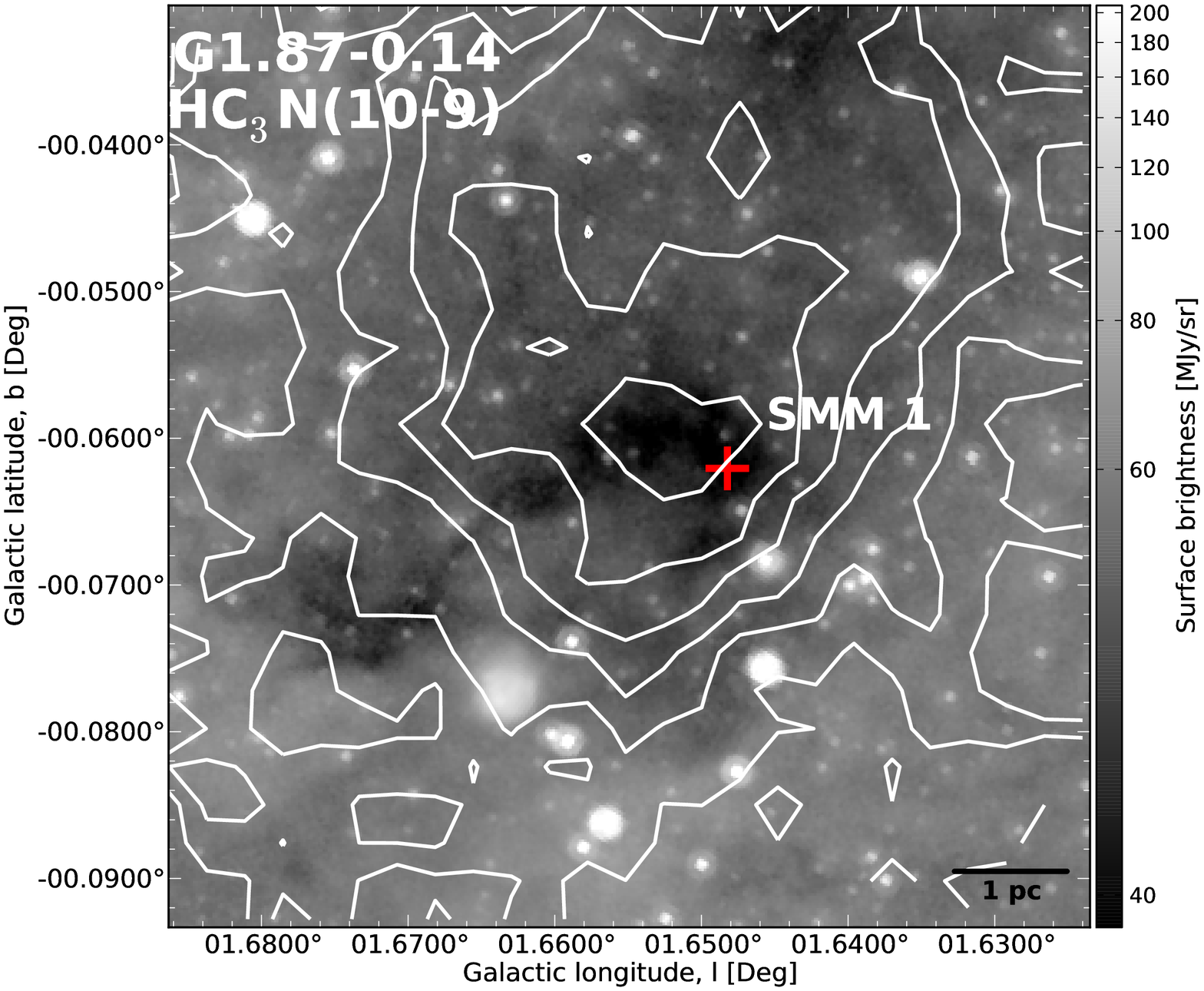}
\includegraphics[width=0.245\textwidth]{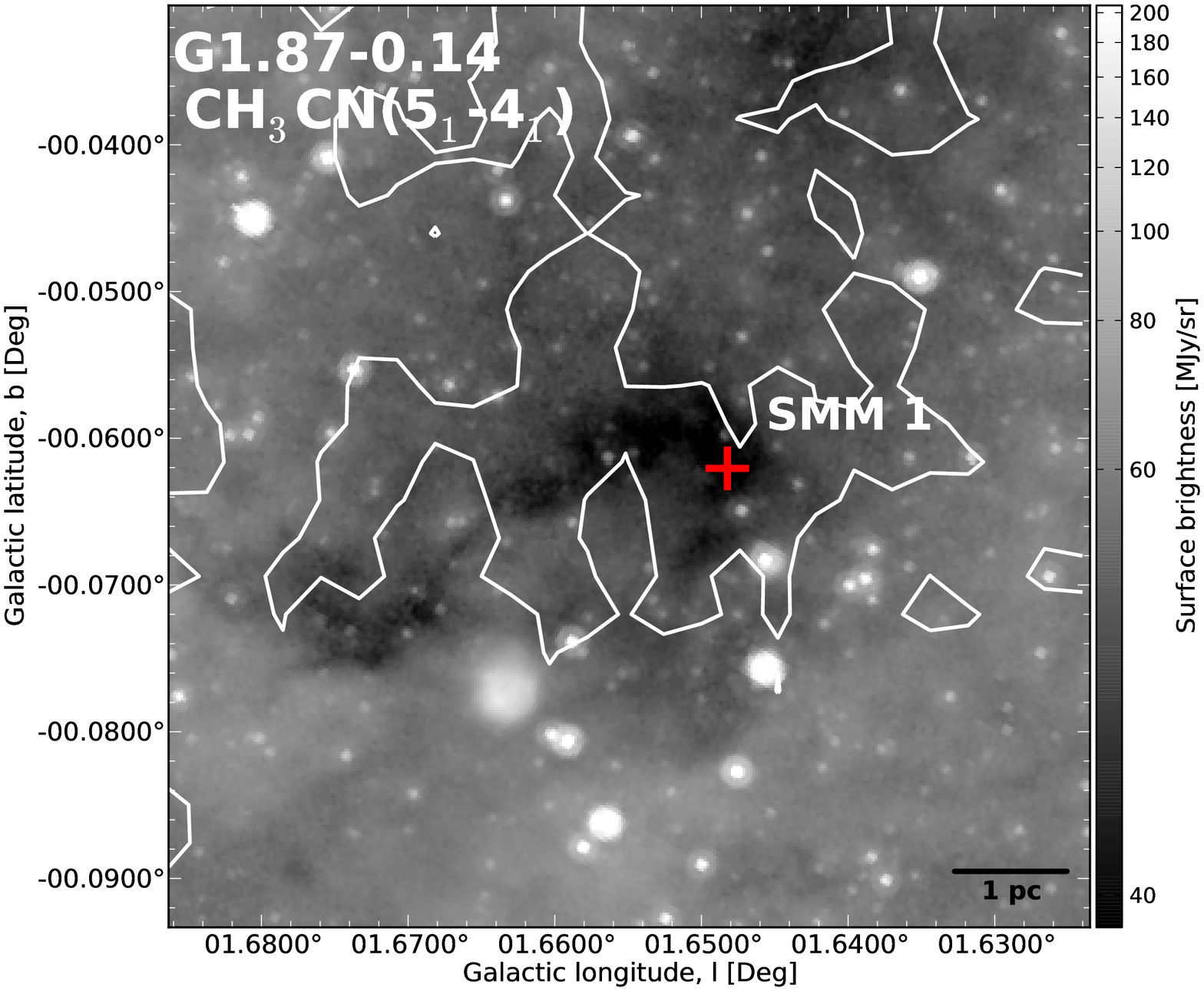}
\includegraphics[width=0.245\textwidth]{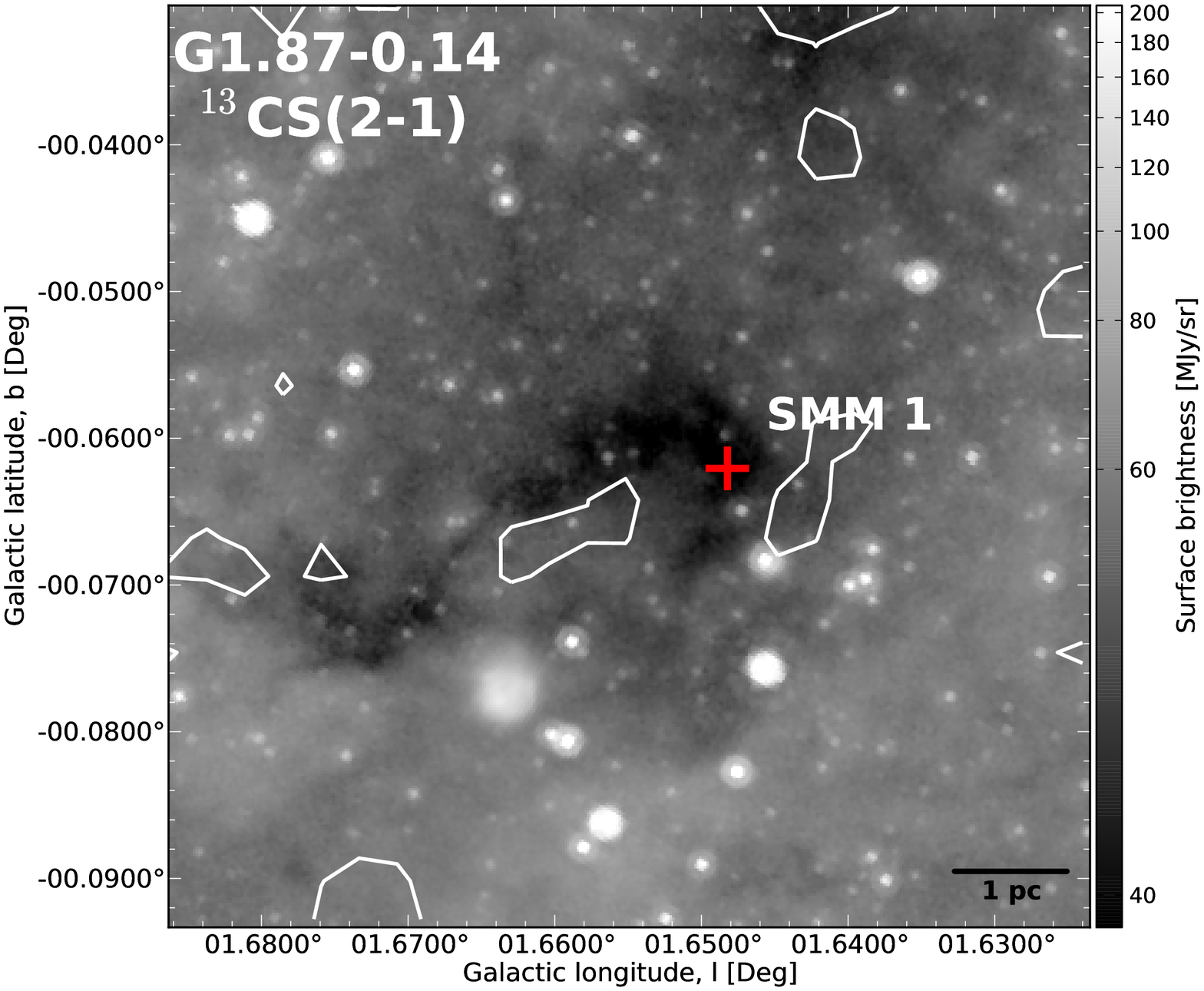}
\includegraphics[width=0.245\textwidth]{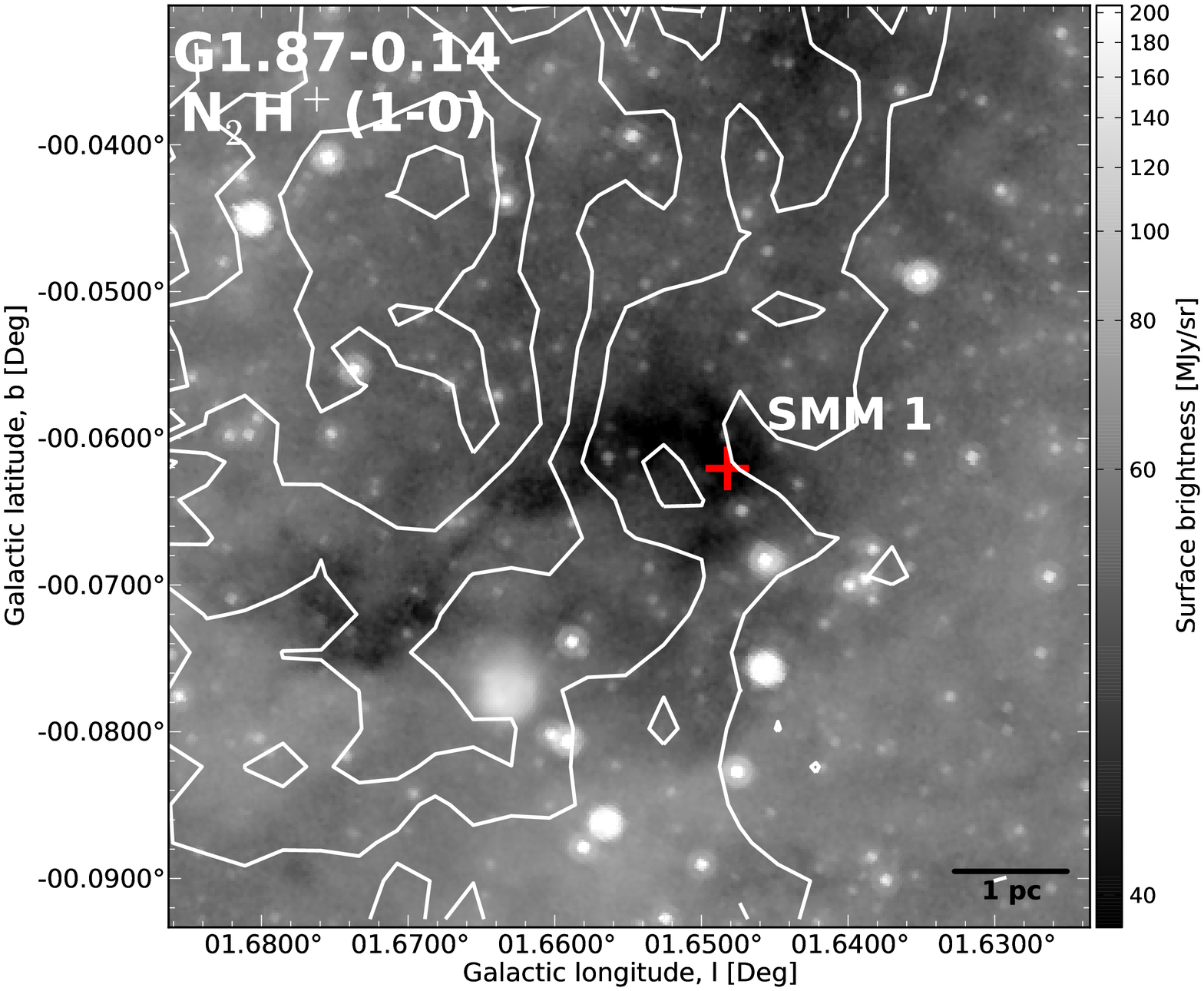}
\caption{Contour maps of integrated intensity of the MALT90 lines detected 
towards G1.87--SMM 1. In each panel, the contours are overlaid on the 
arcsinh-scaled \textit{Spitzer} 8-$\mu$m image (cf.~Fig.~\ref{figure:irac}). 
The contour levels start at $3\sigma$ for H$^{13}$CO$^+$, SiO, HN$^{13}$C, 
C$_2$H, CH$_3$CN, and $^{13}$CS. For HNCO$(4_{0,\,4}-3_{0,\,3})$, HCN, HCO$^+$, 
HNC, HC$_3$N, and N$_2$H$^+$, the contours start at $26\sigma$, $23\sigma$, 
$20\sigma$, $30\sigma$, $6\sigma$, and $15\sigma$, respectively. In all cases, 
the contours go in steps of $3\sigma$. The average $1\sigma$ value in 
$T_{\rm MB}$ units is $\sim0.69$ K~km~s$^{-1}$. The red plus sign marks the 
LABOCA 870-$\mu$m peak position of the clump. A scale bar indicating the 1 pc 
projected length is indicated. The line emission is extended in many 
cases, and the HCN, HCO$^+$, and HNC emissions are well correlated with each 
other. The N$_2$H$^+$ emission also shows some resemblance to these species. 
The spatial distributions of HNCO and HC$_3$N appear to be similar to each 
other, while weak CH$_3$CN emission traces reasonably well the 8-$\mu$m 
absorption feature.}
\label{figure:G187SMM1lines}
\end{center}
\end{figure*}

\begin{figure*}
\begin{center}
\includegraphics[width=0.245\textwidth]{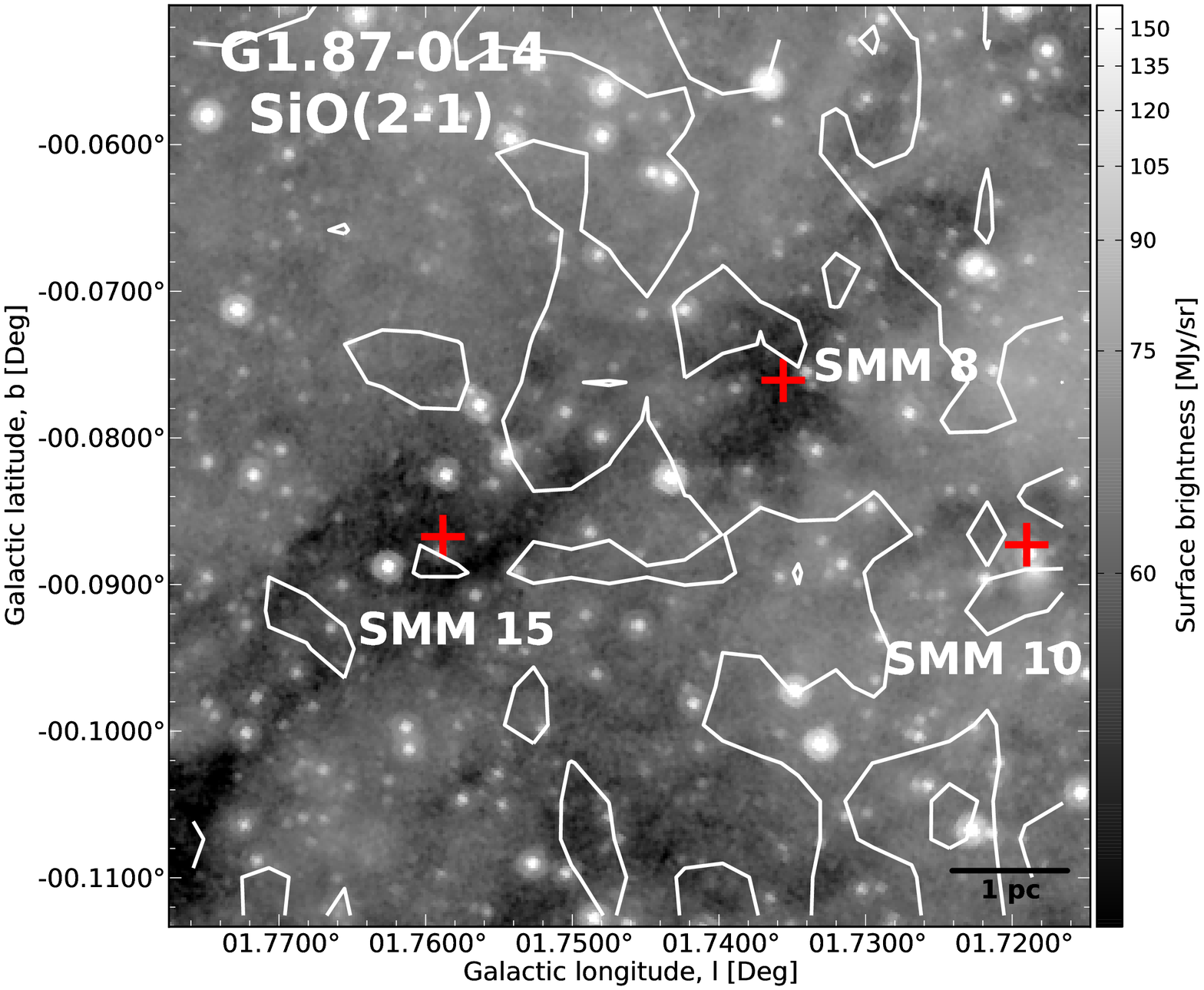}
\includegraphics[width=0.245\textwidth]{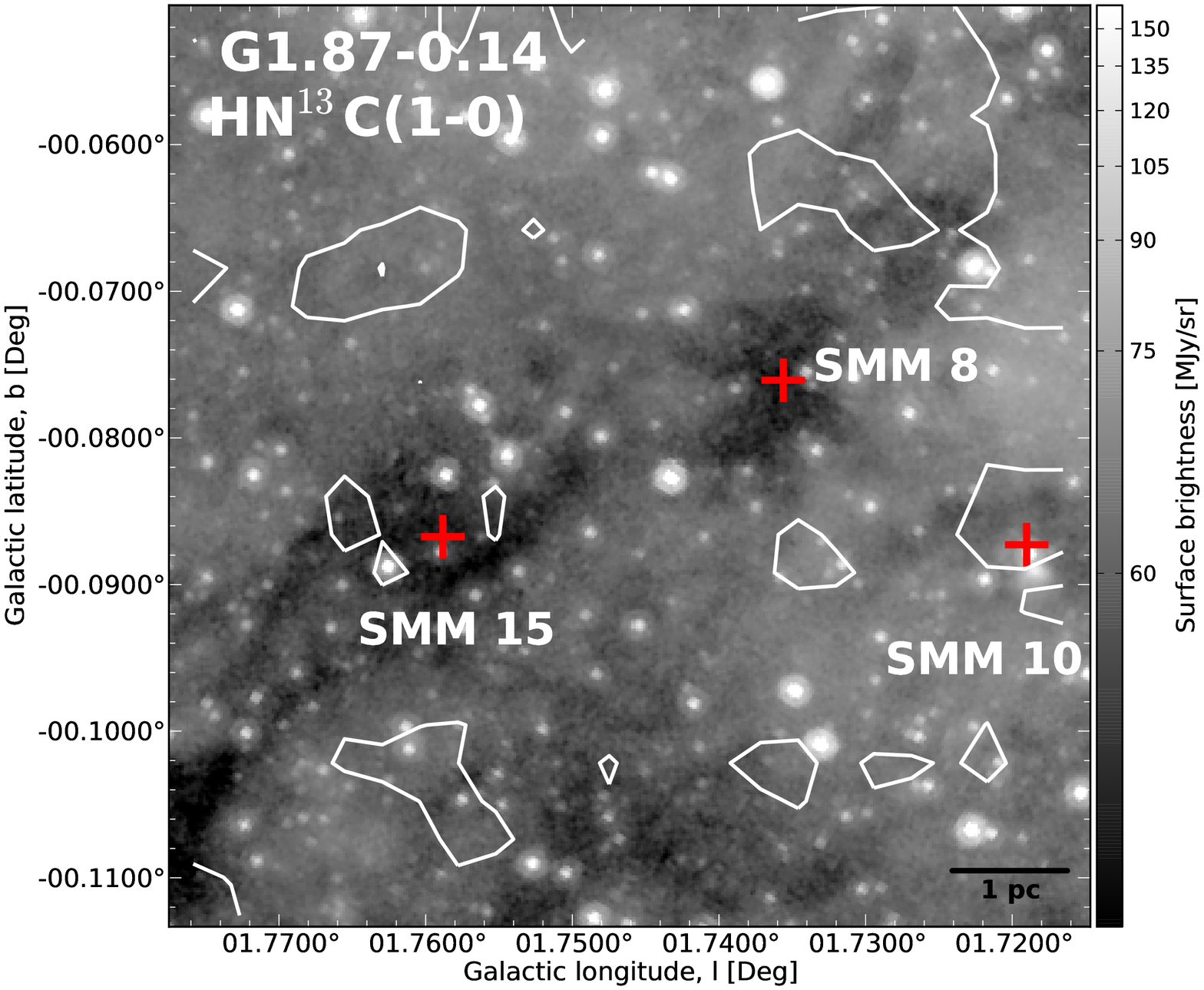}
\includegraphics[width=0.245\textwidth]{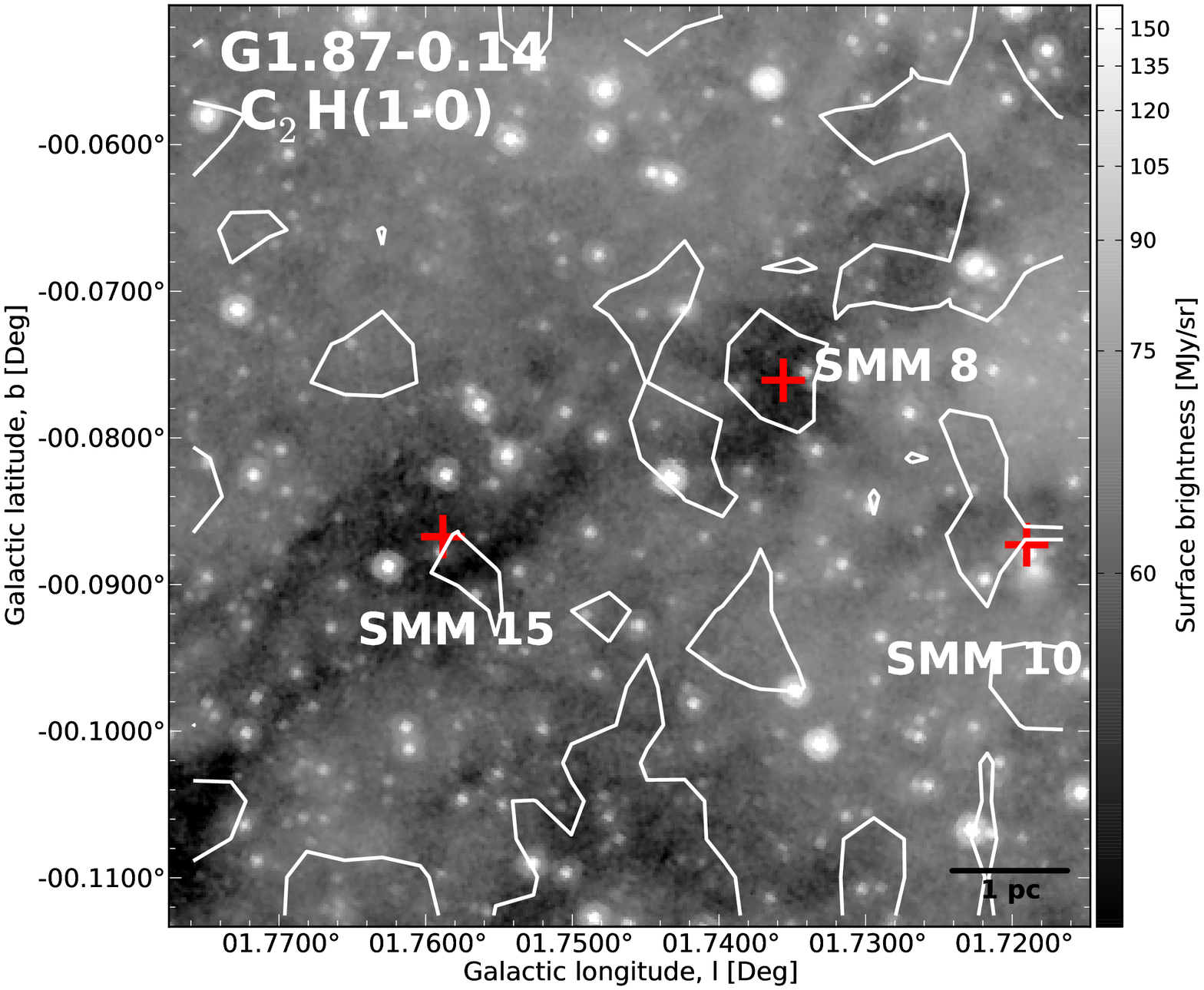}
\includegraphics[width=0.245\textwidth]{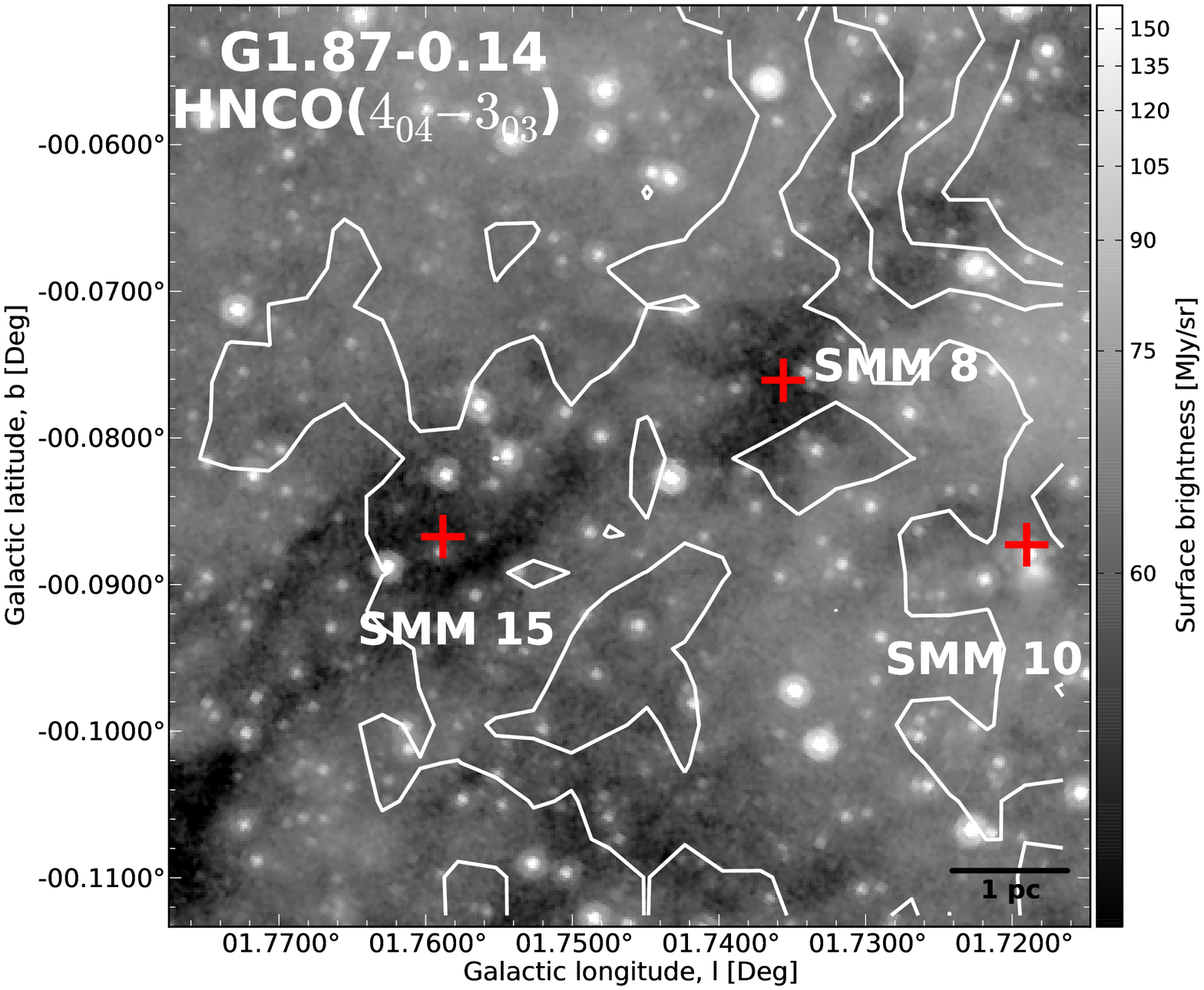}
\includegraphics[width=0.245\textwidth]{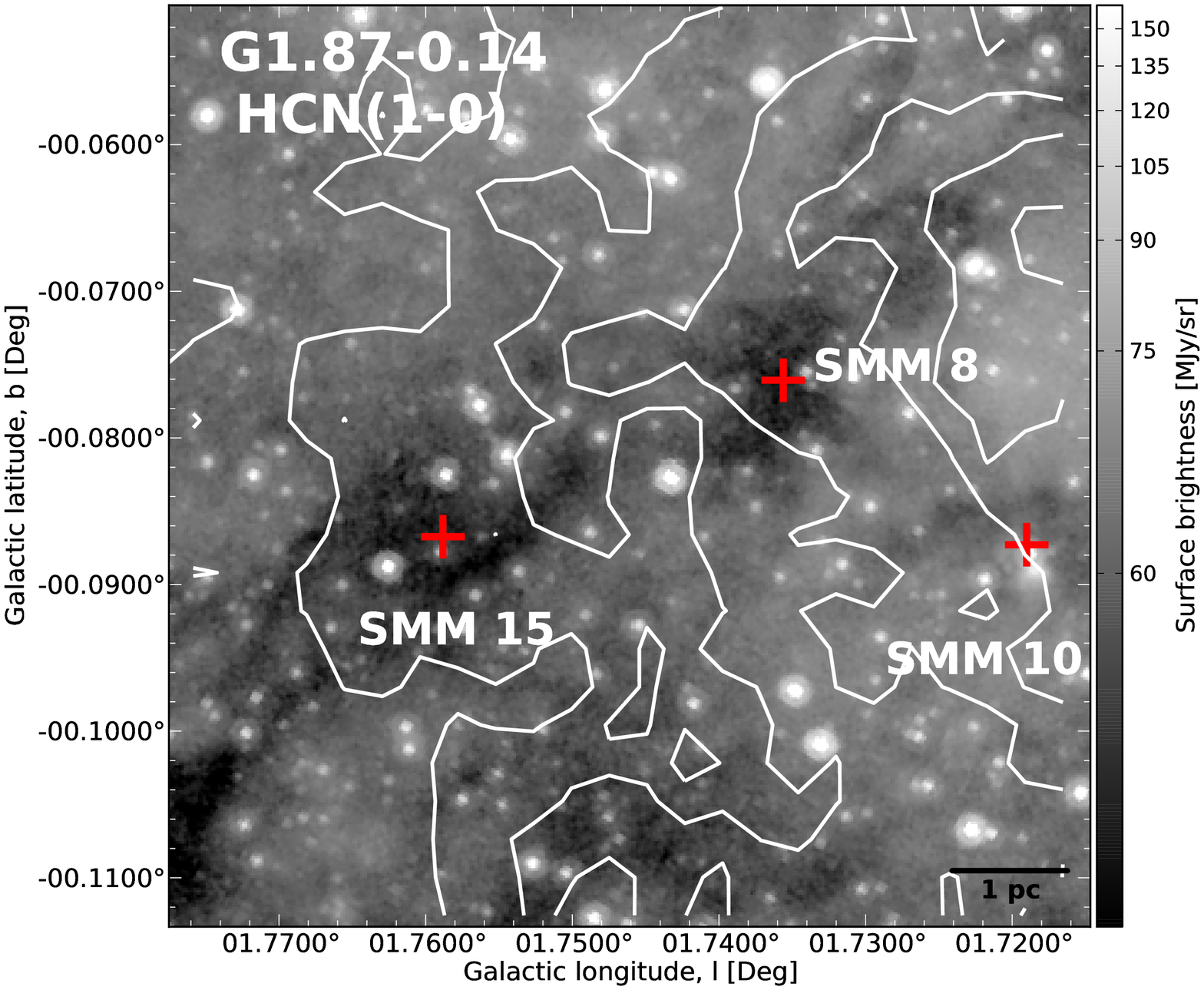}
\includegraphics[width=0.245\textwidth]{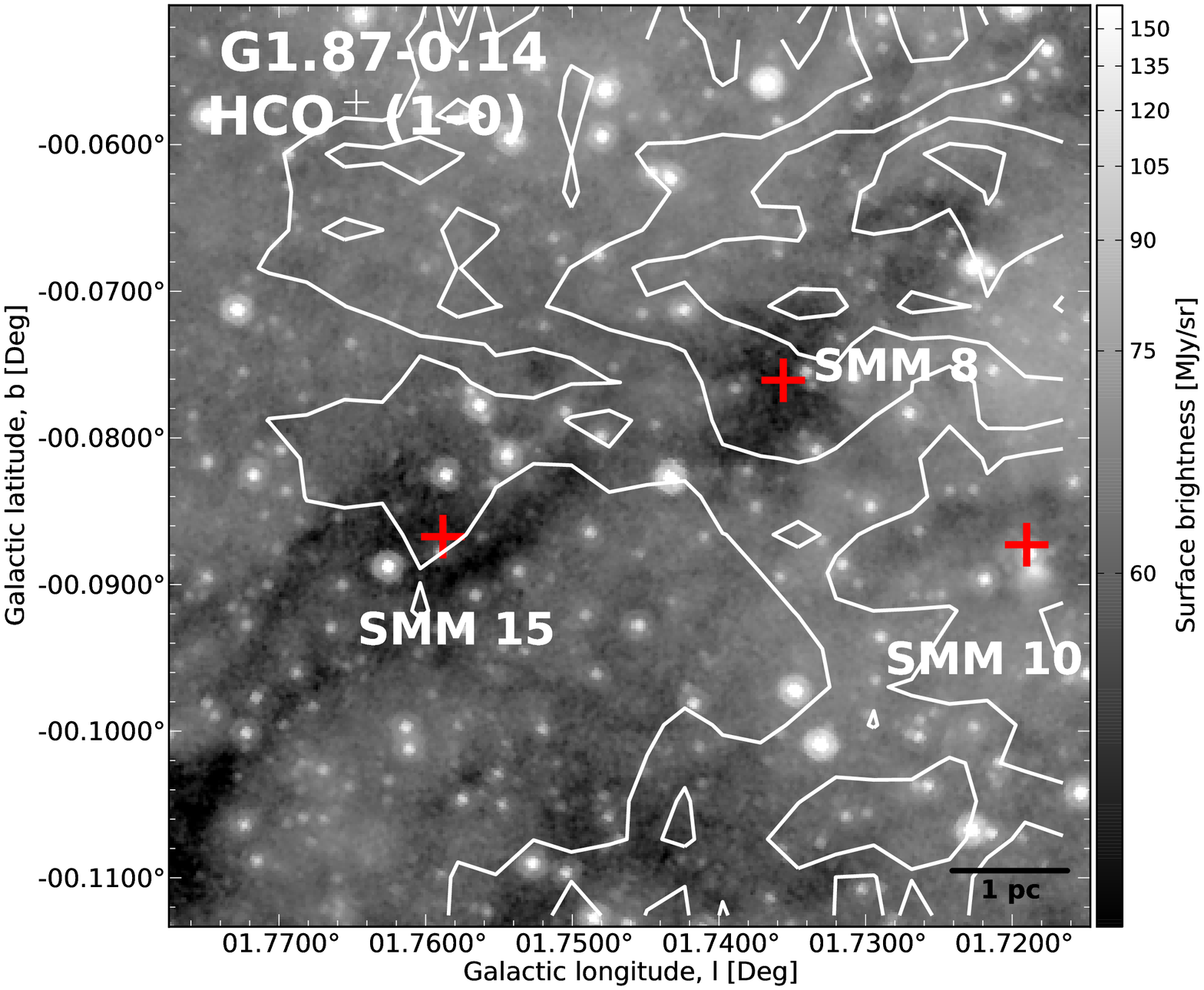}
\includegraphics[width=0.245\textwidth]{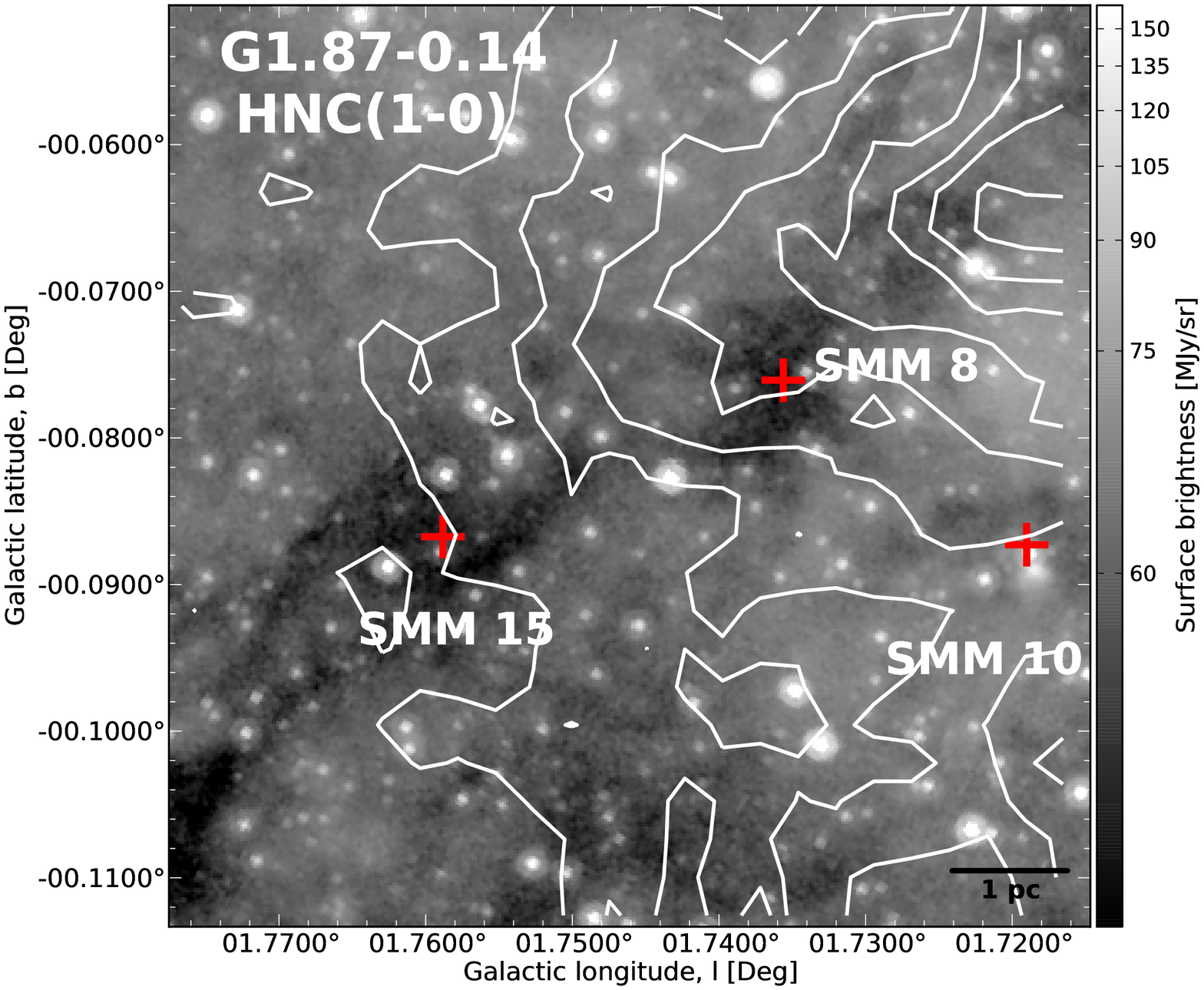}
\includegraphics[width=0.245\textwidth]{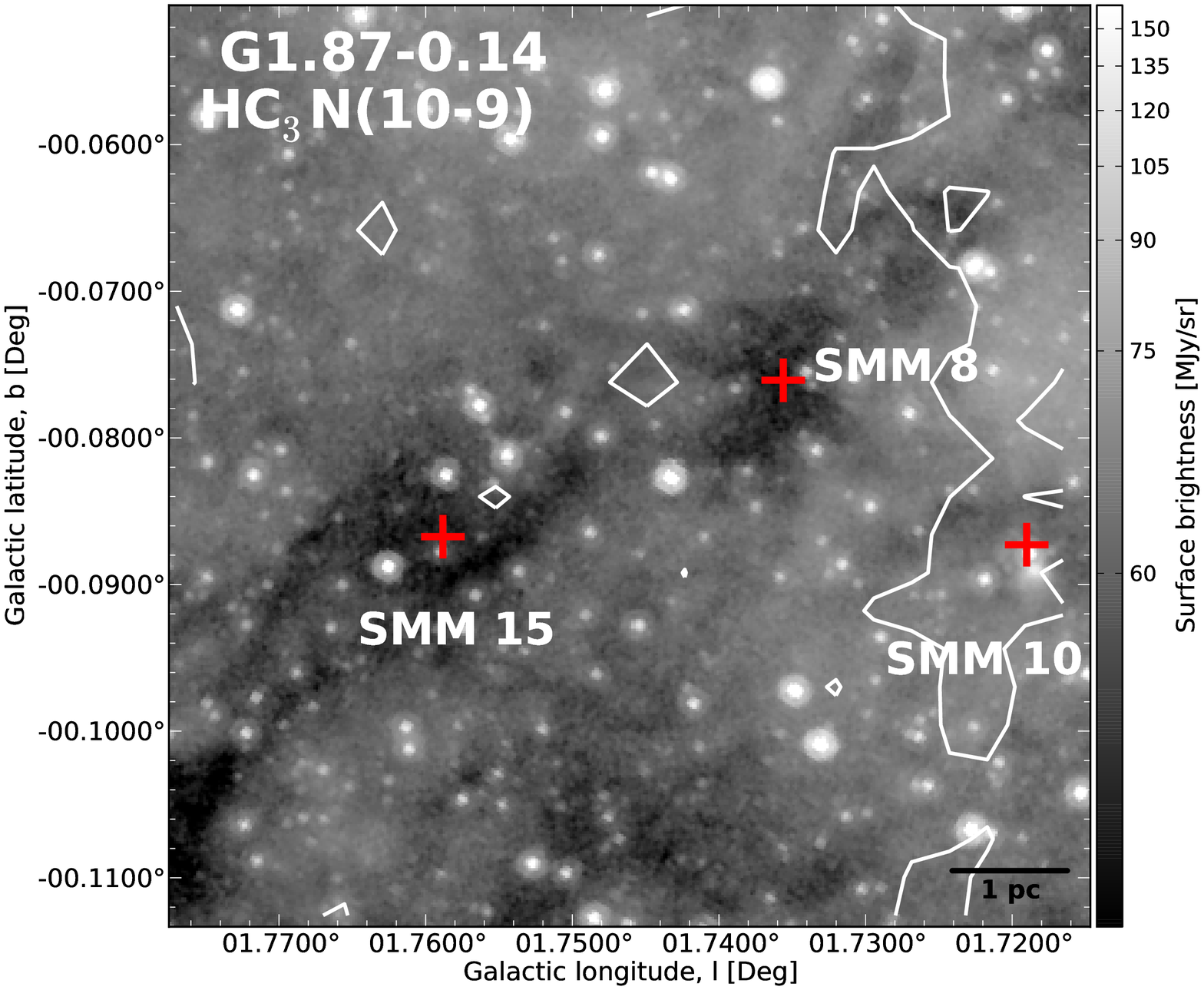}
\includegraphics[width=0.245\textwidth]{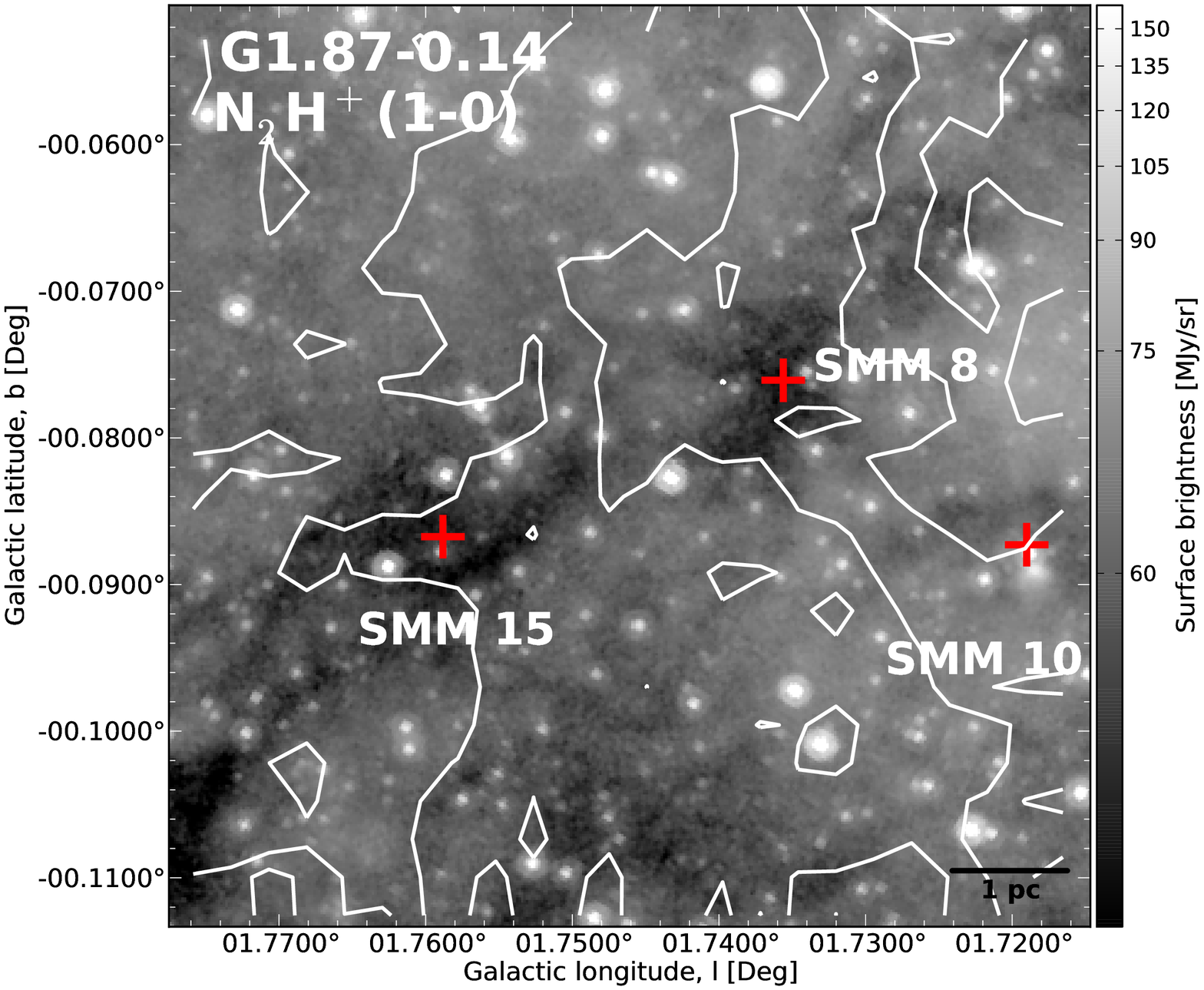}
\caption{Similar to Fig.~\ref{figure:G187SMM1lines} but towards 
G1.87--SMM 8, 10, 15. The contour levels start at 
$3\sigma$ for SiO, HN$^{13}$C, C$_2$H, and HC$_3$N.  
For HNCO$(4_{0,\,4}-3_{0,\,3})$, HCN, HCO$^+$, HNC, and N$_2$H$^+$, 
the contours start at $9\sigma$, $16\sigma$, $15\sigma$, $12\sigma$, 
and $5\sigma$, respectively. In all cases, the contours go in 
steps of $3\sigma$. The average $1\sigma$ value in $T_{\rm MB}$ units is 
$\sim0.70$ K~km~s$^{-1}$. The LABOCA 870-$\mu$m peak positions of the clumps
are marked by red plus signs. A scale bar indicating the 1 pc 
projected length is indicated. The spatial distributions of the HCN, HCO$^+$, 
and HNC emissions appear to be quite similar. Those of HNCO and N$_2$H$^+$ 
show some similarities also.}
\label{figure:G187SMM8lines}
\end{center}
\end{figure*}

\begin{figure*}
\begin{center}
\includegraphics[width=0.245\textwidth]{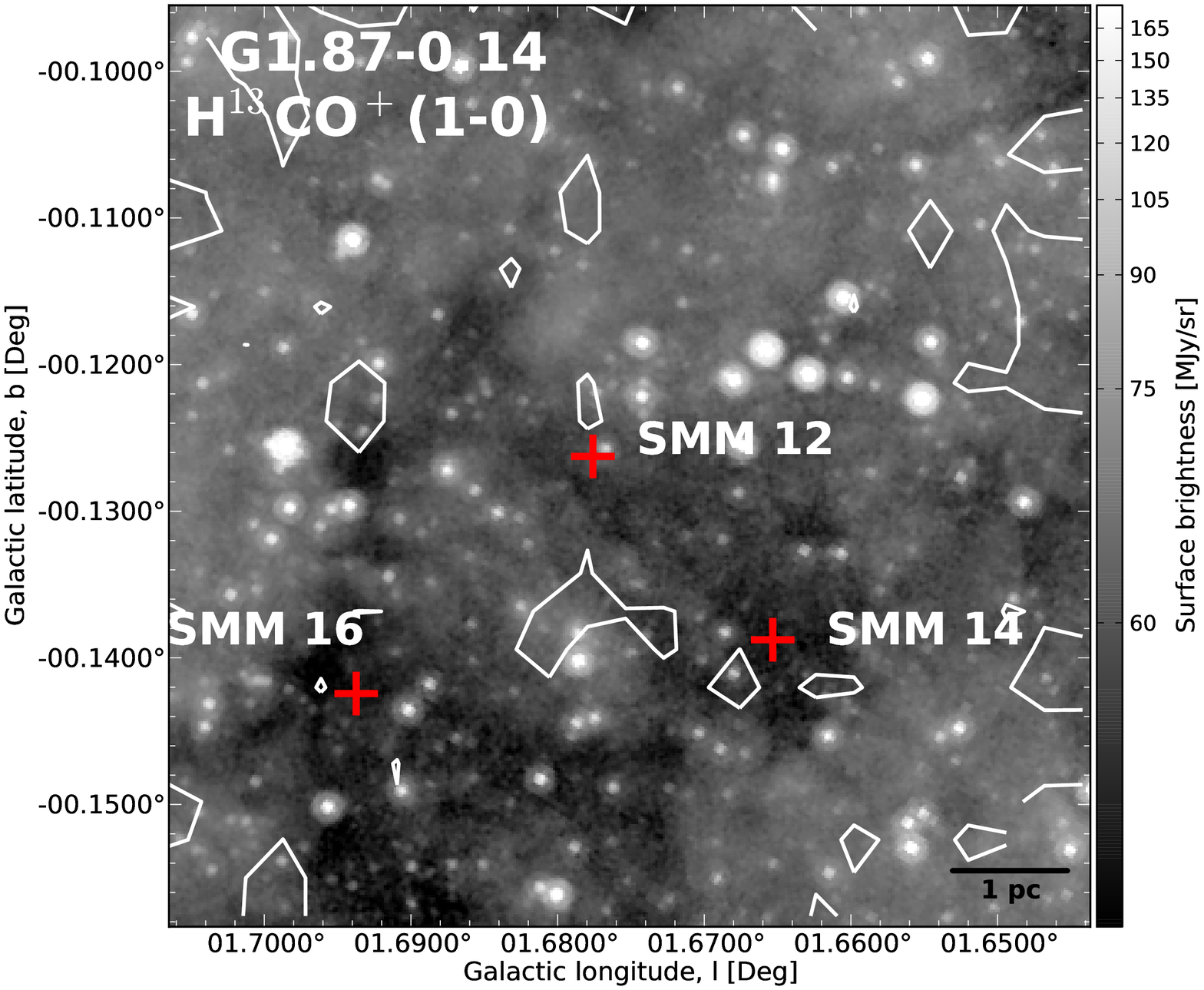}
\includegraphics[width=0.245\textwidth]{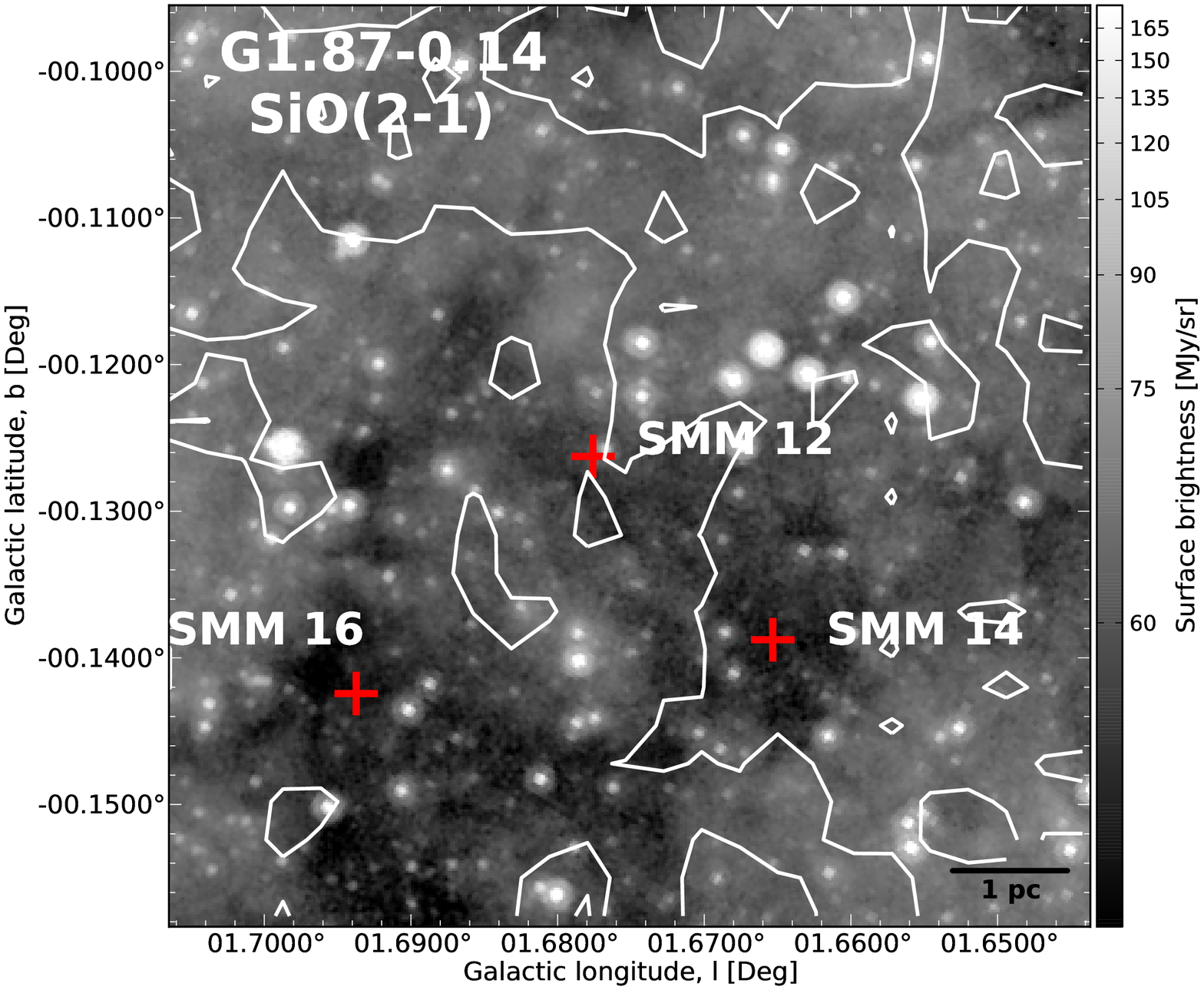}
\includegraphics[width=0.245\textwidth]{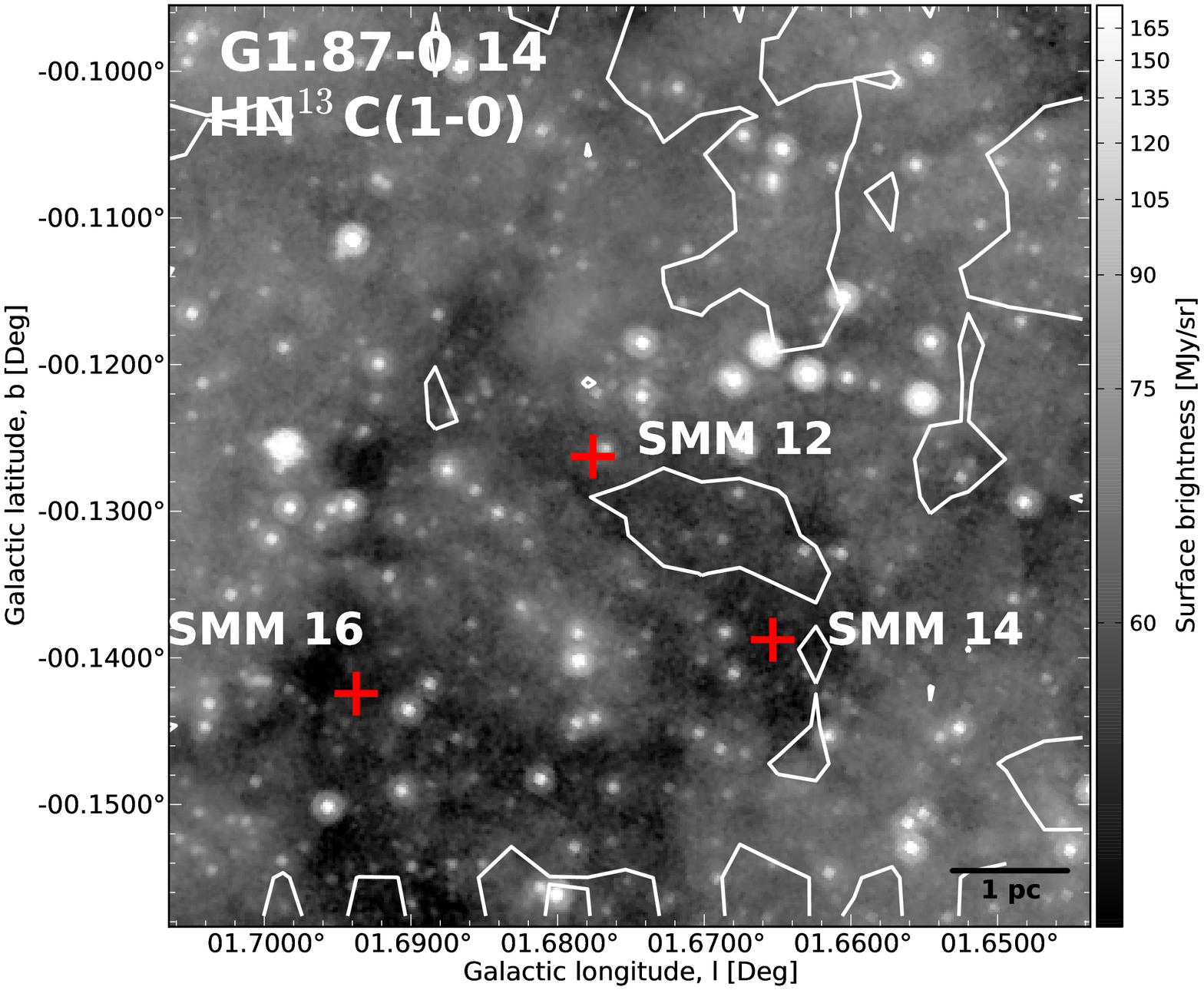}
\includegraphics[width=0.245\textwidth]{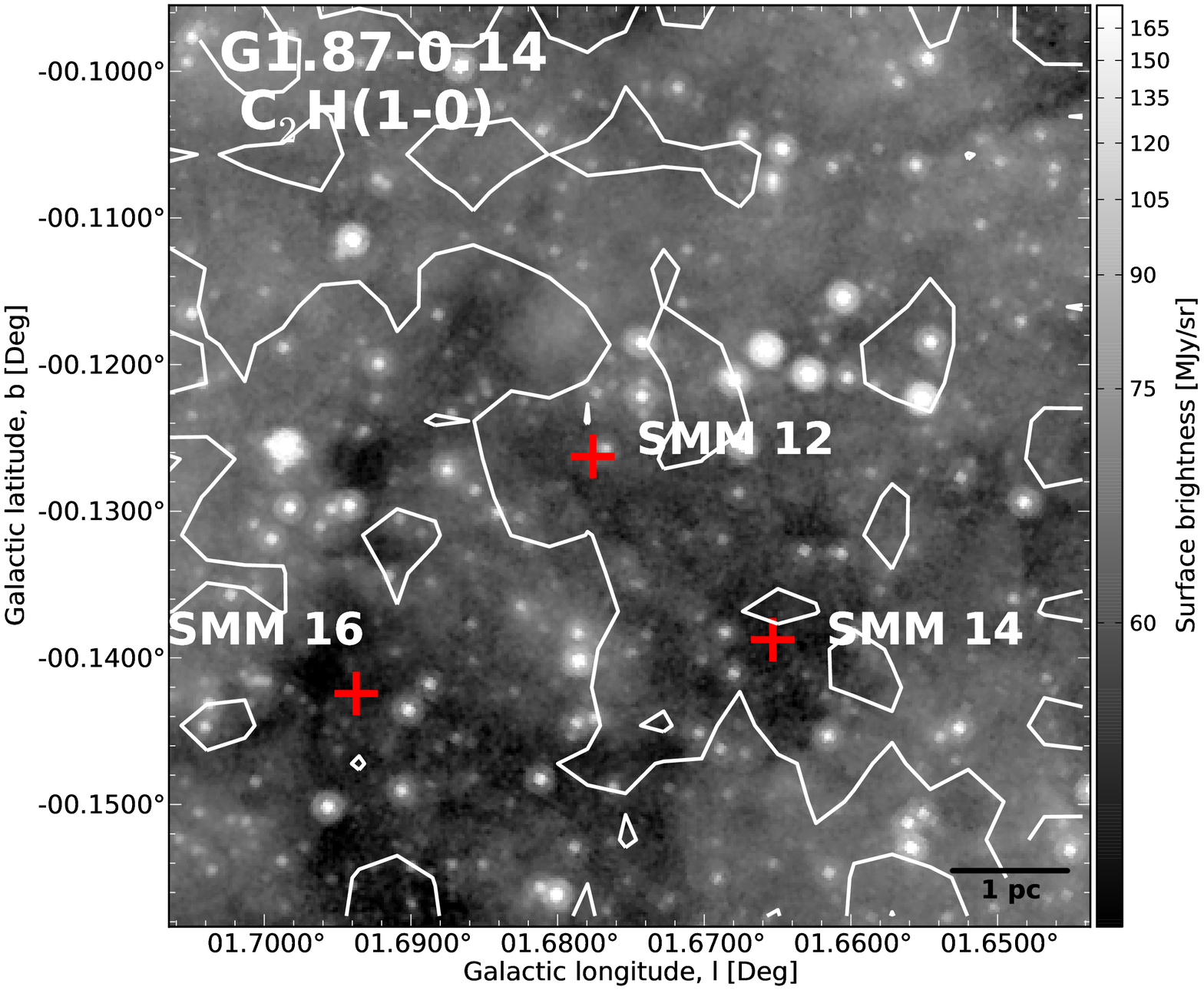}
\includegraphics[width=0.245\textwidth]{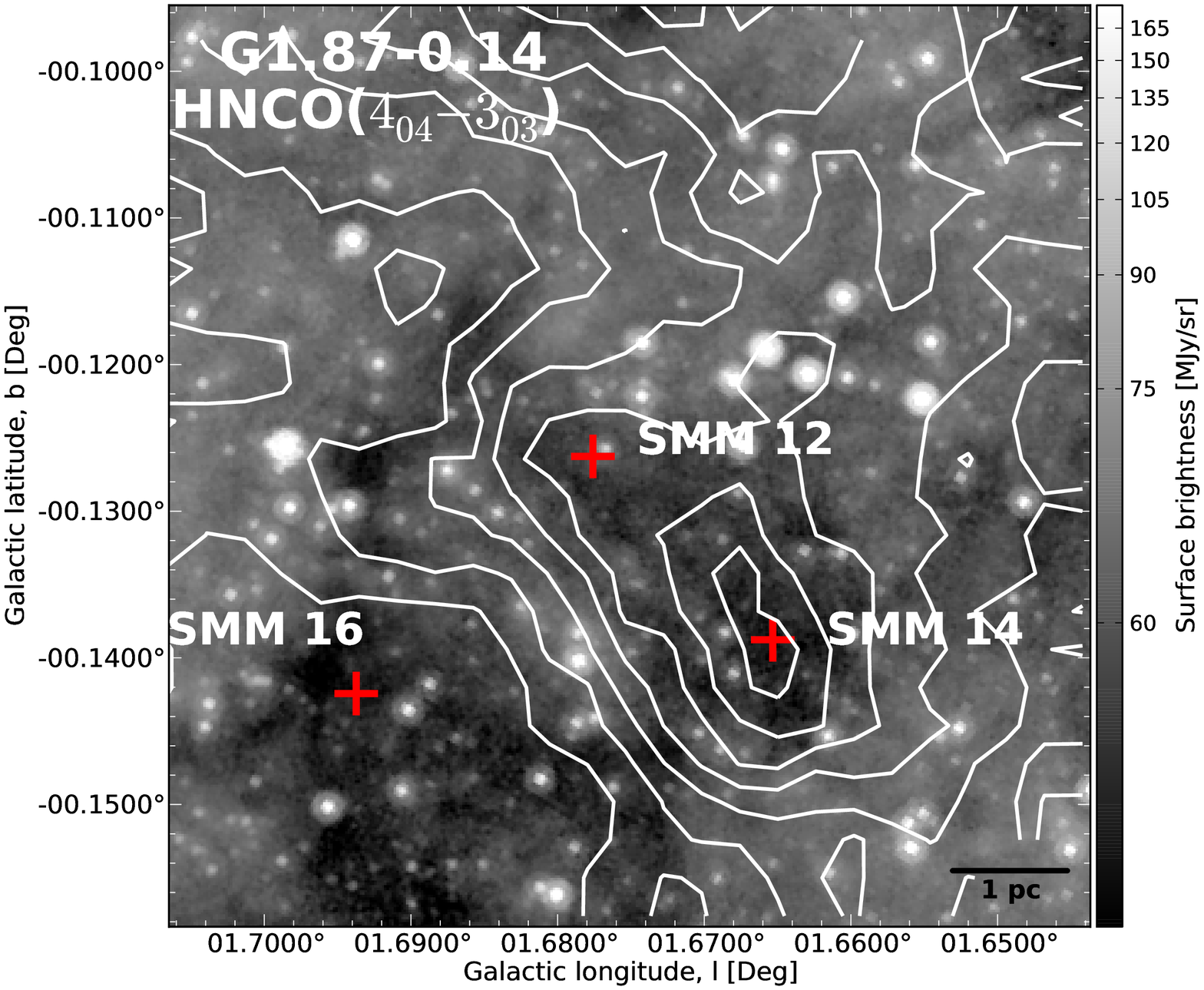}
\includegraphics[width=0.245\textwidth]{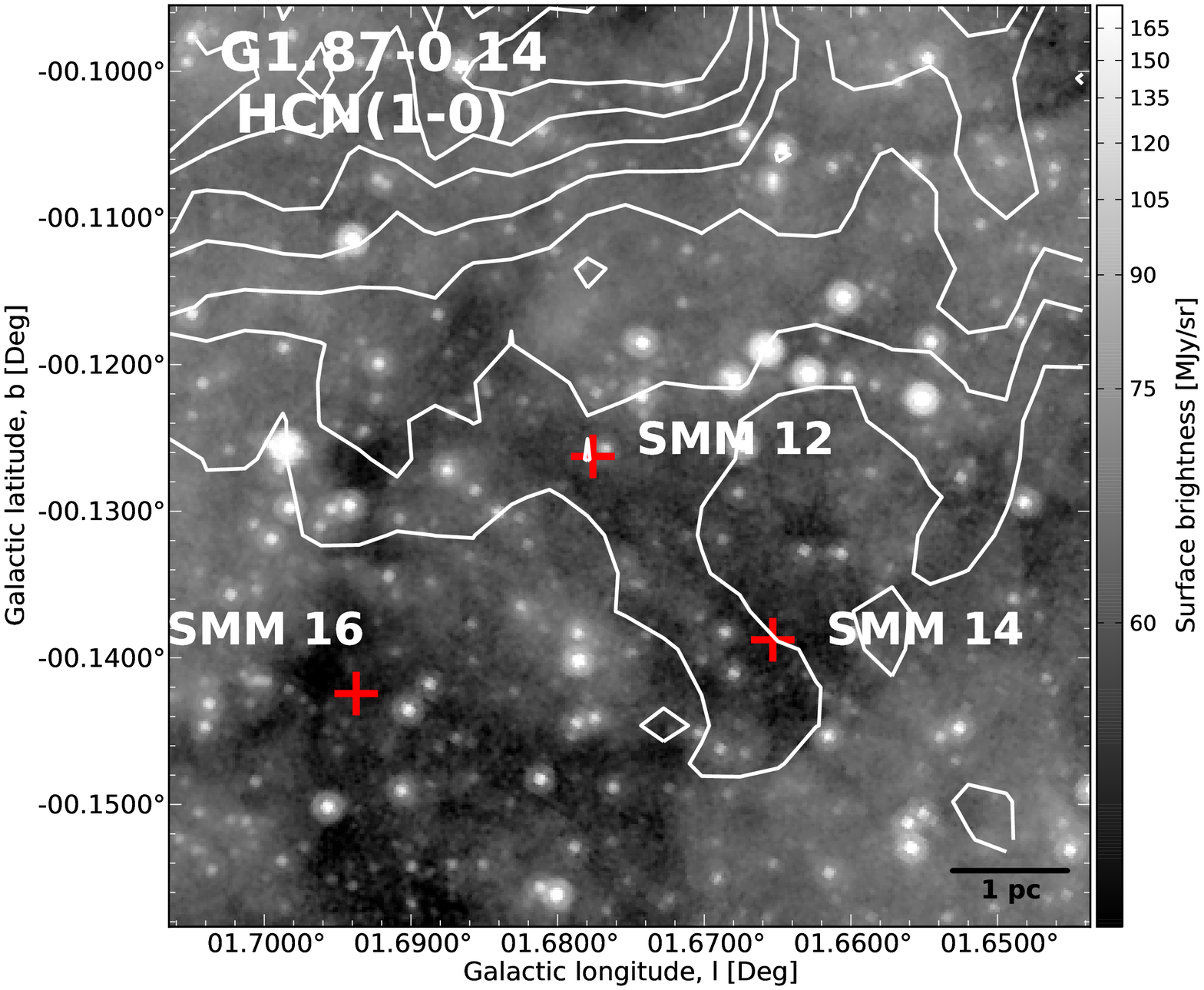}
\includegraphics[width=0.245\textwidth]{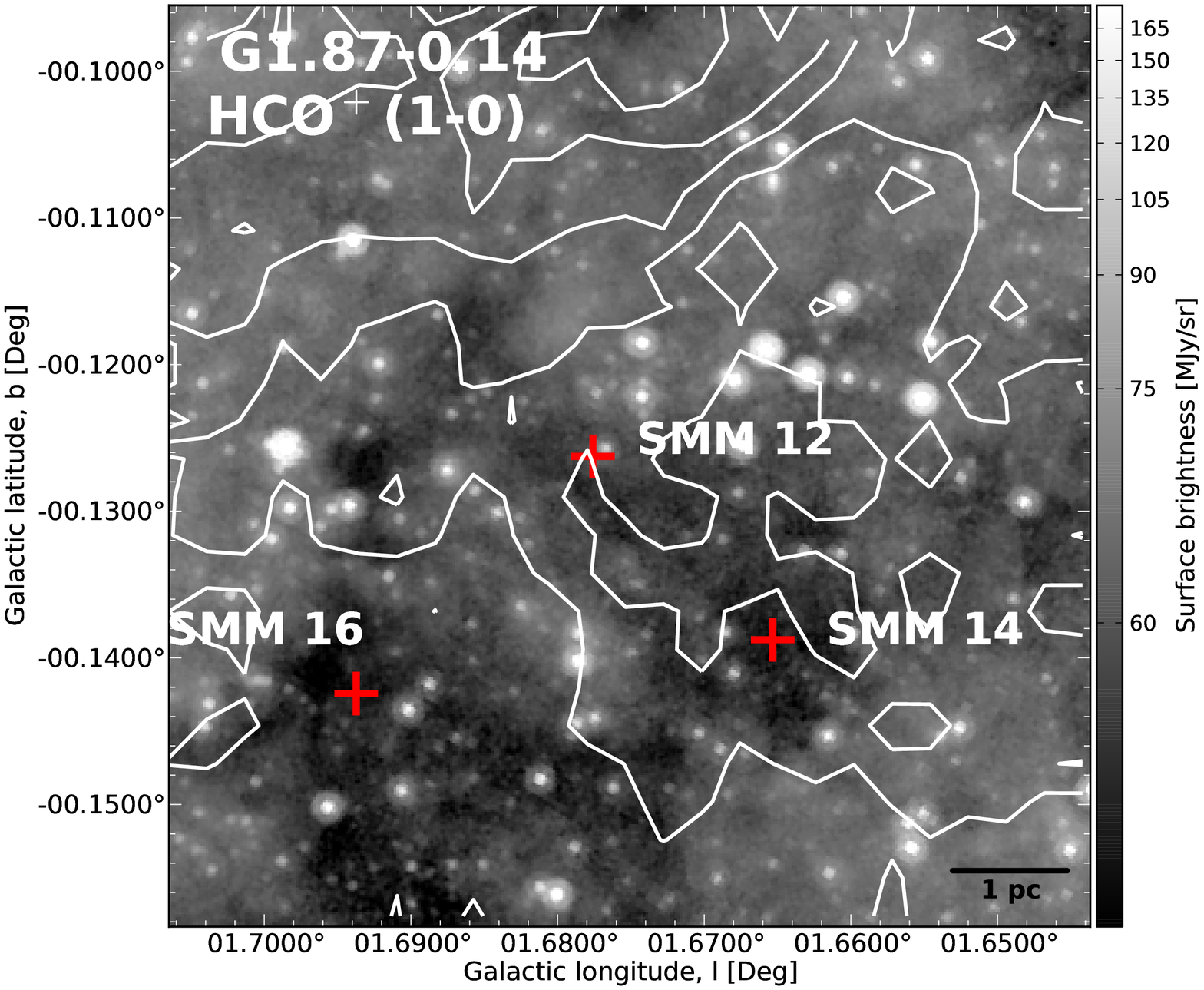}
\includegraphics[width=0.245\textwidth]{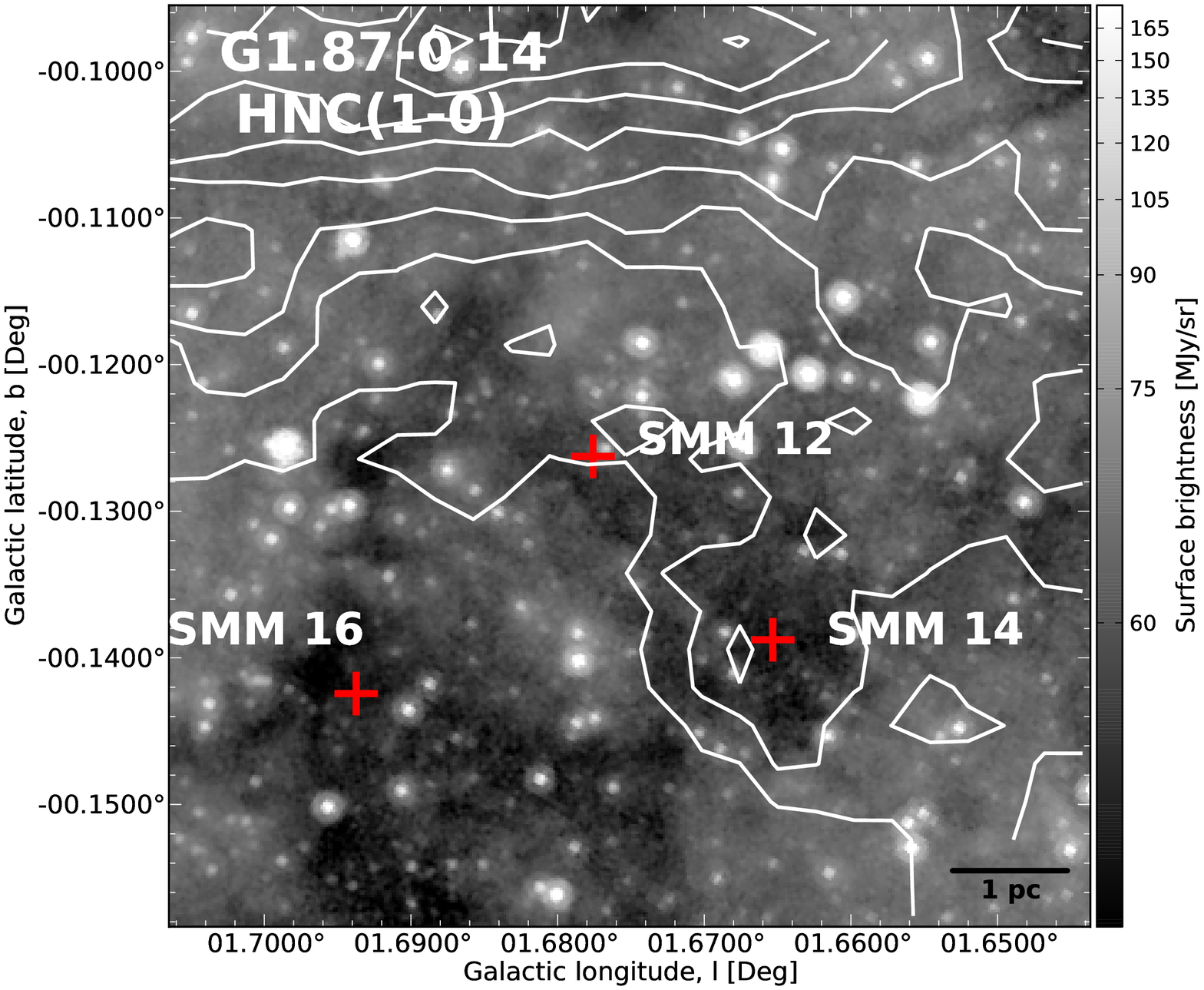}
\includegraphics[width=0.245\textwidth]{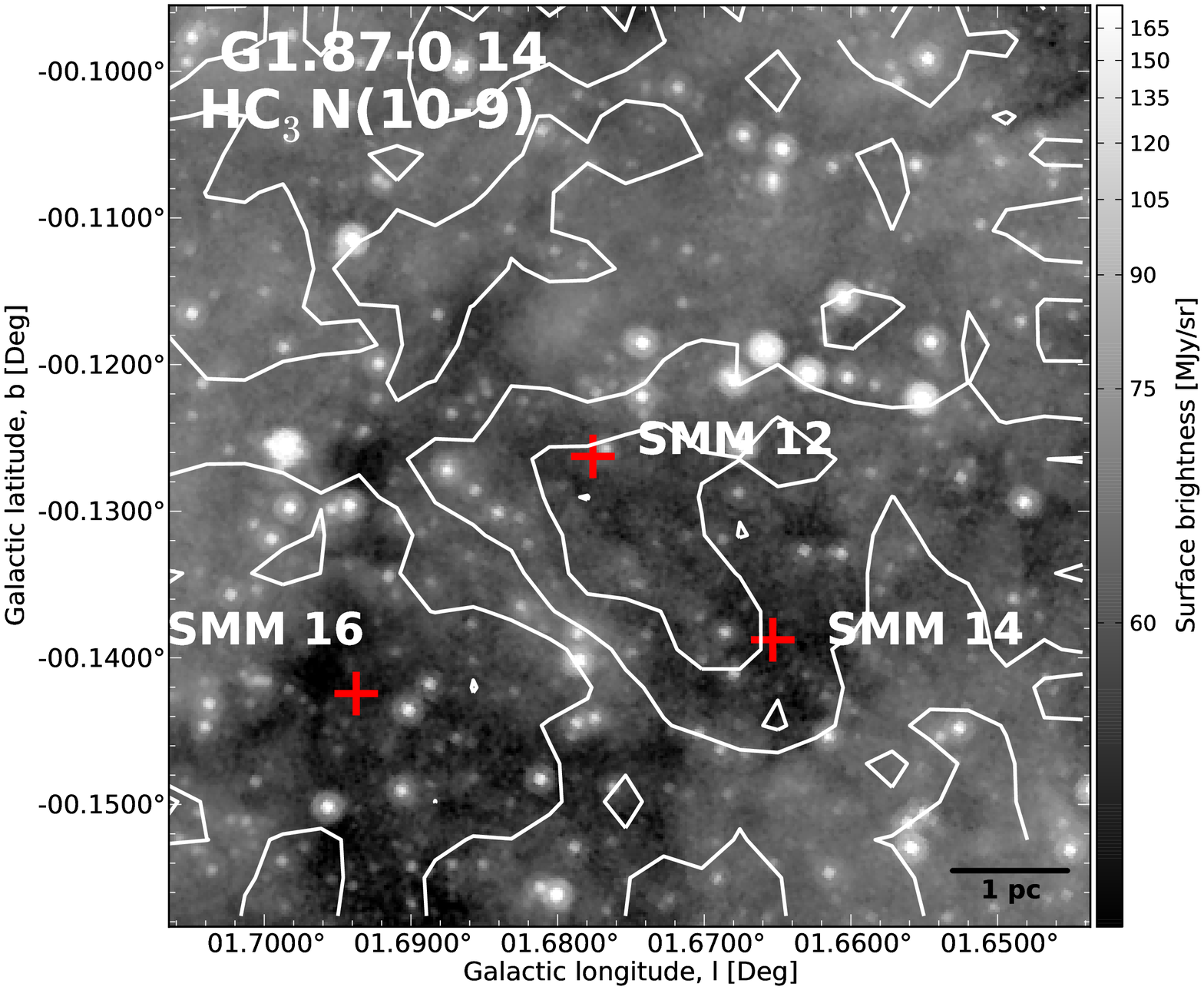}
\includegraphics[width=0.245\textwidth]{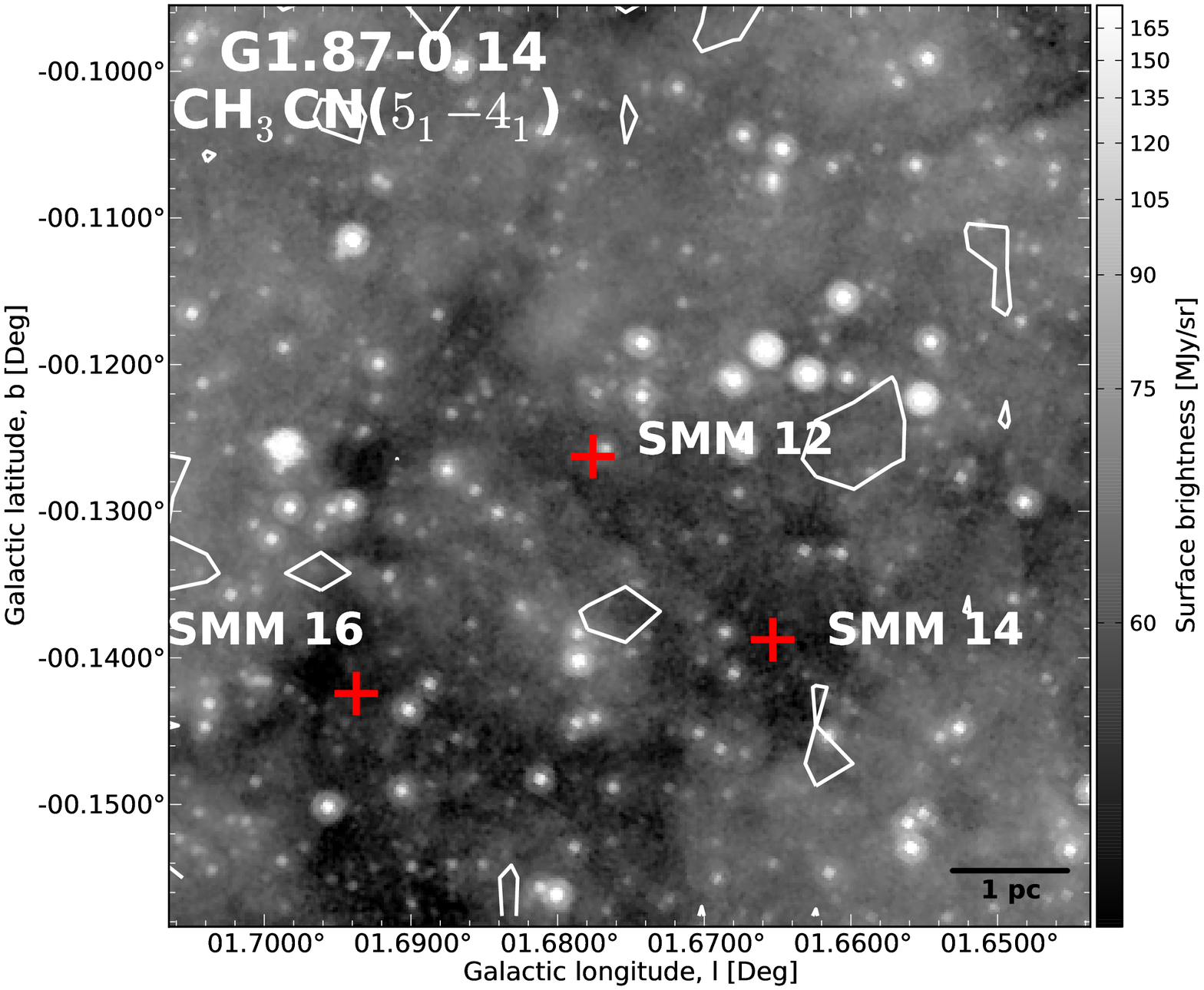}
\includegraphics[width=0.245\textwidth]{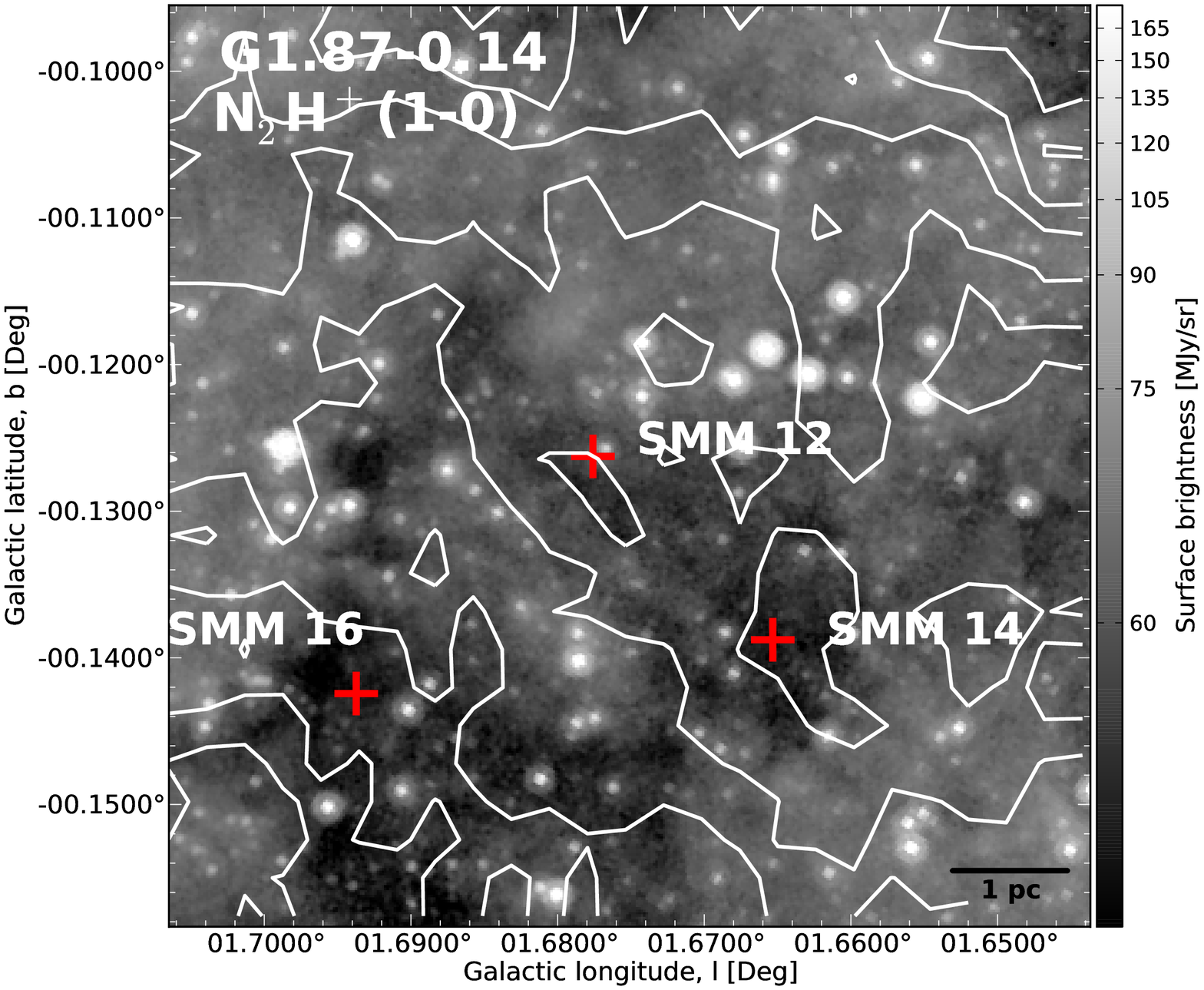}
\caption{Similar to Fig.~\ref{figure:G187SMM1lines} but towards 
G1.87--SMM 12, 14, 16. The contour levels start at 
$3\sigma$ for H$^{13}$CO$^+$, HN$^{13}$C, C$_2$H, HC$_3$N, and CH$_3$CN. 
For SiO, HNCO$(4_{0,\,4}-3_{0,\,3})$, HCN, HCO$^+$, HNC, and N$_2$H$^+$, 
the contours start at $5\sigma$, $15\sigma$, $19\sigma$, $10\sigma$, 
$15\sigma$, and $6\sigma$, respectively. In all cases, the contours go in 
steps of $3\sigma$. The average $1\sigma$ value in $T_{\rm MB}$ units is 
$\sim0.63$ K~km~s$^{-1}$. The LABOCA 870-$\mu$m peak positions of the clumps
are marked by red plus signs. A scale bar indicating the 1 pc 
projected length is indicated. The emissions of HNCO, HCN, HCO$^+$, HNC, 
HC$_3$N, and N$_2$H$^+$ are extended in a similar fashion, but HNCO  
clearly shows the strongest.}
\label{figure:G187SMM12lines}
\end{center}
\end{figure*}

\begin{figure*}
\begin{center}
\includegraphics[width=0.245\textwidth]{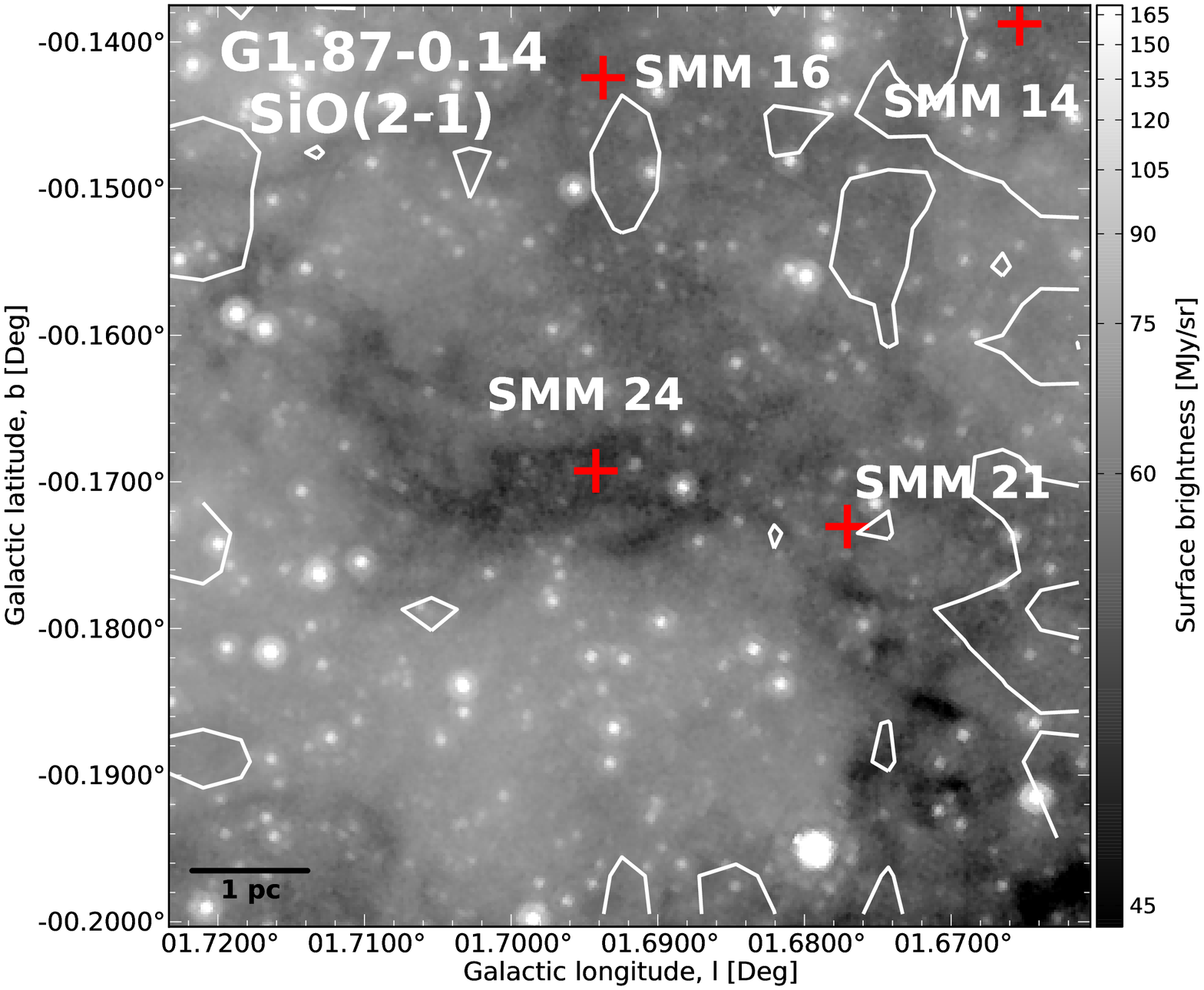}
\includegraphics[width=0.245\textwidth]{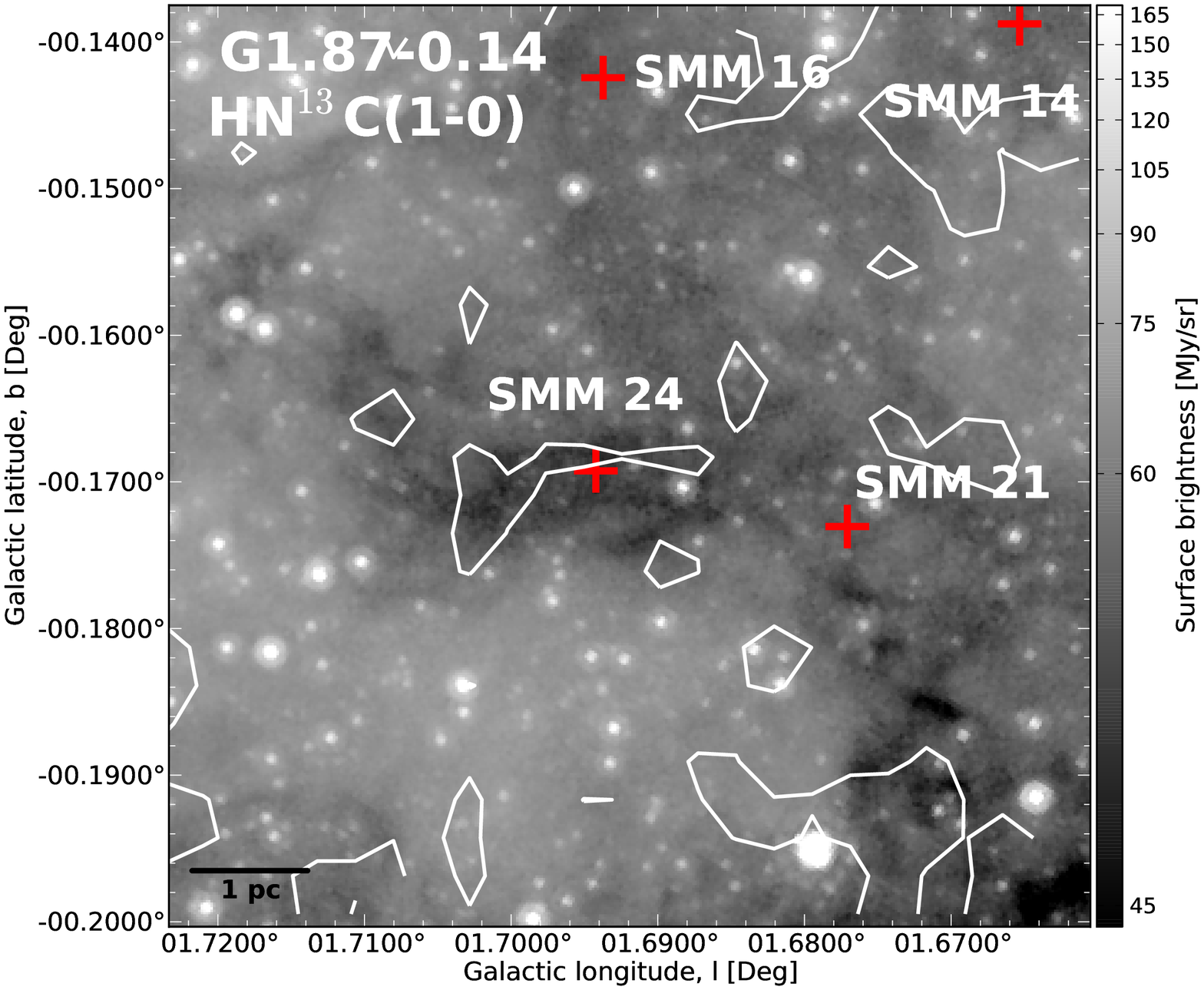}
\includegraphics[width=0.245\textwidth]{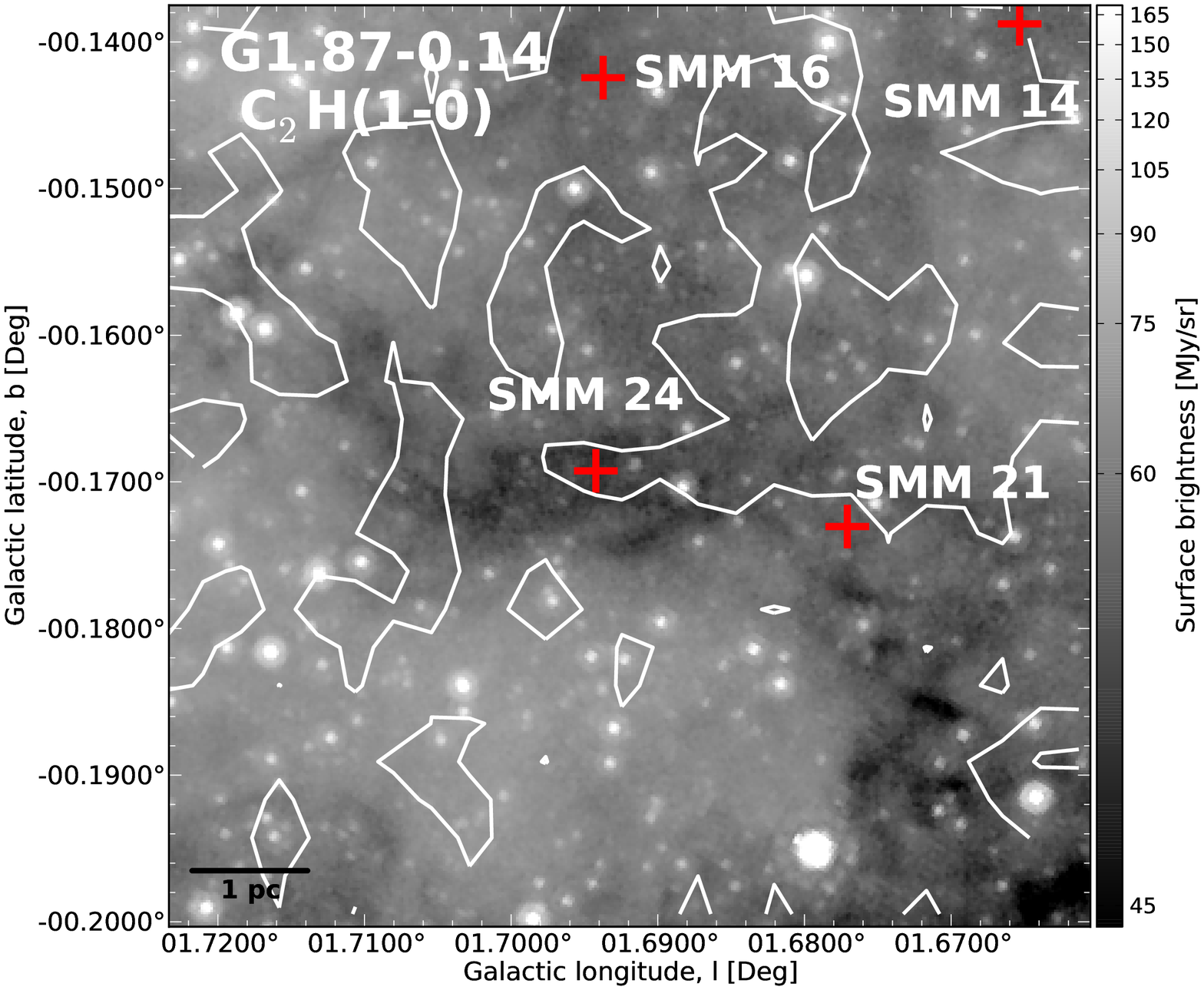}
\includegraphics[width=0.245\textwidth]{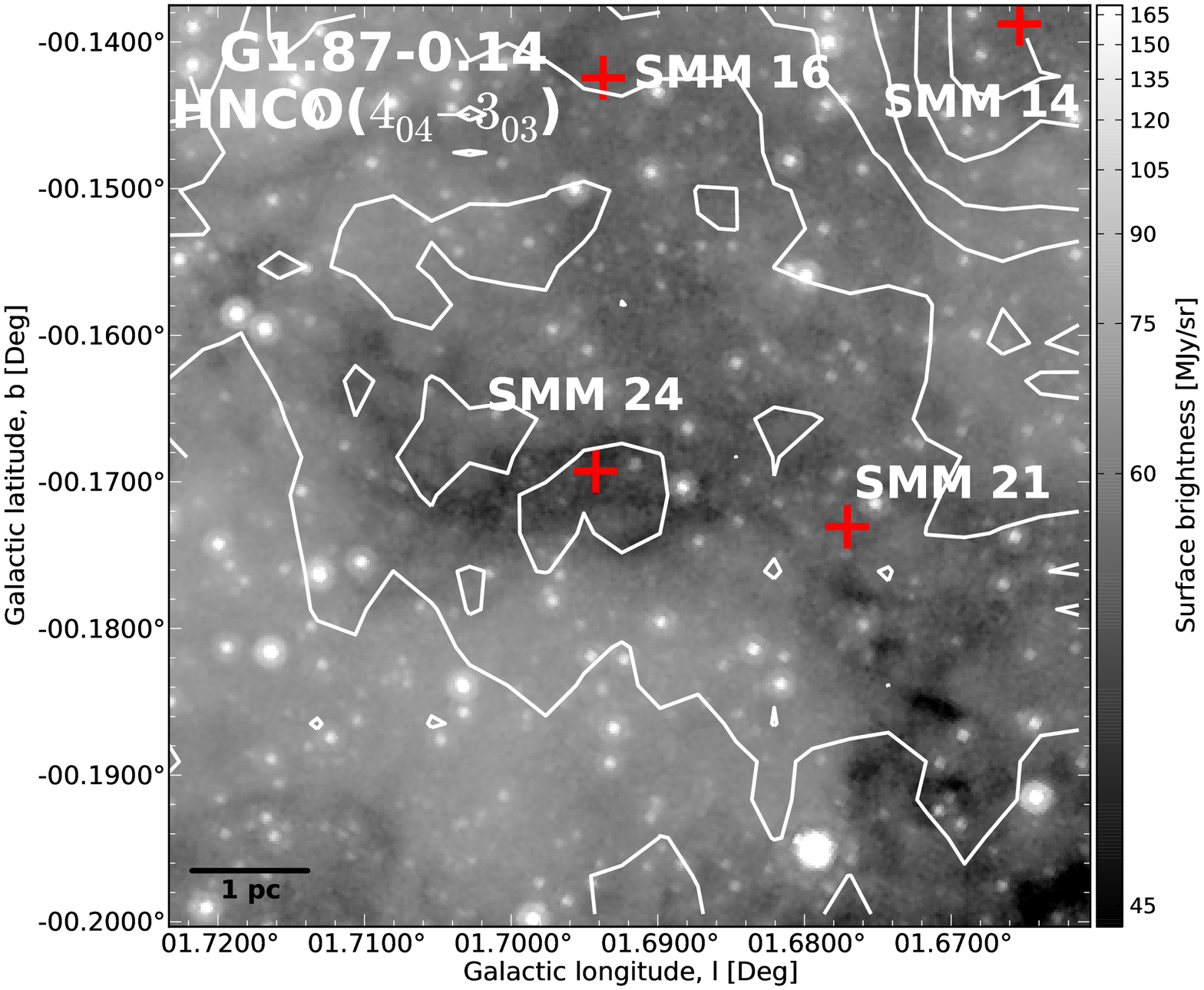}
\includegraphics[width=0.245\textwidth]{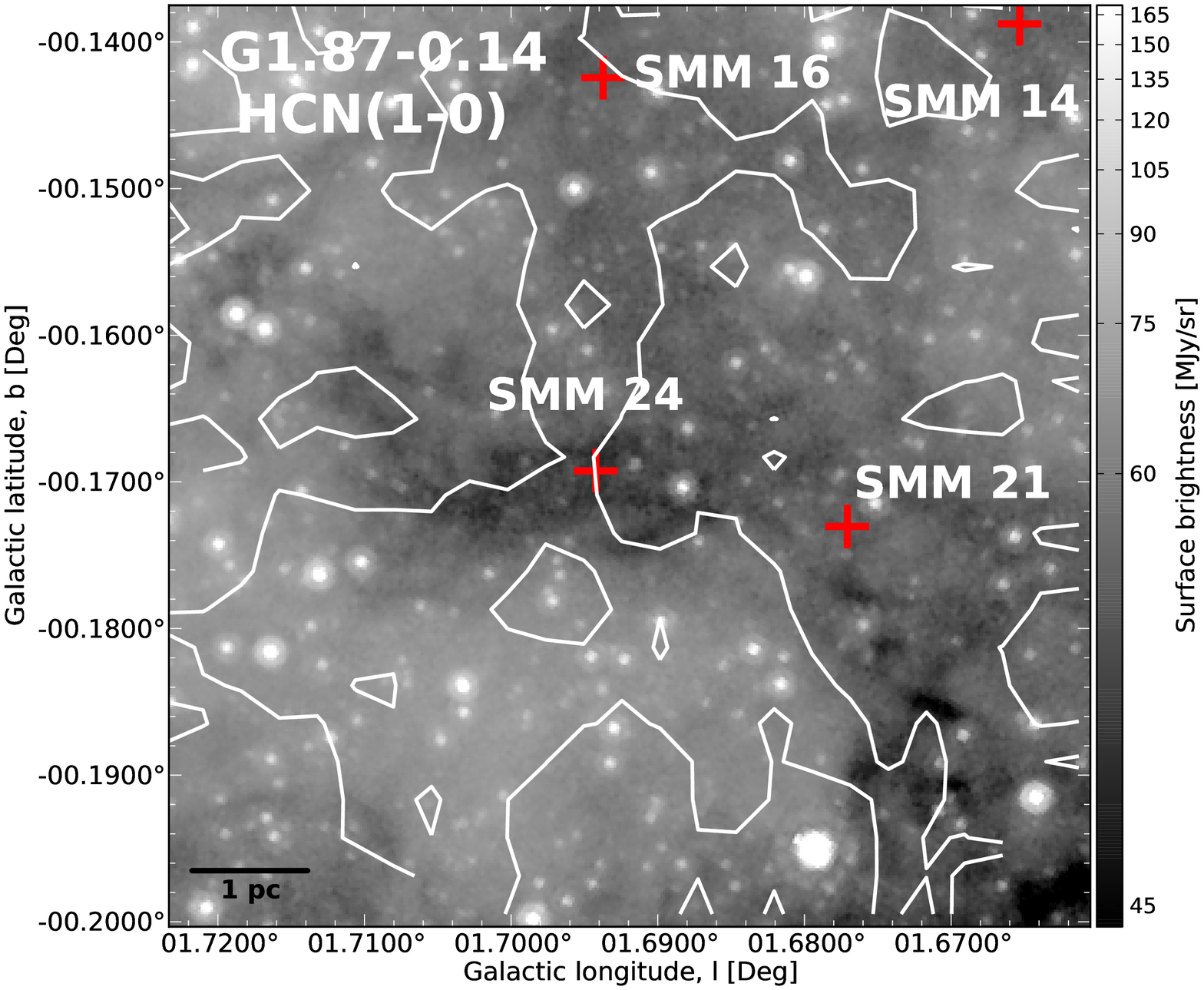}
\includegraphics[width=0.245\textwidth]{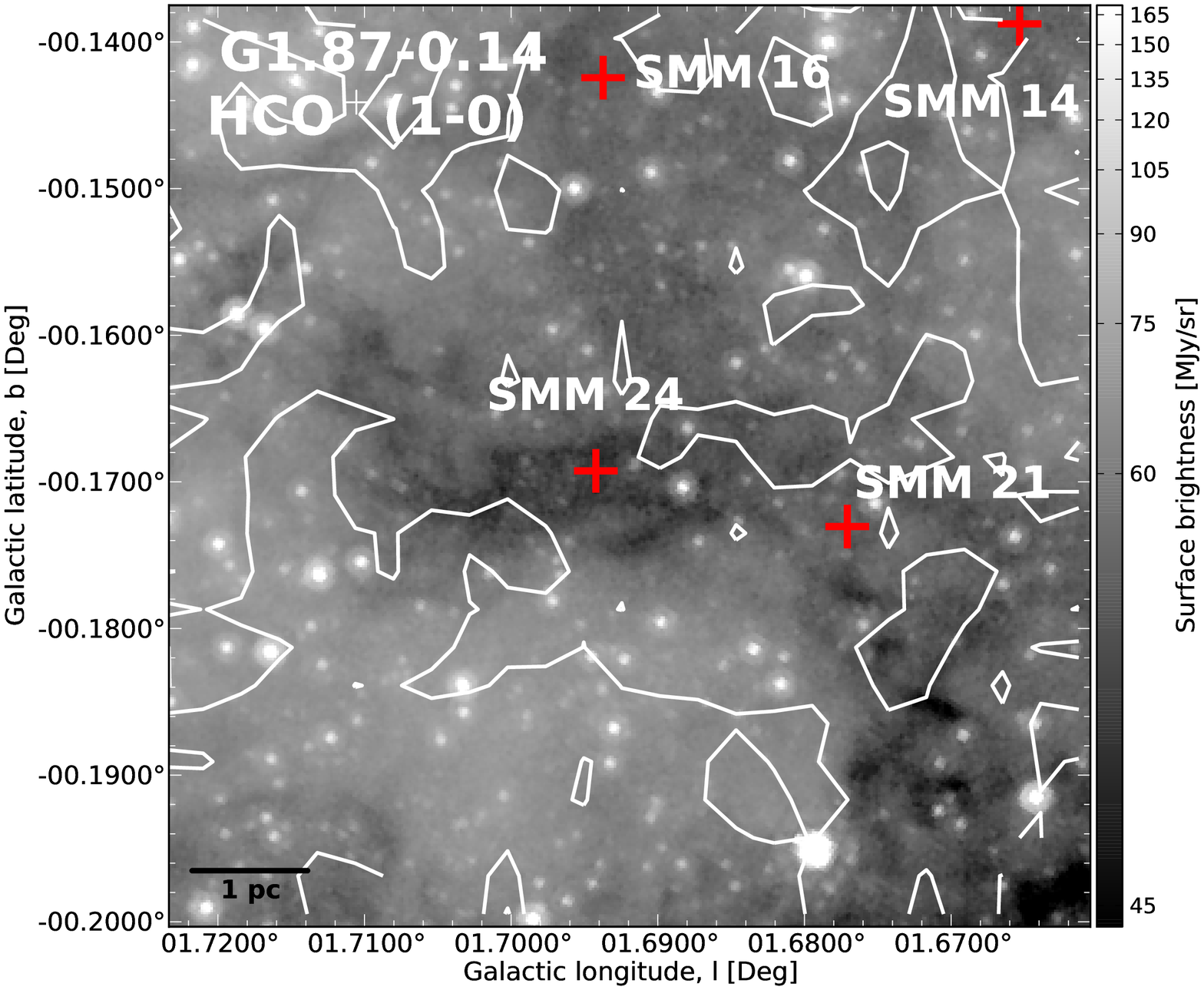}
\includegraphics[width=0.245\textwidth]{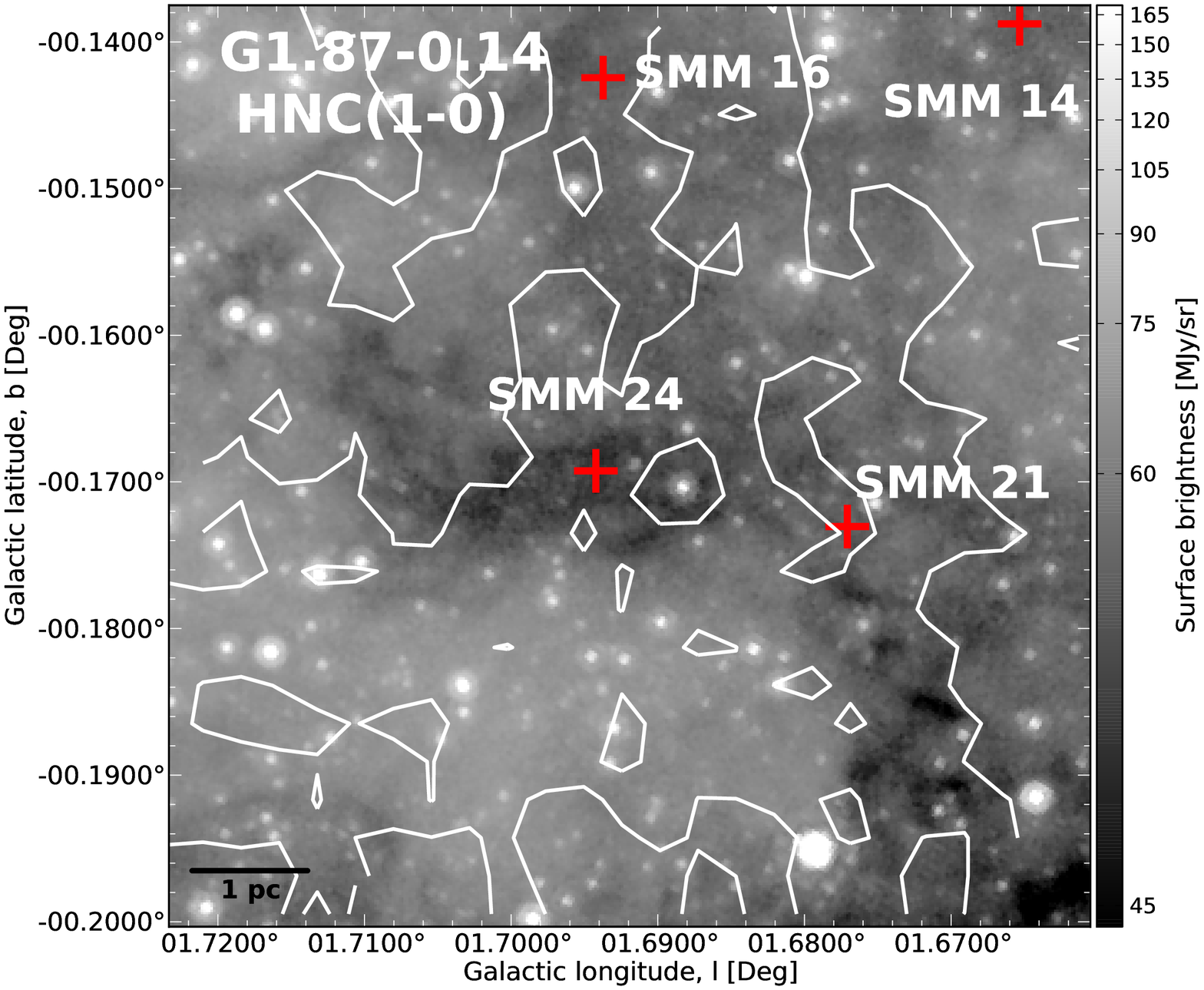}
\includegraphics[width=0.245\textwidth]{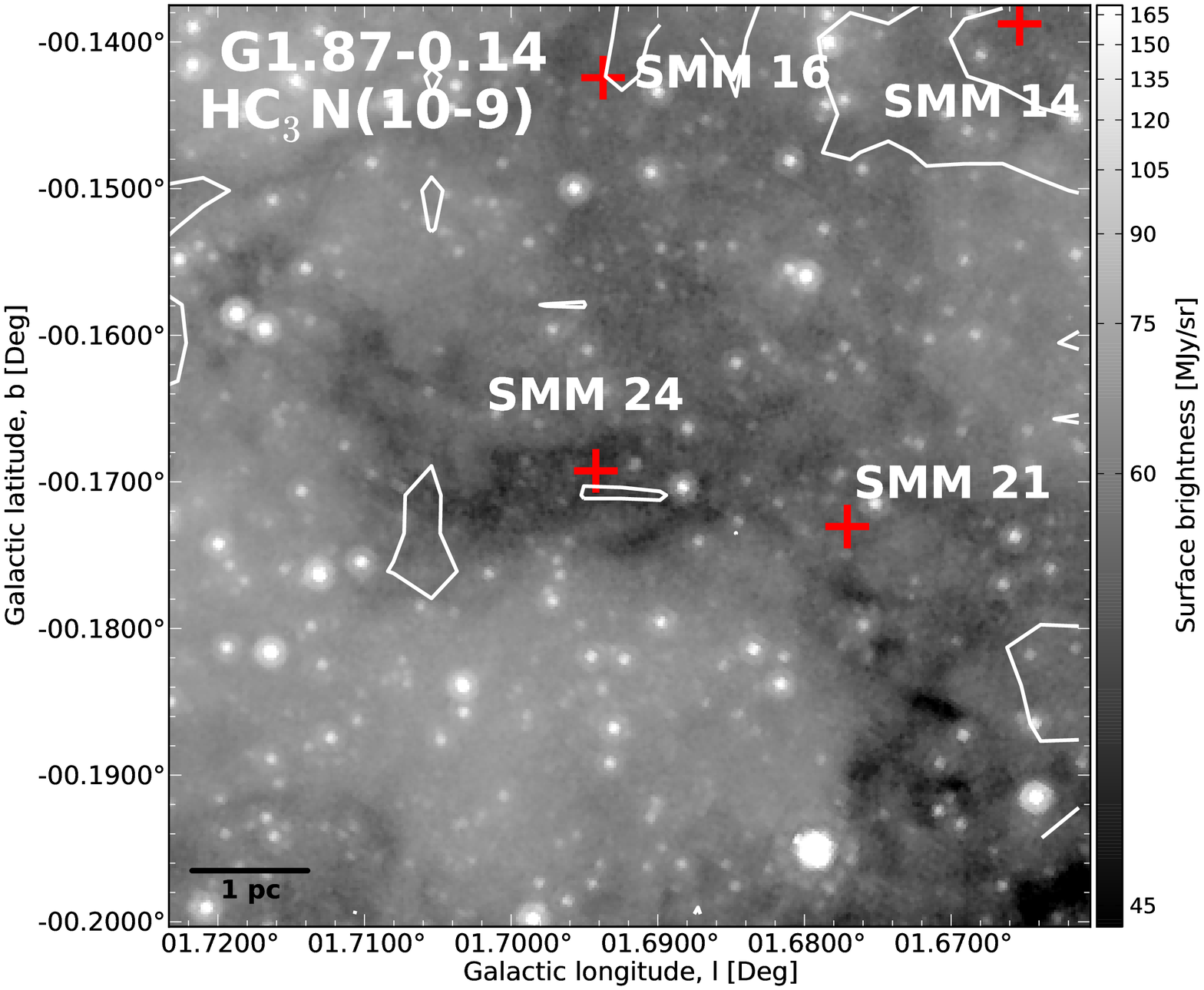}
\includegraphics[width=0.245\textwidth]{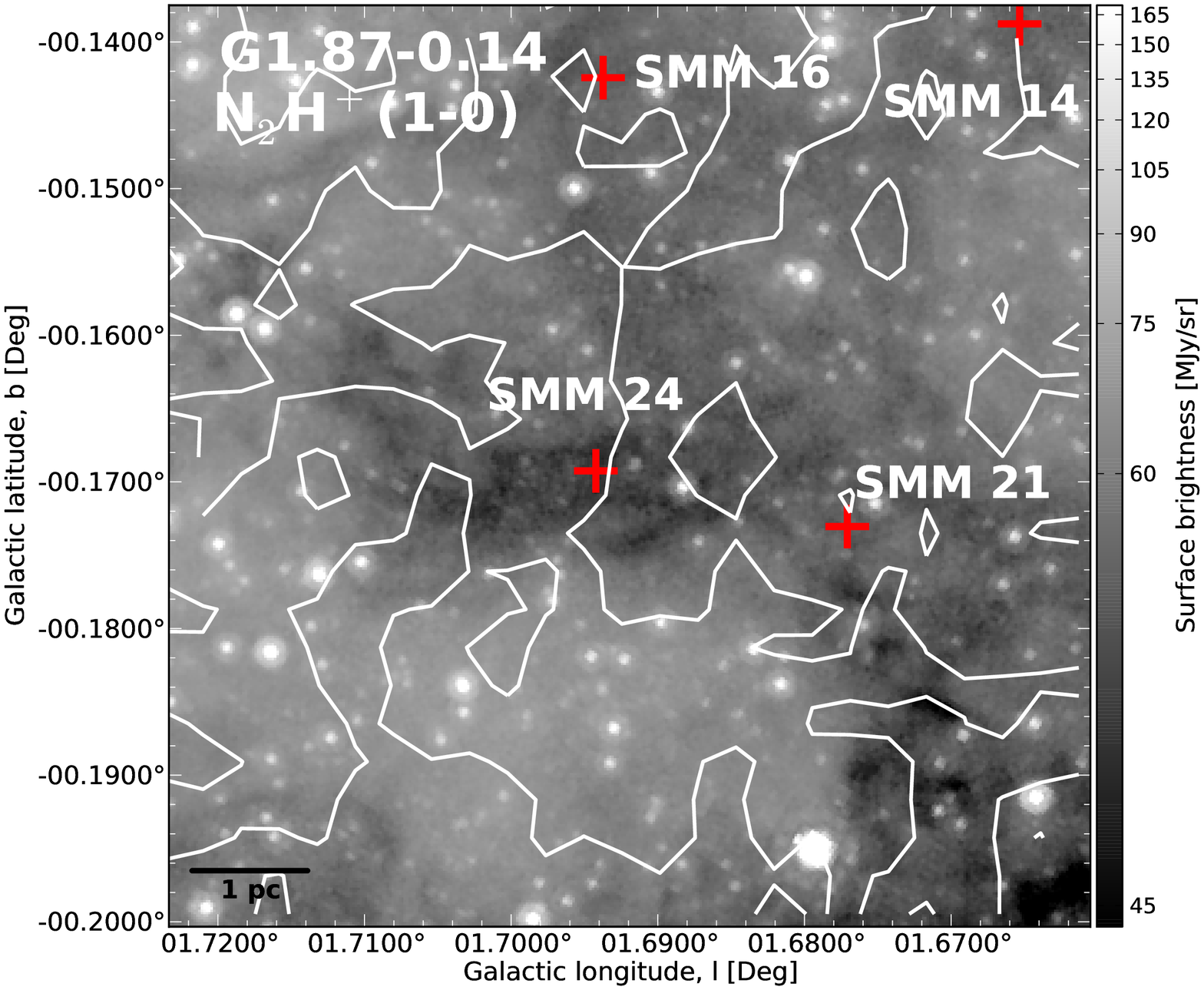}
\caption{Similar to Fig.~\ref{figure:G187SMM1lines} but towards 
G1.87--SMM 14, 16, 17, 21, 24. The contour levels start at 
$3\sigma$ for SiO, HN$^{13}$C, C$_2$H, and HC$_3$N. For 
HNCO$(4_{0,\,4}-3_{0,\,3})$, HCN, HCO$^+$, HNC, and N$_2$H$^+$, the contours 
start at $8\sigma$, $8\sigma$, $5\sigma$, $5\sigma$, and $5\sigma$, 
respectively. In all cases, the contours go in 
steps of $3\sigma$. The average $1\sigma$ value in $T_{\rm MB}$ units is 
$\sim0.78$ K~km~s$^{-1}$. The LABOCA 870-$\mu$m peak positions of the clumps
are marked by red plus signs. The 870-$\mu$m peak of SMM 17 
(west of SMM 21) lies just outside the MALT90 map 
(cf.~Fig.~\ref{figure:irac}, middle left panel). A scale bar 
indicating the 1 pc projected length is indicated. The HCN and HNC emissions 
show some morphological similarities.}
\label{figure:G187SMM24lines}
\end{center}
\end{figure*}

\begin{figure*}
\begin{center}
\includegraphics[width=0.245\textwidth]{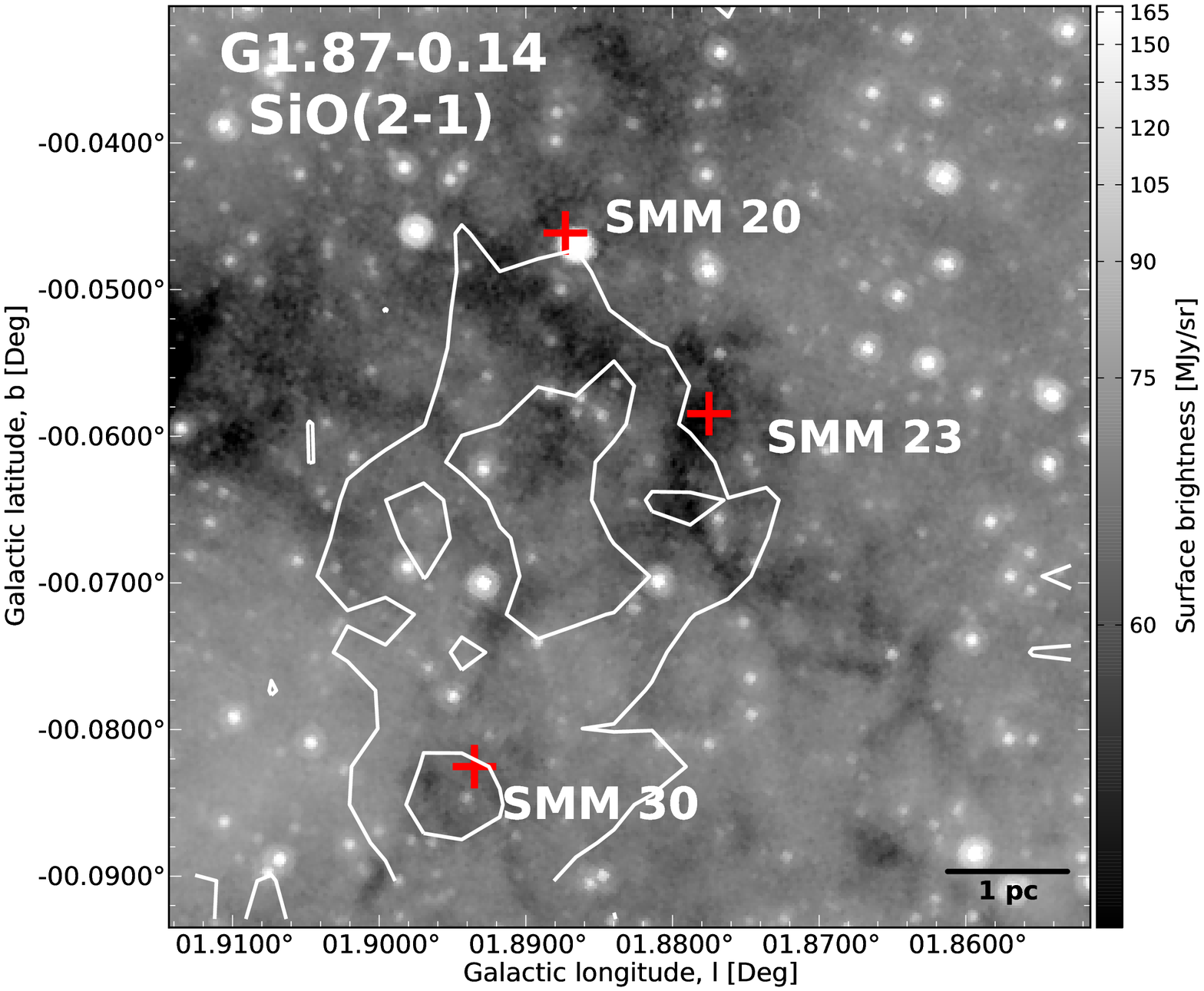}
\includegraphics[width=0.245\textwidth]{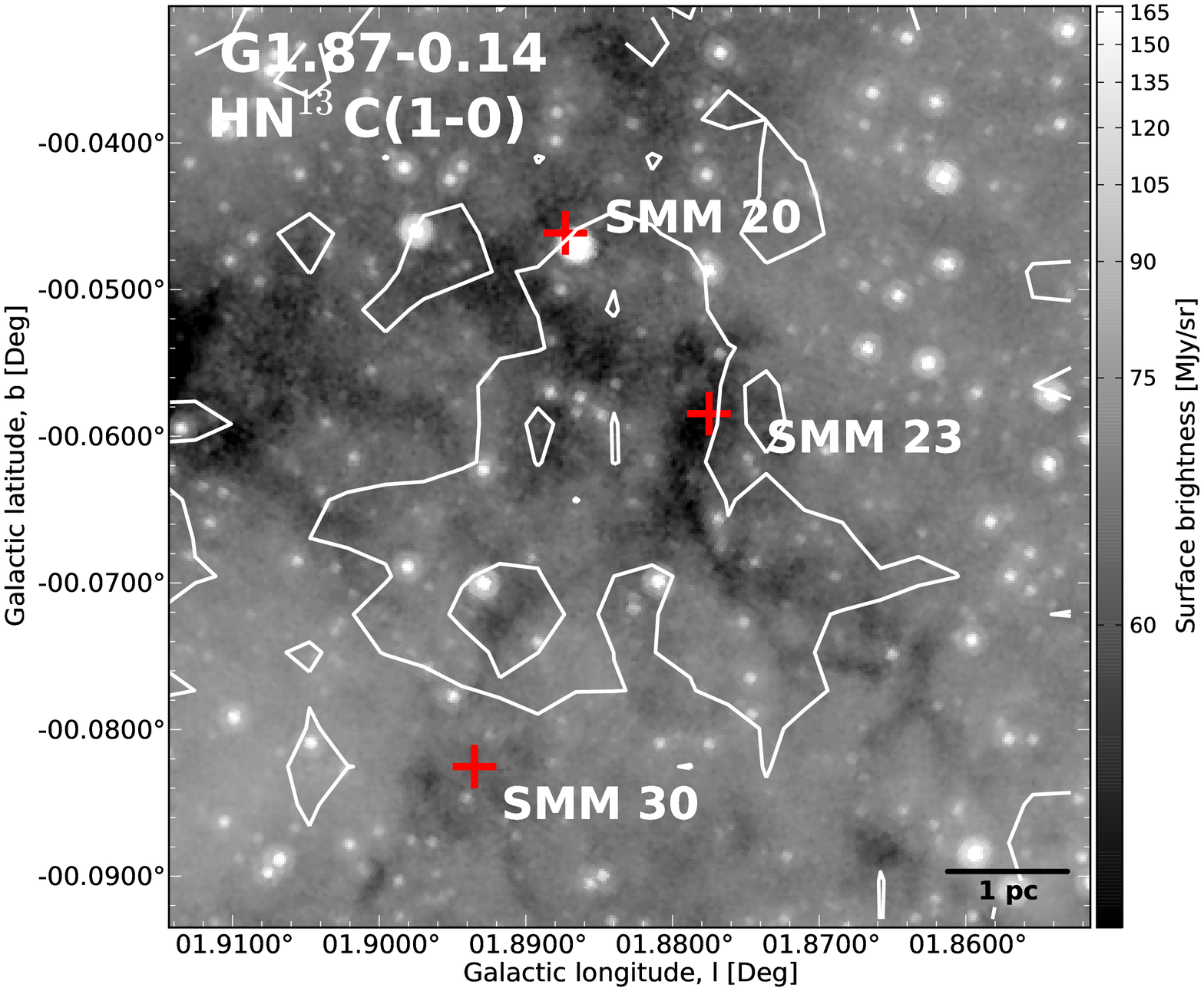}
\includegraphics[width=0.245\textwidth]{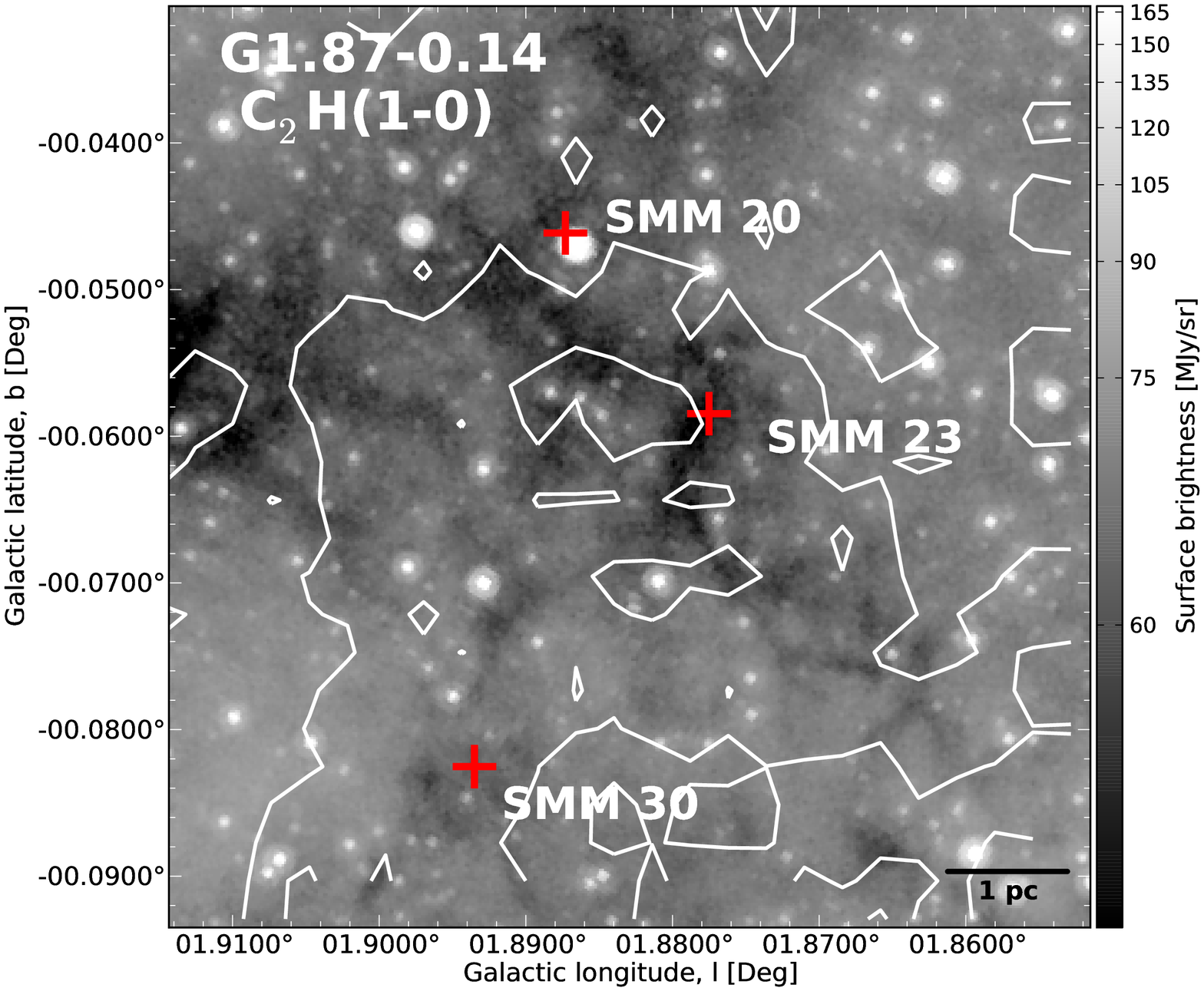}
\includegraphics[width=0.245\textwidth]{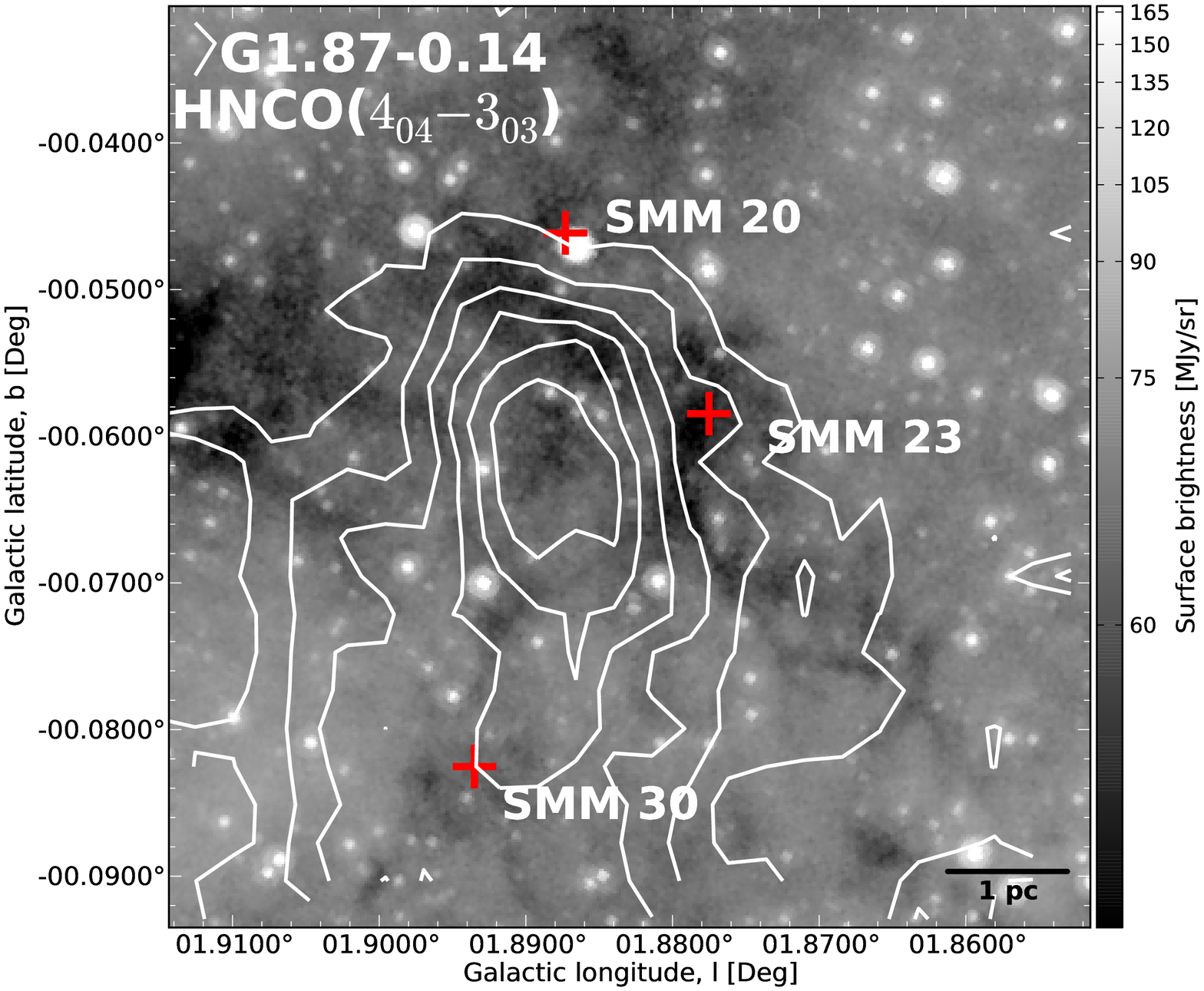}
\includegraphics[width=0.245\textwidth]{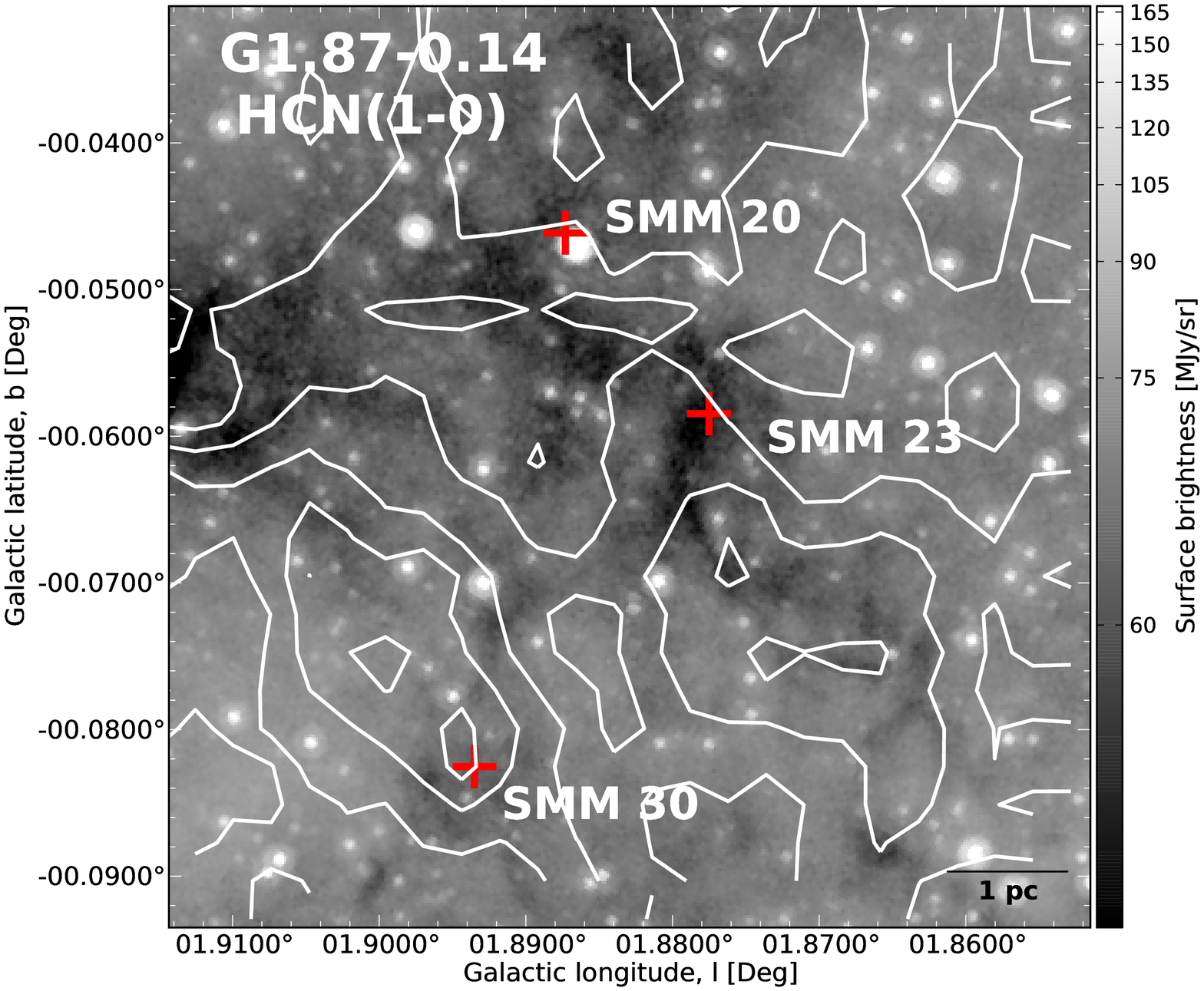}
\includegraphics[width=0.245\textwidth]{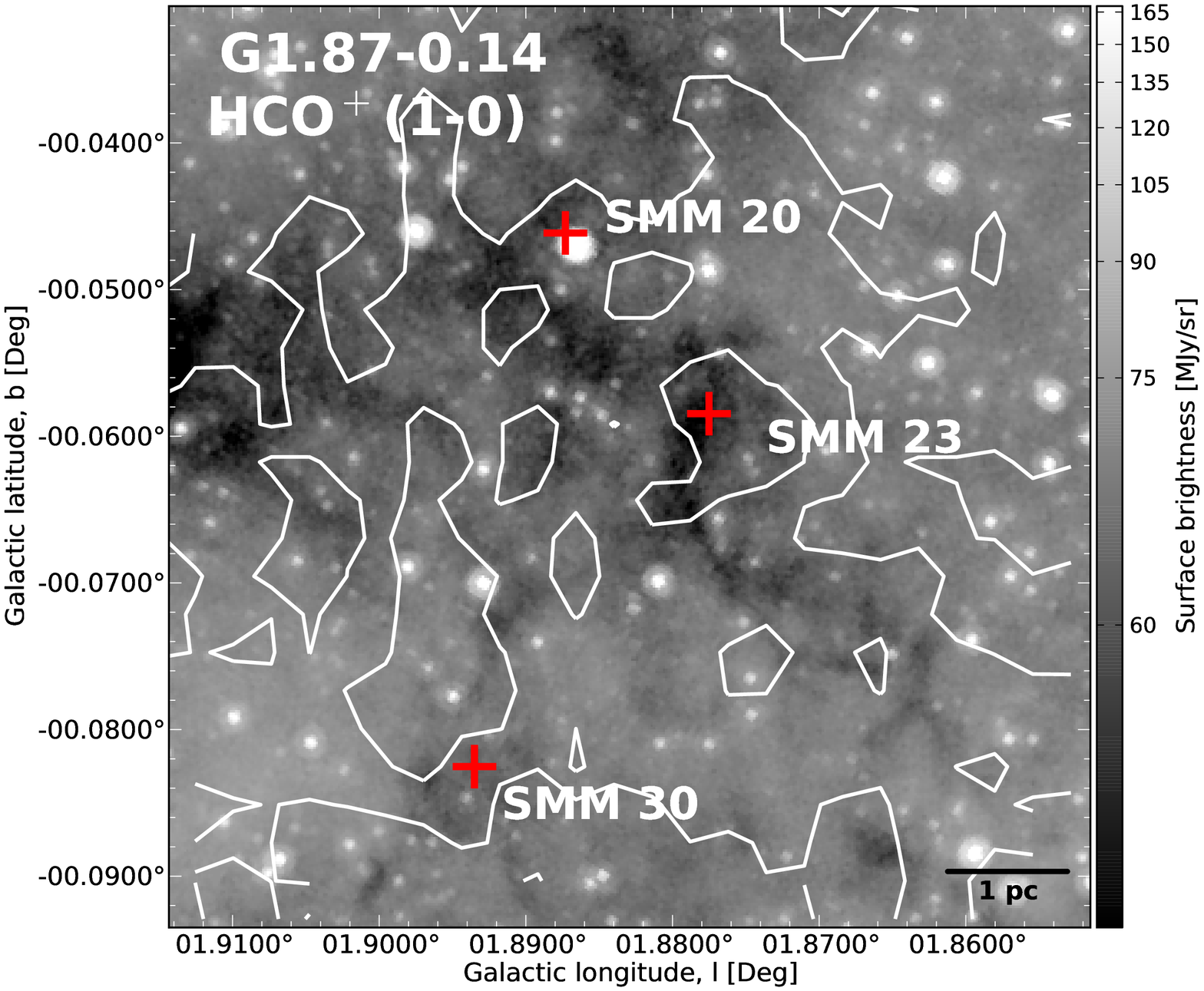}
\includegraphics[width=0.245\textwidth]{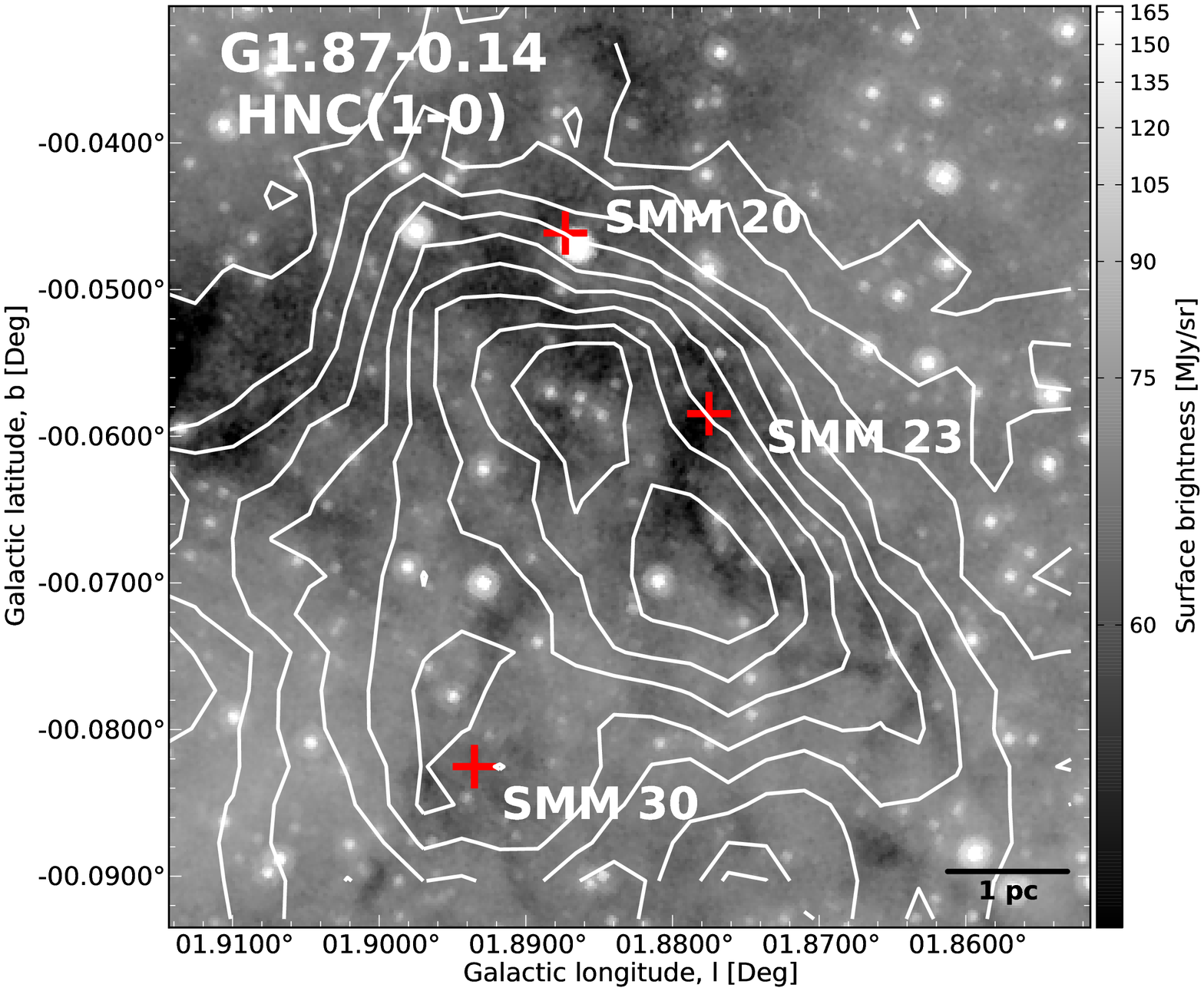}
\includegraphics[width=0.245\textwidth]{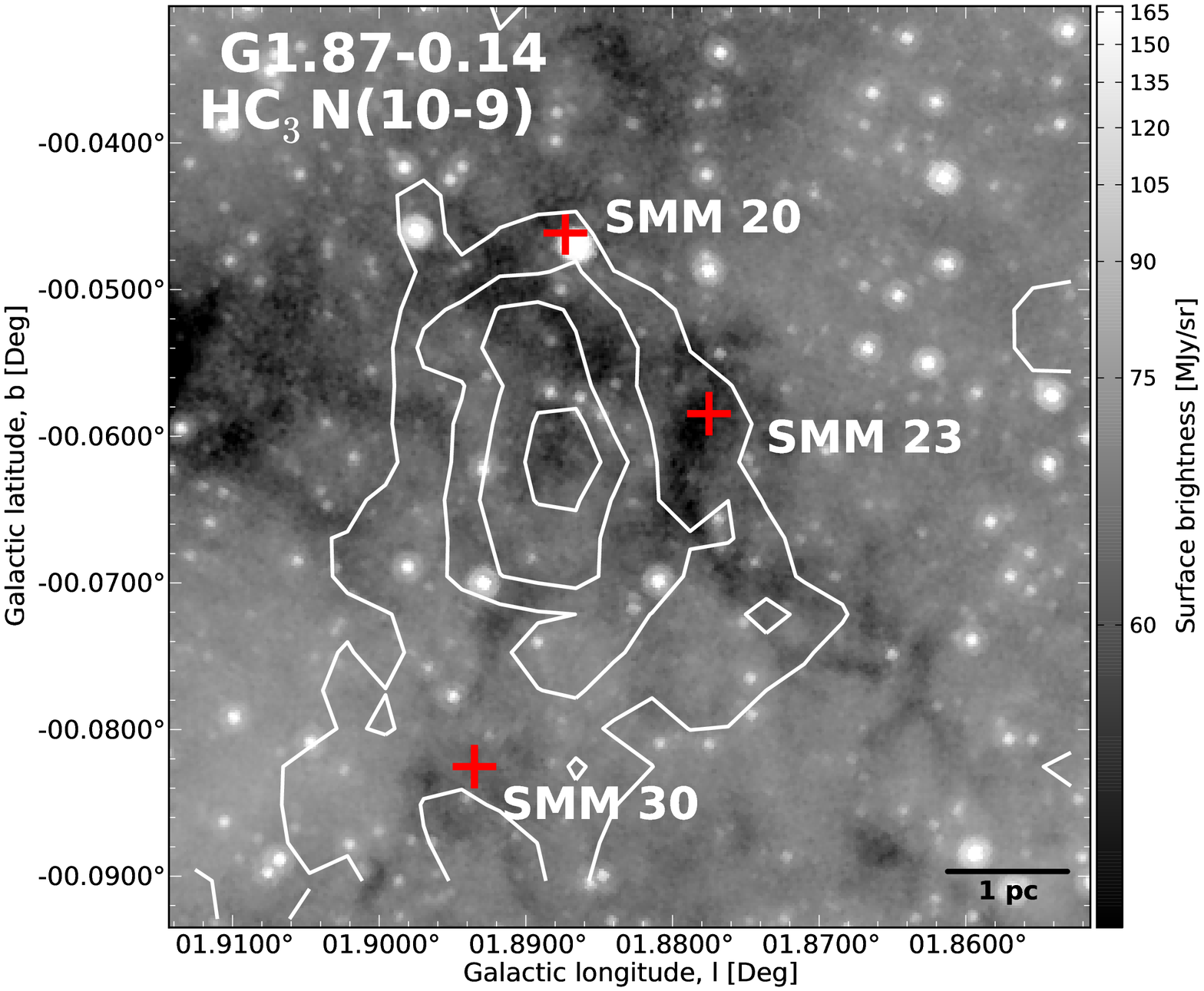}
\includegraphics[width=0.245\textwidth]{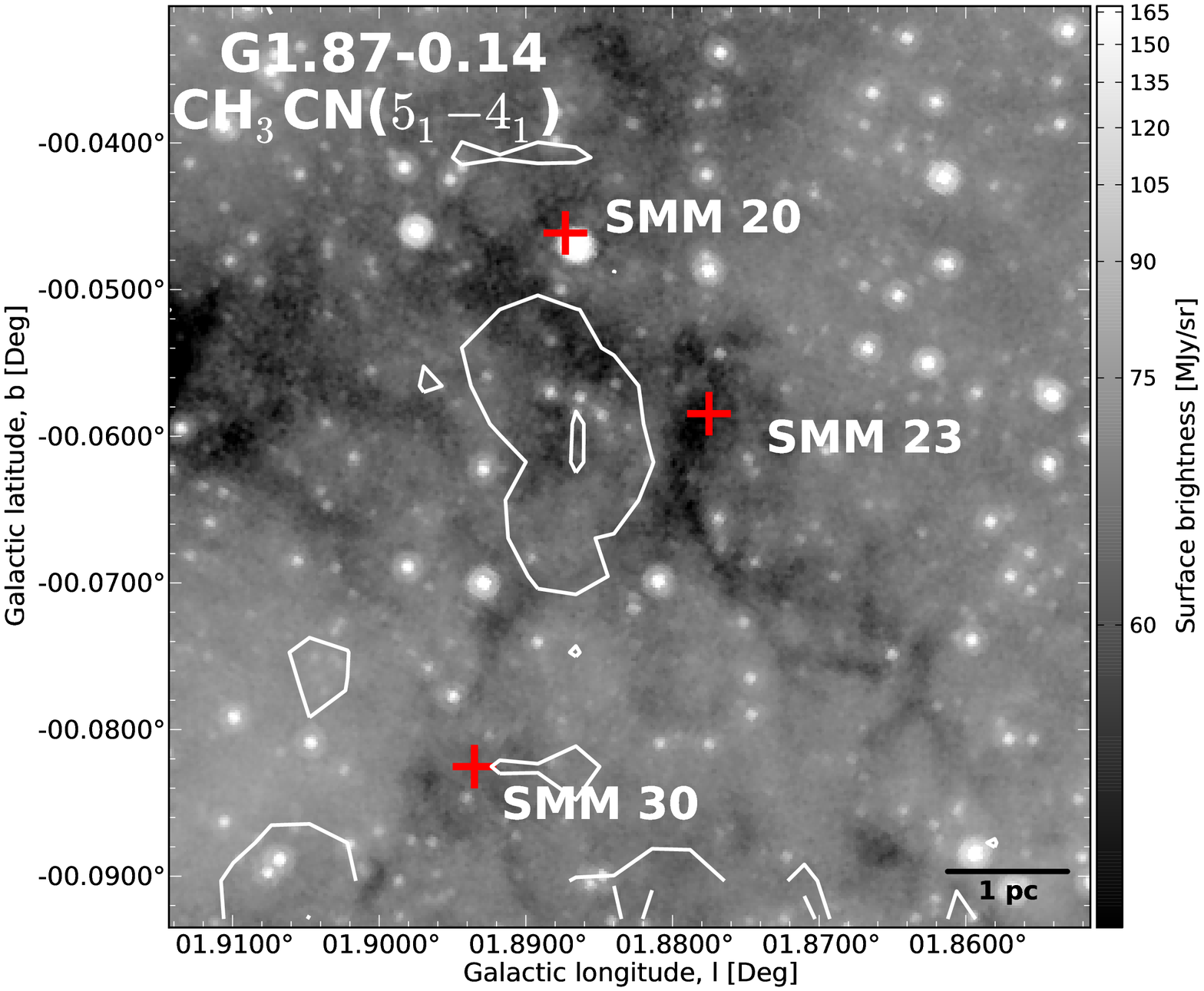}
\includegraphics[width=0.245\textwidth]{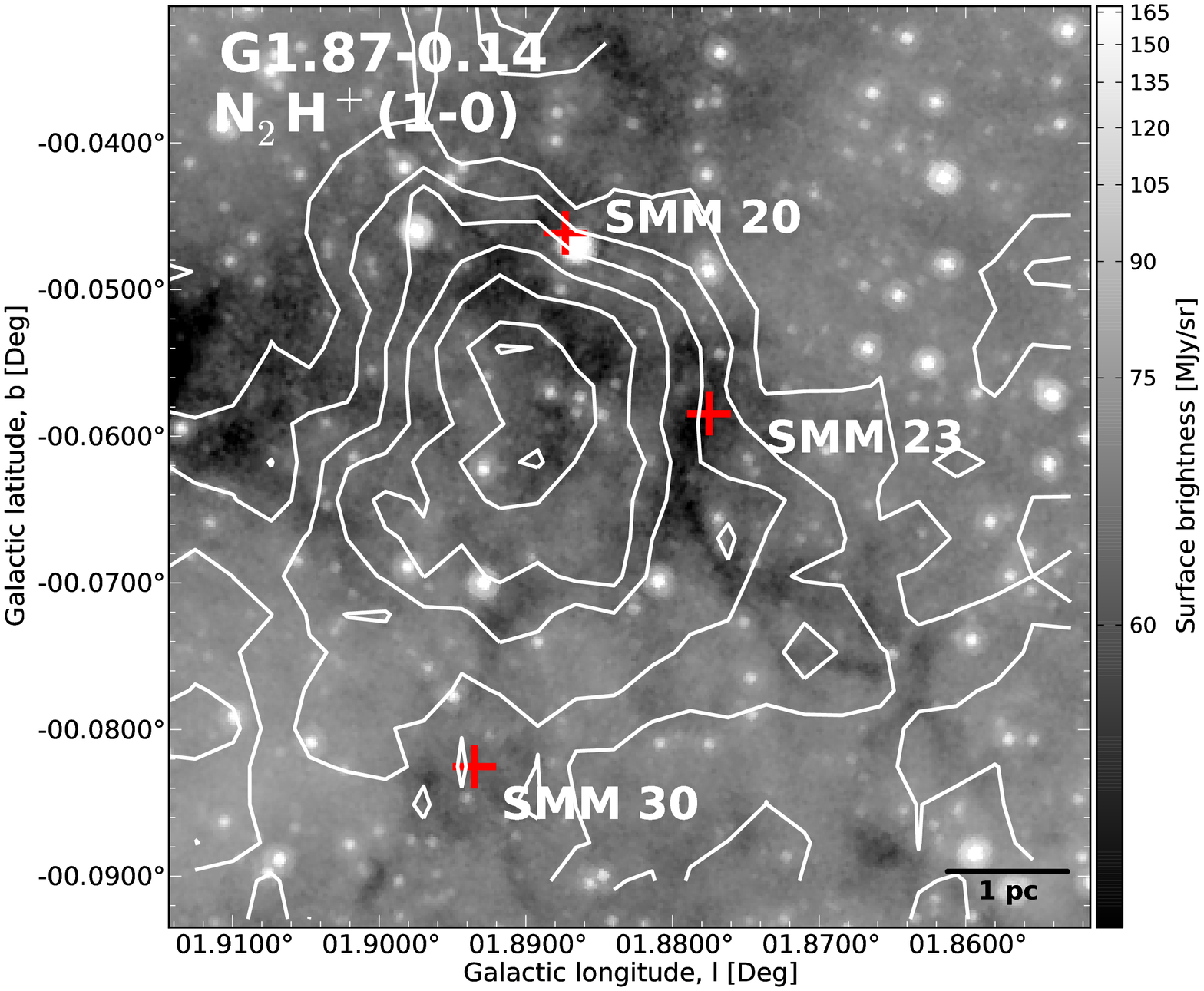}
\caption{Similar to Fig.~\ref{figure:G187SMM1lines} but towards 
G1.87--SMM 20, 23, 30. The contour levels start at 
$3\sigma$ for HN$^{13}$C, C$_2$H, and CH$_3$CN, and $5\sigma$ for
SiO, HNCO$(4_{0,\,4}-3_{0,\,3})$, HCN, HCO$^+$, HNC, HC$_3$N, and N$_2$H$^+$. 
In all cases, the contours go in 
steps of $3\sigma$. The average $1\sigma$ value in $T_{\rm MB}$ units is 
$\sim0.63$ K~km~s$^{-1}$. The LABOCA 870-$\mu$m peak positions of the clumps
are marked by red plus signs. A scale bar indicating the 1 pc 
projected length is indicated. Good correlations are seen between the peak 
emissions of SiO, HNCO, HC$_3$N, CH$_3$CN, and N$_2$H$^+$. The HNC emission 
also shows it maxima in between the submm peaks.}
\label{figure:G187SMM23lines}
\end{center}
\end{figure*}

\begin{figure*}
\begin{center}
\includegraphics[width=0.245\textwidth]{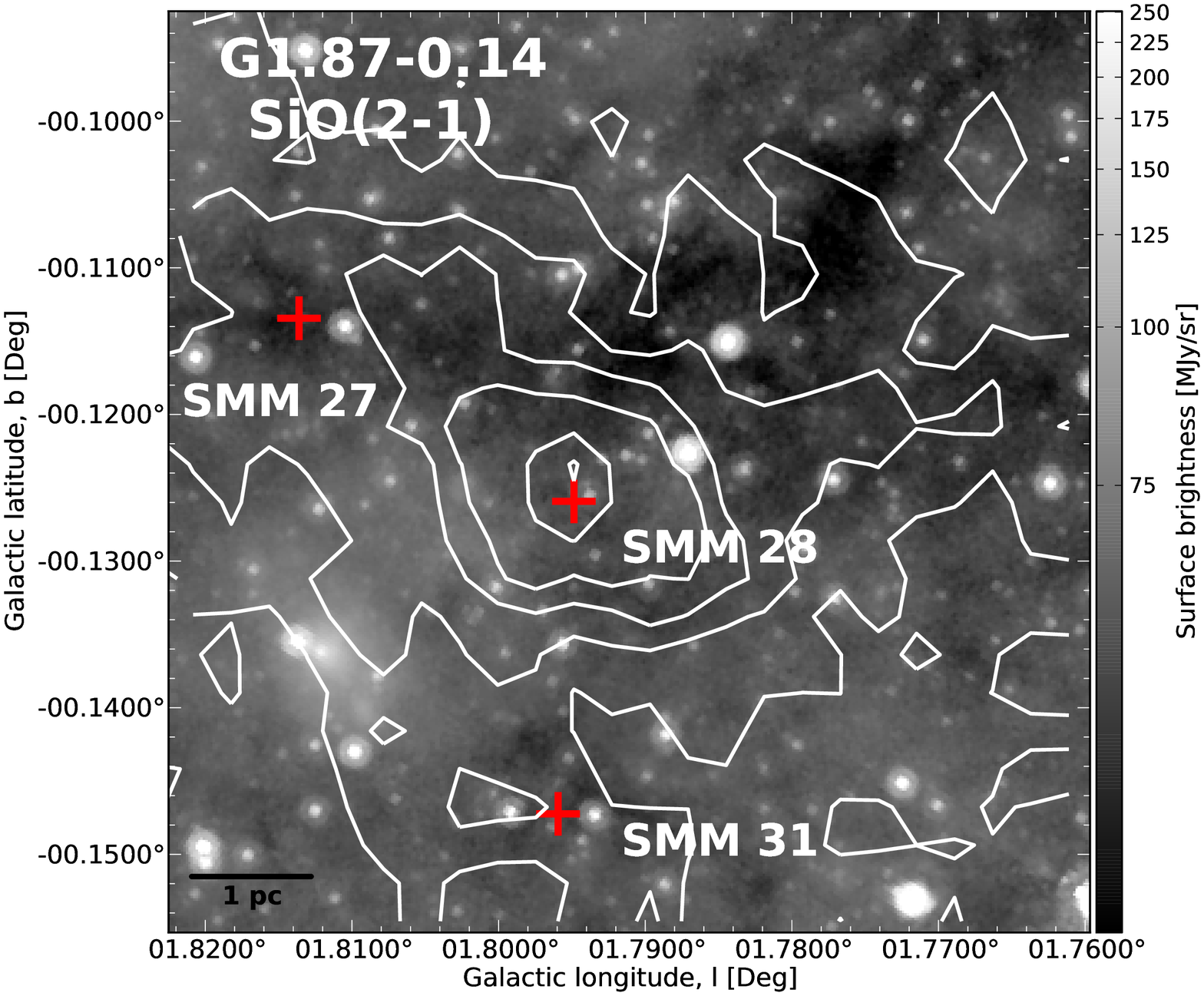}
\includegraphics[width=0.245\textwidth]{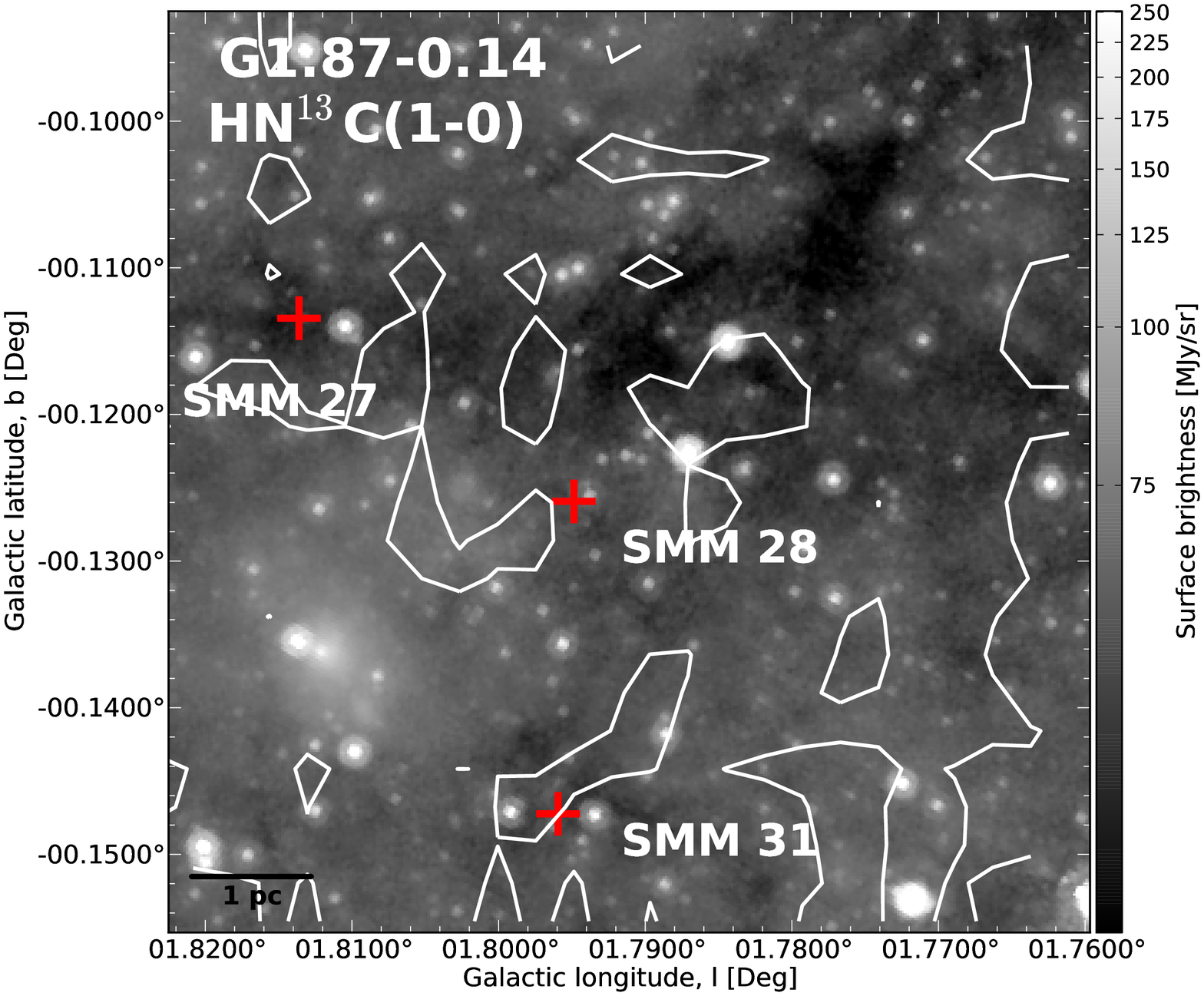}
\includegraphics[width=0.245\textwidth]{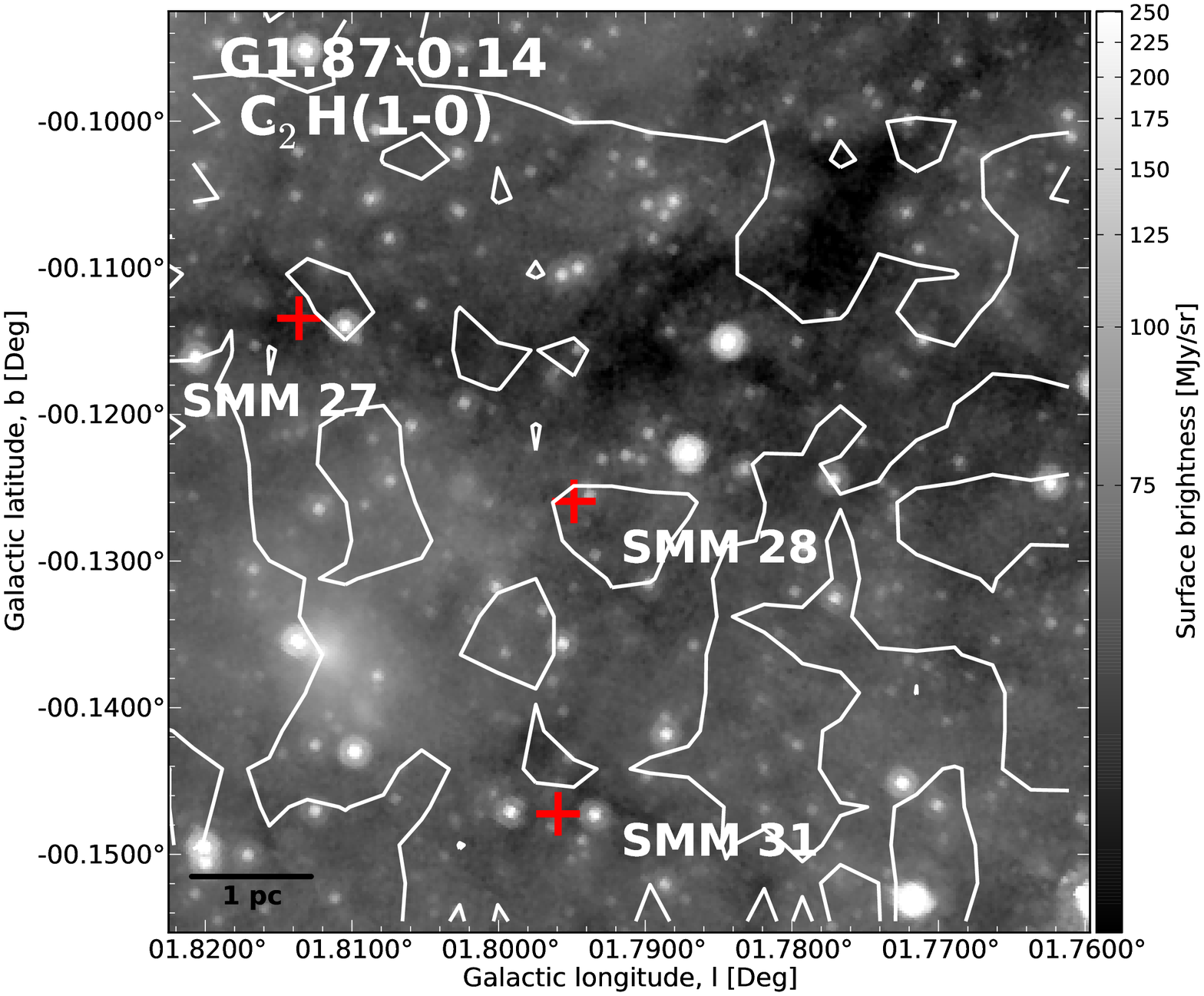}
\includegraphics[width=0.245\textwidth]{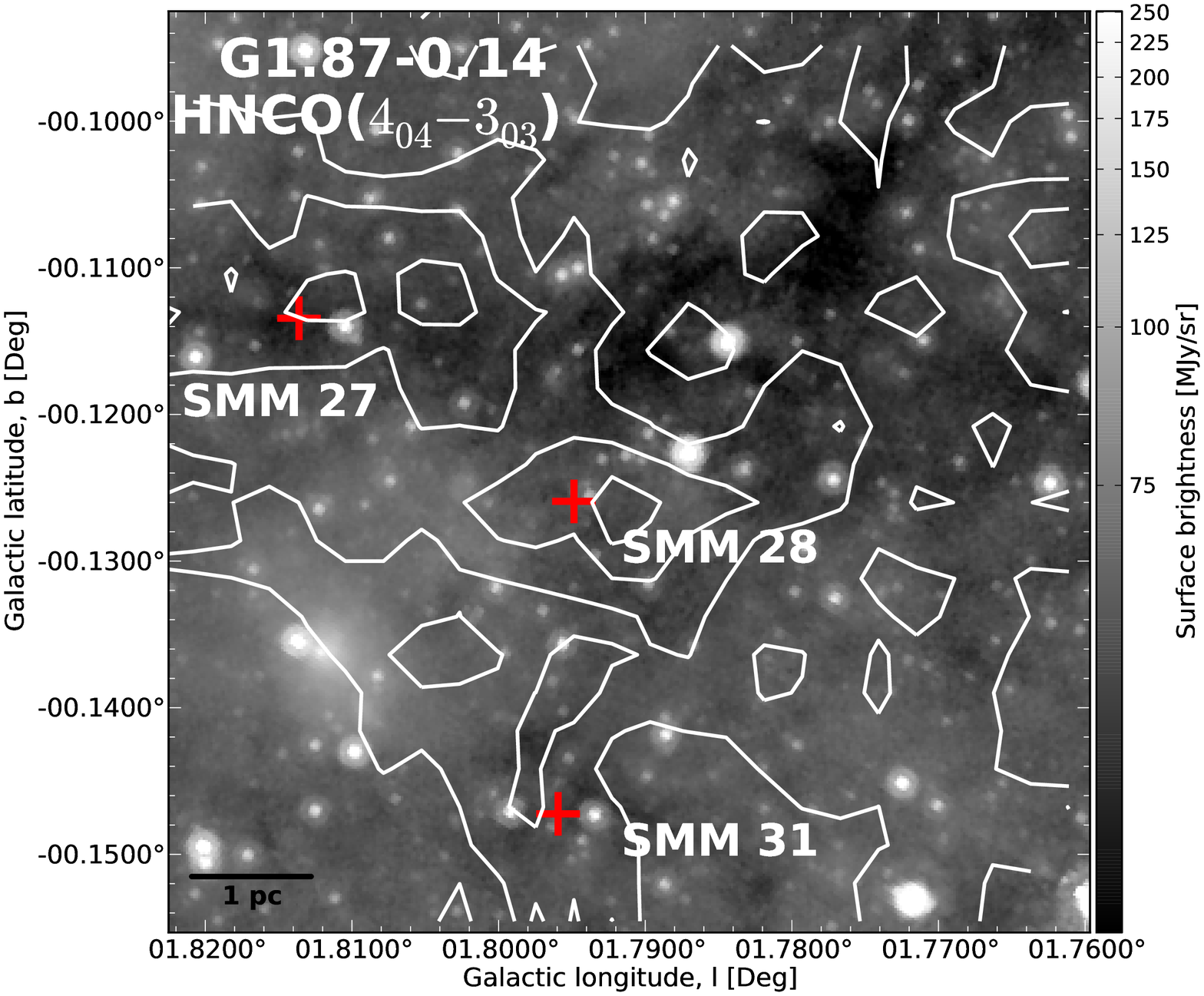}
\includegraphics[width=0.245\textwidth]{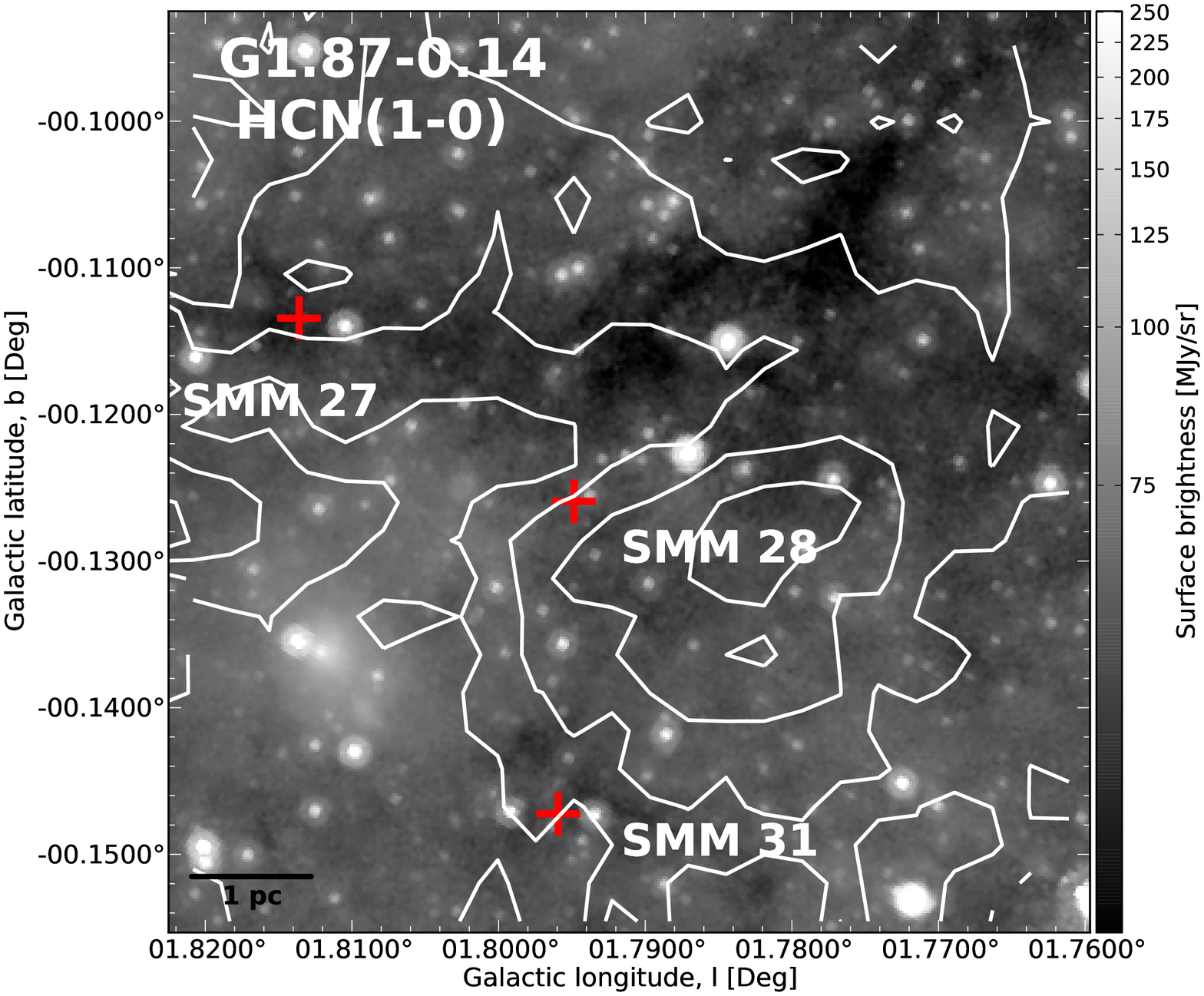}
\includegraphics[width=0.245\textwidth]{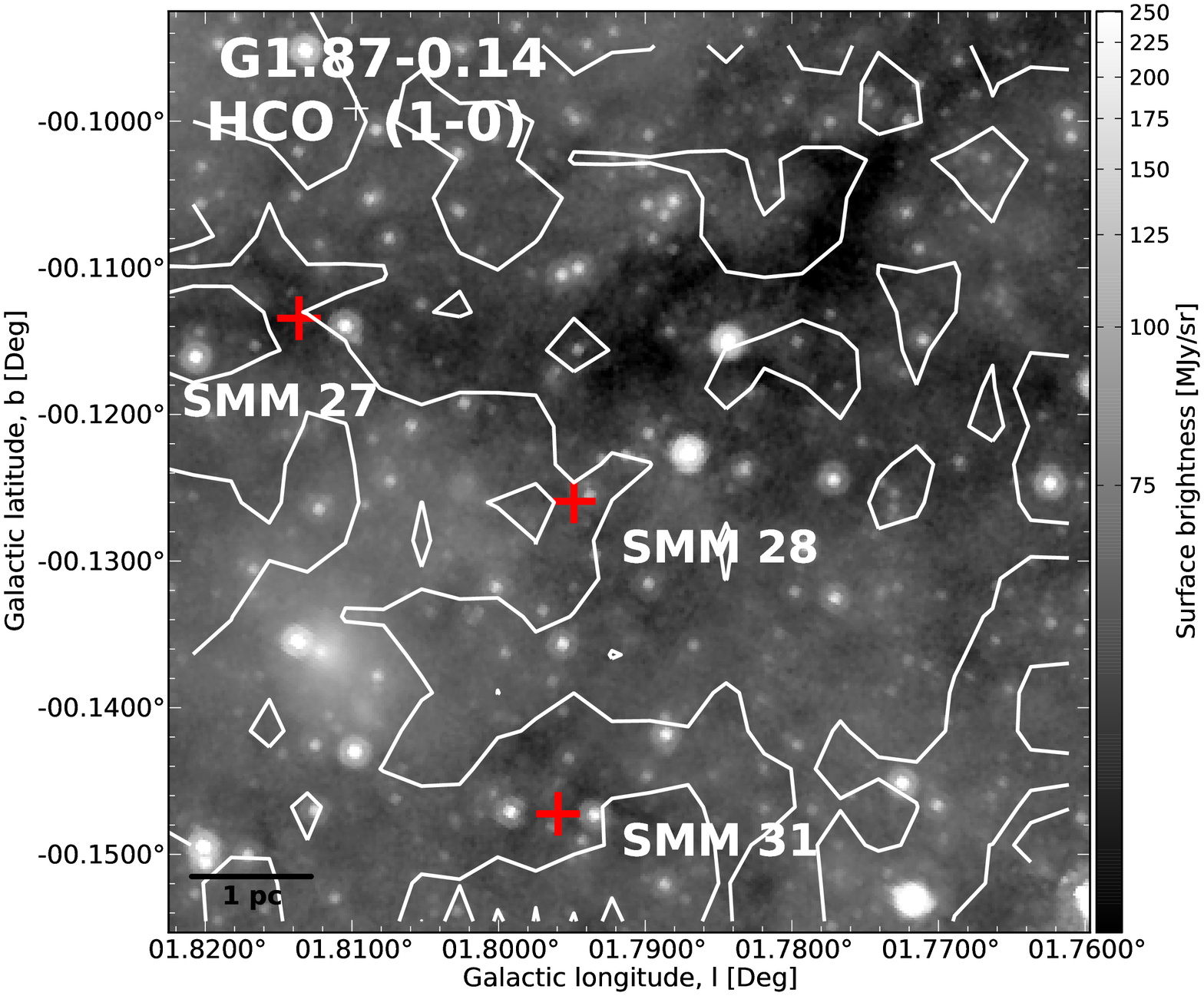}
\includegraphics[width=0.245\textwidth]{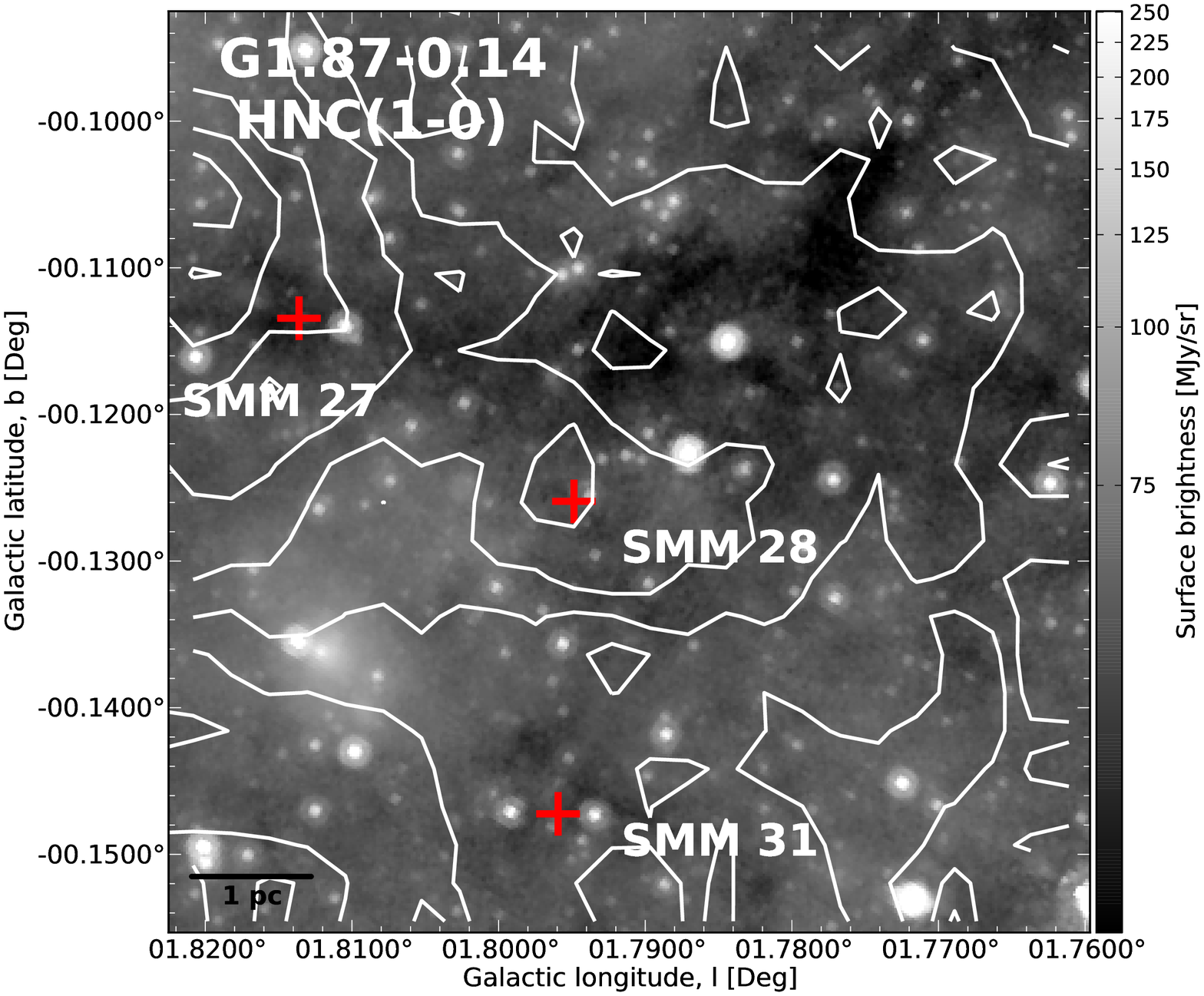}
\includegraphics[width=0.245\textwidth]{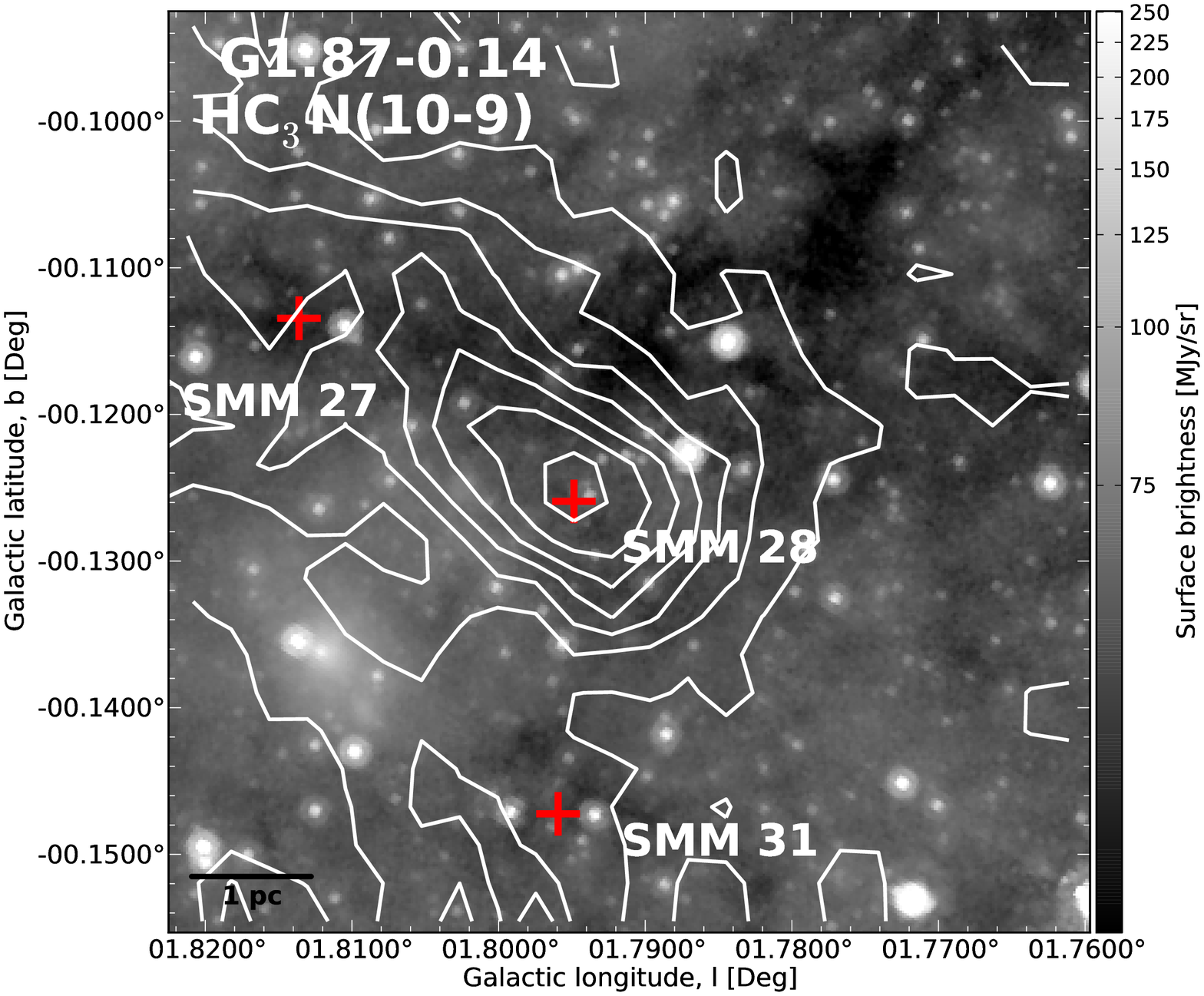}
\includegraphics[width=0.245\textwidth]{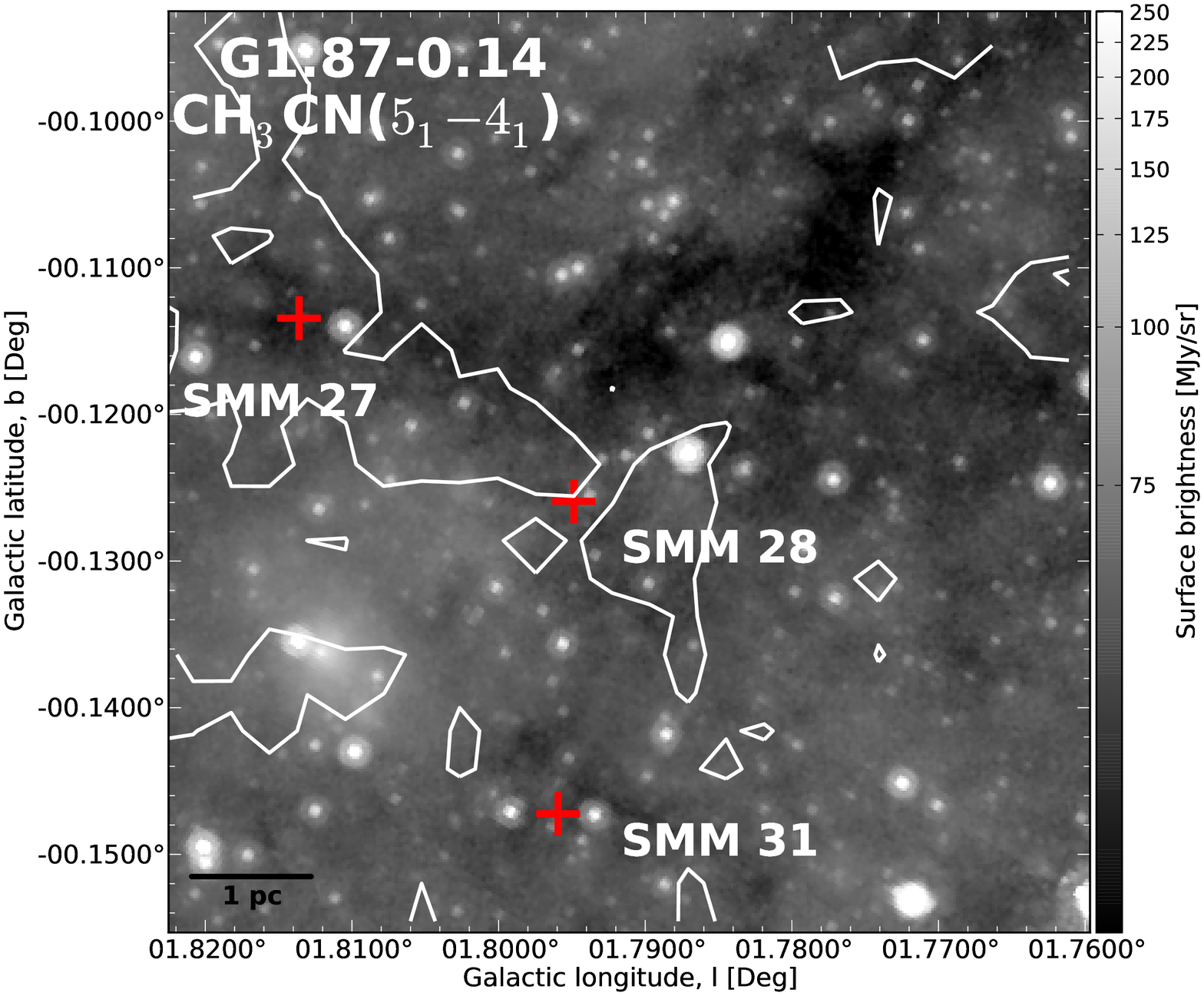}
\includegraphics[width=0.245\textwidth]{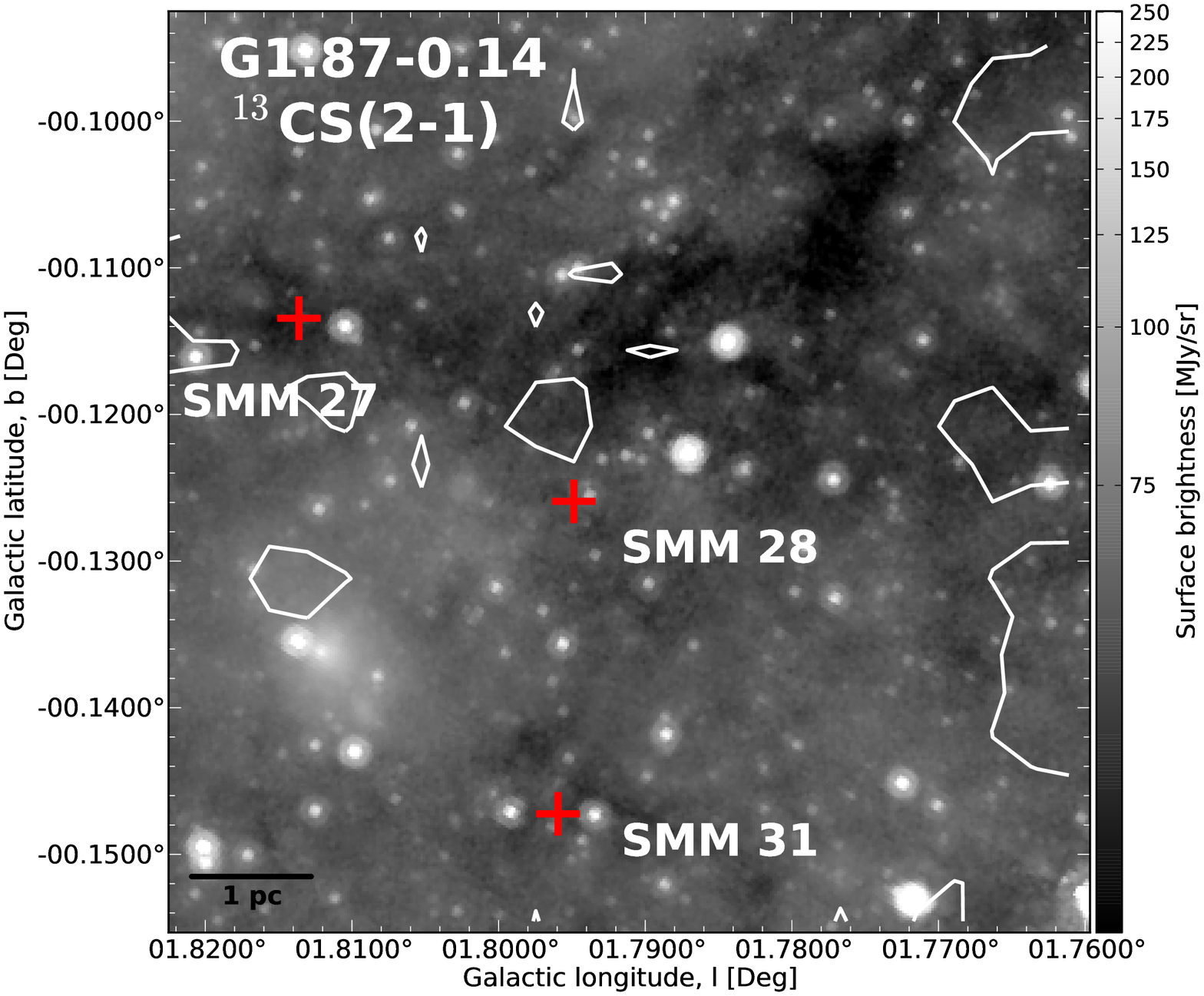}
\includegraphics[width=0.245\textwidth]{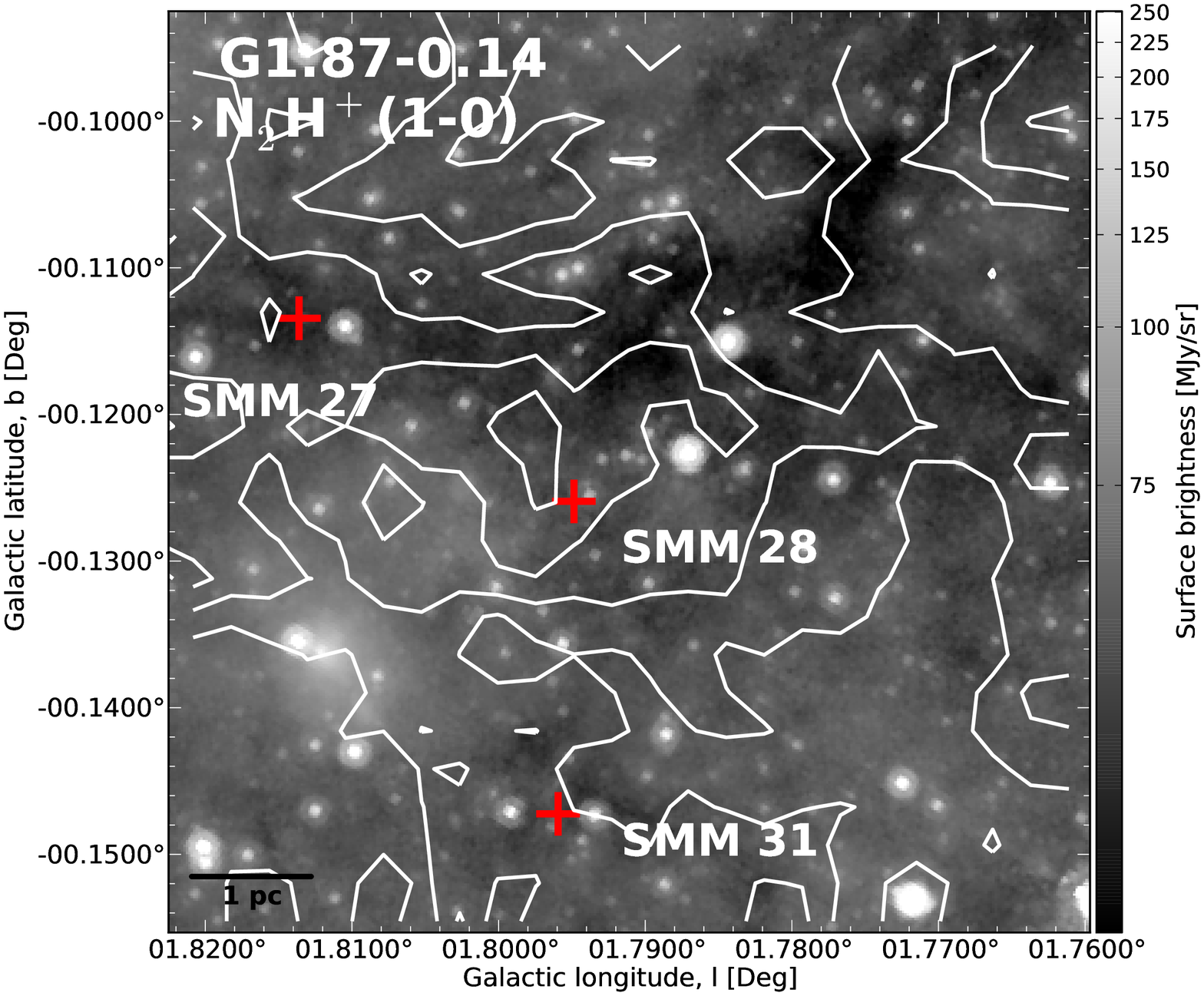}
\caption{Similar to Fig.~\ref{figure:G187SMM1lines} but towards 
G1.87--SMM 27, 28, 31. The contour levels start at $3\sigma$ for SiO, 
HN$^{13}$C, C$_2$H, CH$_3$CN, and $^{13}$CS. For, HNCO$(4_{0,\,4}-3_{0,\,3})$, HNC, 
HCN, HCO$^+$, HC$_3$N, and N$_2$H$^+$, the contours start at $9\sigma$, 
$7\sigma$, $7\sigma$, $4\sigma$, $4\sigma$, and $5\sigma$, respectively. 
In all cases, the contours go in steps of $3\sigma$. The average $1\sigma$ 
value in $T_{\rm MB}$ units is $\sim0.63$ K~km~s$^{-1}$. The LABOCA 870-$\mu$m 
peak positions are marked by red plus signs. A scale bar indicating 
the 1 pc projected length is indicated. The SiO and HC$_3$N emission 
morphologies resemble each other. Also HNC and N$_2$H$^+$ share some common 
features (e.g., peak close to SMM 28).}
\label{figure:G187SMM28lines}
\end{center}
\end{figure*}

\begin{figure*}
\begin{center}
\includegraphics[width=0.245\textwidth]{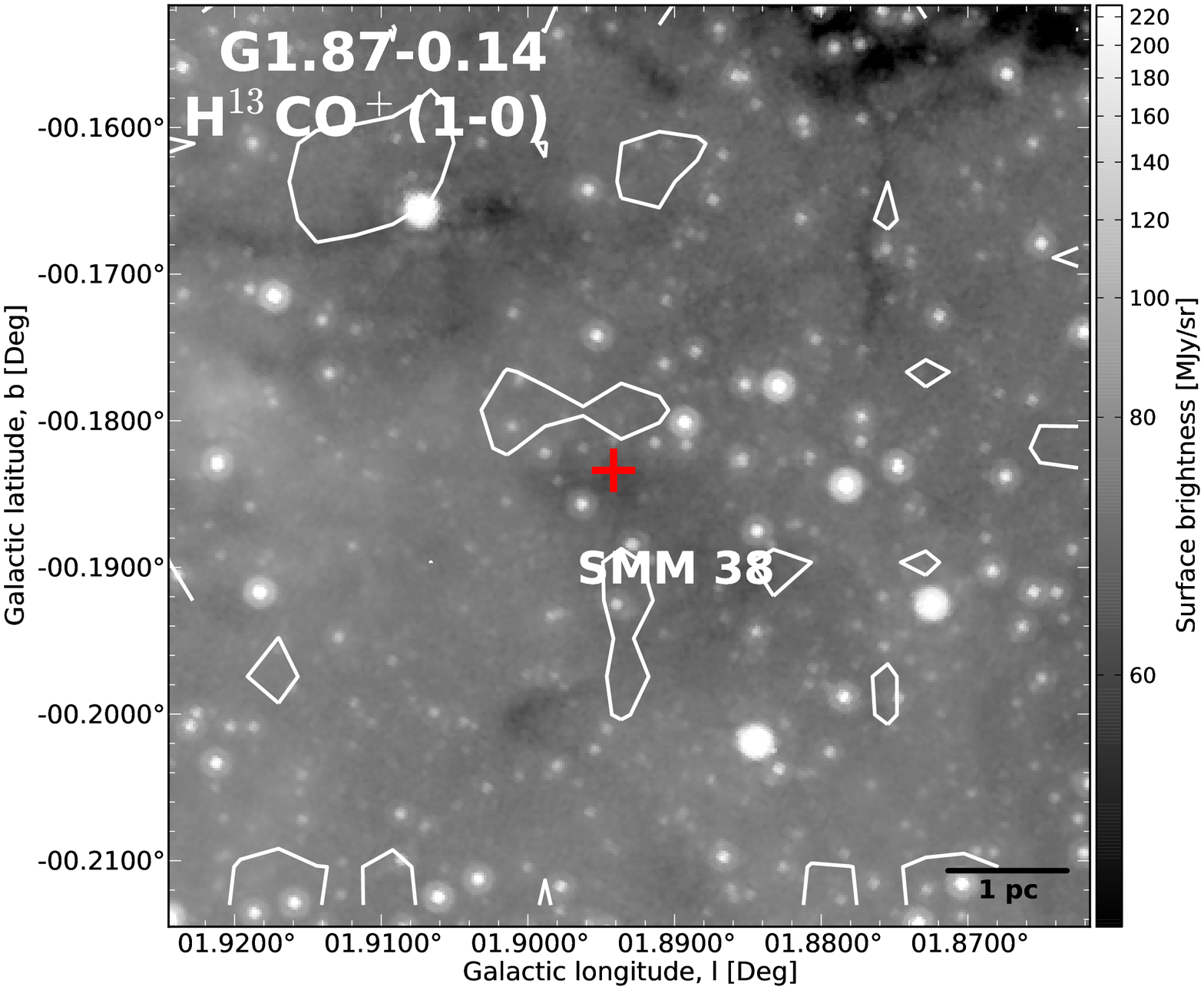}
\includegraphics[width=0.245\textwidth]{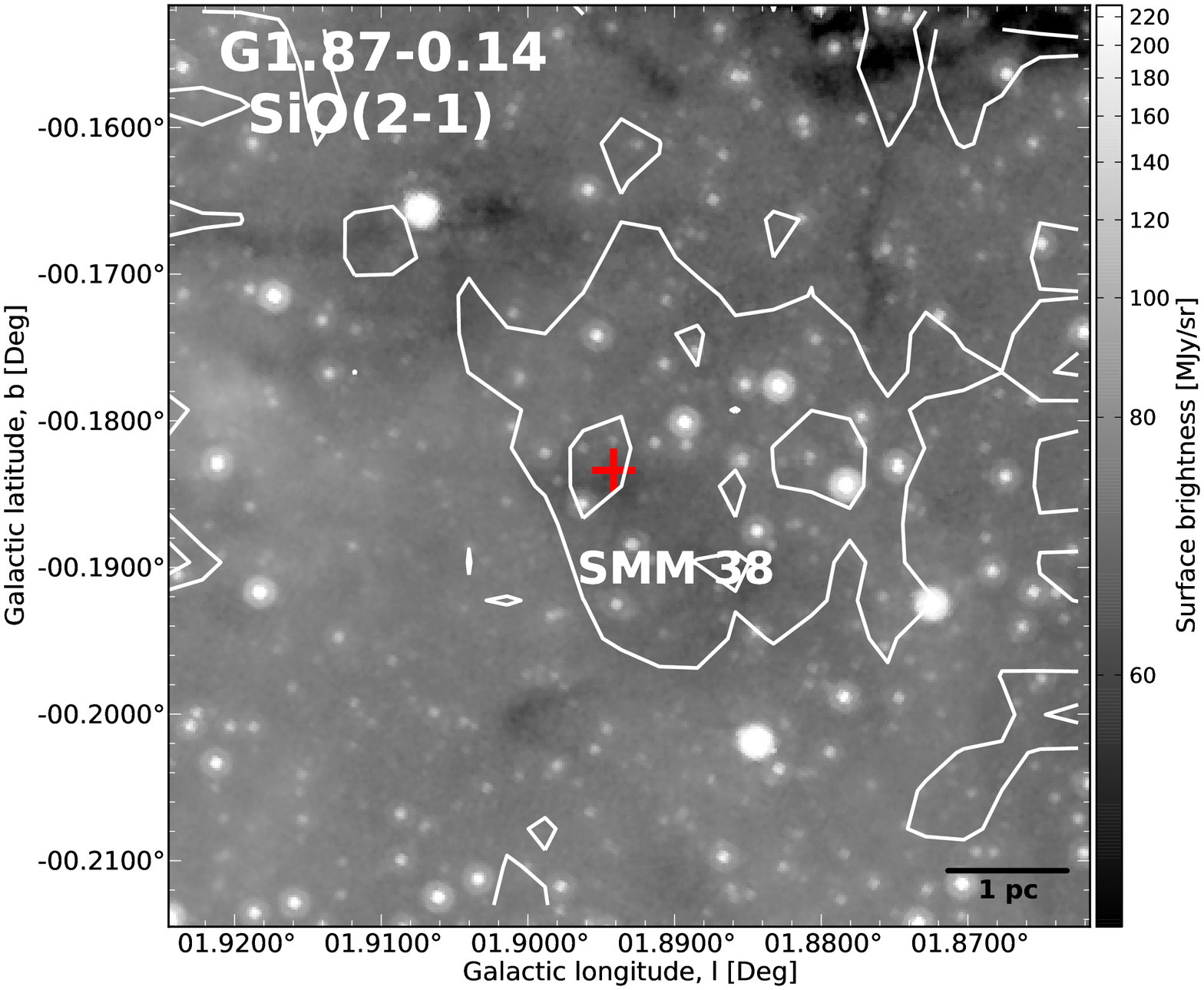}
\includegraphics[width=0.245\textwidth]{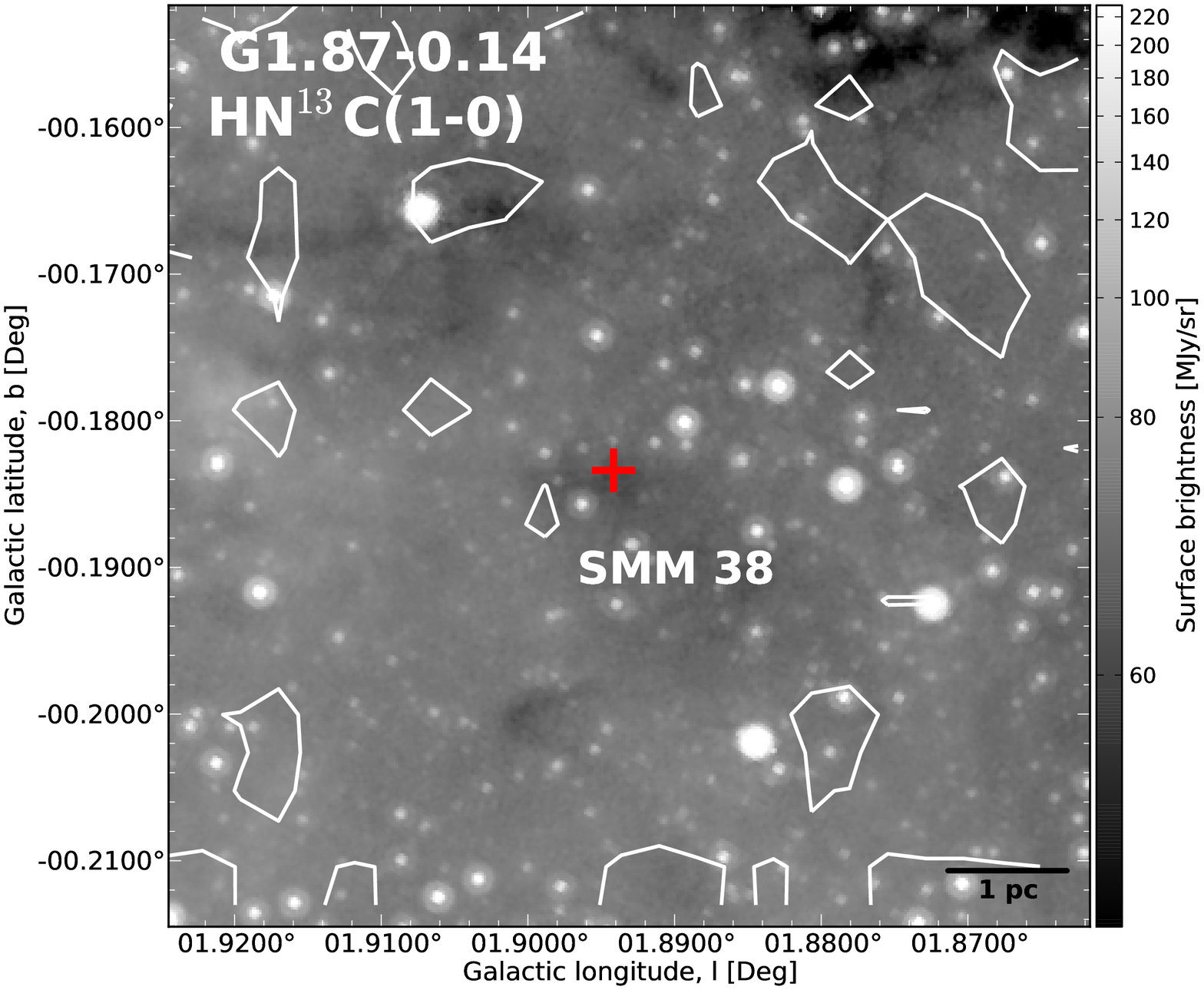}
\includegraphics[width=0.245\textwidth]{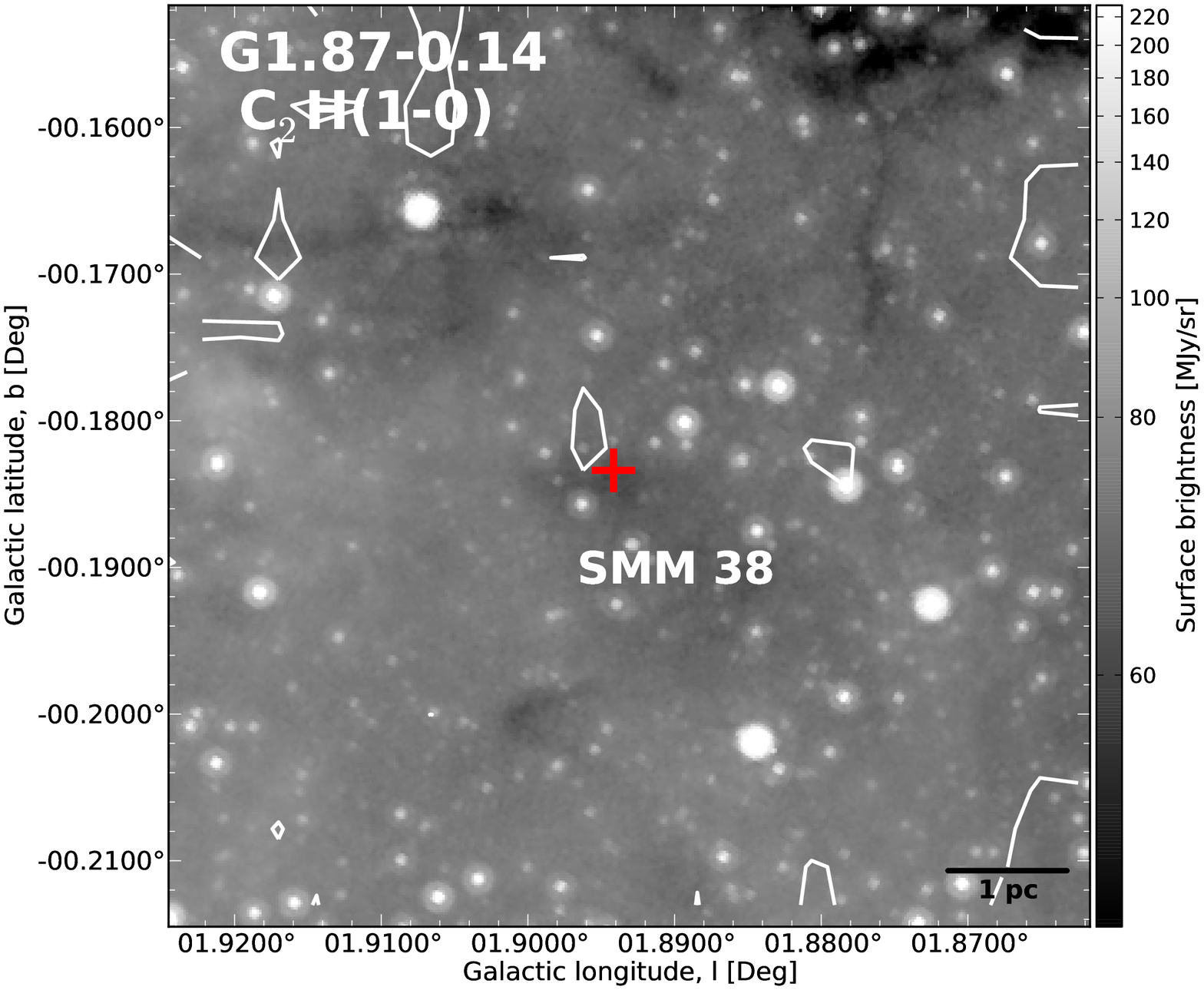}
\includegraphics[width=0.245\textwidth]{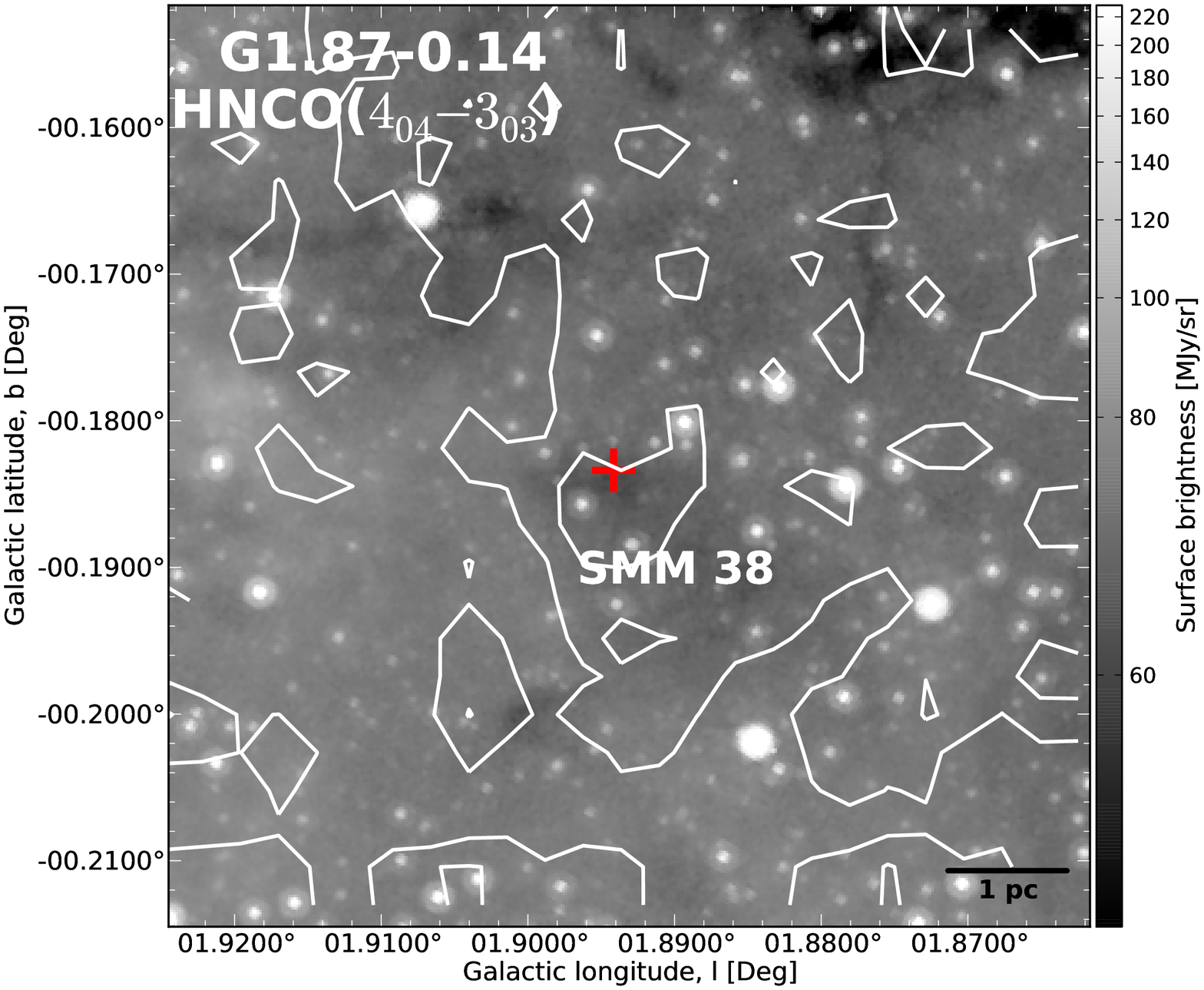}
\includegraphics[width=0.245\textwidth]{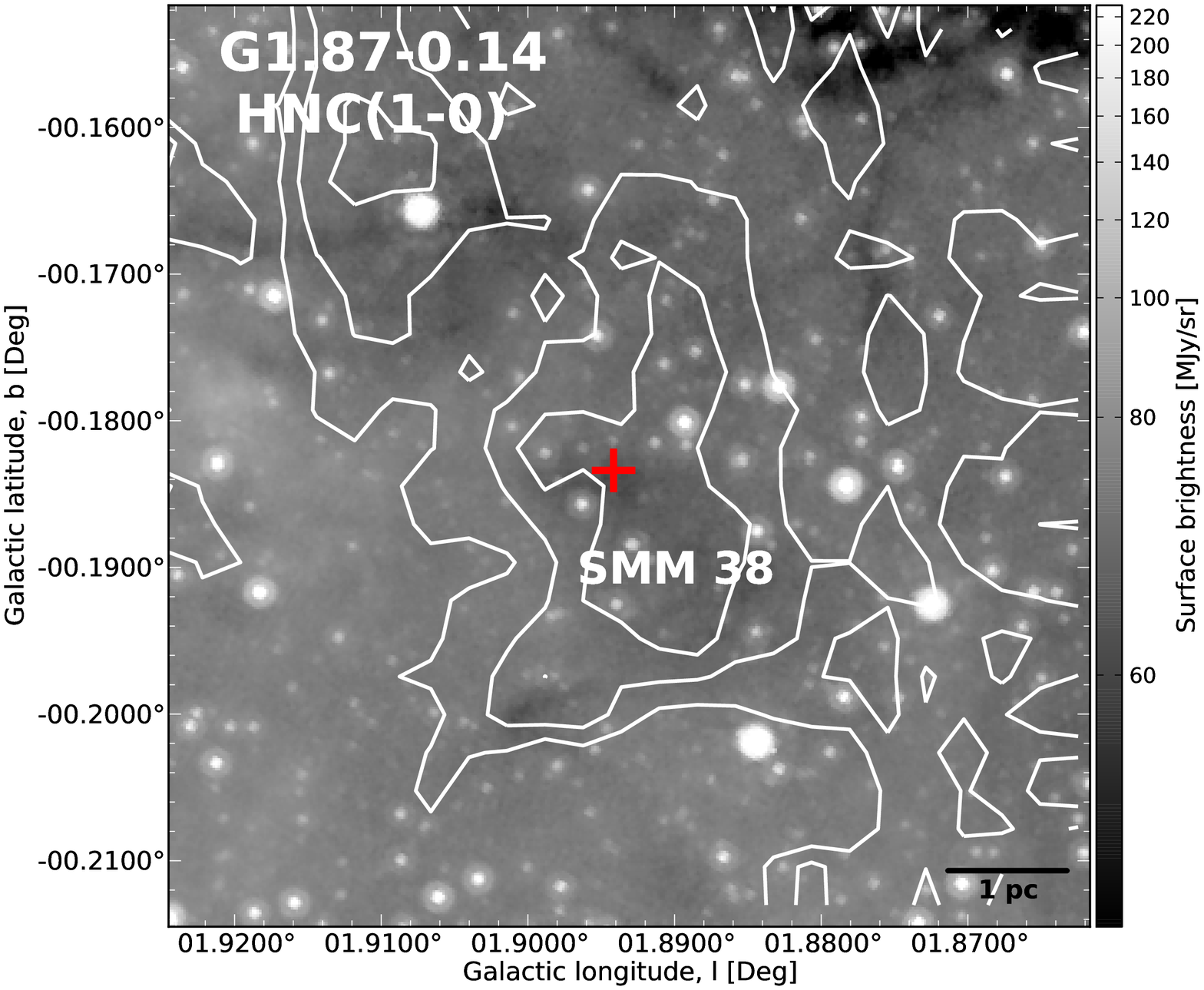}
\includegraphics[width=0.245\textwidth]{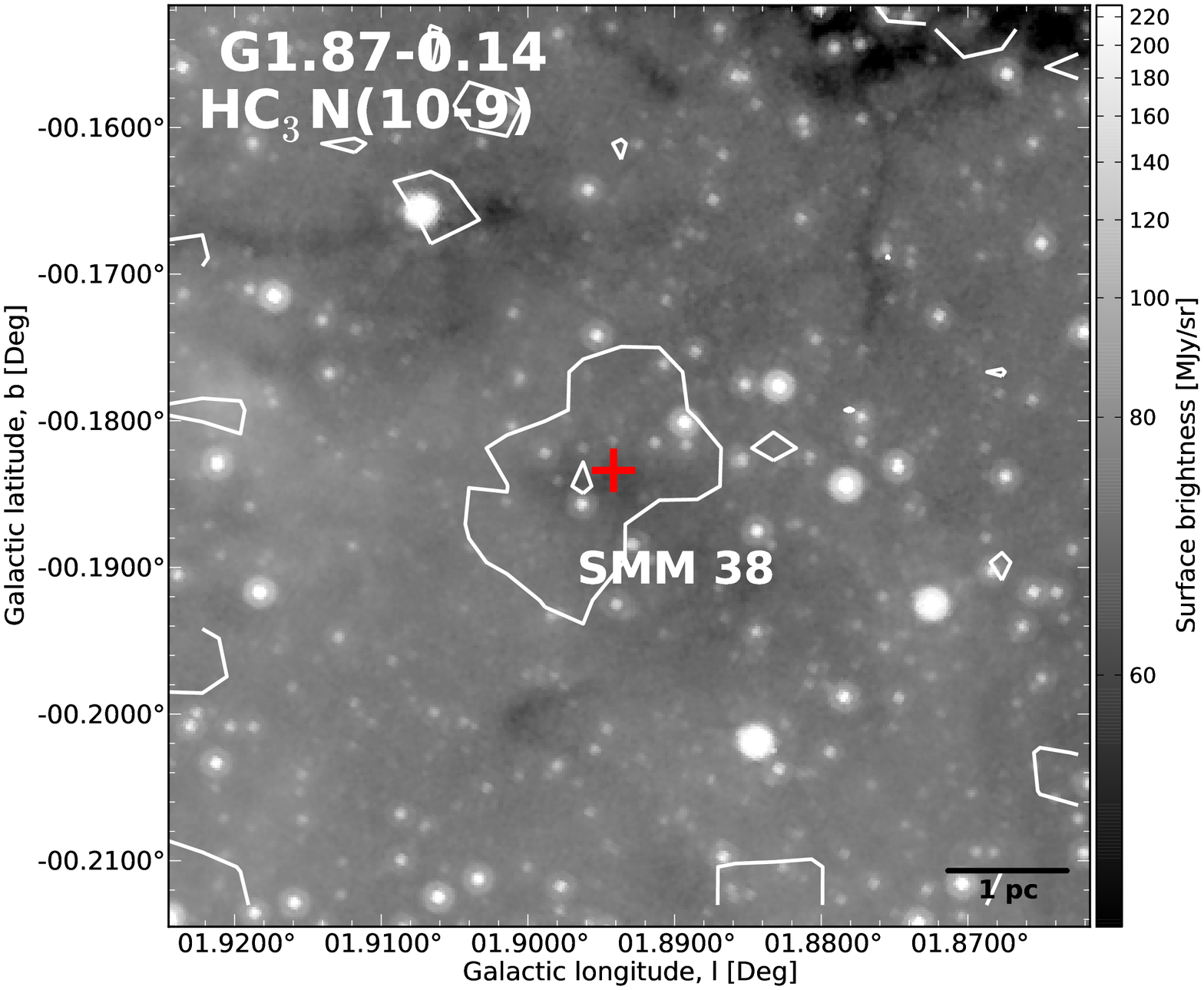}
\includegraphics[width=0.245\textwidth]{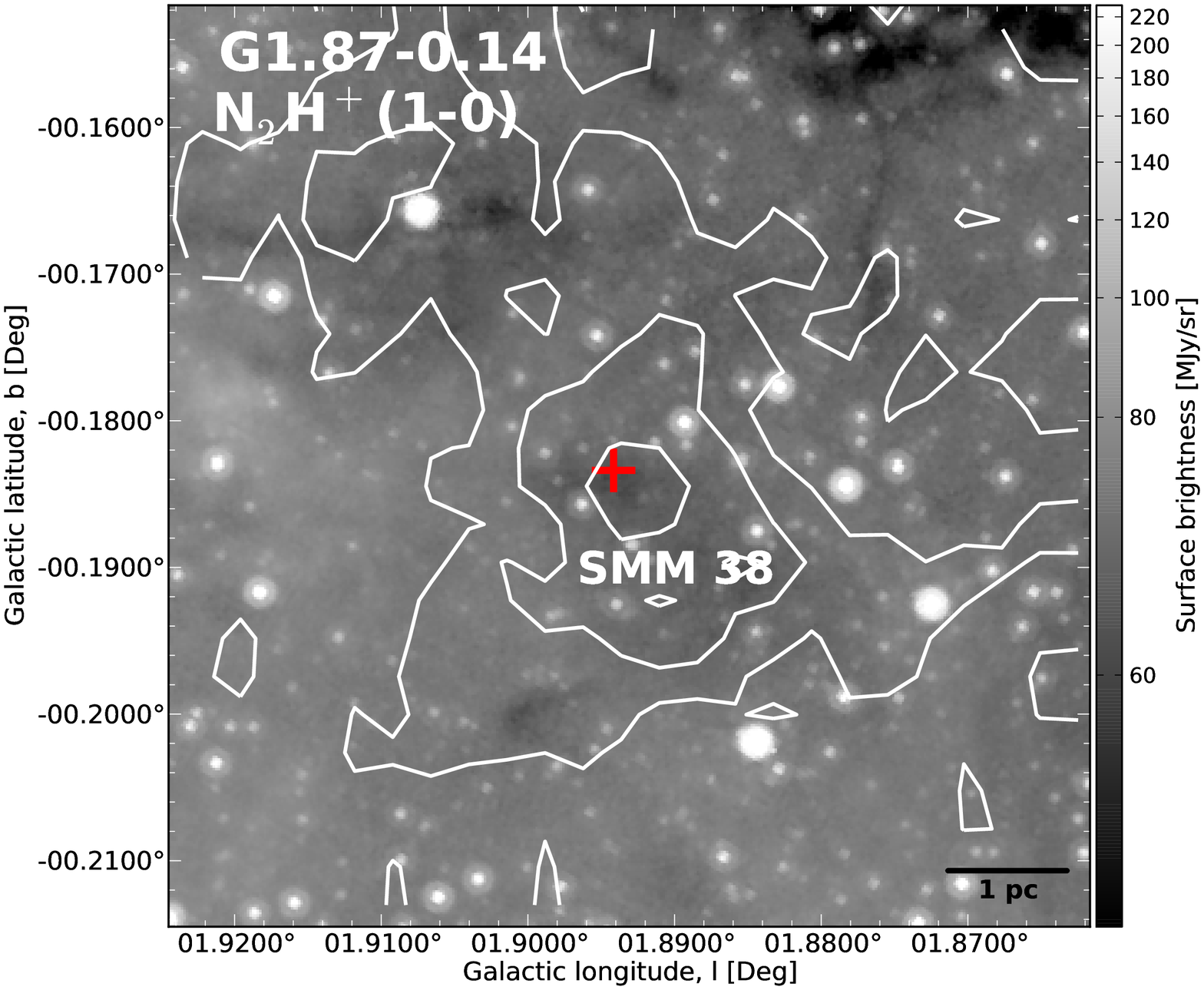}
\caption{Similar to Fig.~\ref{figure:G187SMM1lines} but towards 
G1.87--SMM 38. The contour levels start at $3\sigma$ for H$^{13}$CO$^+$, SiO, 
HN$^{13}$C, C$_2$H, HNCO$(4_{0,\,4}-3_{0,\,3})$, HC$_3$N, and N$_2$H$^+$. For 
HNC, the contours start at $5\sigma$. In all cases, the contours go in 
steps of $3\sigma$. The average $1\sigma$ value in $T_{\rm MB}$ units is 
$\sim0.77$ K~km~s$^{-1}$. The clump's LABOCA 870-$\mu$m peak position is 
marked by a red plus sign. A scale bar indicating the 1 pc 
projected length is indicated. Note that the SiO, HNC, HC$_3$N, and N$_2$H$^+$ 
emissions peak towards the submm maximum. HNC and N$_2$H$^+$ also show  
otherwise comparable spatial distributions.}
\label{figure:G187SMM38lines}
\end{center}
\end{figure*}

\begin{figure*}
\begin{center}
\includegraphics[width=0.245\textwidth]{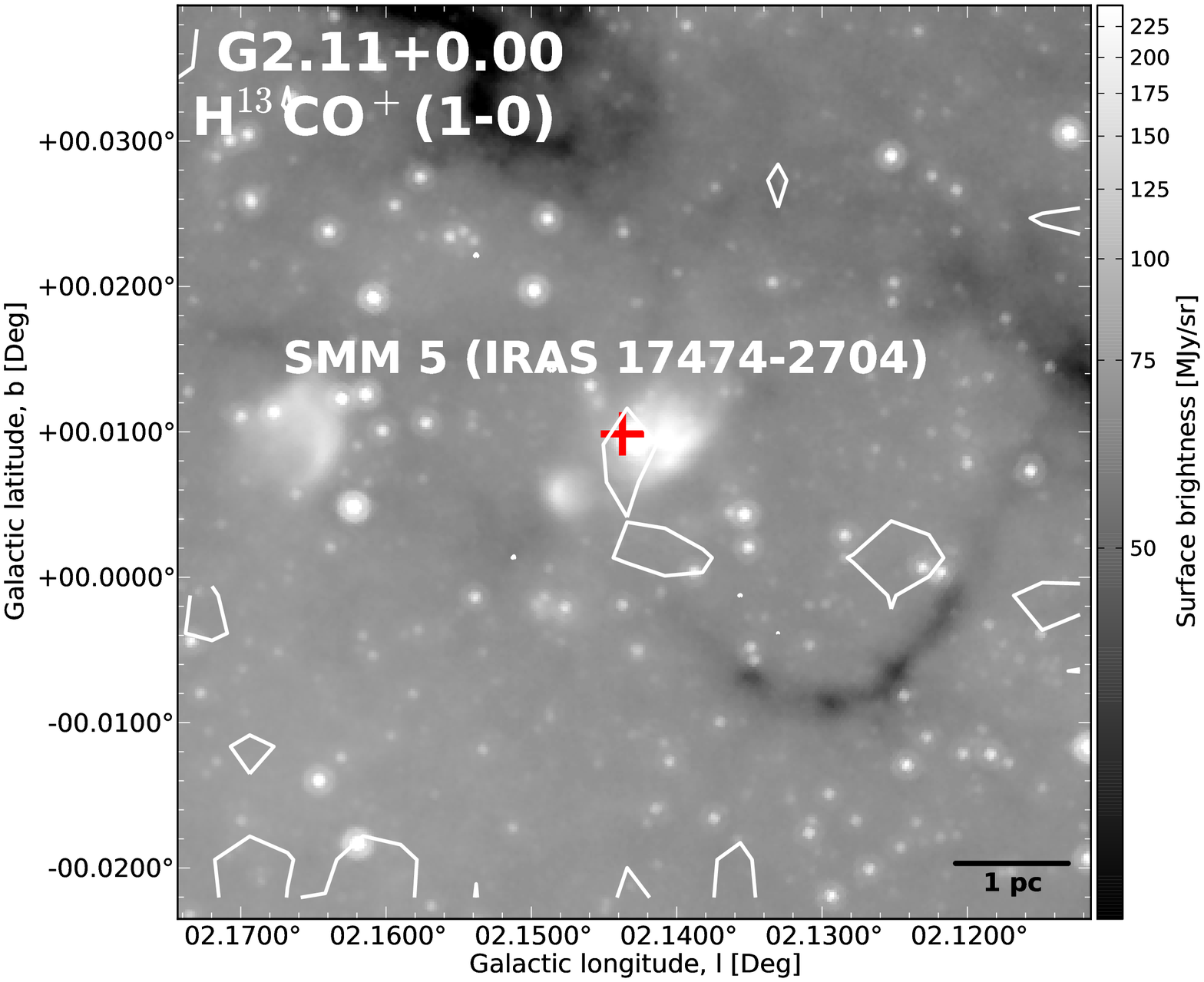}
\includegraphics[width=0.245\textwidth]{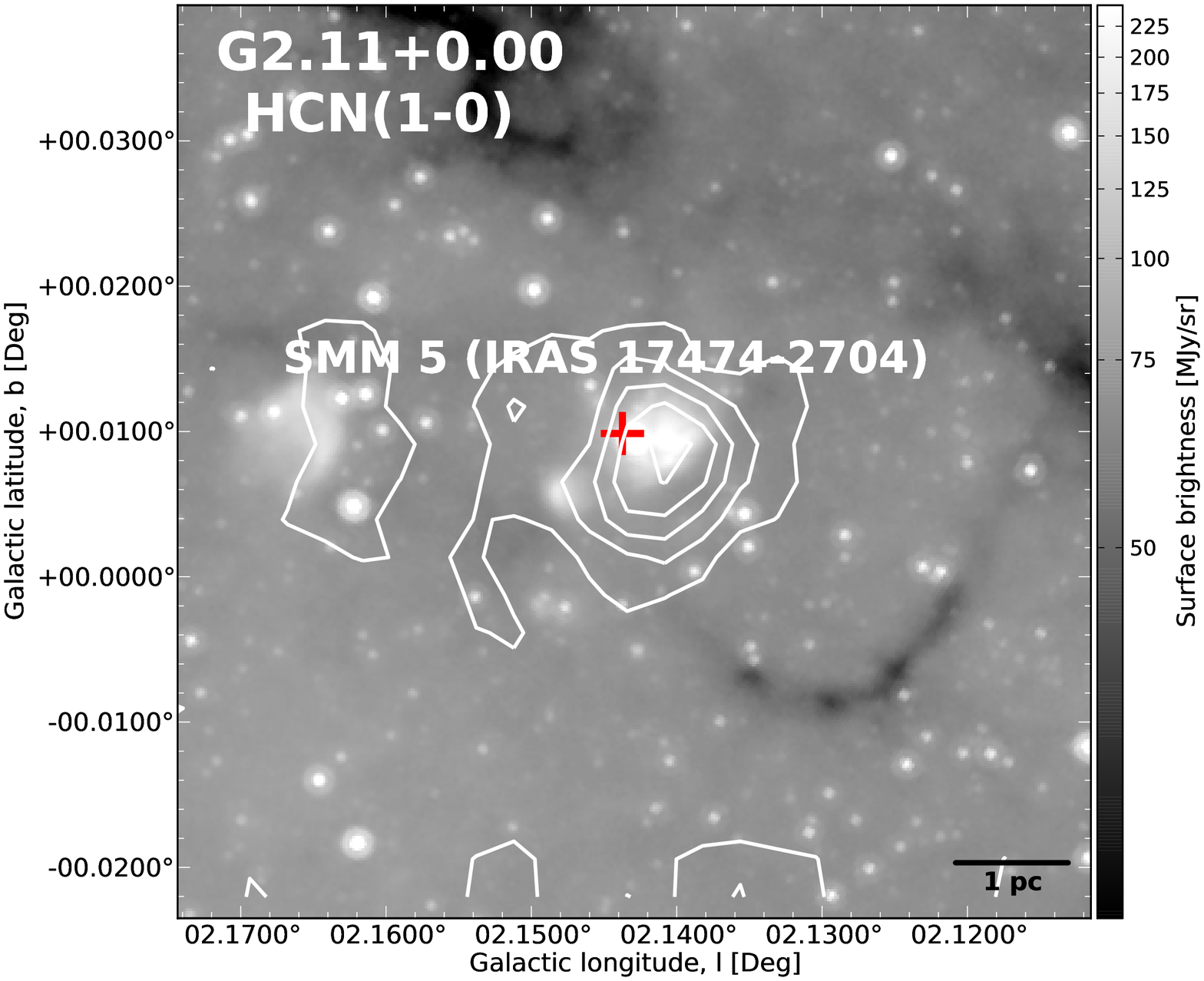}
\includegraphics[width=0.245\textwidth]{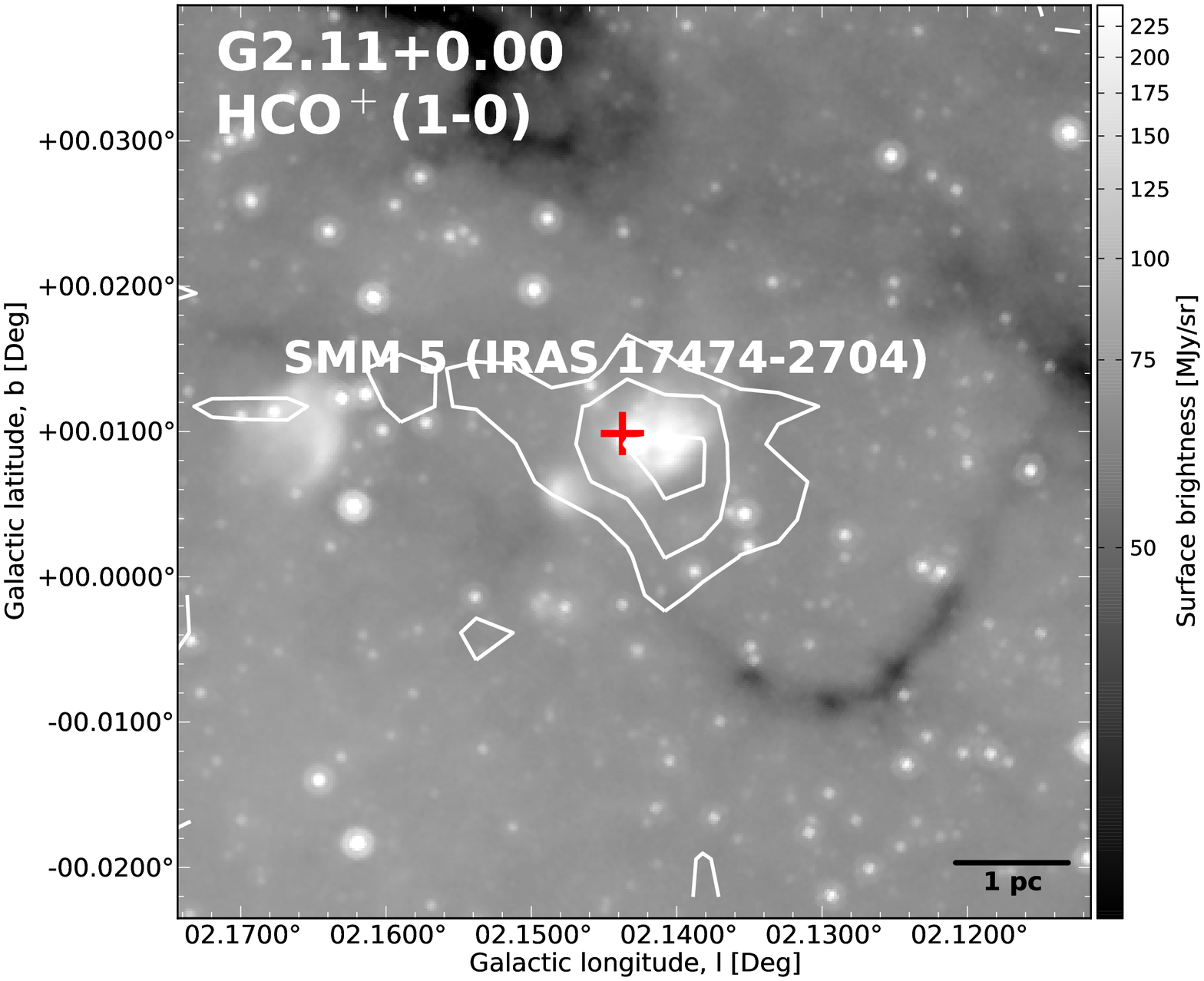}
\includegraphics[width=0.245\textwidth]{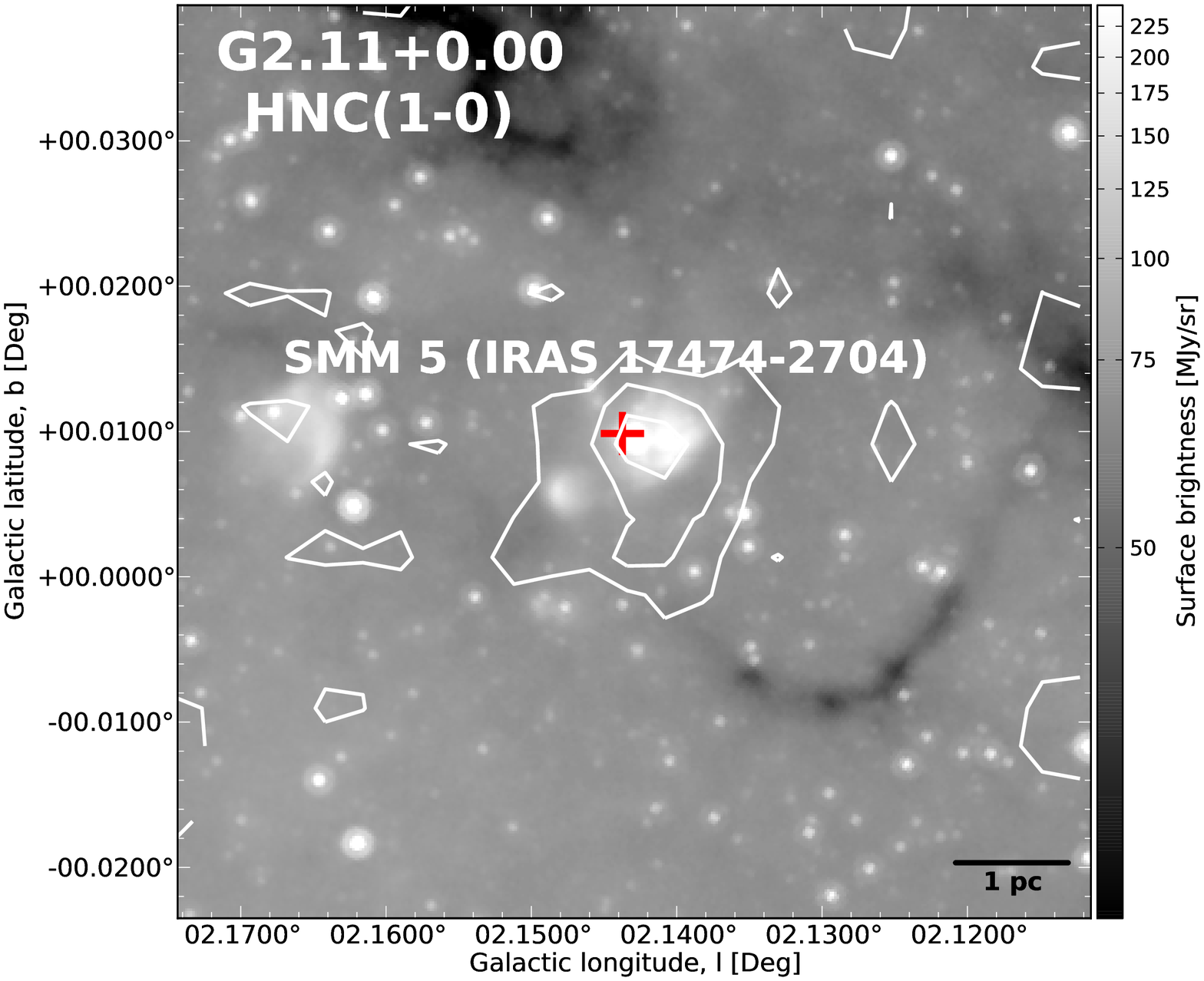}
\includegraphics[width=0.245\textwidth]{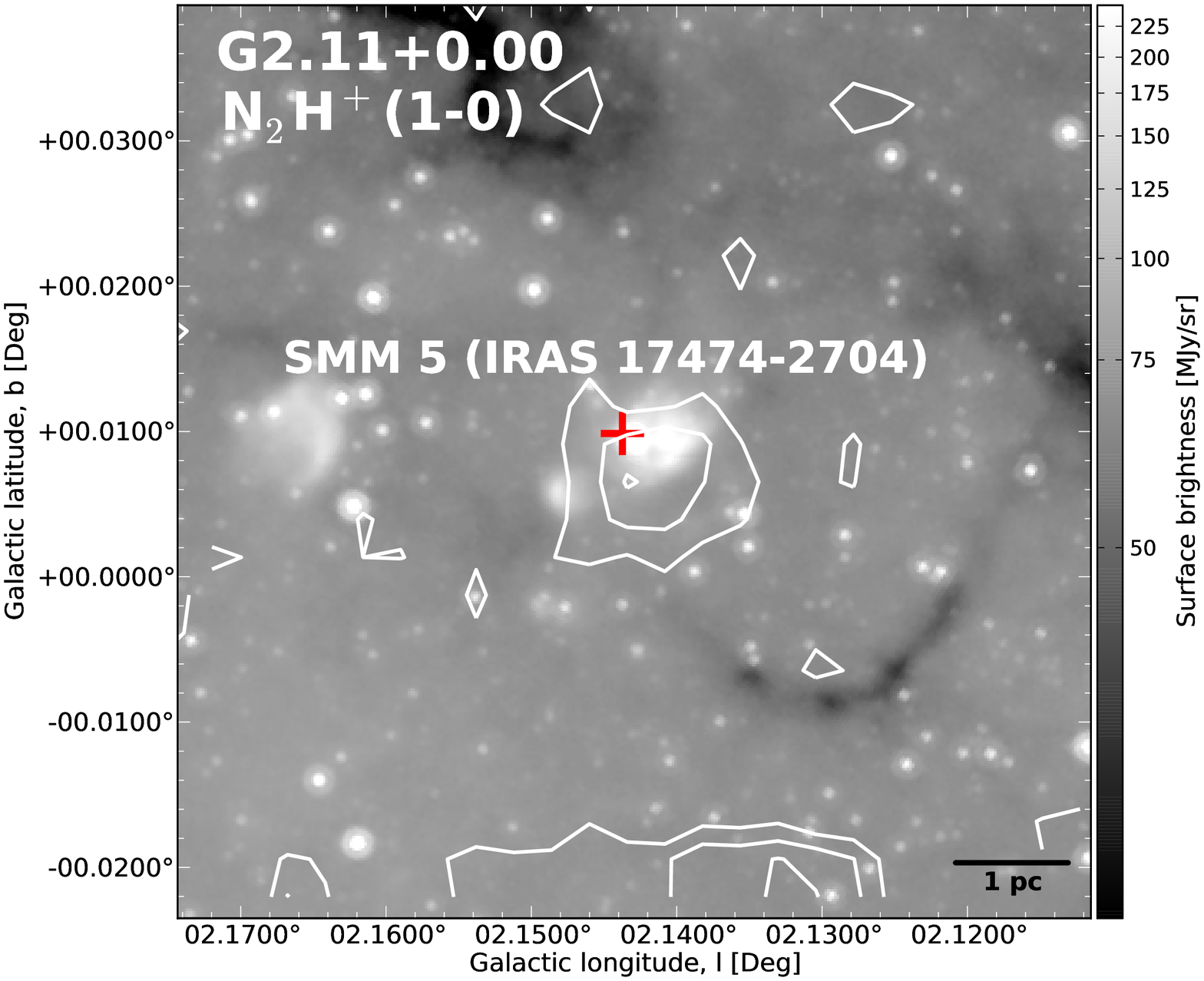}
\caption{Similar to Fig.~\ref{figure:G187SMM1lines} but towards 
G2.11--SMM 5. In each panel, the contour levels start at $3\sigma$, and go in 
steps of $3\sigma$. The average $1\sigma$ value in $T_{\rm MB}$ units is 
$\sim0.68$ K~km~s$^{-1}$. The clump's LABOCA 870-$\mu$m peak position is 
marked by a red plus sign. A scale bar indicating the 1 pc 
projected length is indicated. The HCN, HCO$^+$, HNC, and N$_2$H$^+$ show 
similar clump-like morphologies. The former two species also show an extension 
to the west of the emission peak.}
\label{figure:G211SMM5lines}
\end{center}
\end{figure*}

\begin{figure*}
\begin{center}
\includegraphics[width=0.245\textwidth]{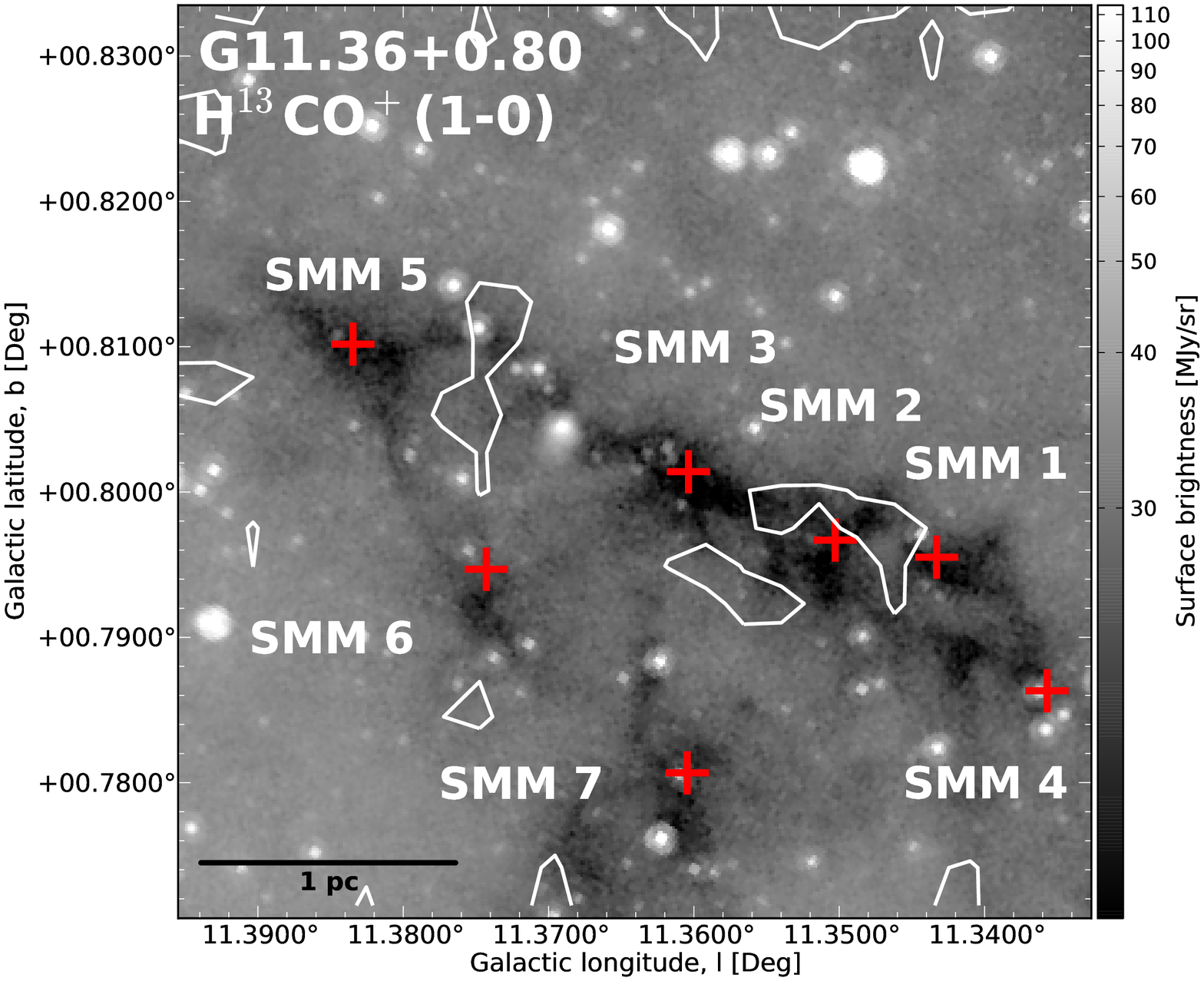}
\includegraphics[width=0.245\textwidth]{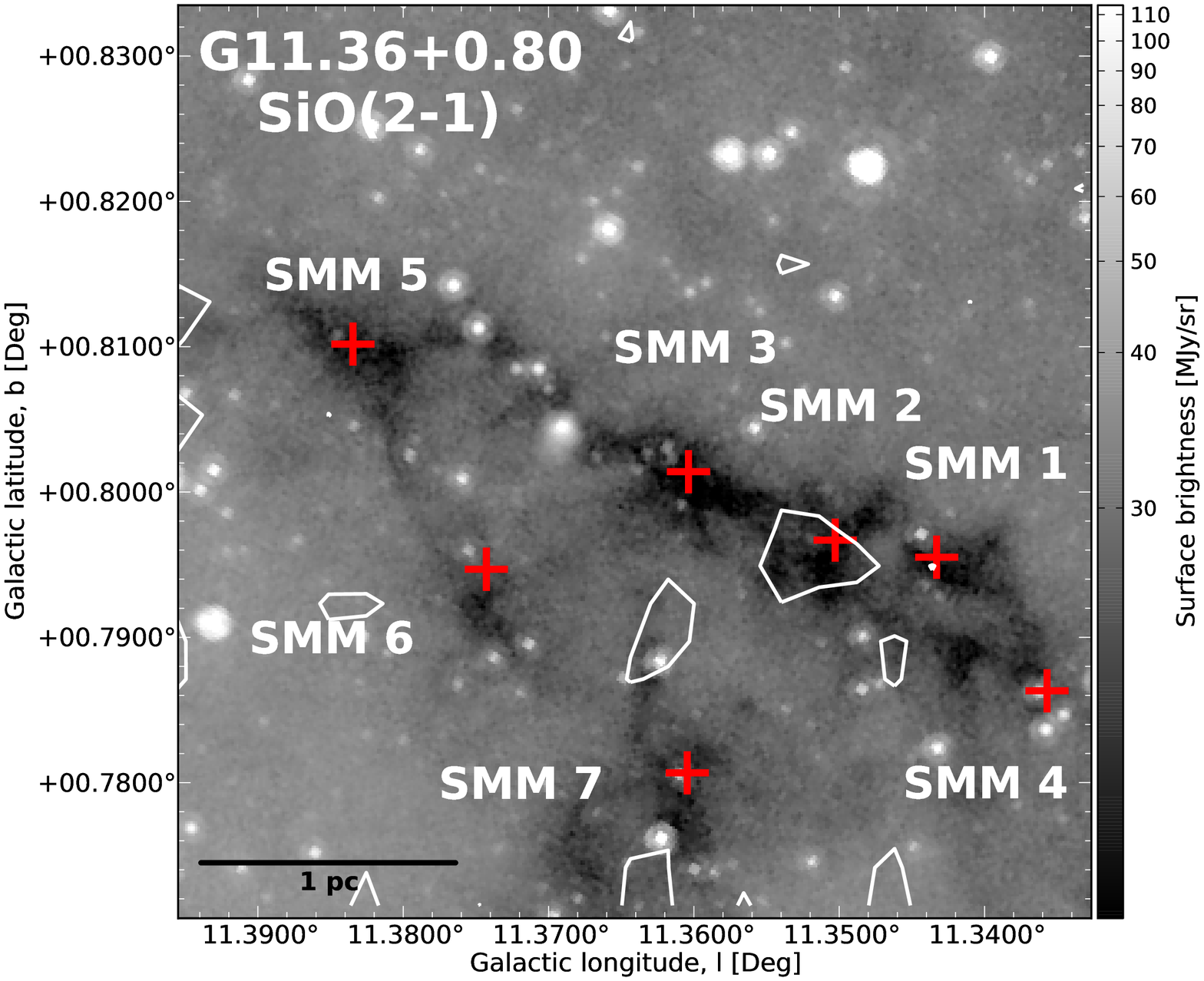}
\includegraphics[width=0.245\textwidth]{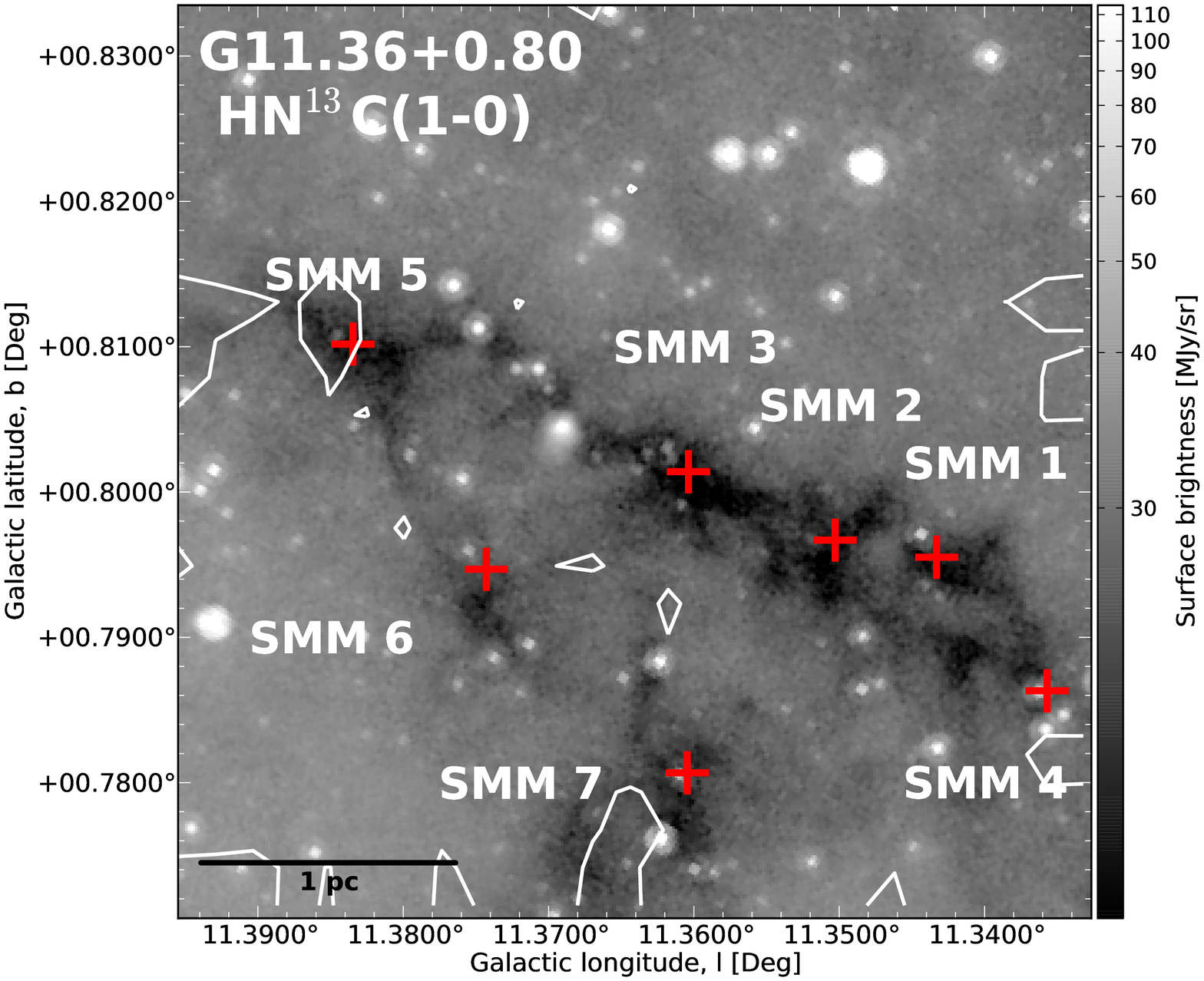}
\includegraphics[width=0.245\textwidth]{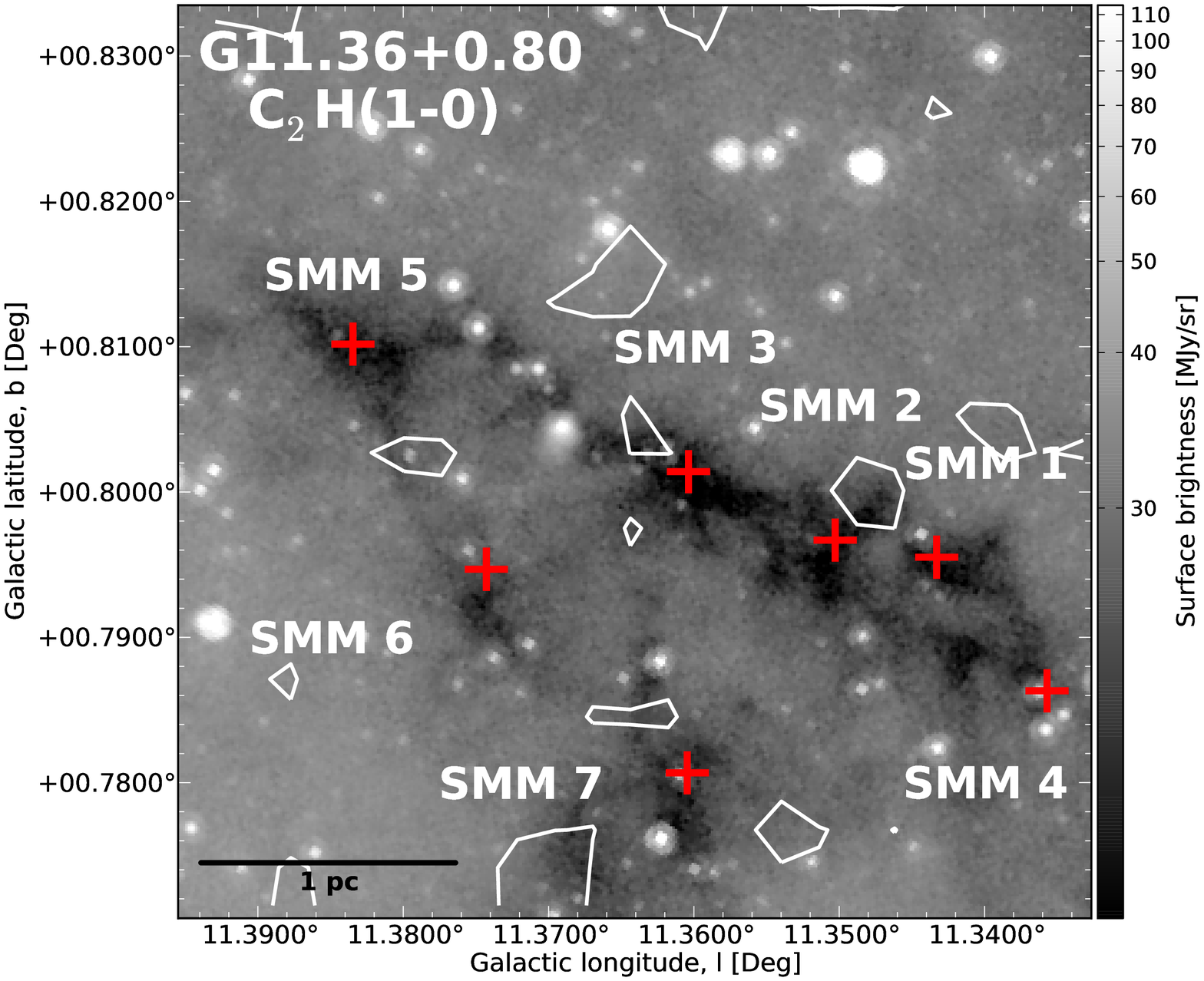}
\includegraphics[width=0.245\textwidth]{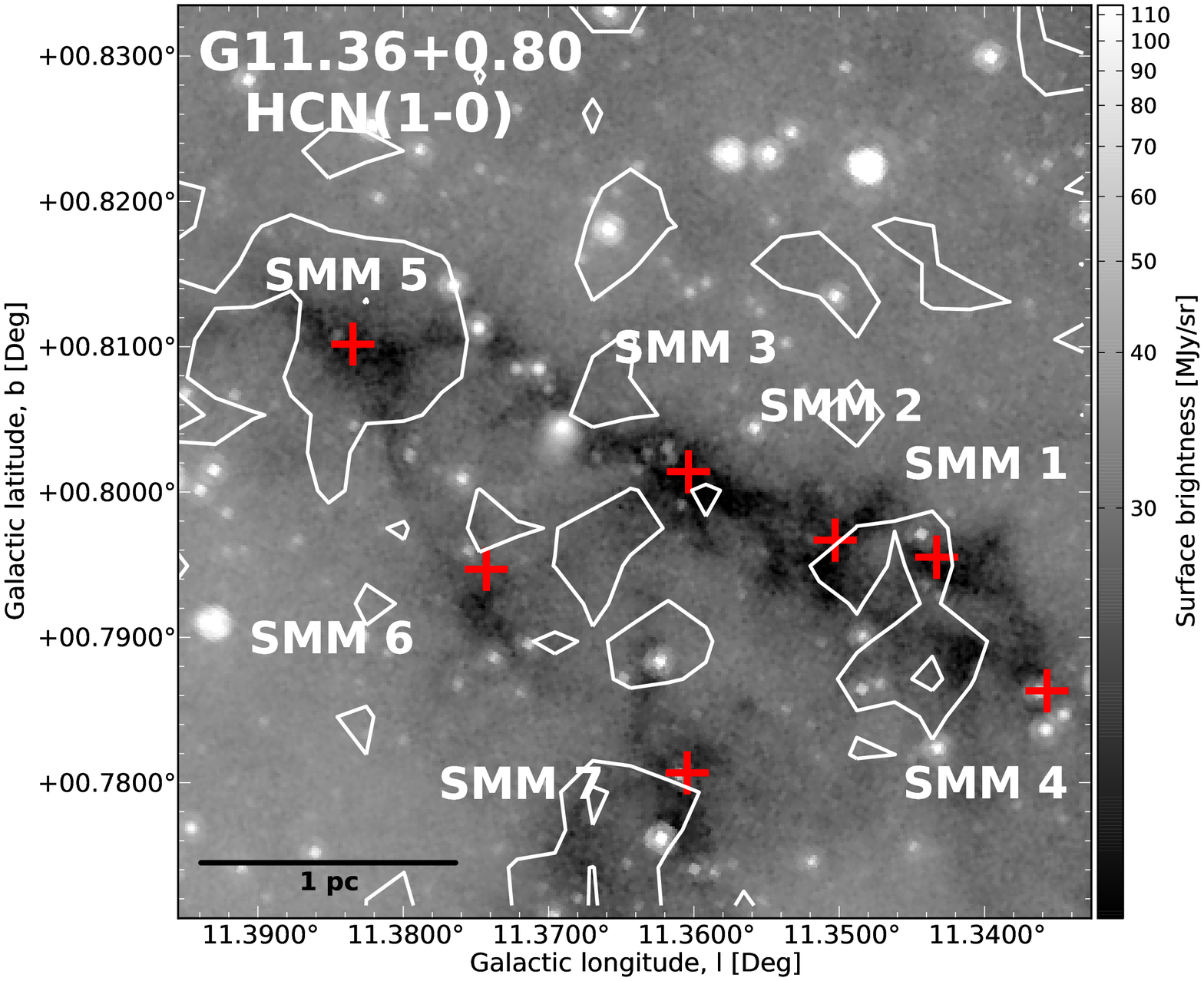}
\includegraphics[width=0.245\textwidth]{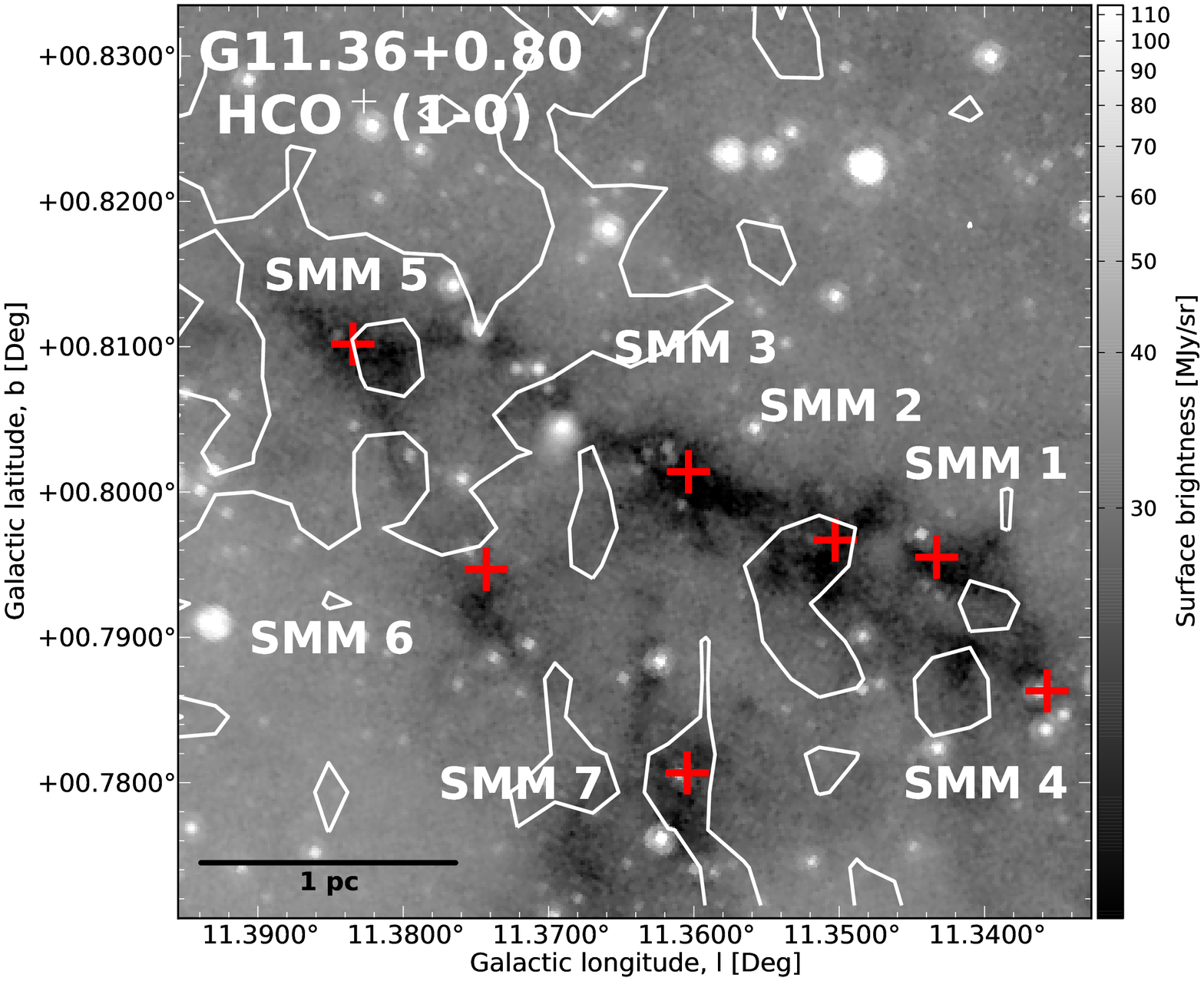}
\includegraphics[width=0.245\textwidth]{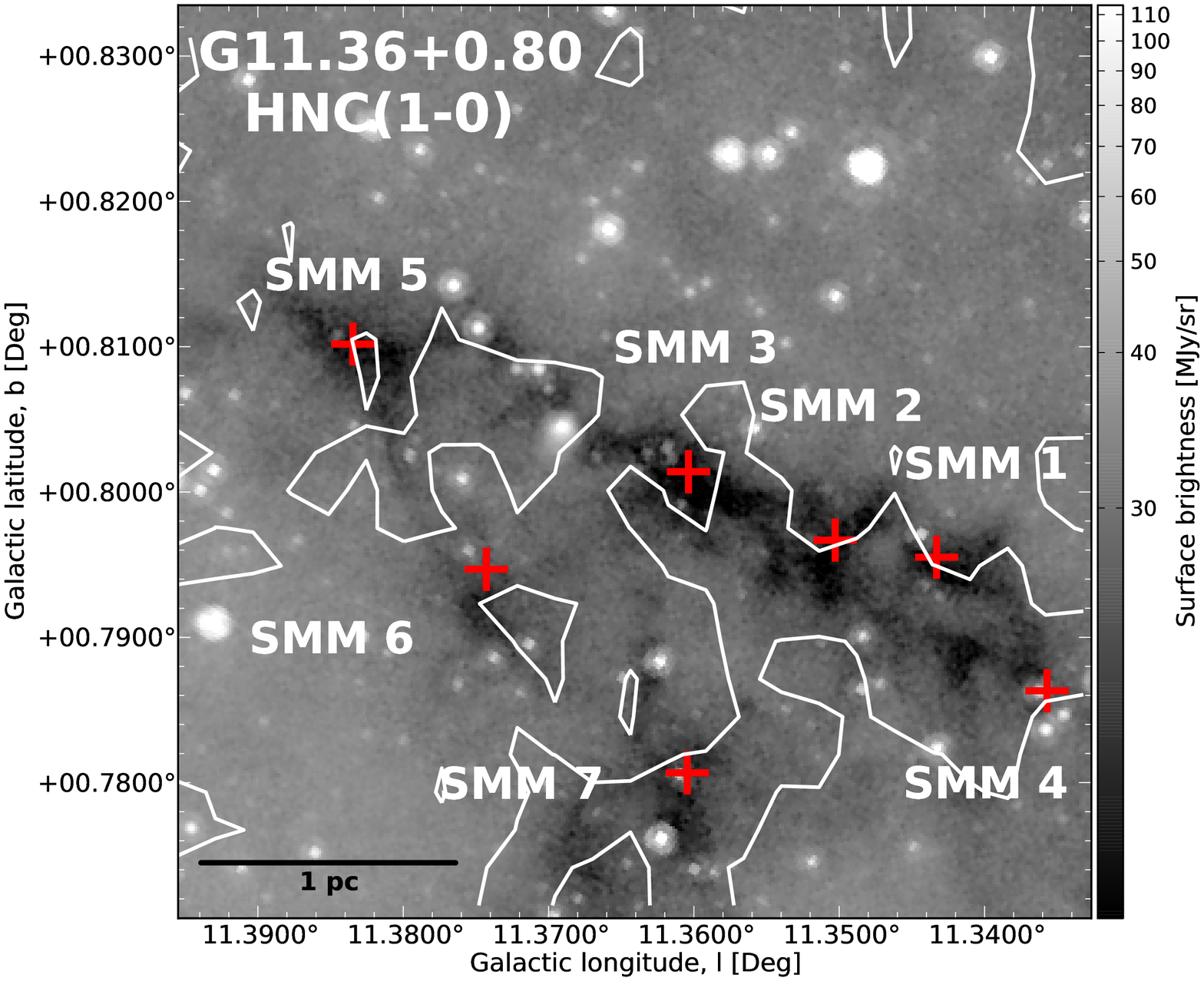}
\includegraphics[width=0.245\textwidth]{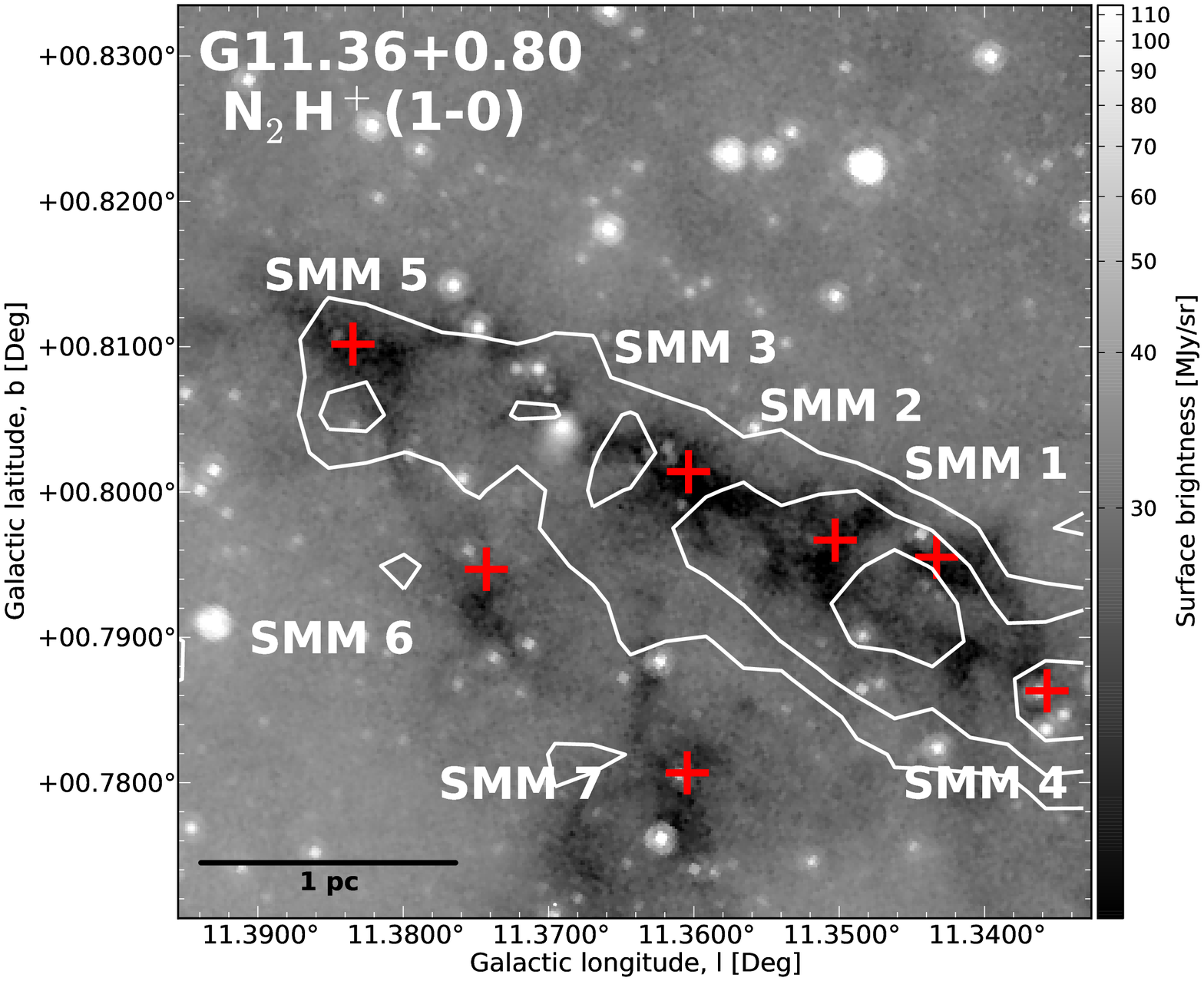}
\caption{Similar to Fig.~\ref{figure:G187SMM1lines} but towards 
the filamentary IRDC G11.36. The contour levels start at $3\sigma$ in all 
cases except for N$_2$H$^+$, where they start at $5\sigma$. In all cases, the 
contours go in steps of $3\sigma$. The average $1\sigma$ value in $T_{\rm MB}$ 
units is $\sim0.74$ K~km~s$^{-1}$. The LABOCA 870-$\mu$m peak positions of the 
clumps are marked by red plus signs. A scale bar indicating the 1 pc 
projected length is indicated. The HCN and HCO$^+$ appear to concentrate 
towards SMM 5, while HNC, and particularly N$_2$H$^+$, are tracing the 
filamentary 8-$\mu$m absorption feature.}
\label{figure:G1136lines}
\end{center}
\end{figure*}

\begin{figure*}
\begin{center}
\includegraphics[width=0.245\textwidth]{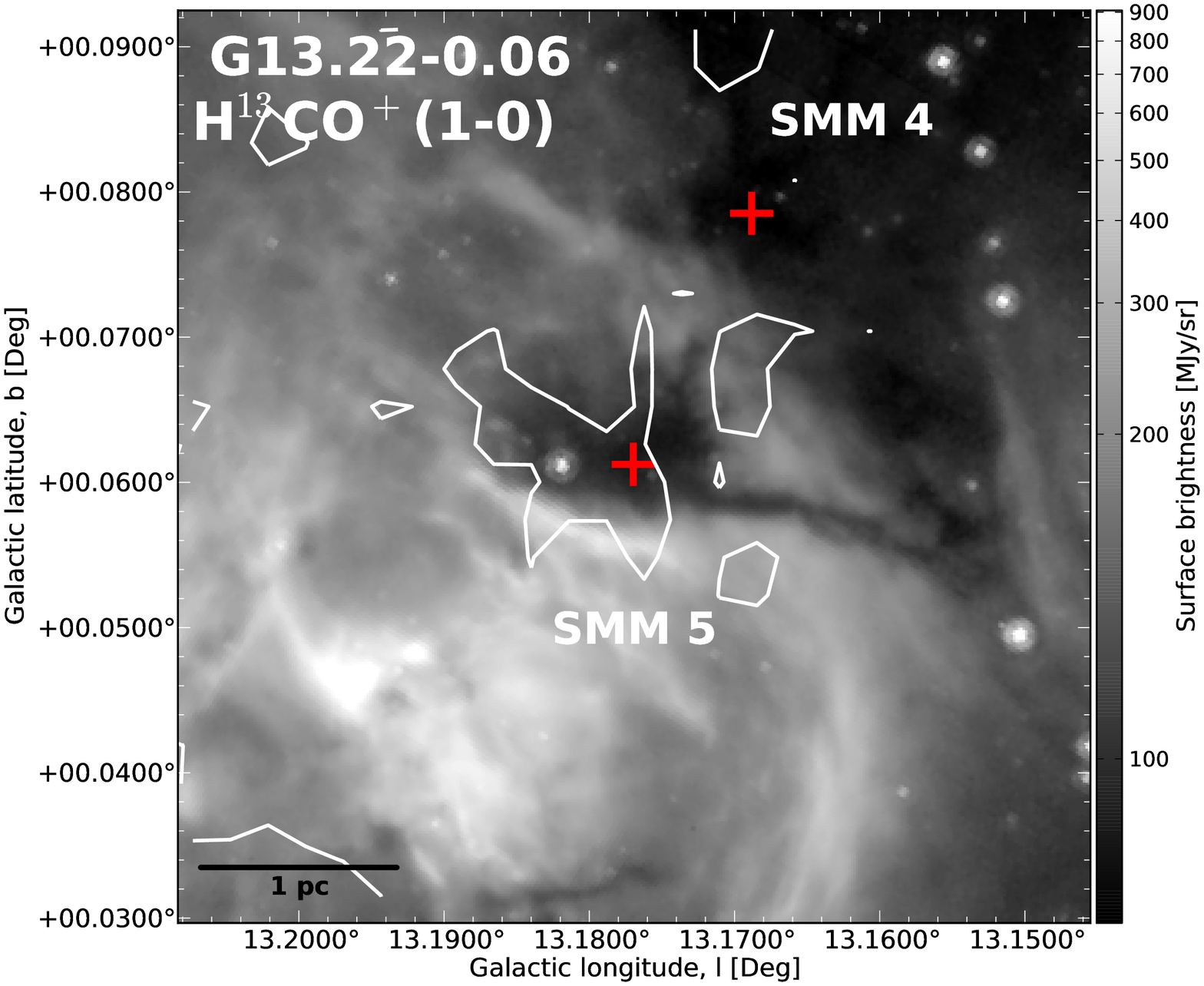}
\includegraphics[width=0.245\textwidth]{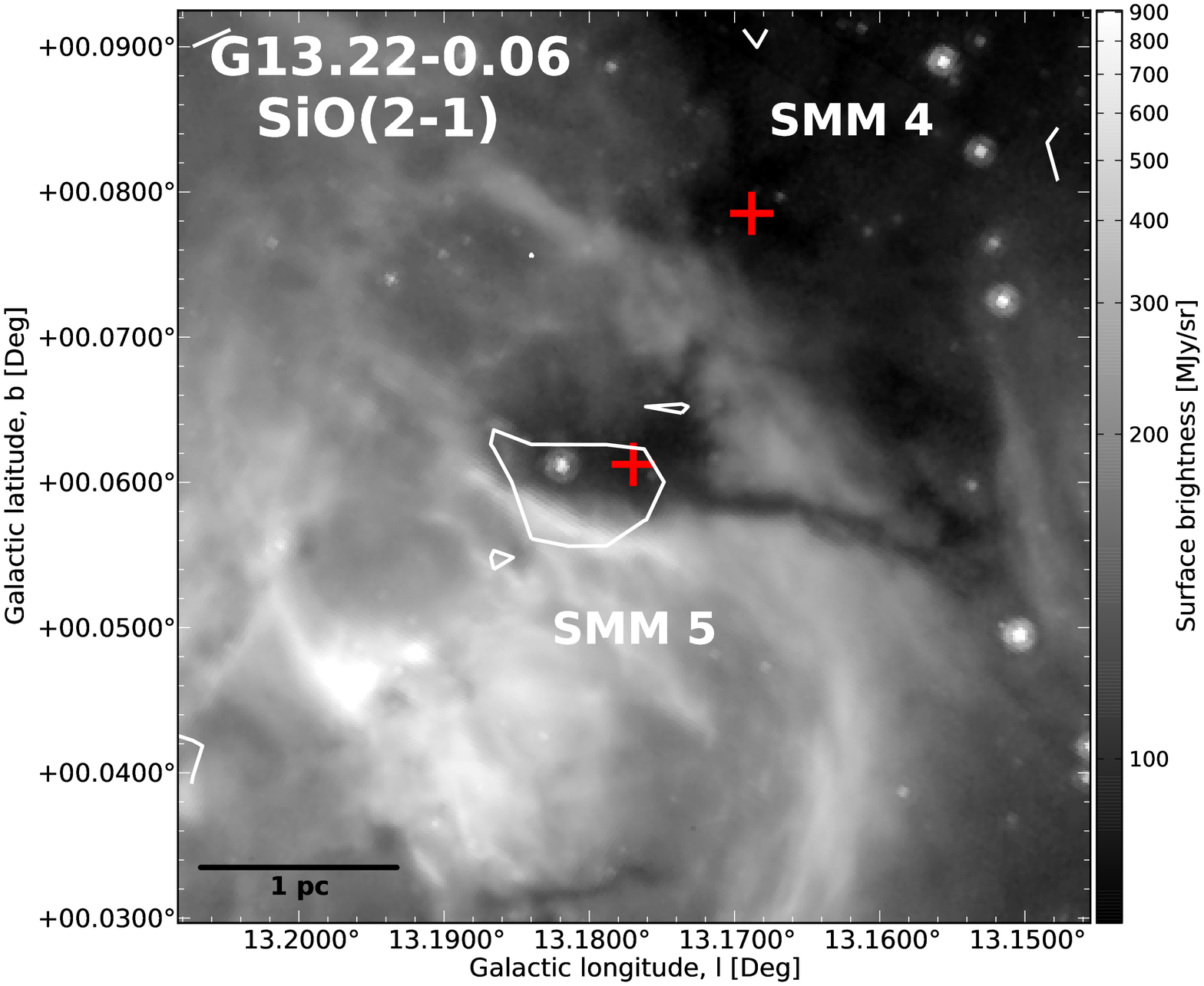}
\includegraphics[width=0.245\textwidth]{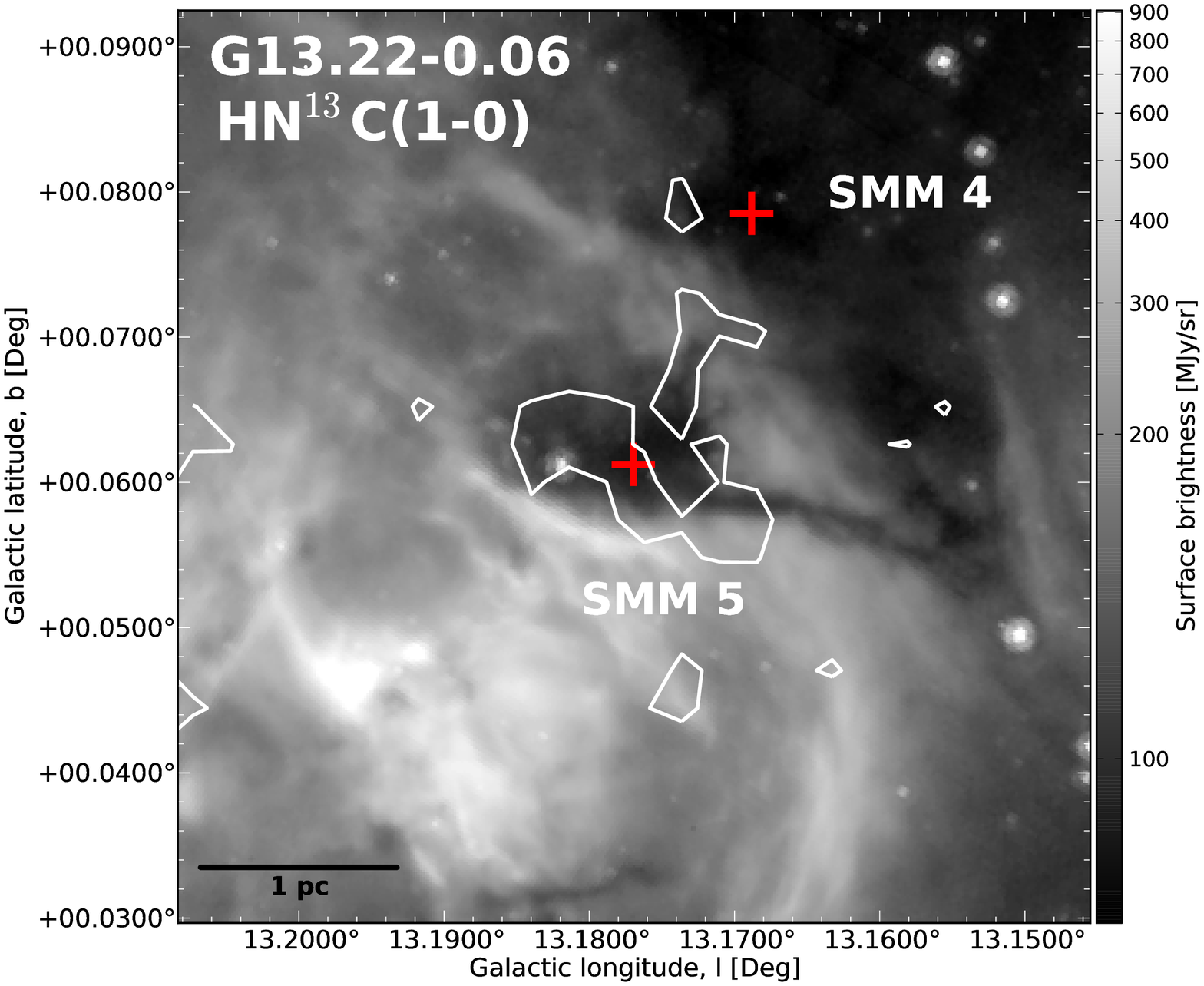}
\includegraphics[width=0.245\textwidth]{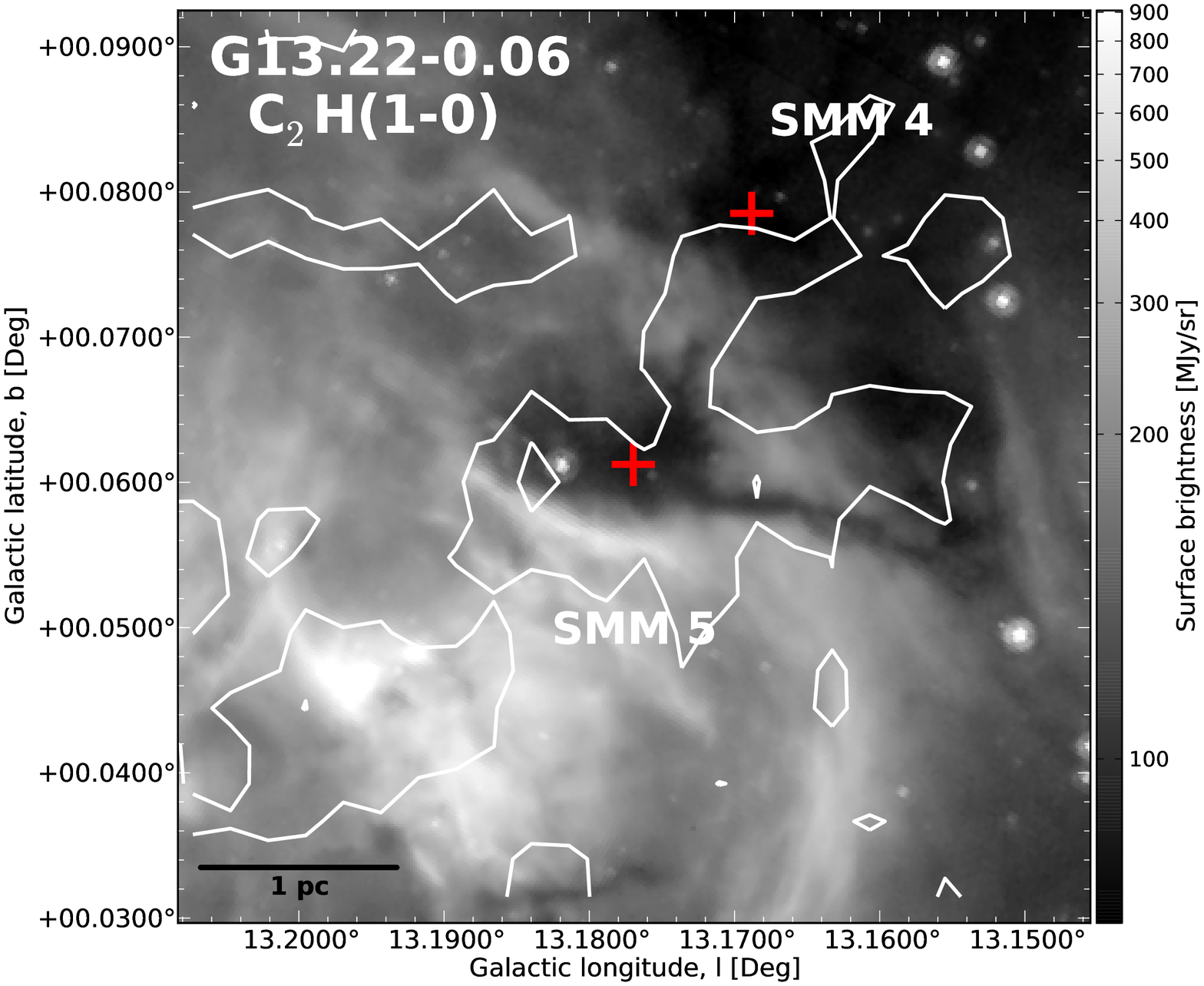}
\includegraphics[width=0.245\textwidth]{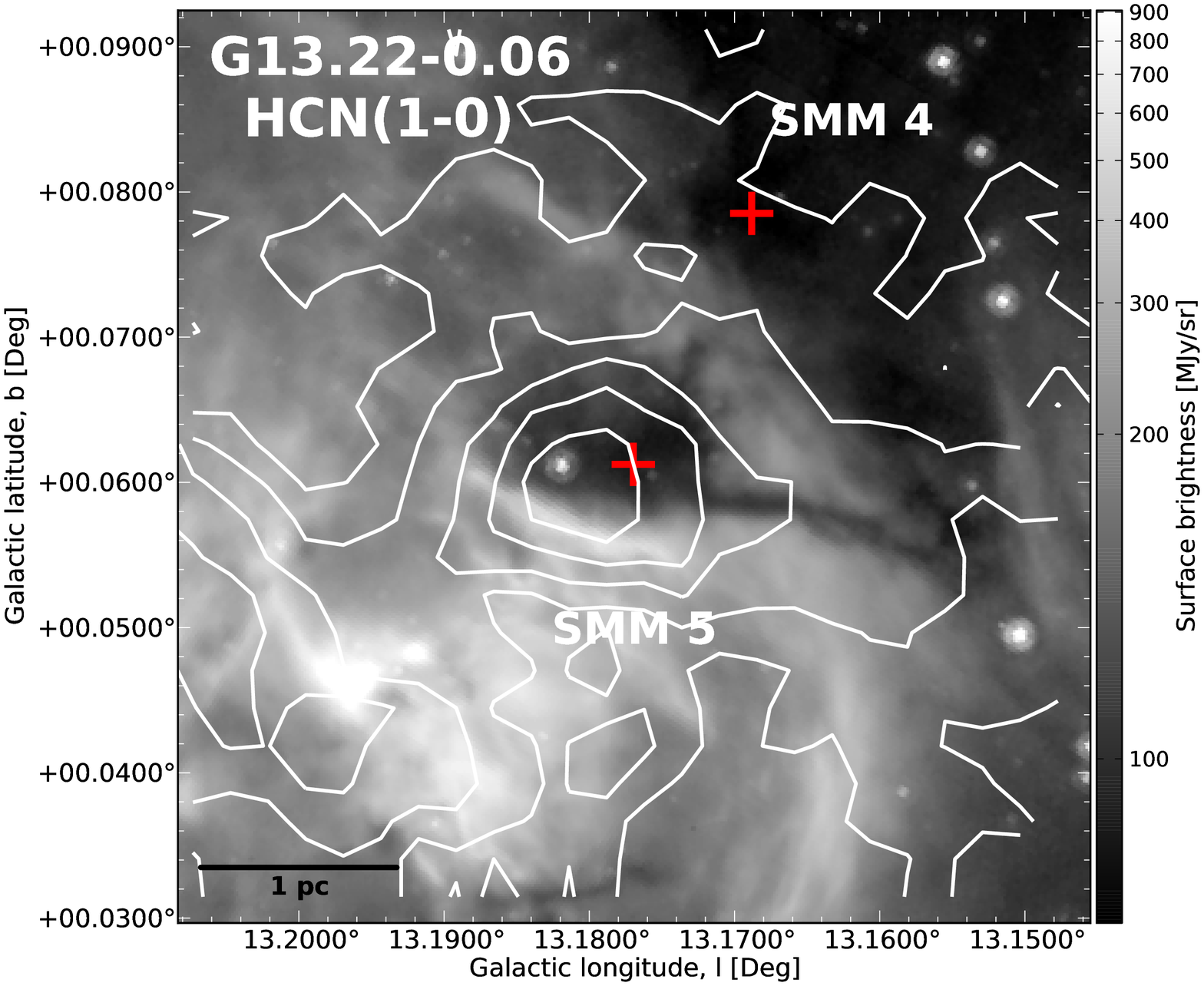}
\includegraphics[width=0.245\textwidth]{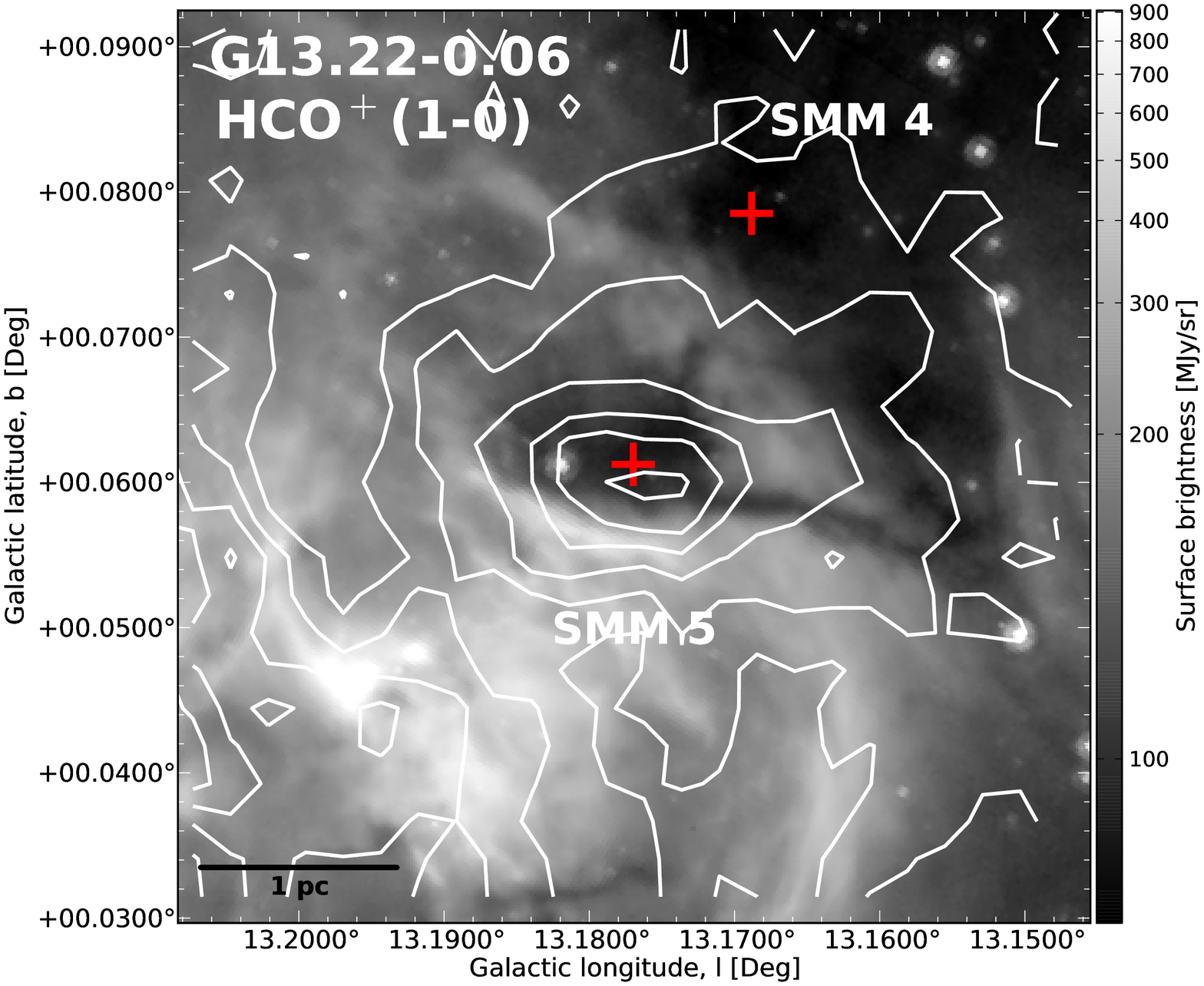}
\includegraphics[width=0.245\textwidth]{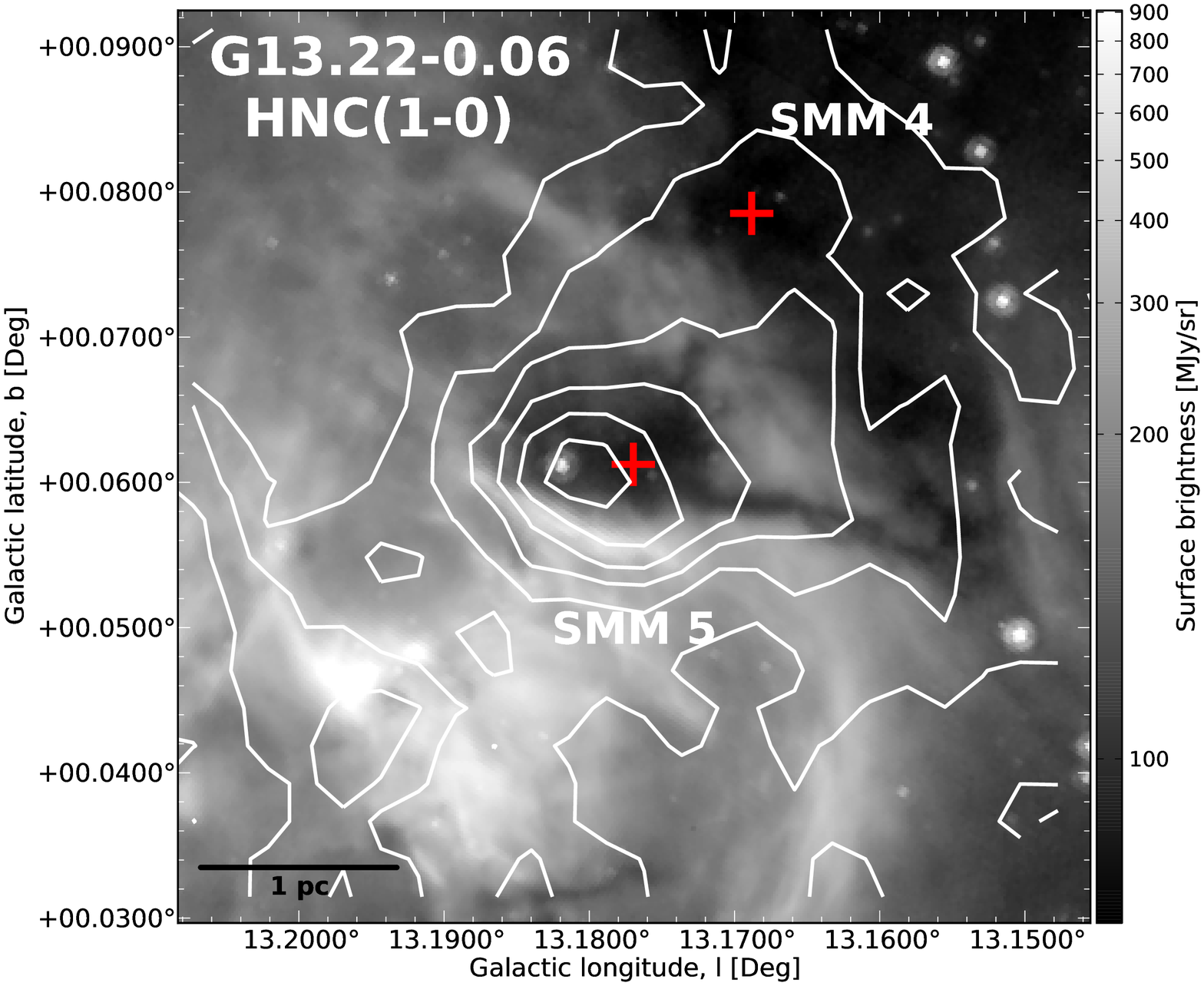}
\includegraphics[width=0.245\textwidth]{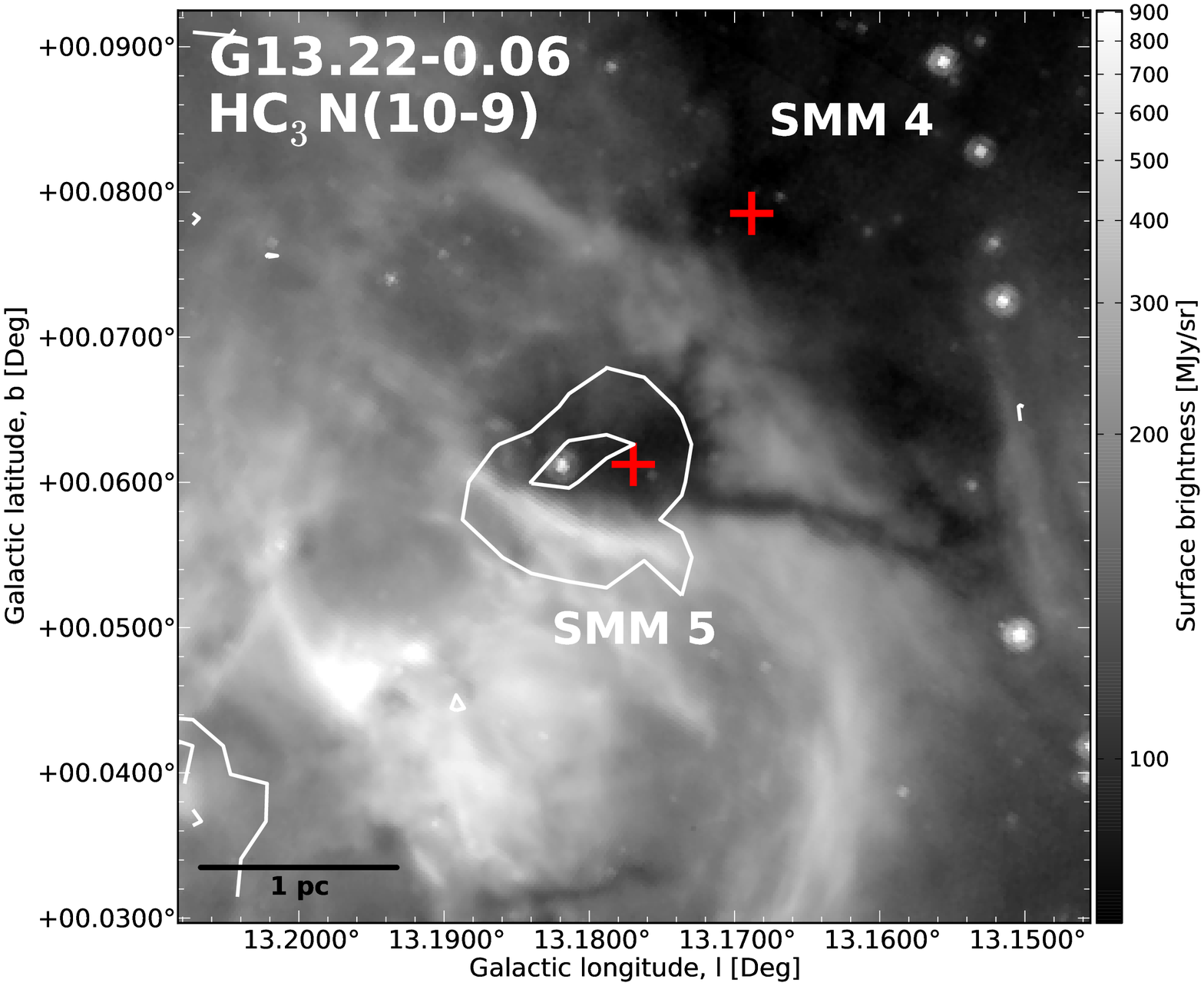}
\includegraphics[width=0.245\textwidth]{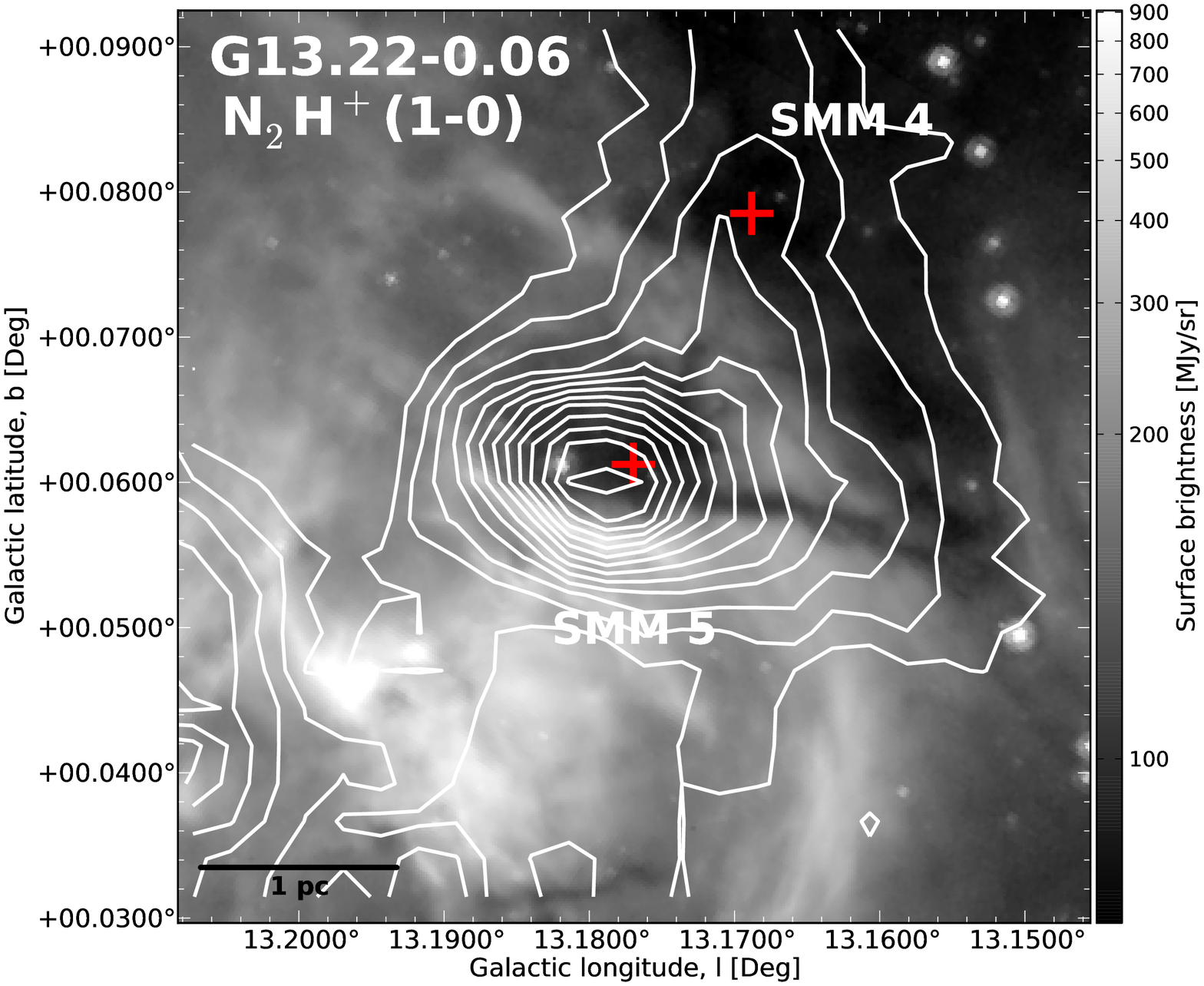}
\caption{Similar to Fig.~\ref{figure:G187SMM1lines} but towards 
G13.22--SMM 4, 5 around the N10/11 IR-bubble pair. The contour levels start at 
$3\sigma$ for H$^{13}$CO$^+$, SiO, HN$^{13}$C, C$_2$H, and HC$_3$N, while for 
HCN, HCO$^+$, HNC, and N$_2$H$^+$ they start at $5\sigma$. In all cases, the 
contours go in steps of $3\sigma$. The average $1\sigma$ value in $T_{\rm MB}$ 
units is $\sim0.79$ K~km~s$^{-1}$. The LABOCA 870-$\mu$m peak positions of the 
clumps are marked by red plus signs. A scale bar indicating the 1 pc 
projected length is indicated. Note the extended morphological similarities 
between HCN, HCO$^+$, HNC, and N$_2$H$^+$. C$_2$H shows a ridge-like emission 
along the northwest-southeast direction. The emission of all the species 
peaks towards SMM 5.}
\label{figure:G1322SMM5lines}
\end{center}
\end{figure*}

\begin{figure*}
\begin{center}
\includegraphics[width=0.245\textwidth]{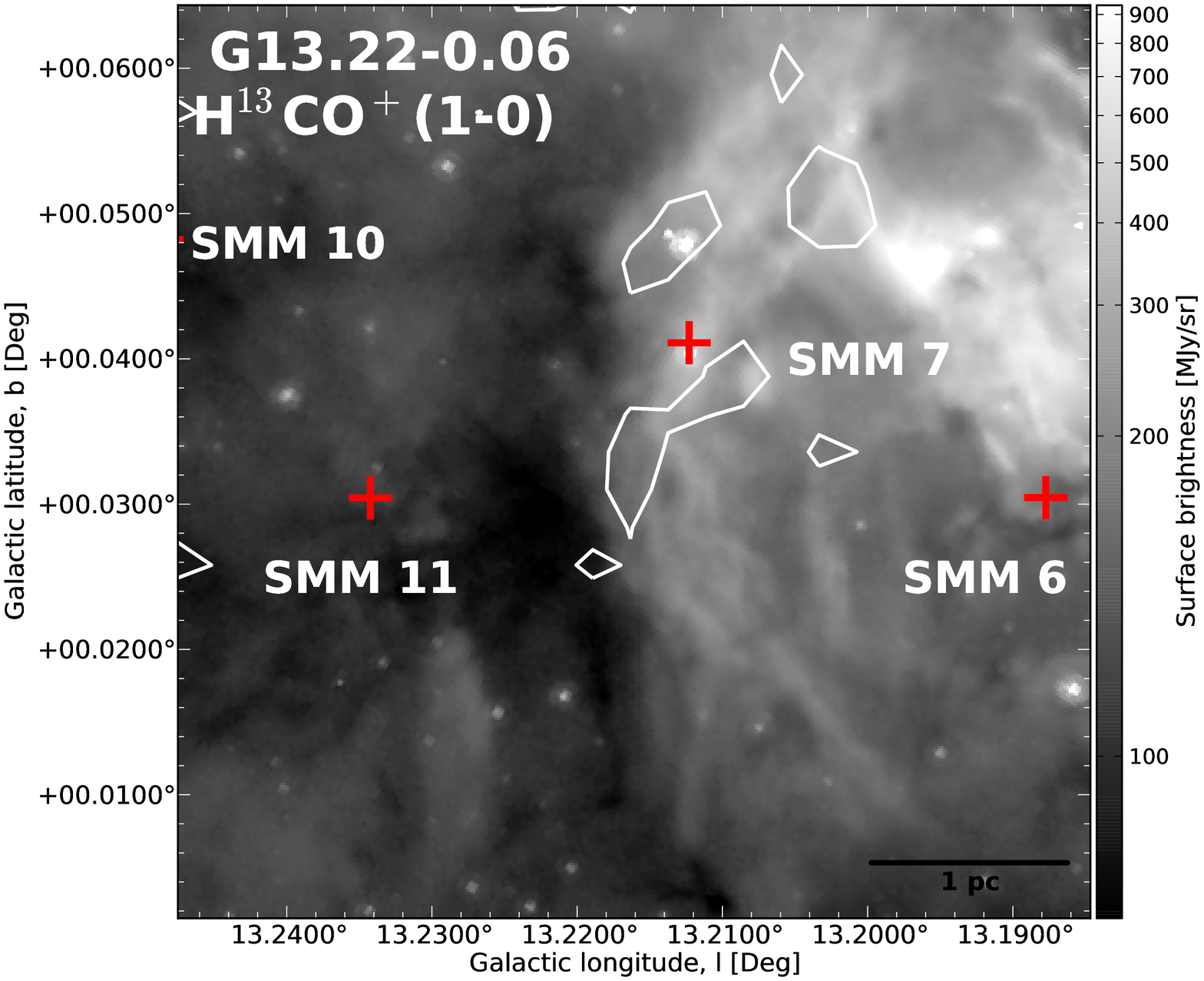}
\includegraphics[width=0.245\textwidth]{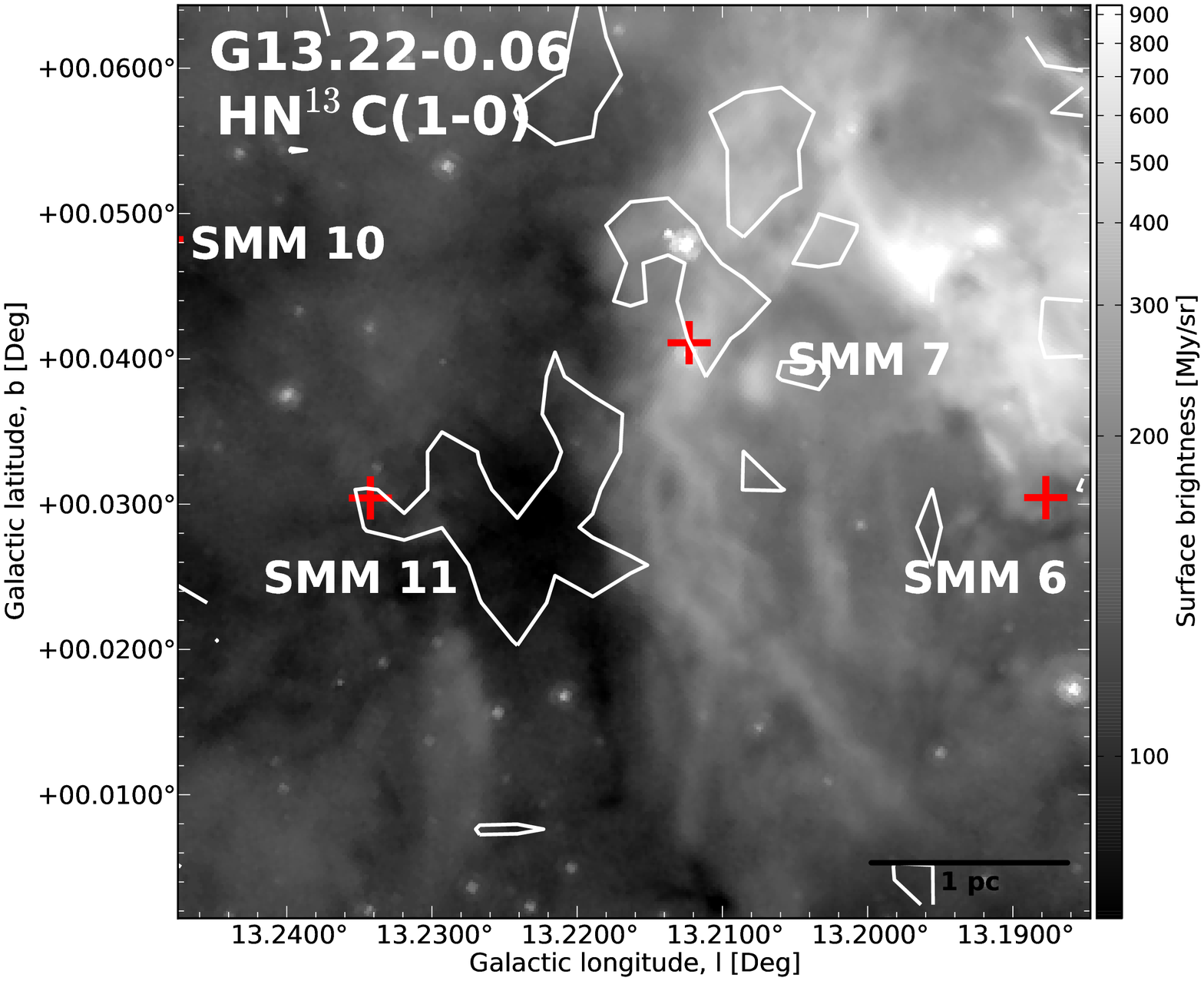}
\includegraphics[width=0.245\textwidth]{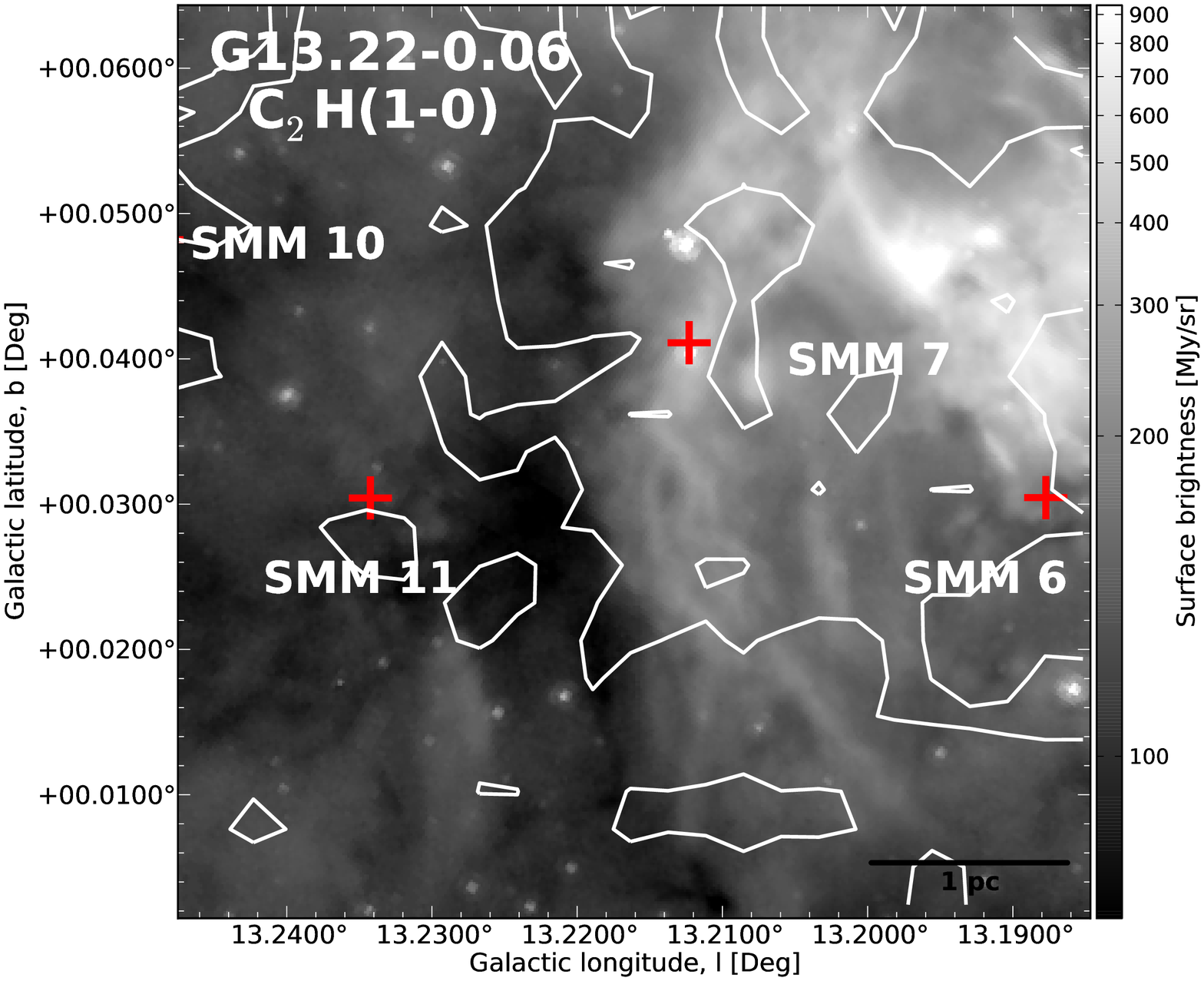}
\includegraphics[width=0.245\textwidth]{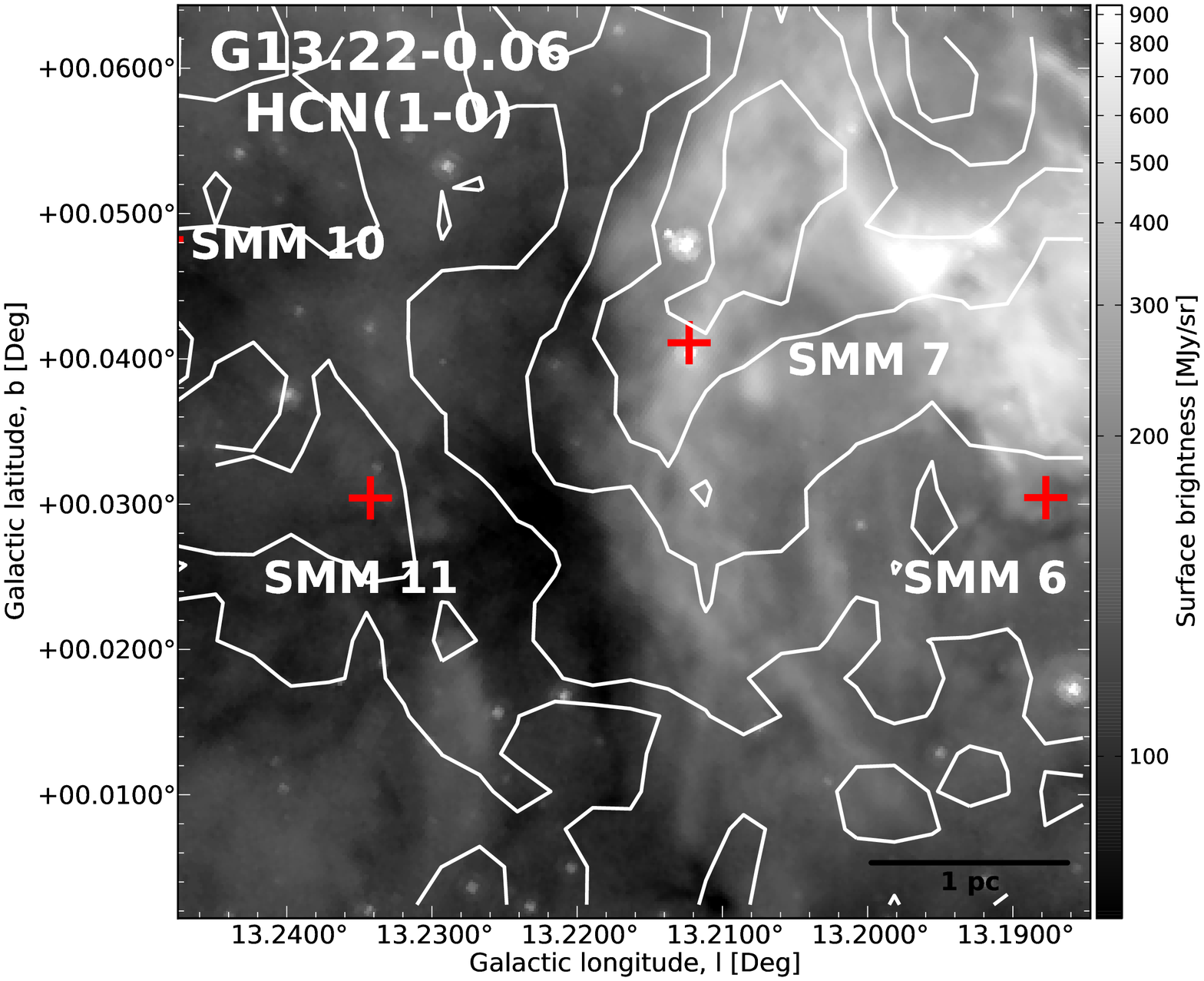}
\includegraphics[width=0.245\textwidth]{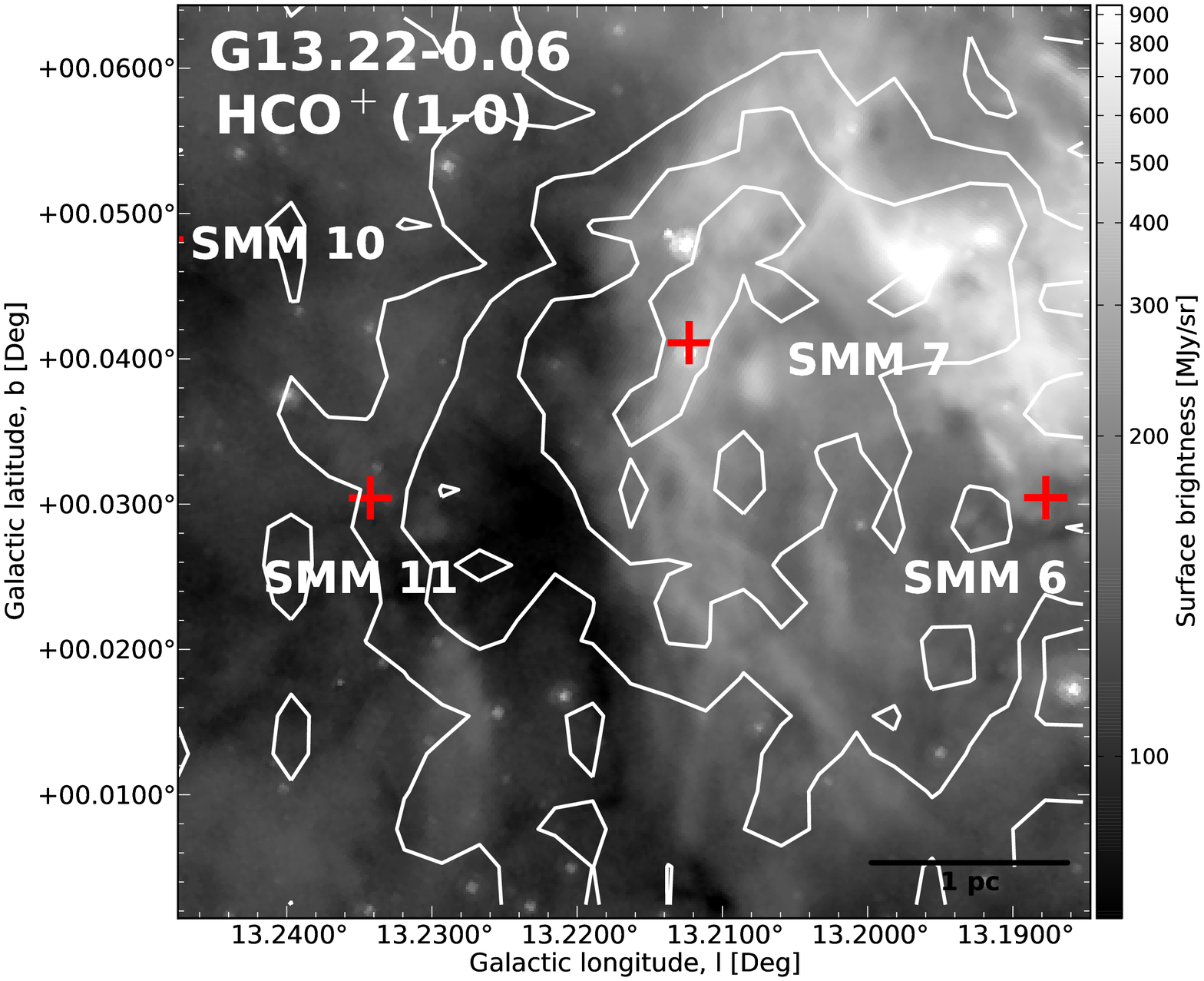}
\includegraphics[width=0.245\textwidth]{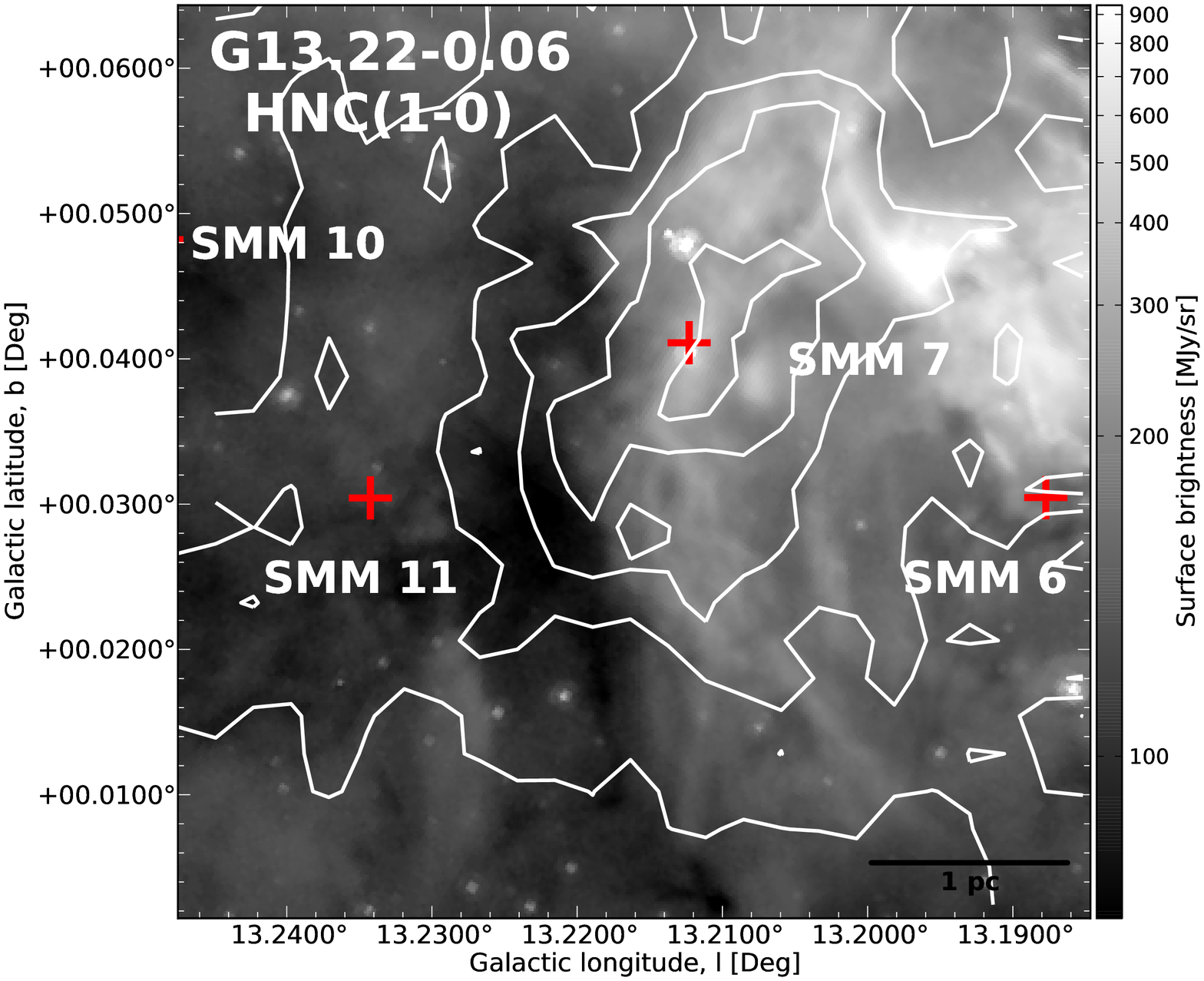}
\includegraphics[width=0.245\textwidth]{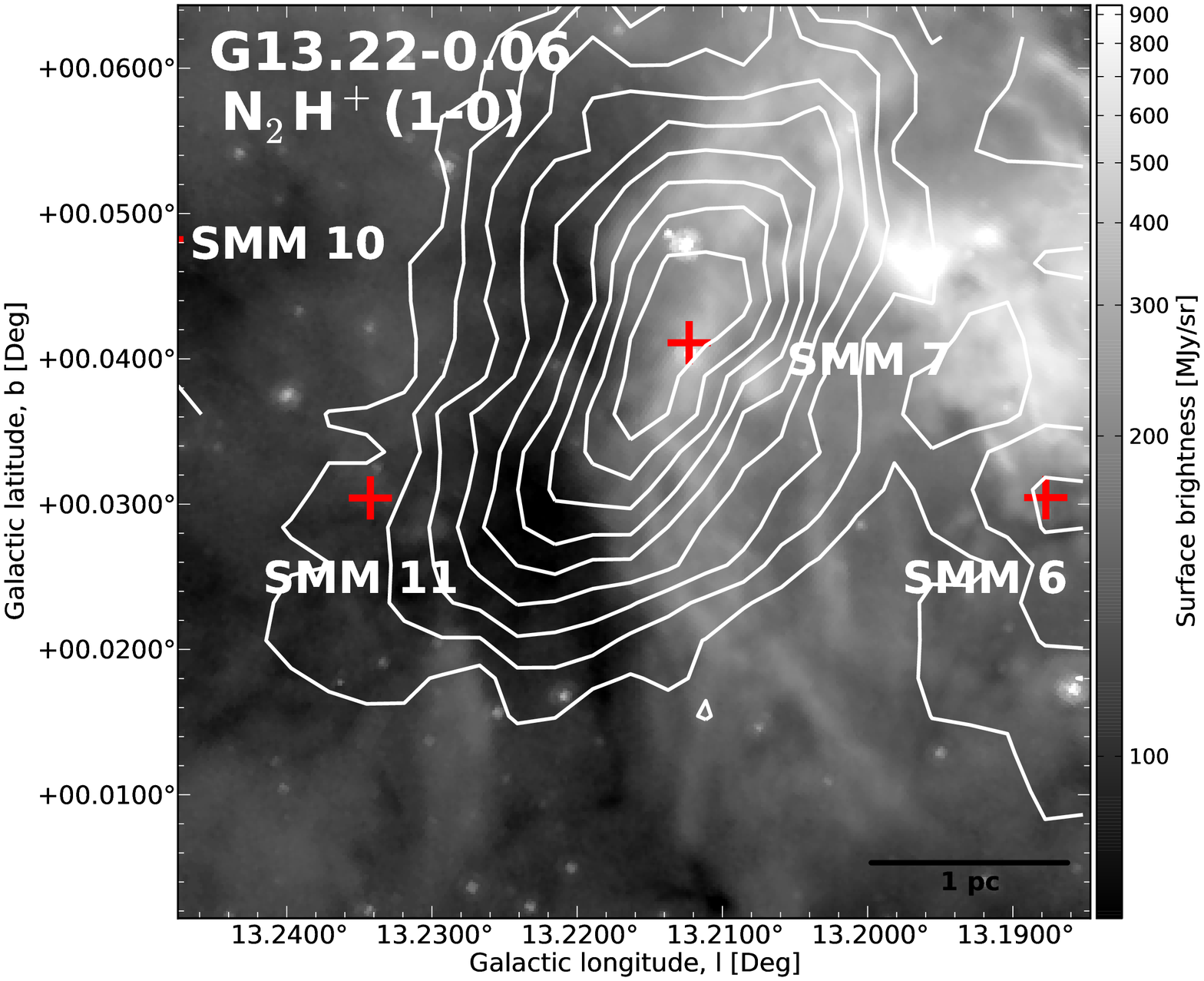}
\caption{Similar to Fig.~\ref{figure:G187SMM1lines} but towards 
G13.22--SMM 6, 7, 10, 11 around the N10/11 bubble. The contour levels 
start at $3\sigma$ for H$^{13}$CO$^+$,  HN$^{13}$C, and C$_2$H, while for HCN, 
HCO$^+$, HNC, and N$_2$H$^+$ they start at $6\sigma$, $6\sigma$, $5\sigma$, 
and $6\sigma$, respectively. In all cases, the contours go in steps of 
$3\sigma$. The average $1\sigma$ value in $T_{\rm MB}$ units is $\sim0.70$ 
K~km~s$^{-1}$. The LABOCA 870-$\mu$m peak positions of the clumps are marked 
by red plus signs. The 870-$\mu$m peak of SMM 10 lies just outside the 
MALT90 map boundary. A scale bar indicating the 1 pc 
projected length is indicated. Emission from C$_2$H, HCN, HCO$^+$, HNC, and 
N$_2$H$^+$ are similarly spatially extended, peaking towards SMM 7.}
\label{figure:G1322SMM7lines}
\end{center}
\end{figure*}

\begin{figure*}
\begin{center}
\includegraphics[width=0.245\textwidth]{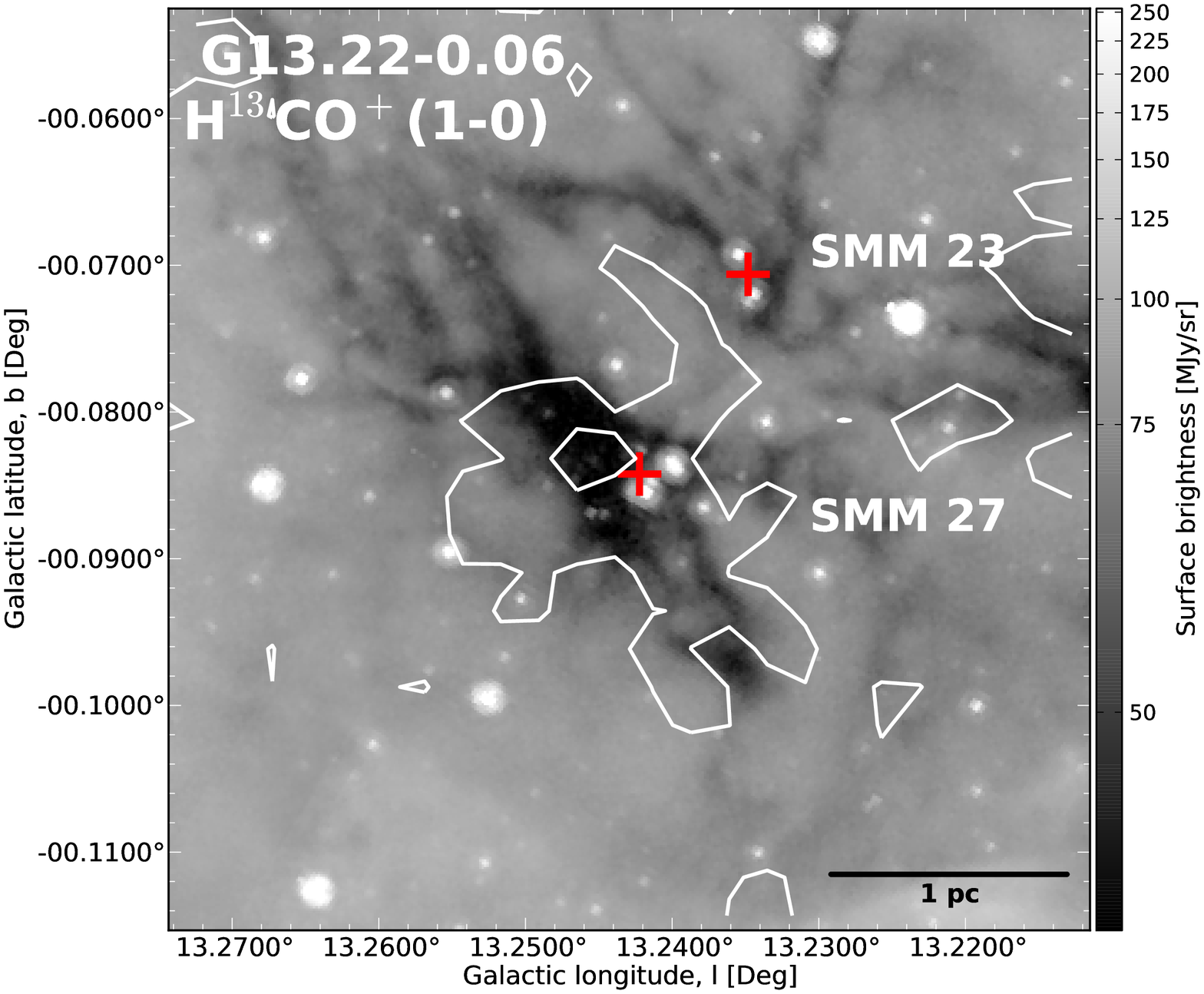}
\includegraphics[width=0.245\textwidth]{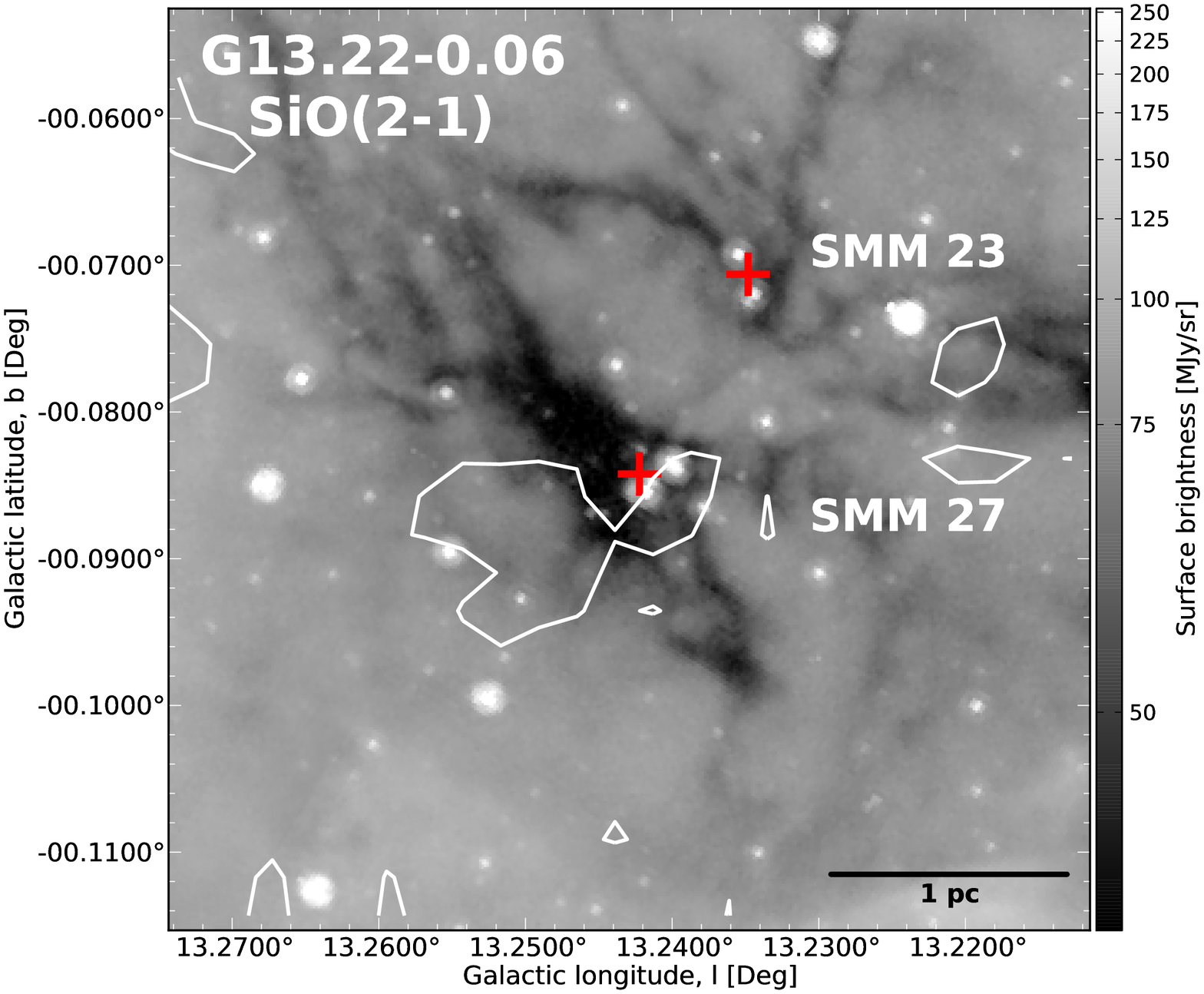}
\includegraphics[width=0.245\textwidth]{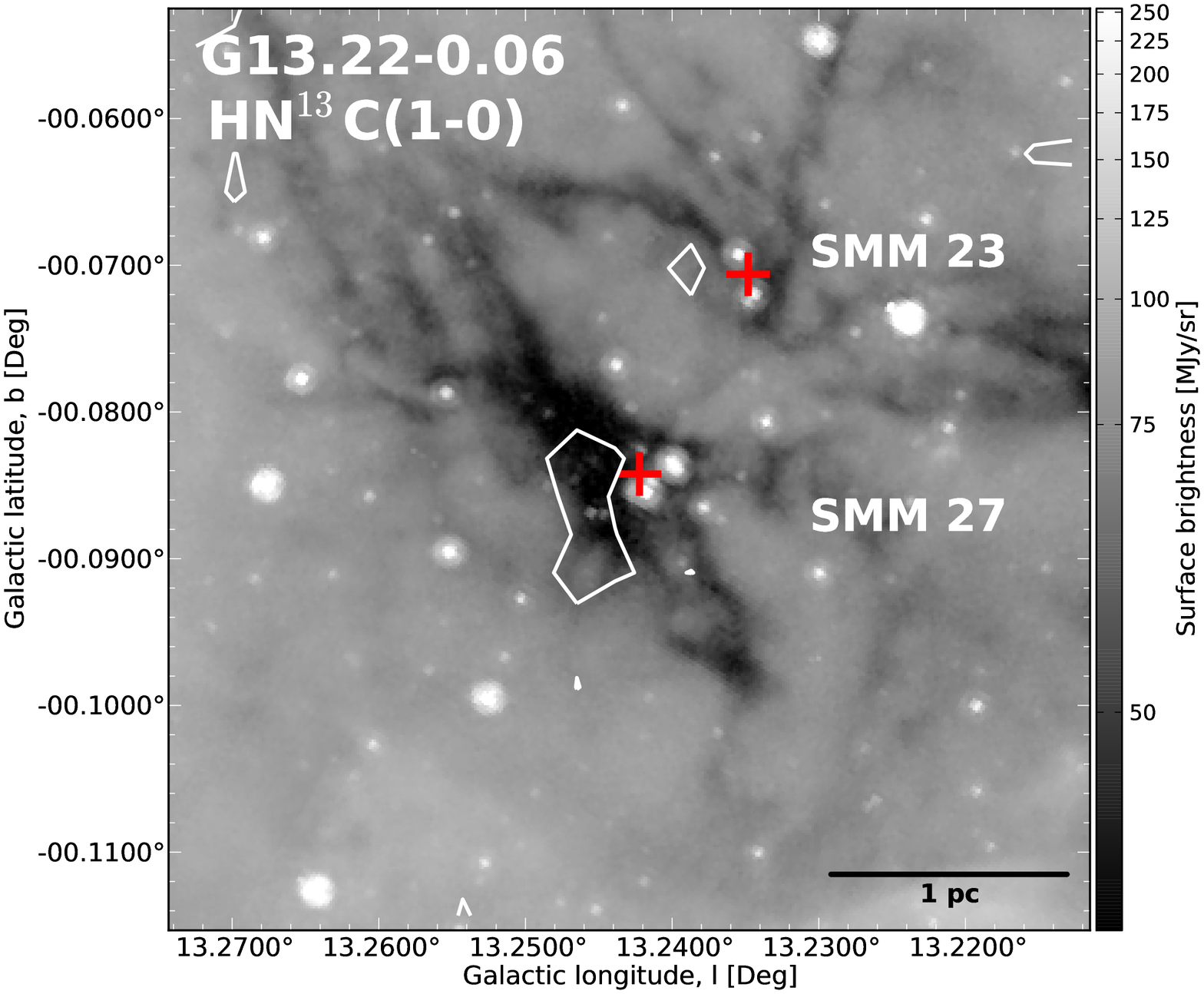}
\includegraphics[width=0.245\textwidth]{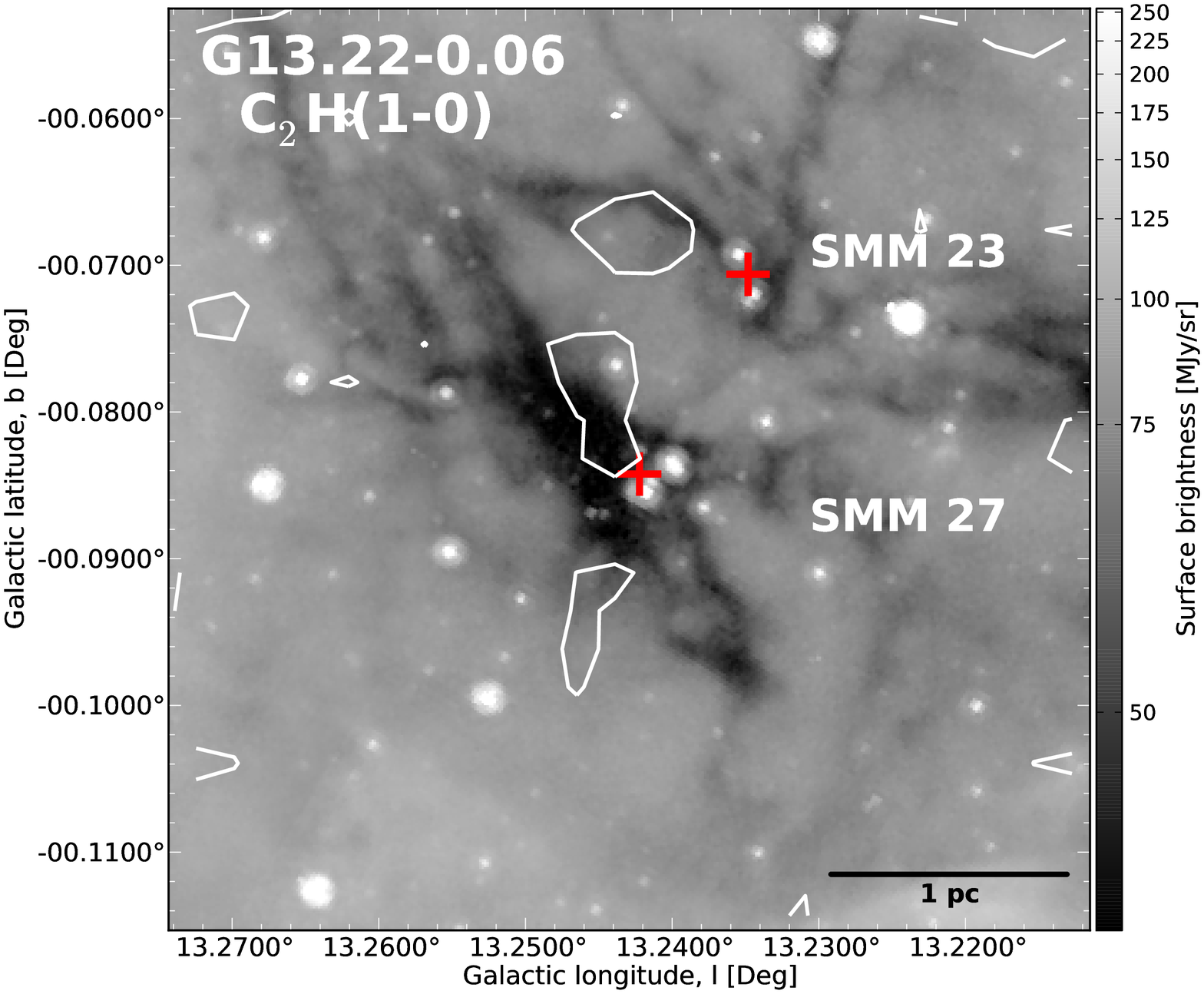}
\includegraphics[width=0.245\textwidth]{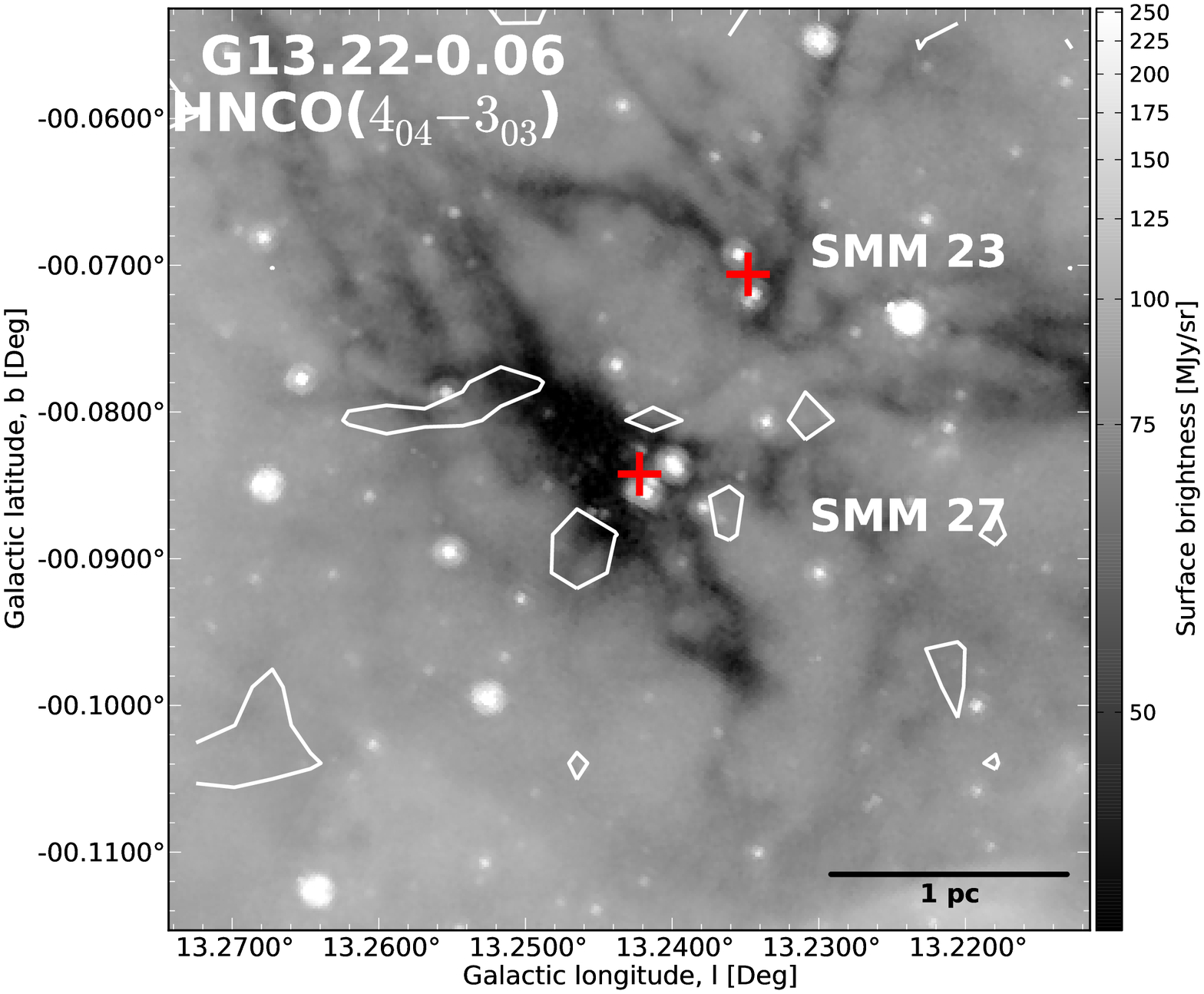}
\includegraphics[width=0.245\textwidth]{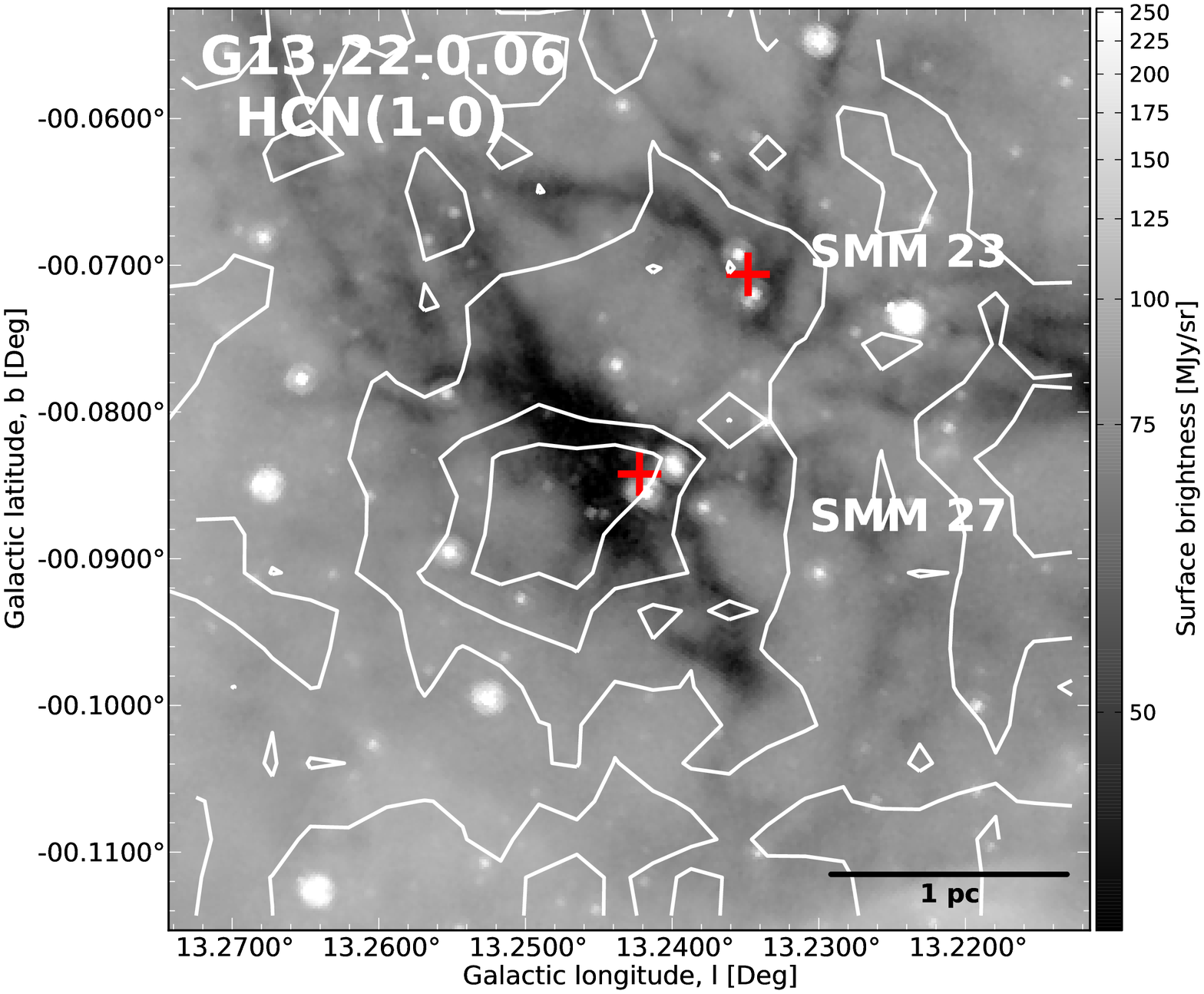}
\includegraphics[width=0.245\textwidth]{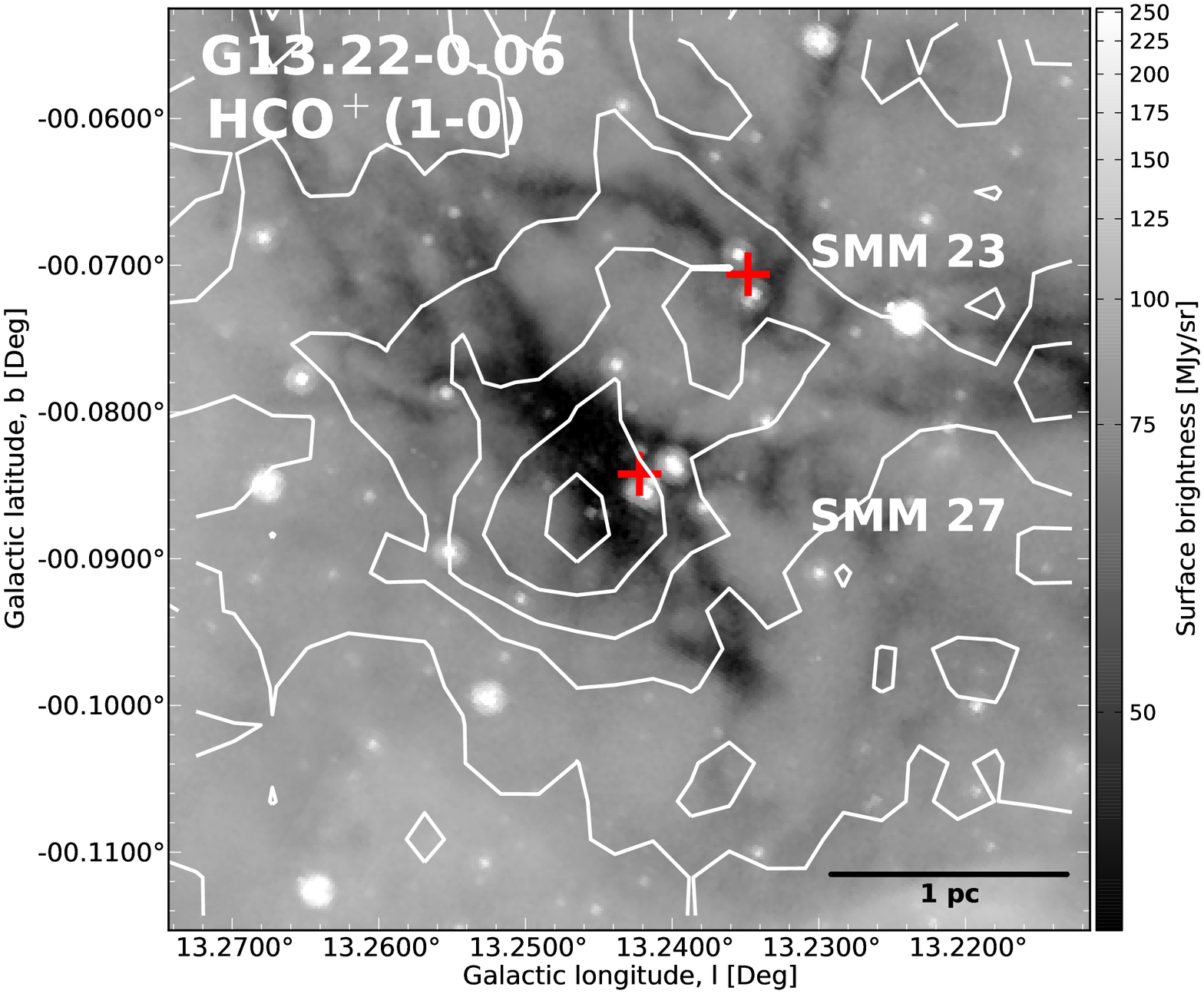}
\includegraphics[width=0.245\textwidth]{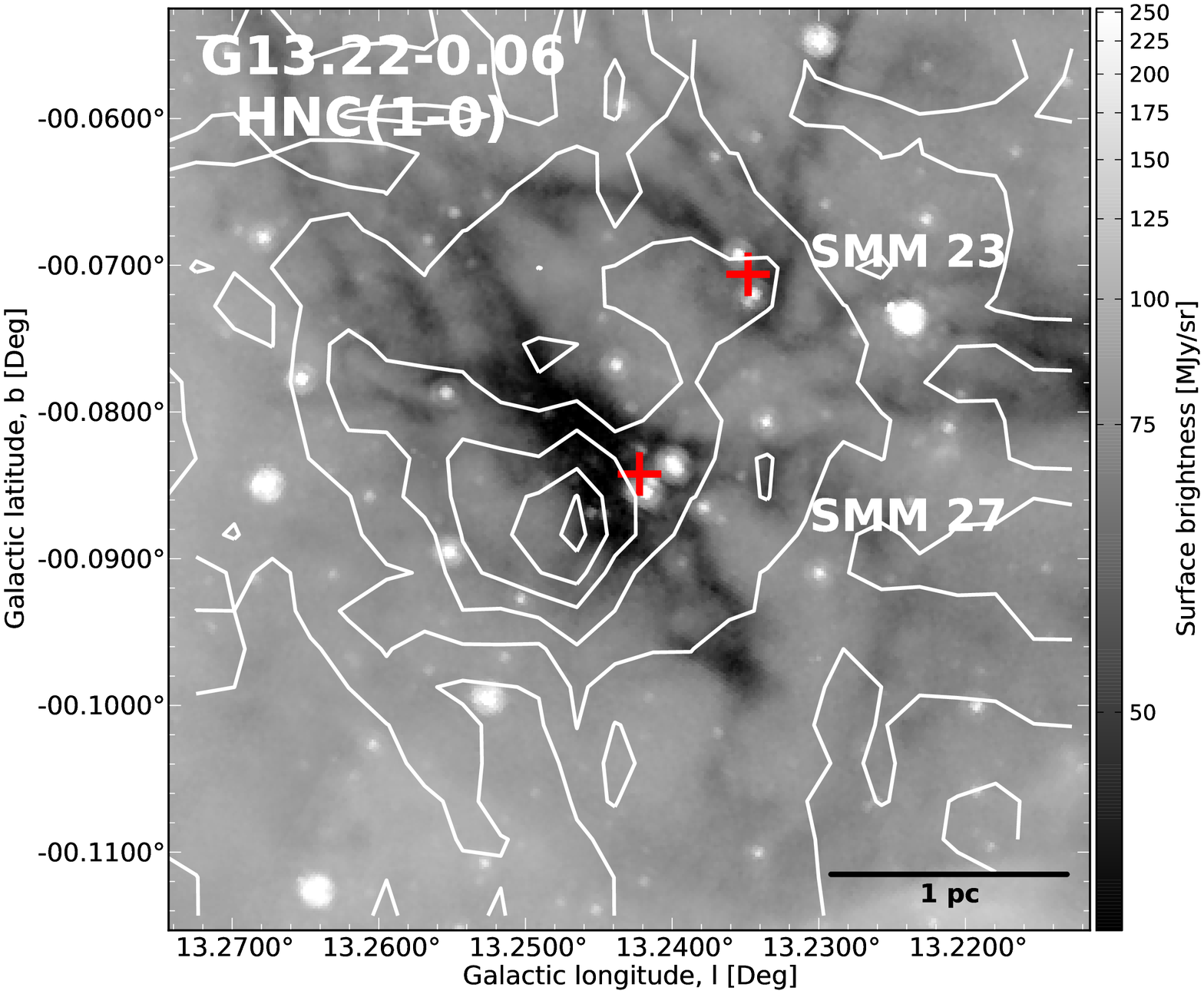}
\includegraphics[width=0.245\textwidth]{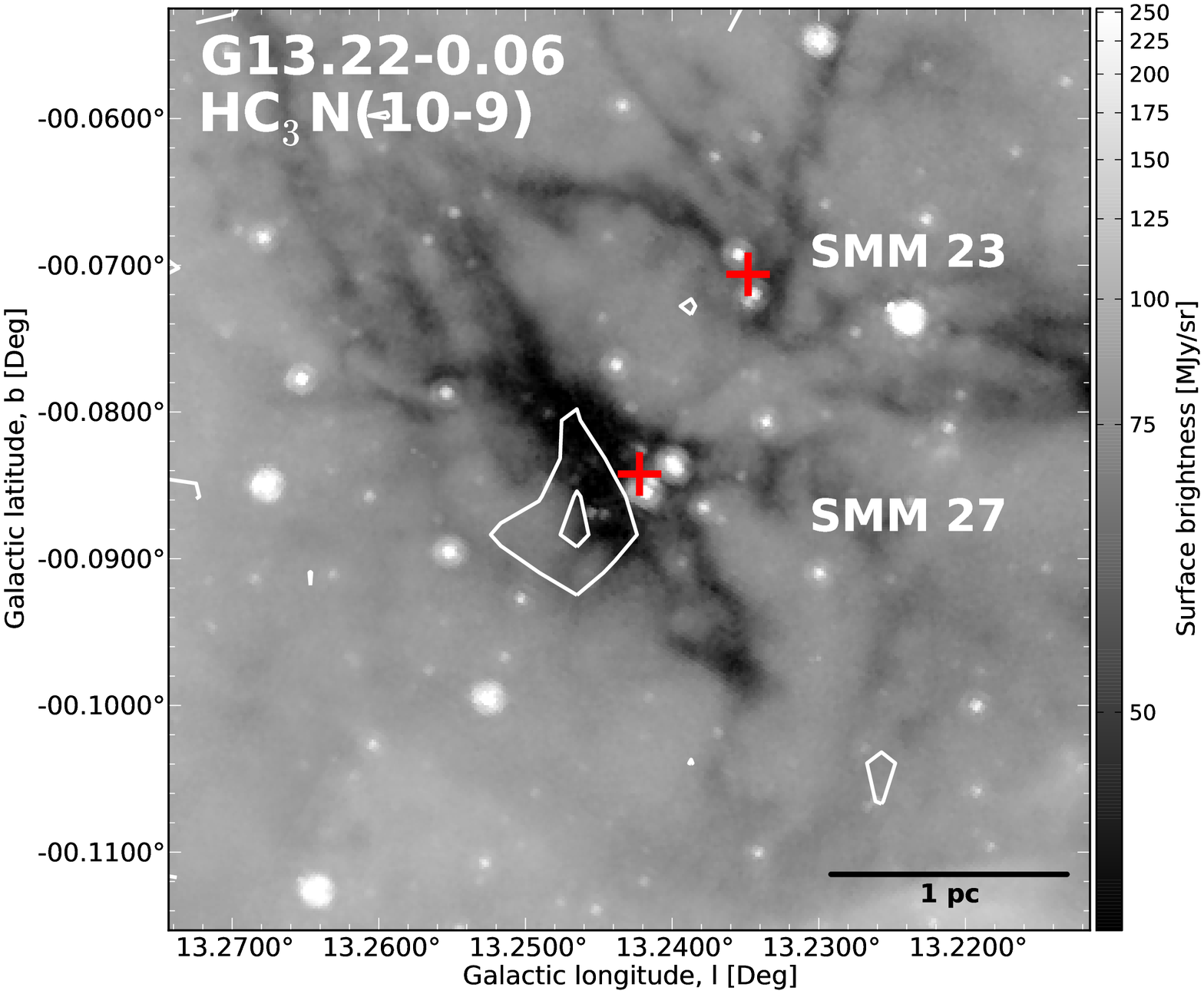}
\includegraphics[width=0.245\textwidth]{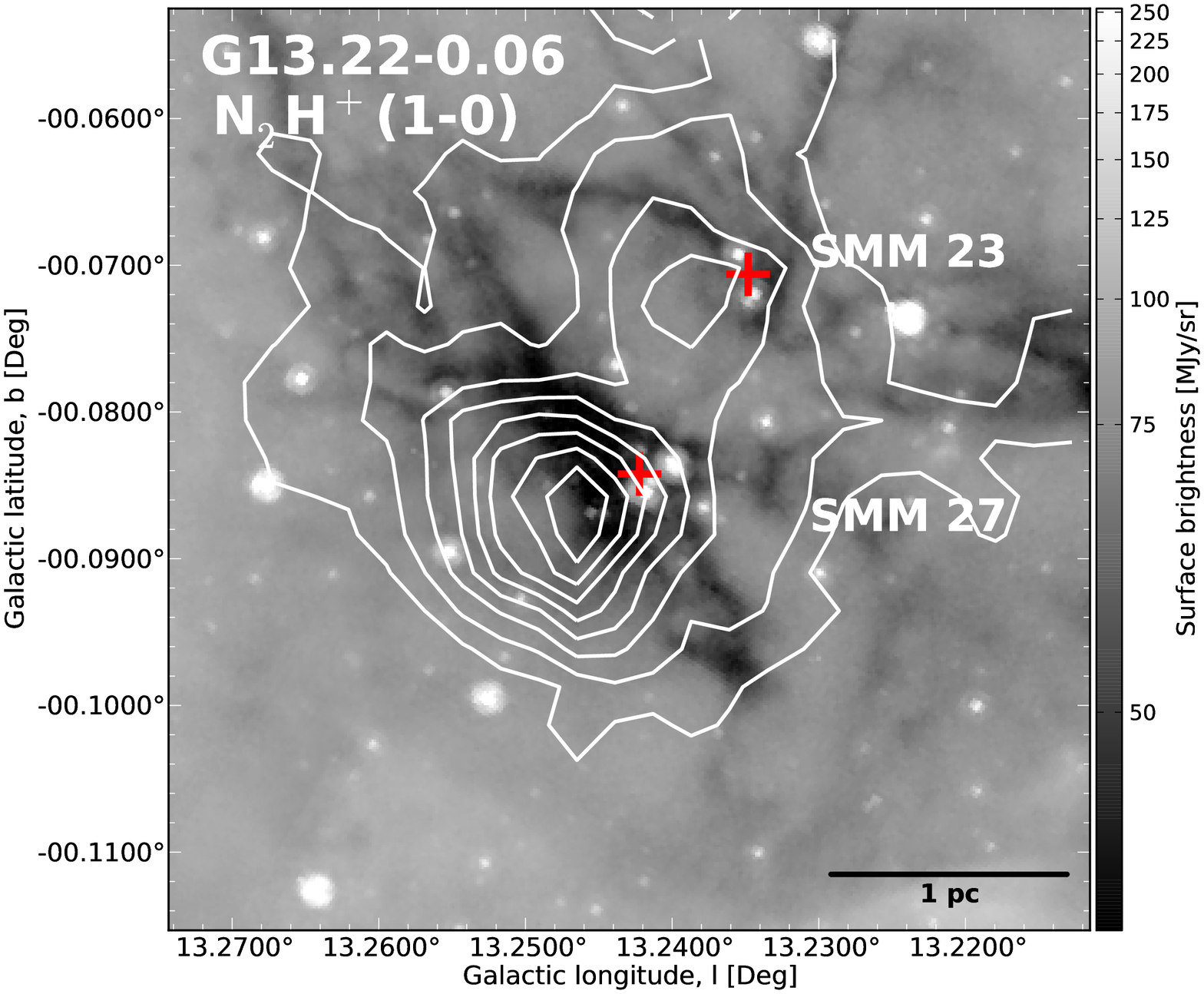}
\caption{Similar to Fig.~\ref{figure:G187SMM1lines} but towards 
G13.22--SMM 23, 27. The contour levels start at $3\sigma$ for 
H$^{13}$CO$^+$, SiO, HN$^{13}$C, C$_2$H, HNCO$(4_{0,\,4}-3_{0,\,3})$, HCN, and 
HC$_3$N, while for HCO$^+$, HNC, and N$_2$H$^+$ they start at $5\sigma$, 
$4\sigma$, and $5\sigma$, respectively. In all cases, the contours go in steps 
of $3\sigma$. The average $1\sigma$ value in $T_{\rm MB}$ units is $\sim0.69$ 
K~km~s$^{-1}$. The LABOCA 870-$\mu$m peak positions of the clumps are marked by 
red plus signs. A scale bar indicating the 1 pc 
projected length is indicated. HCN, HCO$^+$, HNC, and N$_2$H$^+$ show similar 
spatial distributions with the emission peaks being well correlated. 
Although weaker, H$^{13}$CO$^+$ and HC$_3$N also peak near SMM 27.}
\label{figure:G1322SMM23lines}
\end{center}
\end{figure*}

\begin{figure*}
\begin{center}
\includegraphics[width=0.245\textwidth]{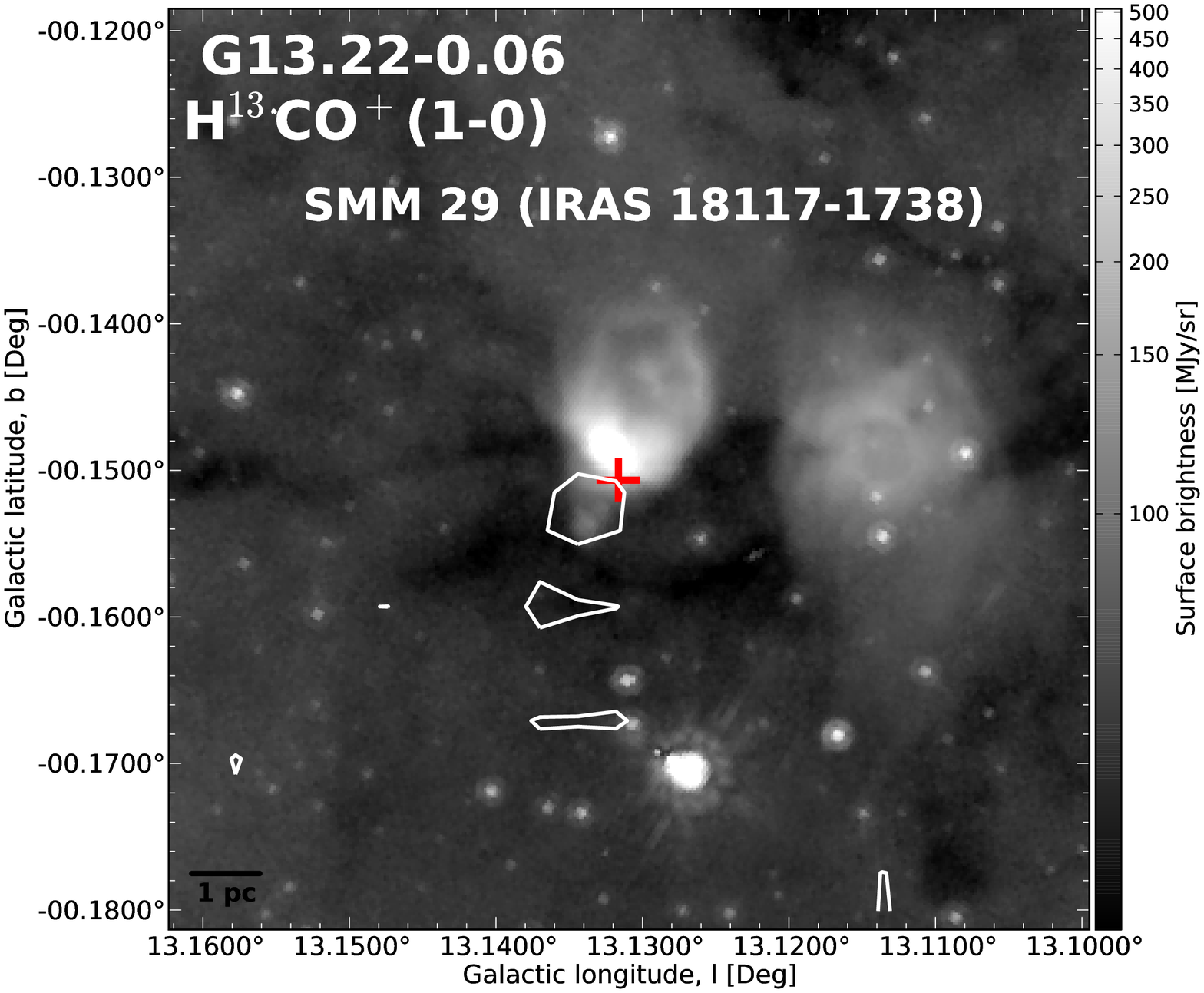}
\includegraphics[width=0.245\textwidth]{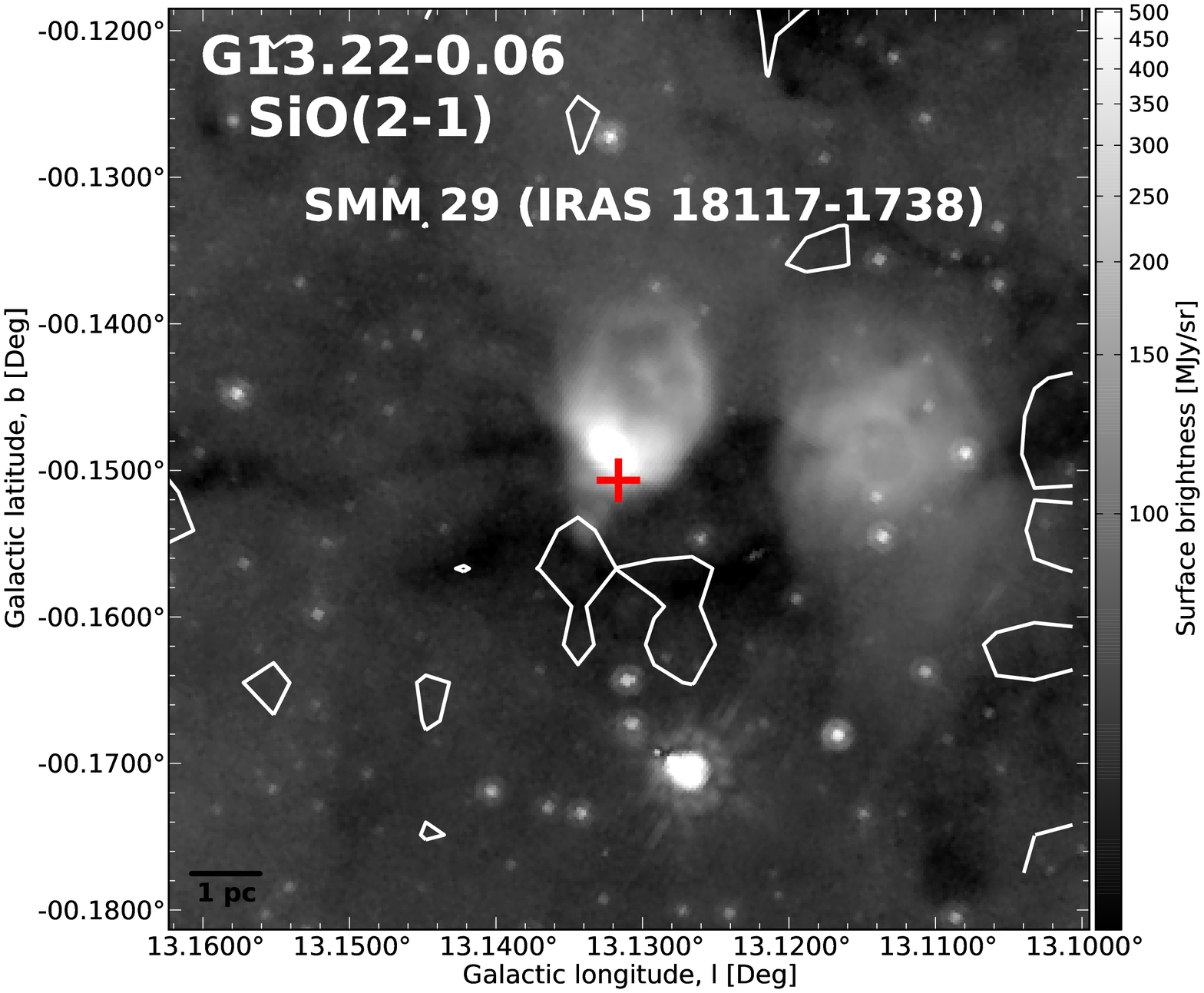}
\includegraphics[width=0.245\textwidth]{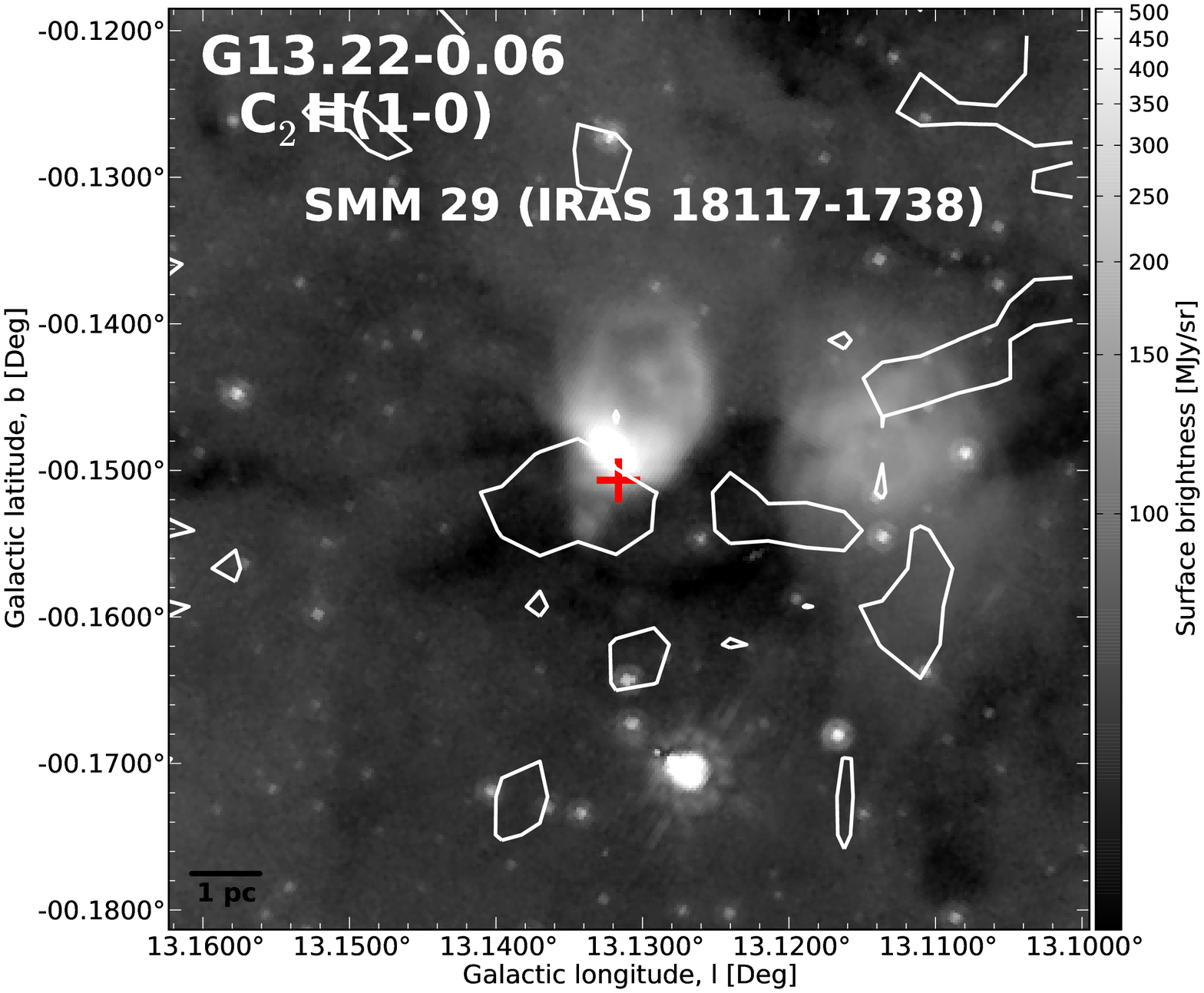}
\includegraphics[width=0.245\textwidth]{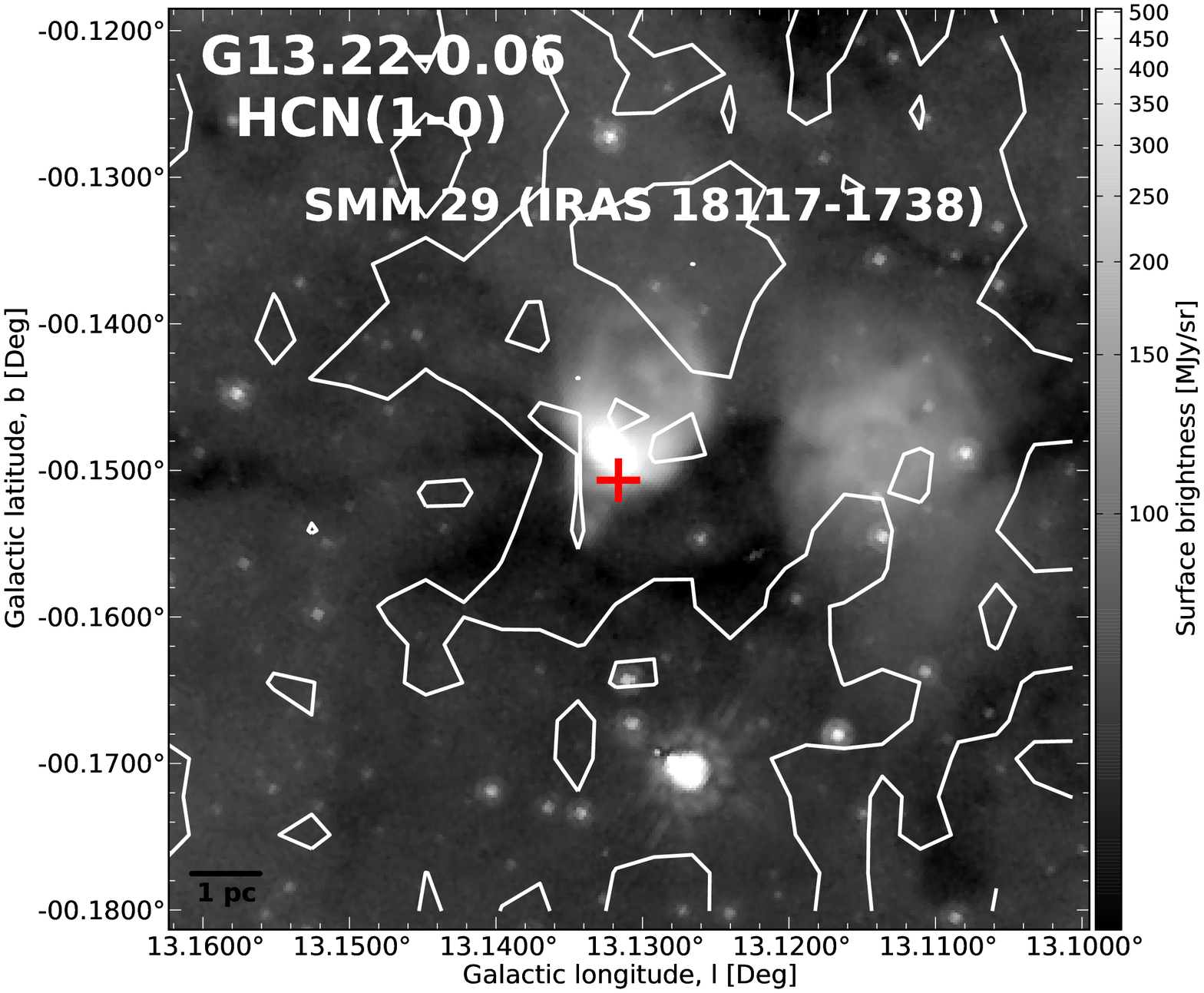}
\includegraphics[width=0.245\textwidth]{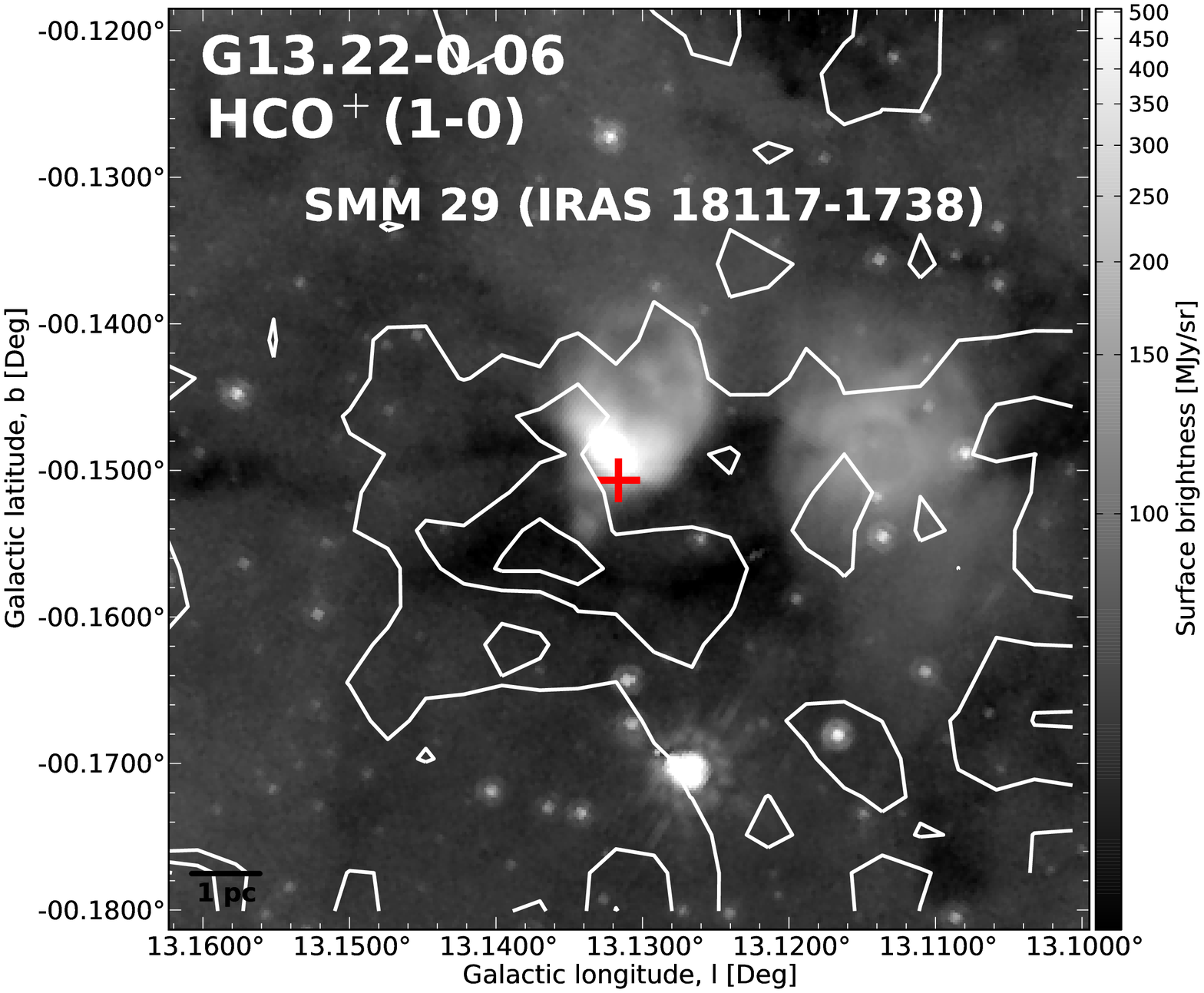}
\includegraphics[width=0.245\textwidth]{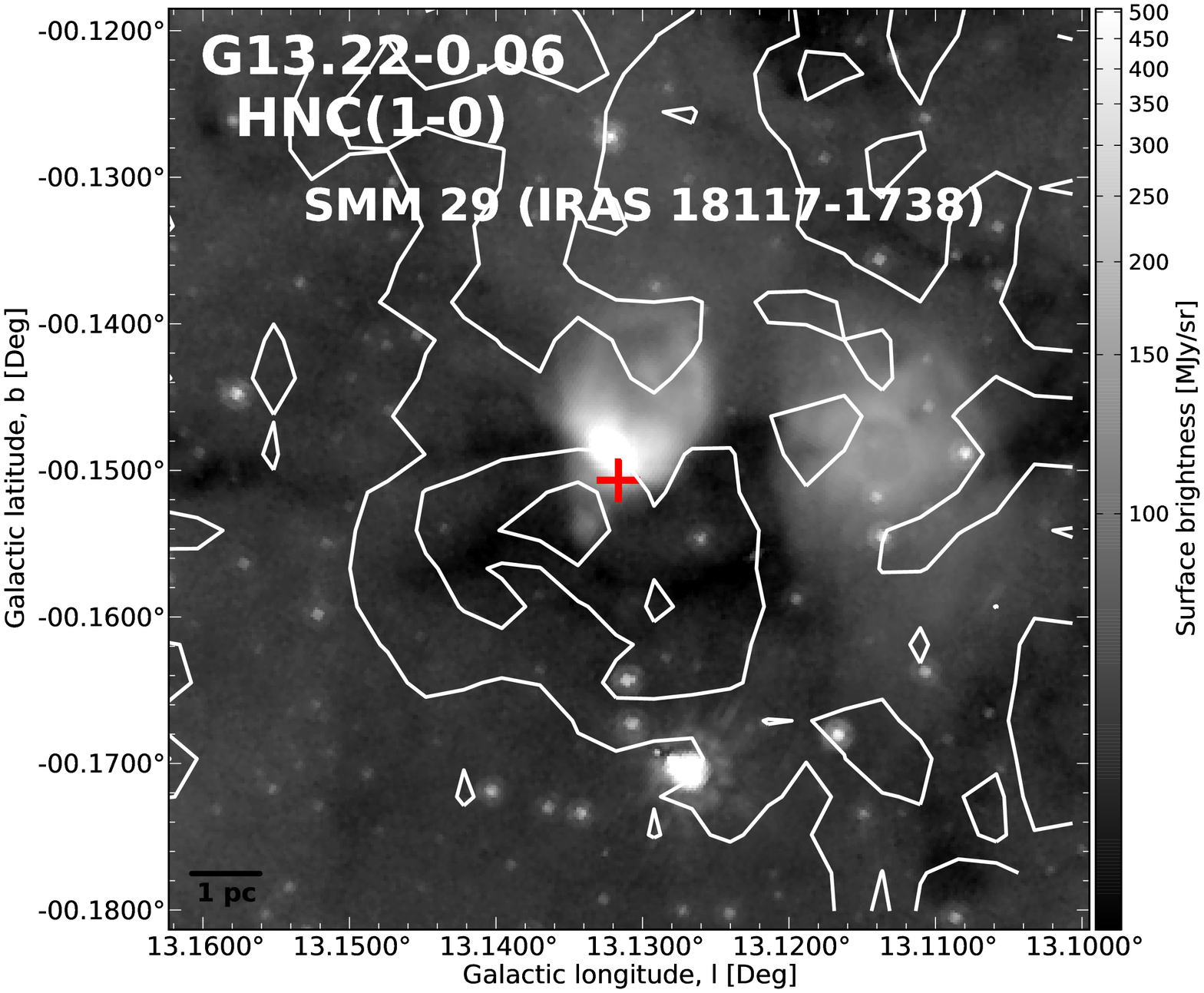}
\includegraphics[width=0.245\textwidth]{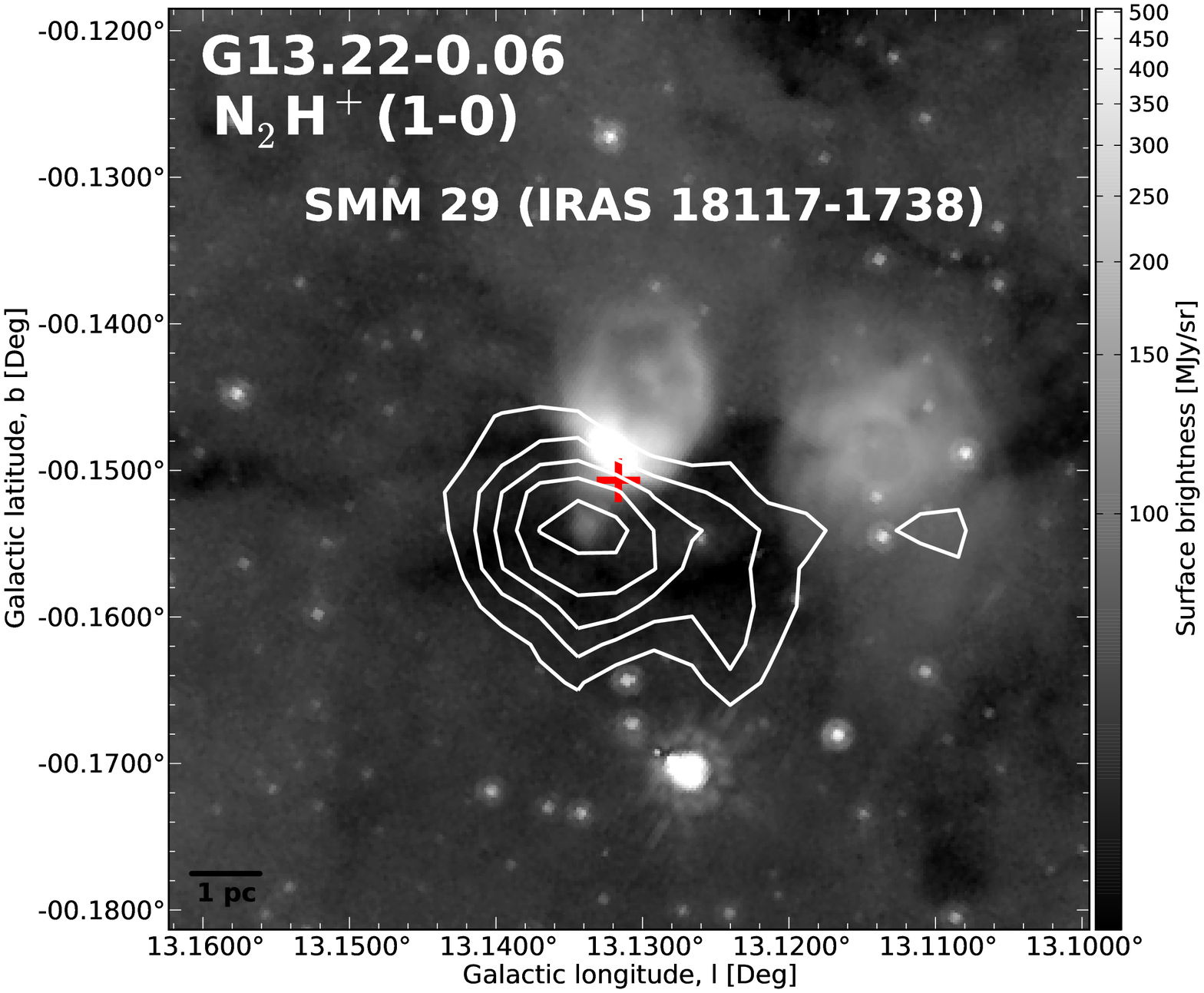}
\caption{Similar to Fig.~\ref{figure:G187SMM1lines} but towards 
G13.22--SMM 29. The contour levels start at $3\sigma$ for H$^{13}$CO$^+$, 
SiO, and C$_2$H, at $4\sigma$ for HCN and HNC, and at $5\sigma$ for HCO$^+$ 
and N$_2$H$^+$. In all cases, the contours go in steps of $3\sigma$. 
The average $1\sigma$ value in $T_{\rm MB}$ units is $\sim0.66$ K~km~s$^{-1}$. 
The LABOCA 870-$\mu$m peak position of the clump is marked by a red plus 
signs. A scale bar indicating the 1 pc projected length is indicated. 
The HCO$^+$ and HNC emissions show similar distributions, with HCN sharing 
some spatial features. The strong N$_2$H$^+$ emission peaks towards the HNC 
maximum.}
\label{figure:G1322SMM29lines}
\end{center}
\end{figure*}

\begin{figure*}
\begin{center}
\includegraphics[width=0.245\textwidth]{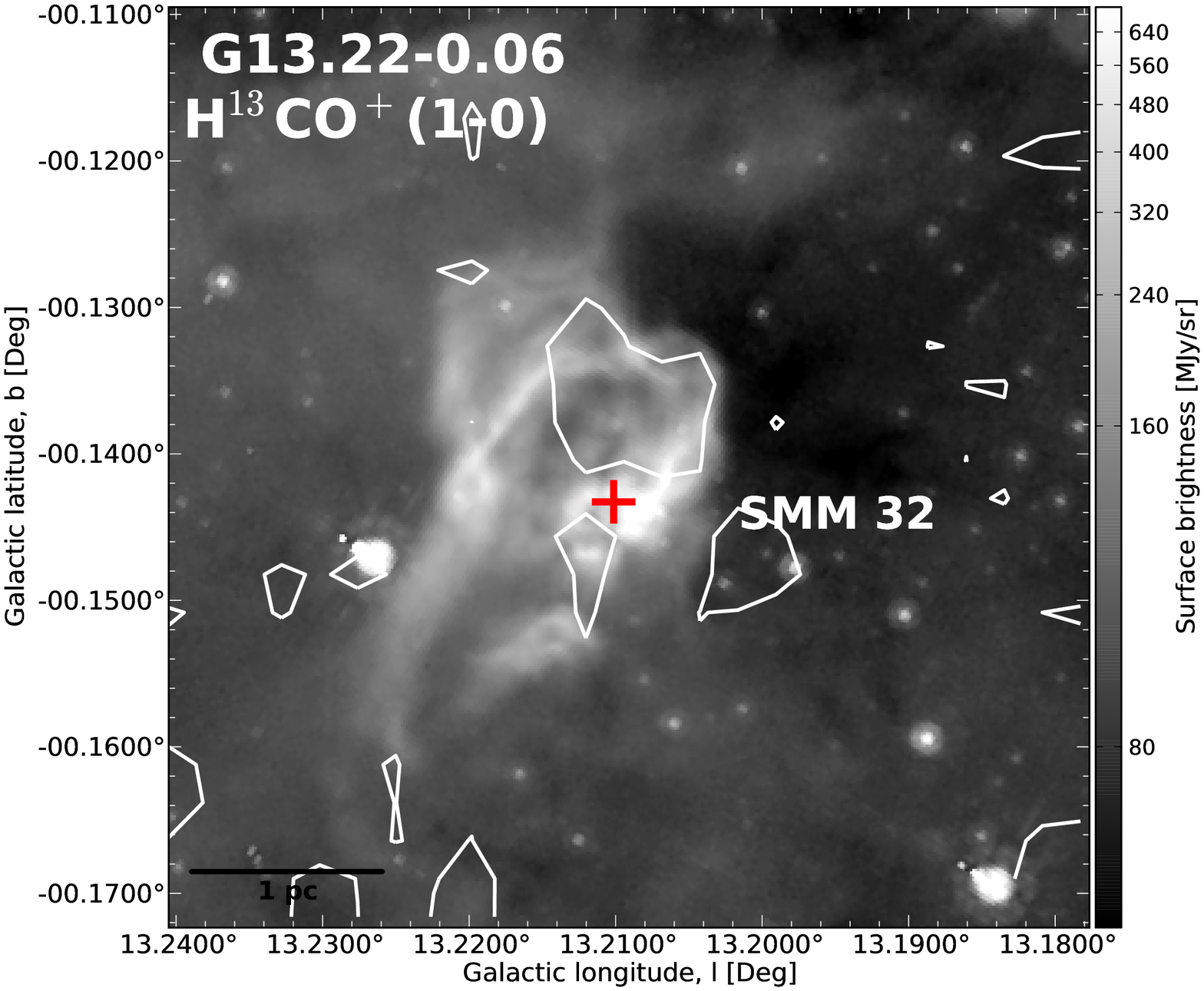}
\includegraphics[width=0.245\textwidth]{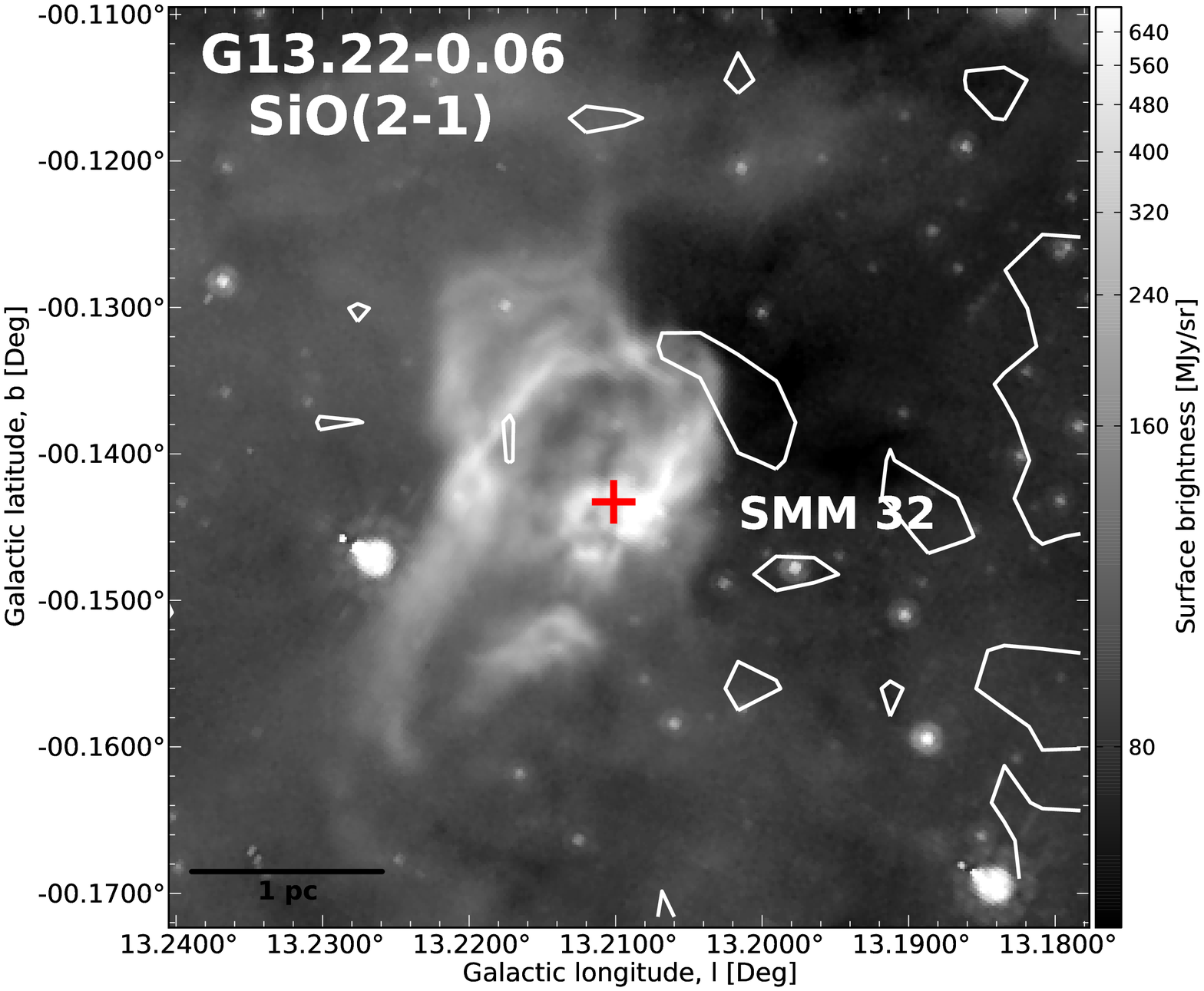}
\includegraphics[width=0.245\textwidth]{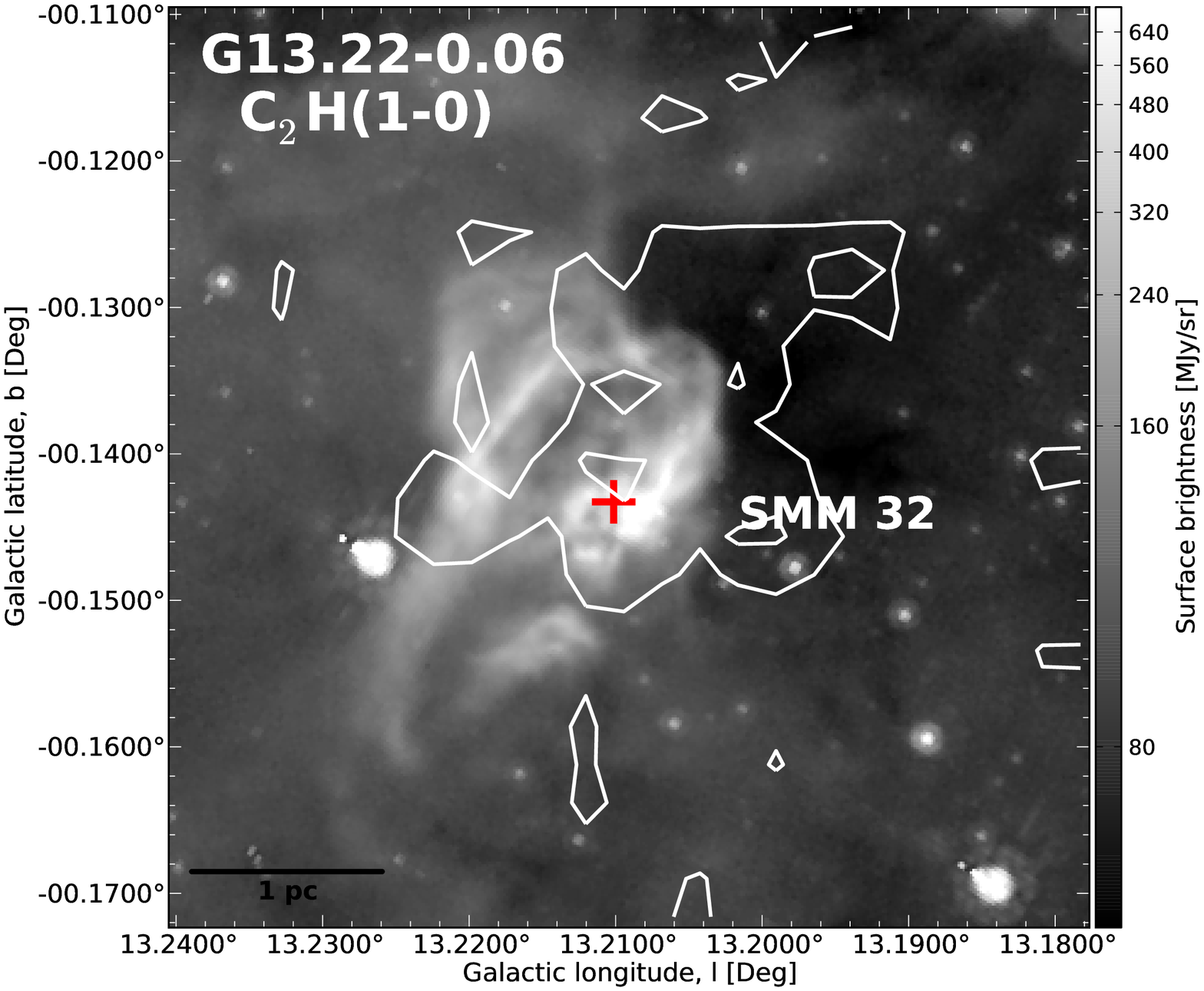}
\includegraphics[width=0.245\textwidth]{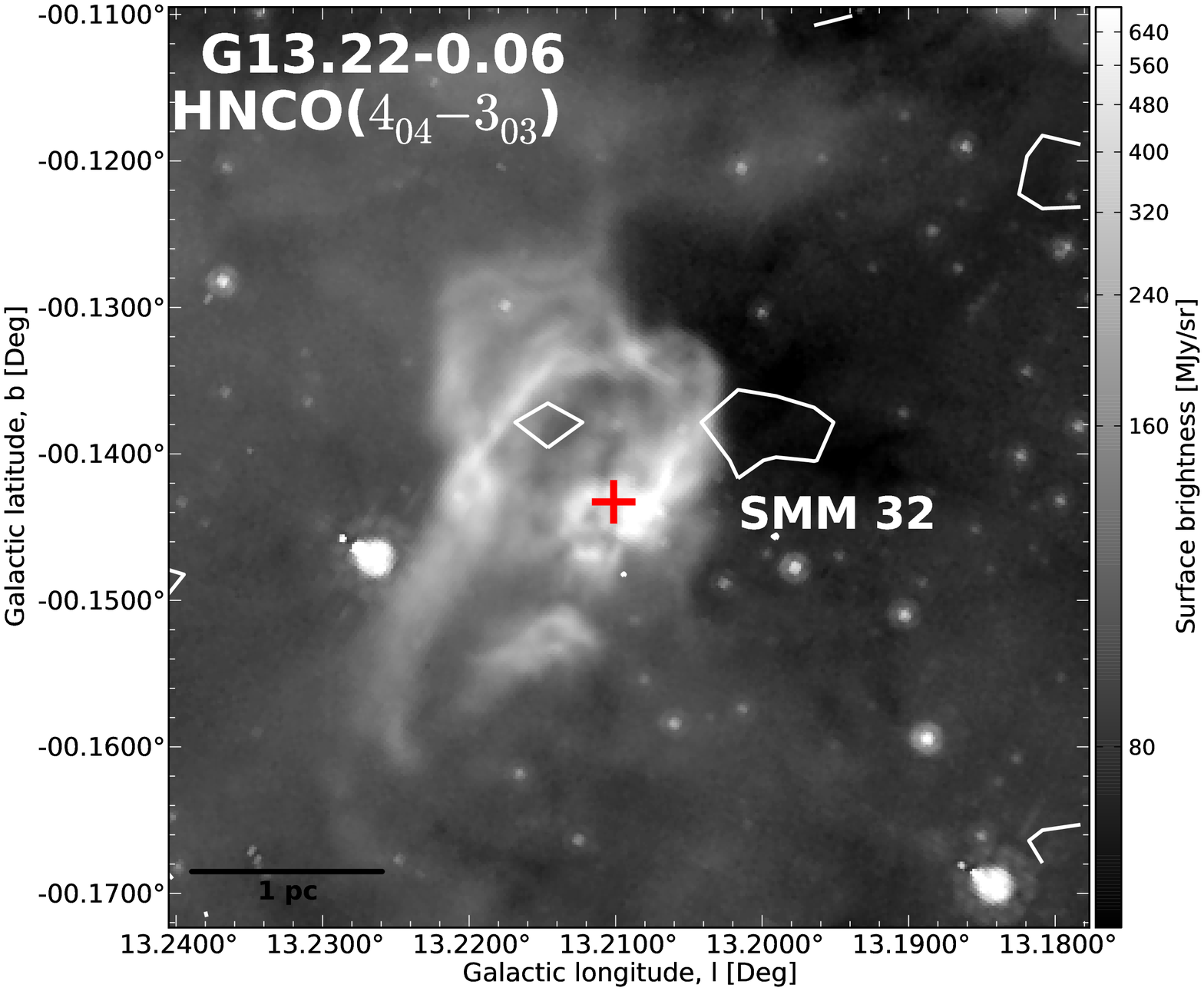}
\includegraphics[width=0.245\textwidth]{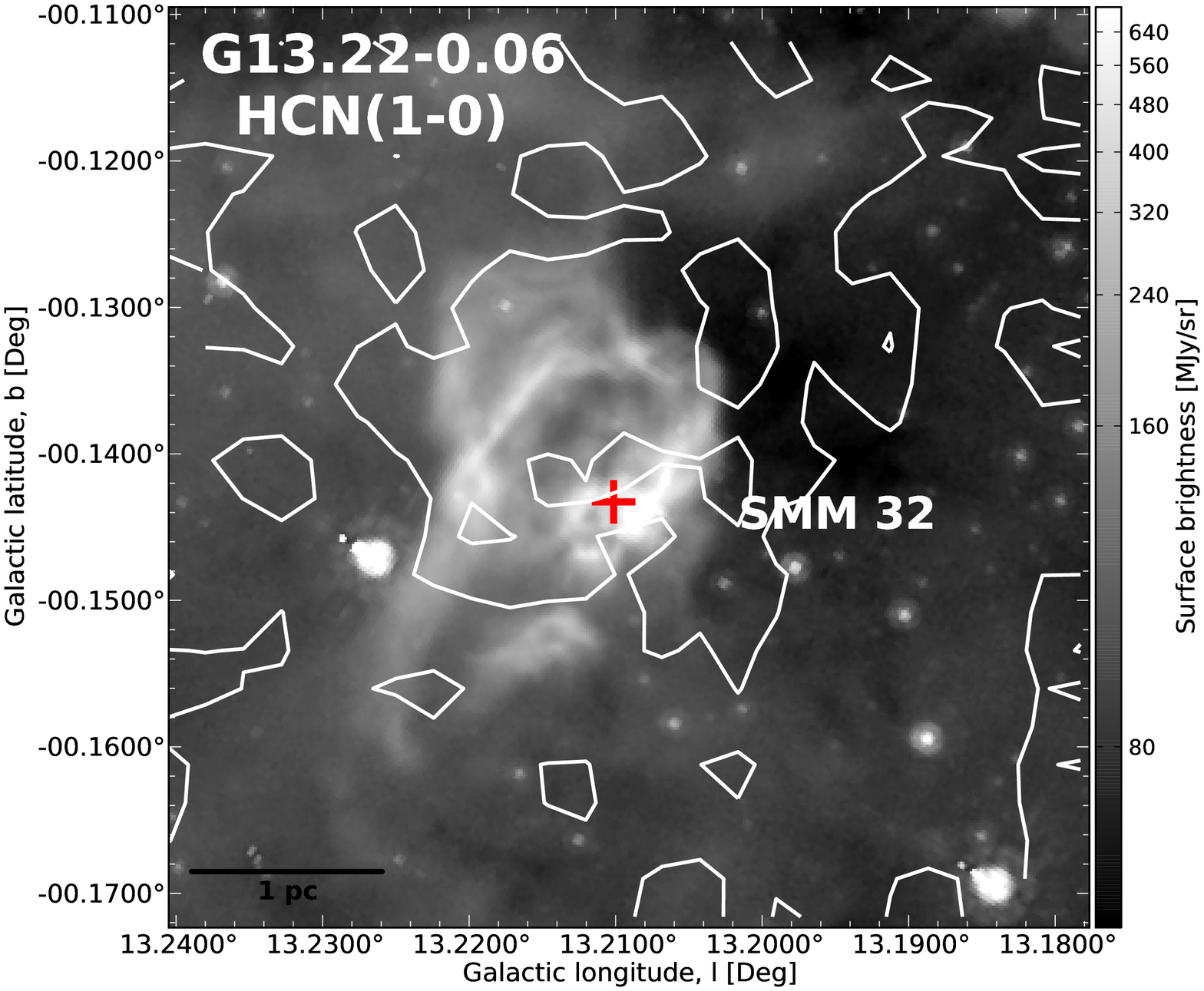}
\includegraphics[width=0.245\textwidth]{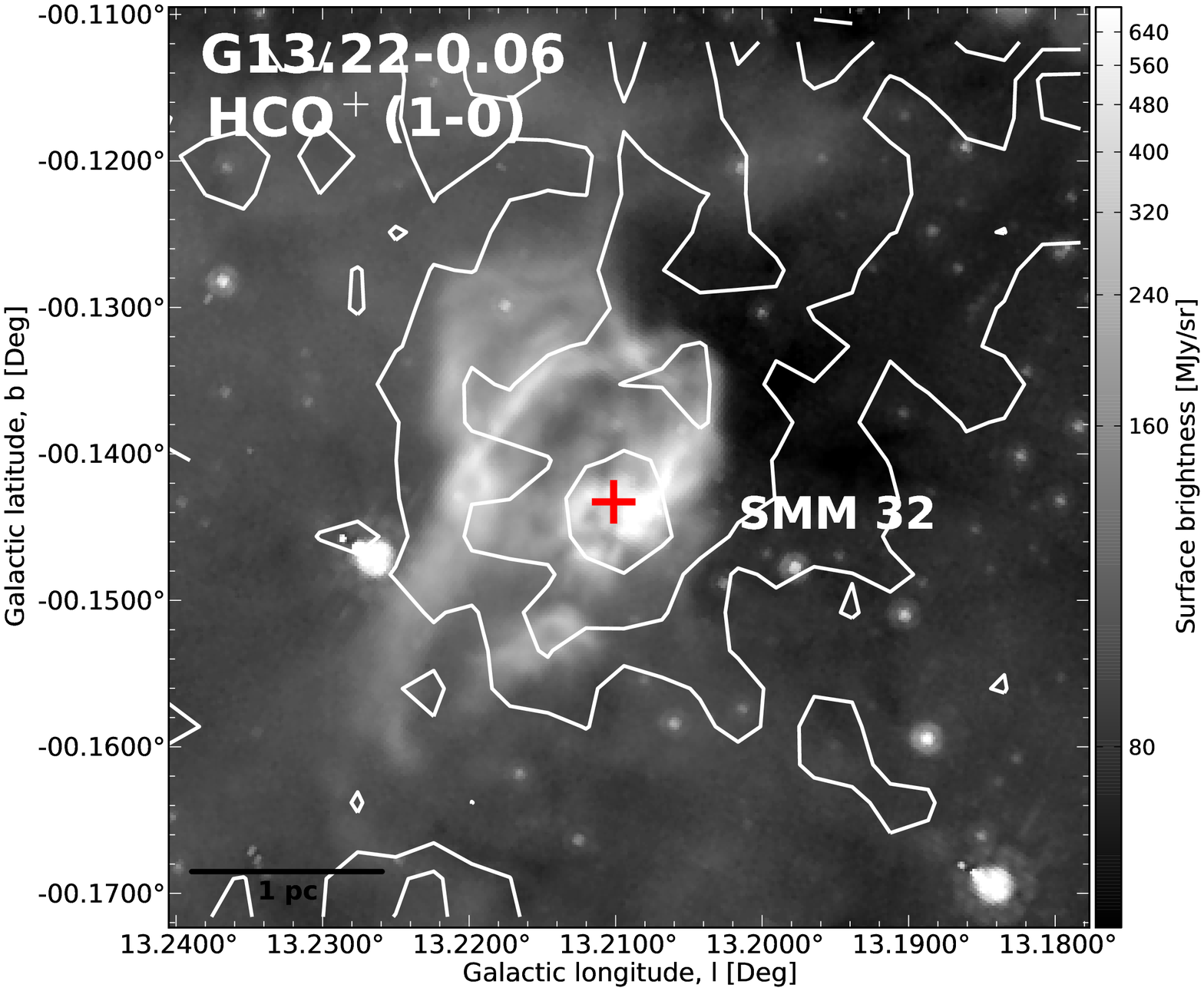}
\includegraphics[width=0.245\textwidth]{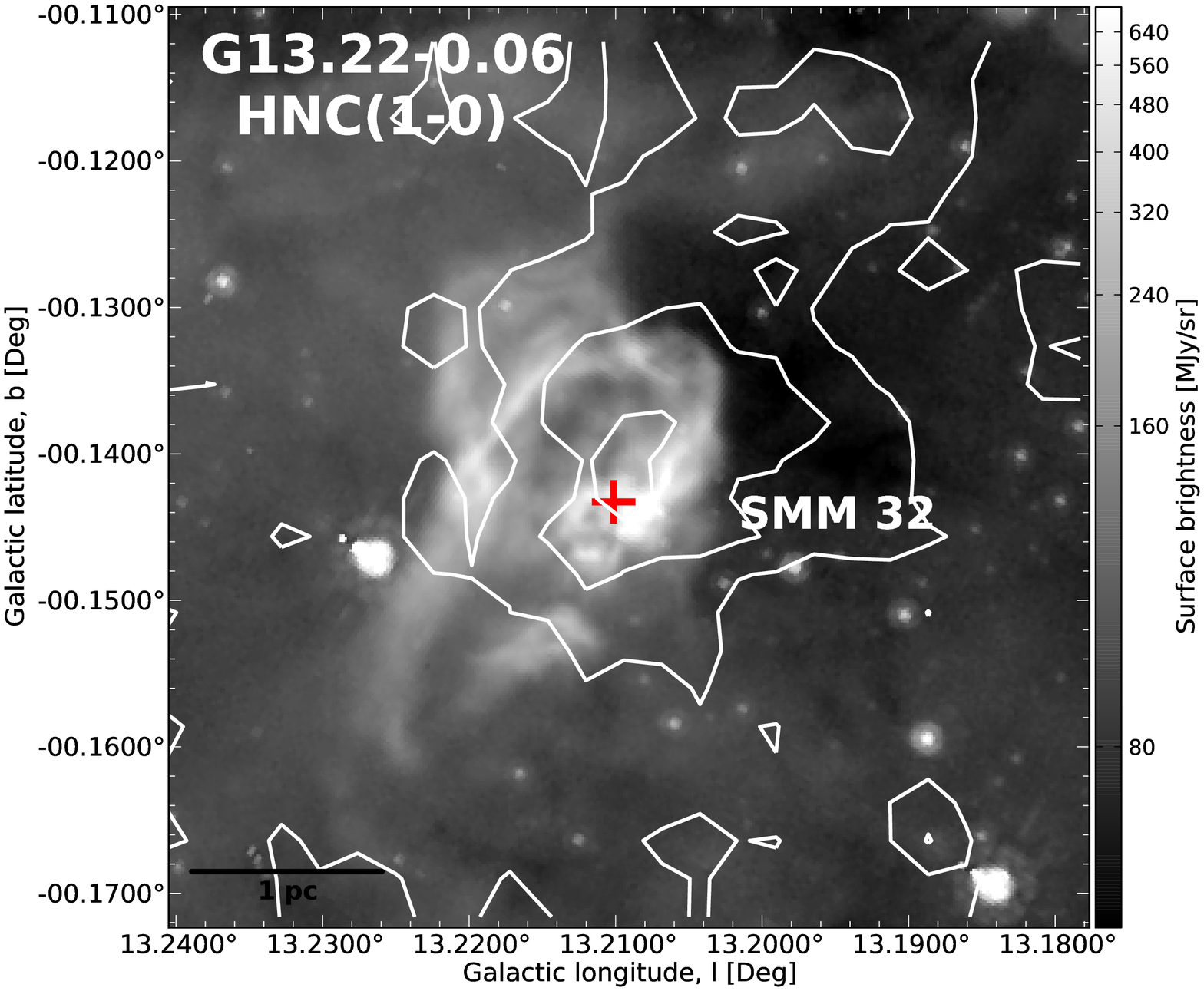}
\includegraphics[width=0.245\textwidth]{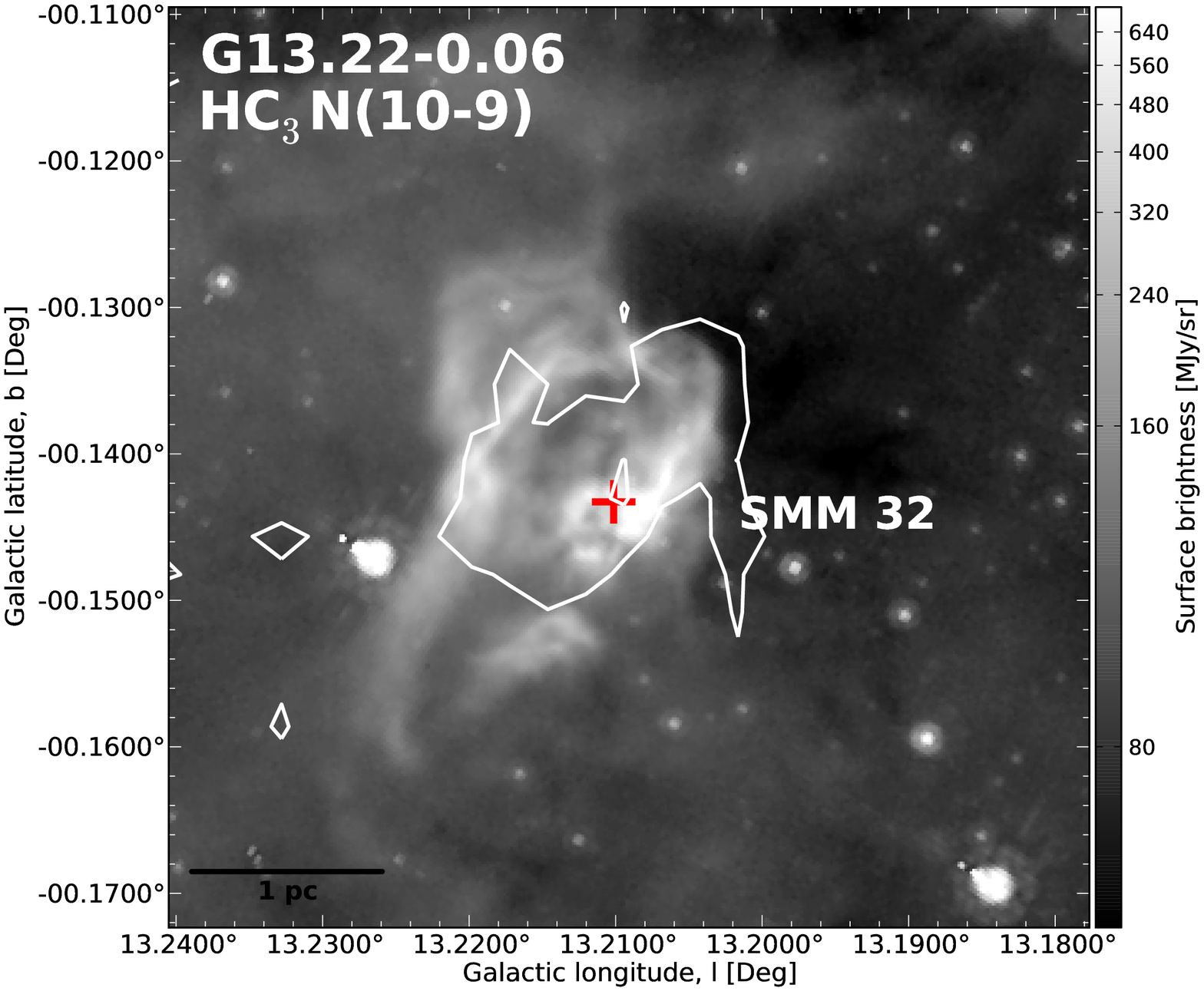}
\includegraphics[width=0.245\textwidth]{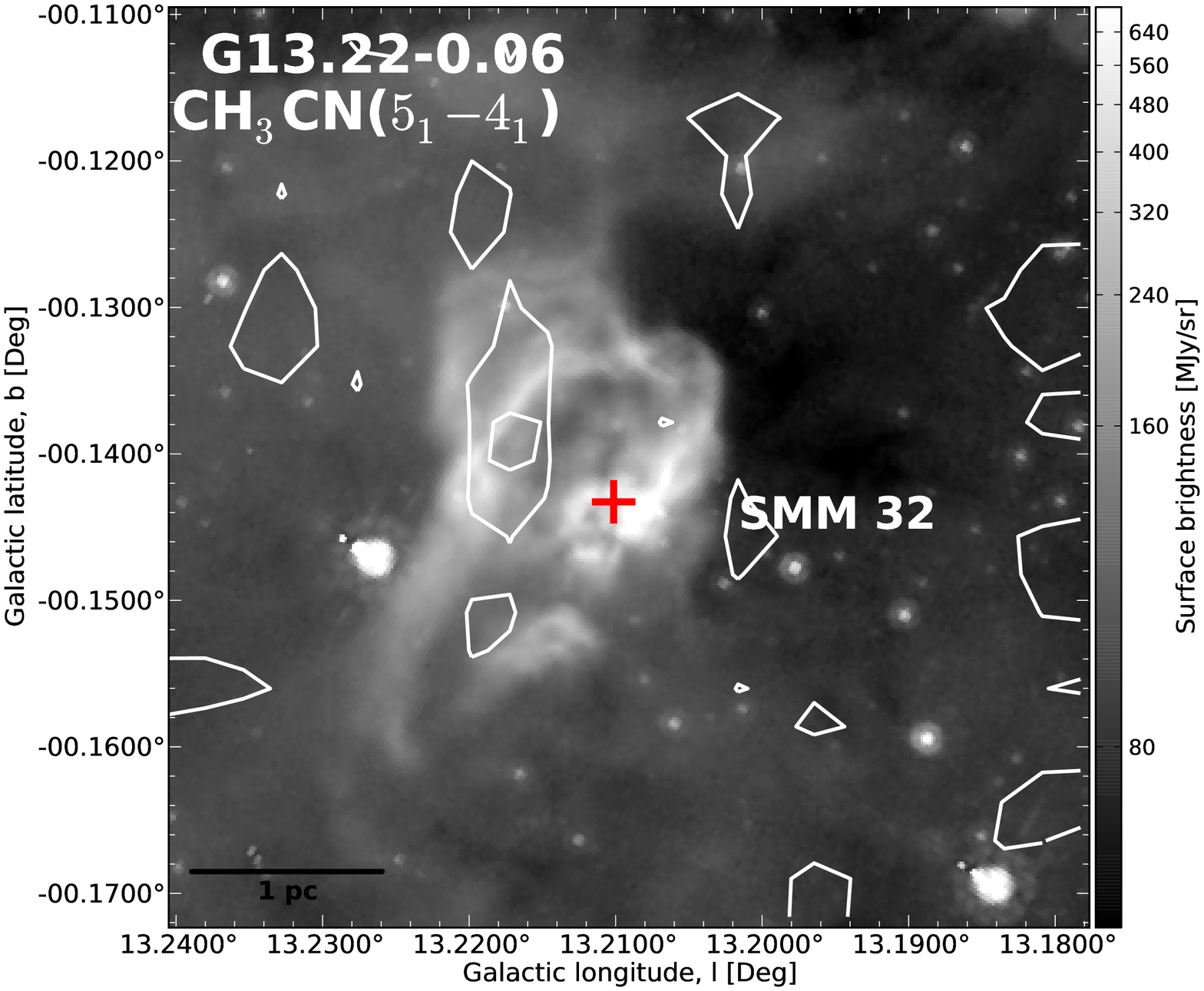}
\includegraphics[width=0.245\textwidth]{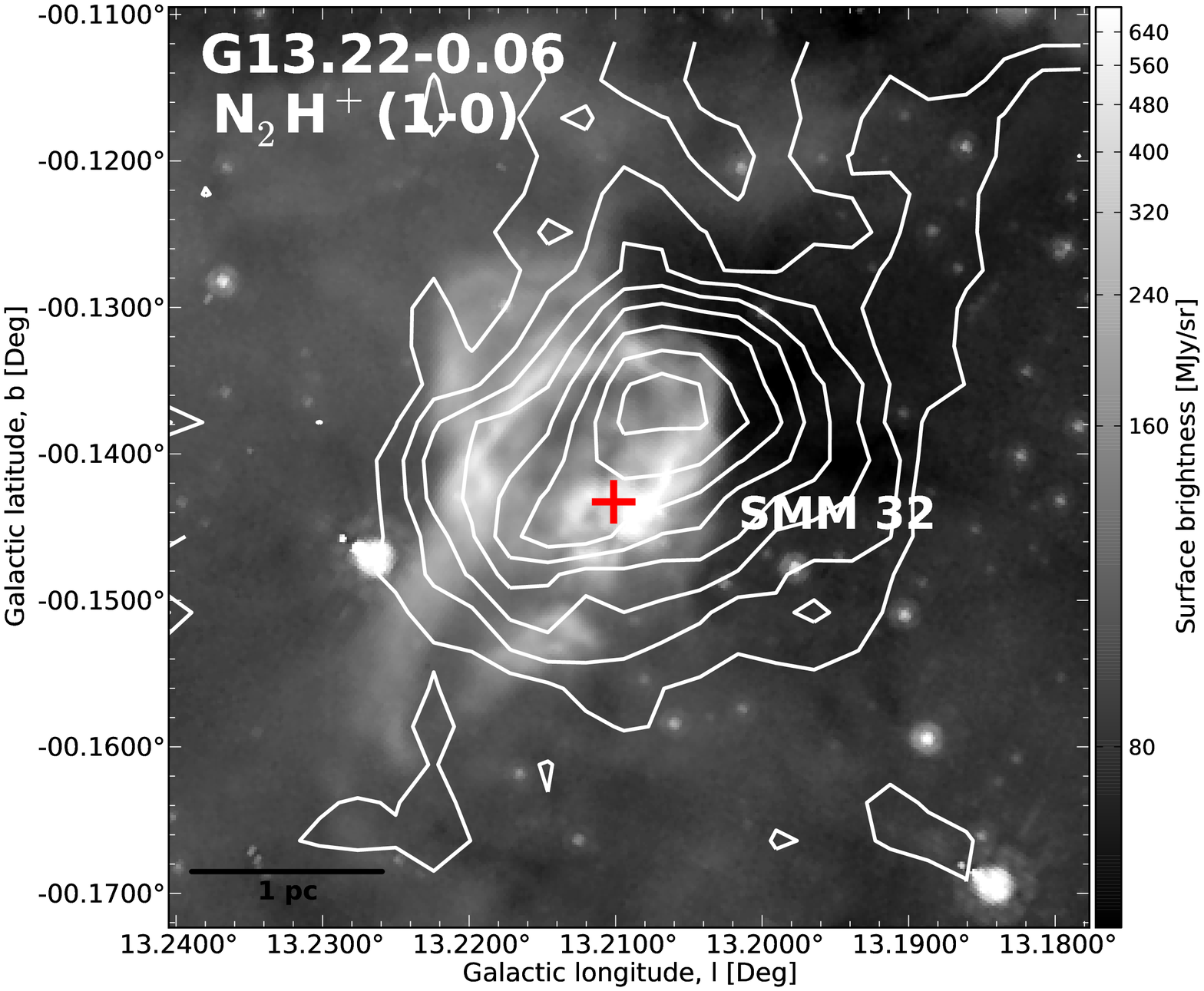}
\caption{Similar to Fig.~\ref{figure:G187SMM1lines} but towards 
G13.22--SMM 32. The contour levels start at $3\sigma$ for H$^{13}$CO$^+$, 
SiO, C$_2$H, HNCO$(4_{0,\,4}-3_{0,\,3})$, HCN, HC$_3$N, and CH$_3$CN, 
while for HCO$^+$, HNC, and  N$_2$H$^+$ they start at $4\sigma$, $5\sigma$, 
and $4\sigma$, respectively. In all cases, the contours go in steps of 
$3\sigma$. The average $1\sigma$ value in $T_{\rm MB}$ units is $\sim0.57$ 
K~km~s$^{-1}$. The LABOCA 870-$\mu$m peak position of the clump is marked by a
red plus sign. A scale bar indicating the 1 pc projected length is 
indicated. The spatial distributions of HCN, HCO$^+$, HNC, and N$_2$H$^+$ are 
quite similar, but note the different peak positions of HCO$^+$ and N$_2$H$^+$ 
(as expected from their chemistry). C$_2$H and HC$_3$N also show weaker 
emission towards SMM 32, with some morphological similarities to the strongly 
emitting species.}
\label{figure:G1322SMM32lines}
\end{center}
\end{figure*}

The integrated intensity maps of the detected spectral lines are presented in 
Figs.~\ref{figure:G187SMM1lines}--\ref{figure:G1322SMM32lines}. In each panel, 
the line emission is shown as contours overlaid on the \textit{Spitzer} 
8-$\mu$m image.

\clearpage

\section{Spectra}

\begin{figure*}
\begin{center}
\includegraphics[width=0.245\textwidth]{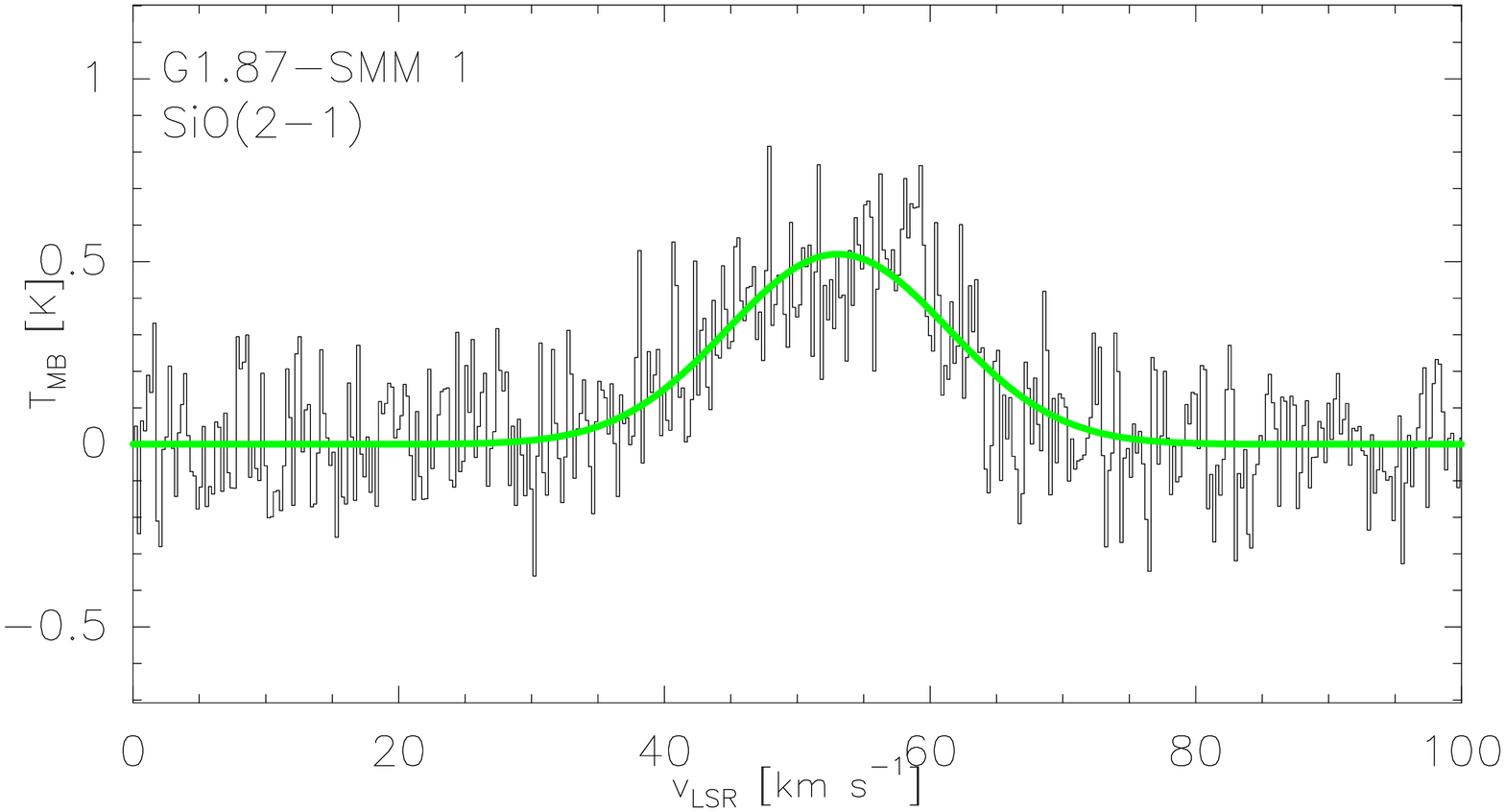}
\includegraphics[width=0.245\textwidth]{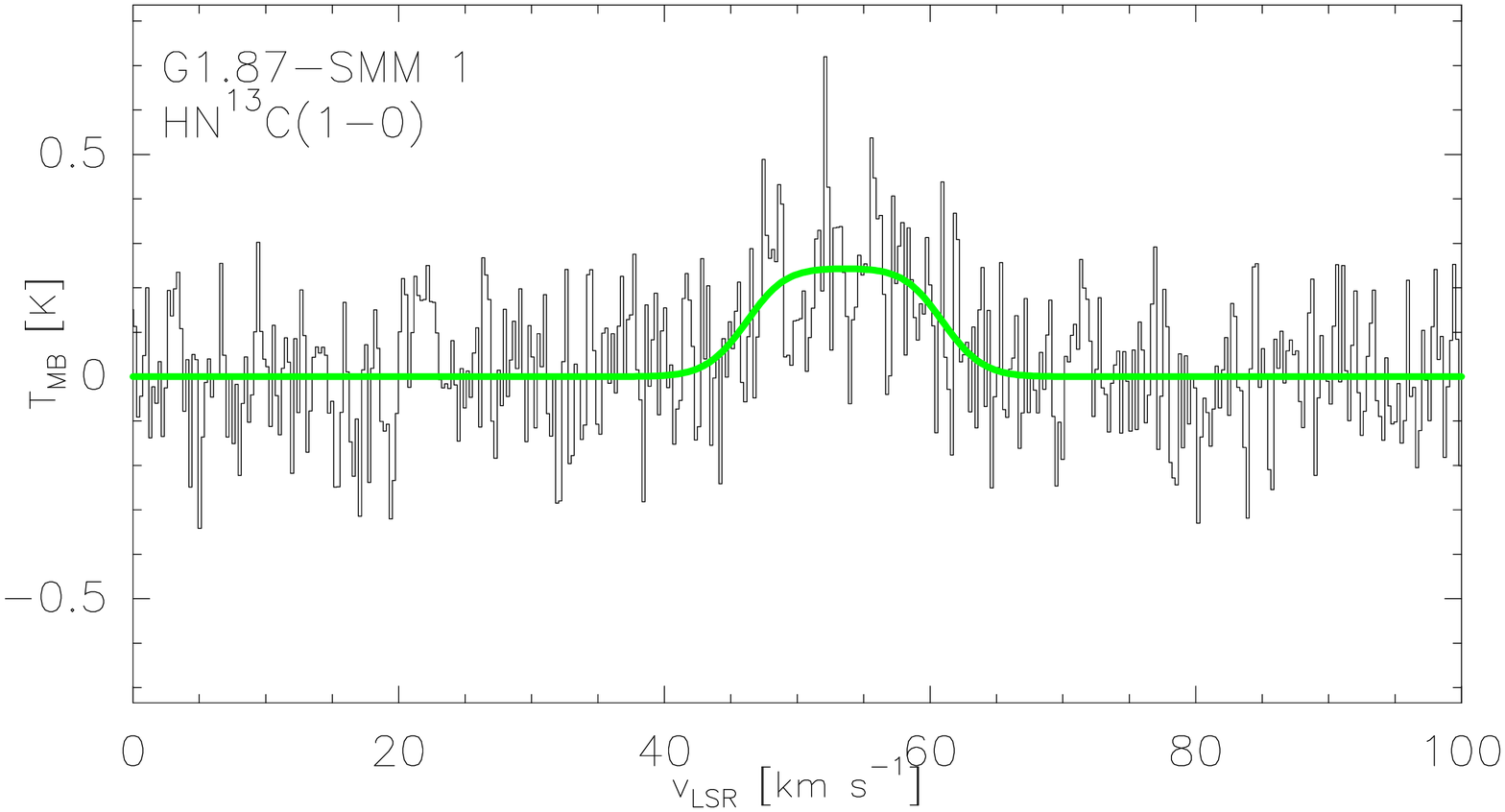}
\includegraphics[width=0.245\textwidth]{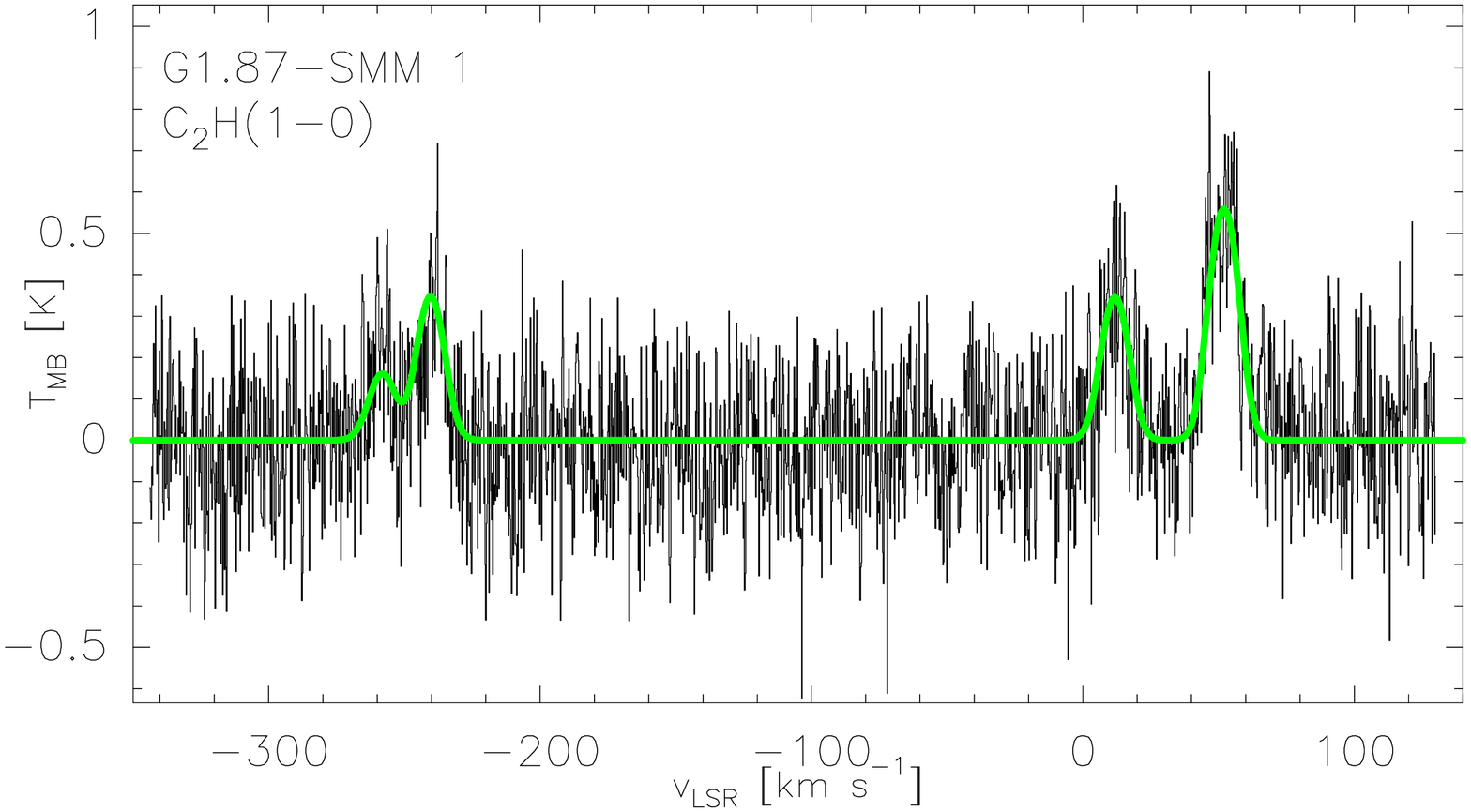}
\includegraphics[width=0.245\textwidth]{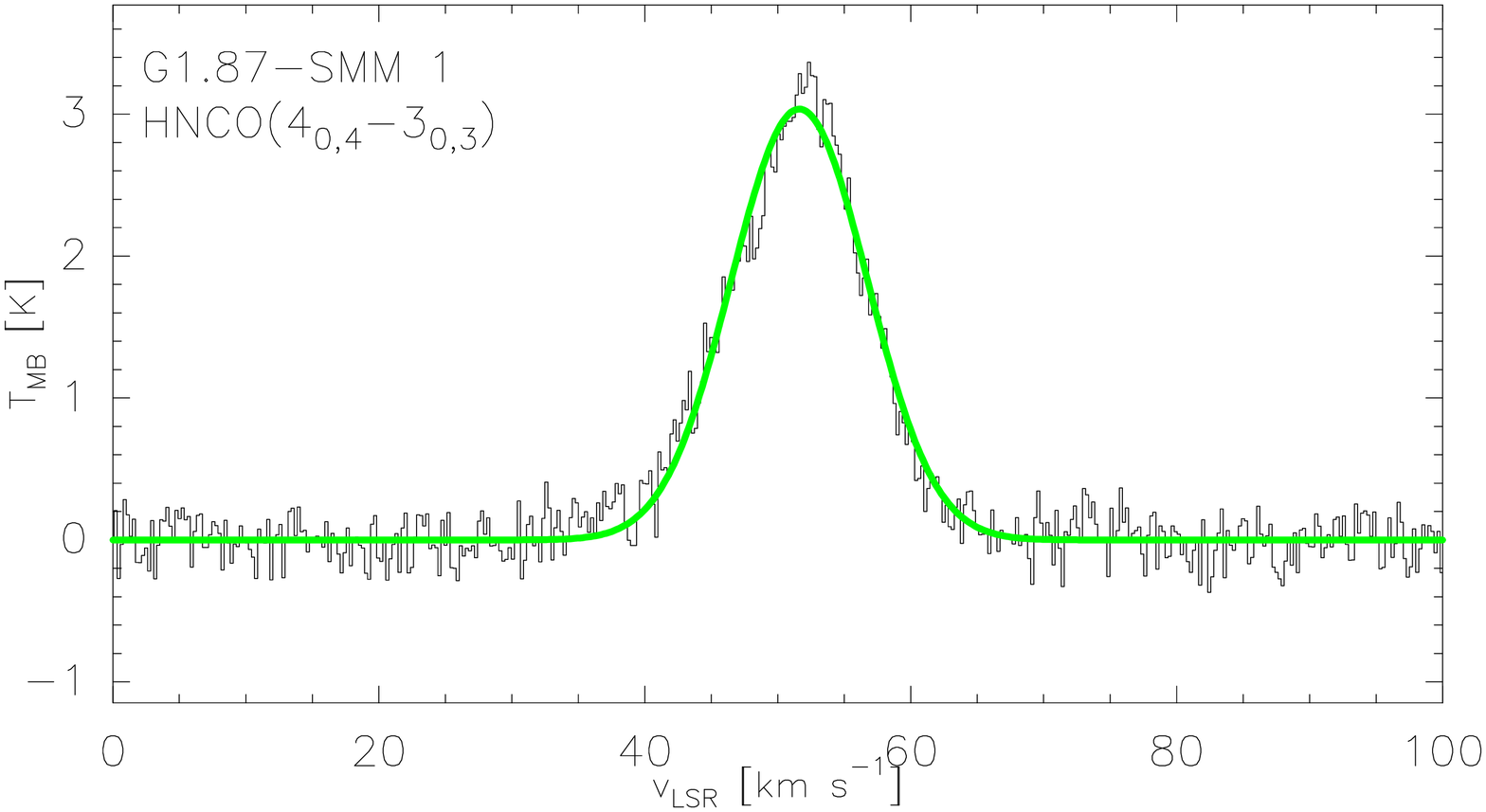}
\includegraphics[width=0.245\textwidth]{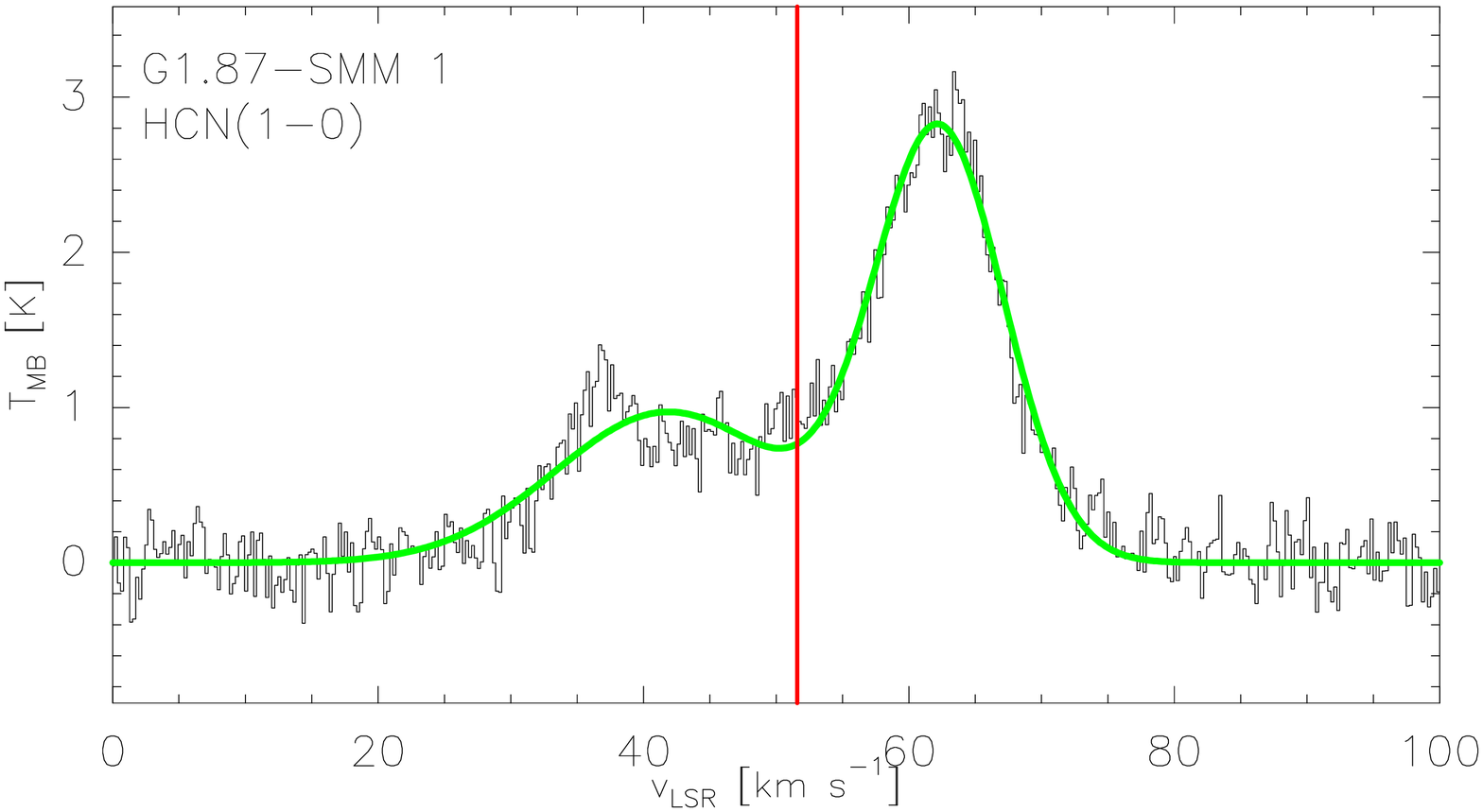}
\includegraphics[width=0.245\textwidth]{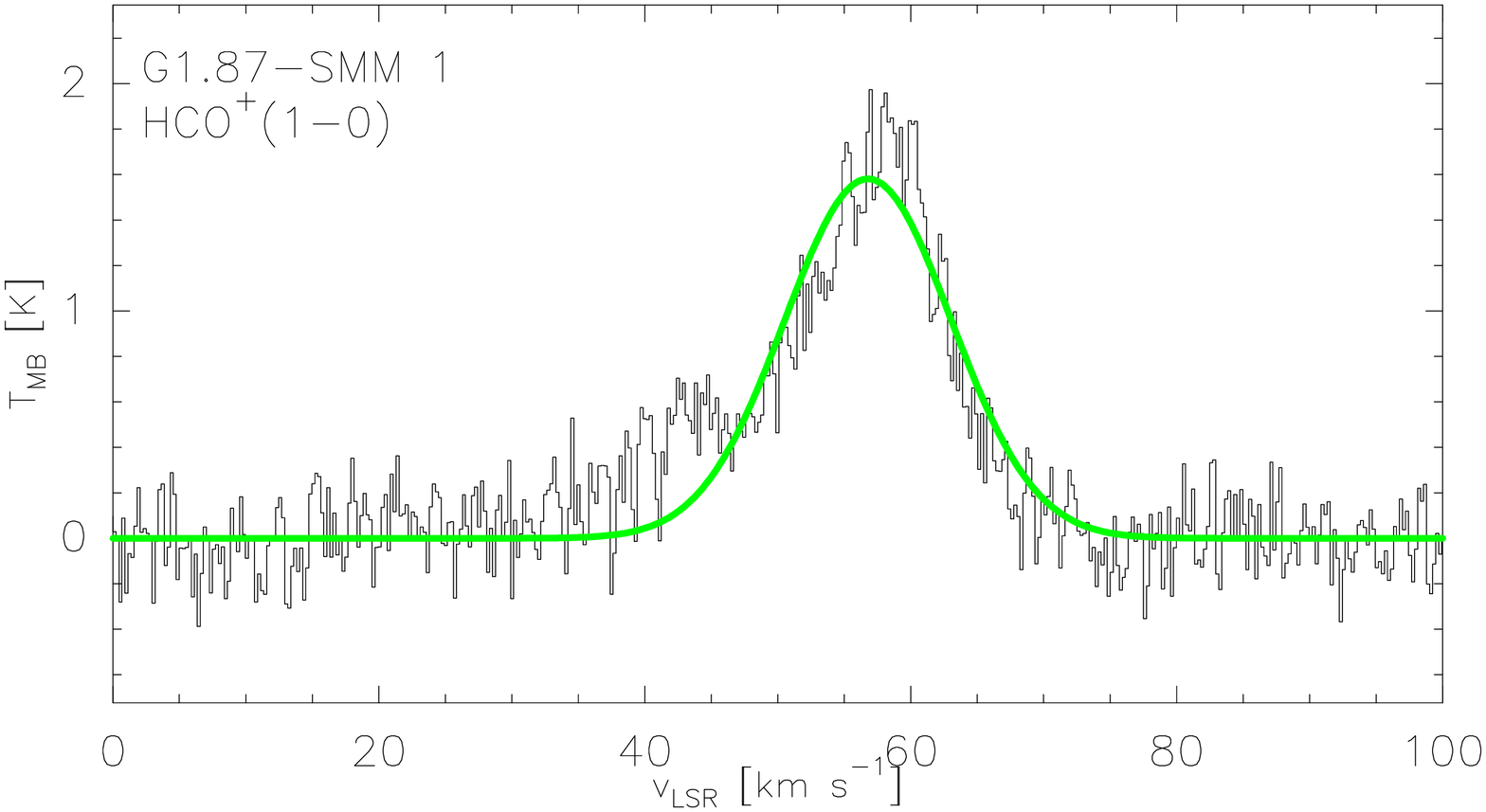}
\includegraphics[width=0.245\textwidth]{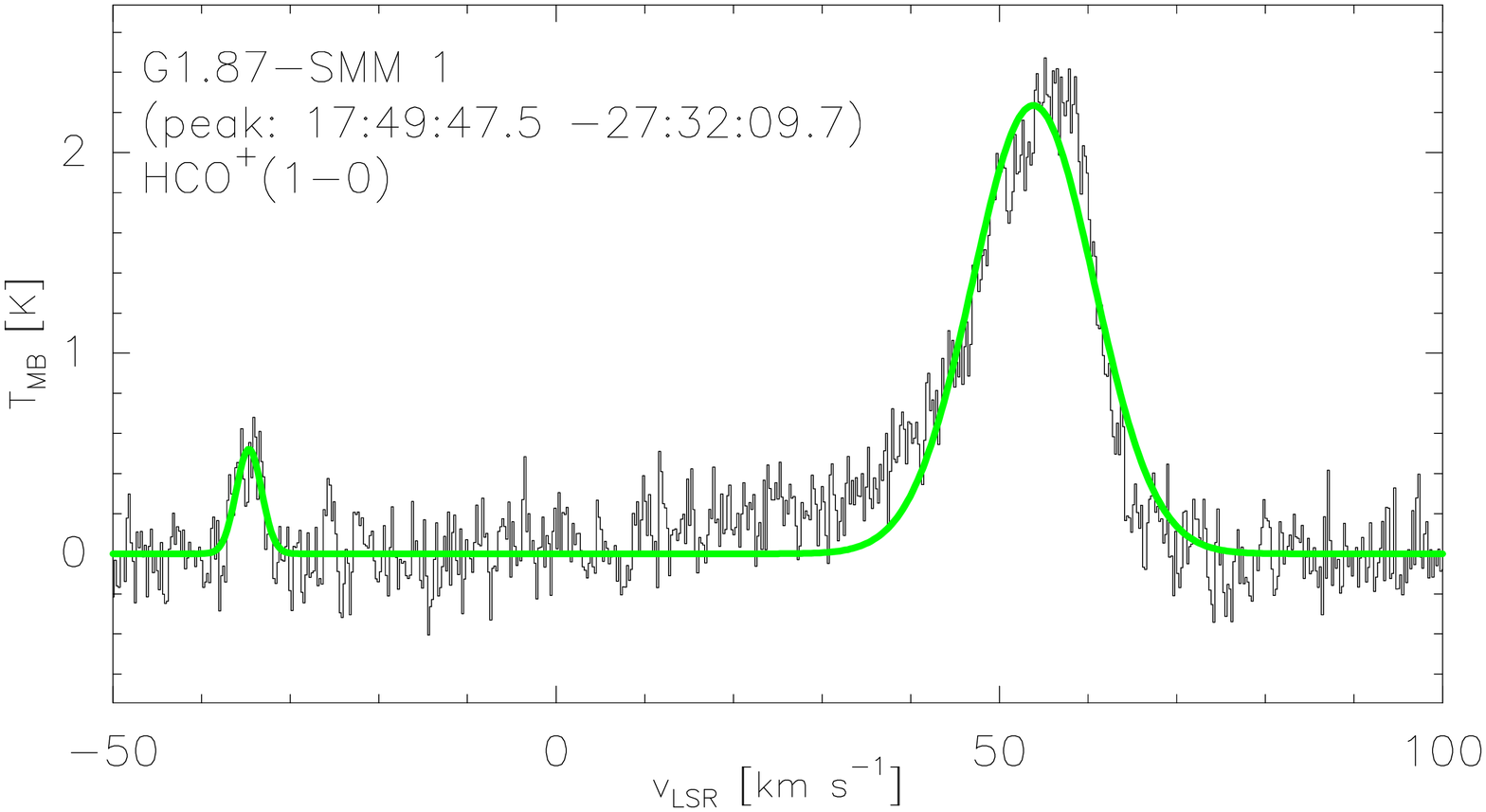}
\includegraphics[width=0.245\textwidth]{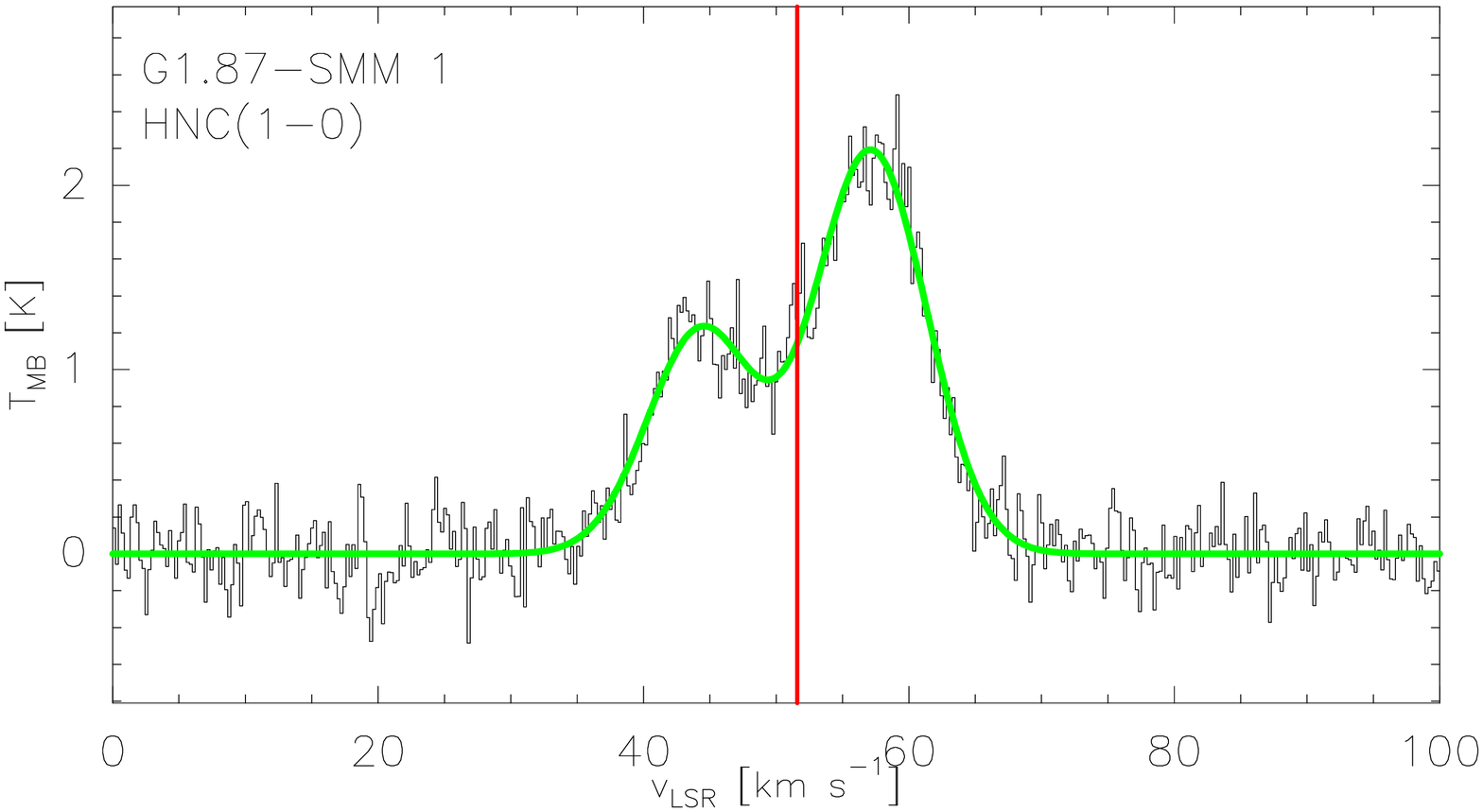}
\includegraphics[width=0.245\textwidth]{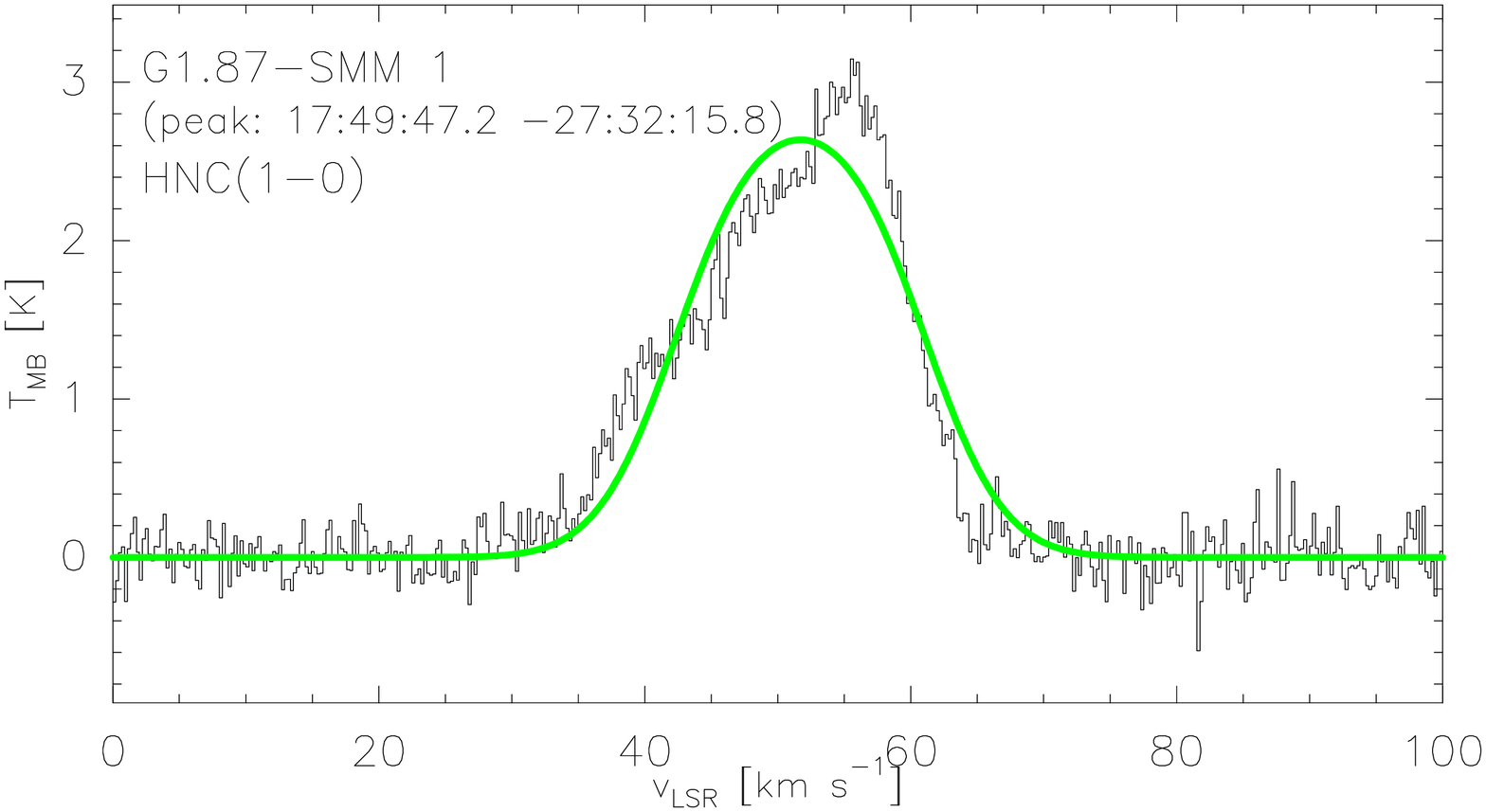}
\includegraphics[width=0.245\textwidth]{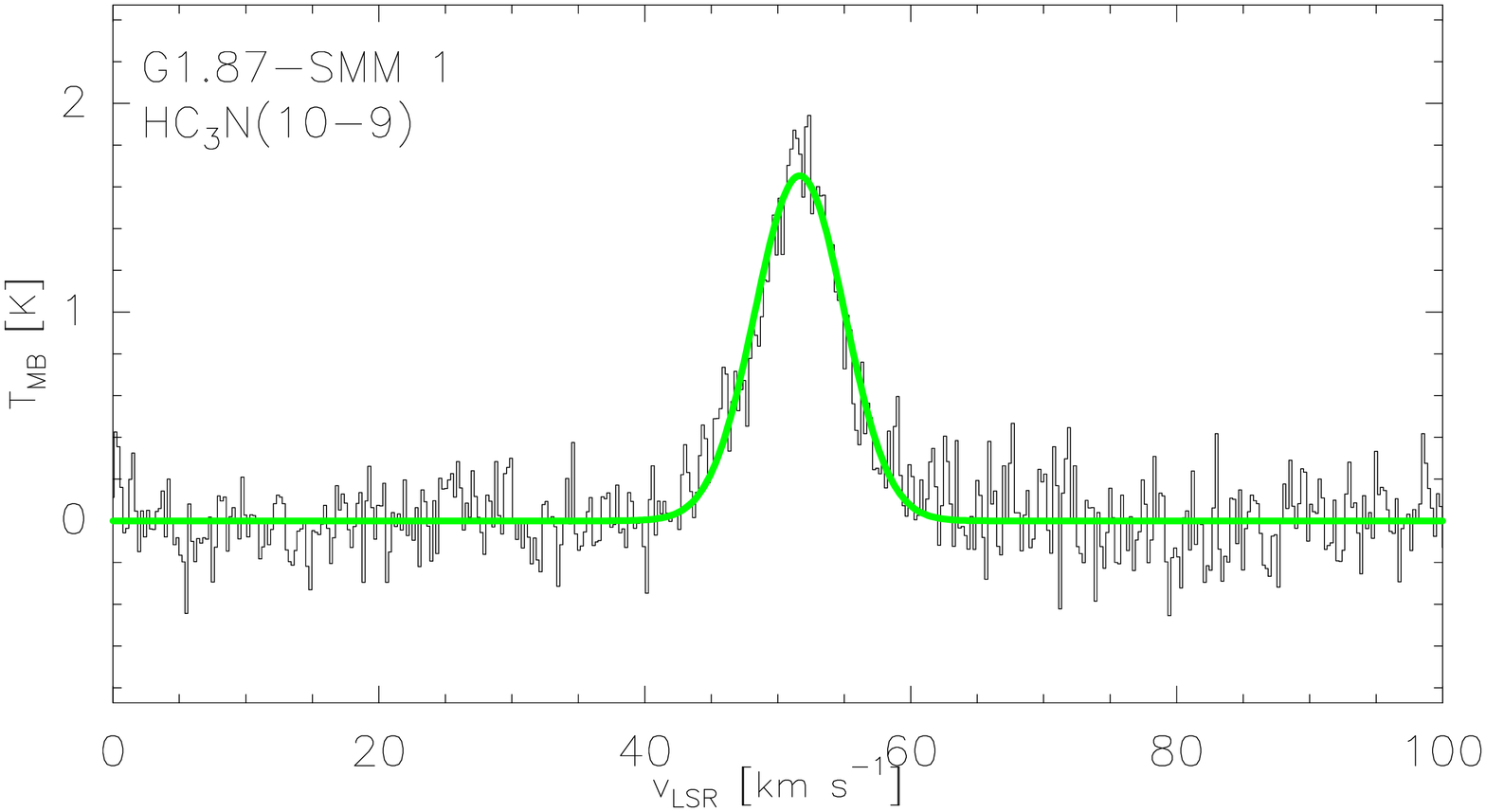}
\includegraphics[width=0.245\textwidth]{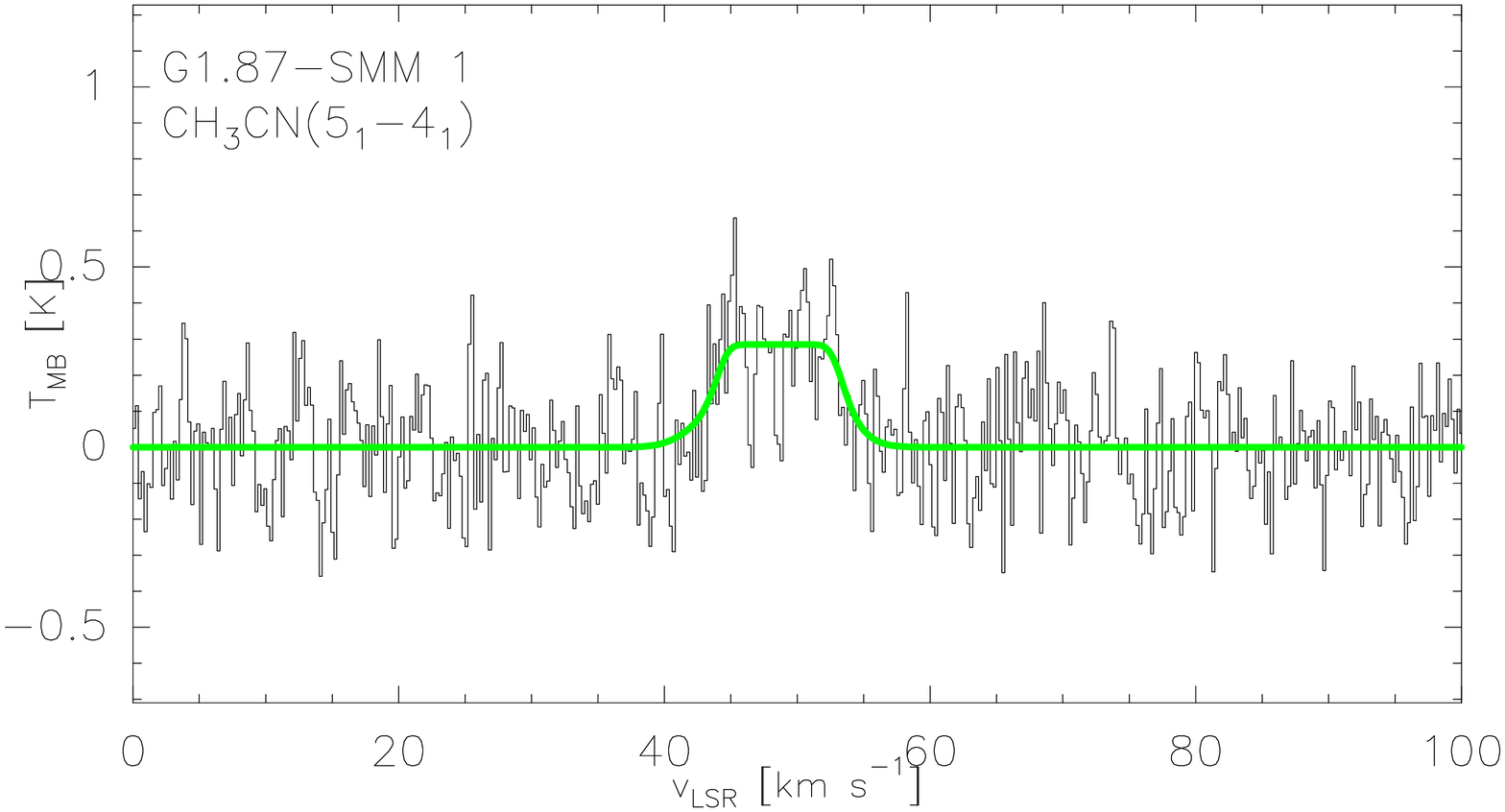}
\includegraphics[width=0.245\textwidth]{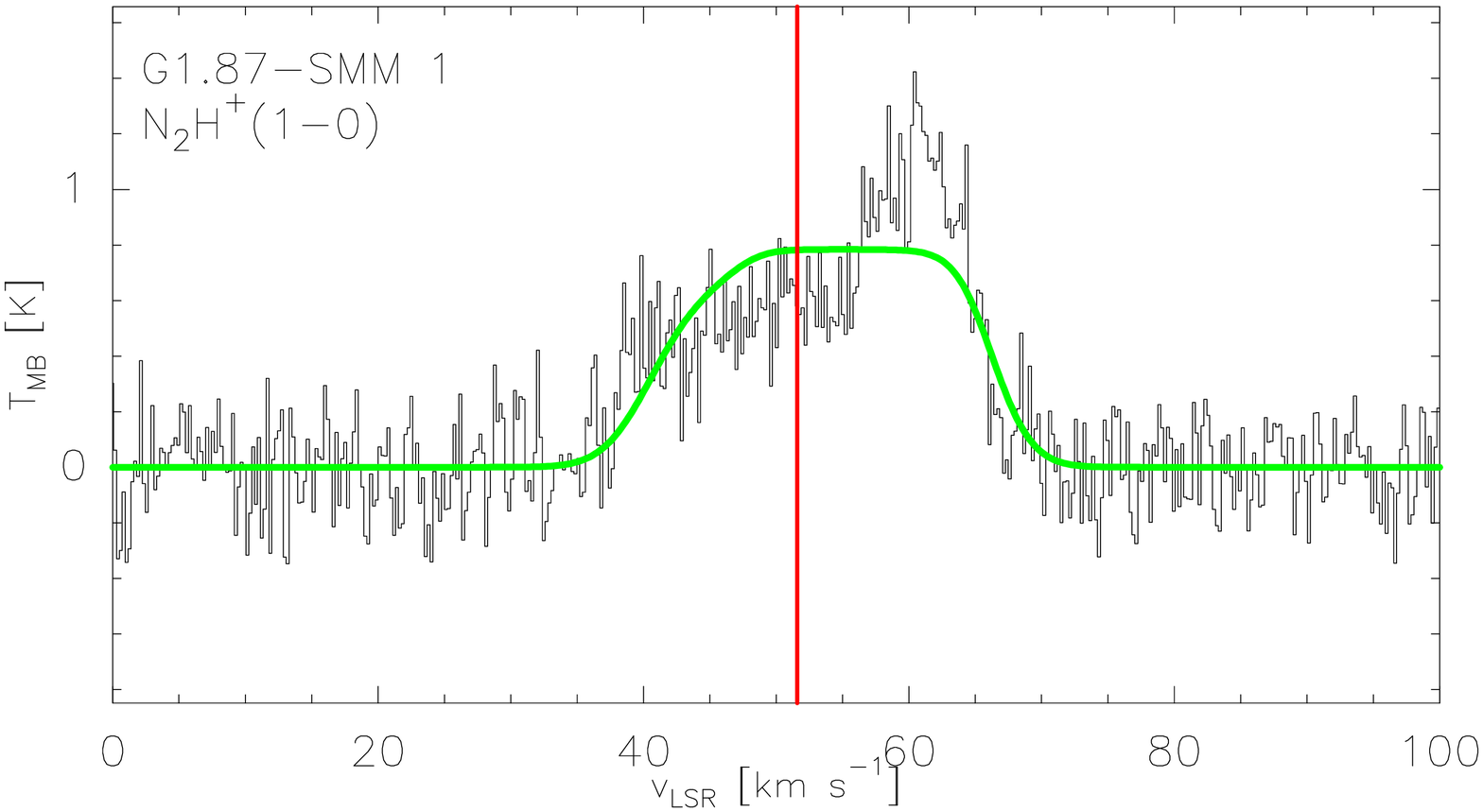}
\caption{Hanning-smoothed spectra towards G1.87--SMM 1. The single Gaussian 
and hf structure fits are shown with green lines. The red vertical line plotted 
on double-peaked profiles indicates the radial velocity of the optically thin 
HC$_3$N line. The velocity range is wider in the C$_2$H and HCO$^+$ (line 
emission peak) spectra than in the other panels to show all the detected lines 
(additional velocity component in the latter case). }
\label{figure:G187SMM1_spectra}
\end{center}
\end{figure*}

\begin{figure*}
\begin{center}
\includegraphics[width=0.245\textwidth]{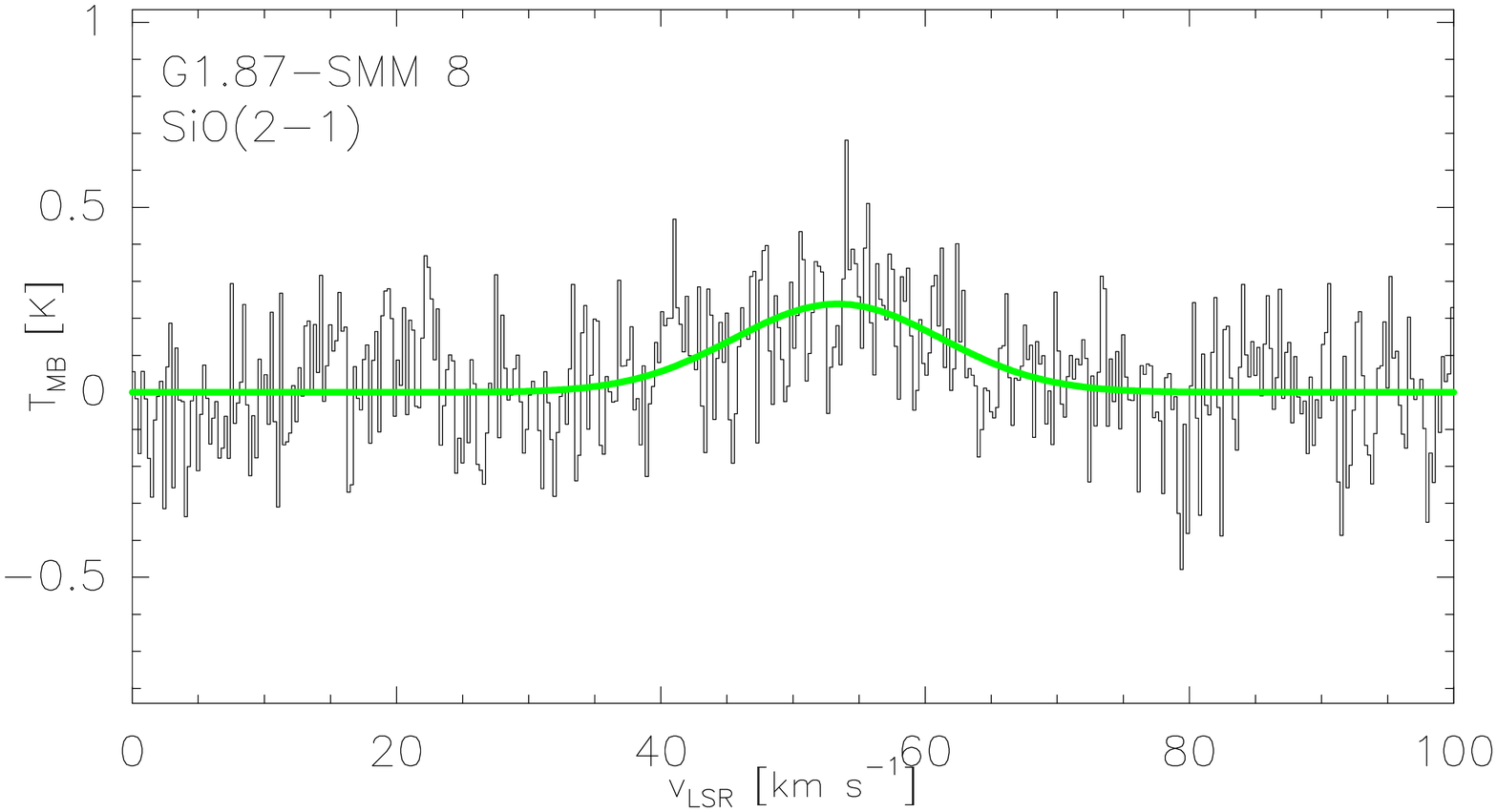}
\includegraphics[width=0.245\textwidth]{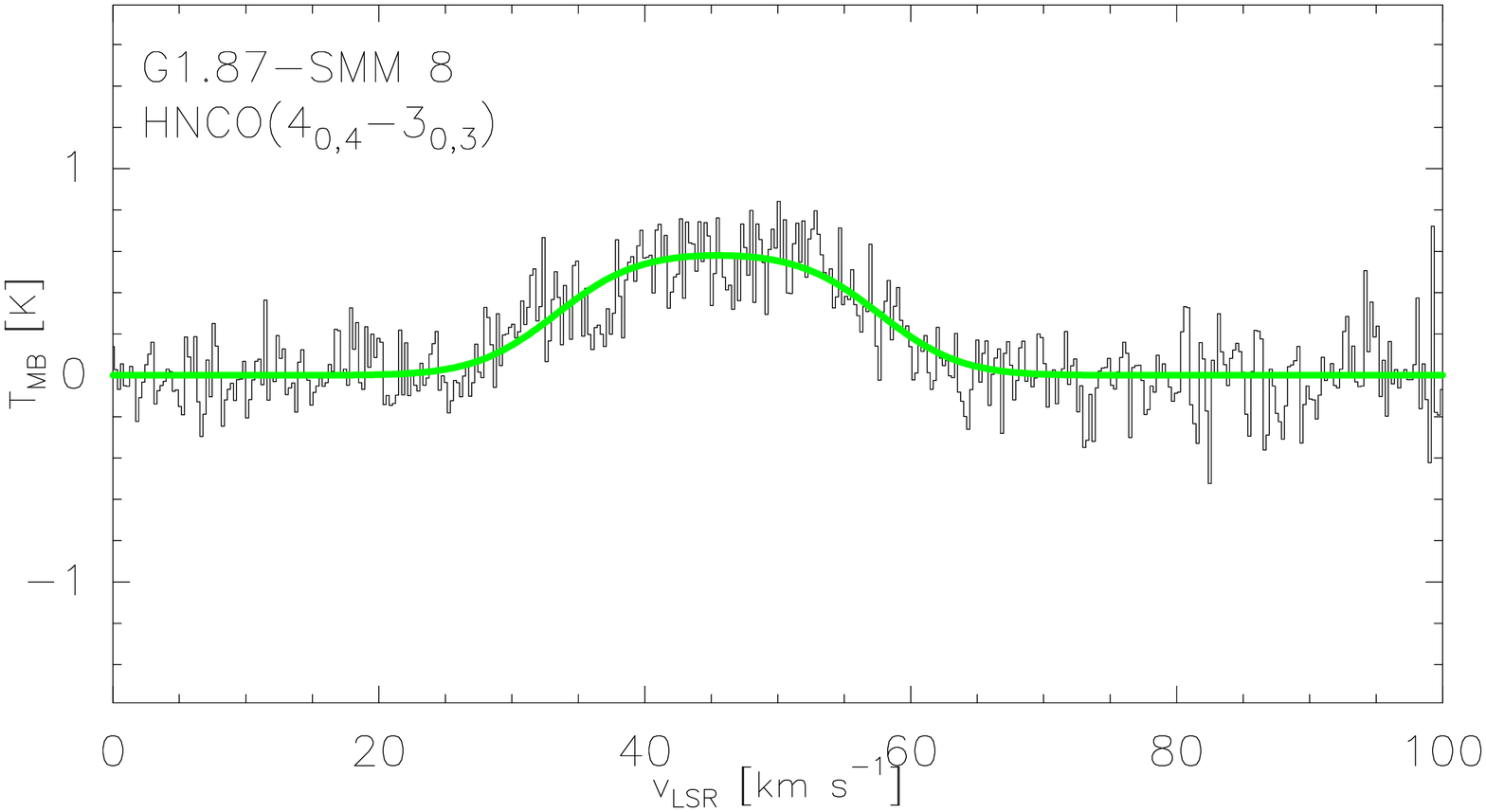}
\includegraphics[width=0.245\textwidth]{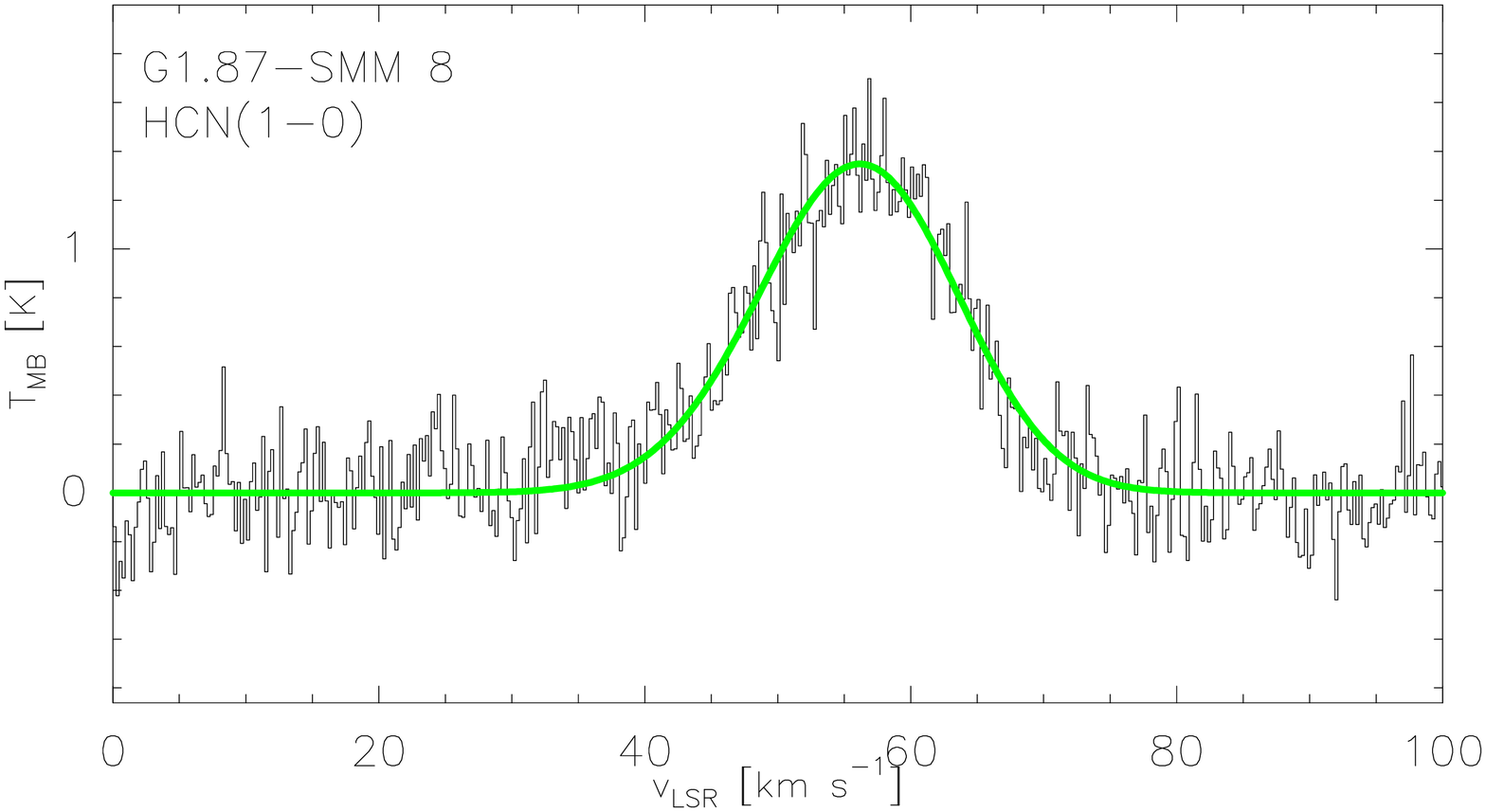}
\includegraphics[width=0.245\textwidth]{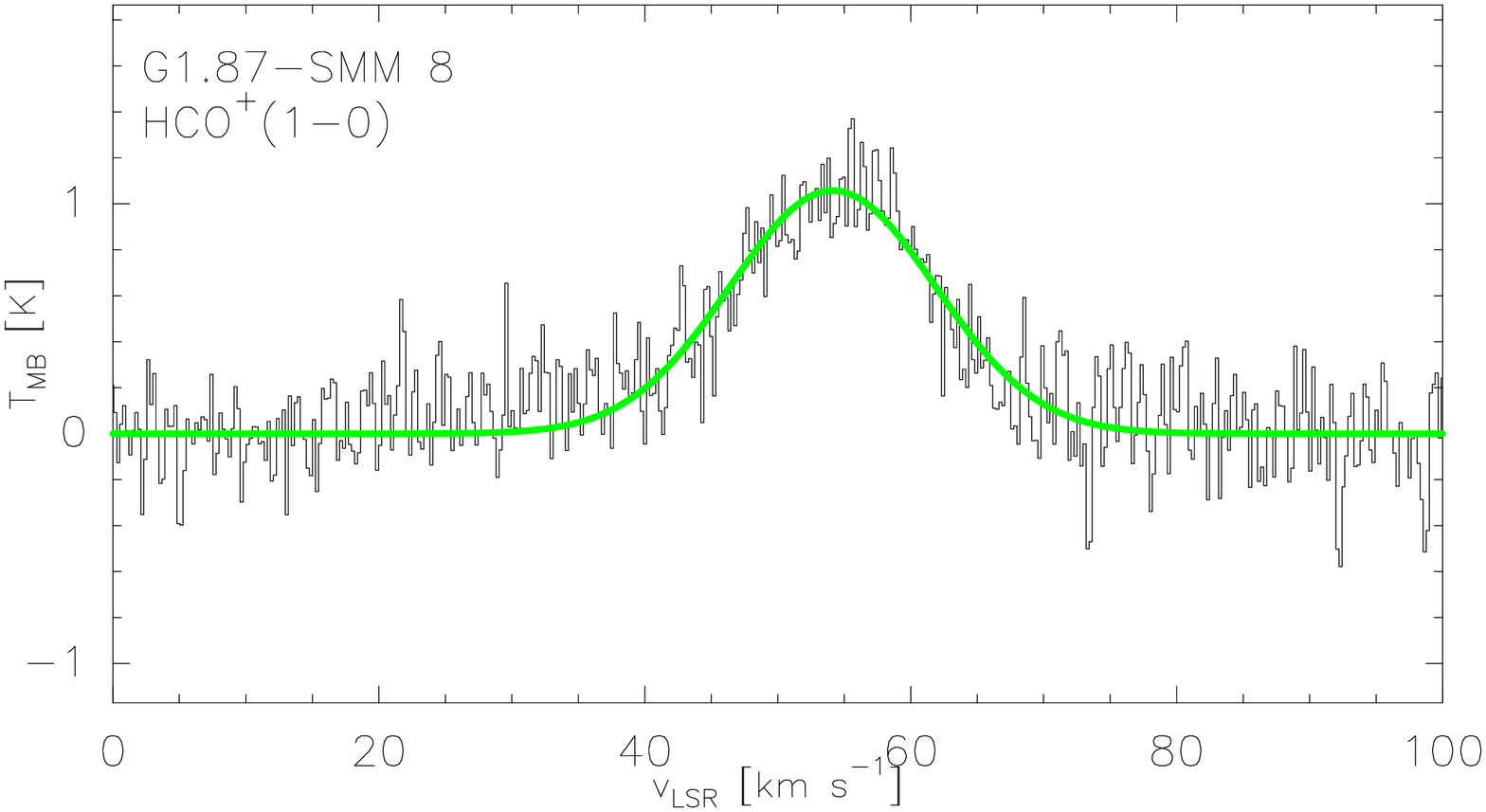}
\includegraphics[width=0.245\textwidth]{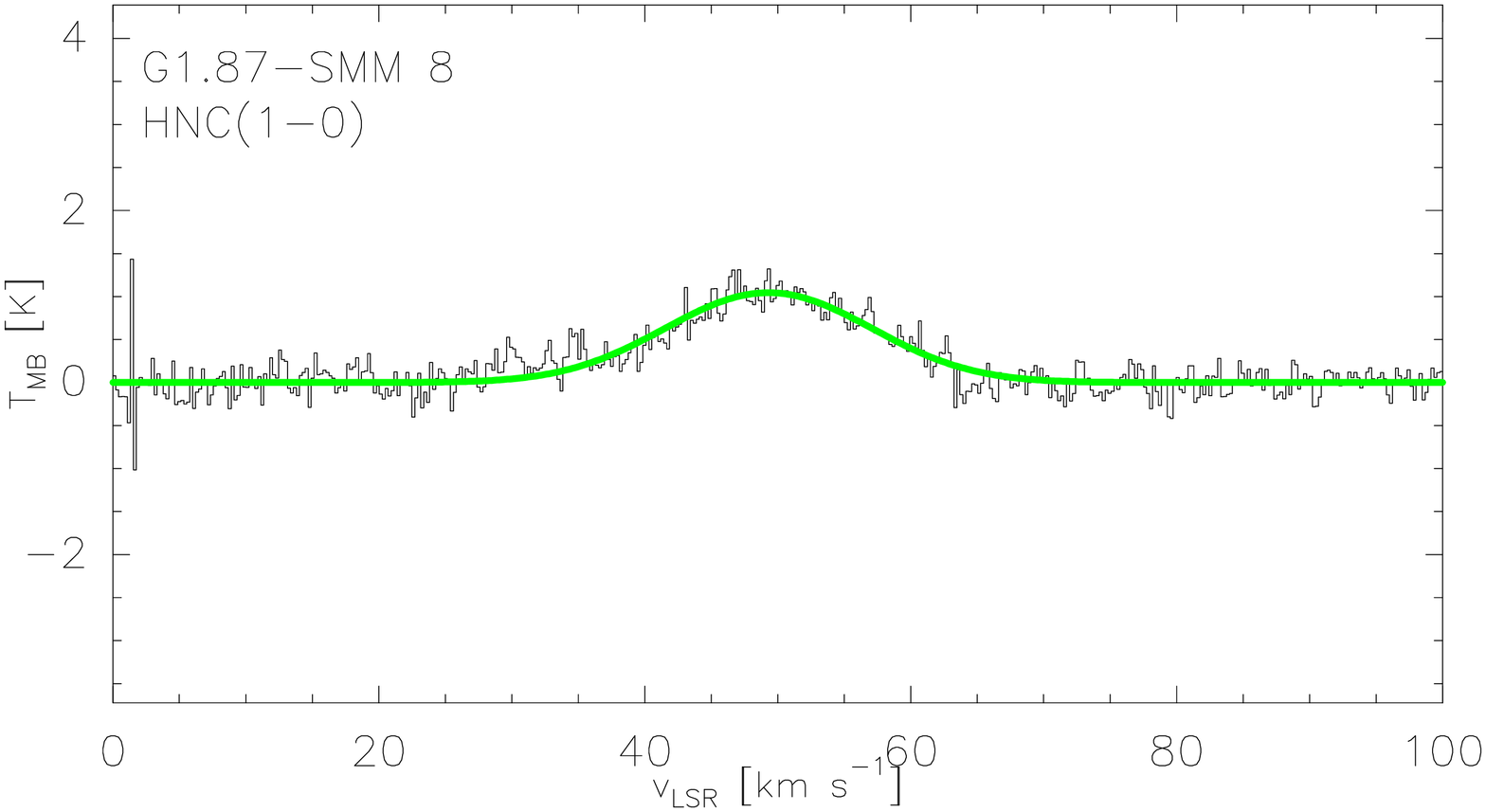}
\includegraphics[width=0.245\textwidth]{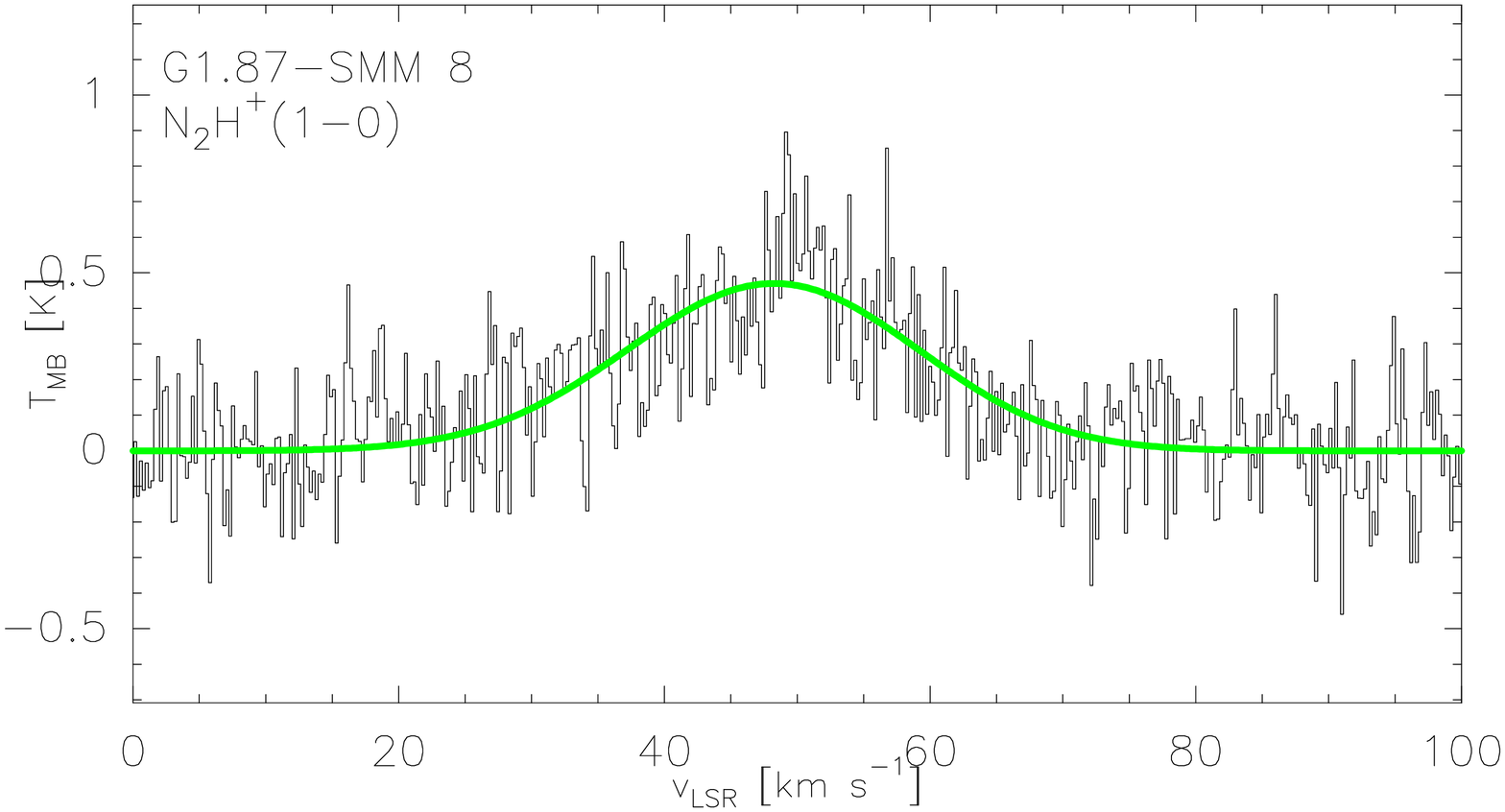}
\caption{Same as Fig.~\ref{figure:G187SMM1_spectra} but towards G1.87--SMM 8.}
\label{figure:G187SMM8_spectra}
\end{center}
\end{figure*}

\begin{figure*}
\begin{center}
\includegraphics[width=0.245\textwidth]{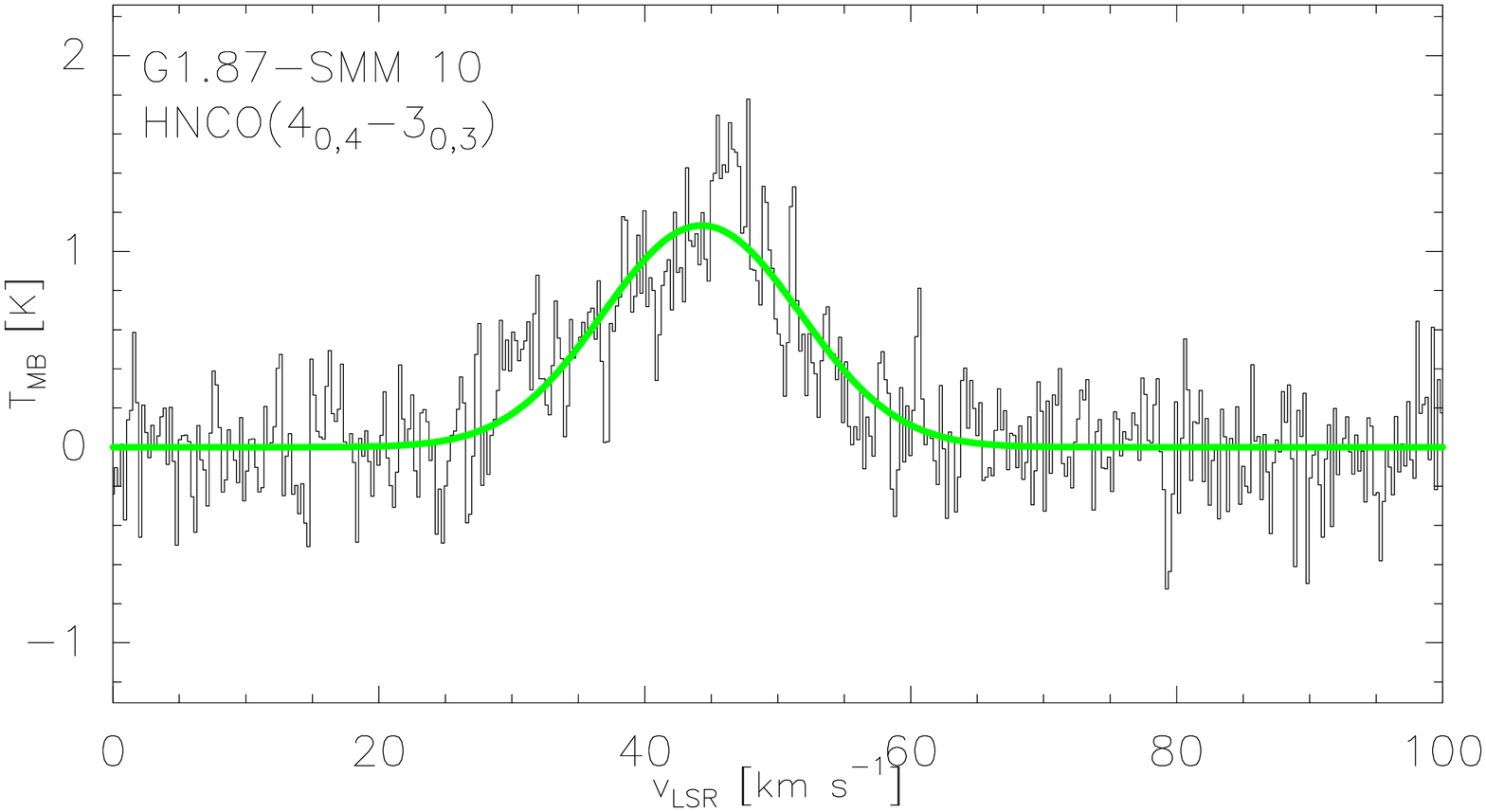}
\includegraphics[width=0.245\textwidth]{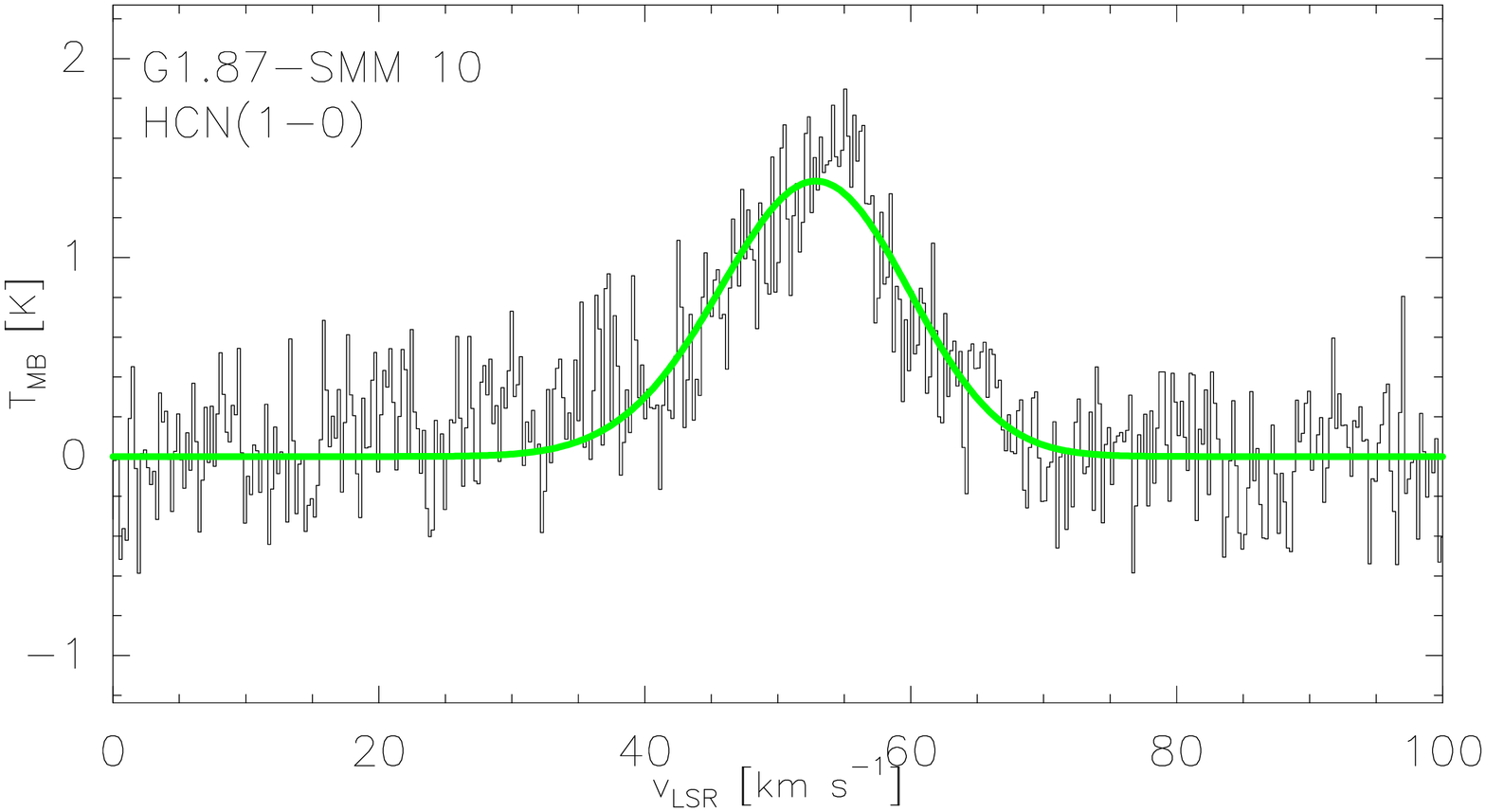}
\includegraphics[width=0.245\textwidth]{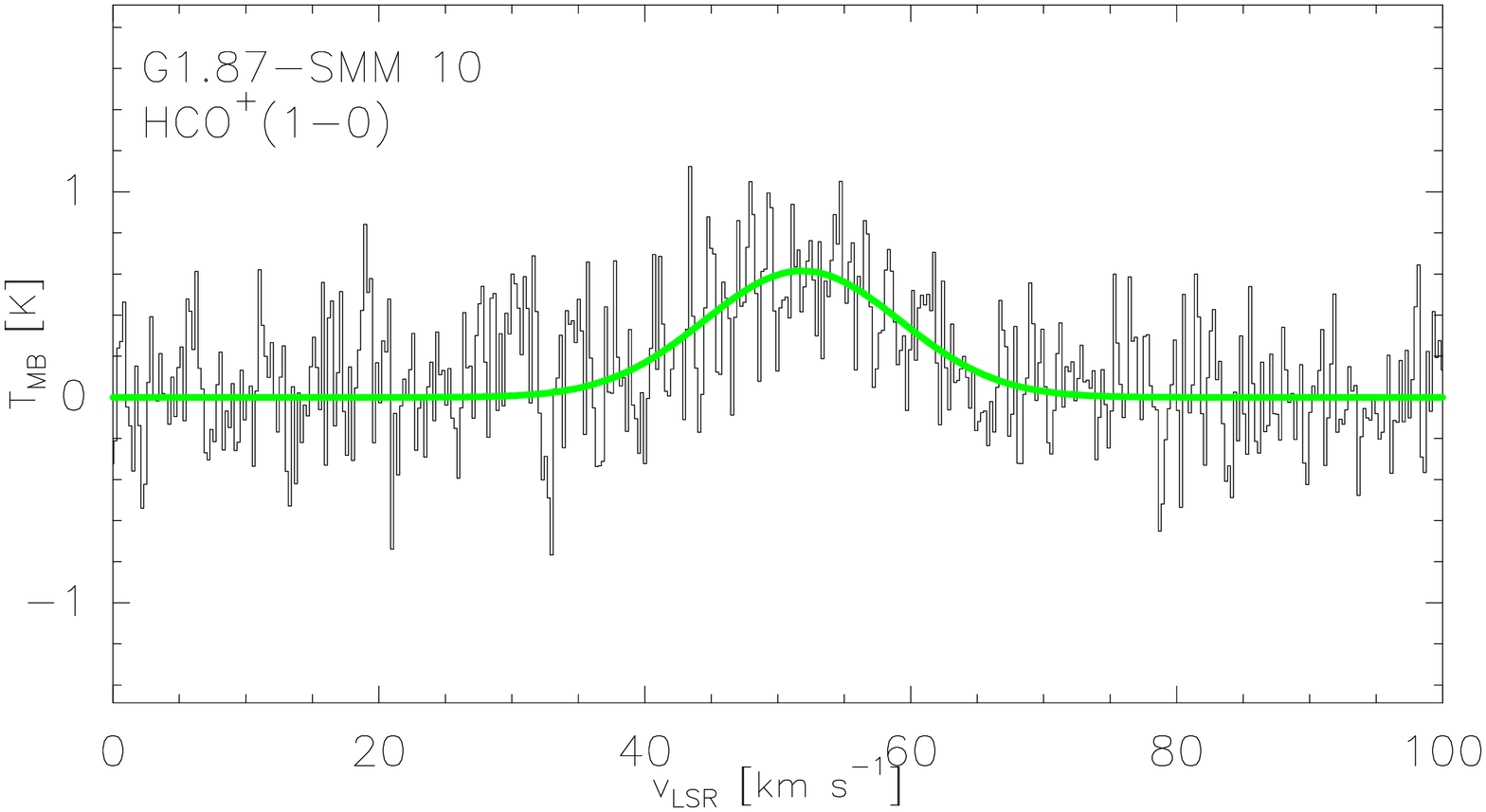}
\includegraphics[width=0.245\textwidth]{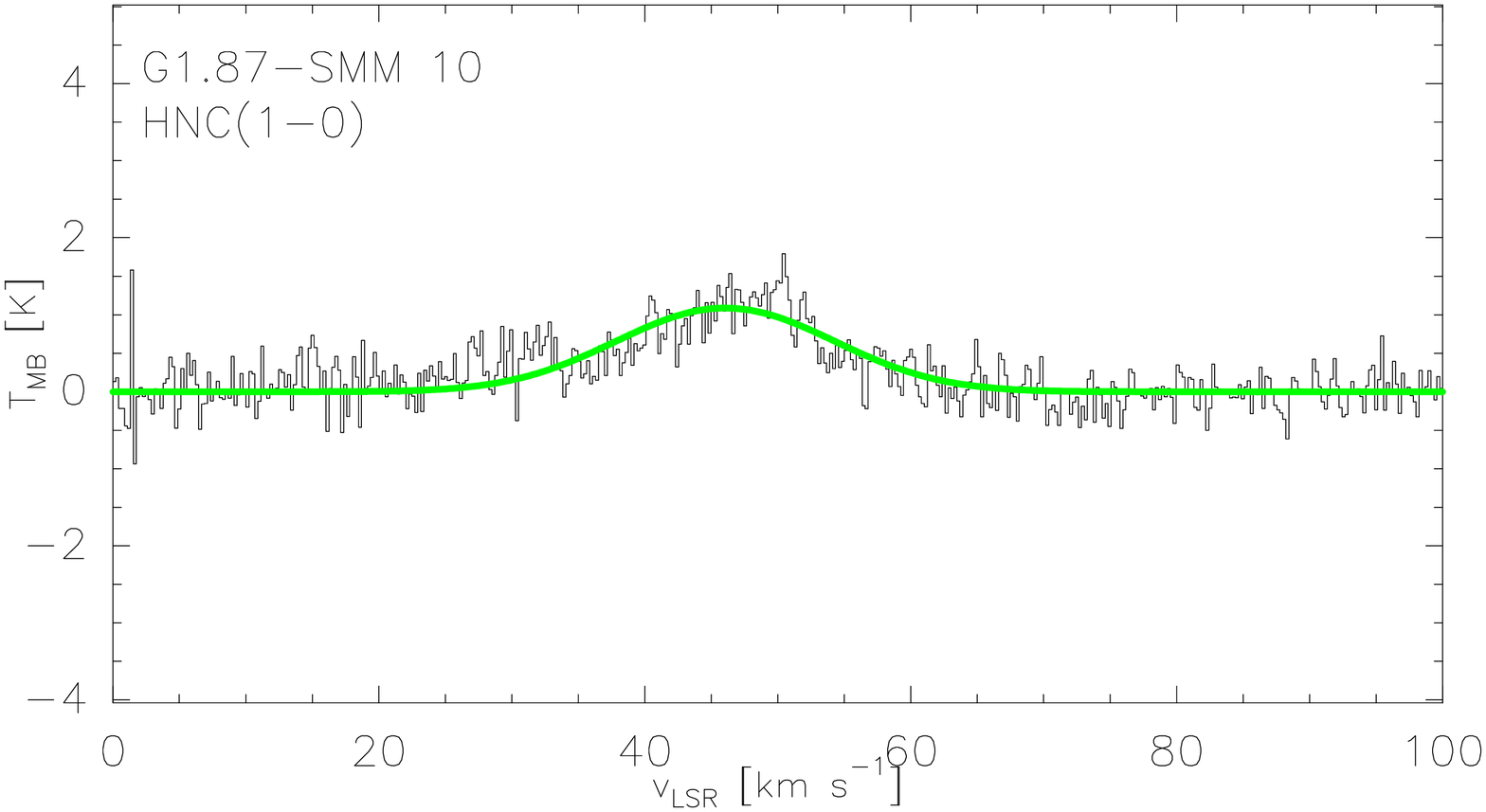}
\includegraphics[width=0.245\textwidth]{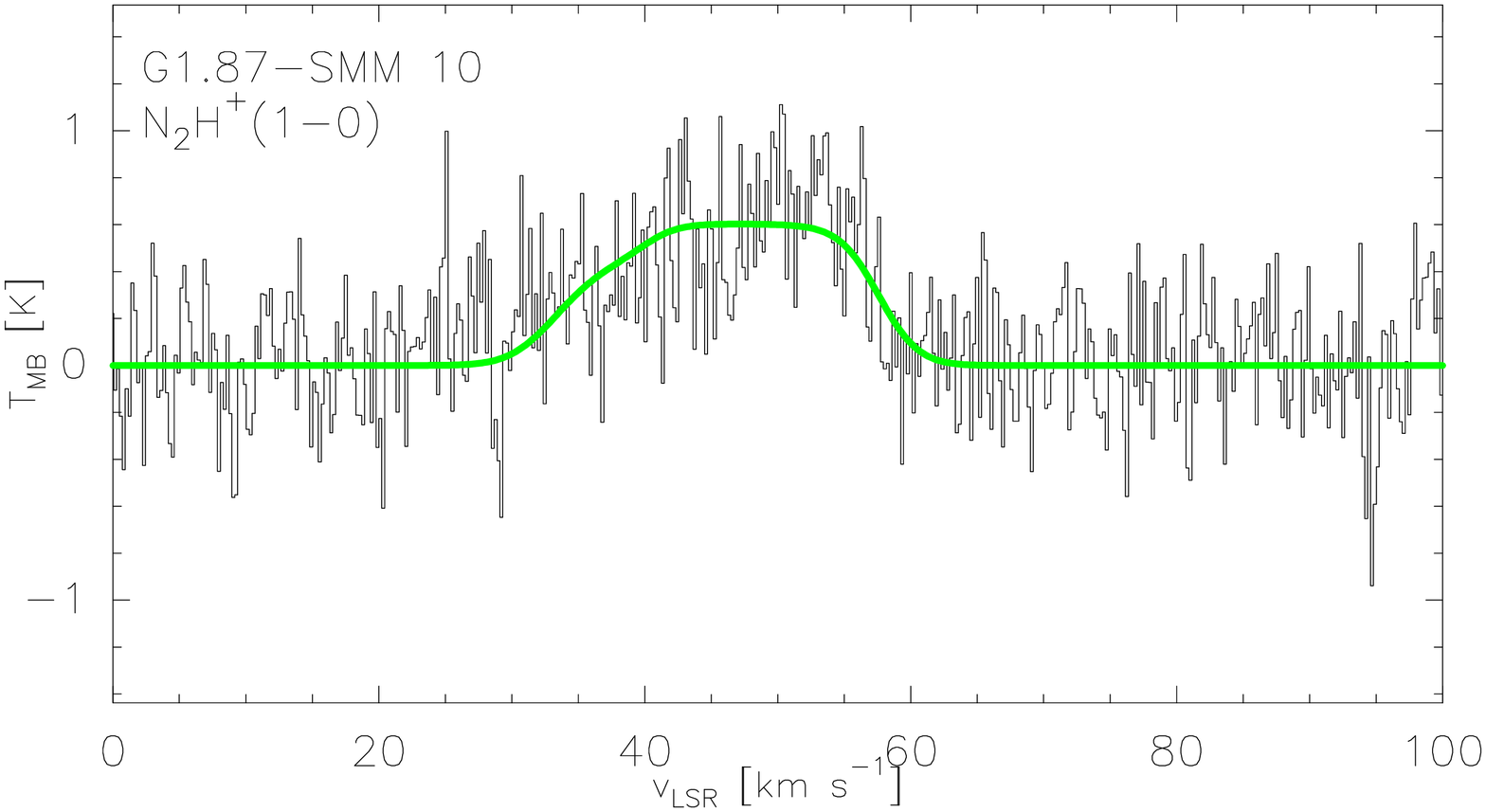}
\caption{Same as Fig.~\ref{figure:G187SMM1_spectra} but towards G1.87--SMM 10.}
\label{figure:G187SMM10_spectra}
\end{center}
\end{figure*}

\begin{figure*}
\begin{center}
\includegraphics[width=0.245\textwidth]{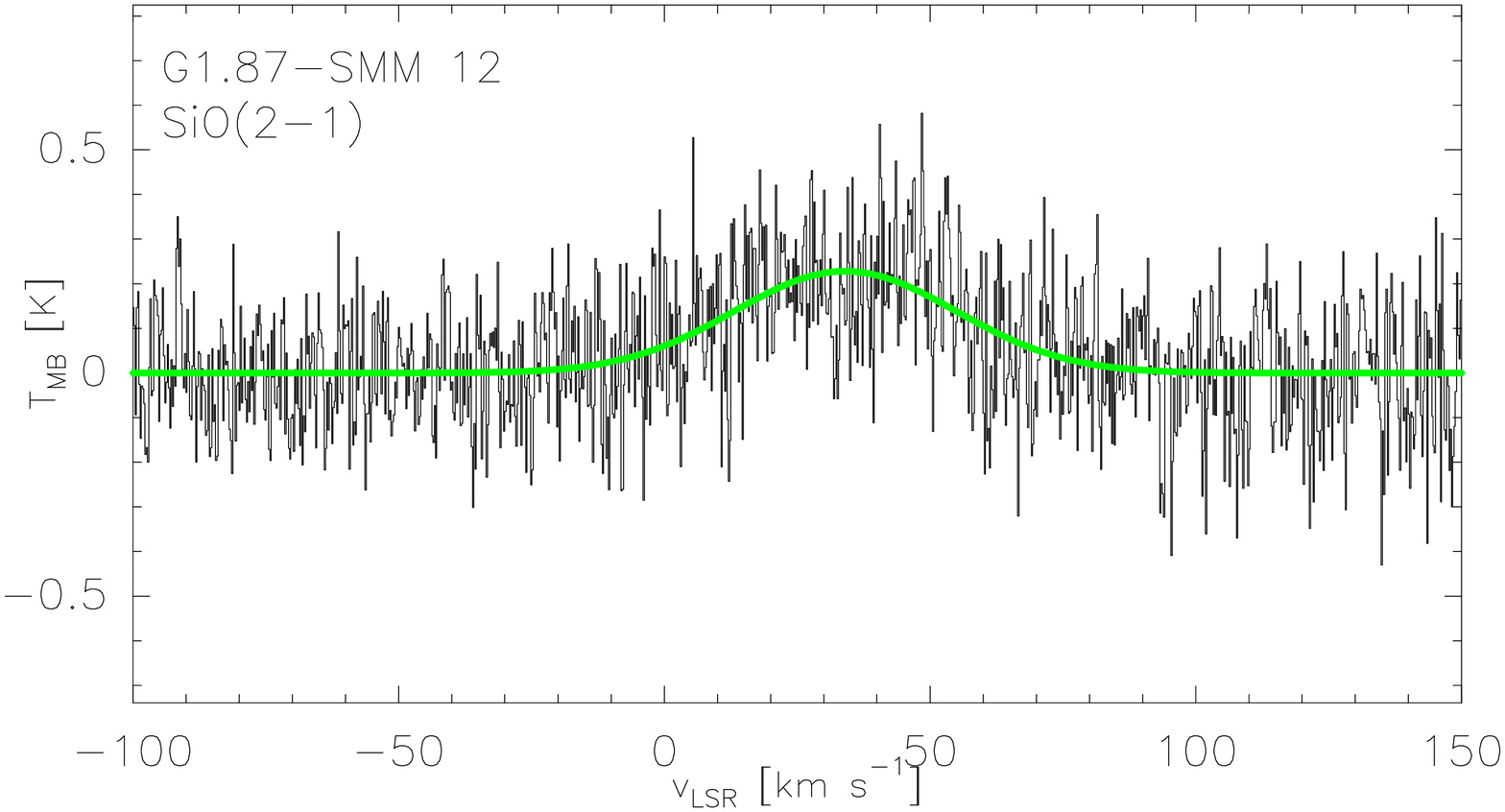}
\includegraphics[width=0.245\textwidth]{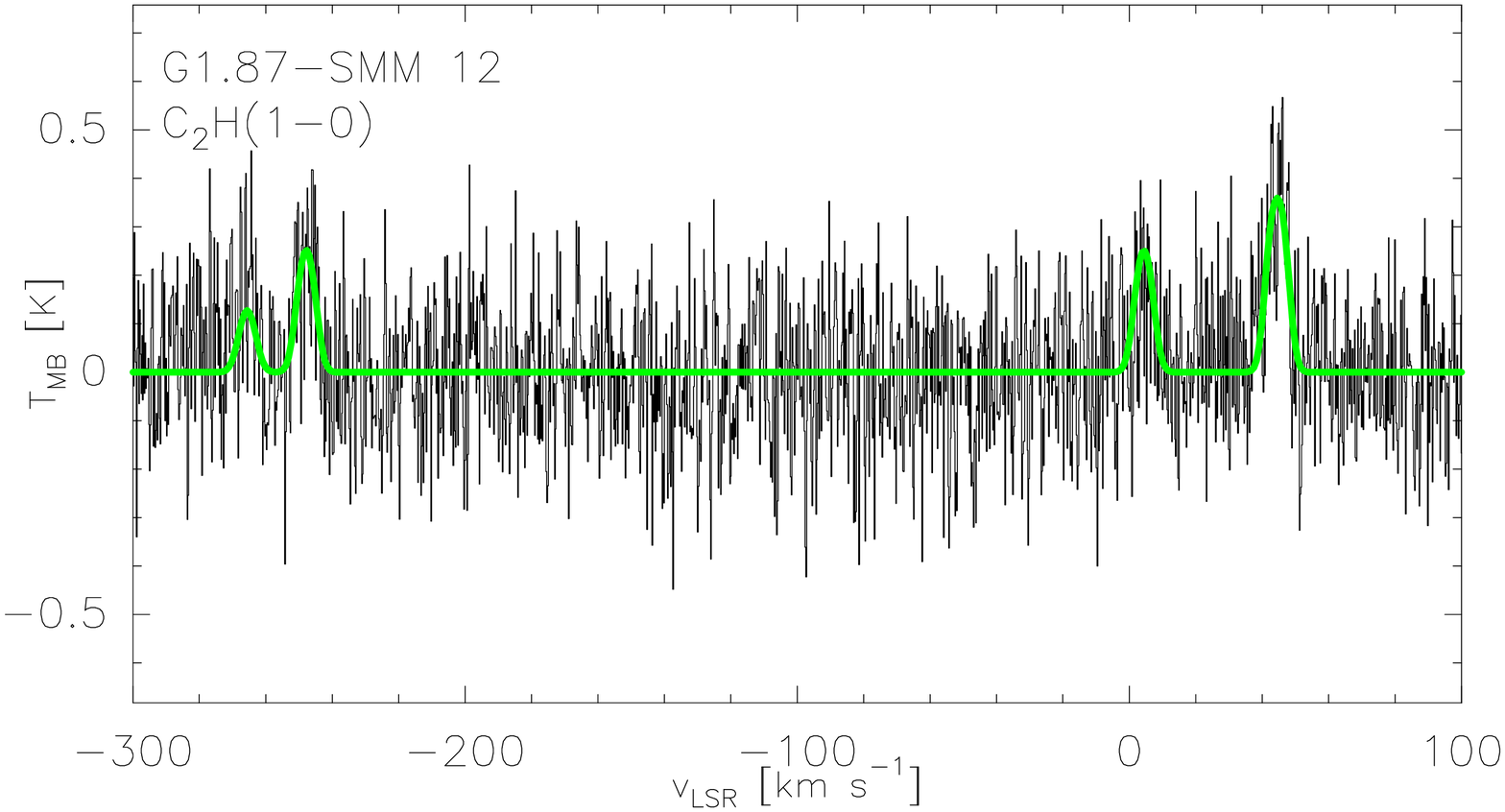}
\includegraphics[width=0.245\textwidth]{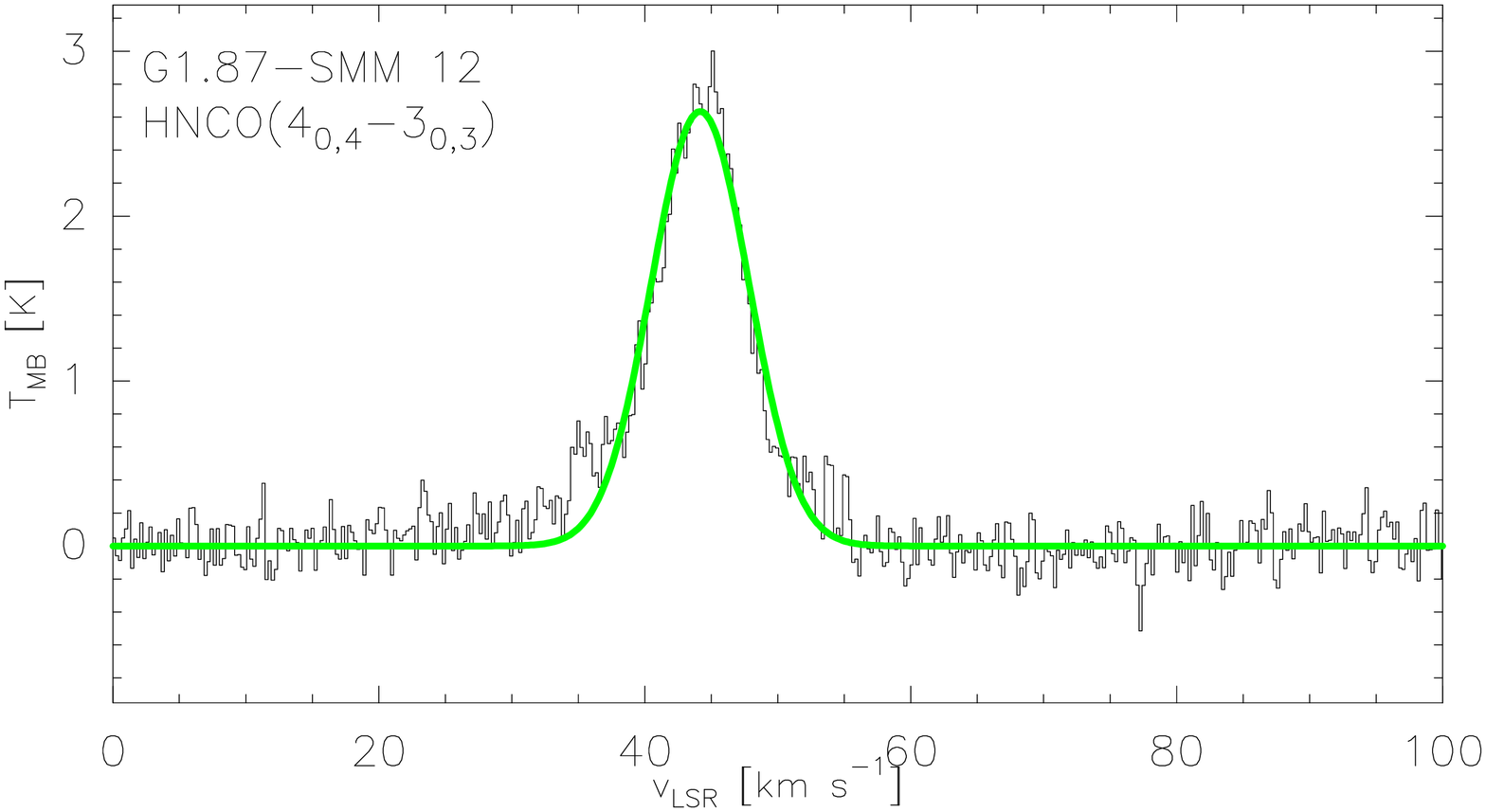}
\includegraphics[width=0.245\textwidth]{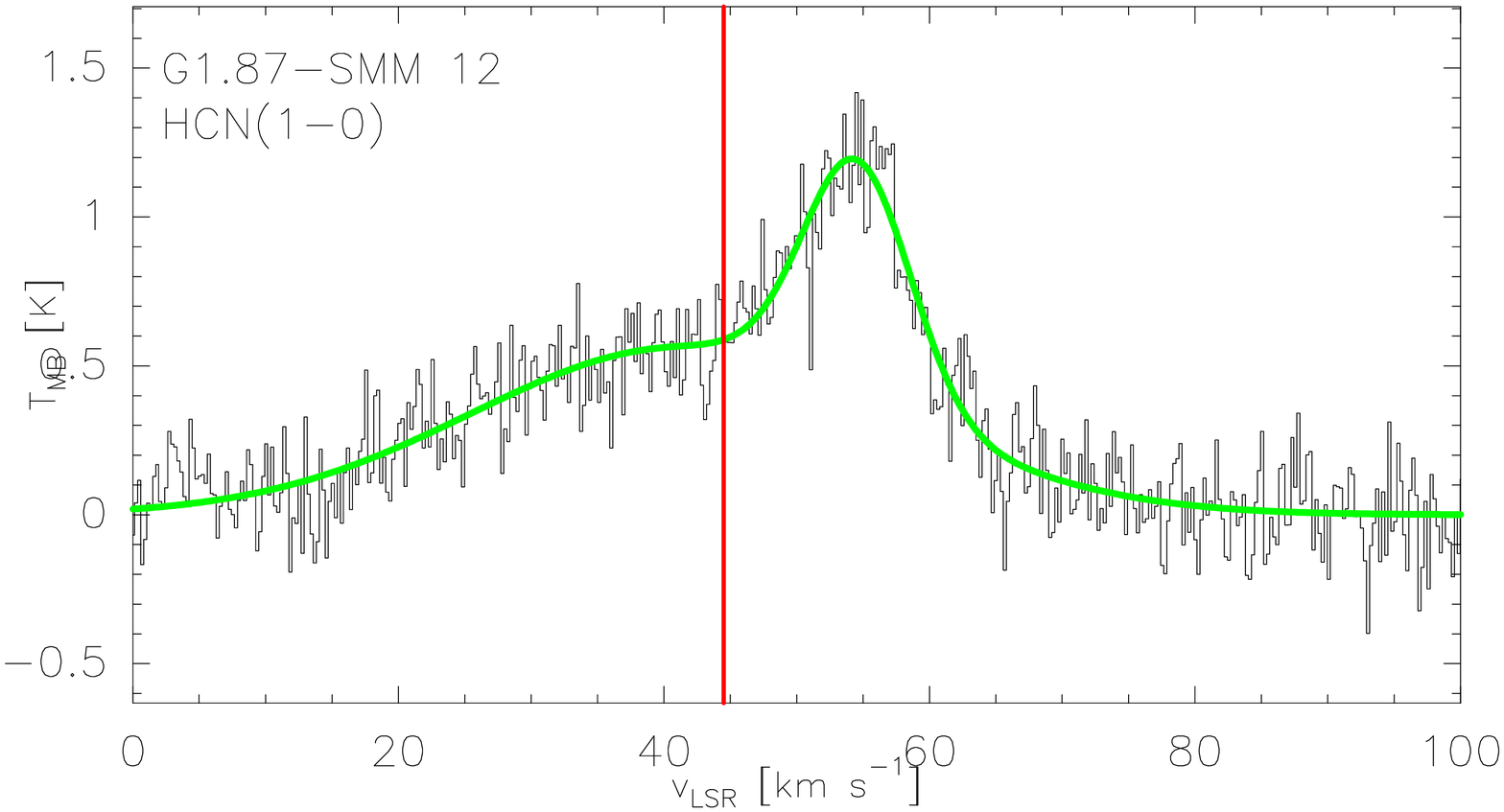}
\includegraphics[width=0.245\textwidth]{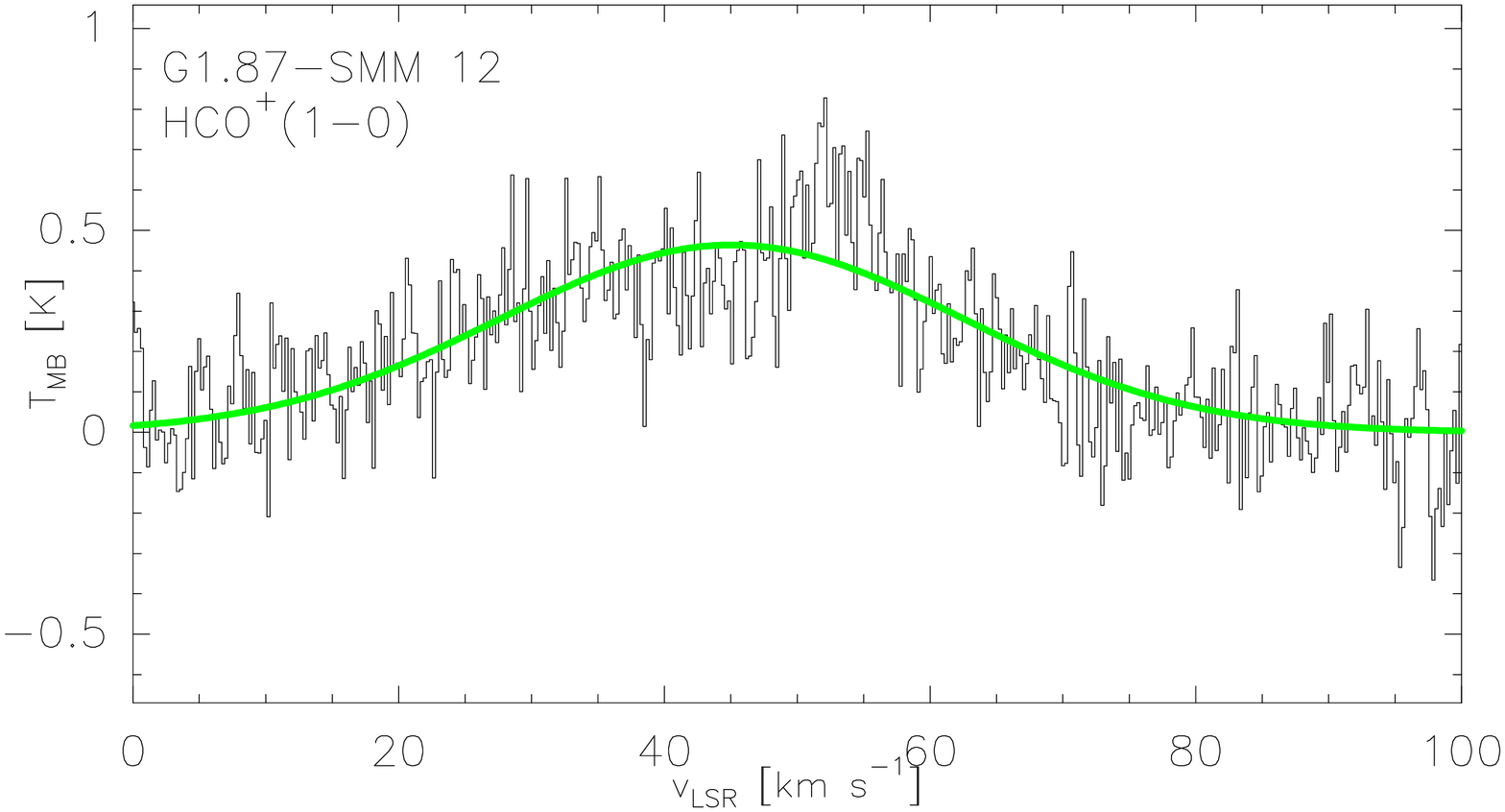}
\includegraphics[width=0.245\textwidth]{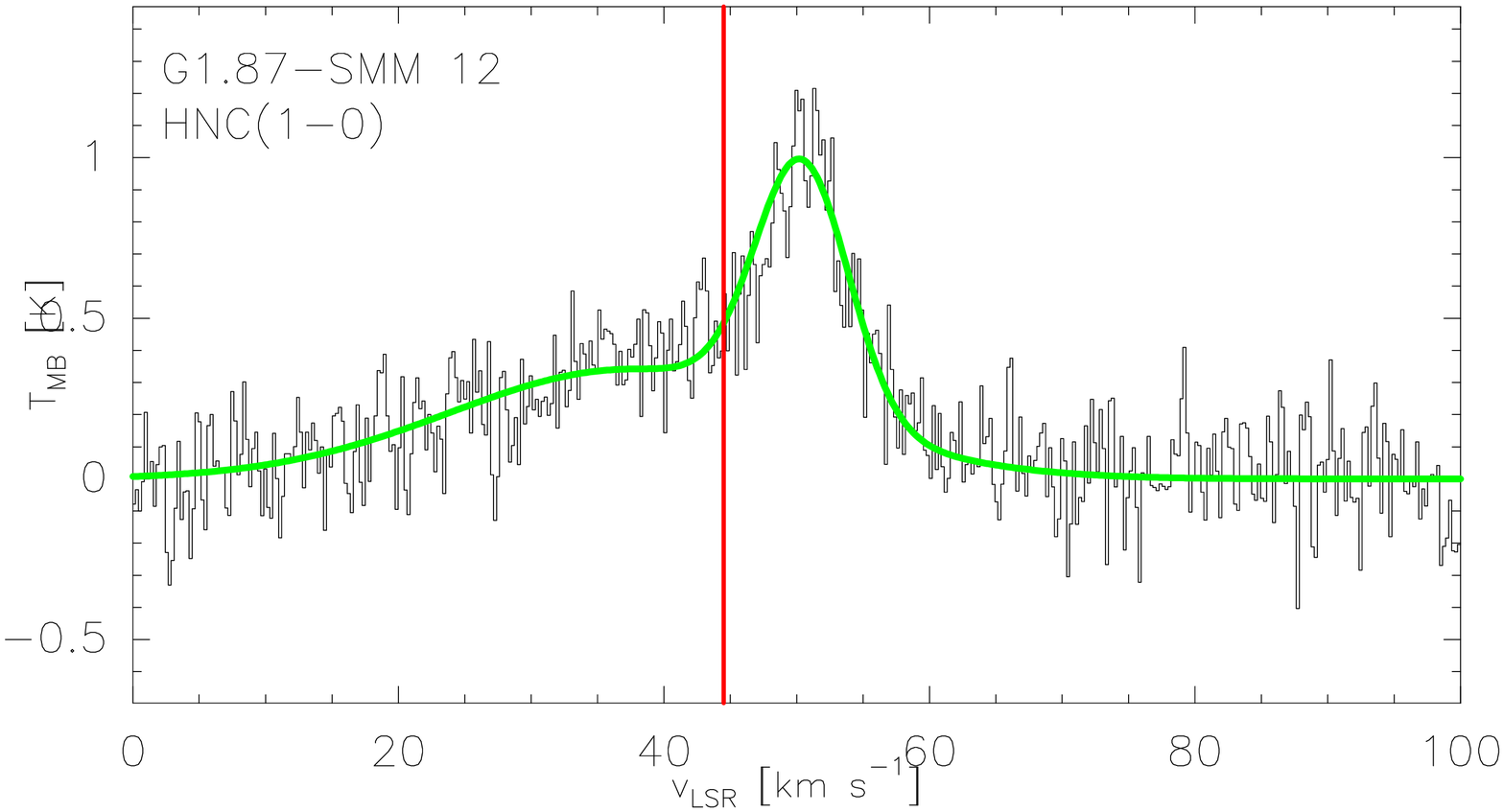}
\includegraphics[width=0.245\textwidth]{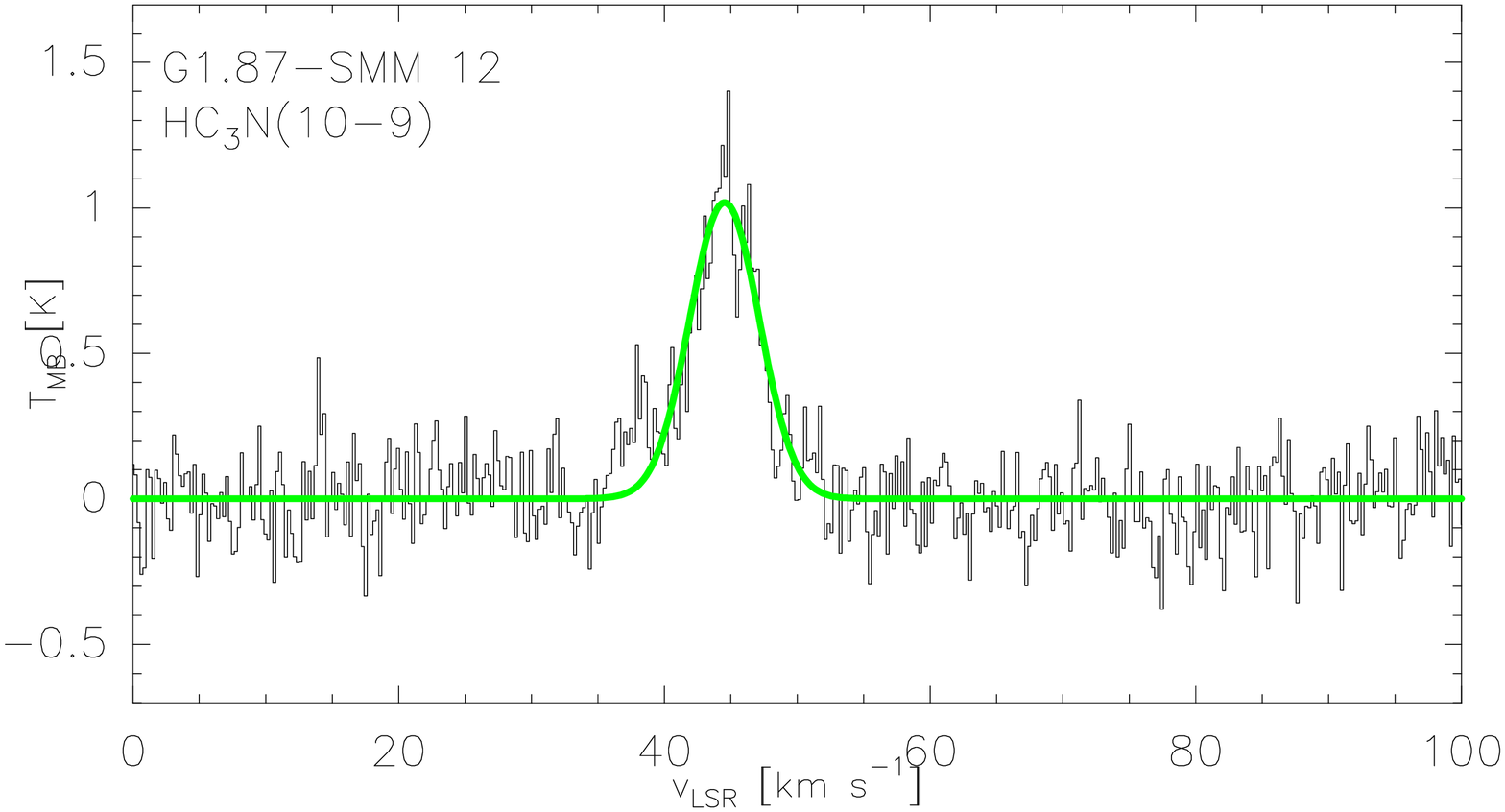}
\includegraphics[width=0.245\textwidth]{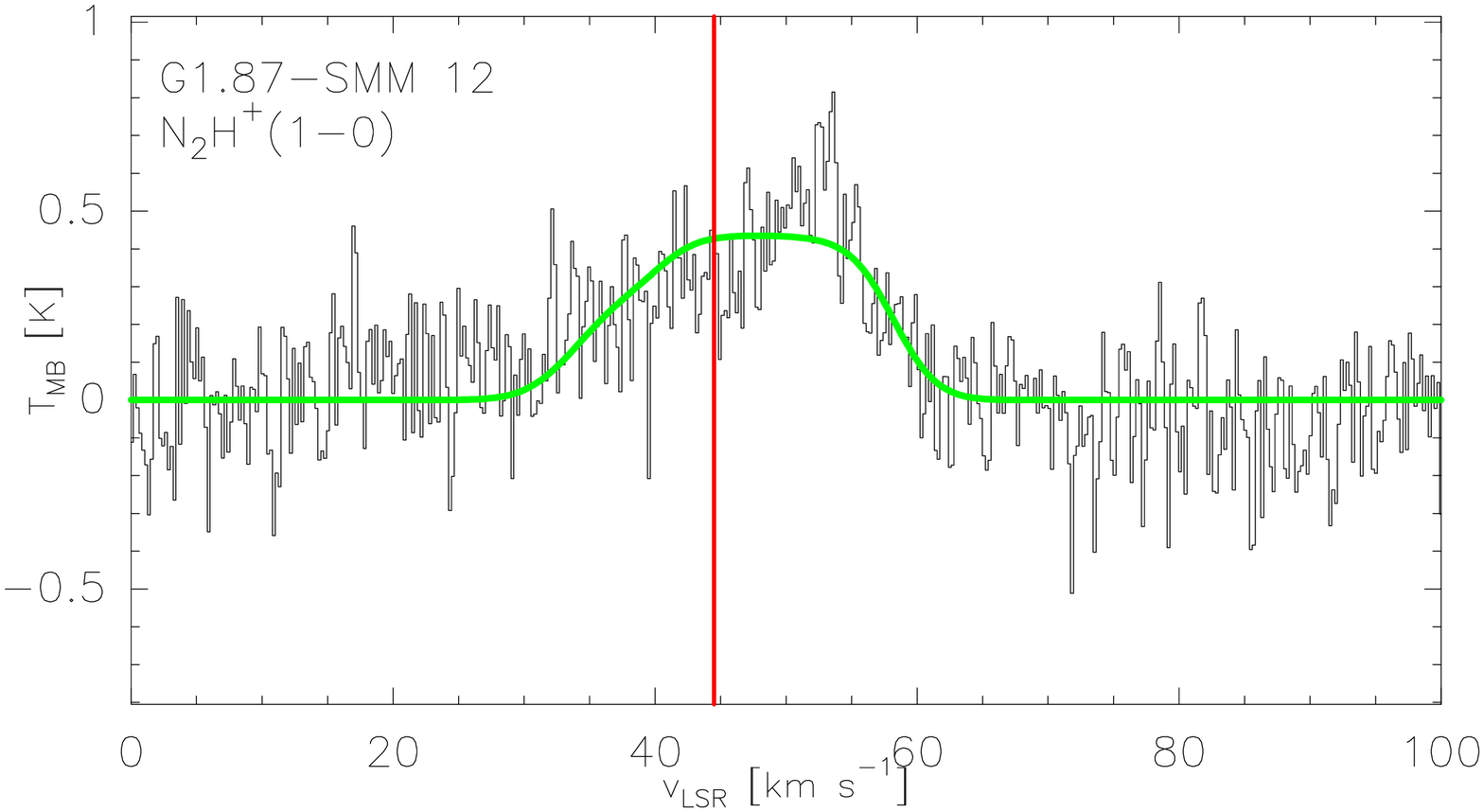}
\caption{Same as Fig.~\ref{figure:G187SMM1_spectra} but towards G1.87--SMM 12. 
Note that the velocity range for the SiO and C$_2$H spectra is wider for 
illustrative purposes. The red vertical line marks the radial velocity of the 
optically thin HC$_3$N line.}
\label{figure:G187SMM12_spectra}
\end{center}
\end{figure*}

\begin{figure*}
\begin{center}
\includegraphics[width=0.245\textwidth]{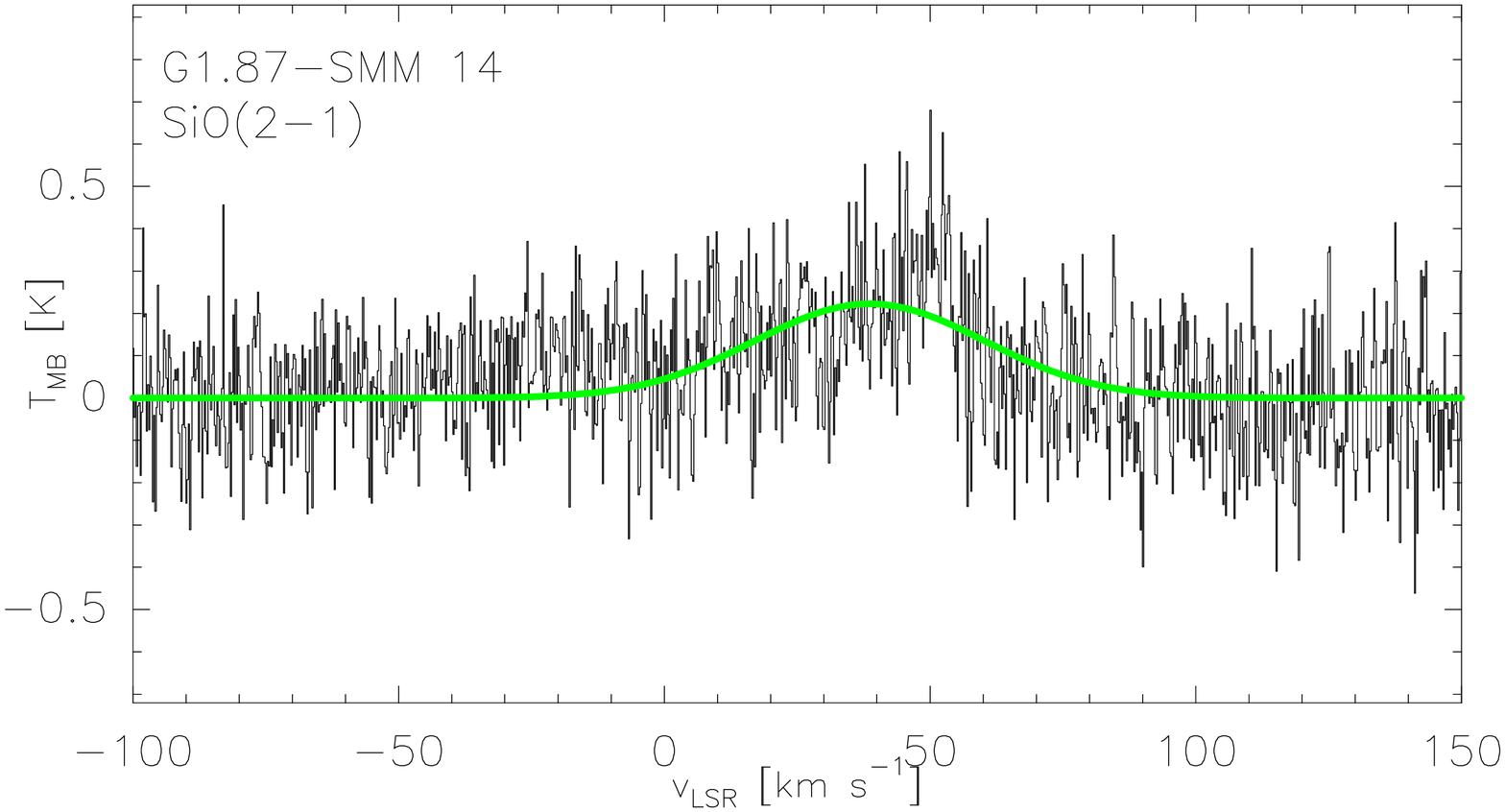}
\includegraphics[width=0.245\textwidth]{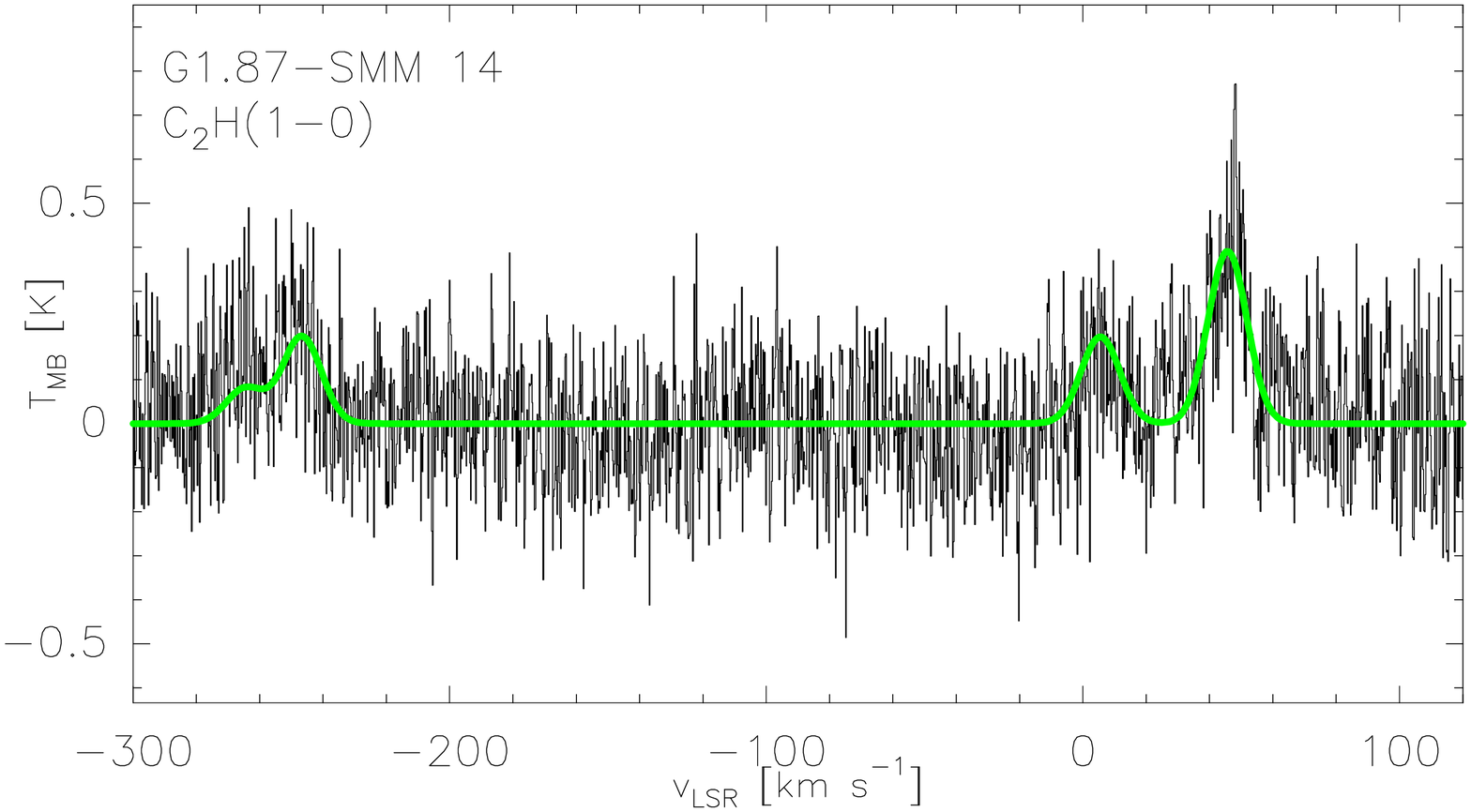}
\includegraphics[width=0.245\textwidth]{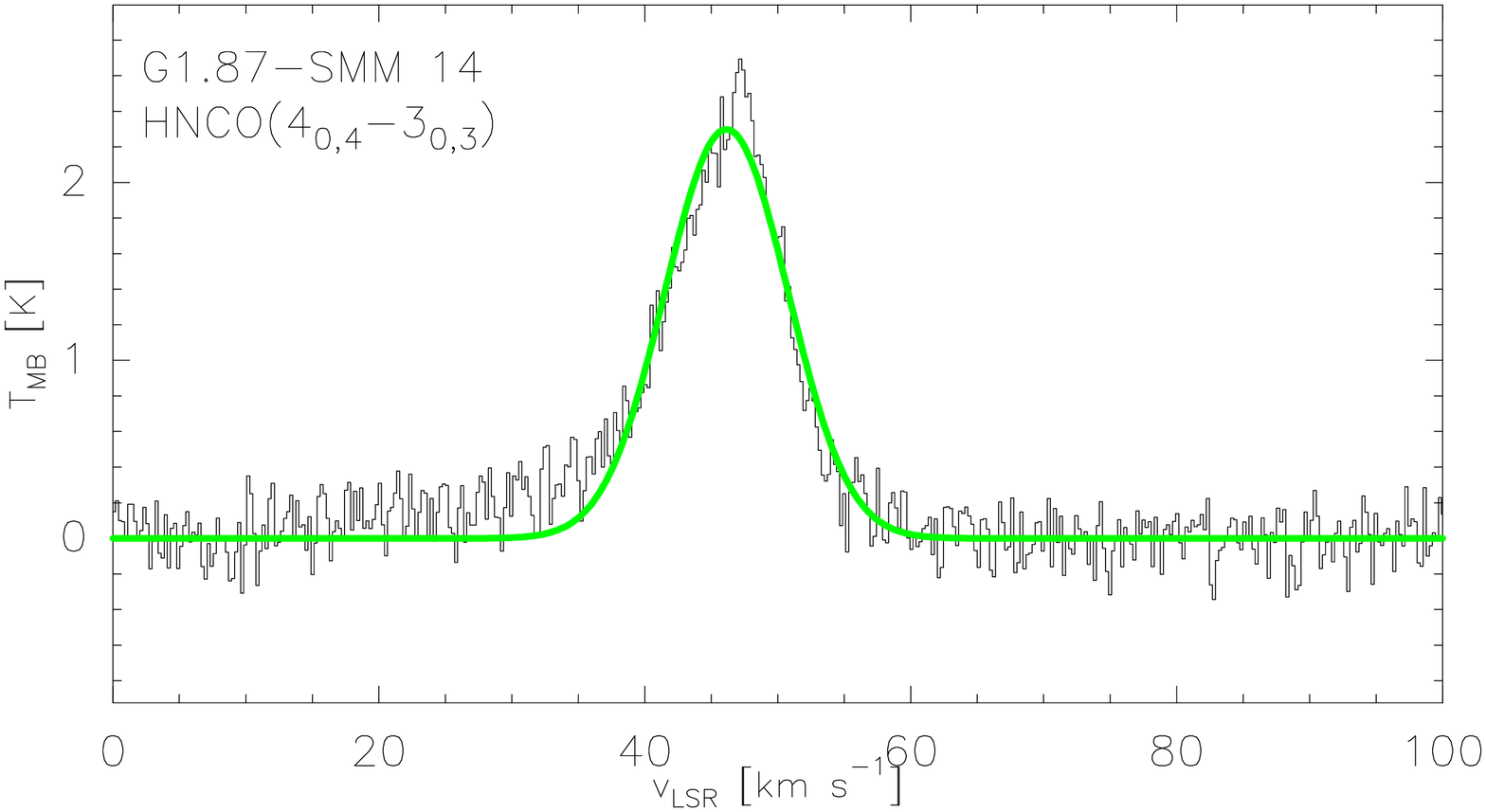}
\includegraphics[width=0.245\textwidth]{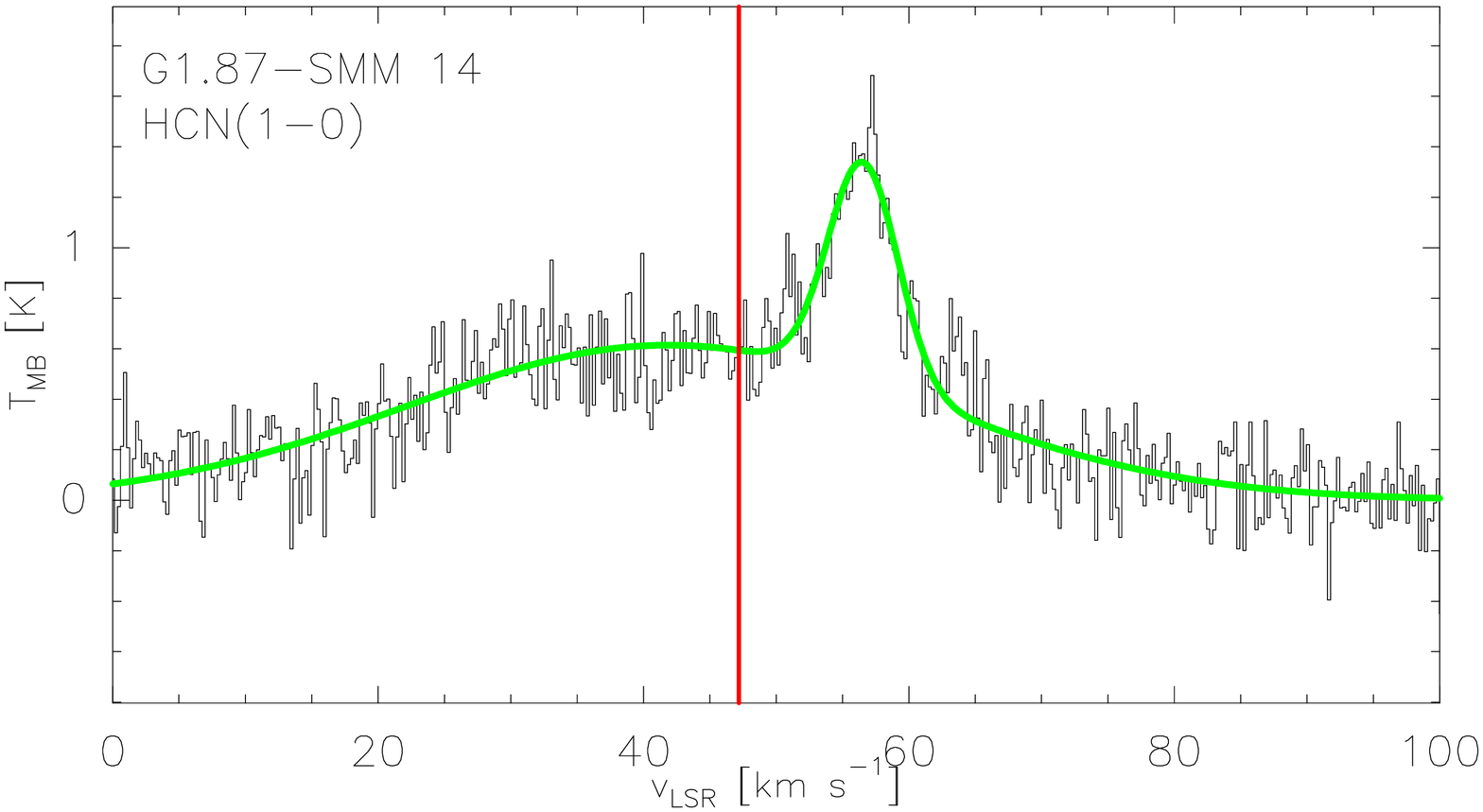}
\includegraphics[width=0.245\textwidth]{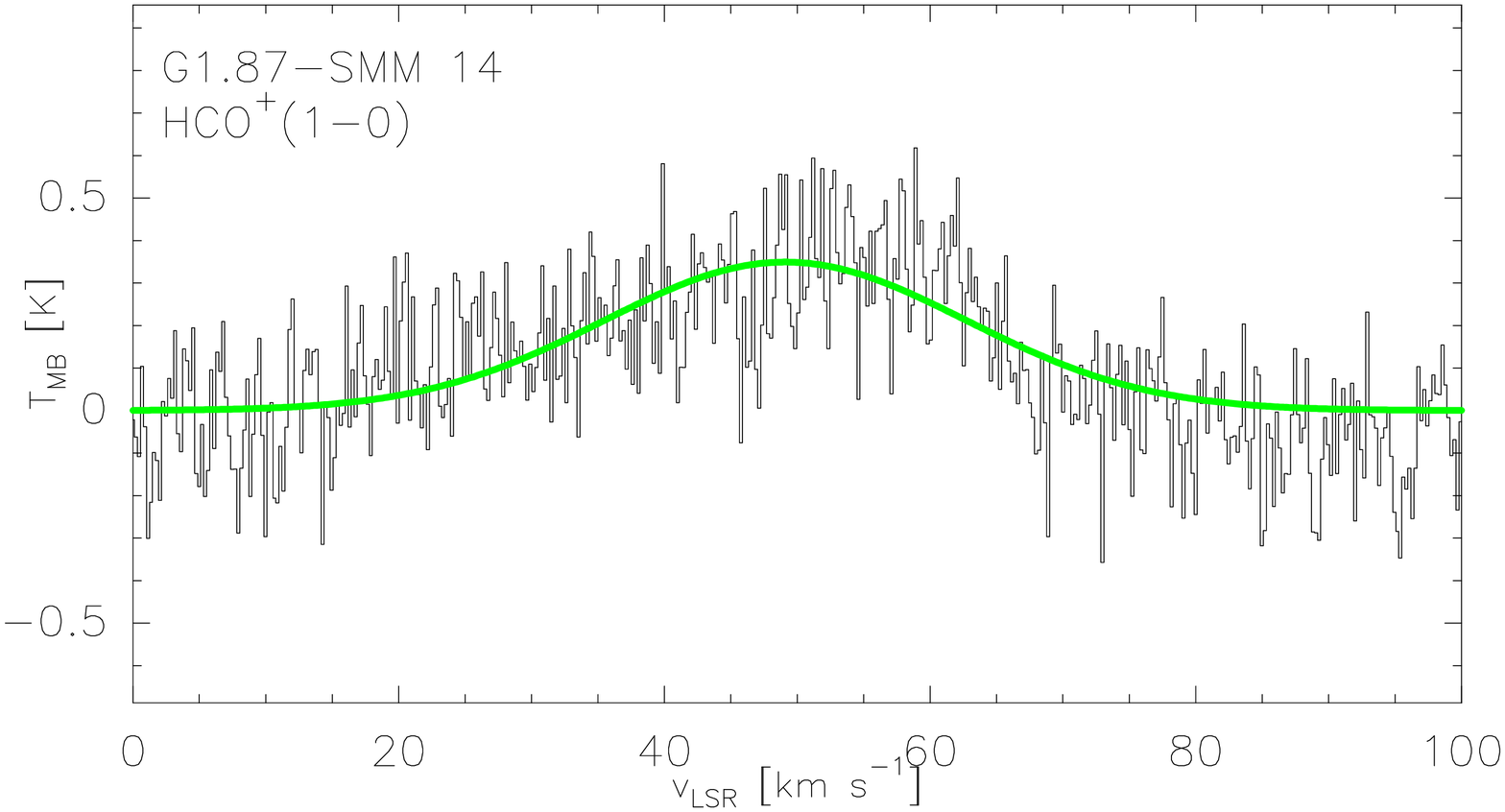}
\includegraphics[width=0.245\textwidth]{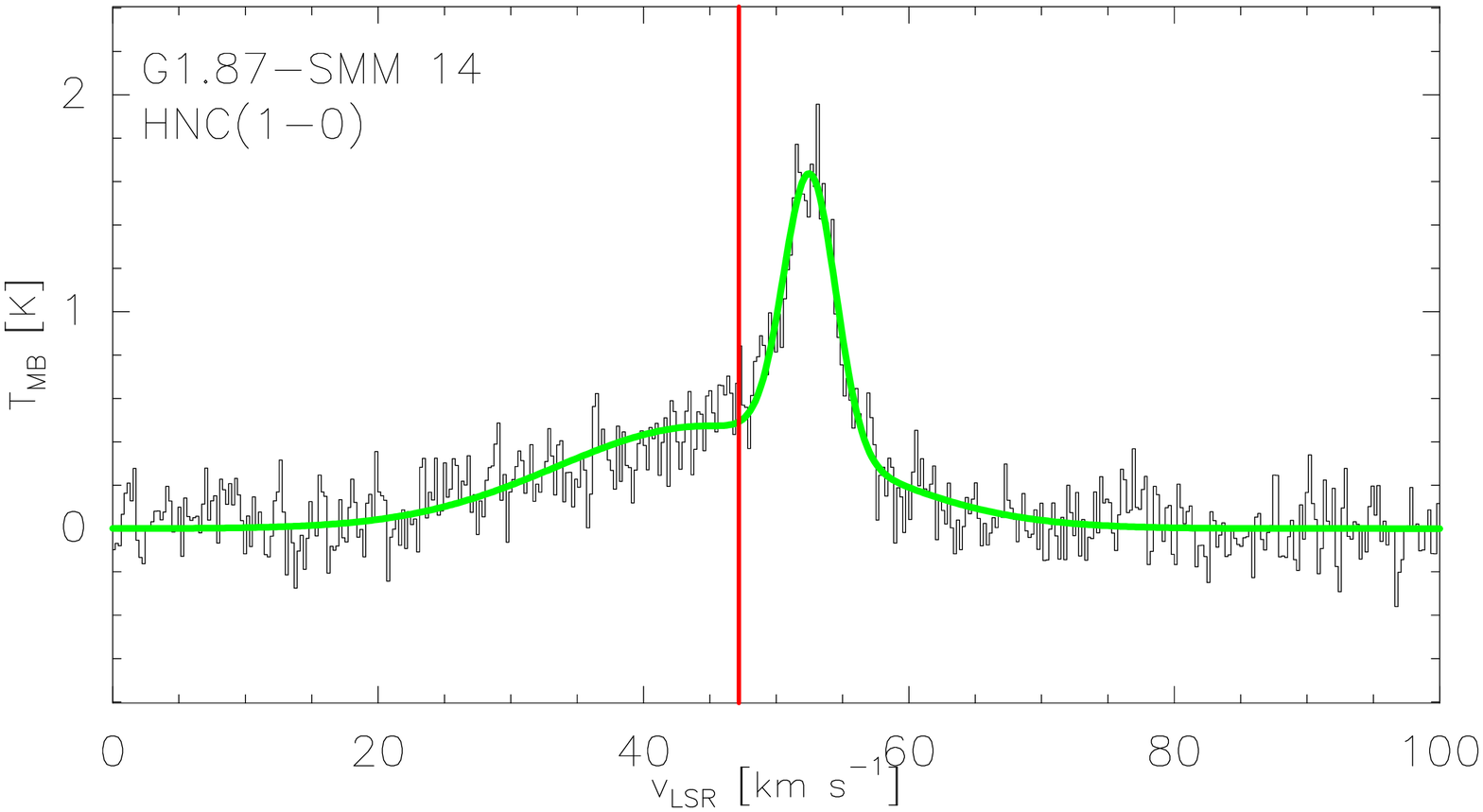}
\includegraphics[width=0.245\textwidth]{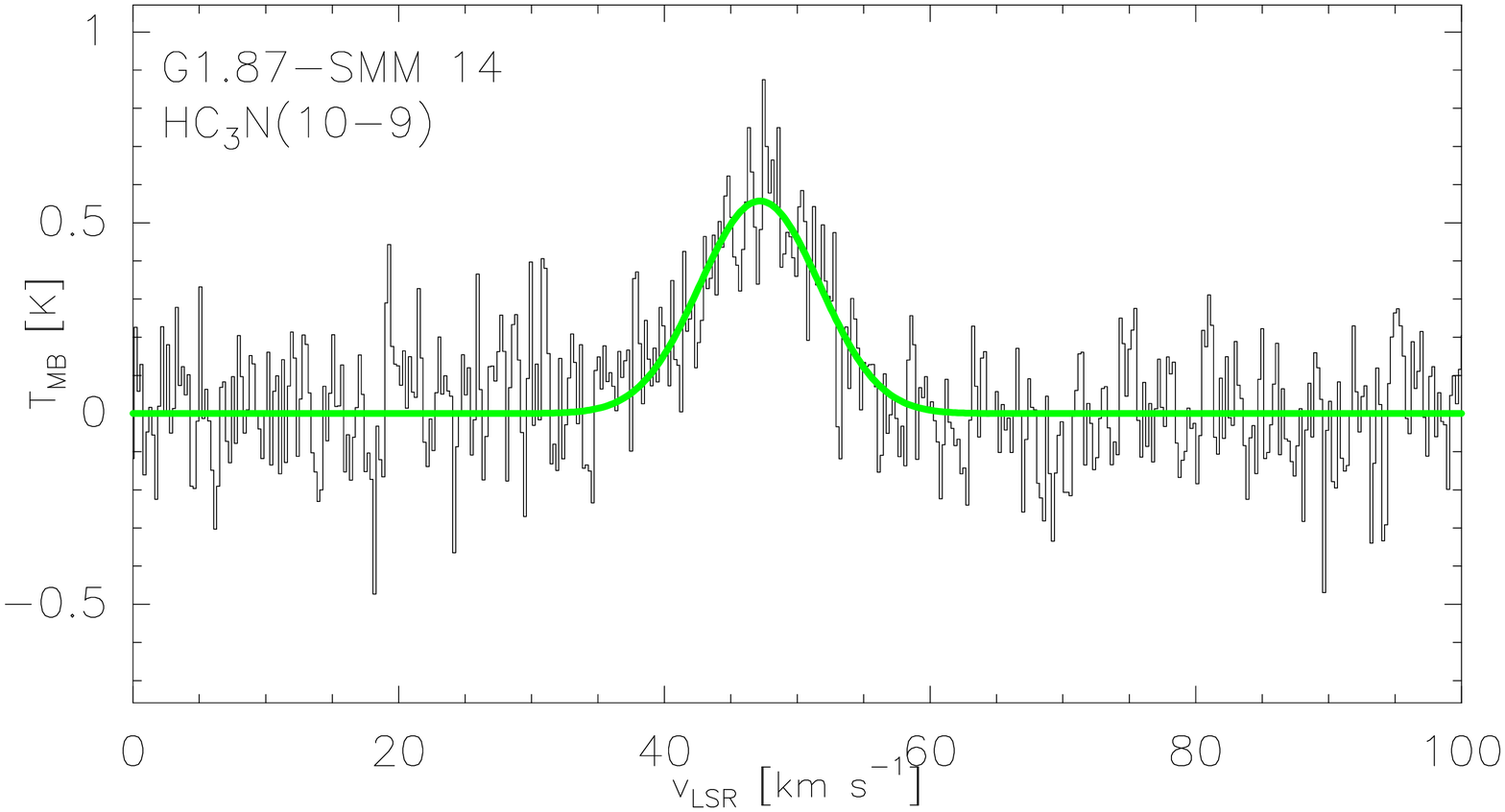}
\includegraphics[width=0.245\textwidth]{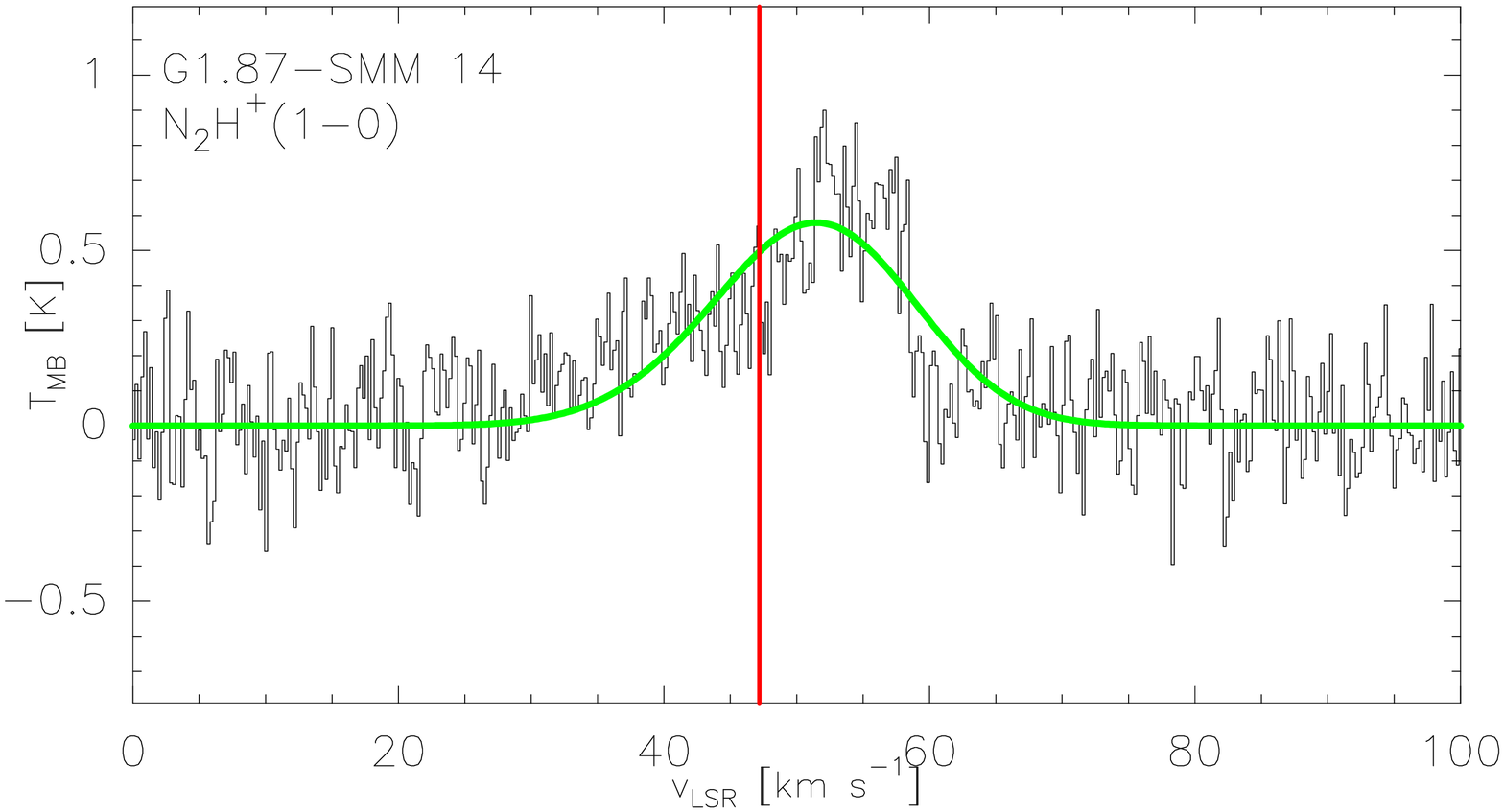}
\caption{Same as Fig.~\ref{figure:G187SMM1_spectra} but towards G1.87--SMM 14. 
Note that the velocity range for the SiO and C$_2$H spectra is wider for 
illustrative purposes. The red vertical line marks the radial velocity of the 
optically thin HC$_3$N line. }
\label{figure:G187SMM14_spectra}
\end{center}
\end{figure*}

\begin{figure*}
\begin{center}
\includegraphics[width=0.245\textwidth]{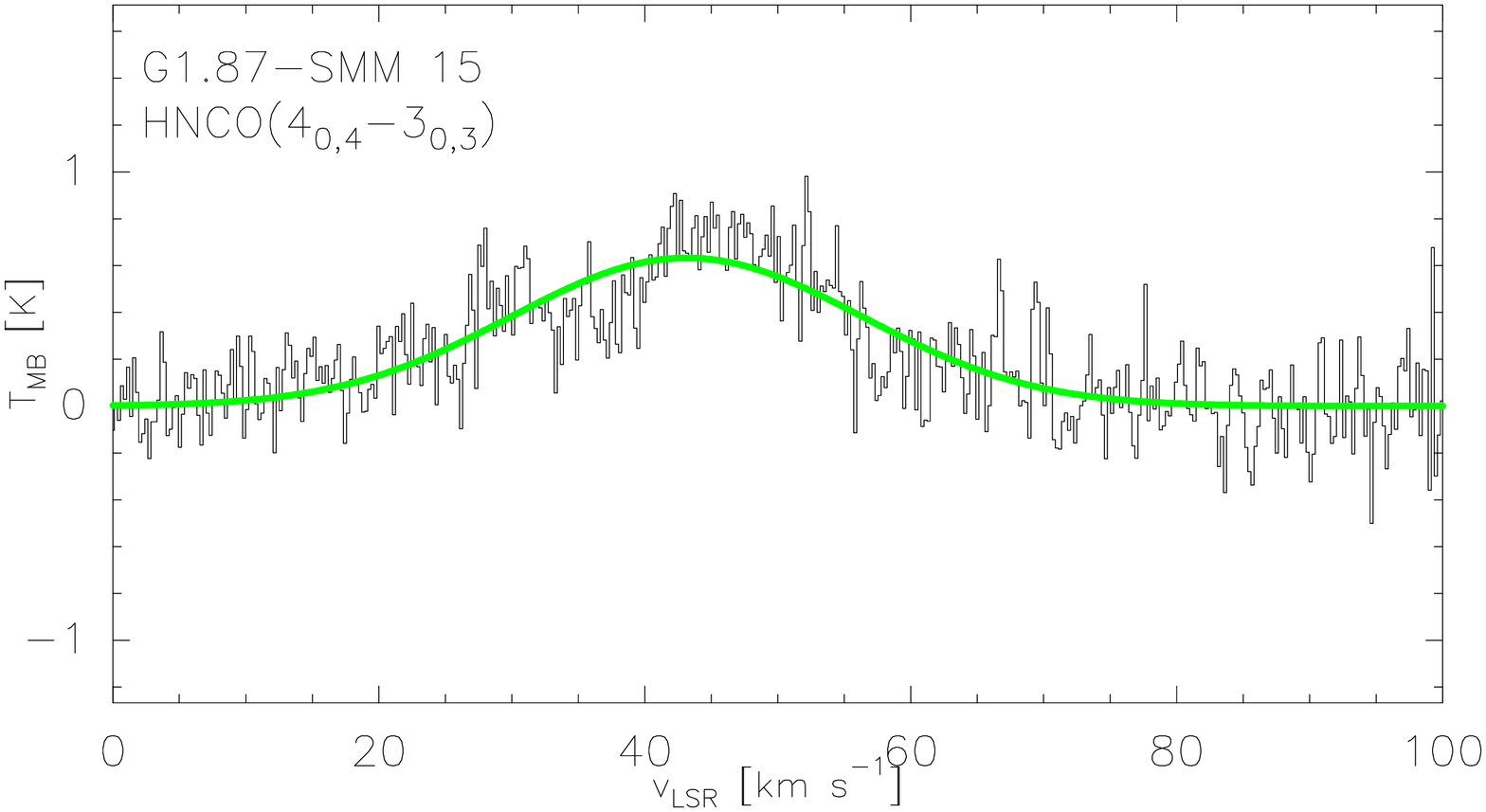}
\includegraphics[width=0.245\textwidth]{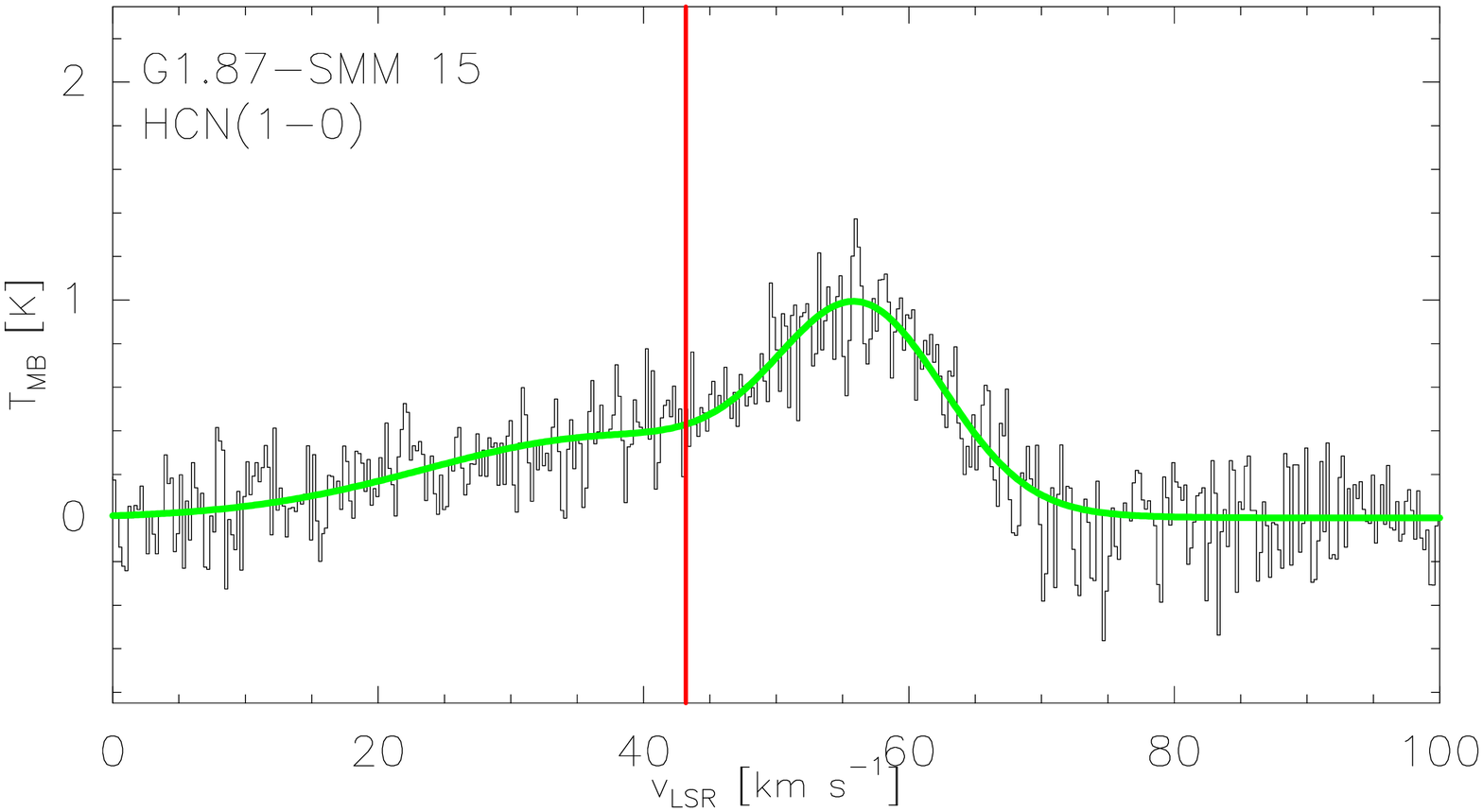}
\includegraphics[width=0.245\textwidth]{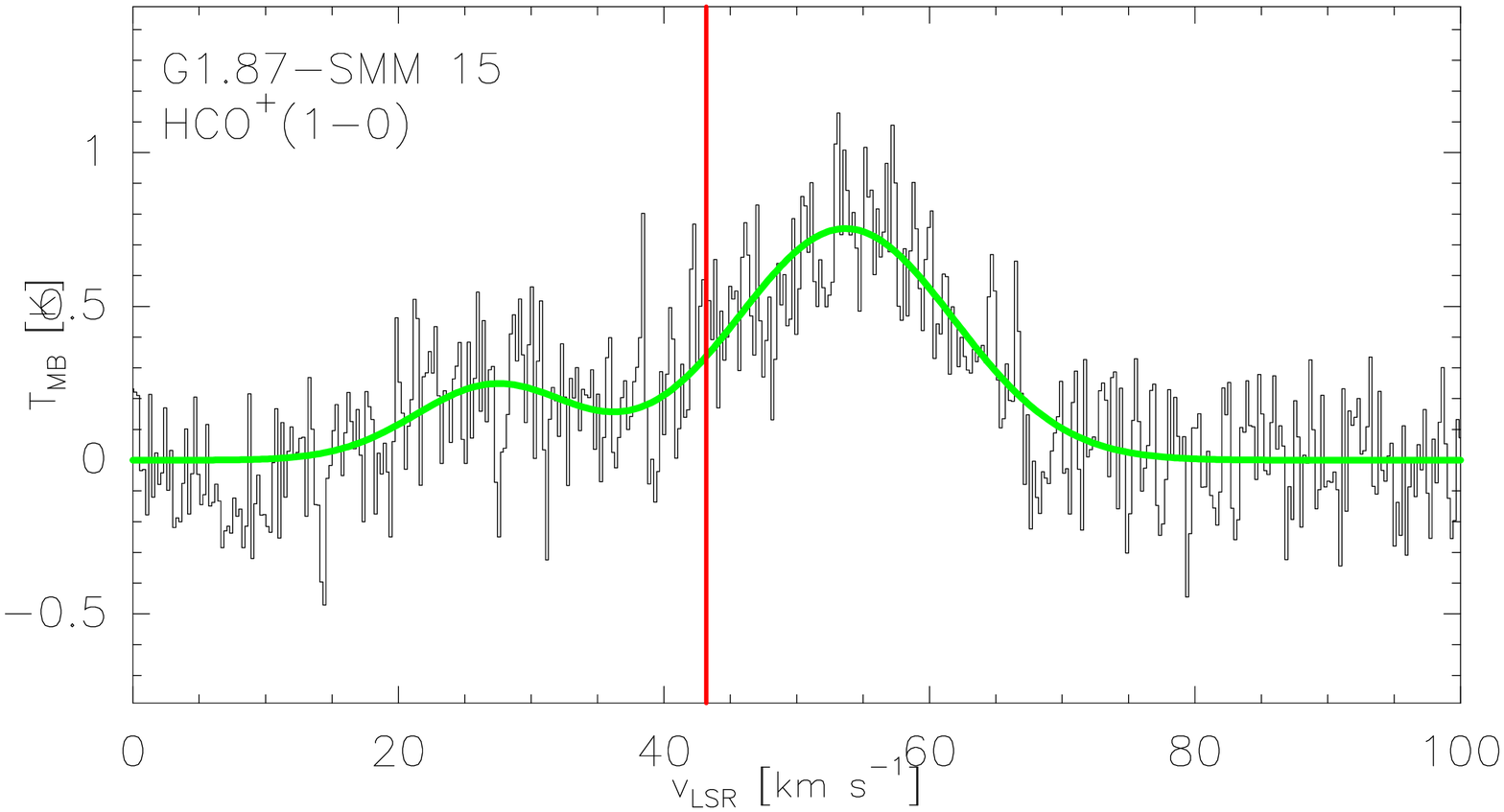}
\includegraphics[width=0.245\textwidth]{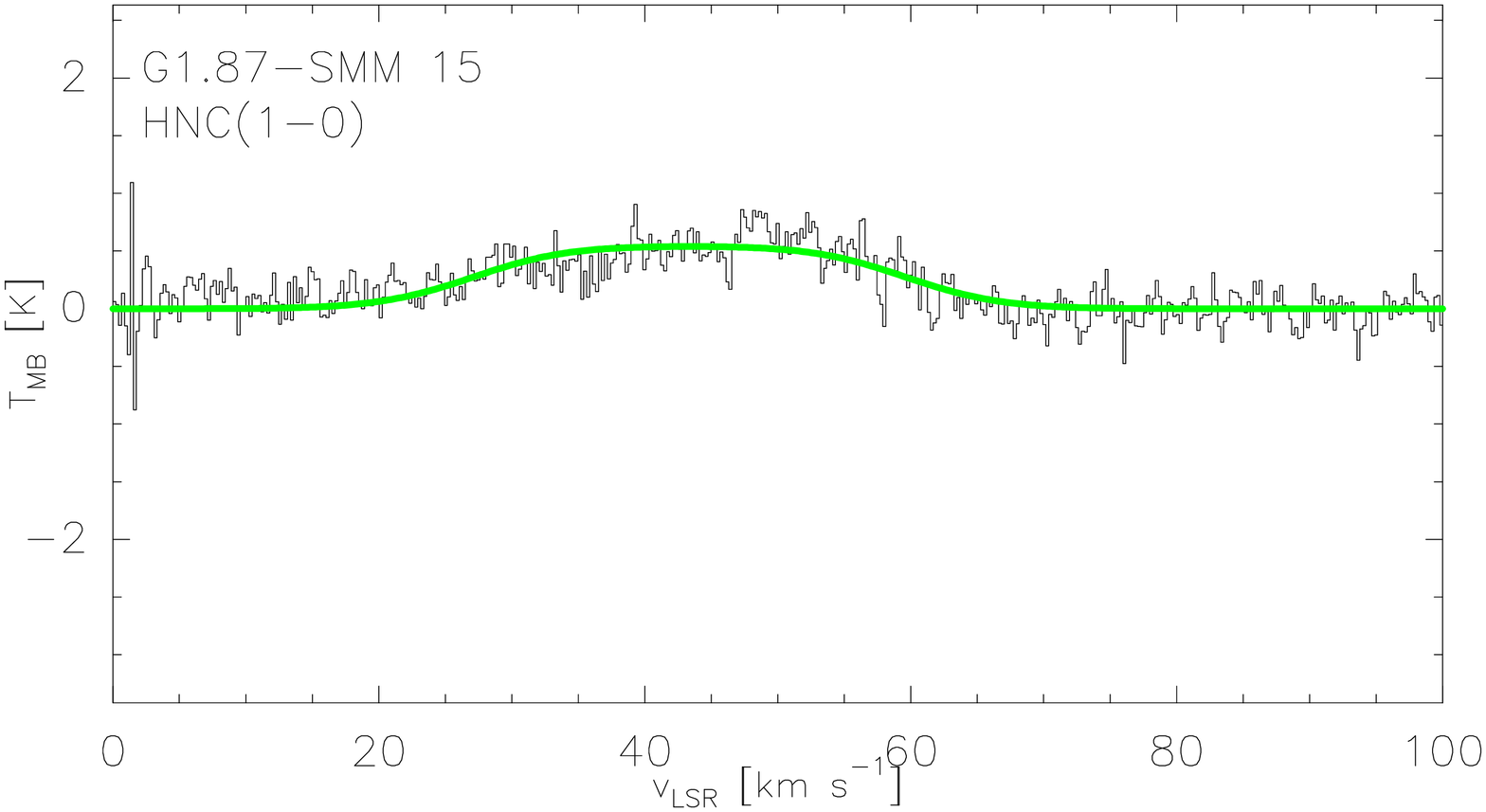}
\includegraphics[width=0.245\textwidth]{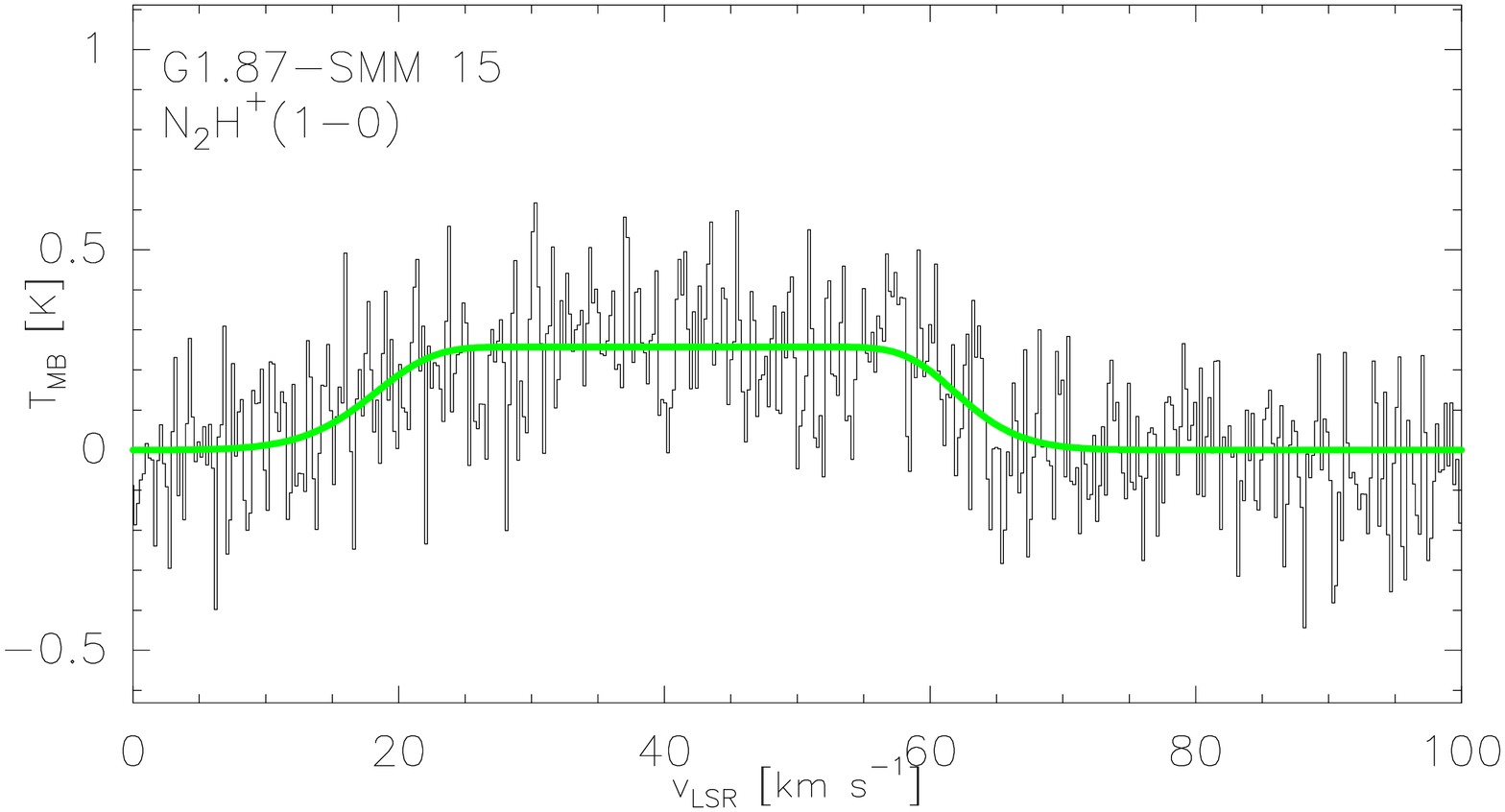}
\caption{Same as Fig.~\ref{figure:G187SMM1_spectra} but towards G1.87--SMM 15. 
The red vertical line indicates the radial velocity of the optically thin 
HNCO line.}
\label{figure:G187SMM15_spectra}
\end{center}
\end{figure*}

\begin{figure*}
\begin{center}
\includegraphics[width=0.245\textwidth]{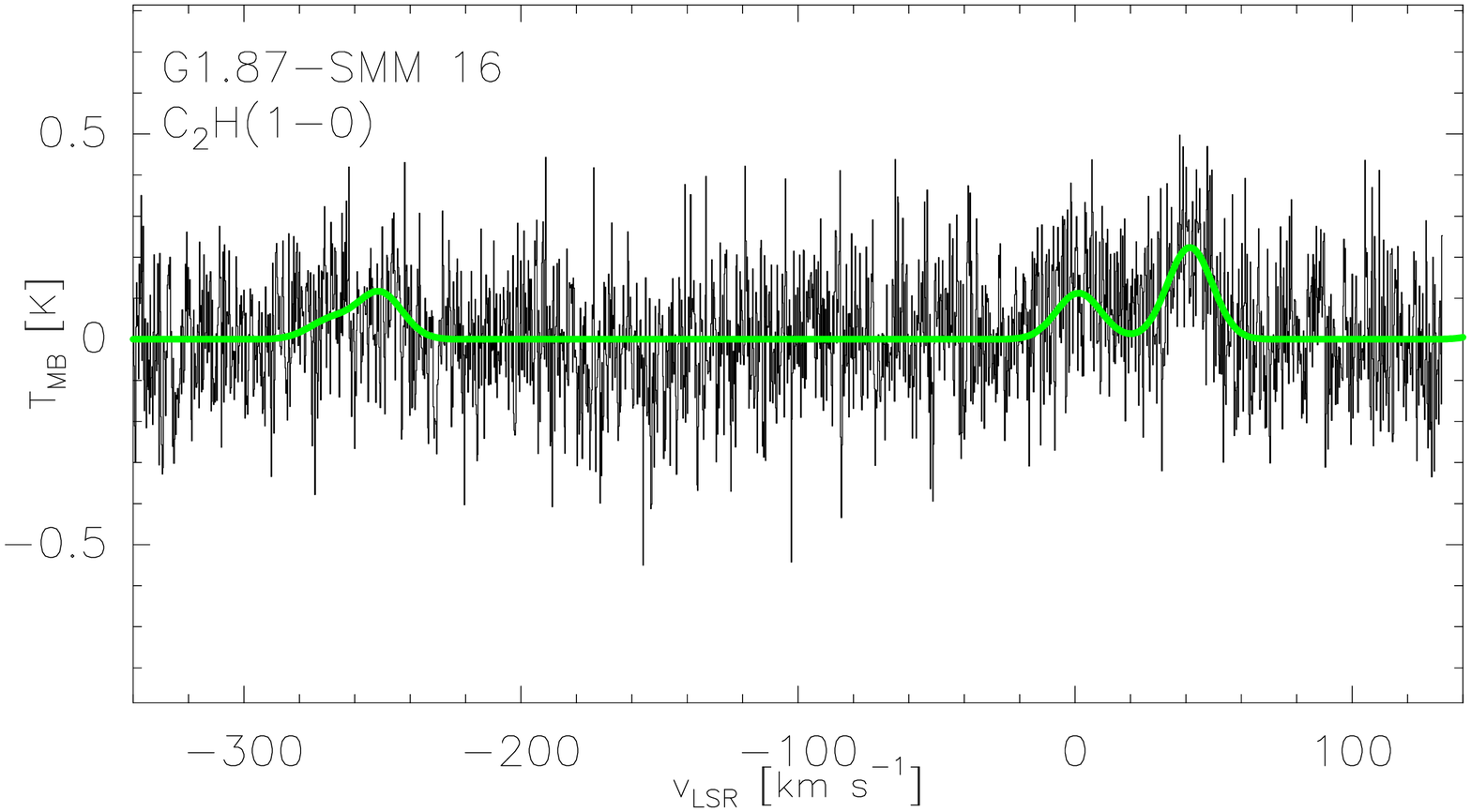}
\includegraphics[width=0.245\textwidth]{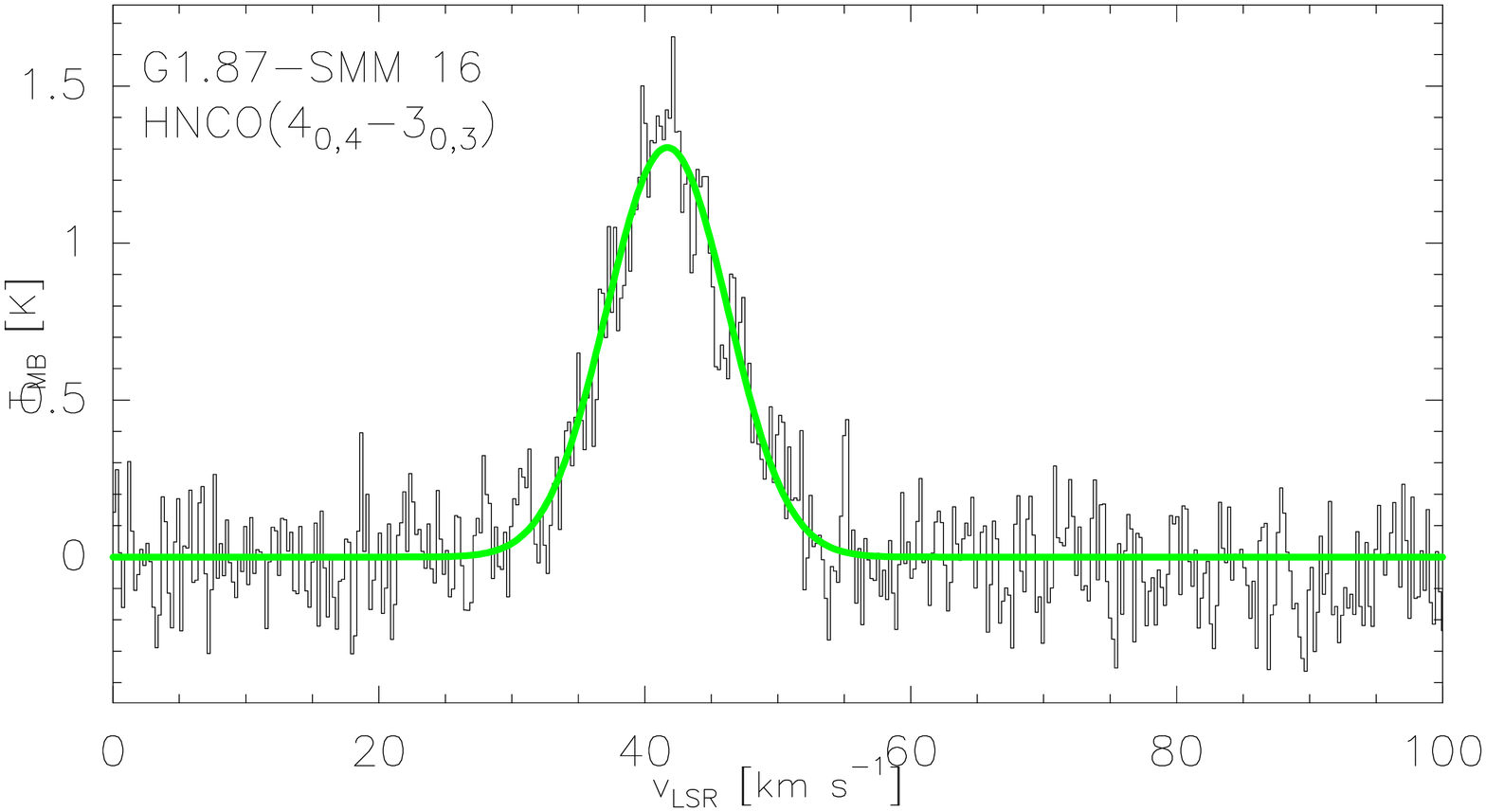}
\includegraphics[width=0.245\textwidth]{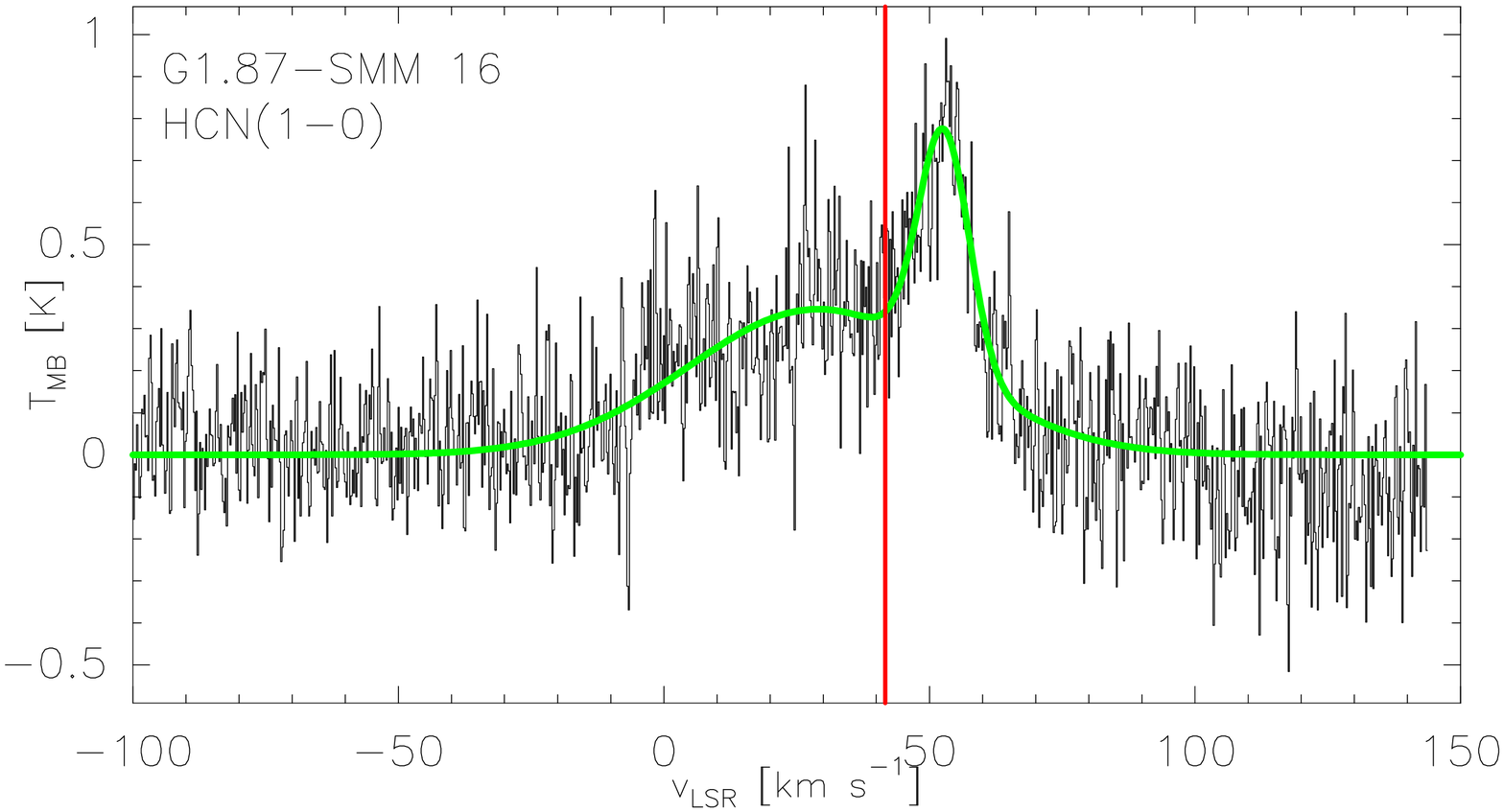}
\includegraphics[width=0.245\textwidth]{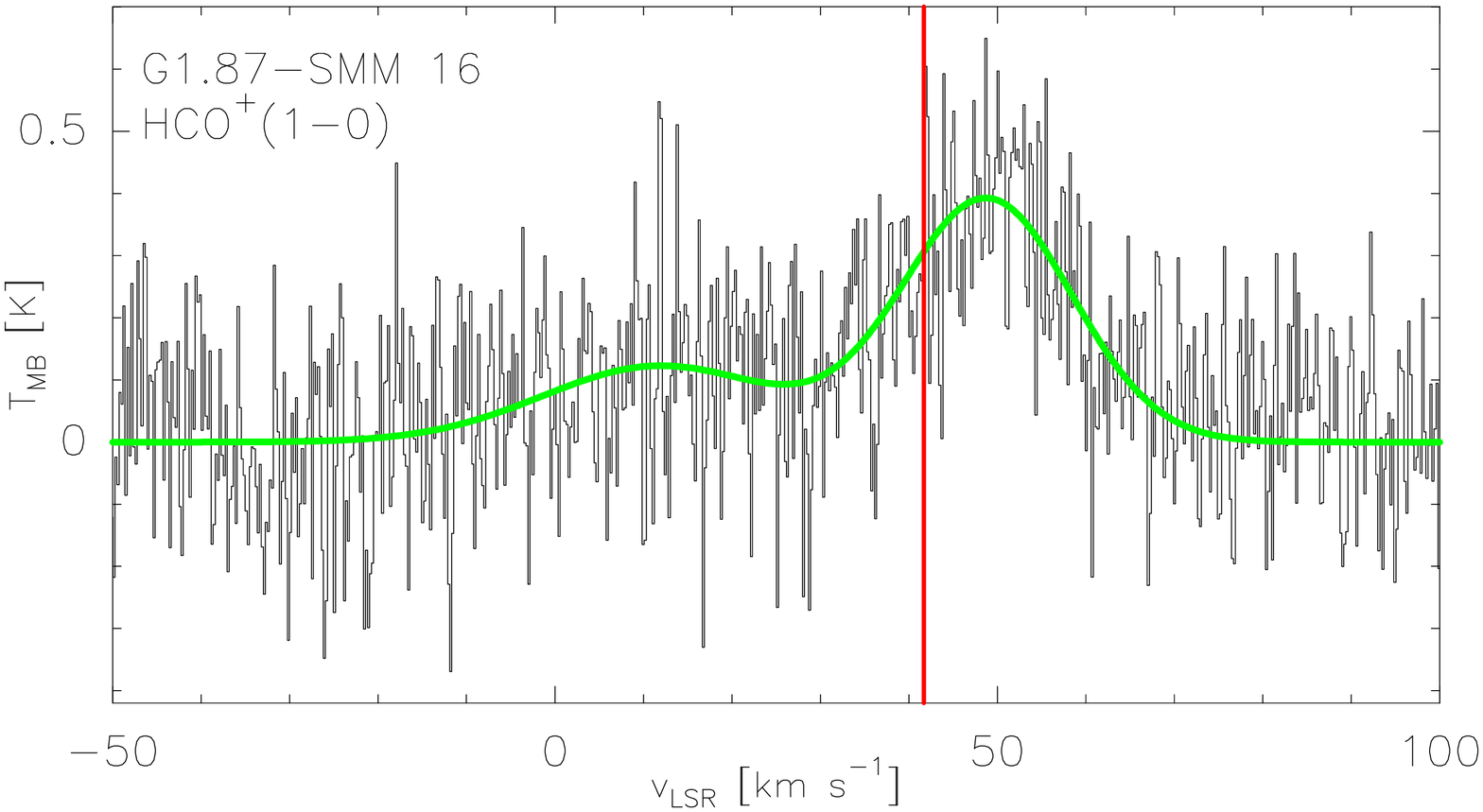}
\includegraphics[width=0.245\textwidth]{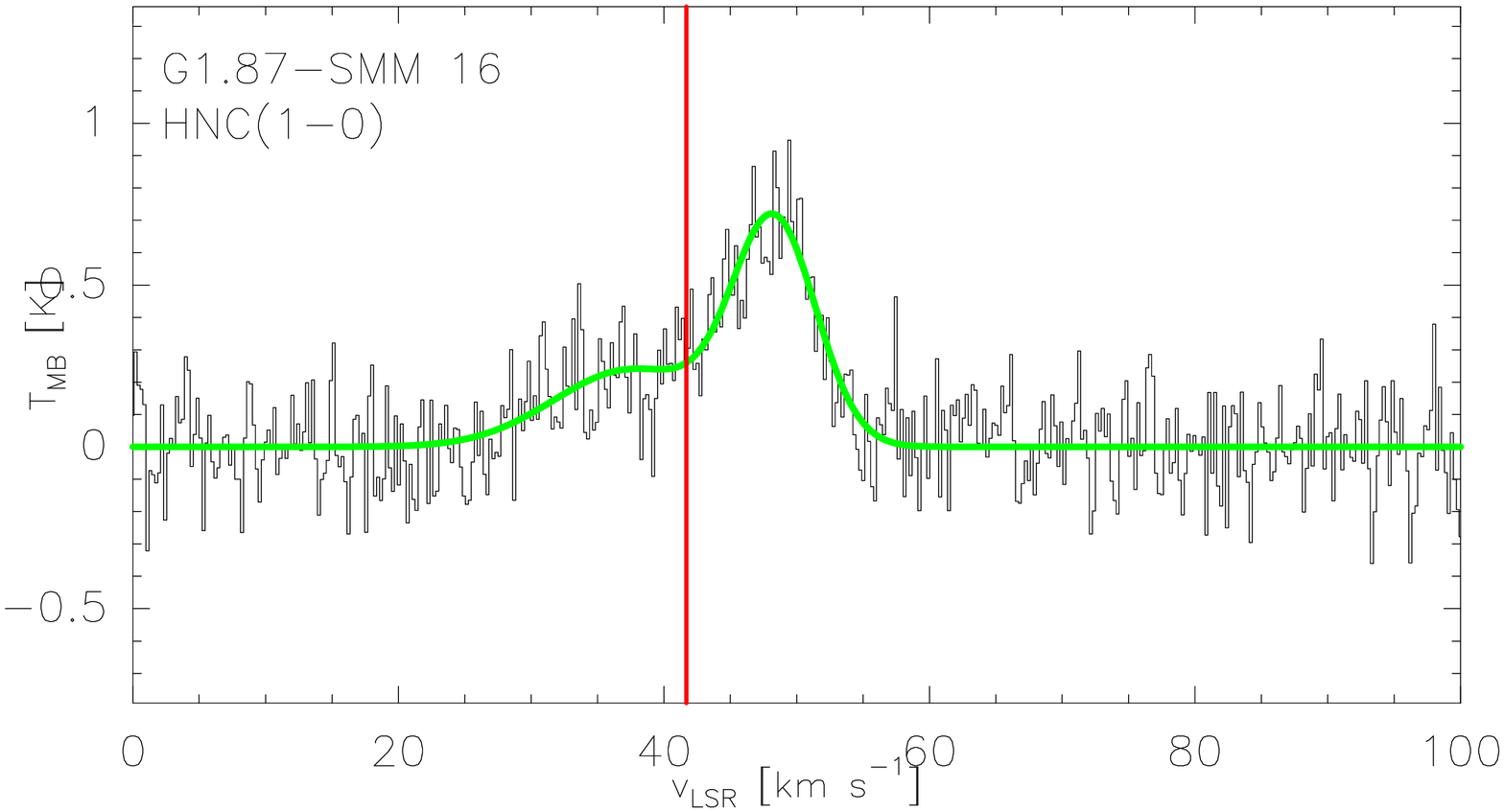}
\caption{Same as Fig.~\ref{figure:G187SMM1_spectra} but towards G1.87--SMM 16. 
In some of the spectra, the velocity range shown is wider for illustrative 
purposes. The red vertical line mark the velocity of the HNCO line.}
\label{figure:G187SMM16_spectra}
\end{center}
\end{figure*}

\begin{figure*}
\begin{center}
\includegraphics[width=0.245\textwidth]{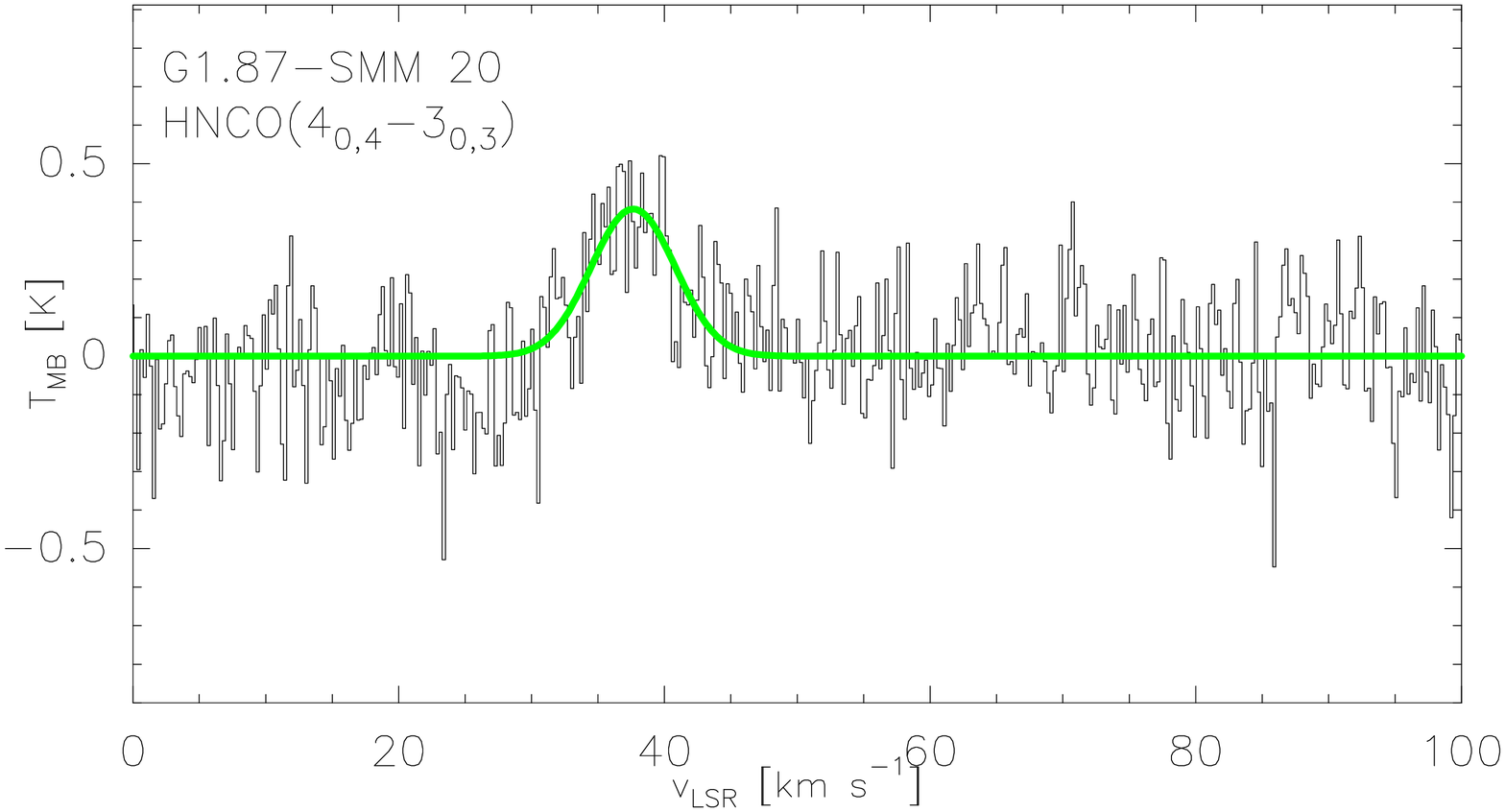}
\includegraphics[width=0.245\textwidth]{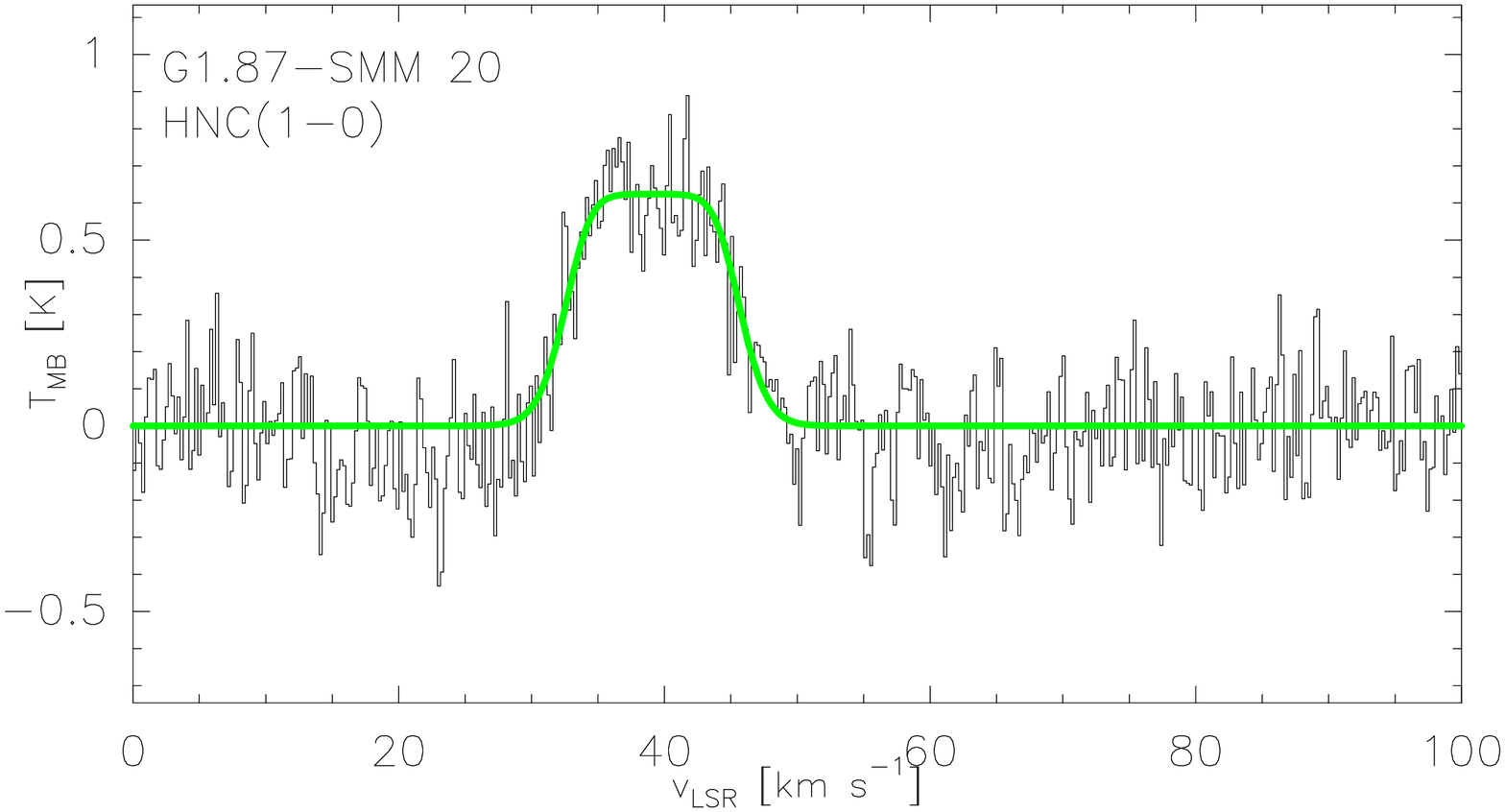}
\includegraphics[width=0.245\textwidth]{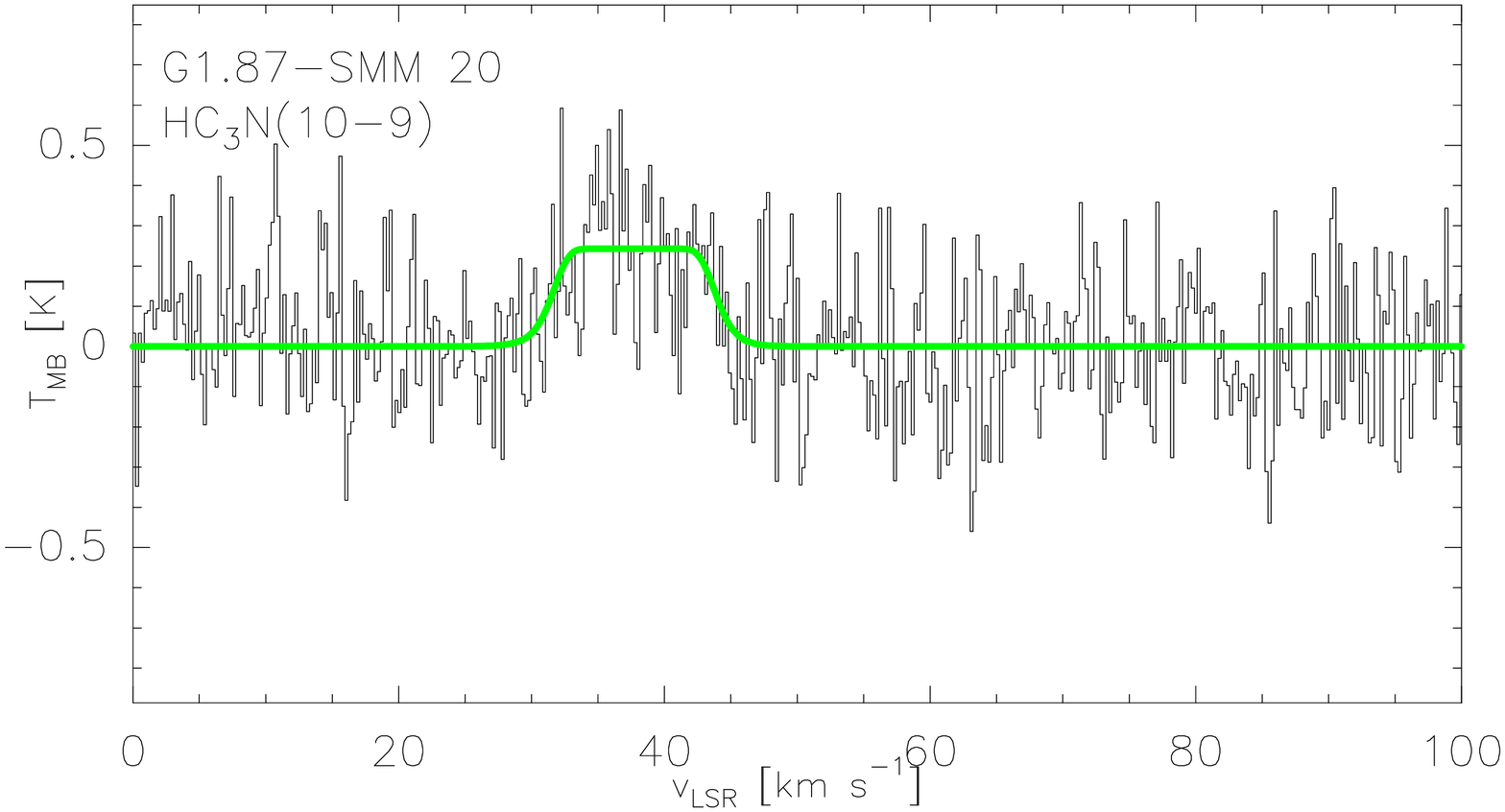}
\includegraphics[width=0.245\textwidth]{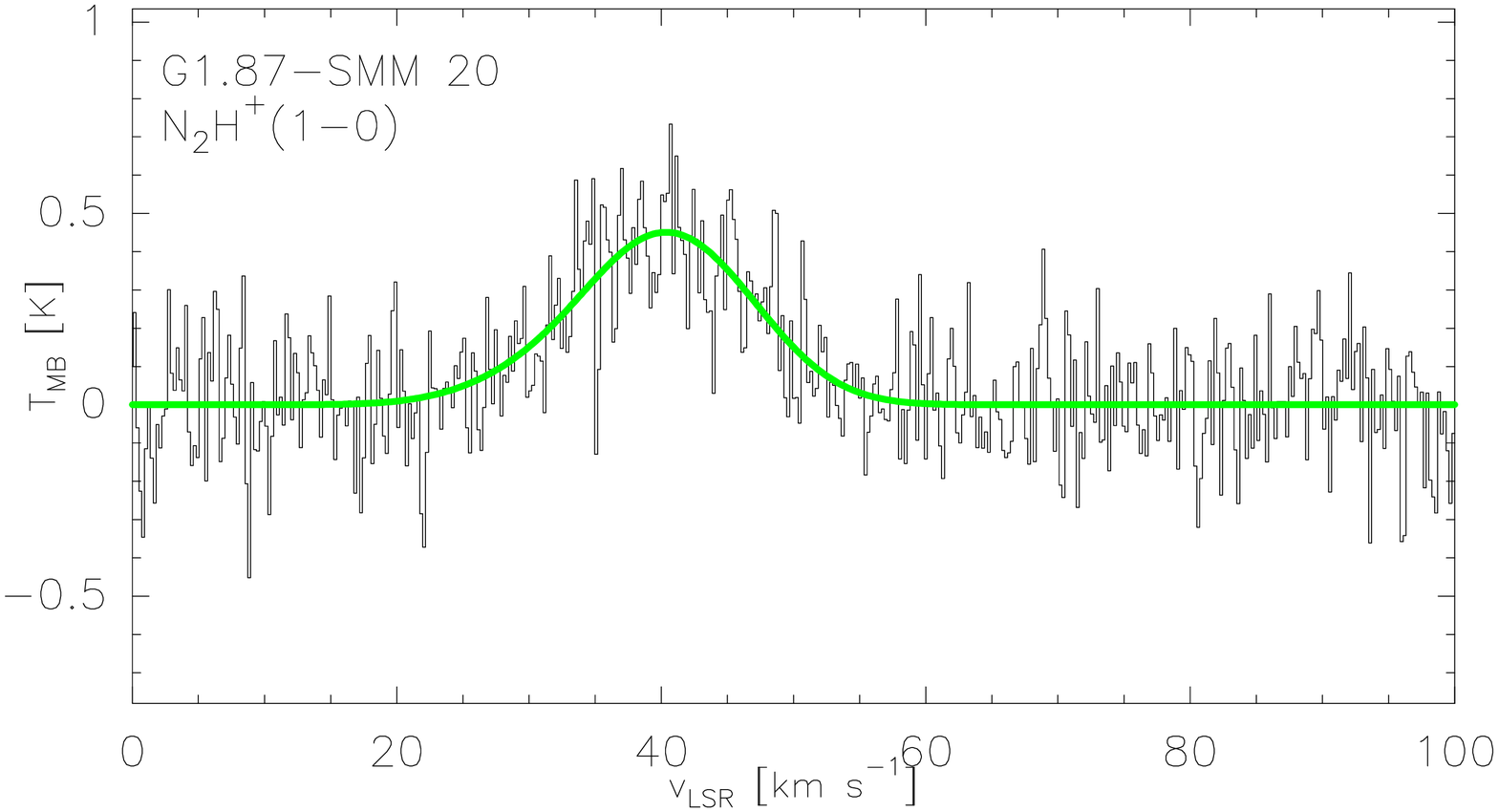}
\caption{Same as Fig.~\ref{figure:G187SMM1_spectra} but towards G1.87--SMM 20.}
\label{figure:G187SMM20_spectra}
\end{center}
\end{figure*}

\begin{figure*}
\begin{center}
\includegraphics[width=0.245\textwidth]{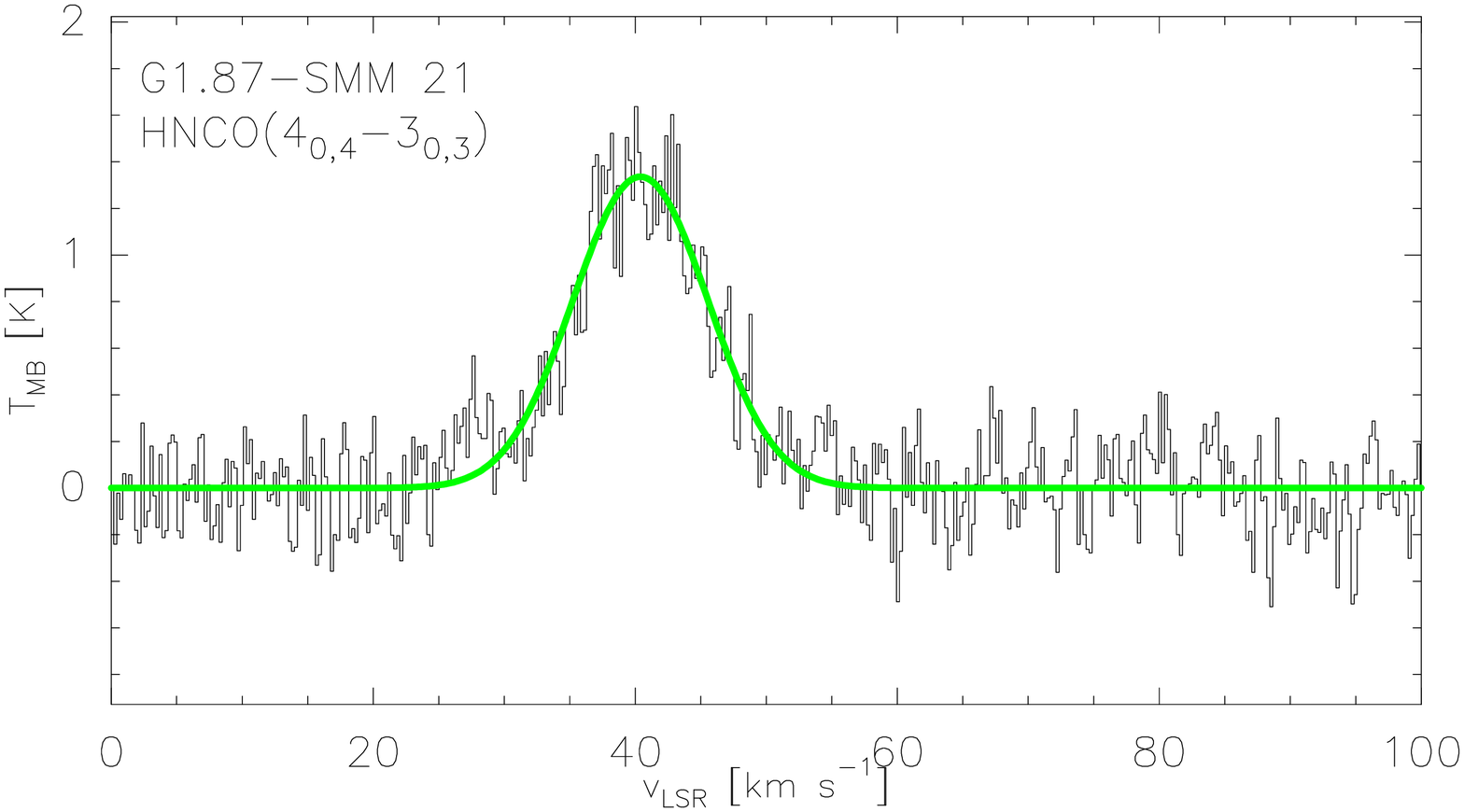}
\includegraphics[width=0.245\textwidth]{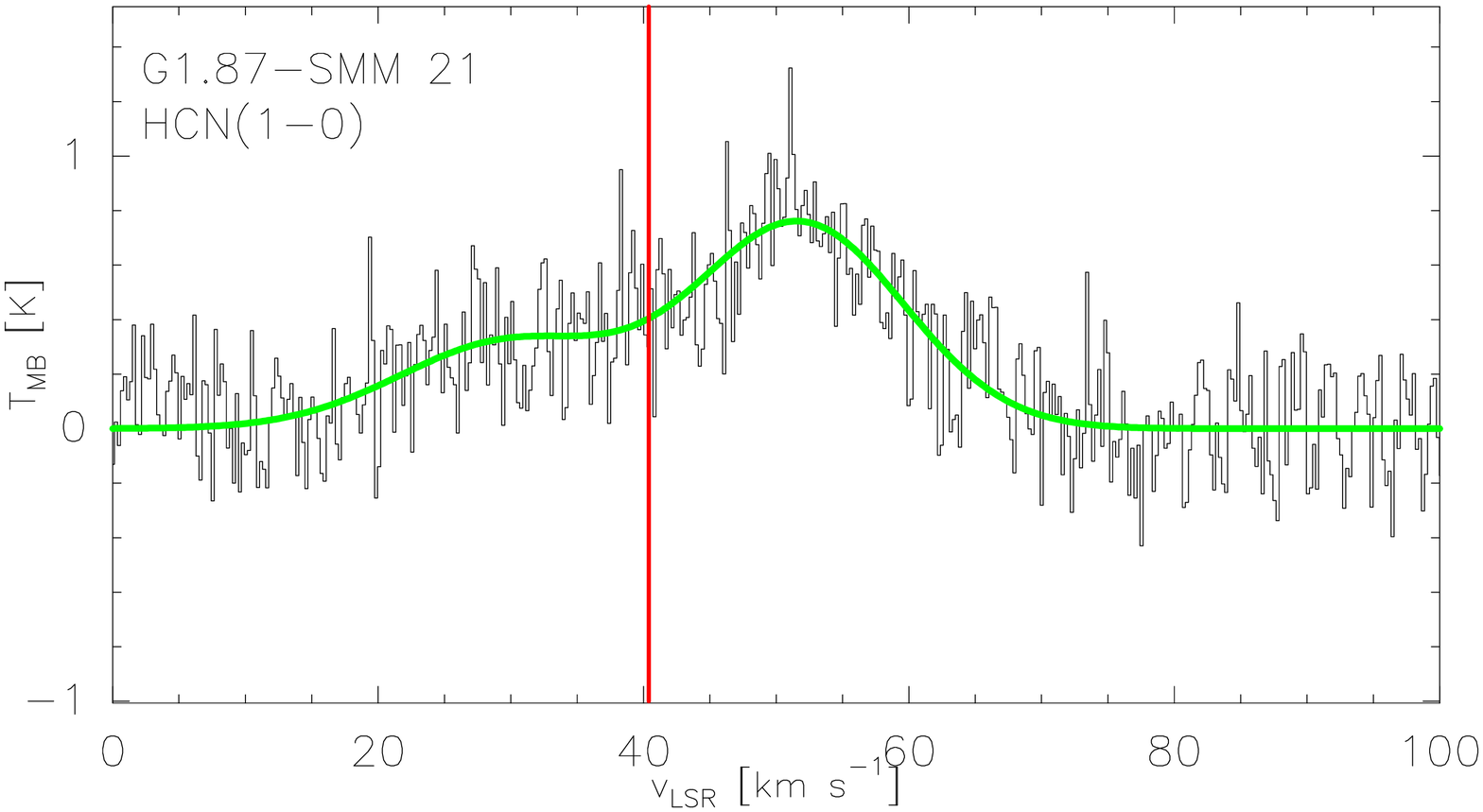}
\includegraphics[width=0.245\textwidth]{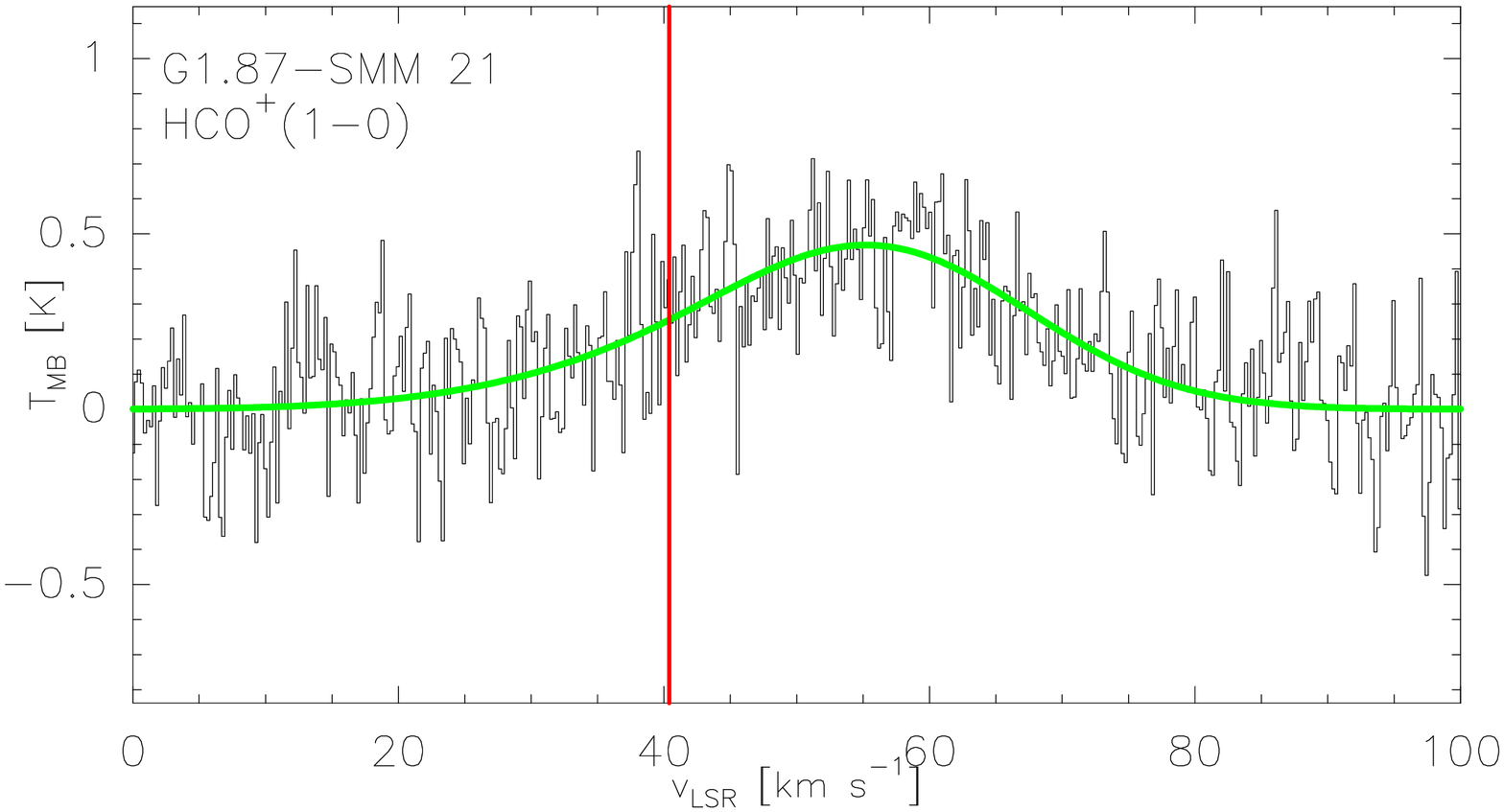}
\includegraphics[width=0.245\textwidth]{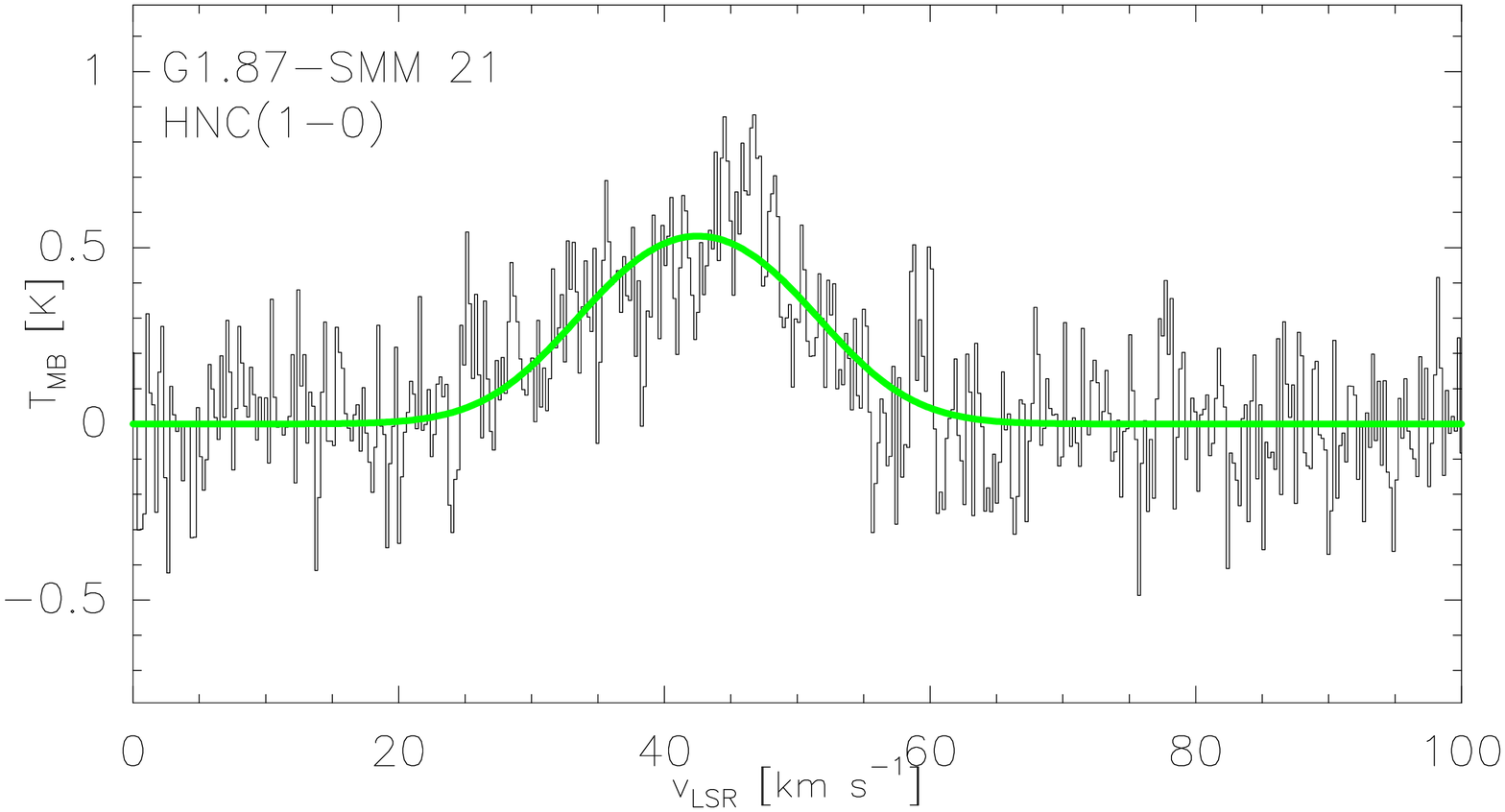}
\includegraphics[width=0.245\textwidth]{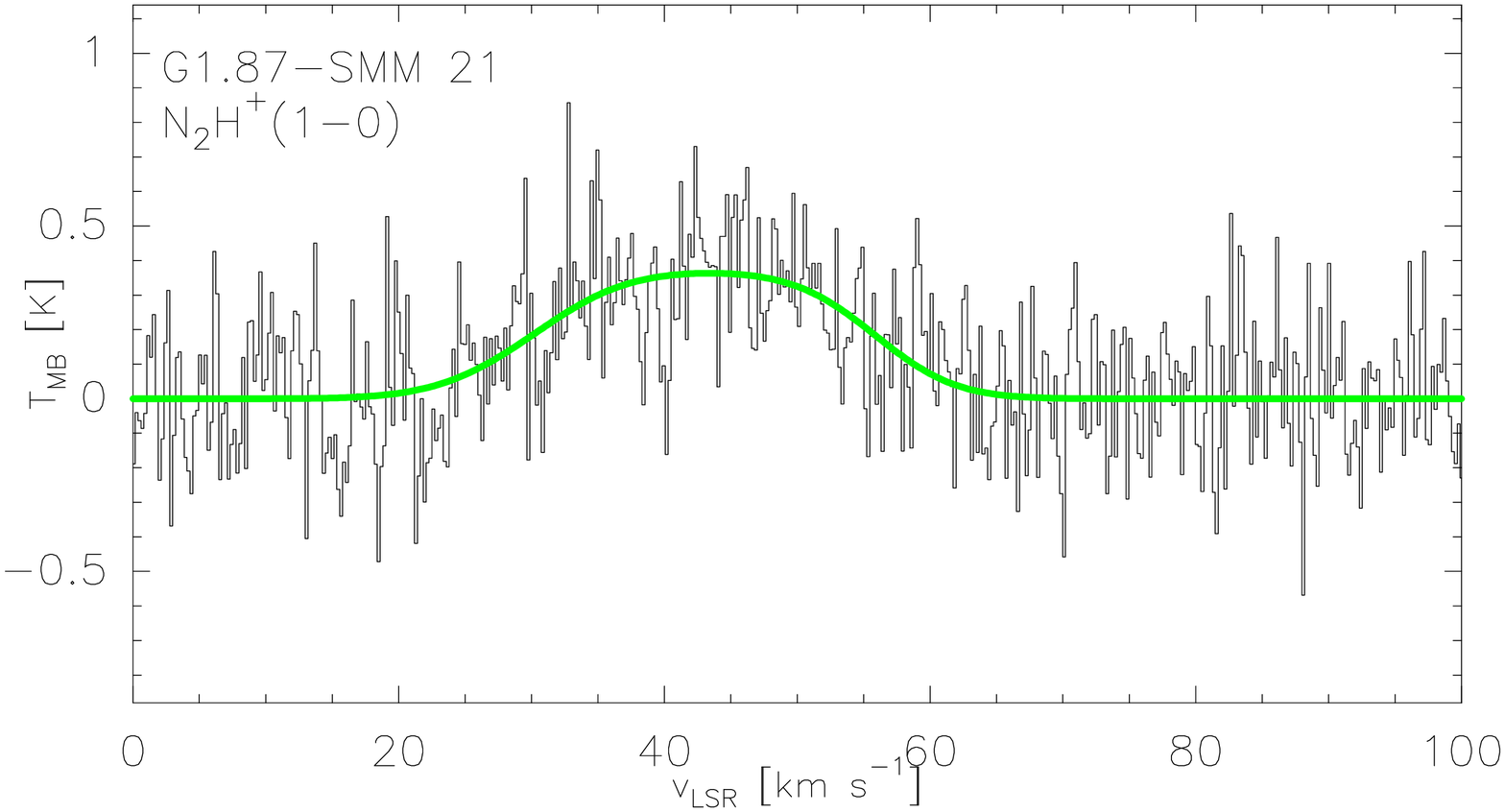}
\caption{Same as Fig.~\ref{figure:G187SMM1_spectra} but towards G1.87--SMM 21. 
The red vertical line indicates the radial velocity of the HNCO line.}
\label{figure:G187SMM21_spectra}
\end{center}
\end{figure*}

\begin{figure*}
\begin{center}
\includegraphics[width=0.245\textwidth]{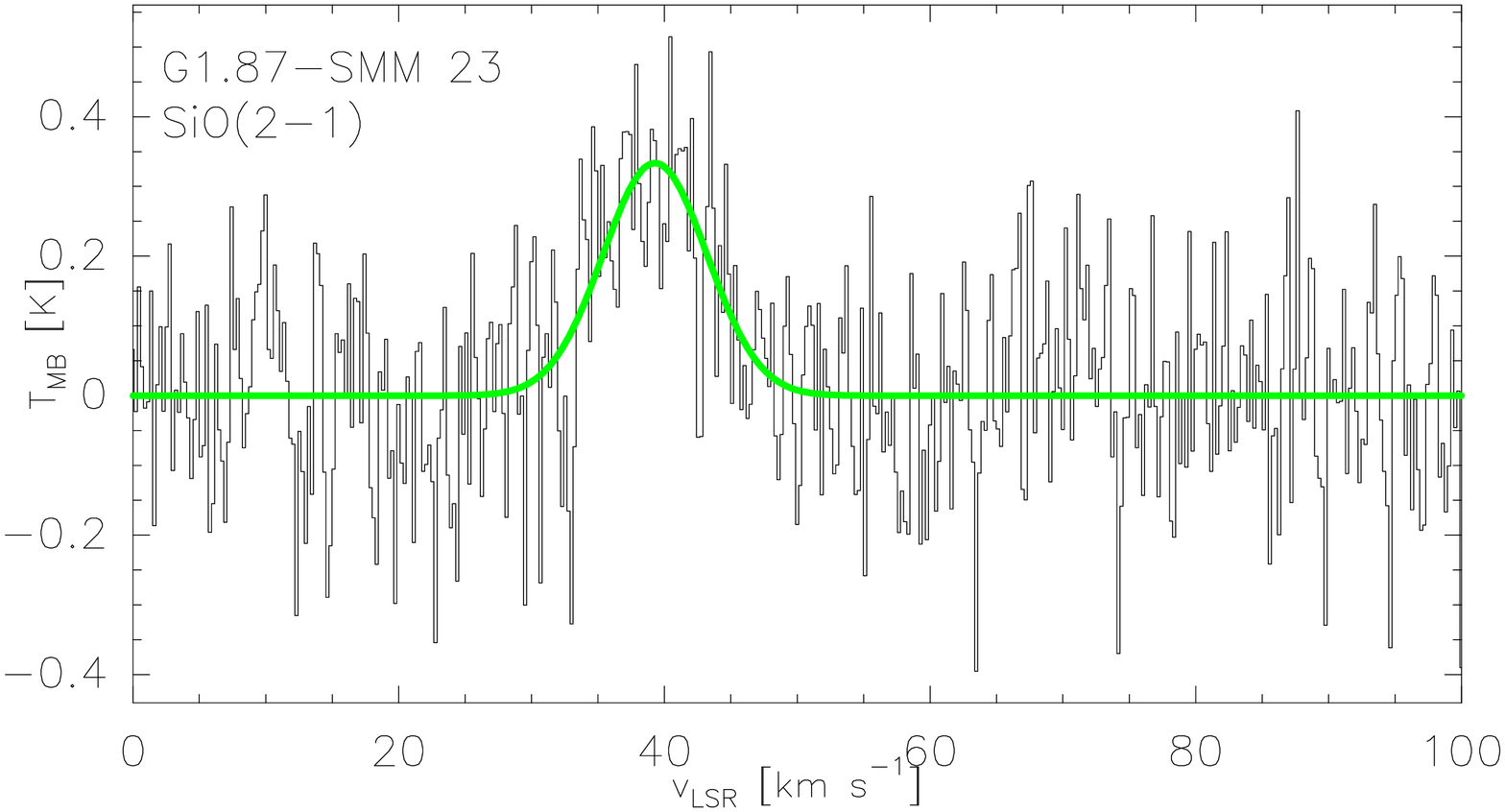}
\includegraphics[width=0.245\textwidth]{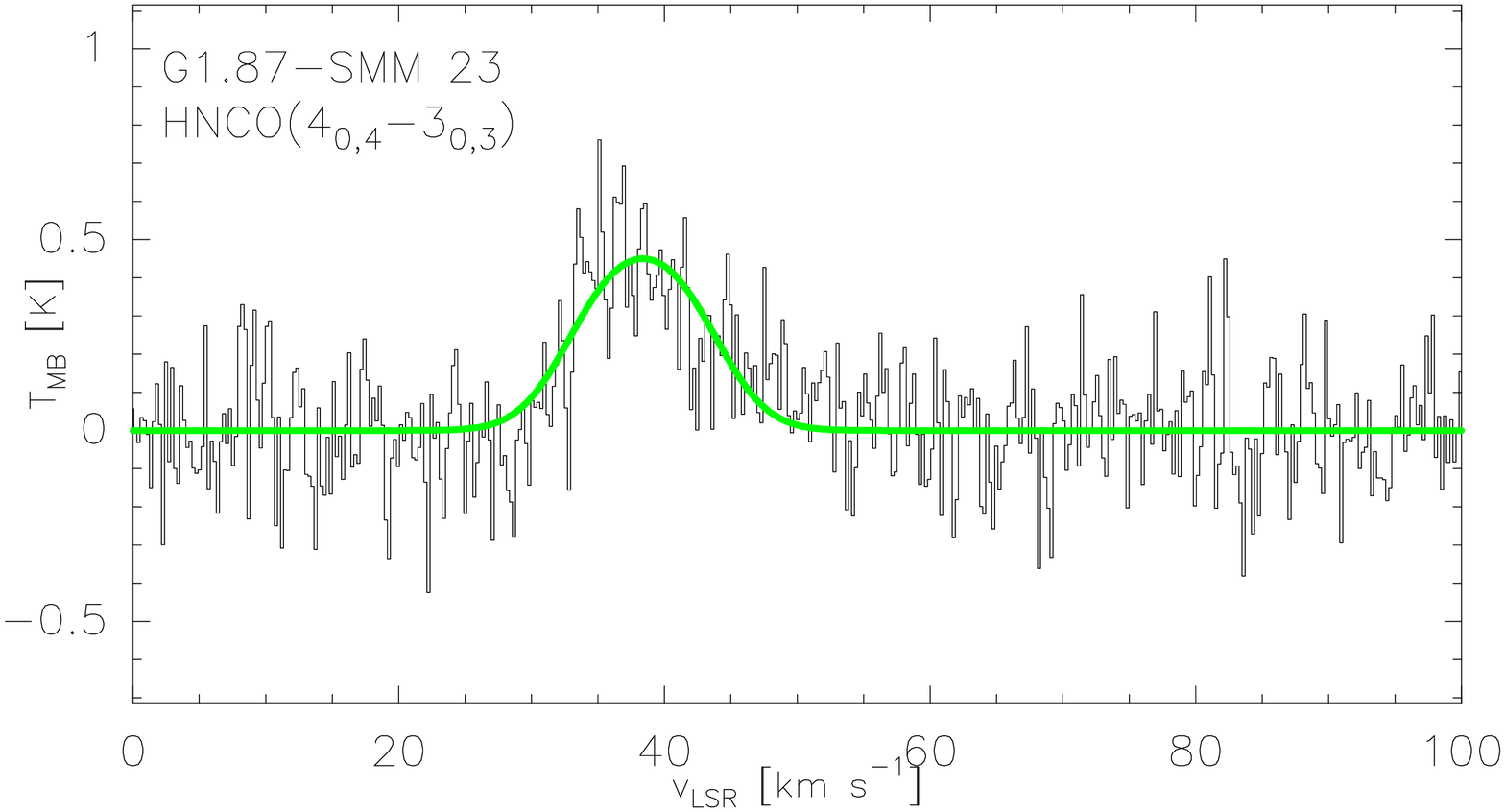}
\includegraphics[width=0.245\textwidth]{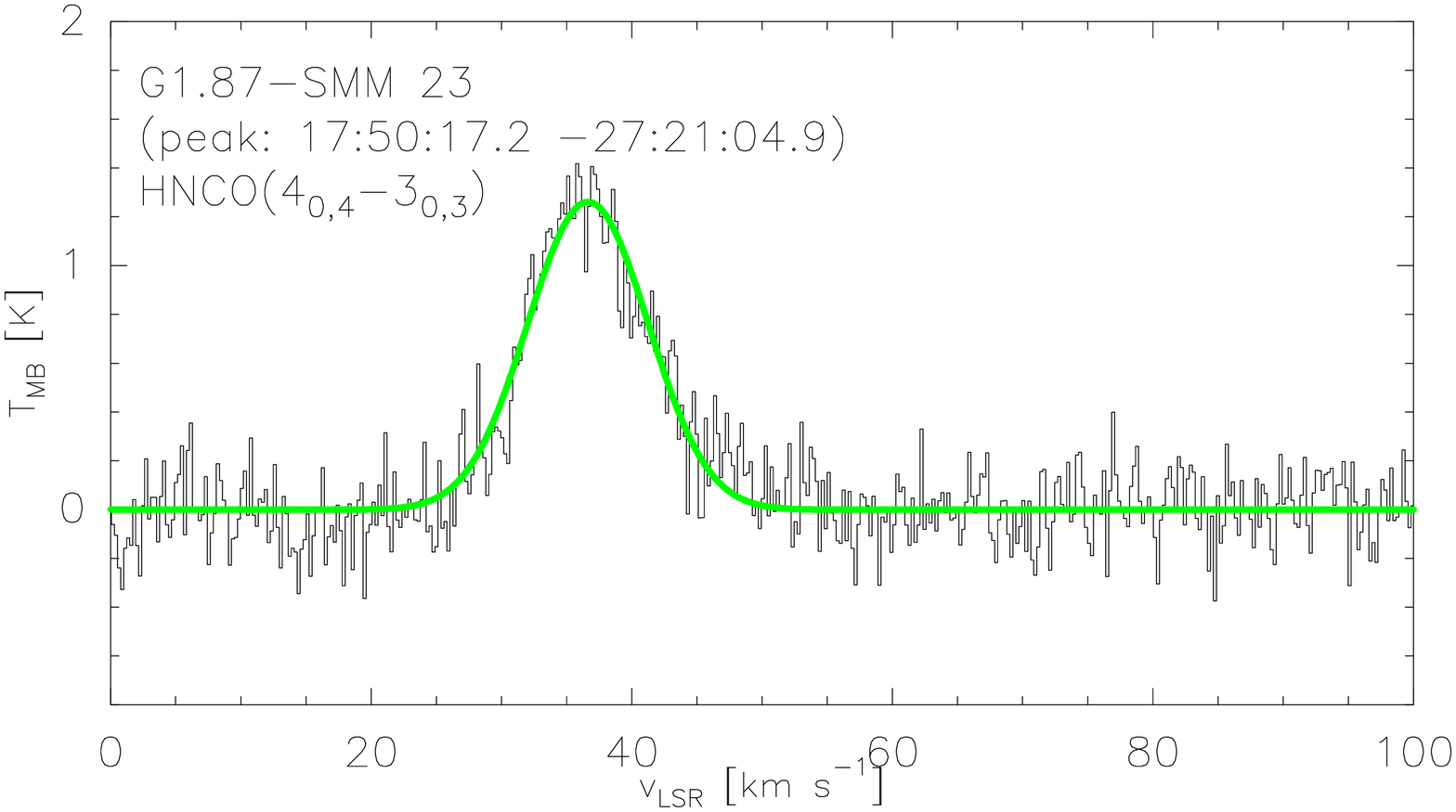}
\includegraphics[width=0.245\textwidth]{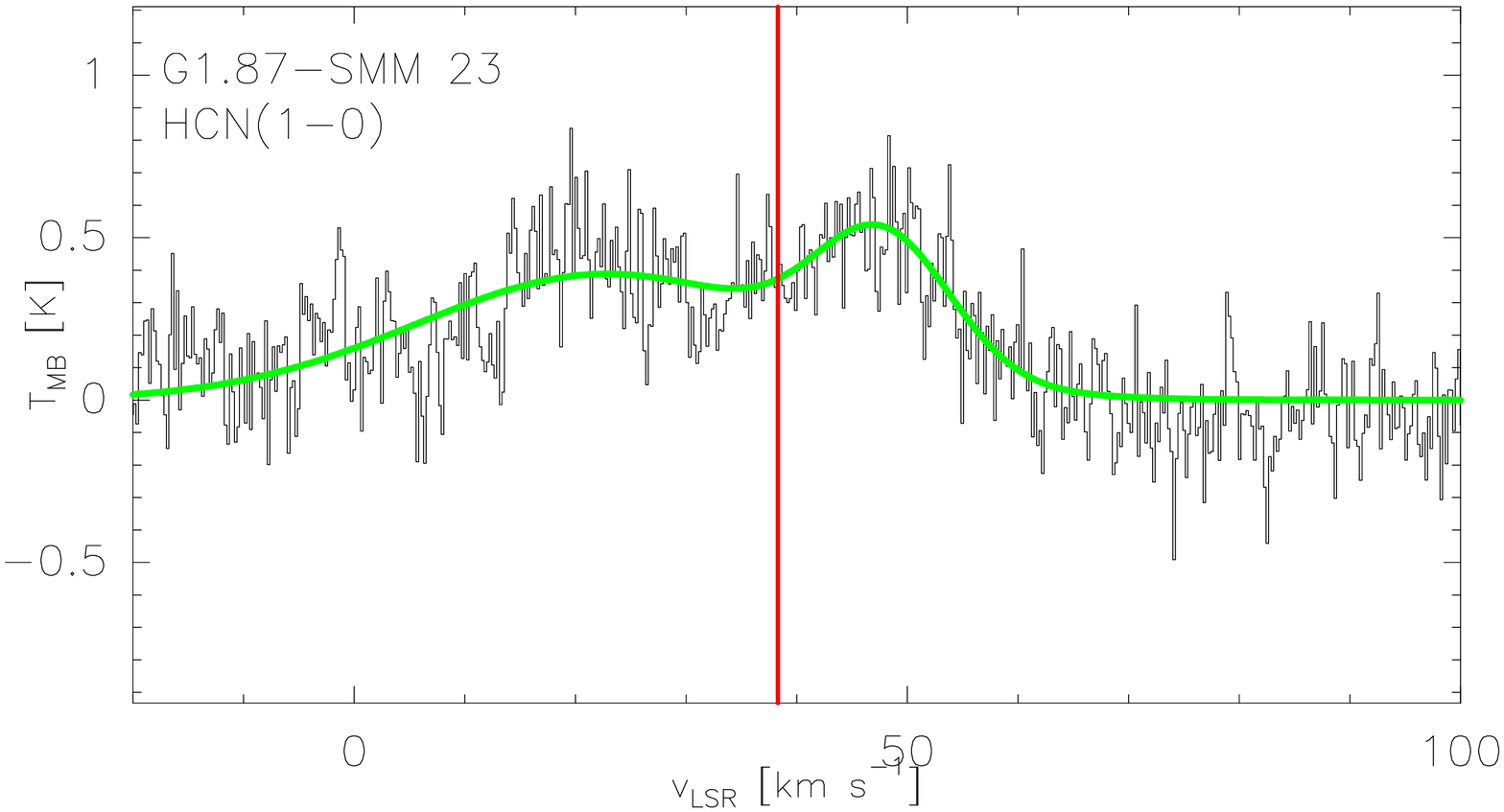}
\includegraphics[width=0.245\textwidth]{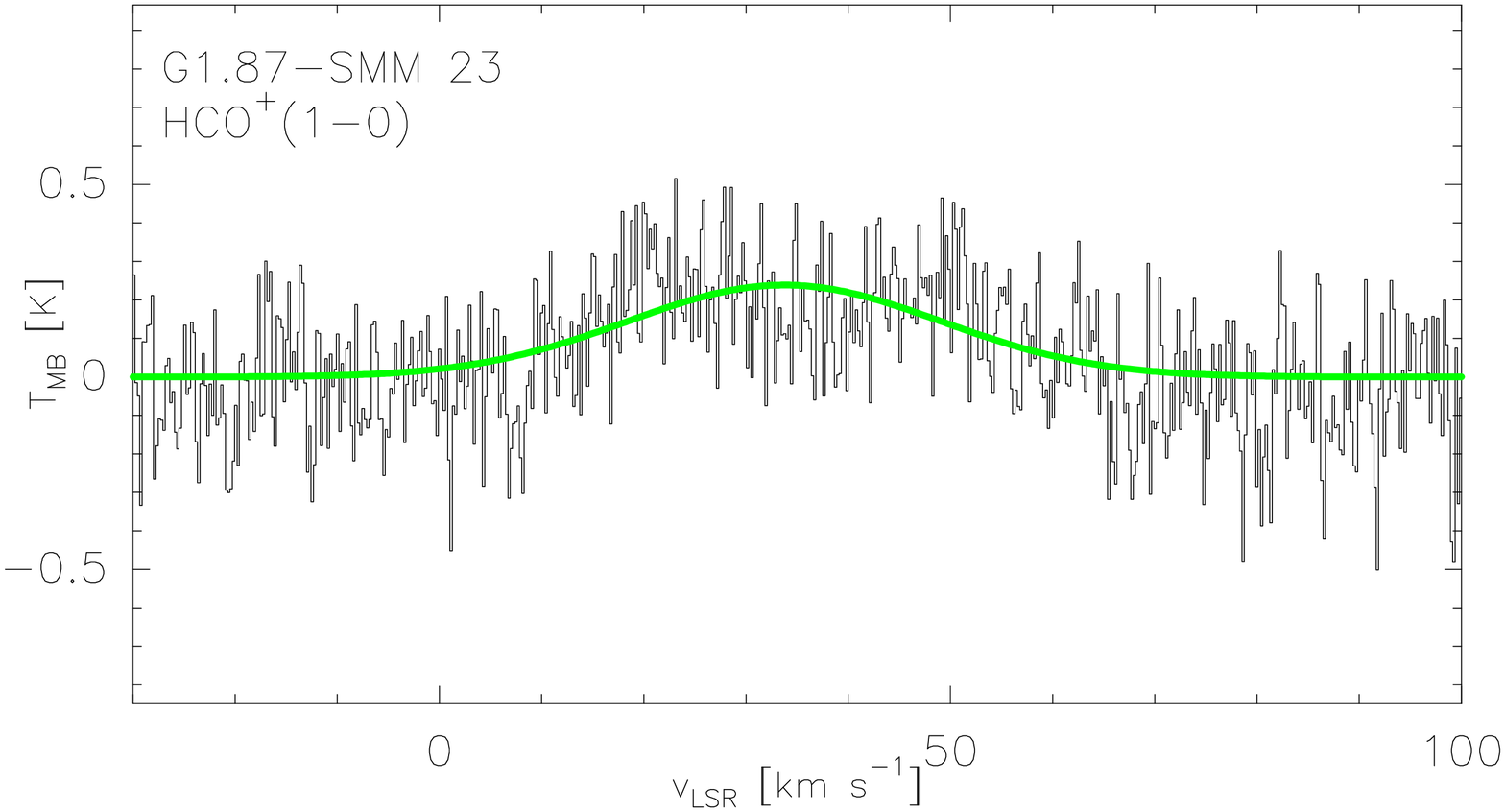}
\includegraphics[width=0.245\textwidth]{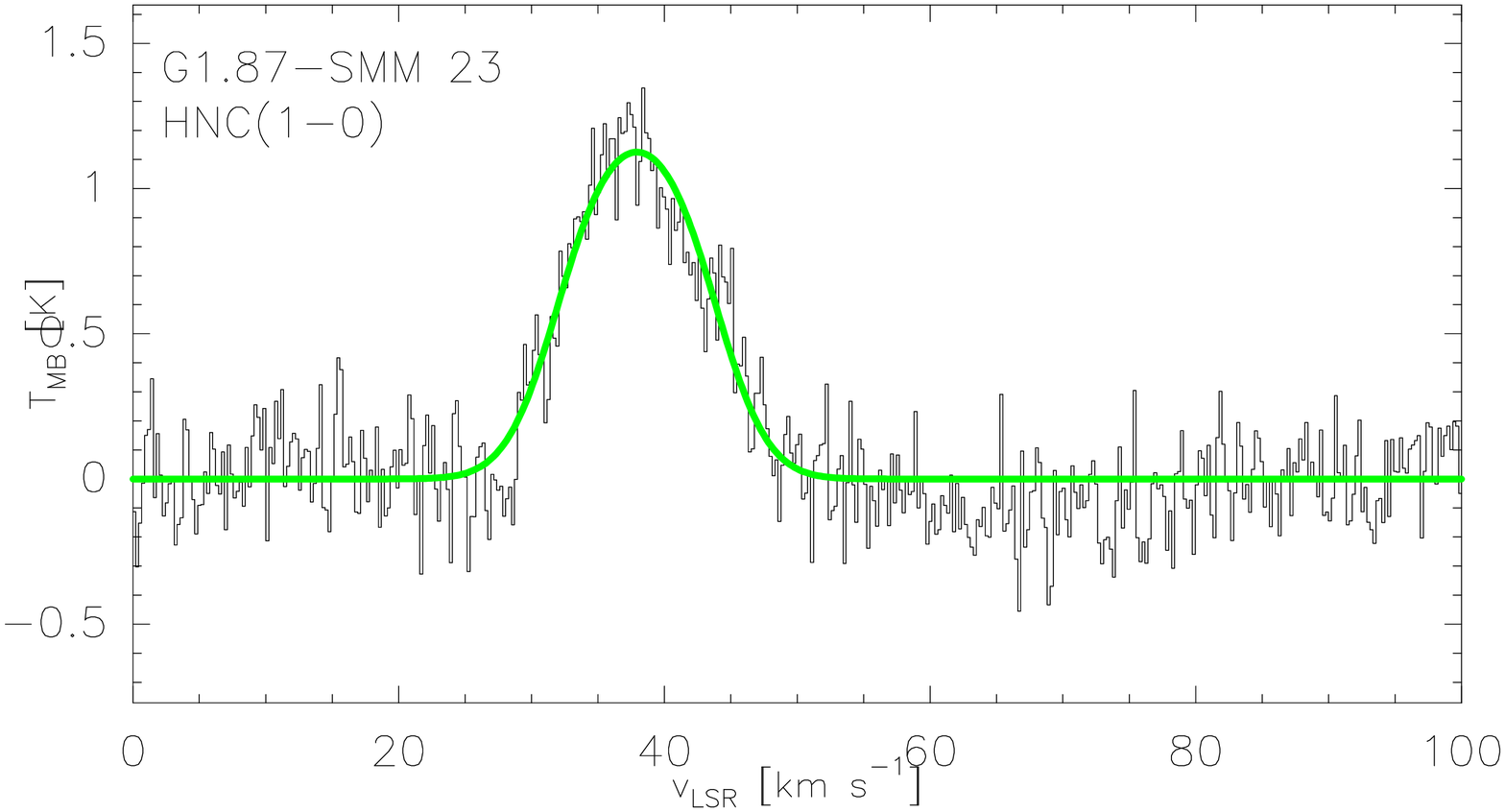}
\includegraphics[width=0.245\textwidth]{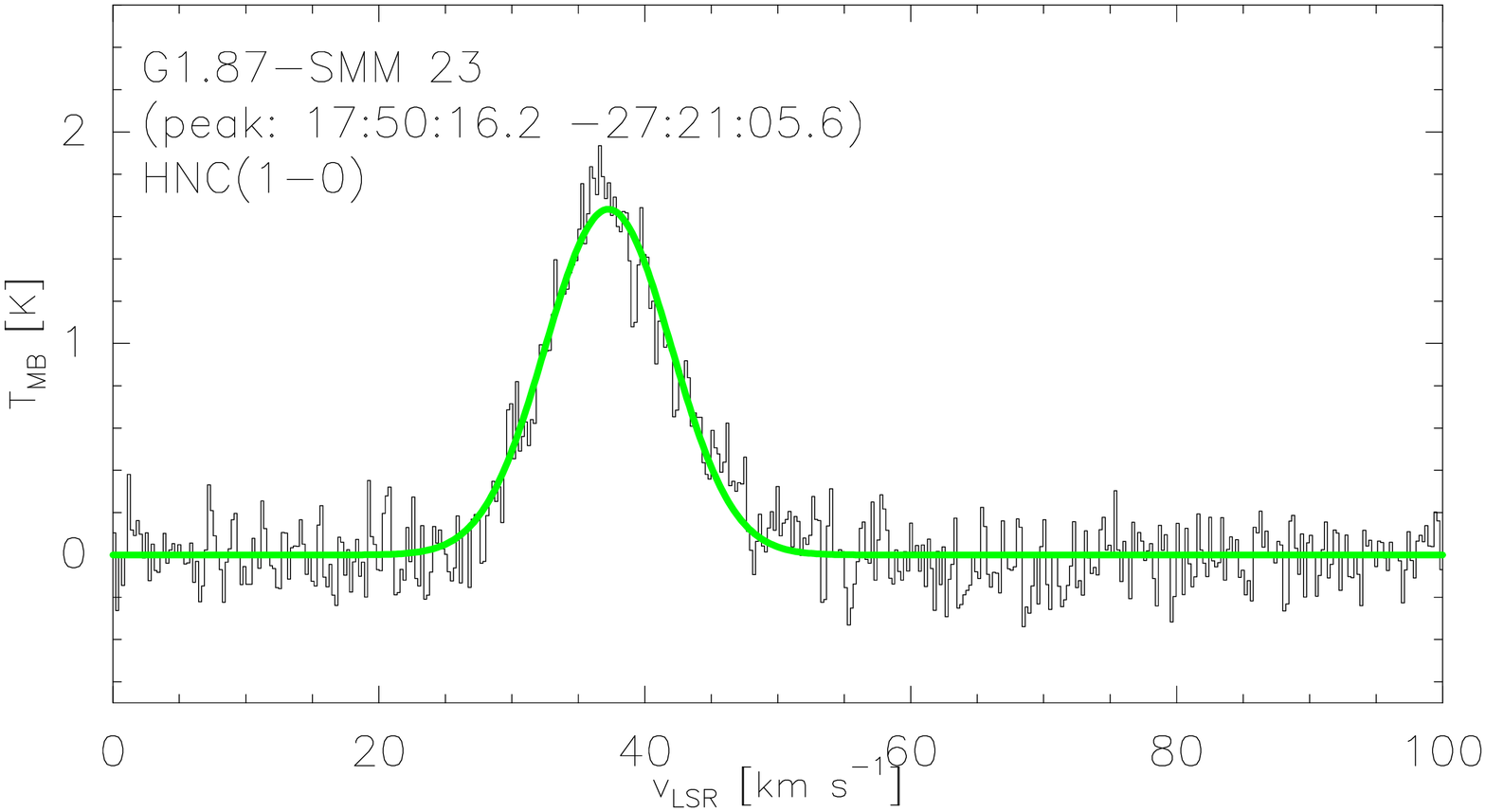}
\includegraphics[width=0.245\textwidth]{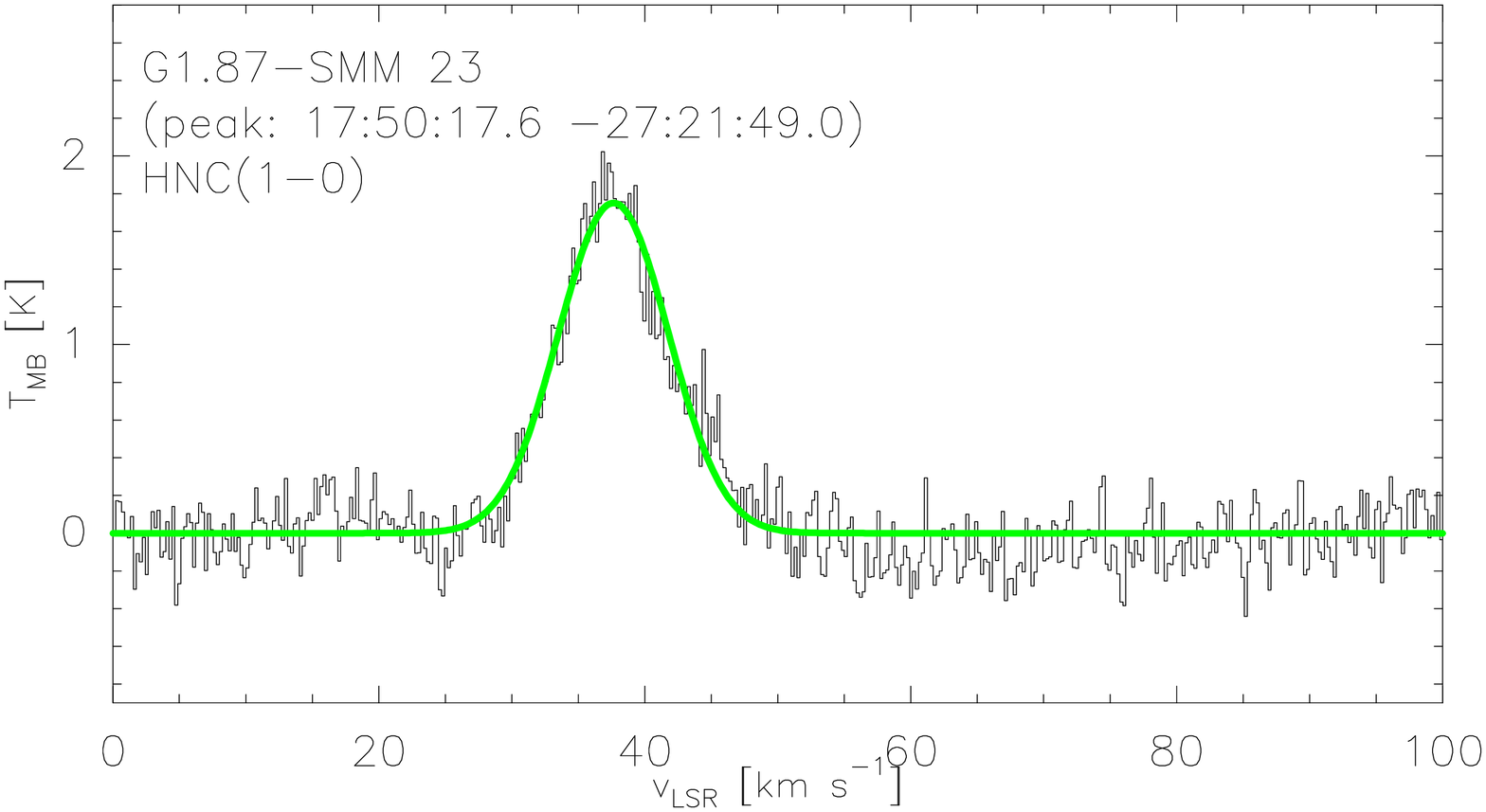}
\includegraphics[width=0.245\textwidth]{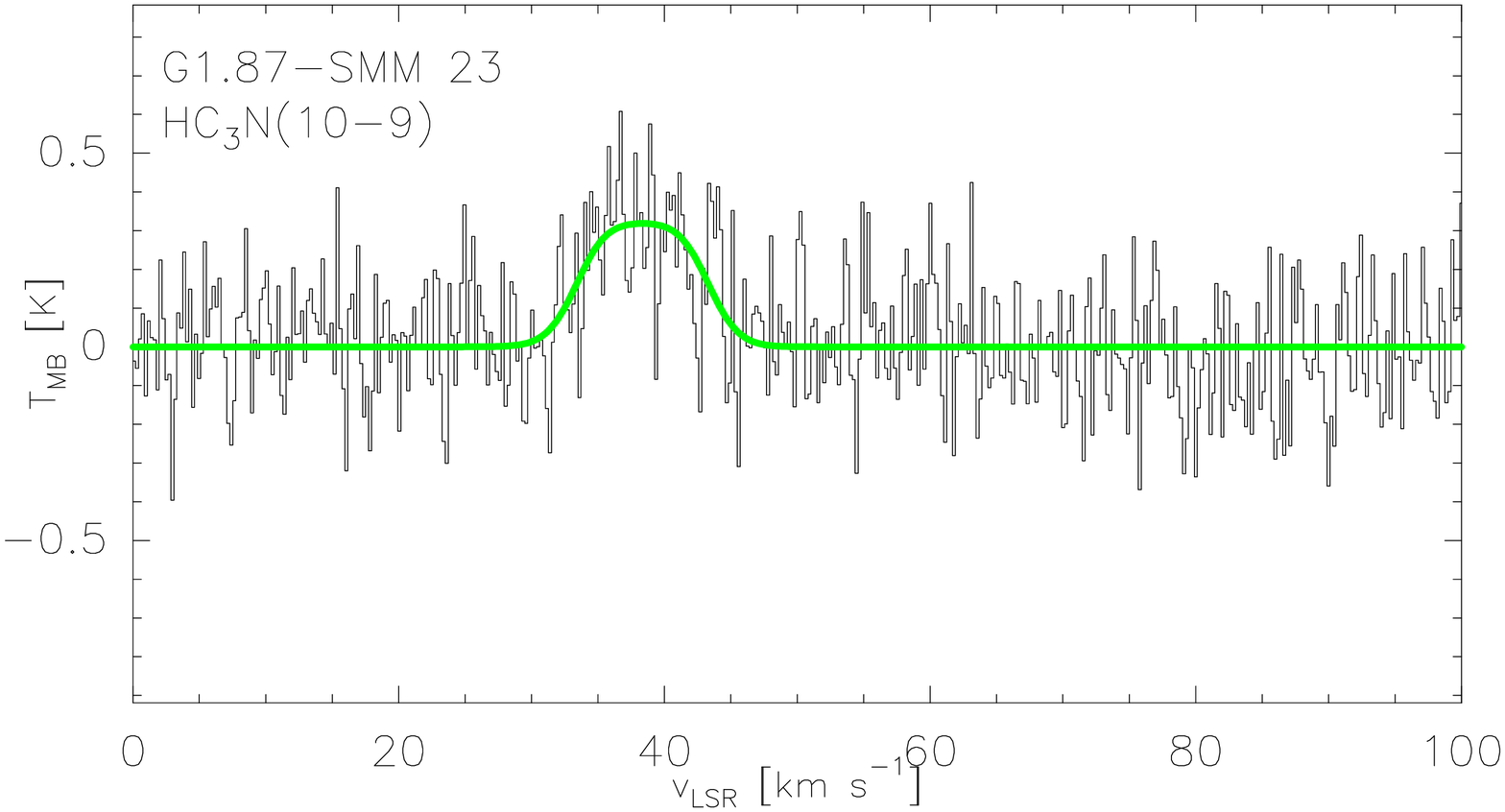}
\includegraphics[width=0.245\textwidth]{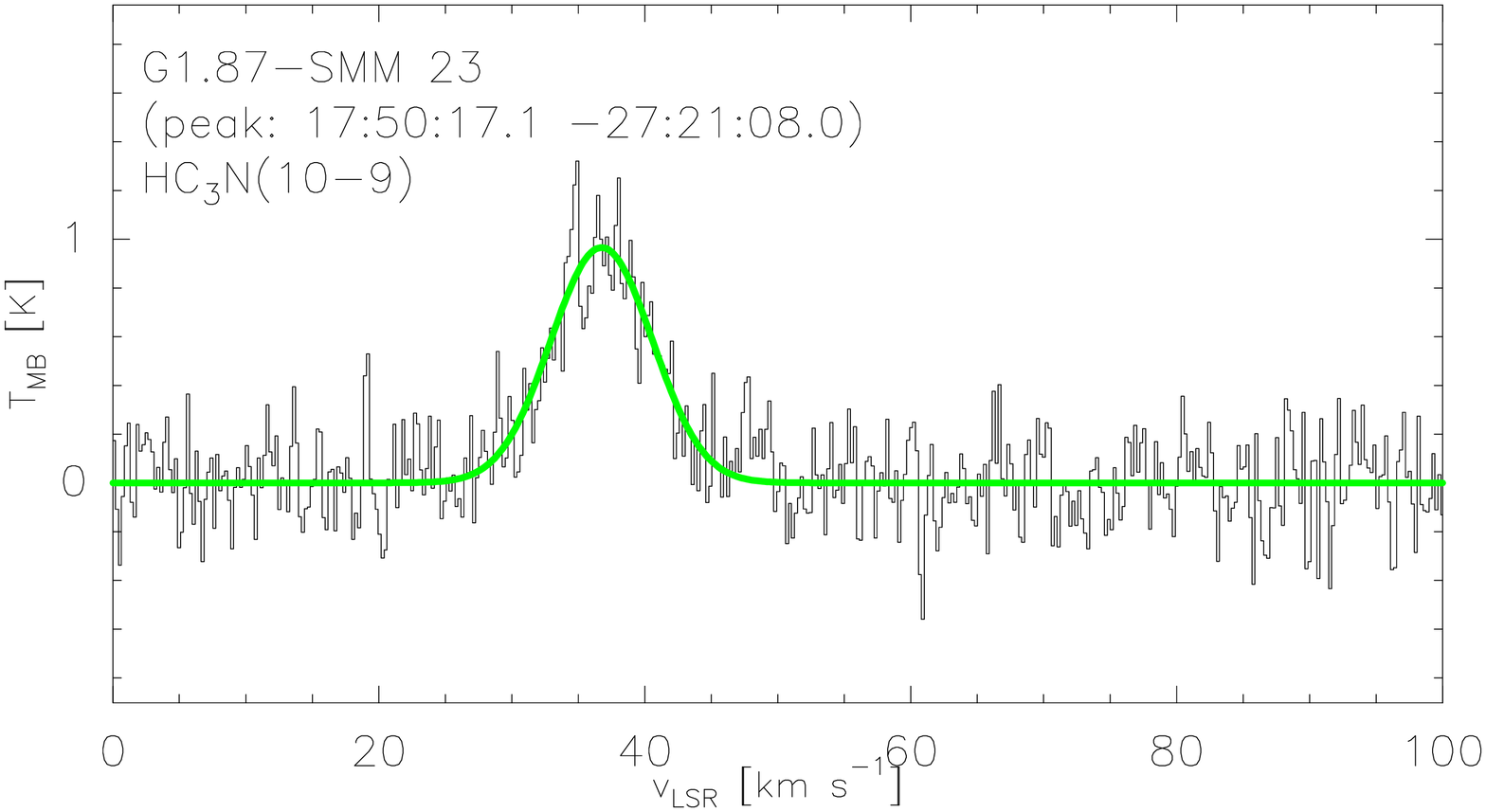}
\includegraphics[width=0.245\textwidth]{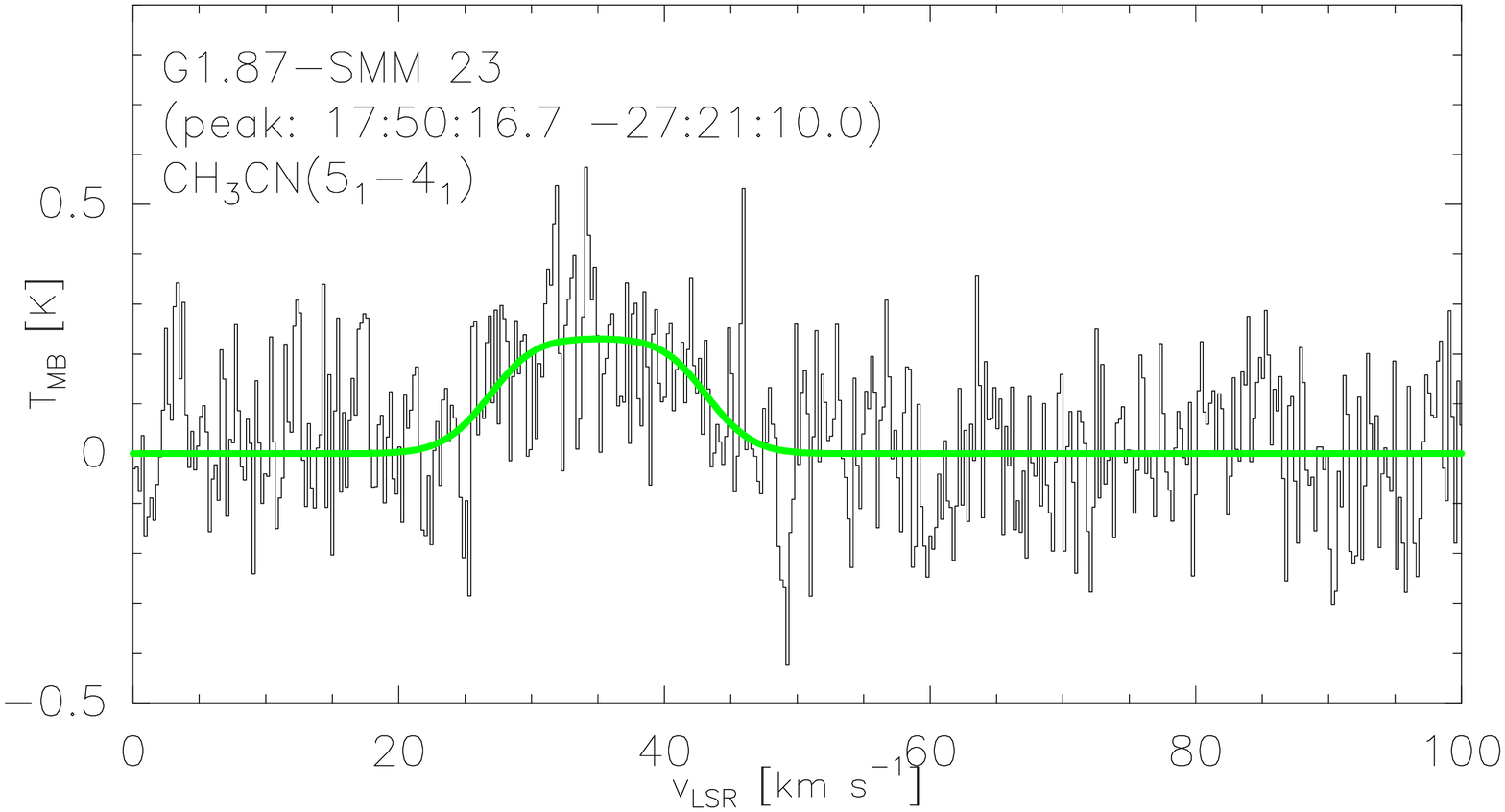}
\includegraphics[width=0.245\textwidth]{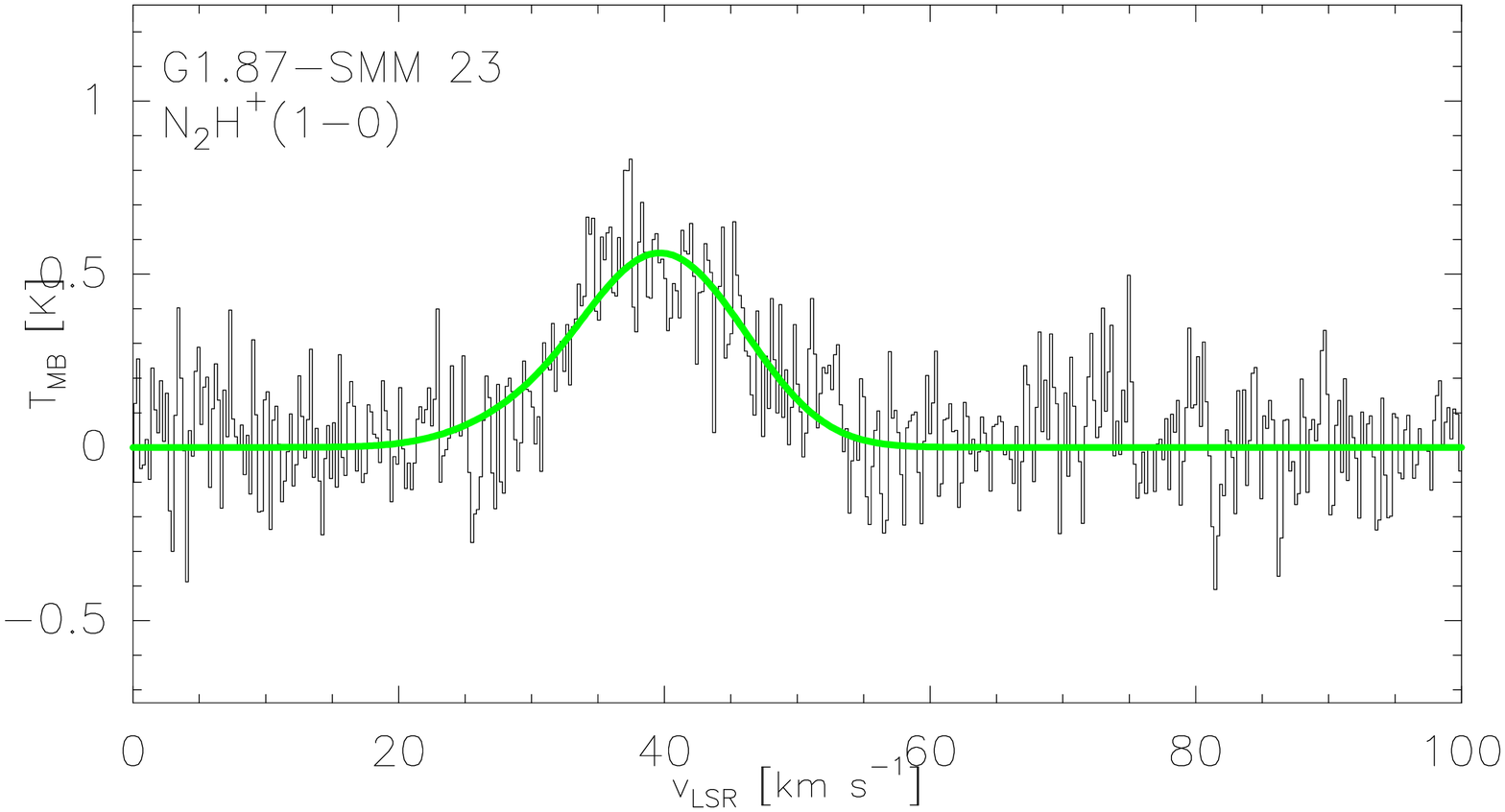}
\includegraphics[width=0.245\textwidth]{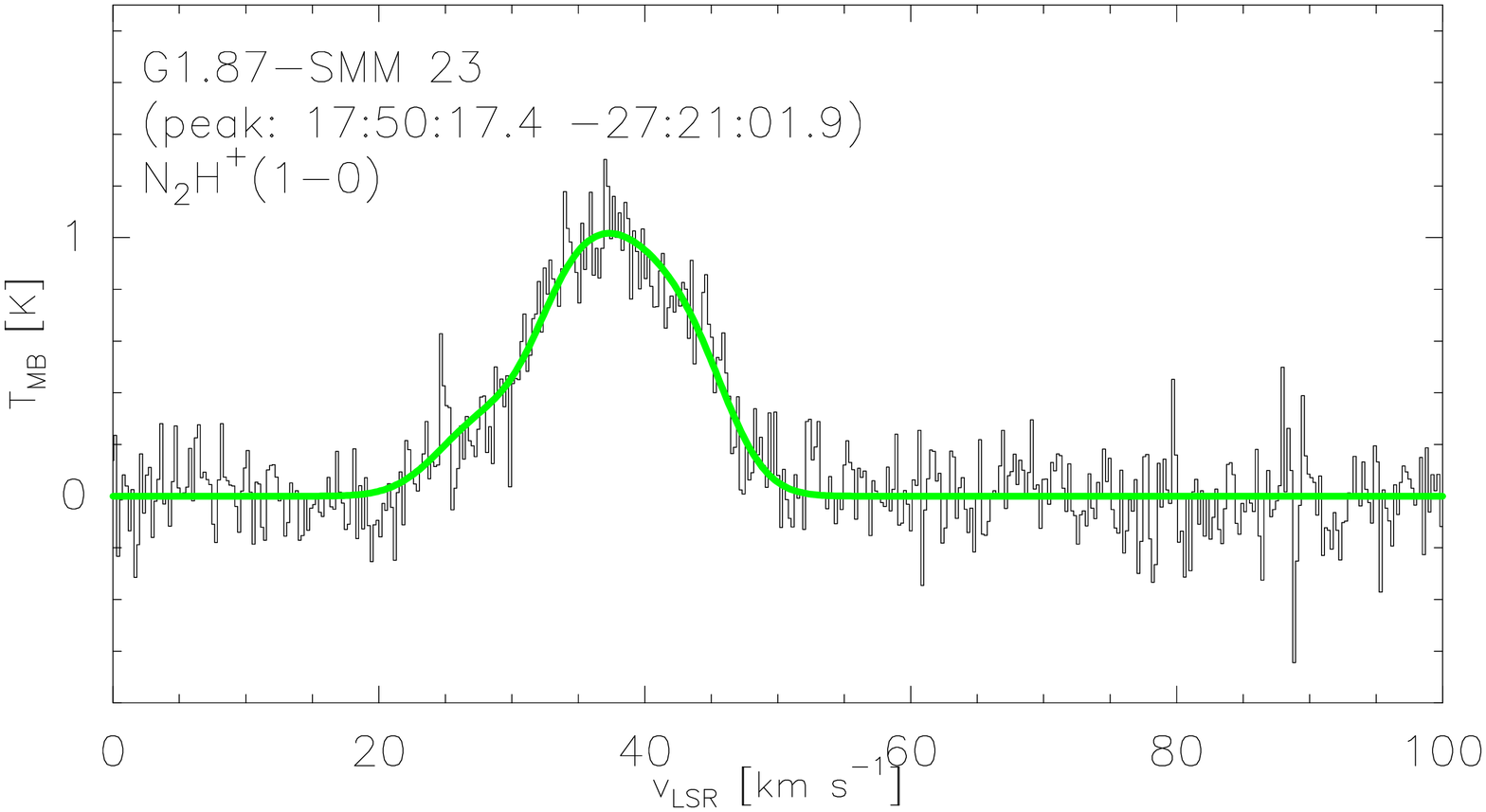}
\caption{Same as Fig.~\ref{figure:G187SMM1_spectra} but towards G1.87--SMM 23. 
Note that the velocity range for the HCN and HCO$^+$ spectra is wider for 
illustrative purposes. The red vertical line indicates the radial velocity of 
the optically thin HC$_3$N line.}
\label{figure:G187SMM23_spectra}
\end{center}
\end{figure*}

\begin{figure*}
\begin{center}
\includegraphics[width=0.245\textwidth]{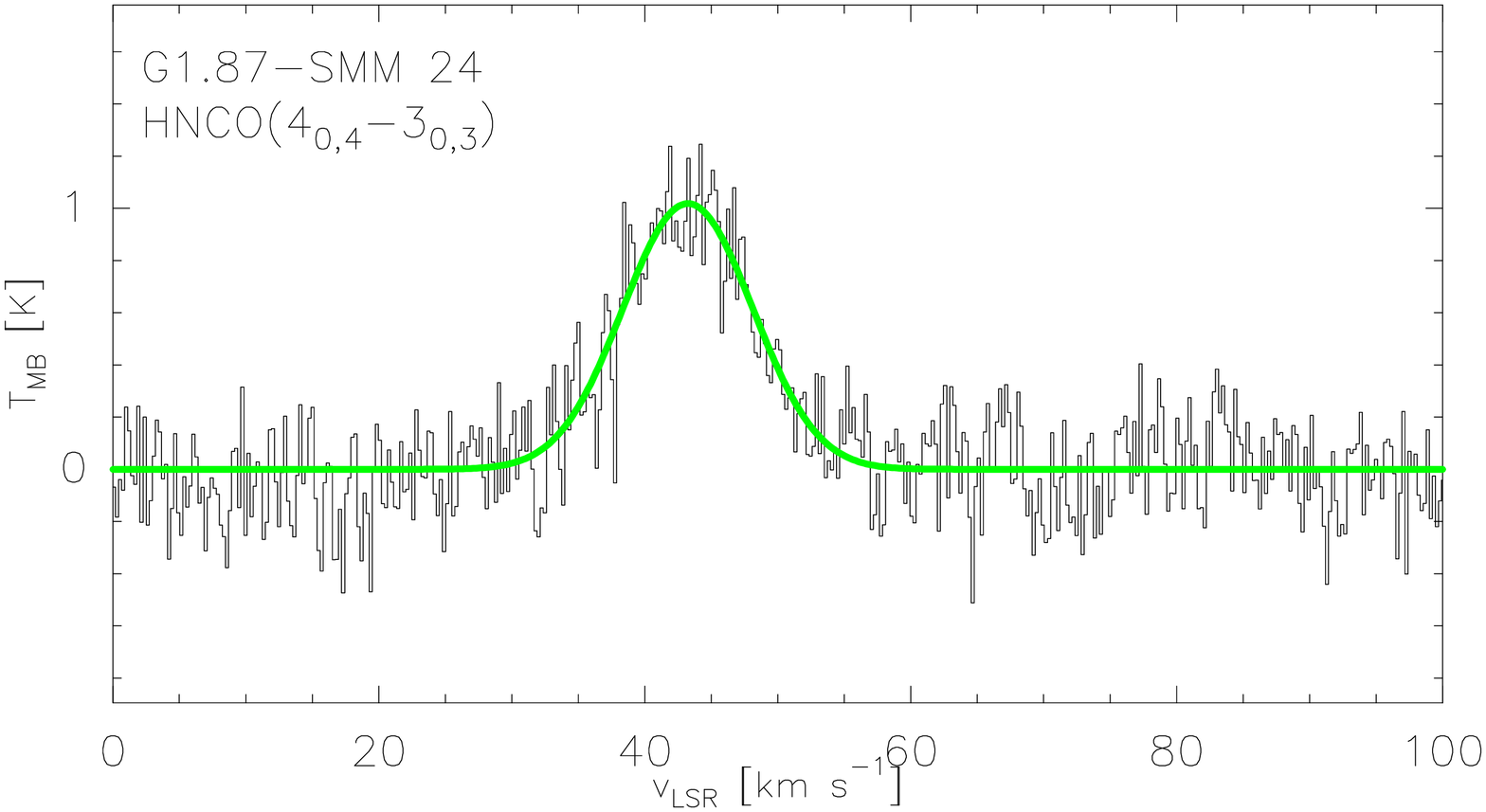}
\includegraphics[width=0.245\textwidth]{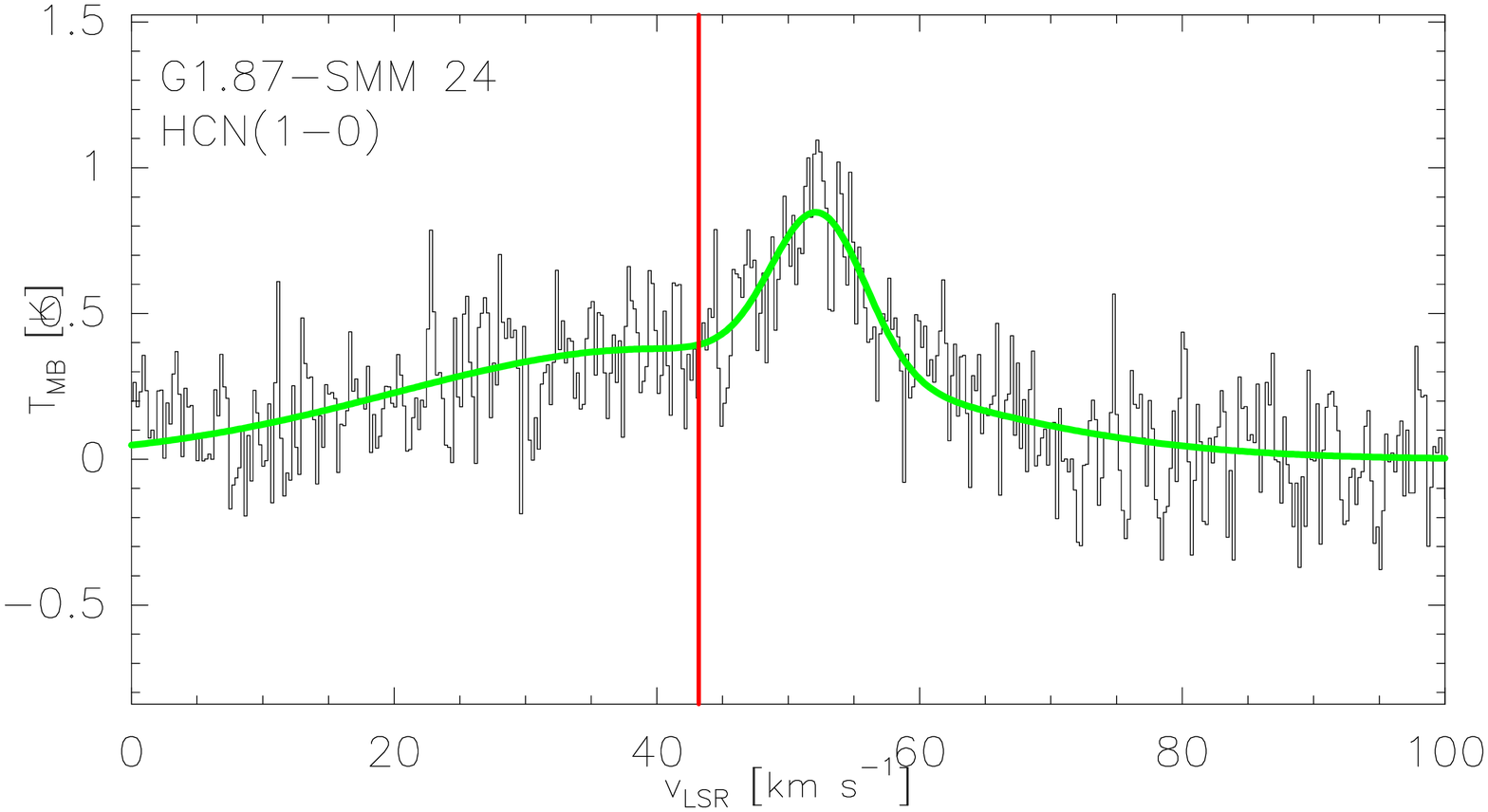}
\includegraphics[width=0.245\textwidth]{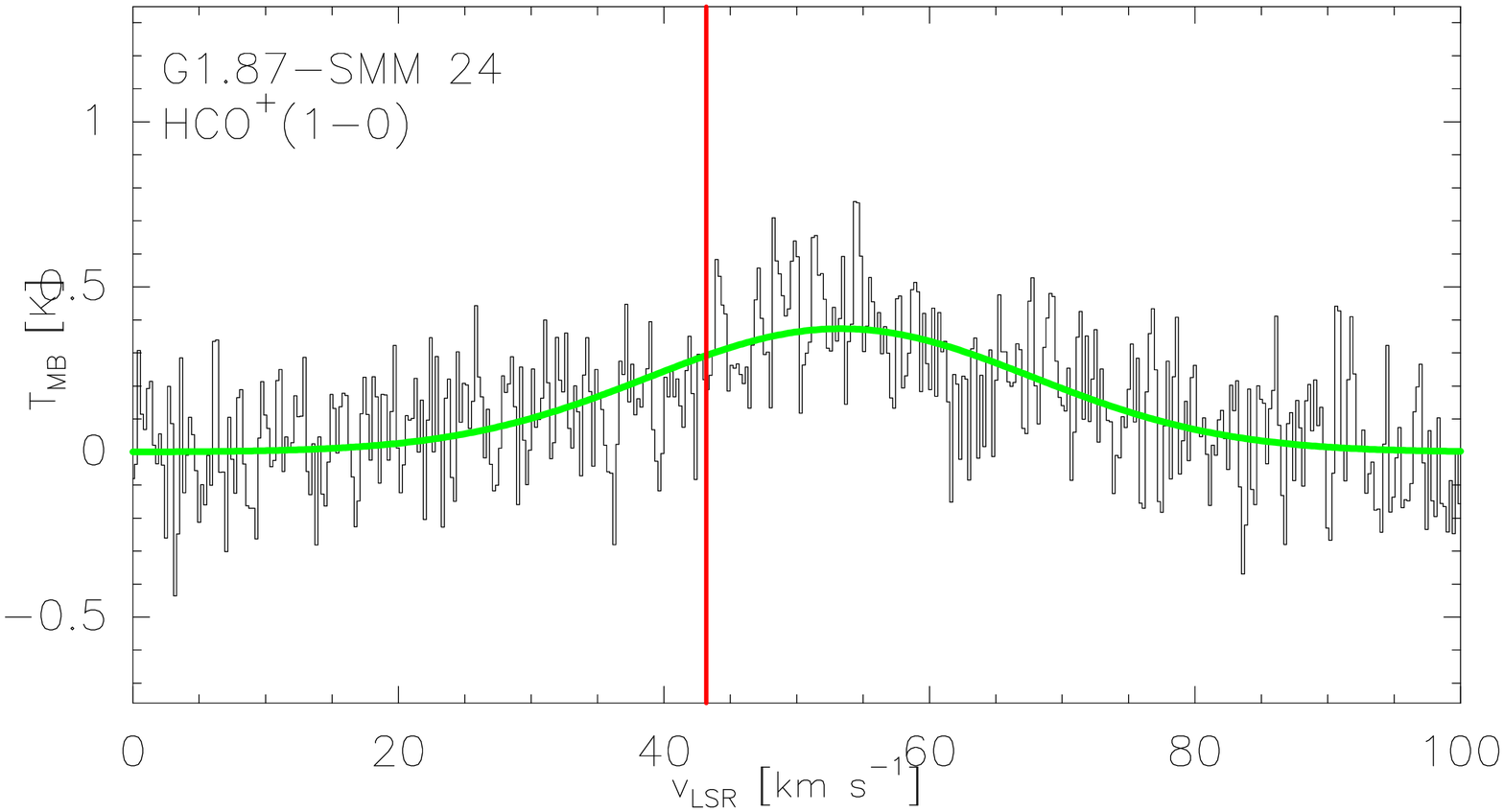}
\includegraphics[width=0.245\textwidth]{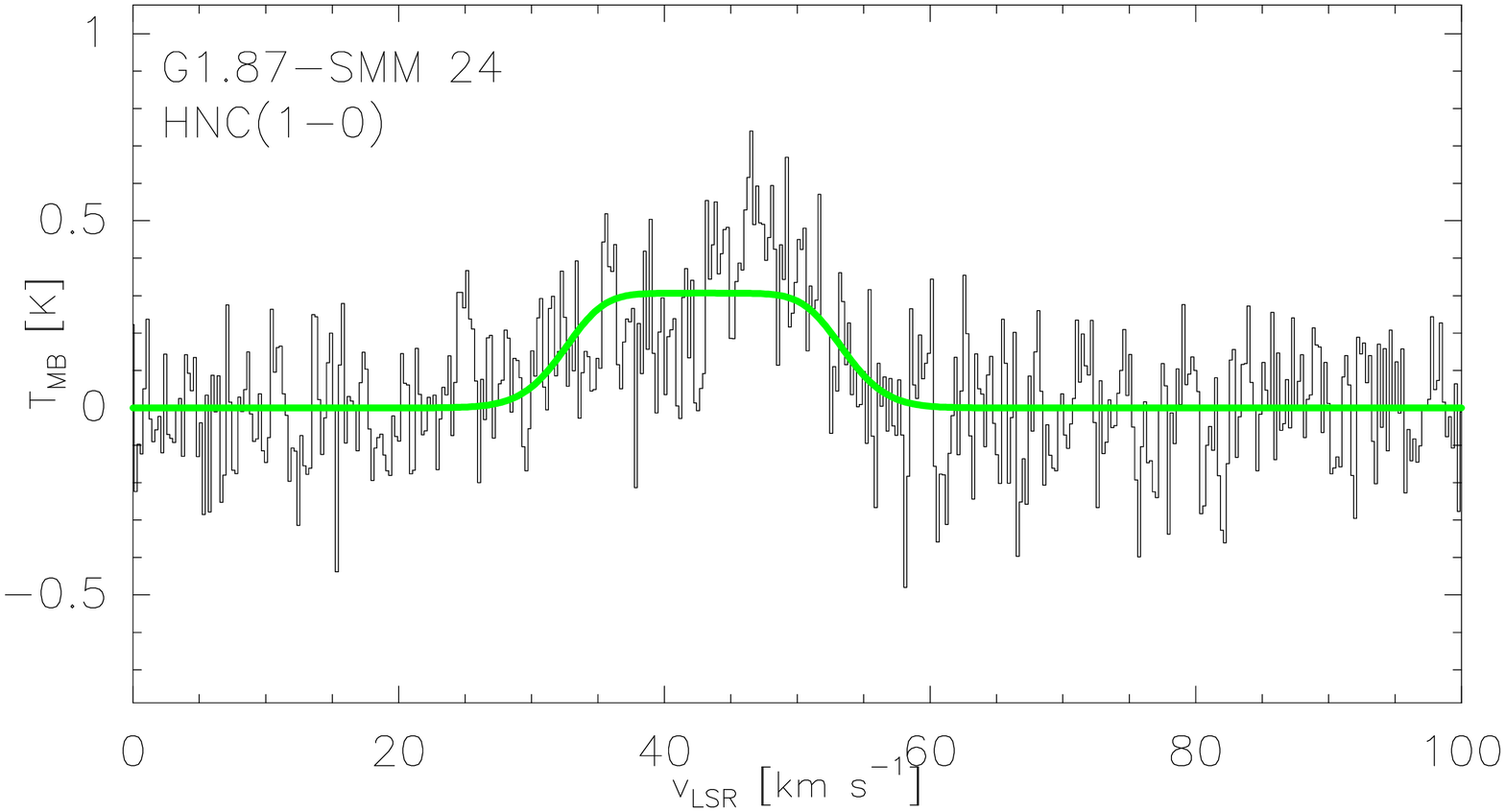}
\includegraphics[width=0.245\textwidth]{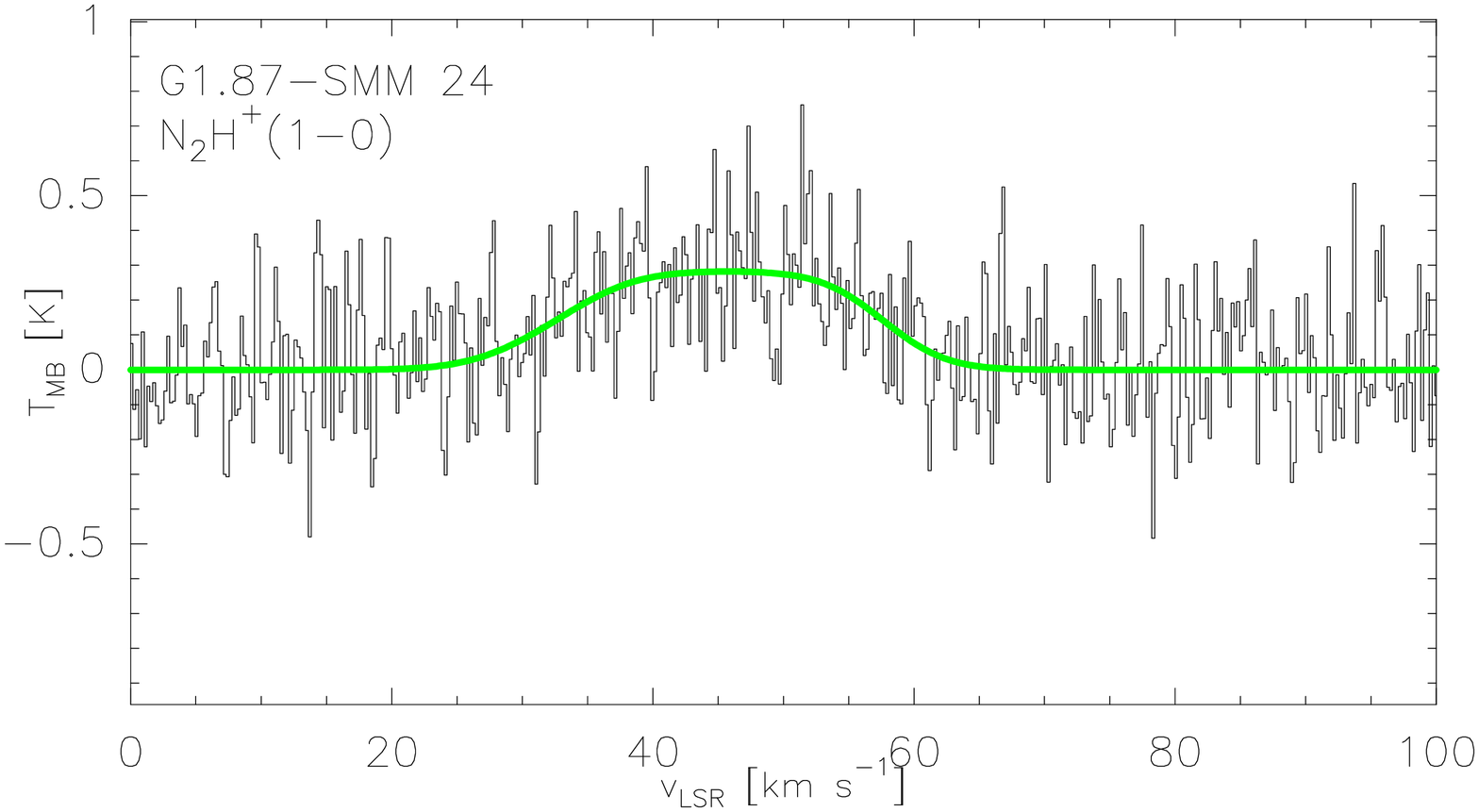}
\caption{Same as Fig.~\ref{figure:G187SMM1_spectra} but towards G1.87--SMM 24. 
The red vertical line indicates the radial velocity of the HNCO line.}
\label{figure:G187SMM24_spectra}
\end{center}
\end{figure*}

\begin{figure*}
\begin{center}
\includegraphics[width=0.245\textwidth]{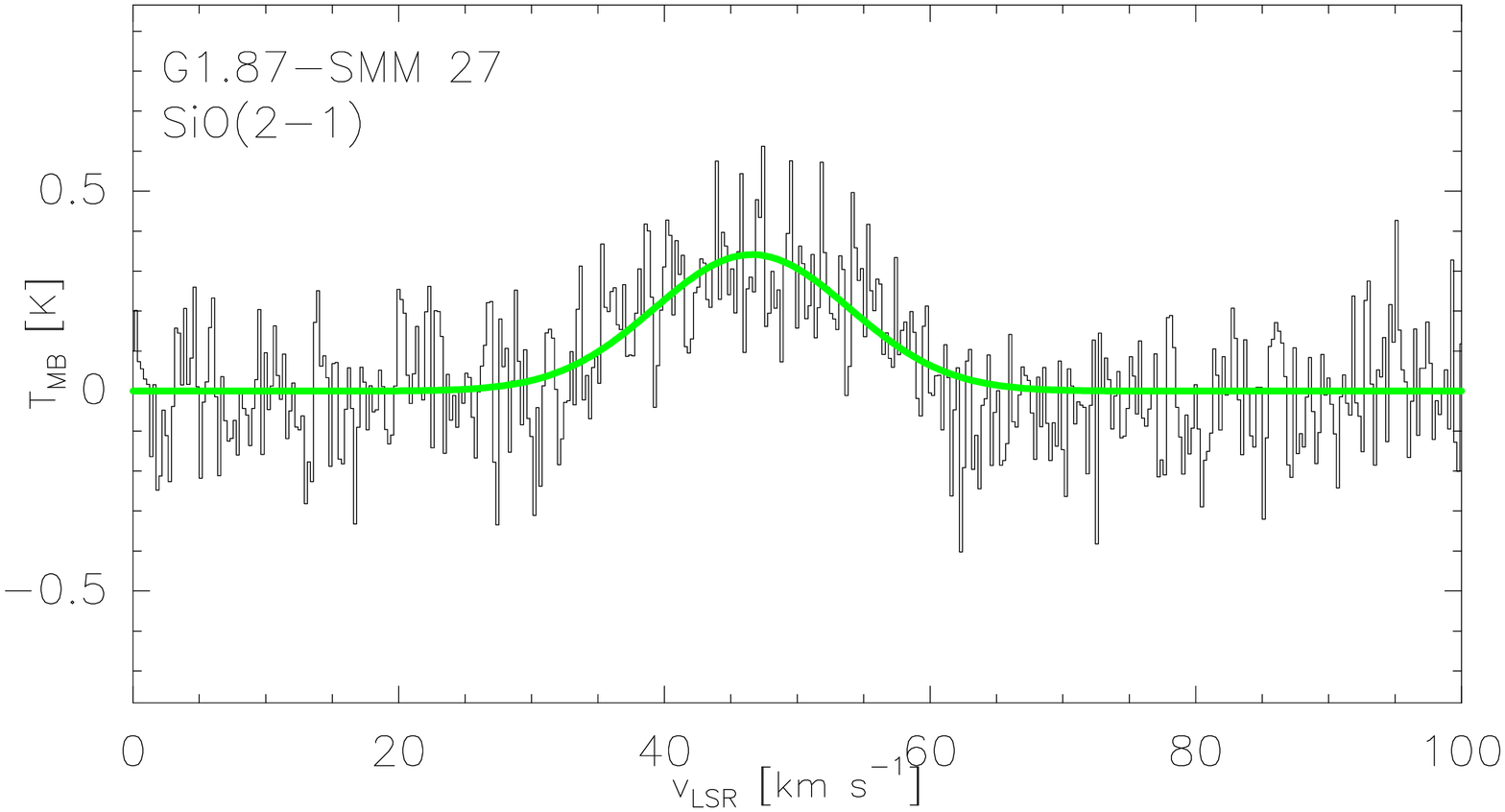}
\includegraphics[width=0.245\textwidth]{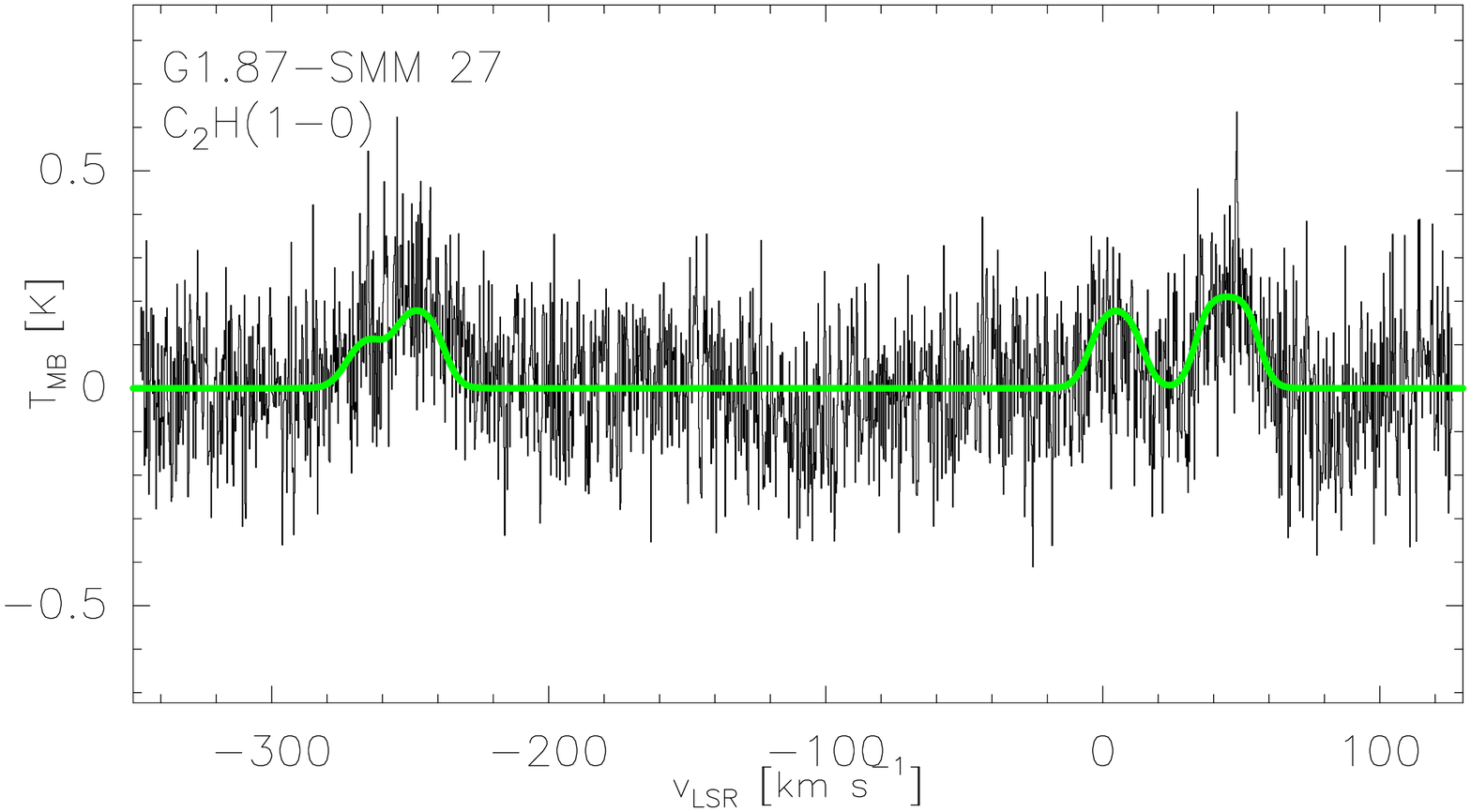}
\includegraphics[width=0.245\textwidth]{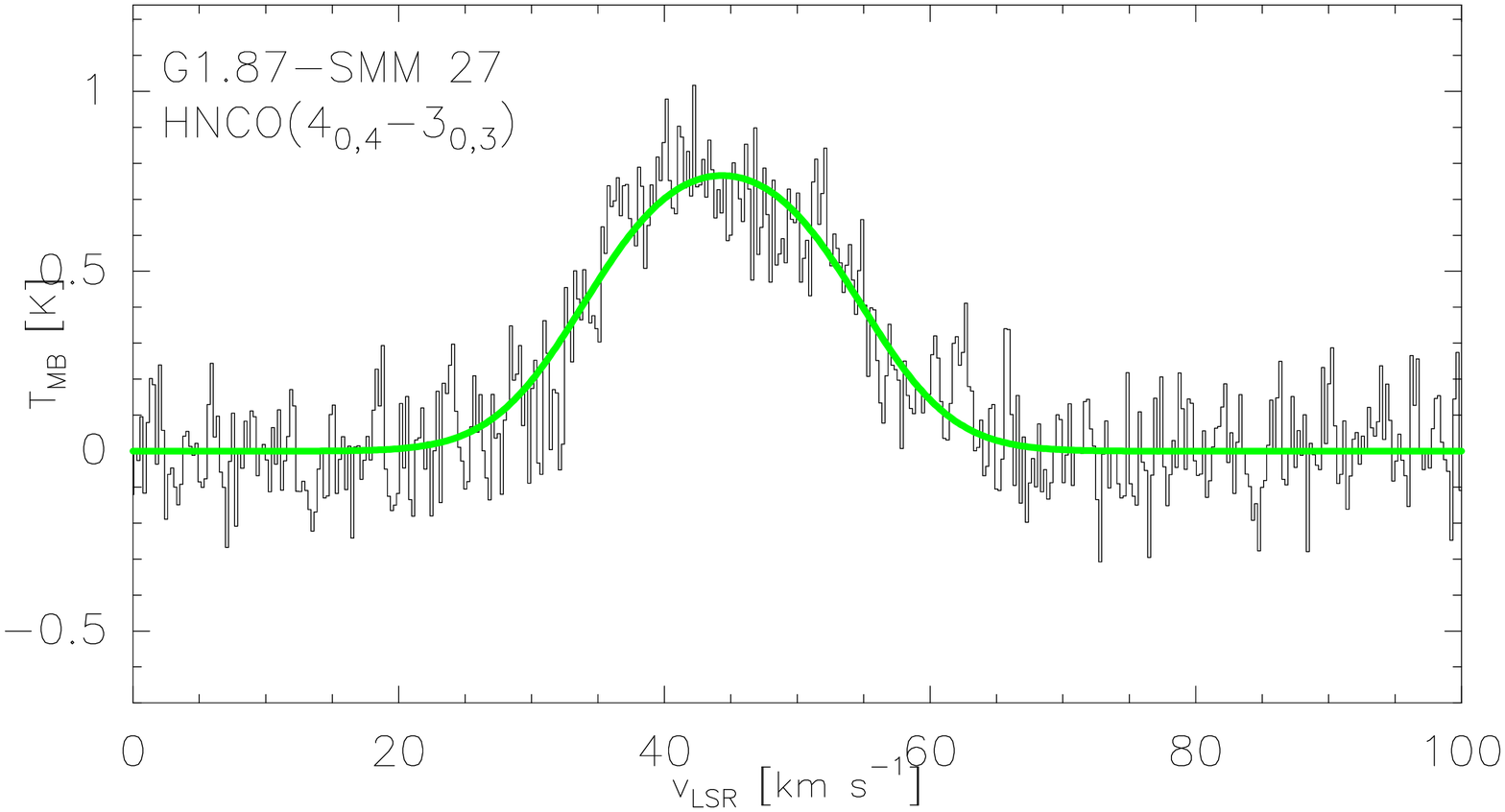}
\includegraphics[width=0.245\textwidth]{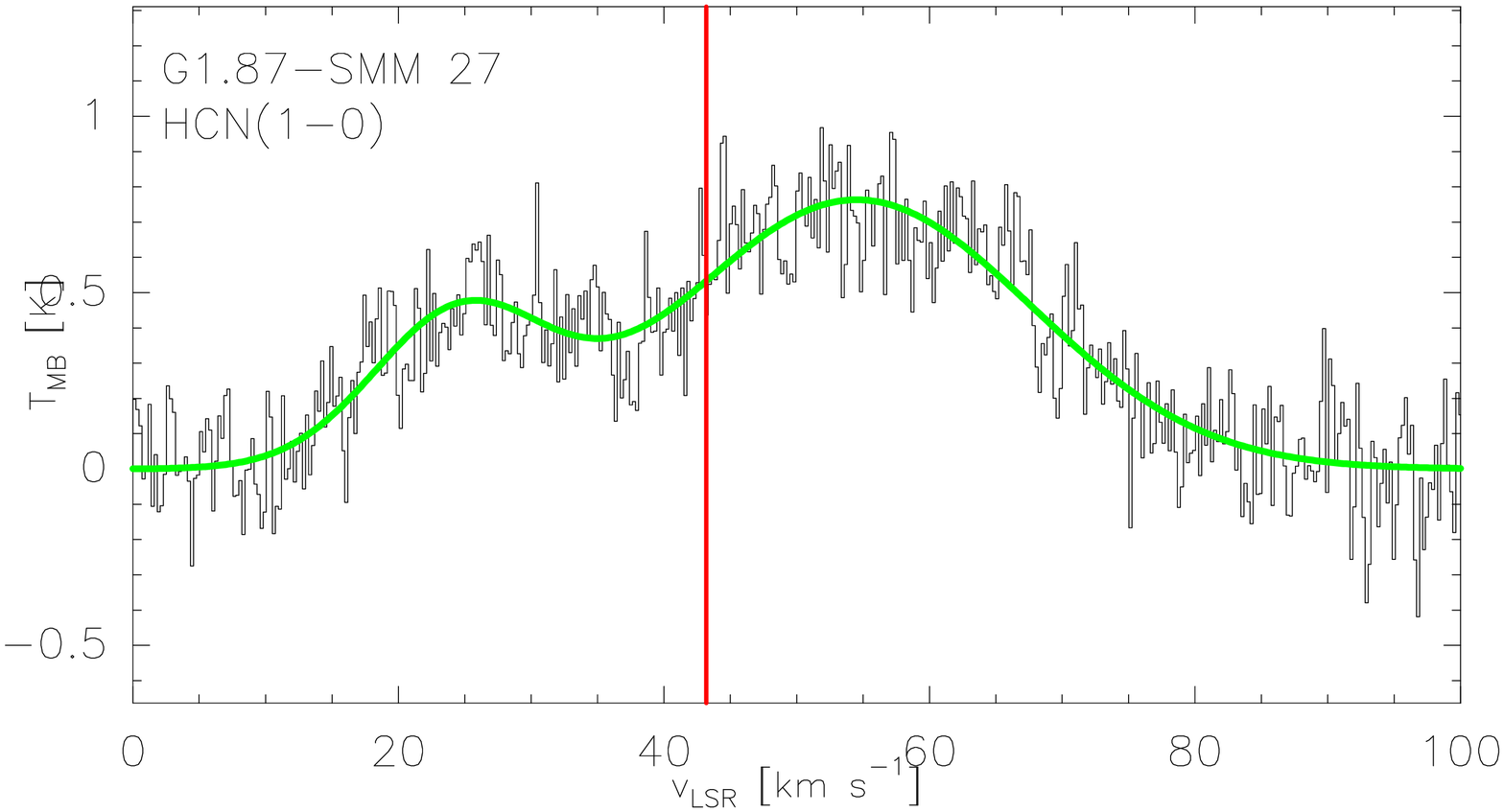}
\includegraphics[width=0.245\textwidth]{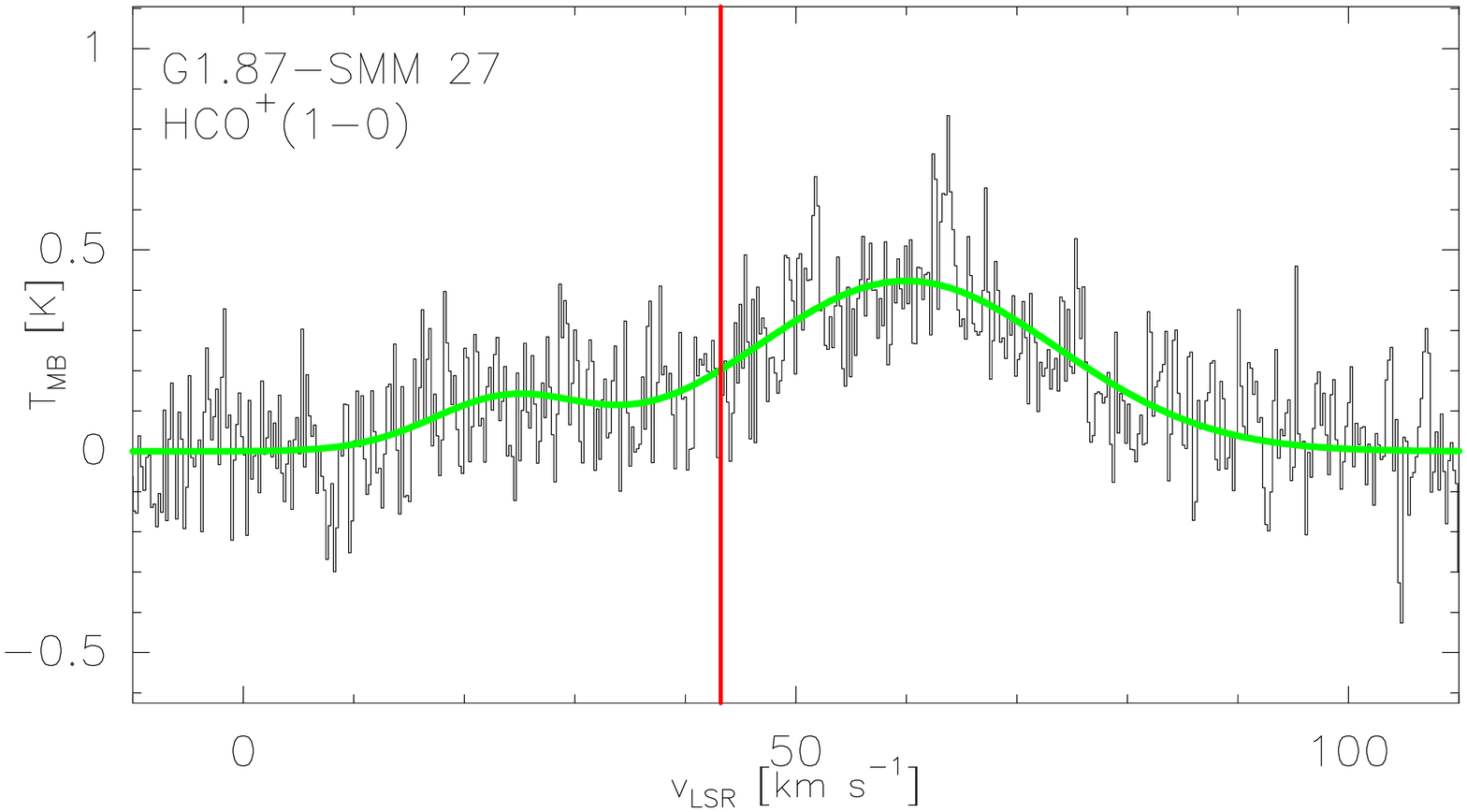}
\includegraphics[width=0.245\textwidth]{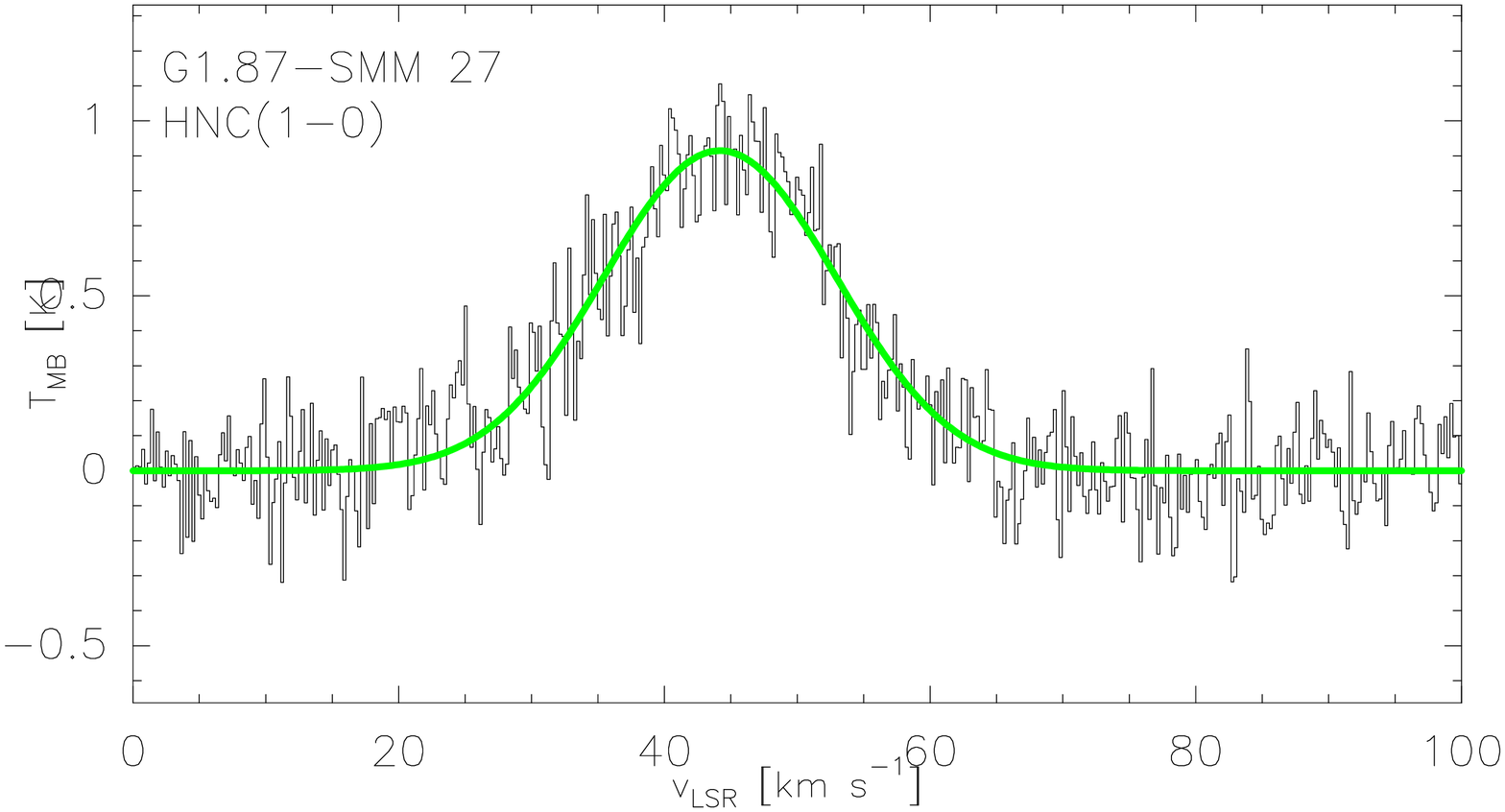}
\includegraphics[width=0.245\textwidth]{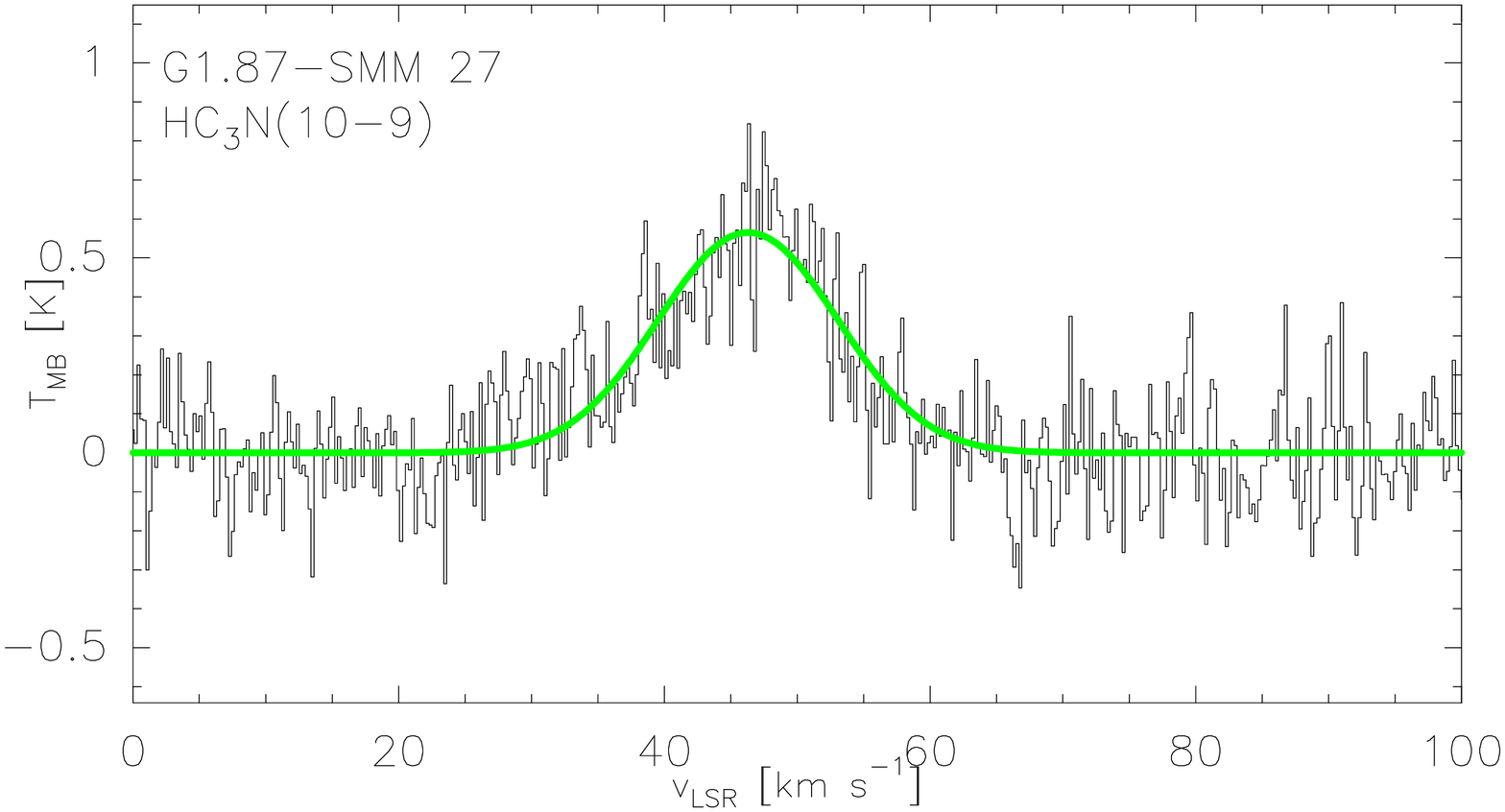}
\includegraphics[width=0.245\textwidth]{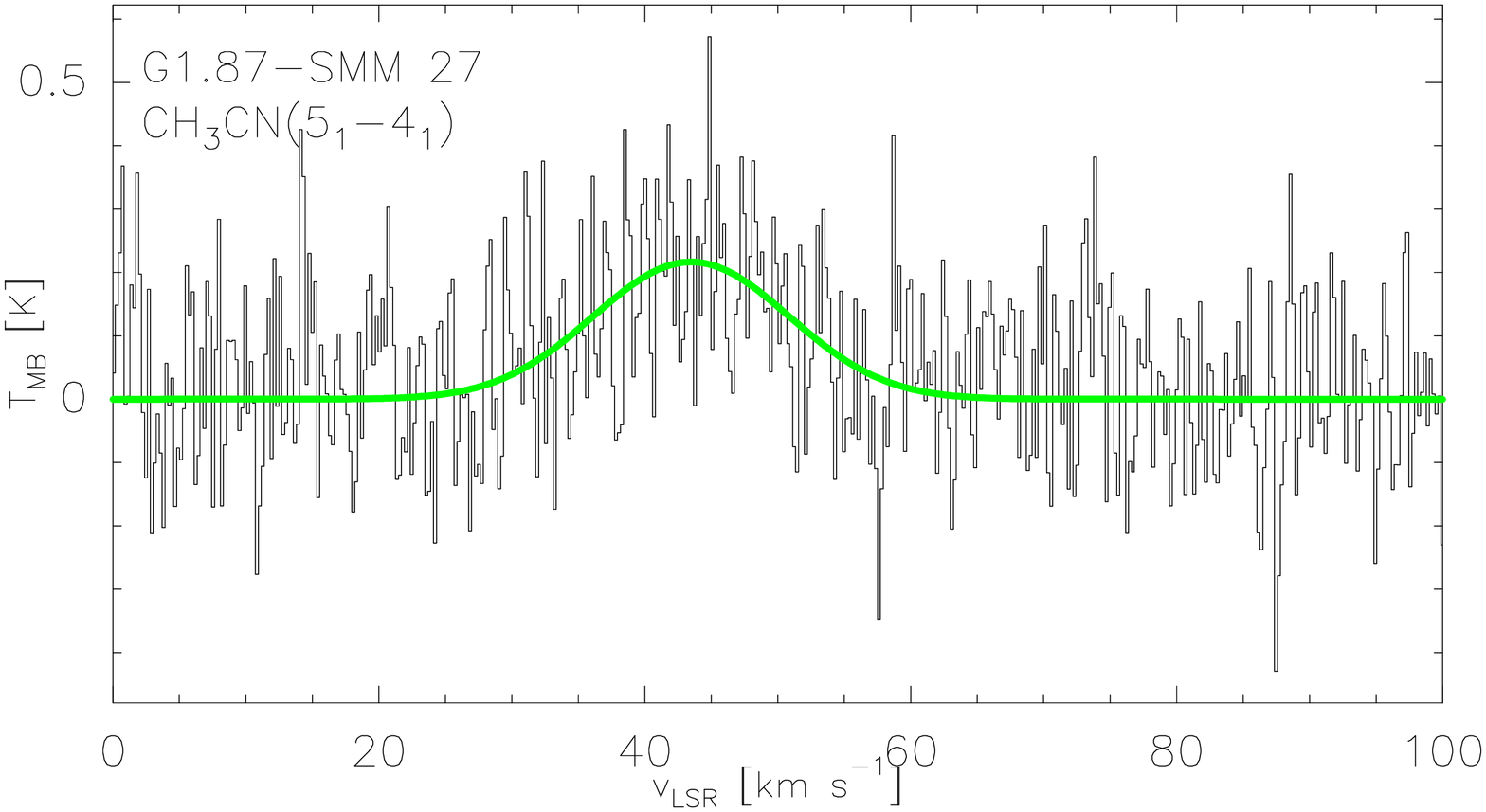}
\includegraphics[width=0.245\textwidth]{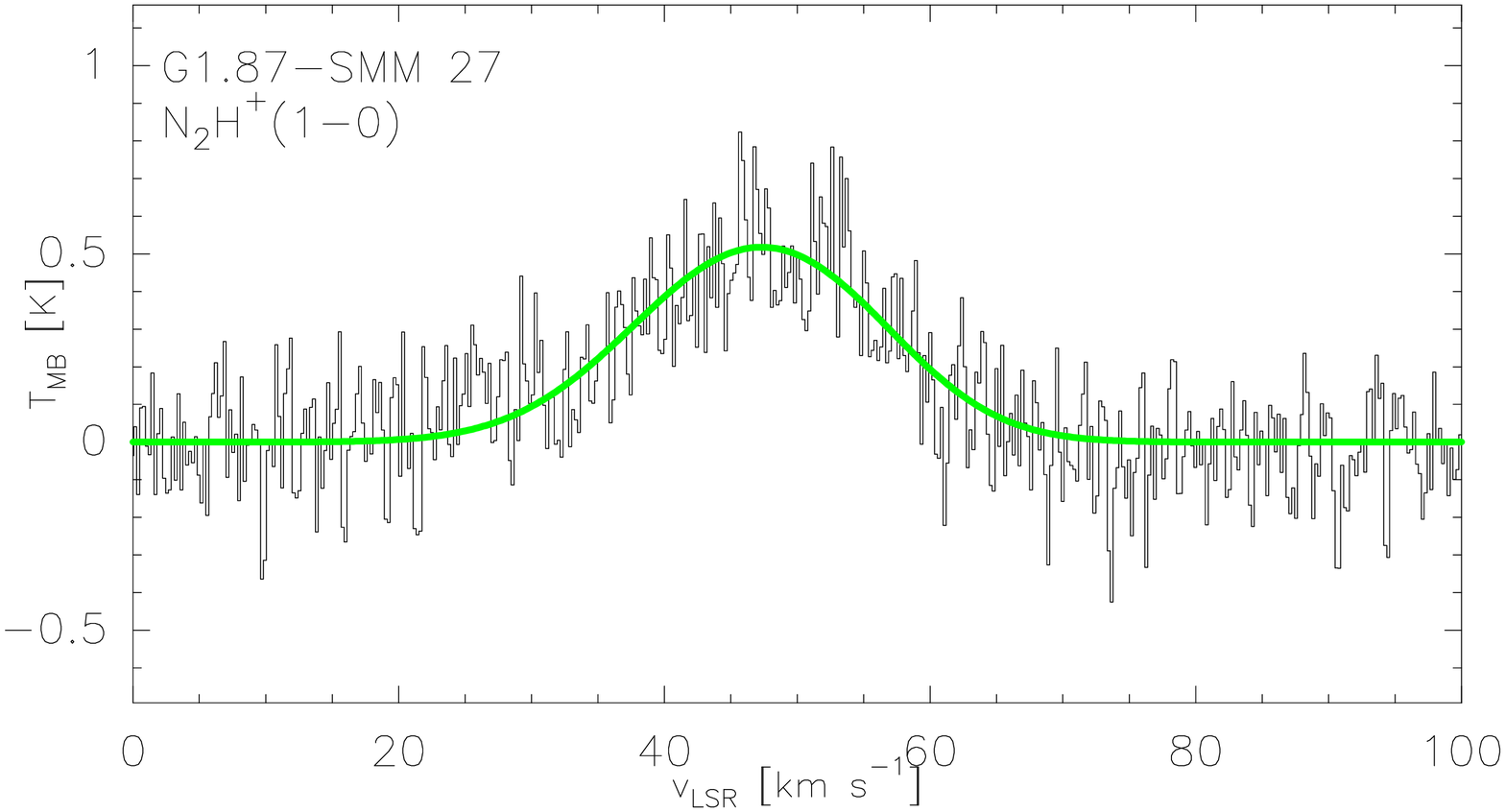}
\caption{Same as Fig.~\ref{figure:G187SMM1_spectra} but towards G1.87--SMM 27. 
The velocity range shown is wider for the C$_2$H and HCO$^+$ spectra. 
The red vertical line indicates the radial velocity of the optically thin 
HNCO line.}
\label{figure:G187SMM27_spectra}
\end{center}
\end{figure*}

\begin{figure*}
\begin{center}
\includegraphics[width=0.245\textwidth]{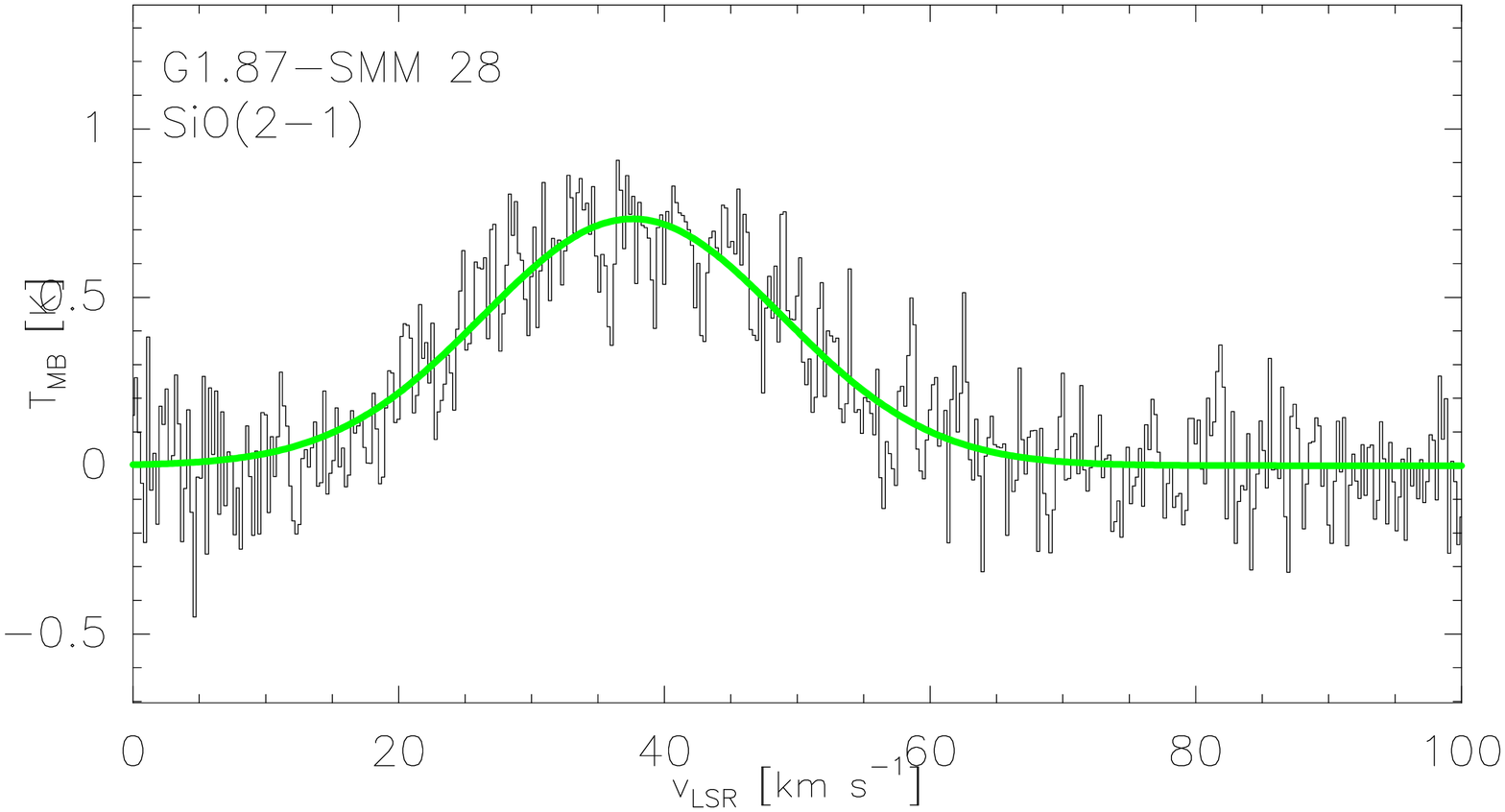}
\includegraphics[width=0.245\textwidth]{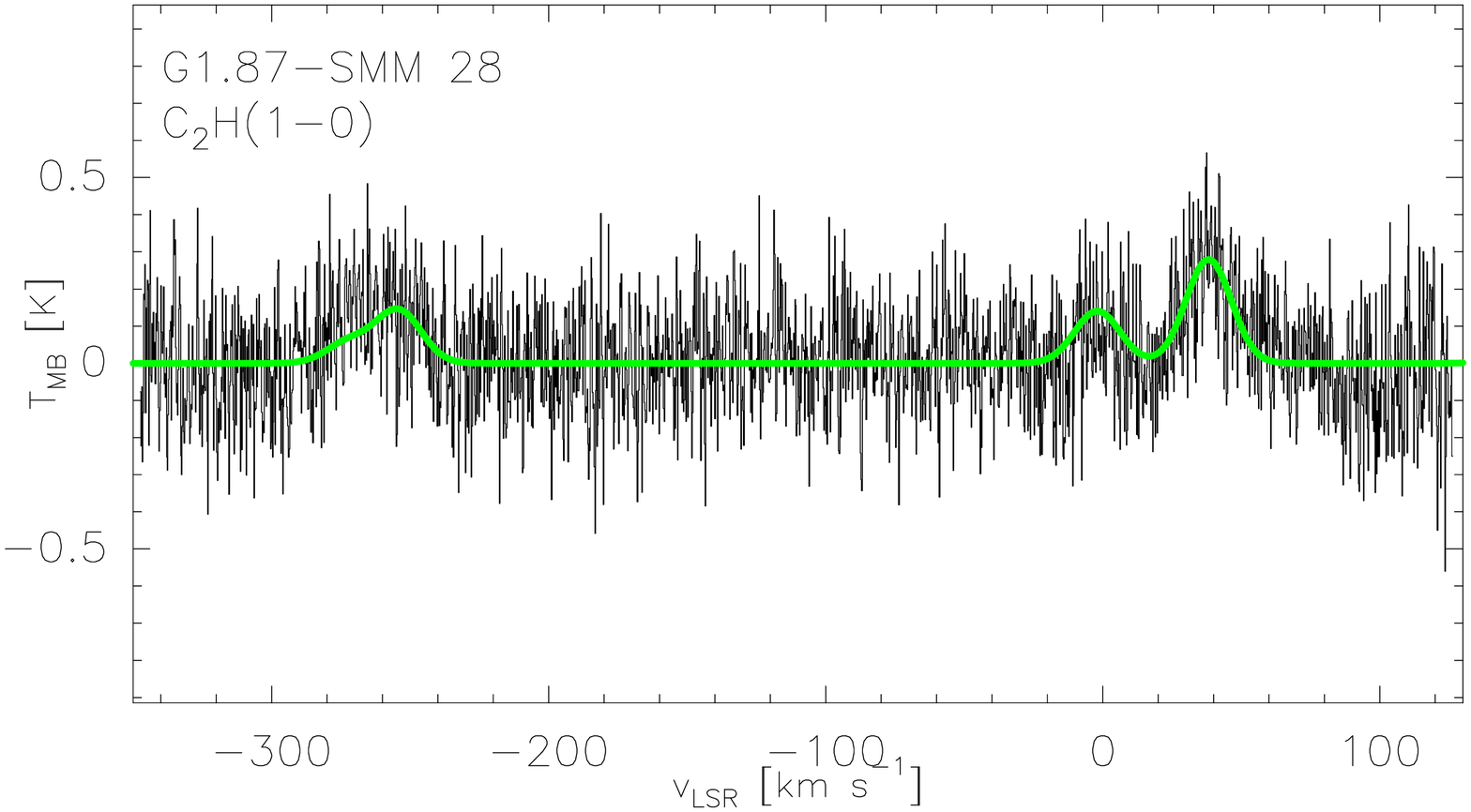}
\includegraphics[width=0.245\textwidth]{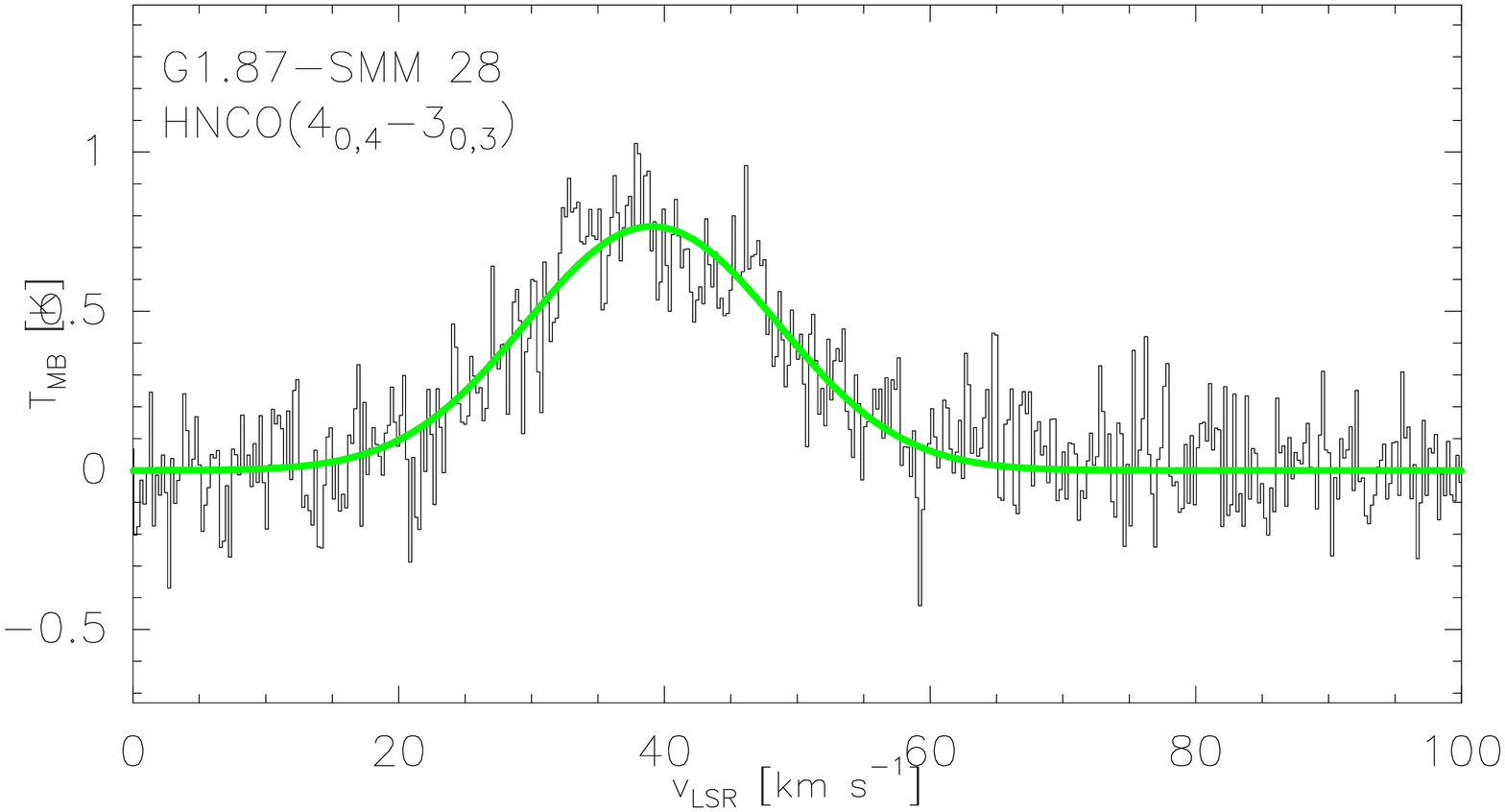}
\includegraphics[width=0.245\textwidth]{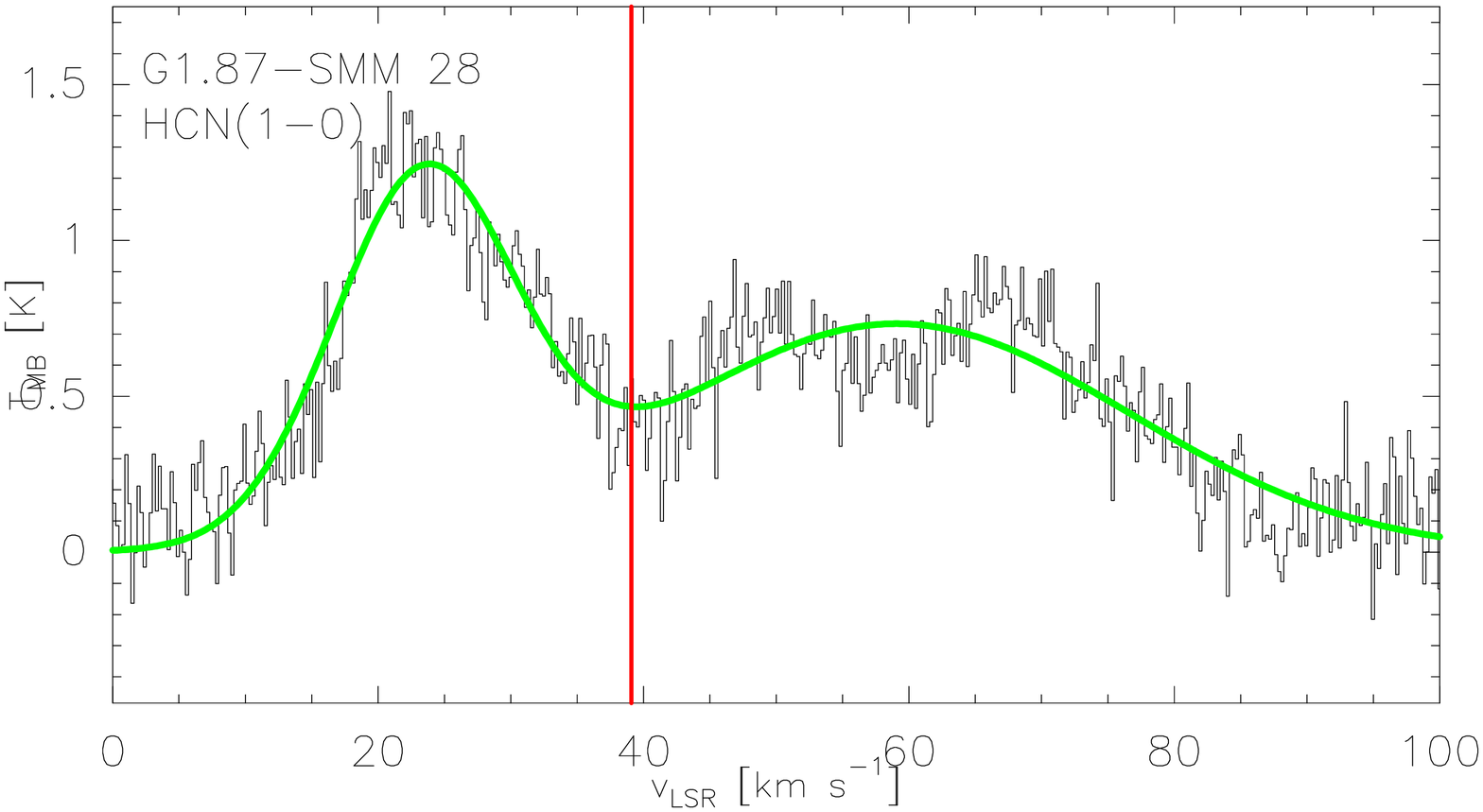}
\includegraphics[width=0.245\textwidth]{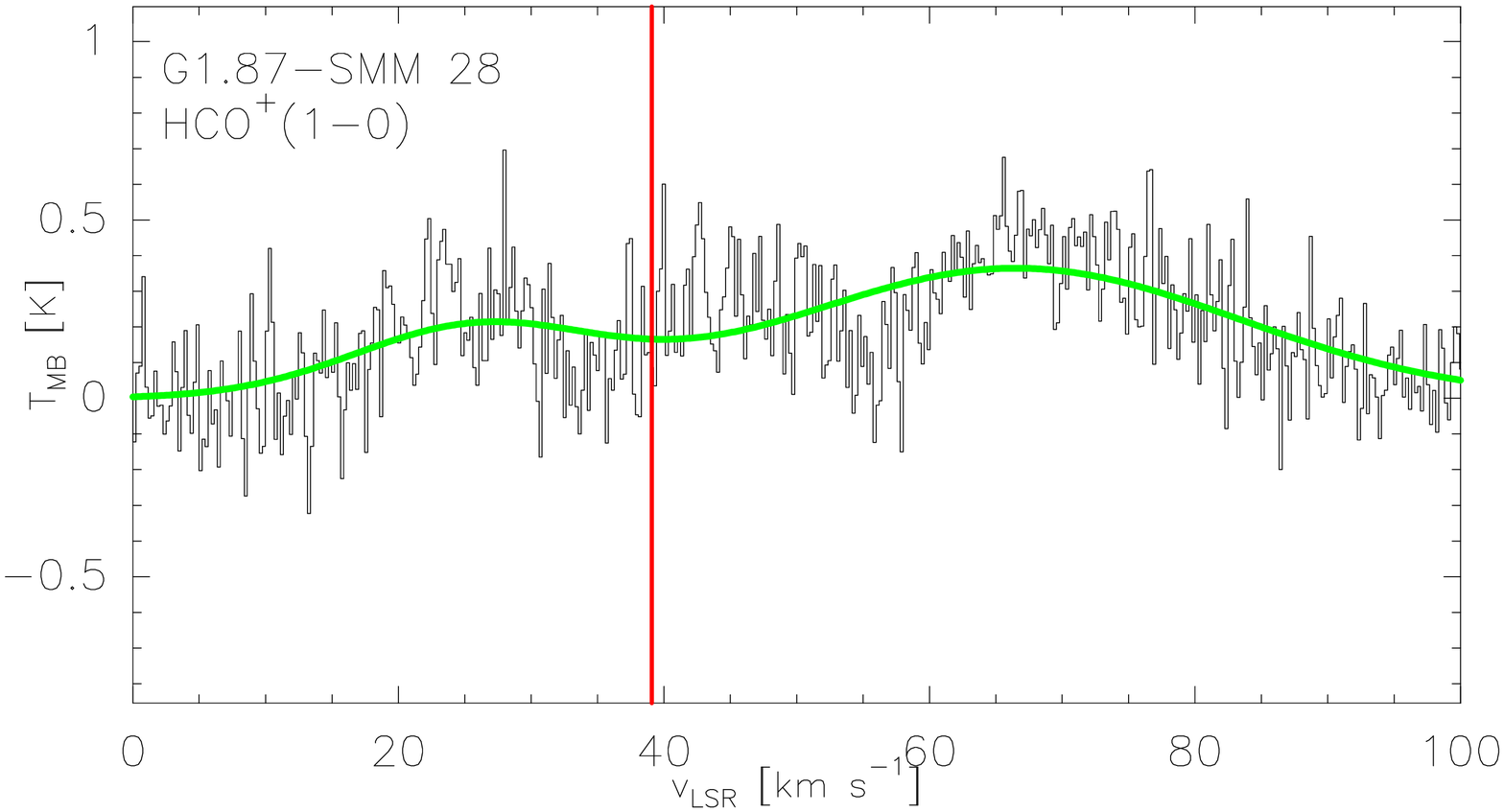}
\includegraphics[width=0.245\textwidth]{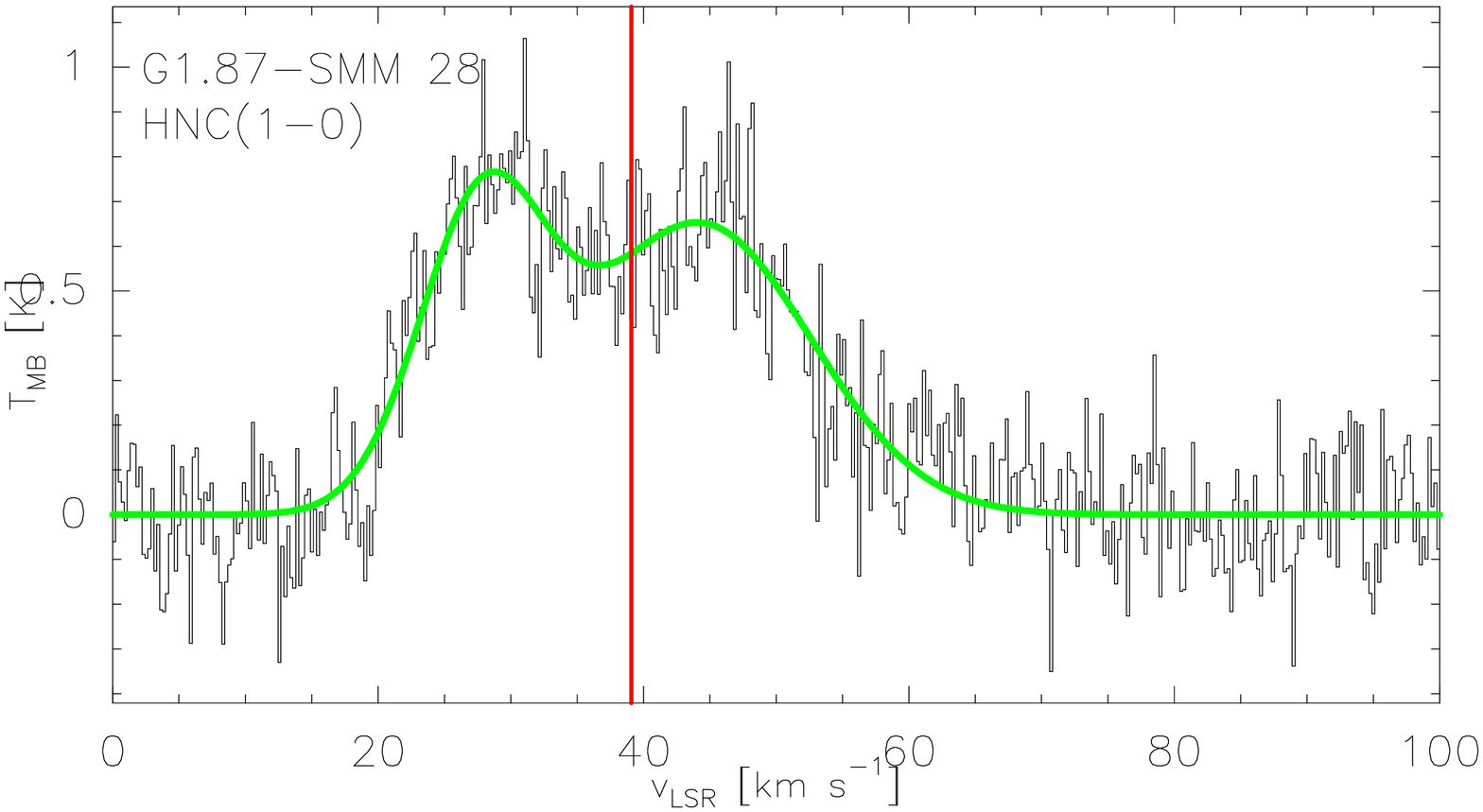}
\includegraphics[width=0.245\textwidth]{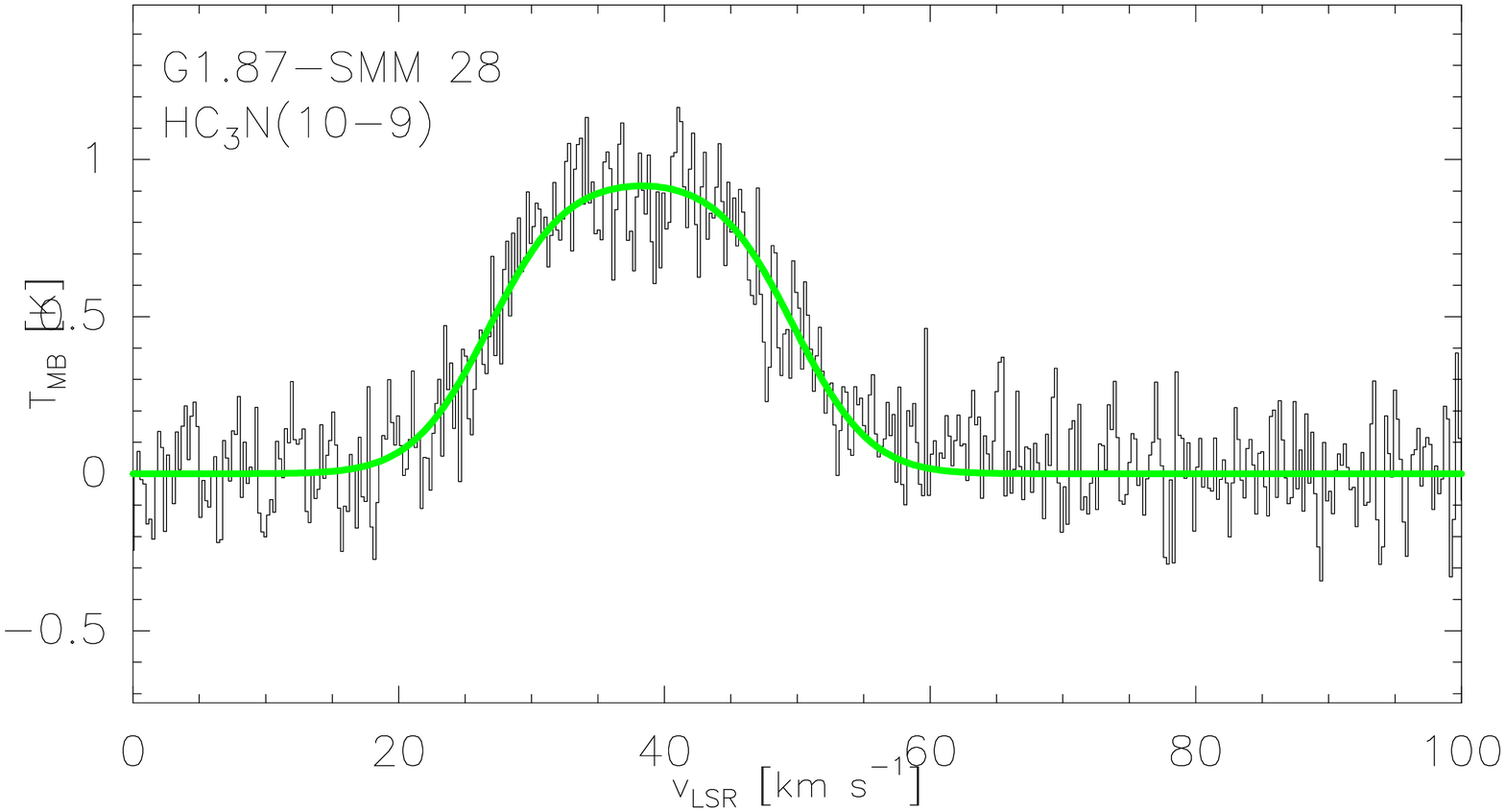}
\includegraphics[width=0.245\textwidth]{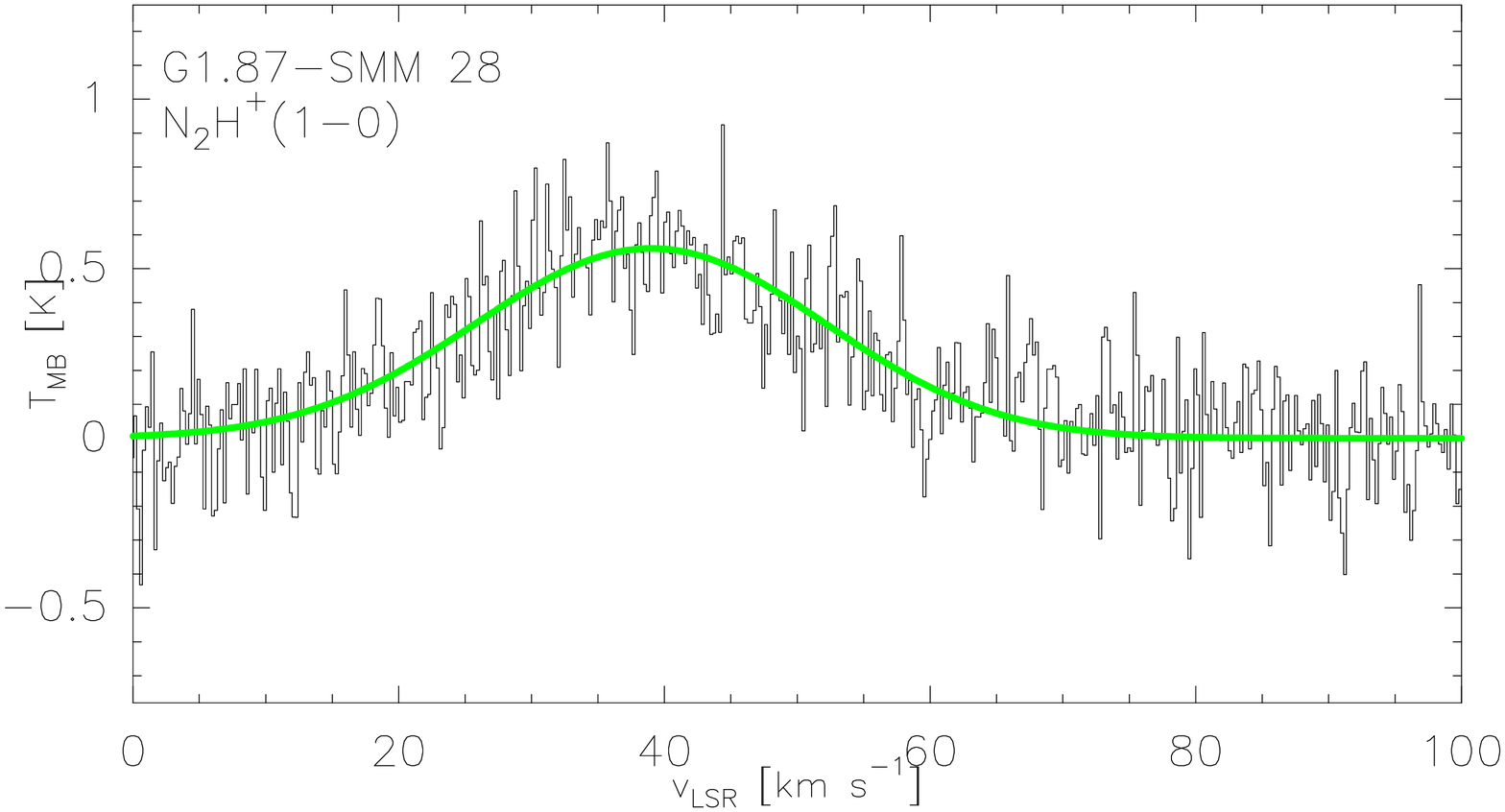}
\caption{Same as Fig.~\ref{figure:G187SMM1_spectra} but towards G1.87--SMM 28. 
The velocity range shown is wider for the C$_2$H spectrum. The red vertical 
line indicates the radial velocity of the optically thin HNCO line.}
\label{figure:G187SMM28_spectra}
\end{center}
\end{figure*}

\begin{figure*}
\begin{center}
\includegraphics[width=0.245\textwidth]{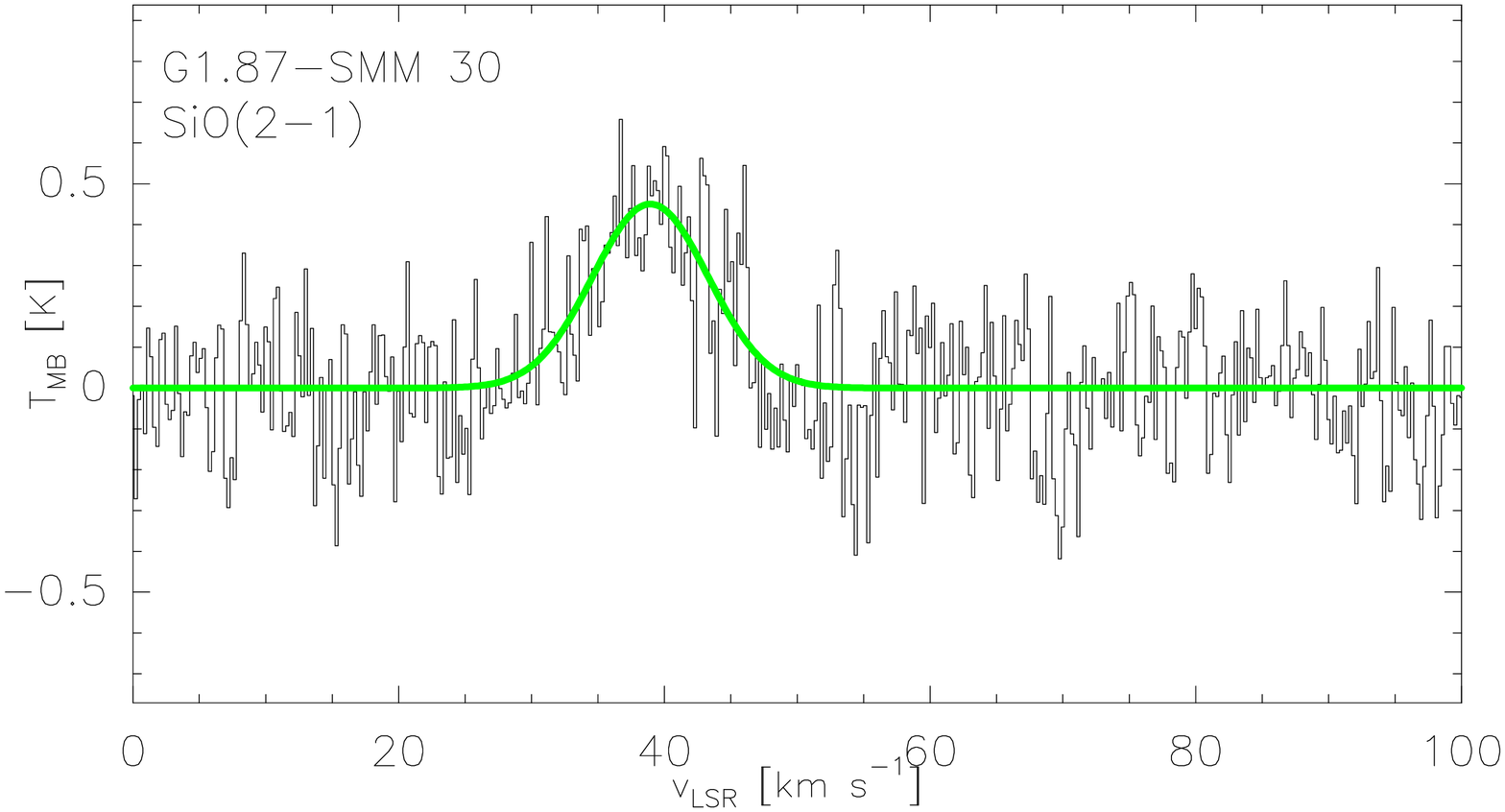}
\includegraphics[width=0.245\textwidth]{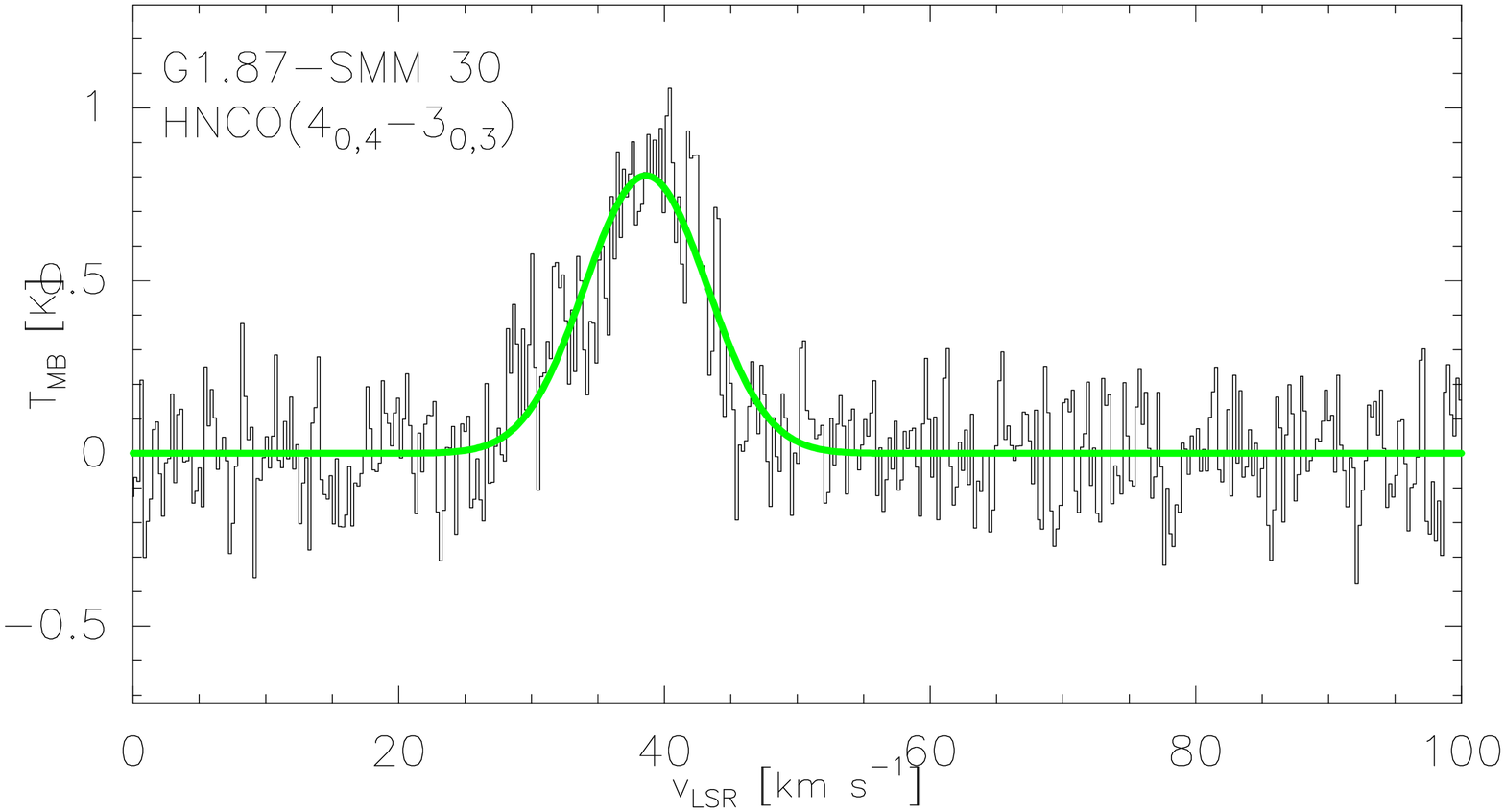}
\includegraphics[width=0.245\textwidth]{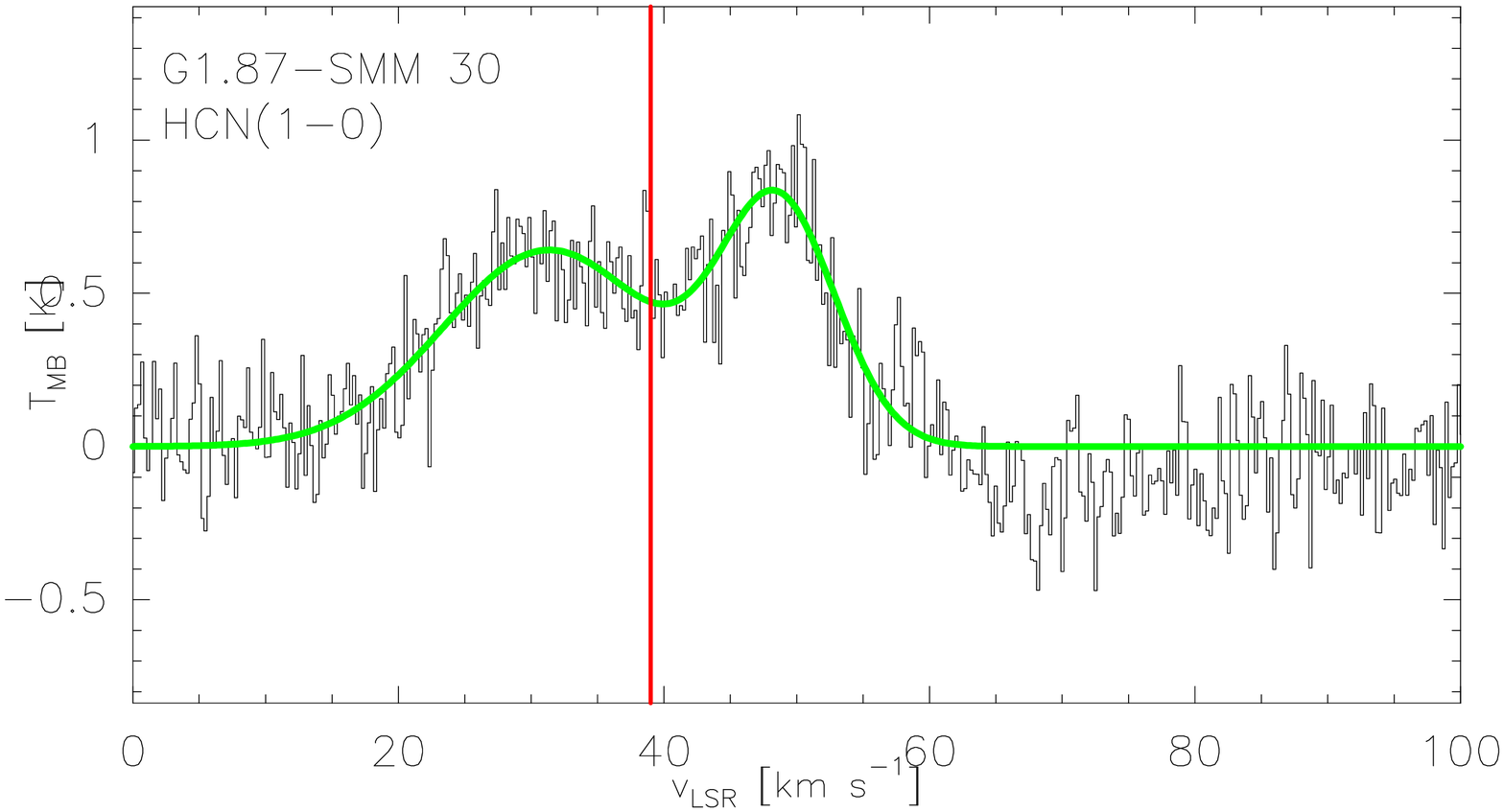}
\includegraphics[width=0.245\textwidth]{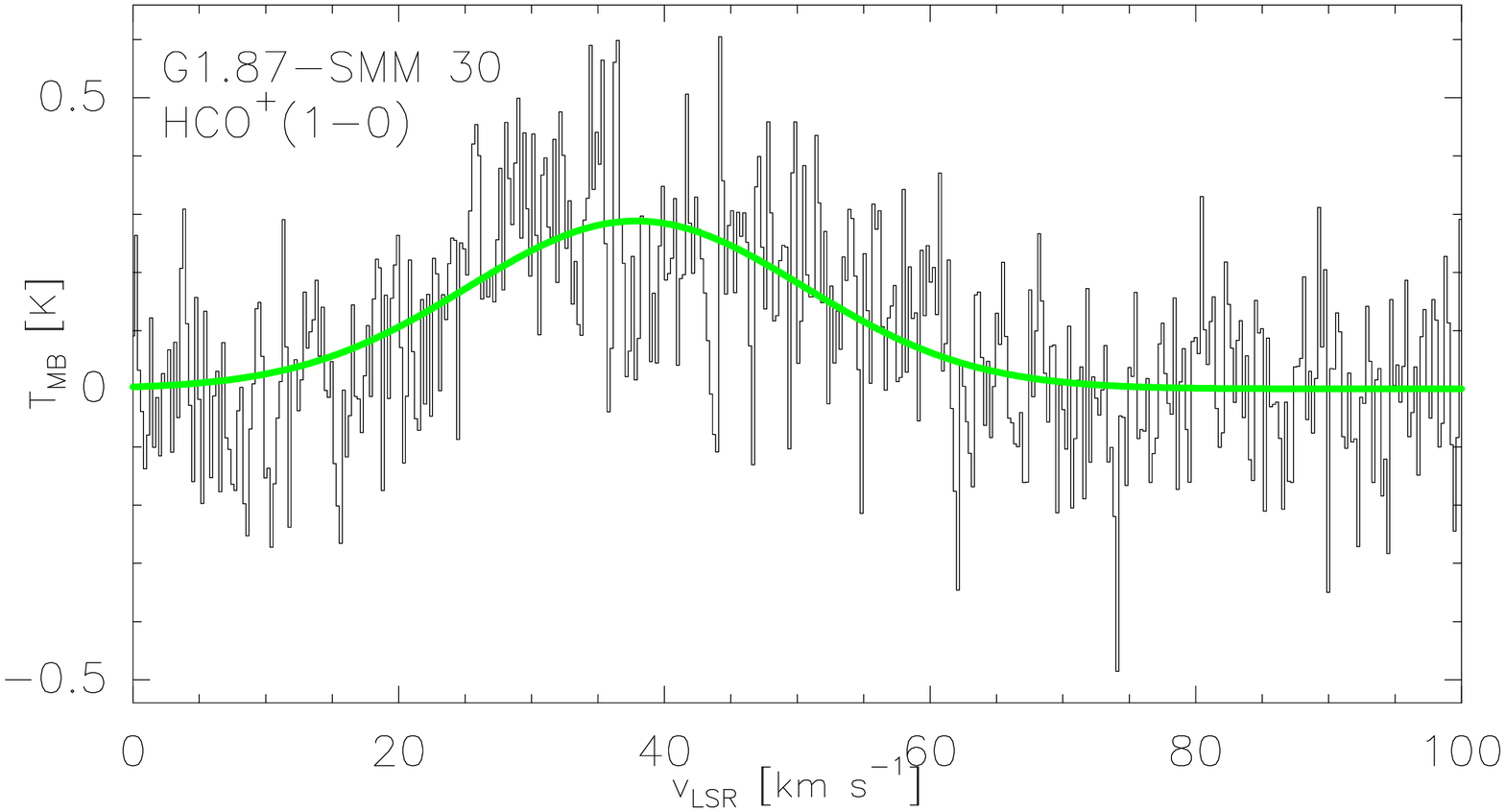}
\includegraphics[width=0.245\textwidth]{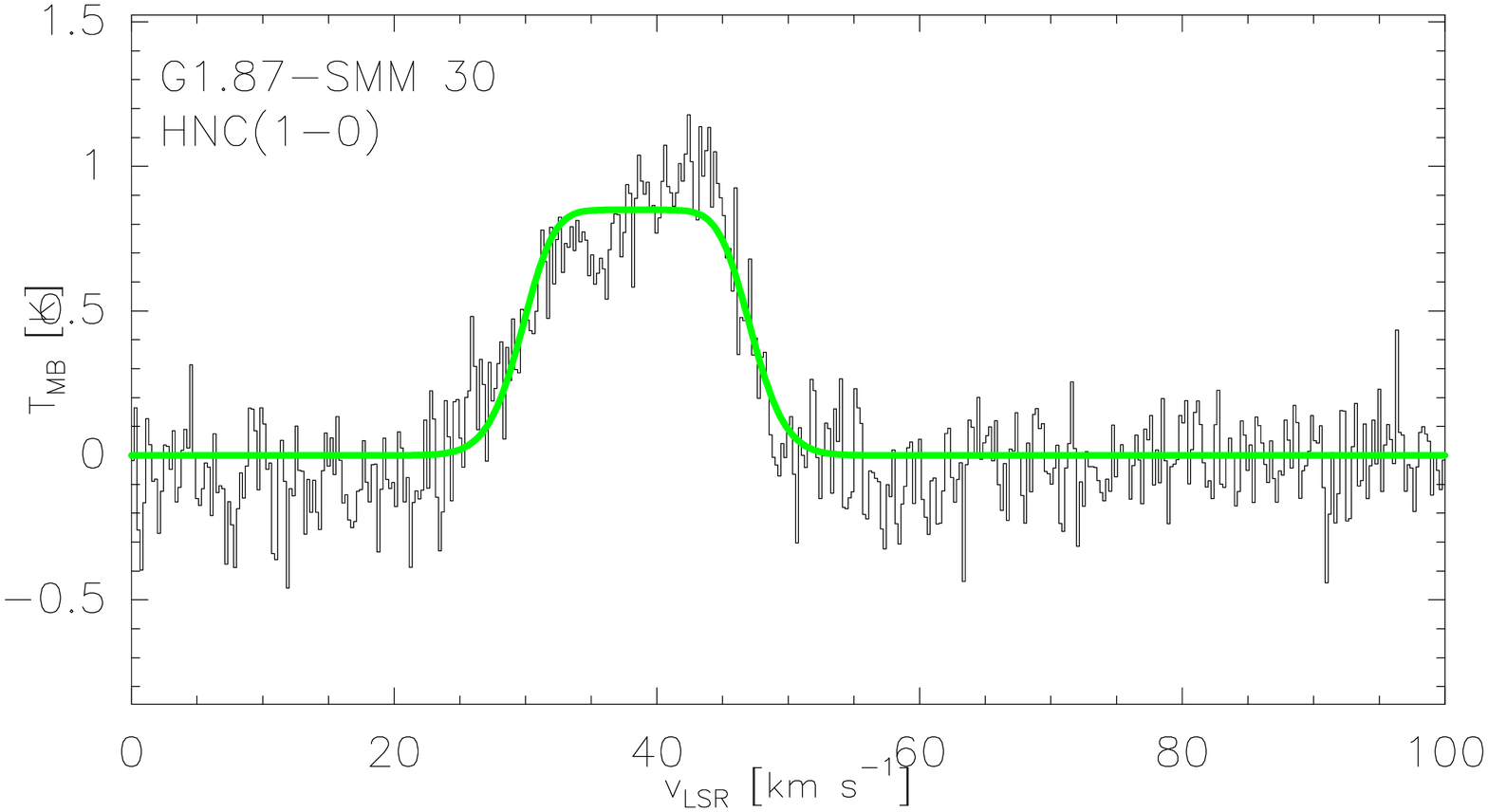}
\includegraphics[width=0.245\textwidth]{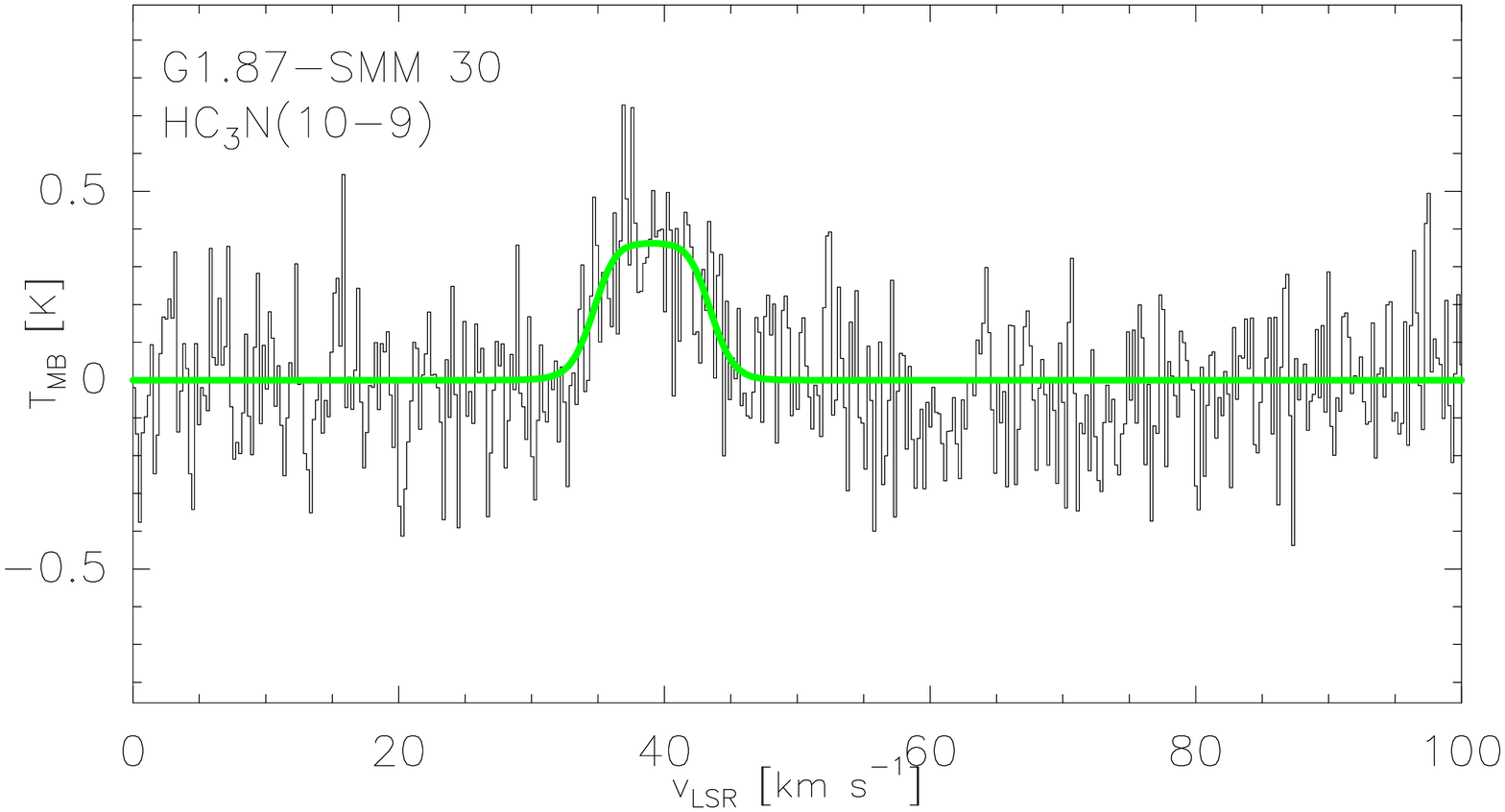}
\includegraphics[width=0.245\textwidth]{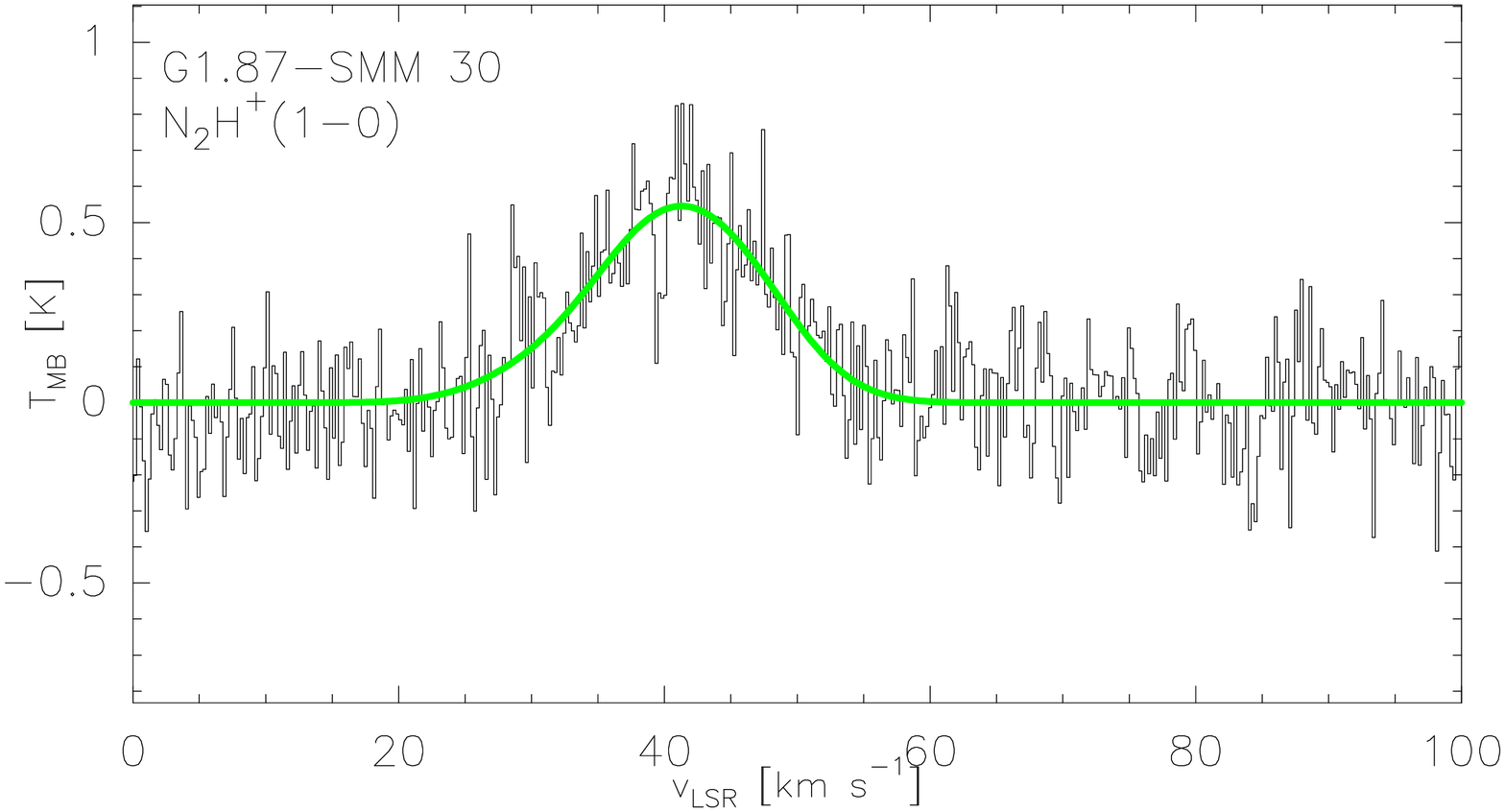}
\caption{Same as Fig.~\ref{figure:G187SMM1_spectra} but towards G1.87--SMM 30. 
The red vertical line indicates the radial velocity of the optically thin 
HC$_3$N line.}
\label{figure:G187SMM30_spectra}
\end{center}
\end{figure*}

\begin{figure*}
\begin{center}
\includegraphics[width=0.245\textwidth]{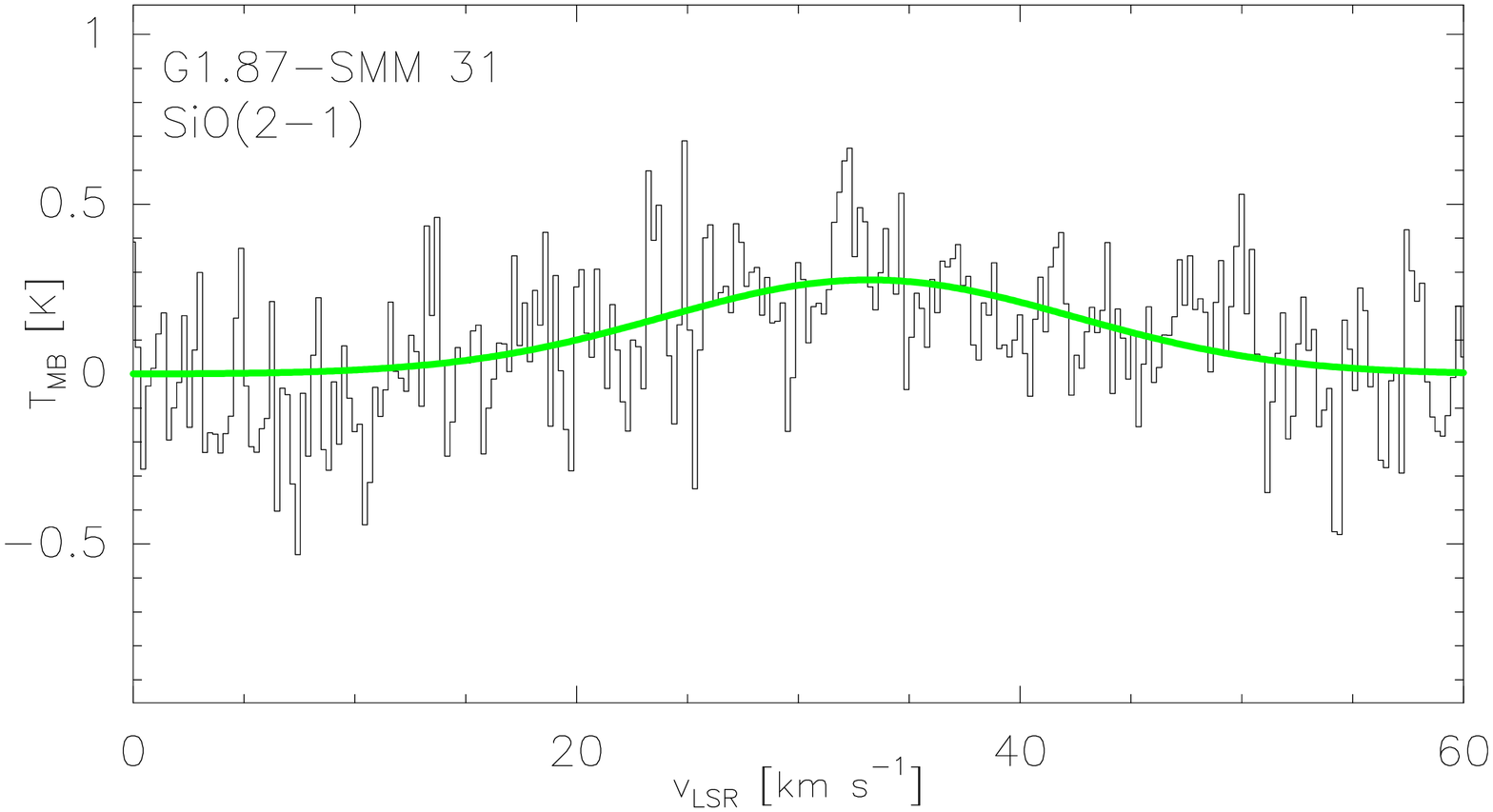}
\includegraphics[width=0.245\textwidth]{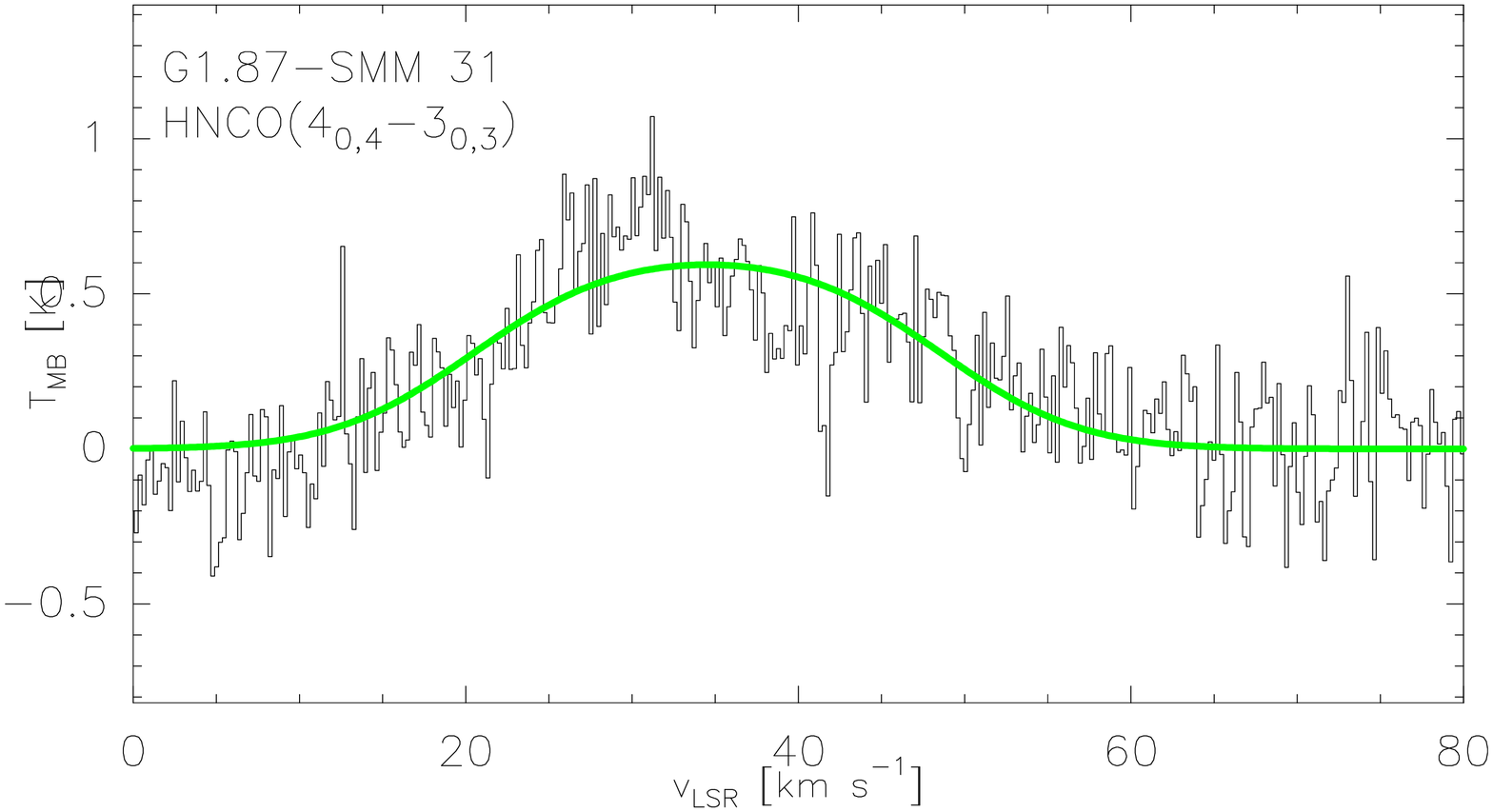}
\includegraphics[width=0.245\textwidth]{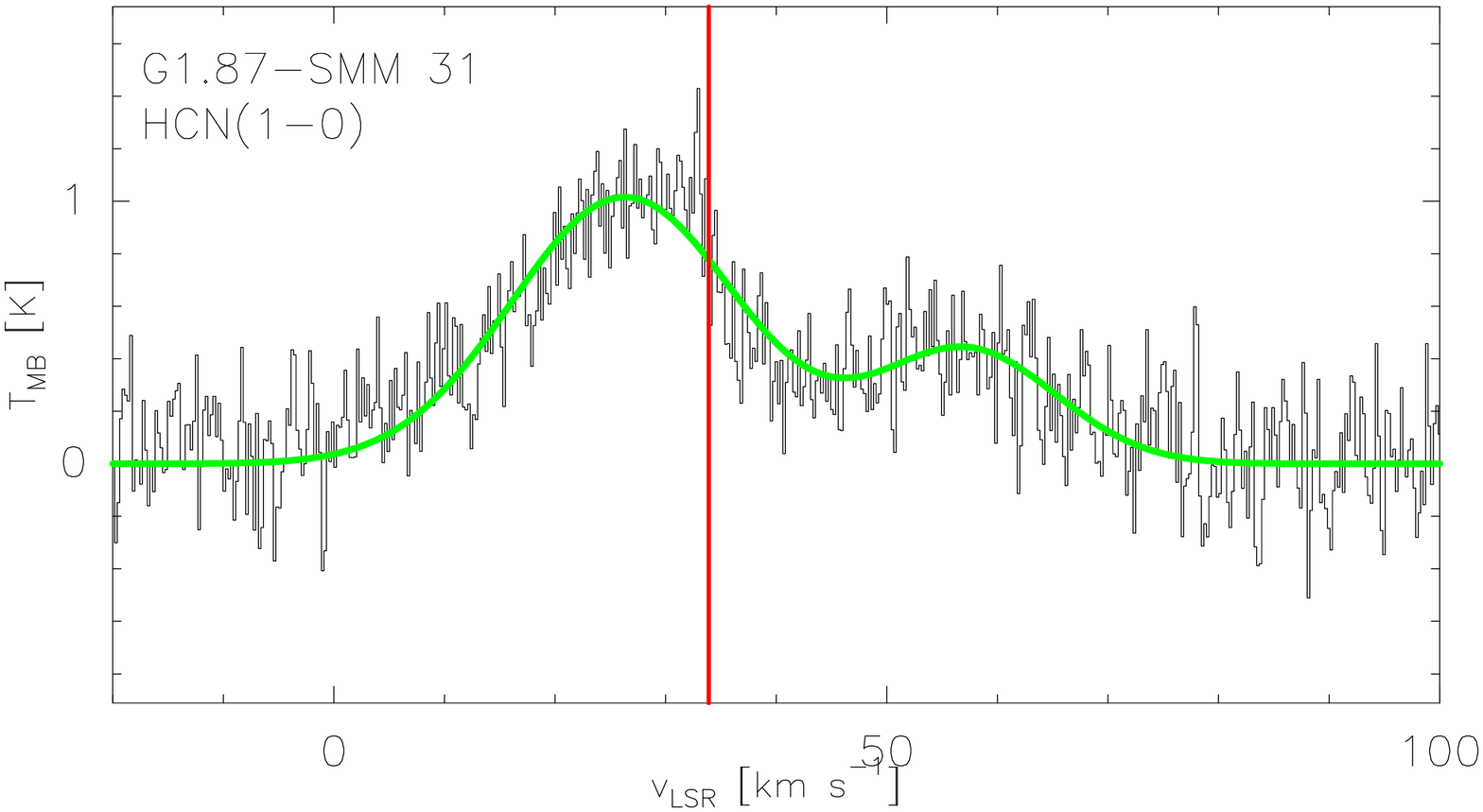}
\includegraphics[width=0.245\textwidth]{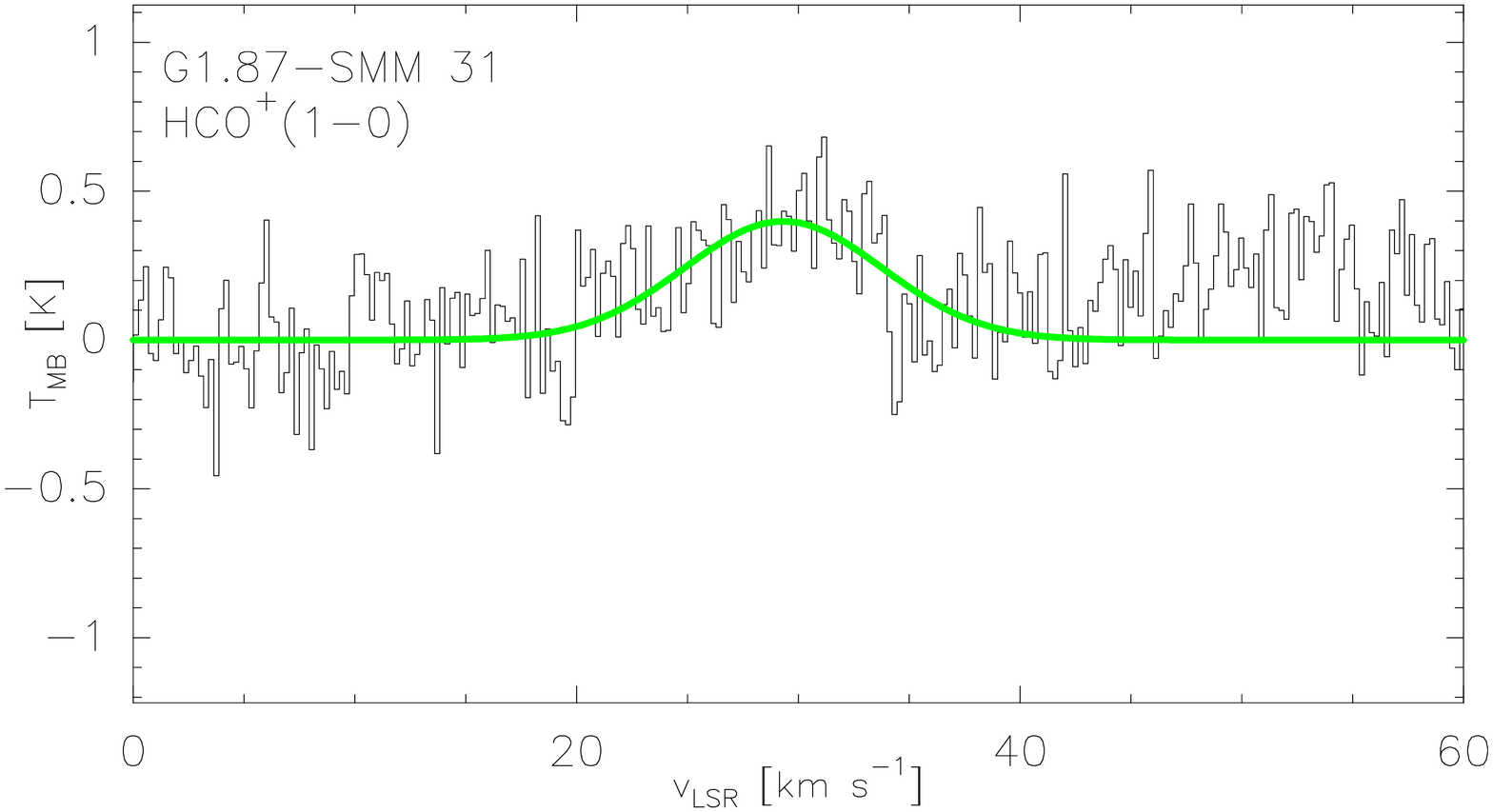}
\includegraphics[width=0.245\textwidth]{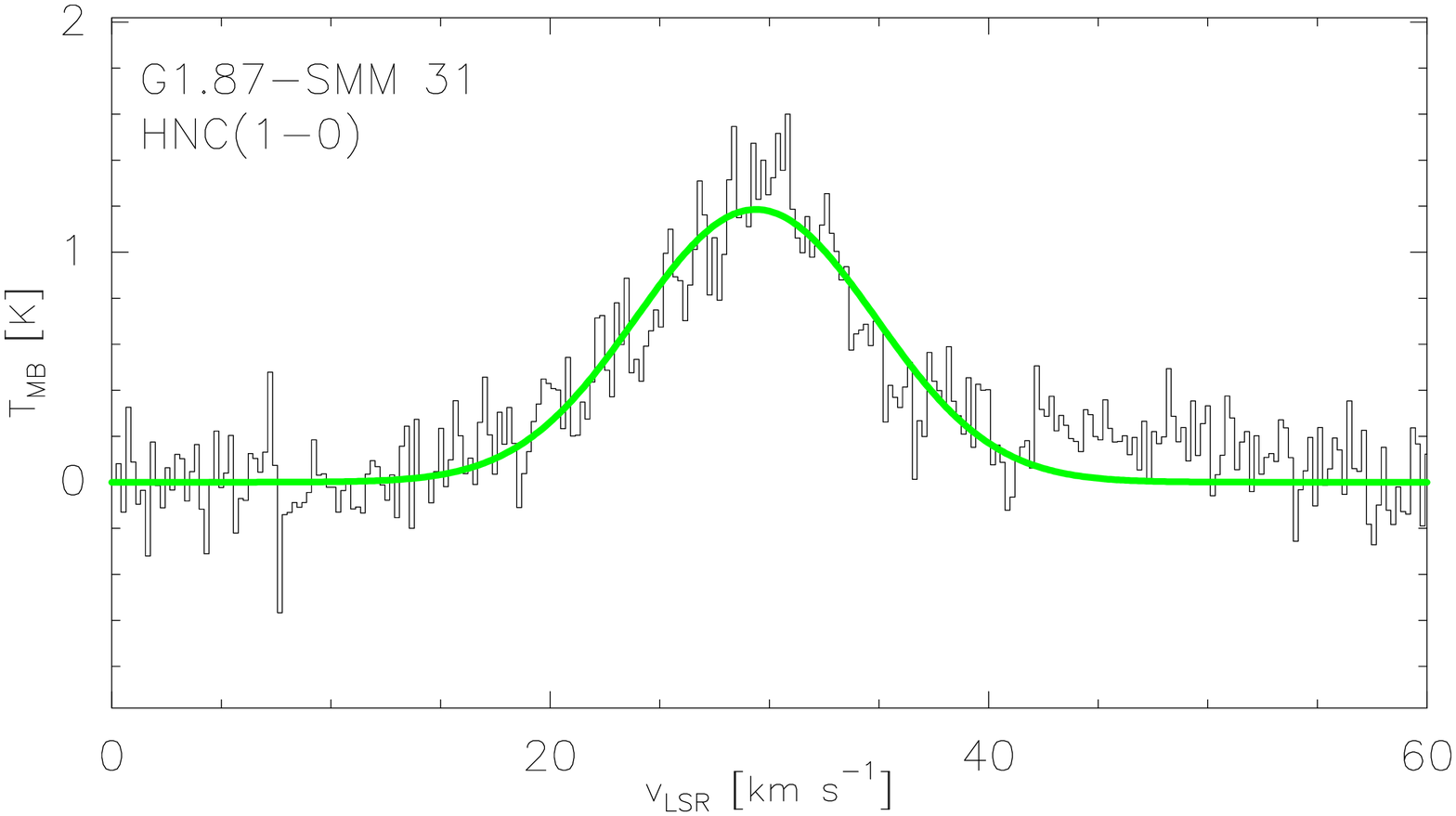}
\includegraphics[width=0.245\textwidth]{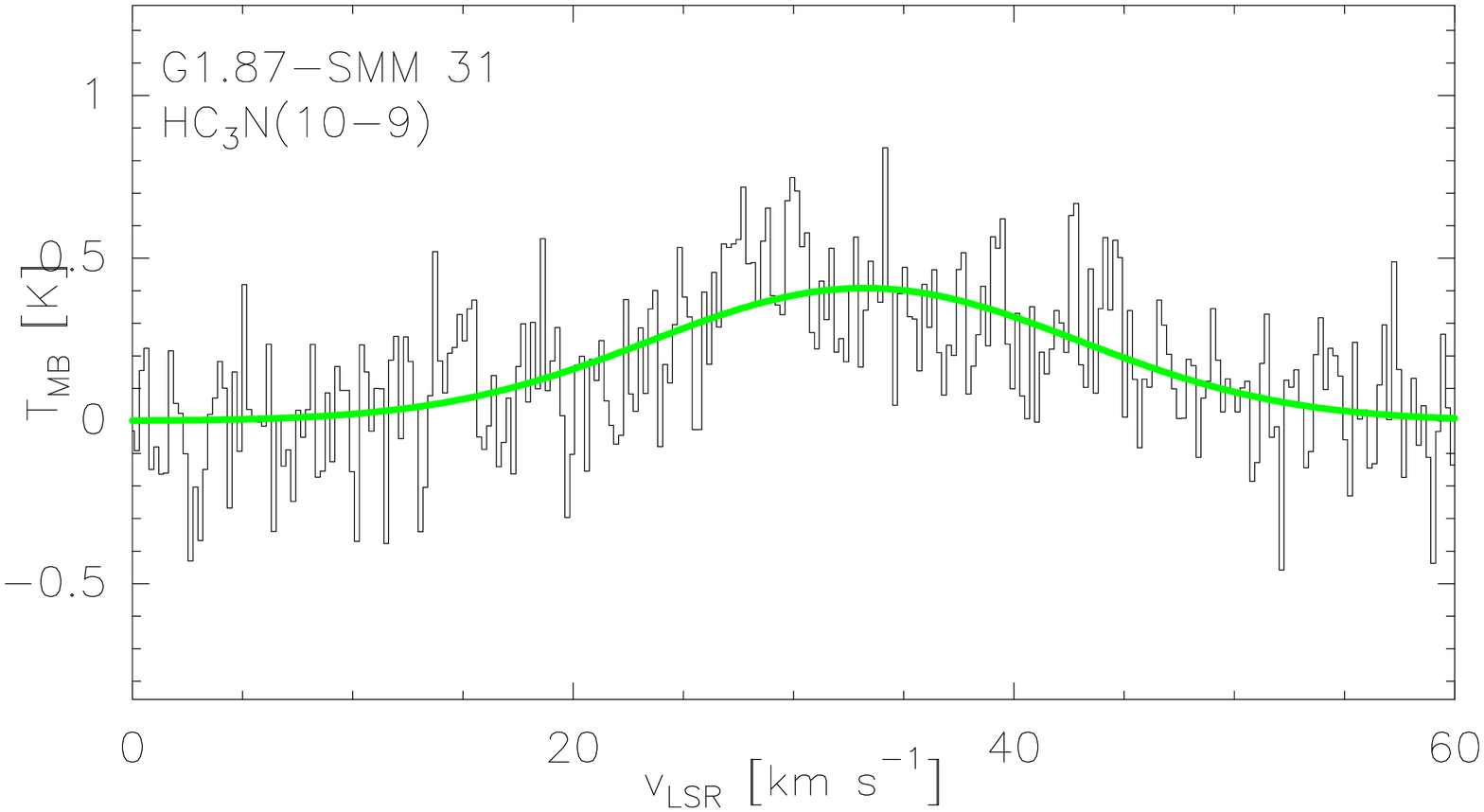}
\includegraphics[width=0.245\textwidth]{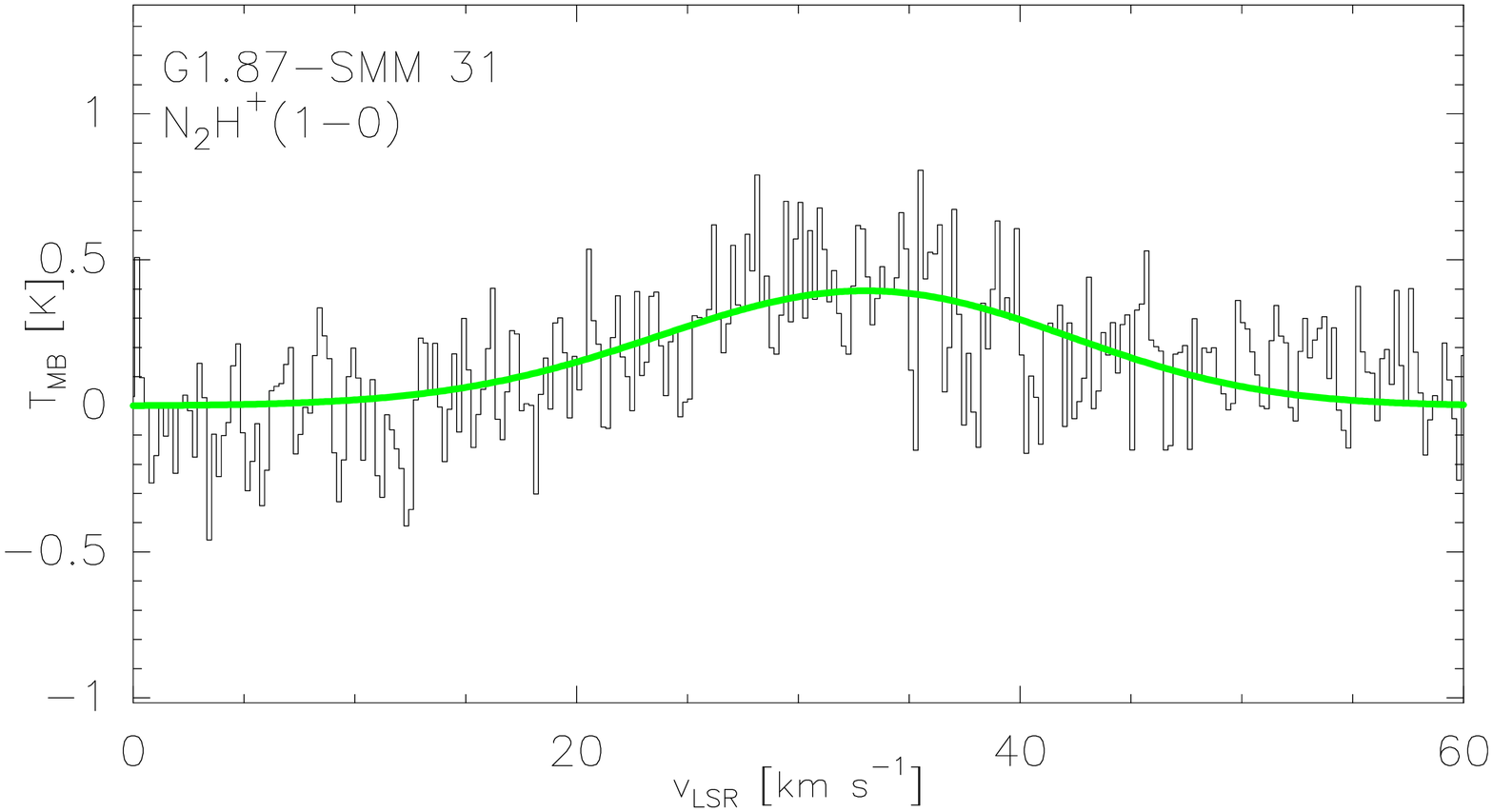}
\caption{Same as Fig.~\ref{figure:G187SMM1_spectra} but towards G1.87--SMM 31.
The velocity range is wider for the HCN spectrum. The red vertical 
line indicates the radial velocity of the optically thin HNCO line. }
\label{figure:G187SMM31_spectra}
\end{center}
\end{figure*}

\begin{figure*}
\begin{center}
\includegraphics[width=0.245\textwidth]{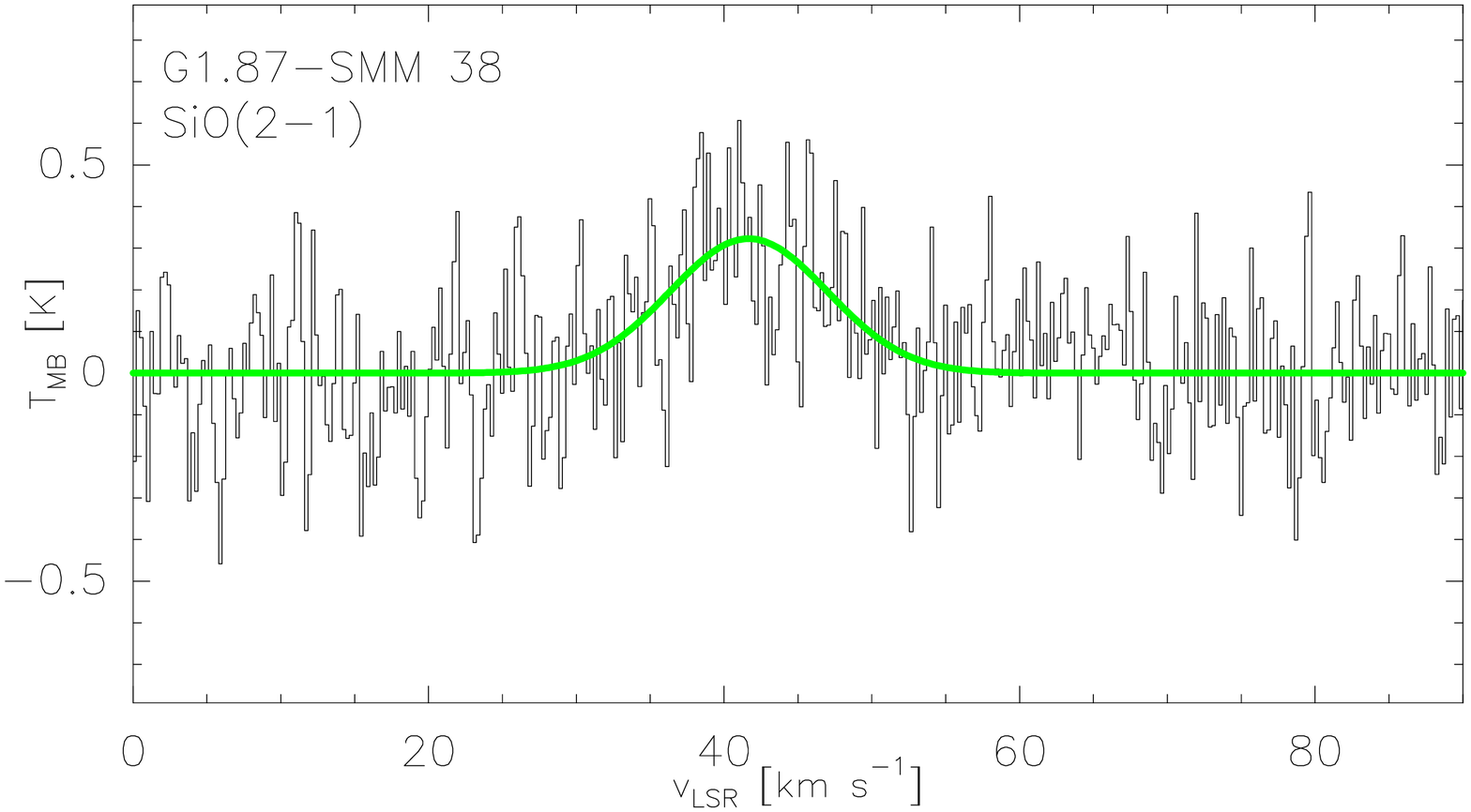}
\includegraphics[width=0.245\textwidth]{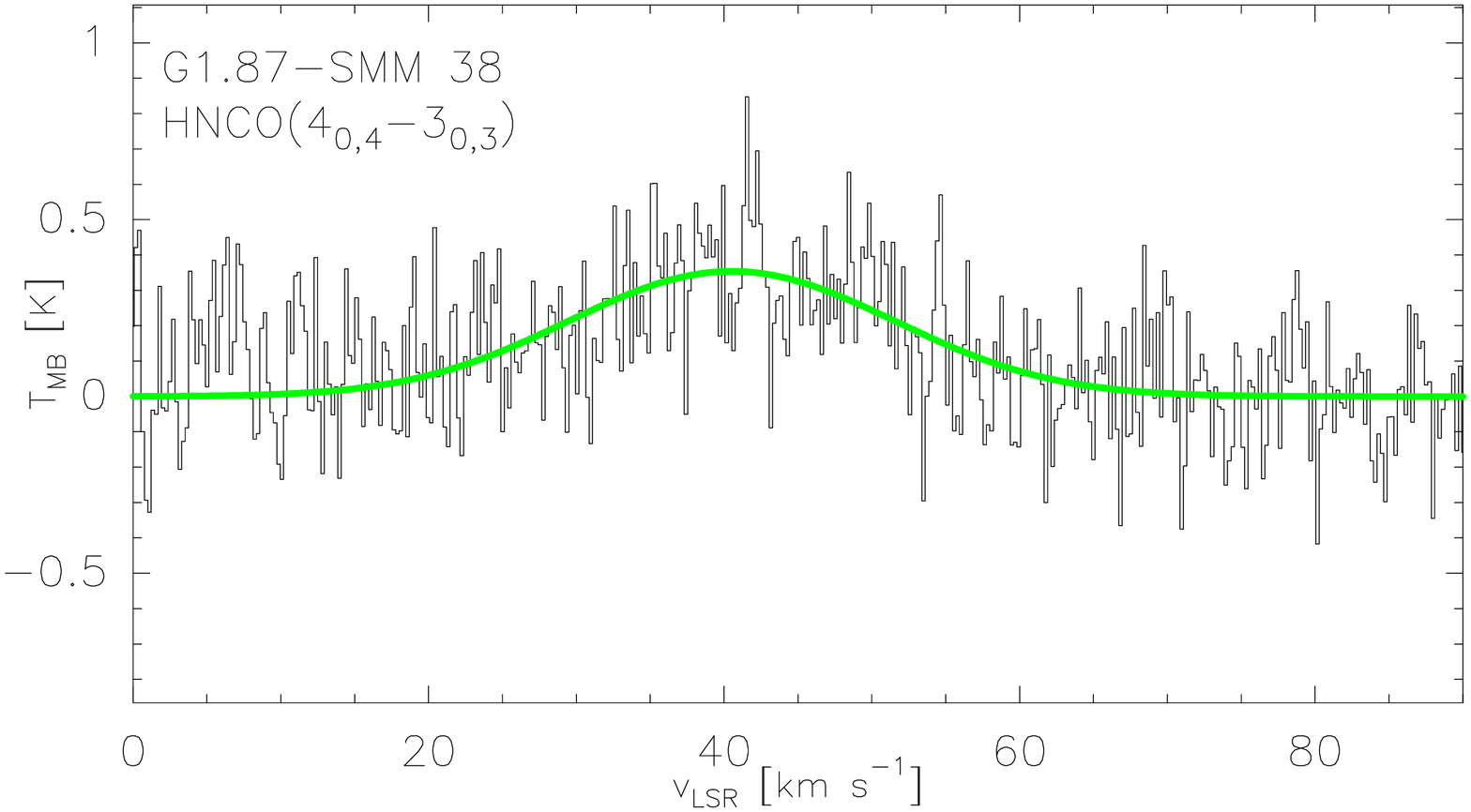}
\includegraphics[width=0.245\textwidth]{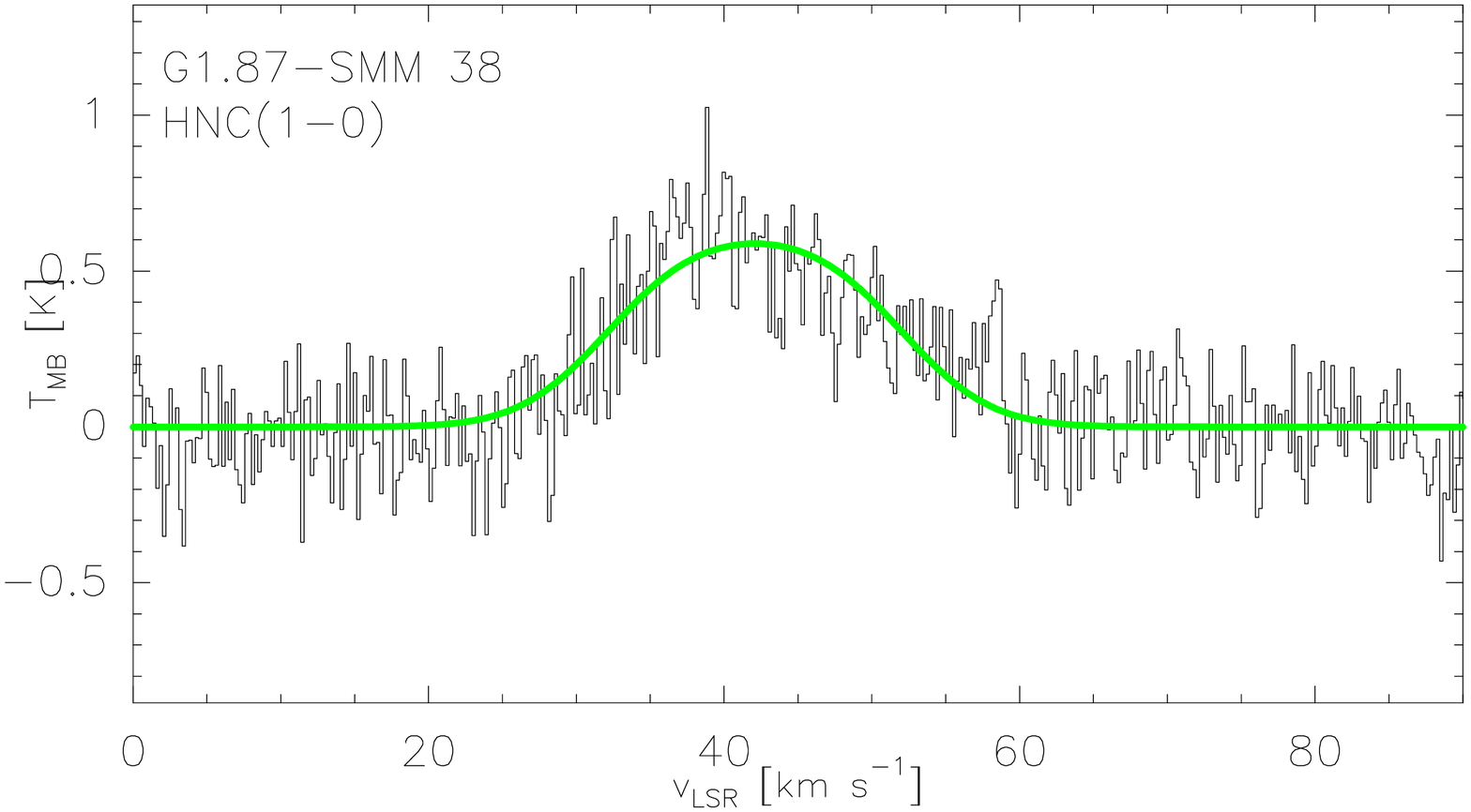}
\includegraphics[width=0.245\textwidth]{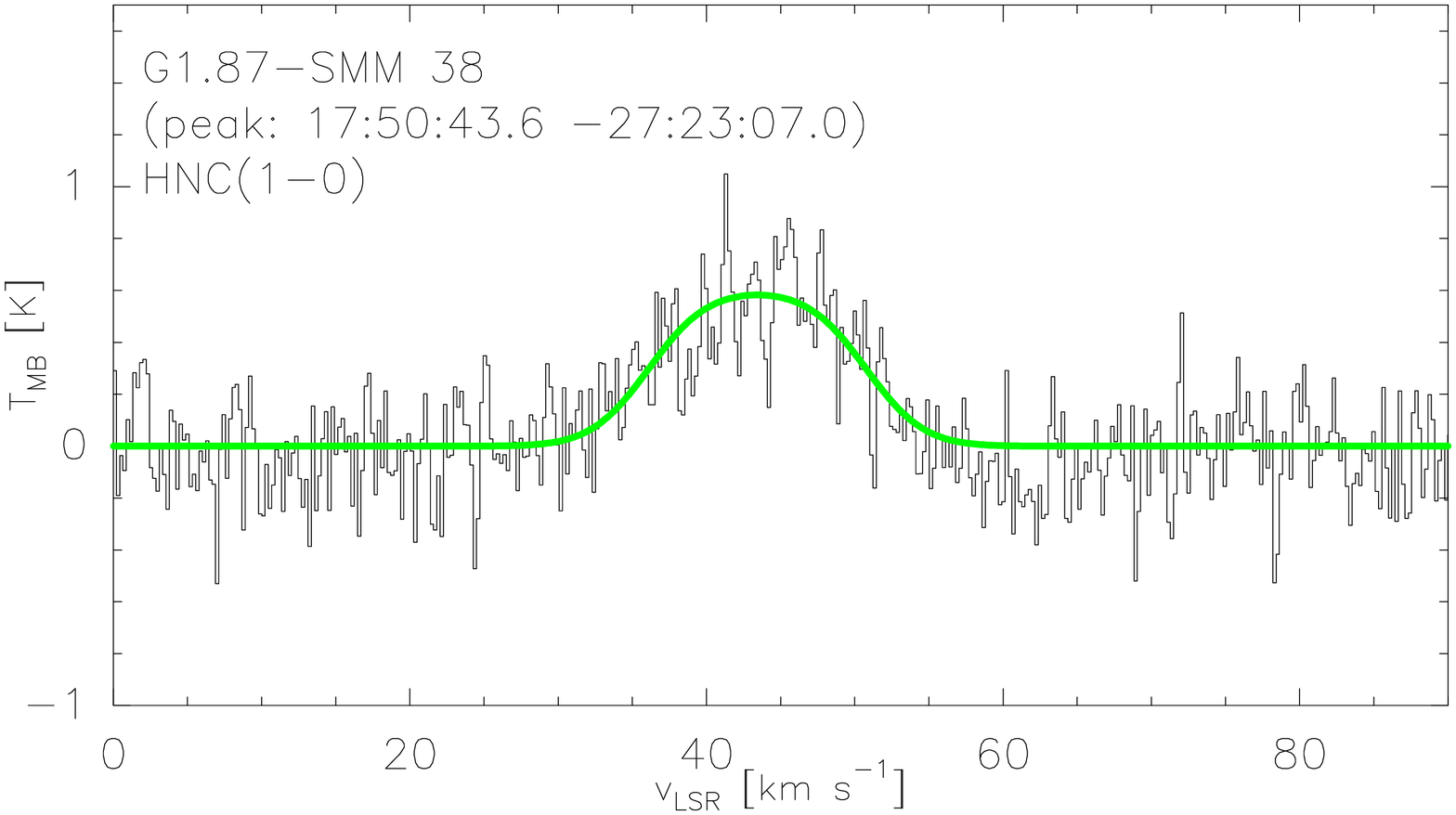}
\includegraphics[width=0.245\textwidth]{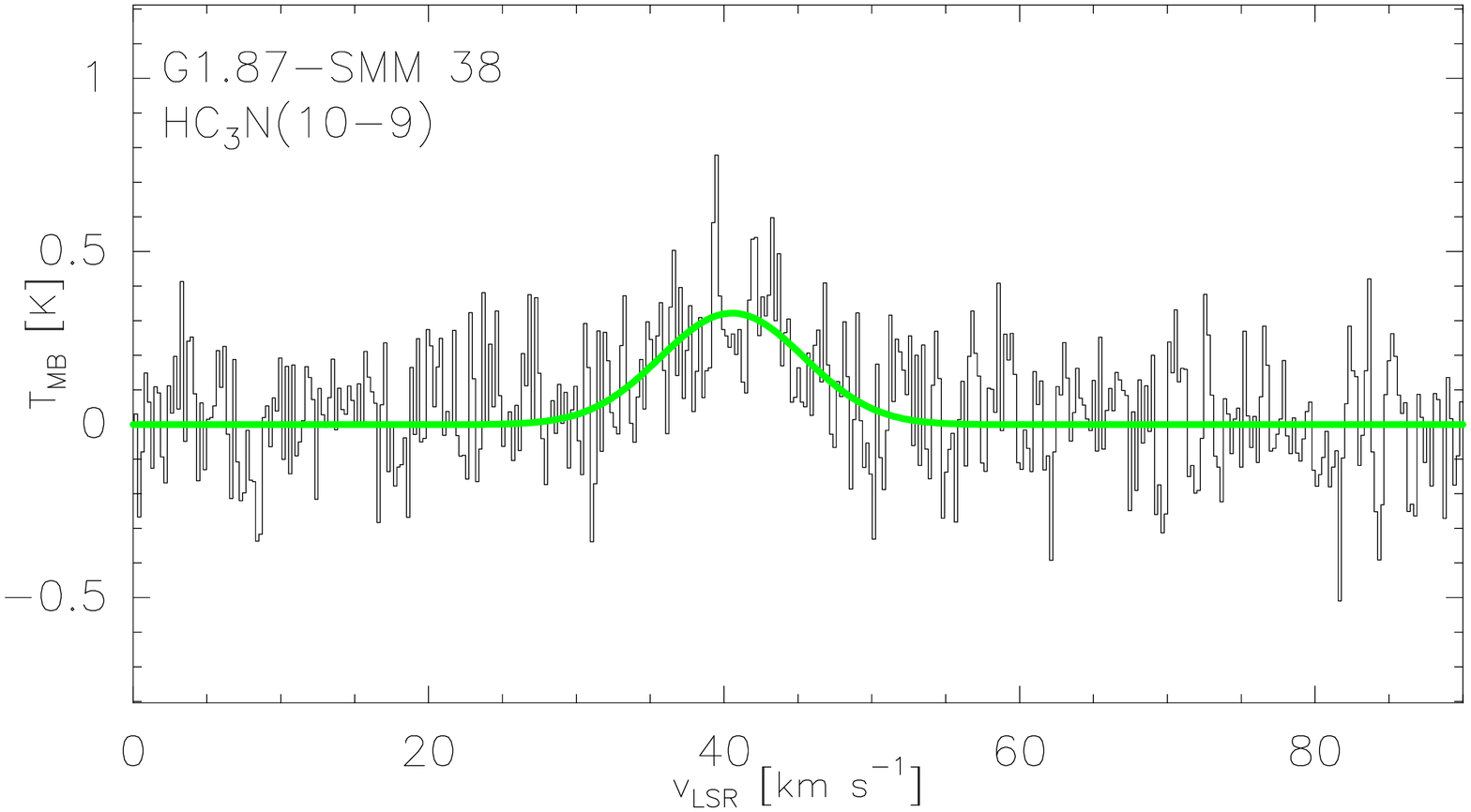}
\includegraphics[width=0.245\textwidth]{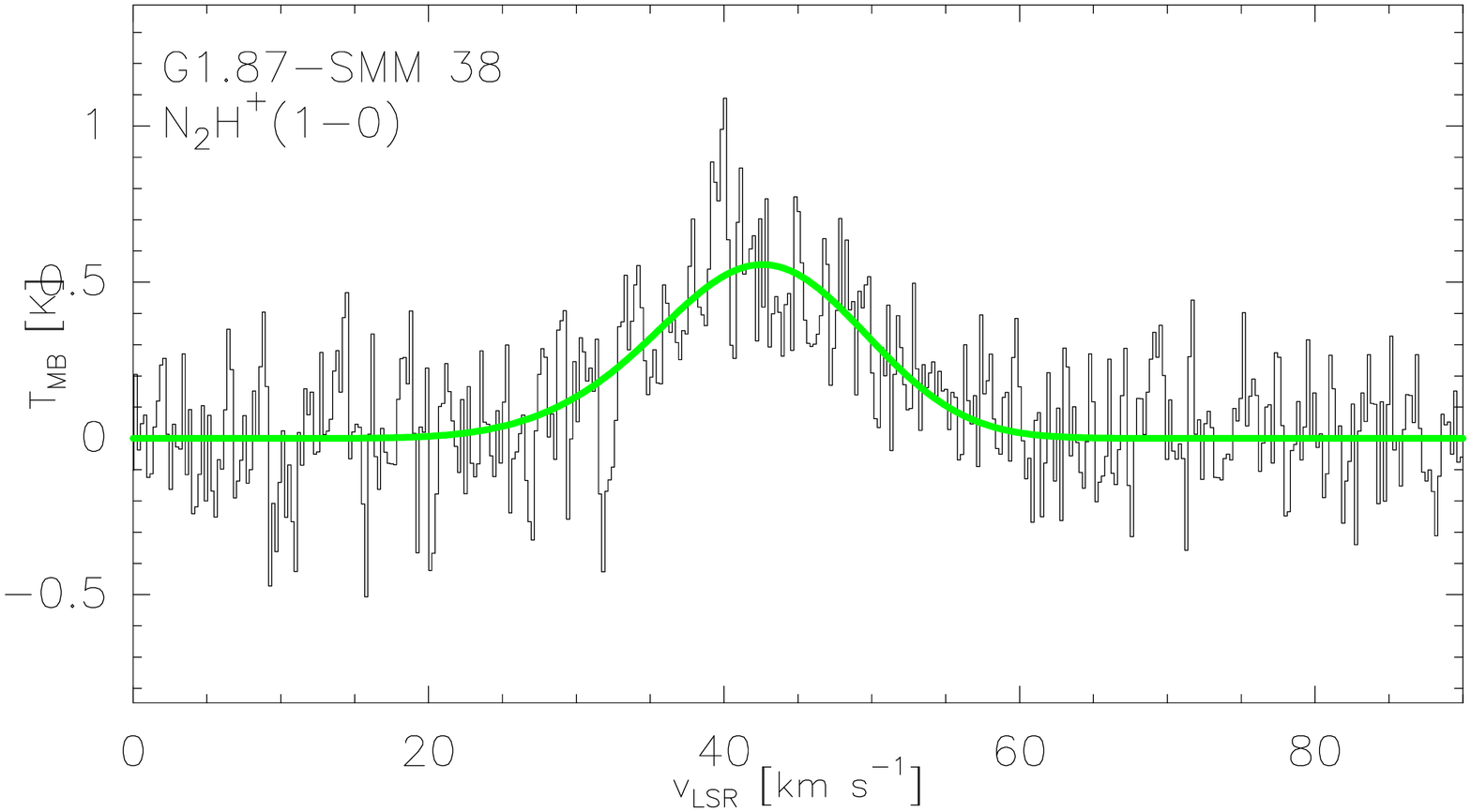}
\includegraphics[width=0.245\textwidth]{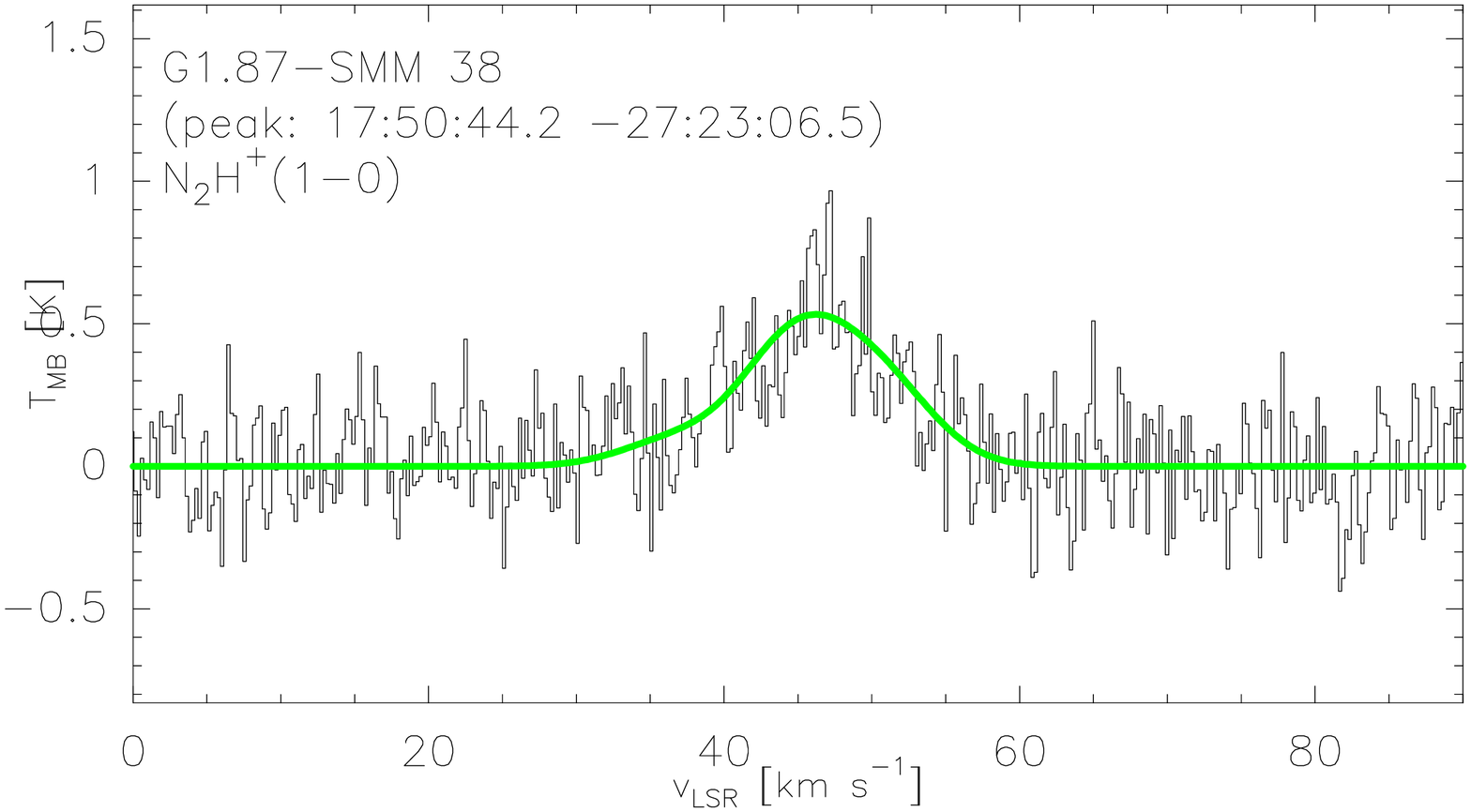}
\caption{Same as Fig.~\ref{figure:G187SMM1_spectra} but towards G1.87--SMM 38.}
\label{figure:G187SMM38_spectra}
\end{center}
\end{figure*}

\begin{figure*}
\begin{center}
\includegraphics[width=0.245\textwidth]{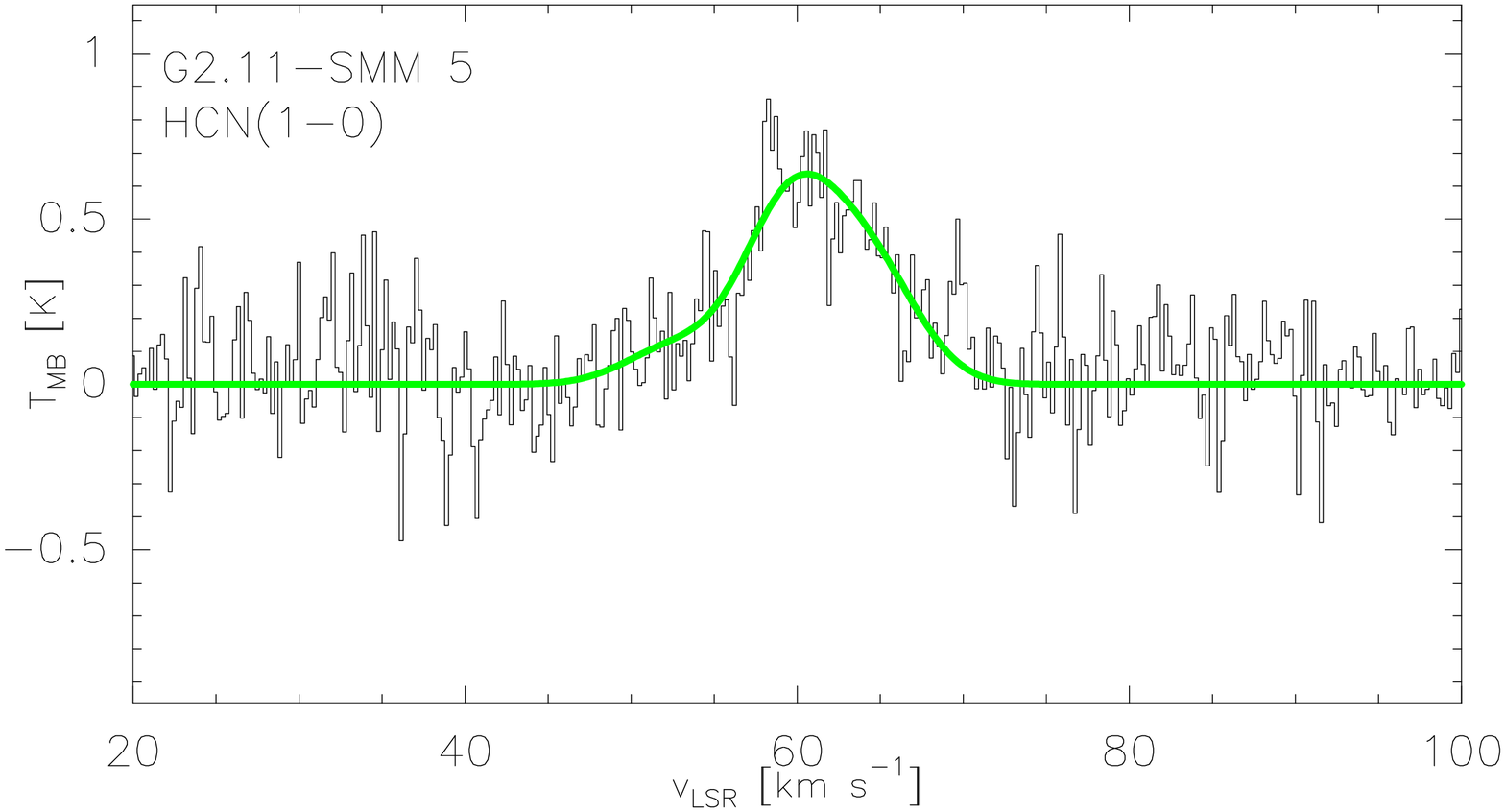}
\includegraphics[width=0.245\textwidth]{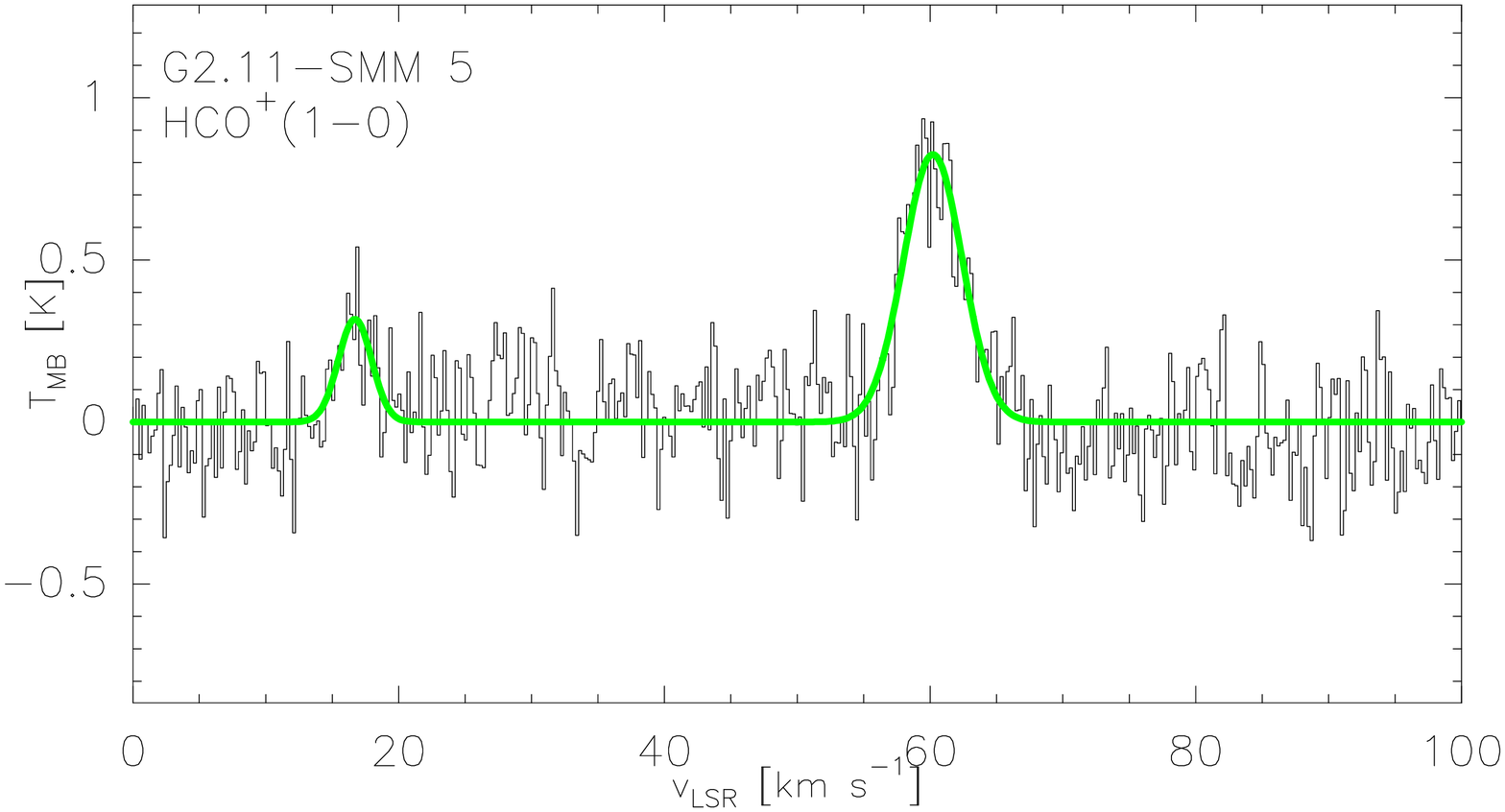}
\includegraphics[width=0.245\textwidth]{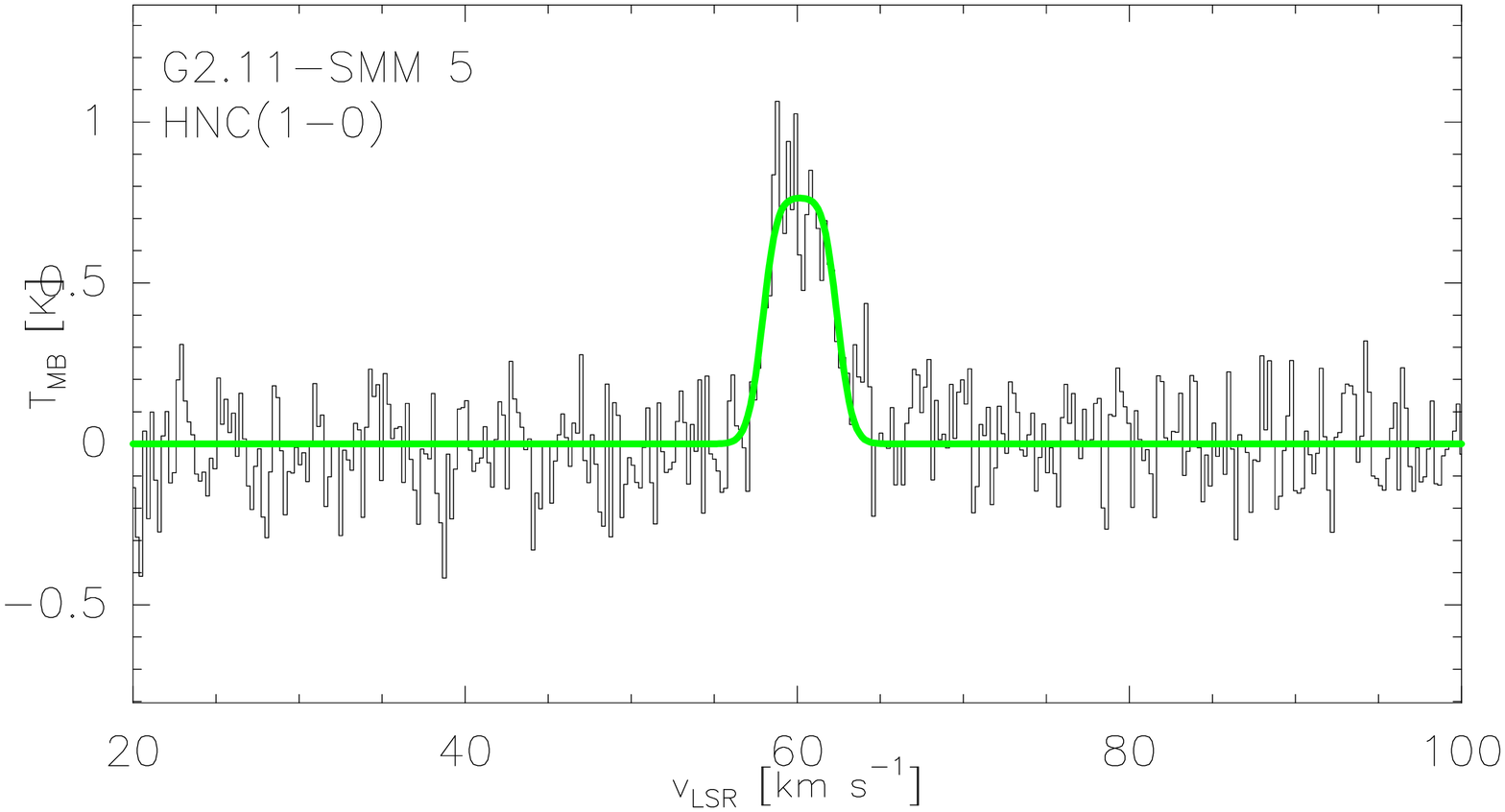}
\includegraphics[width=0.245\textwidth]{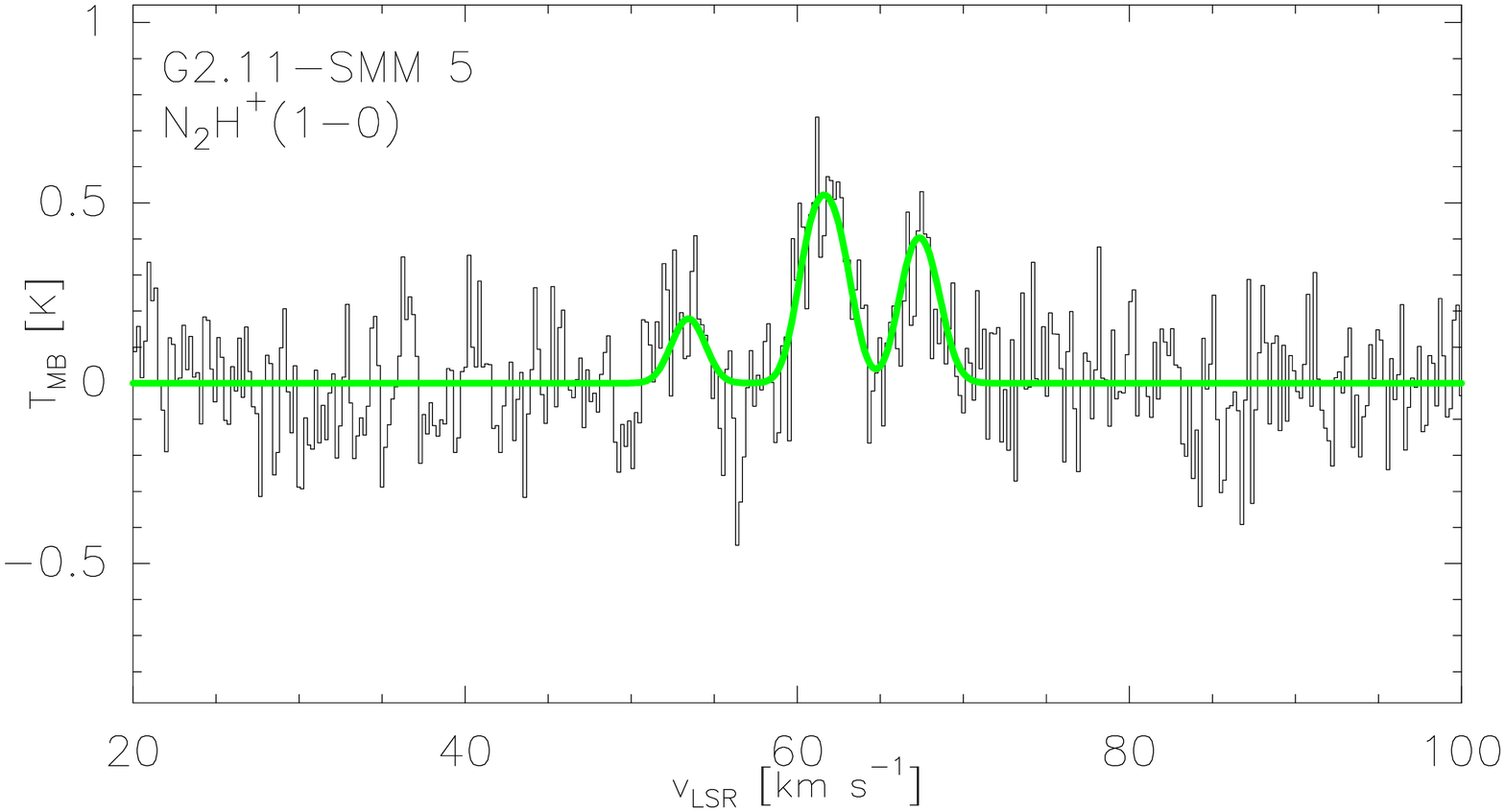}
\caption{Same as Fig.~\ref{figure:G187SMM1_spectra} but towards G2.11--SMM 5. 
A wider velocity range for the HCO$^+$ spectrum is shown due to the additional 
velocity component.}
\label{figure:G211SMM5_spectra}
\end{center}
\end{figure*}

\begin{figure*}
\begin{center}
\includegraphics[width=0.245\textwidth]{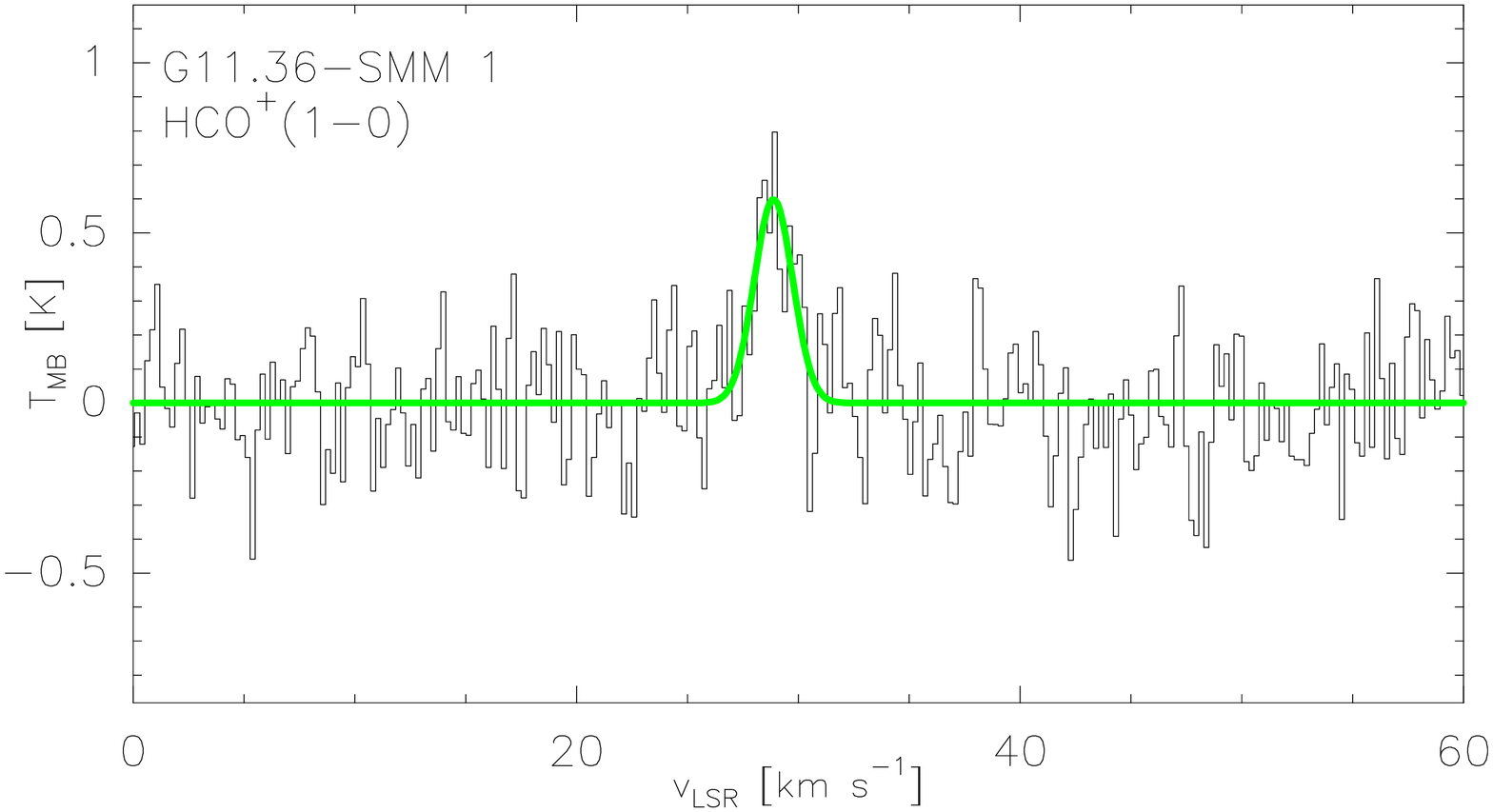}
\includegraphics[width=0.245\textwidth]{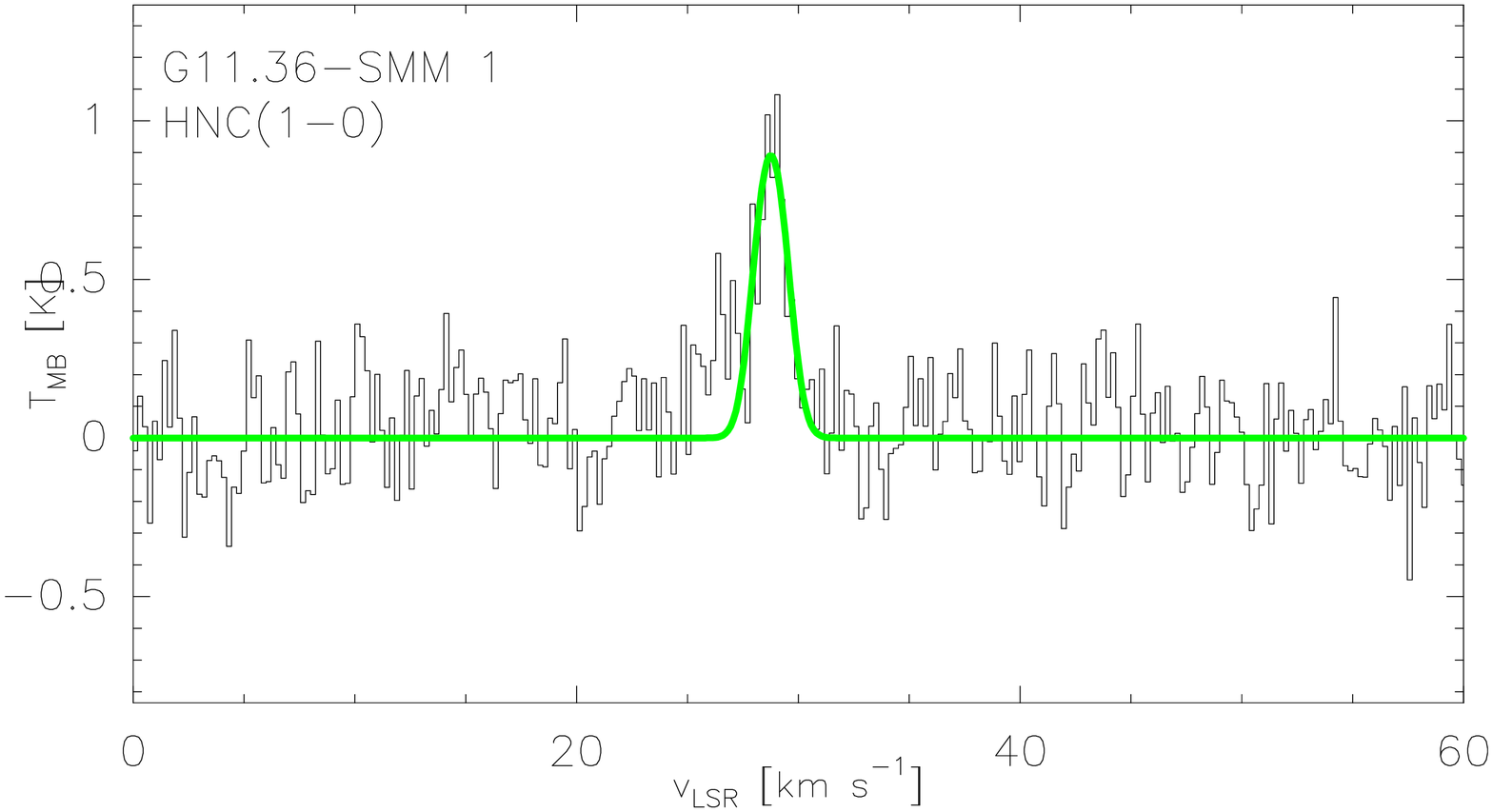}
\includegraphics[width=0.245\textwidth]{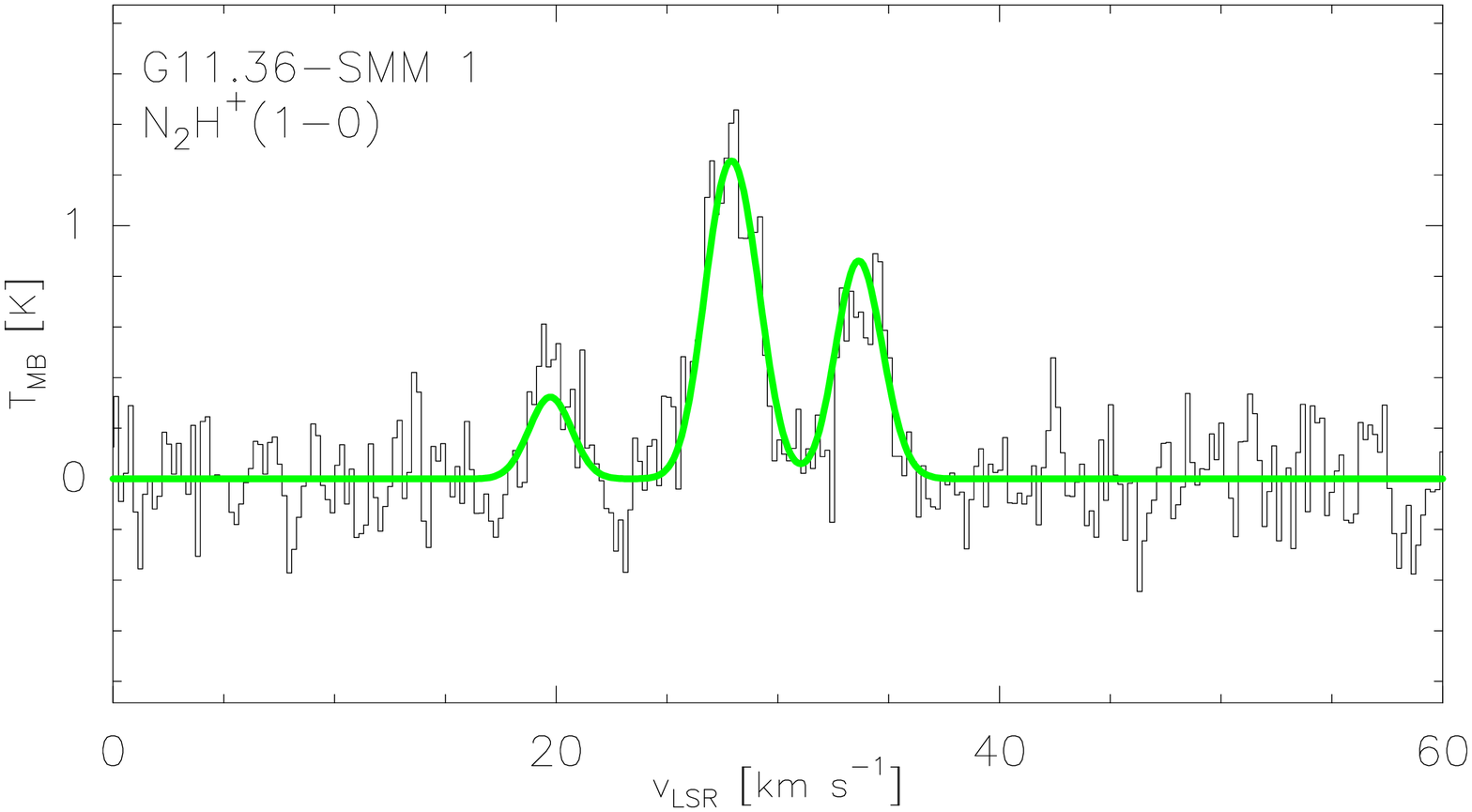}
\caption{Same as Fig.~\ref{figure:G187SMM1_spectra} but towards G11.36--SMM 1.}
\label{figure:G1136SMM1_spectra}
\end{center}
\end{figure*}

\begin{figure*}
\begin{center}
\includegraphics[width=0.245\textwidth]{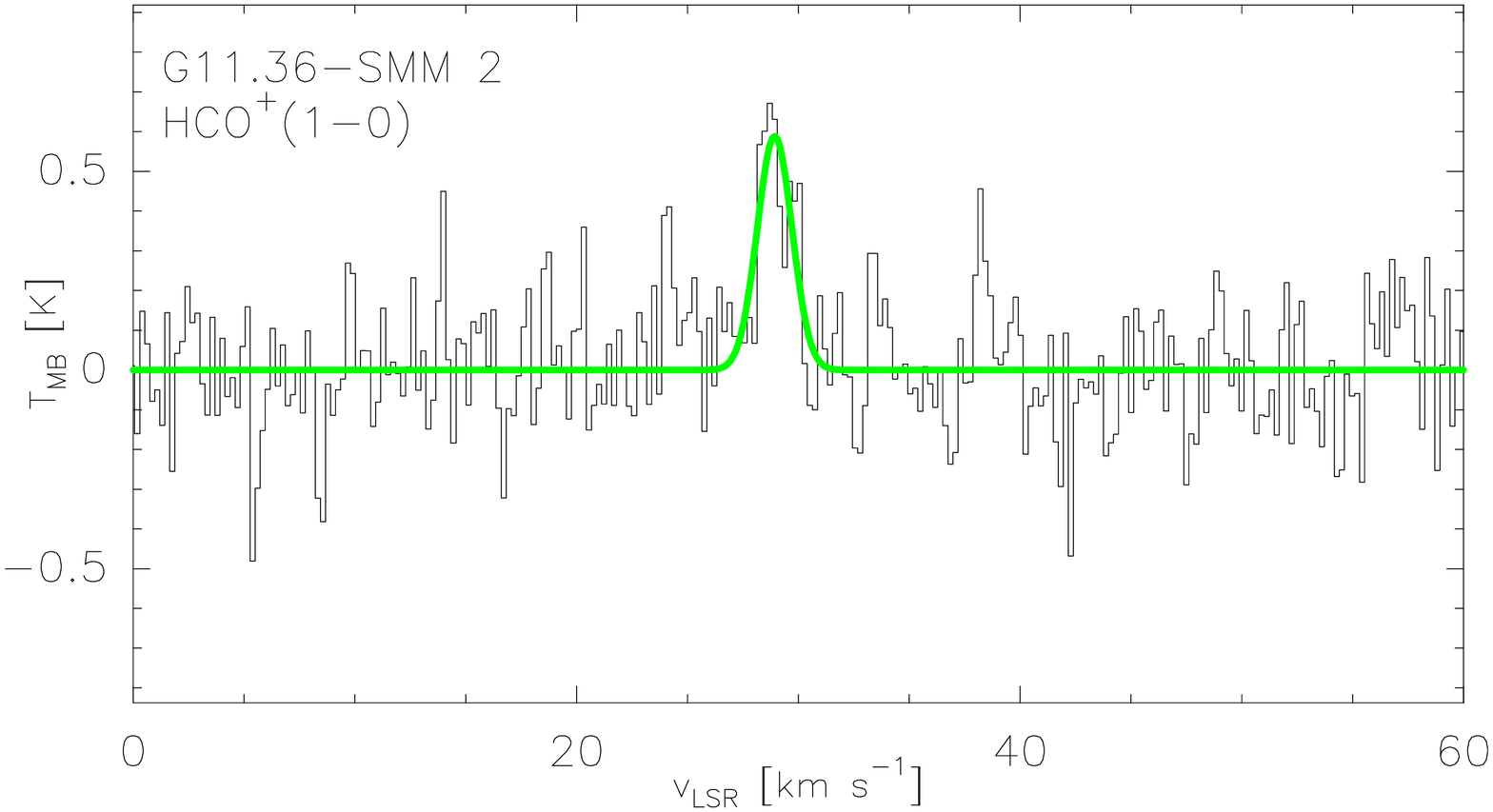}
\includegraphics[width=0.245\textwidth]{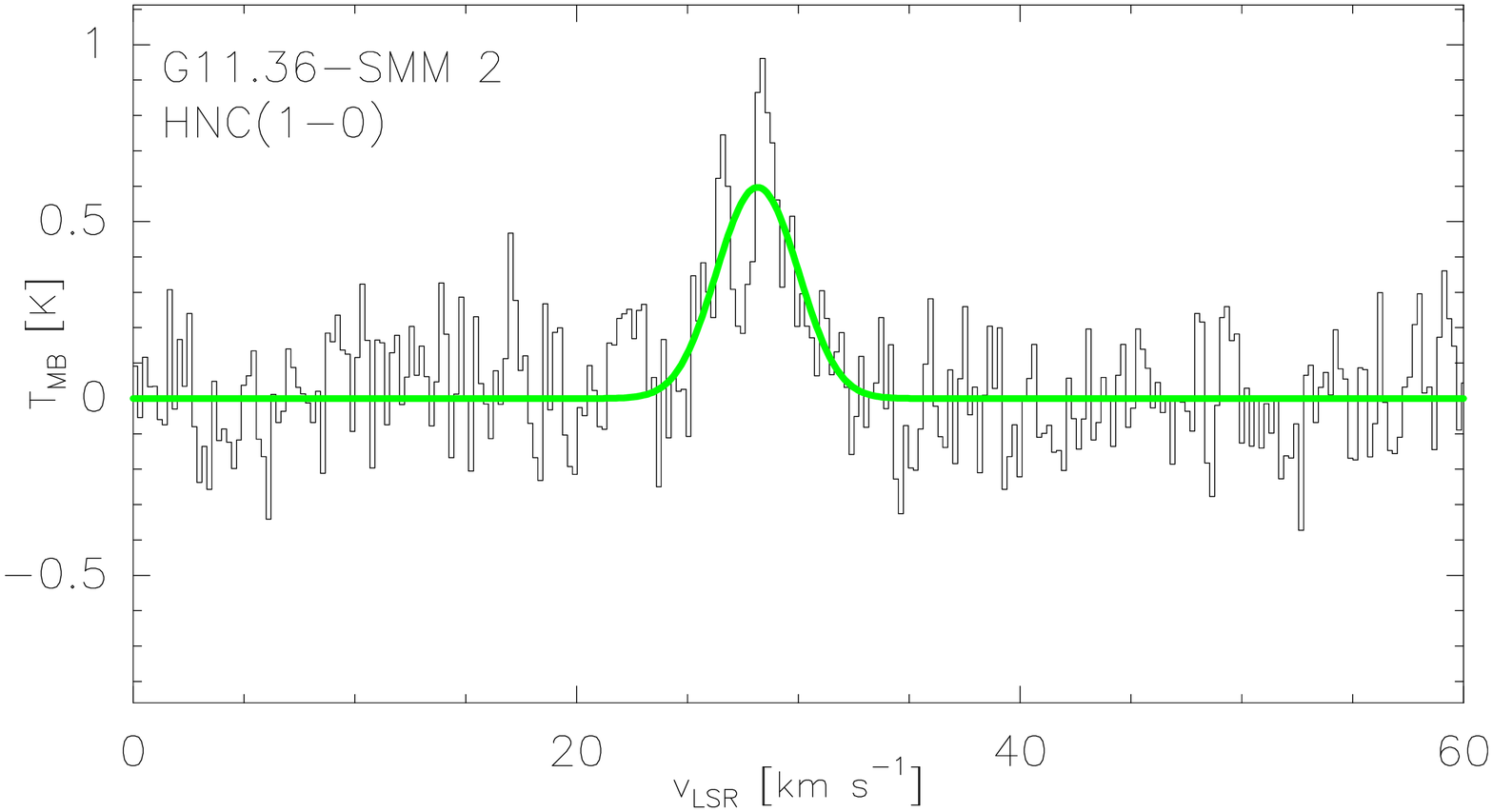}
\includegraphics[width=0.245\textwidth]{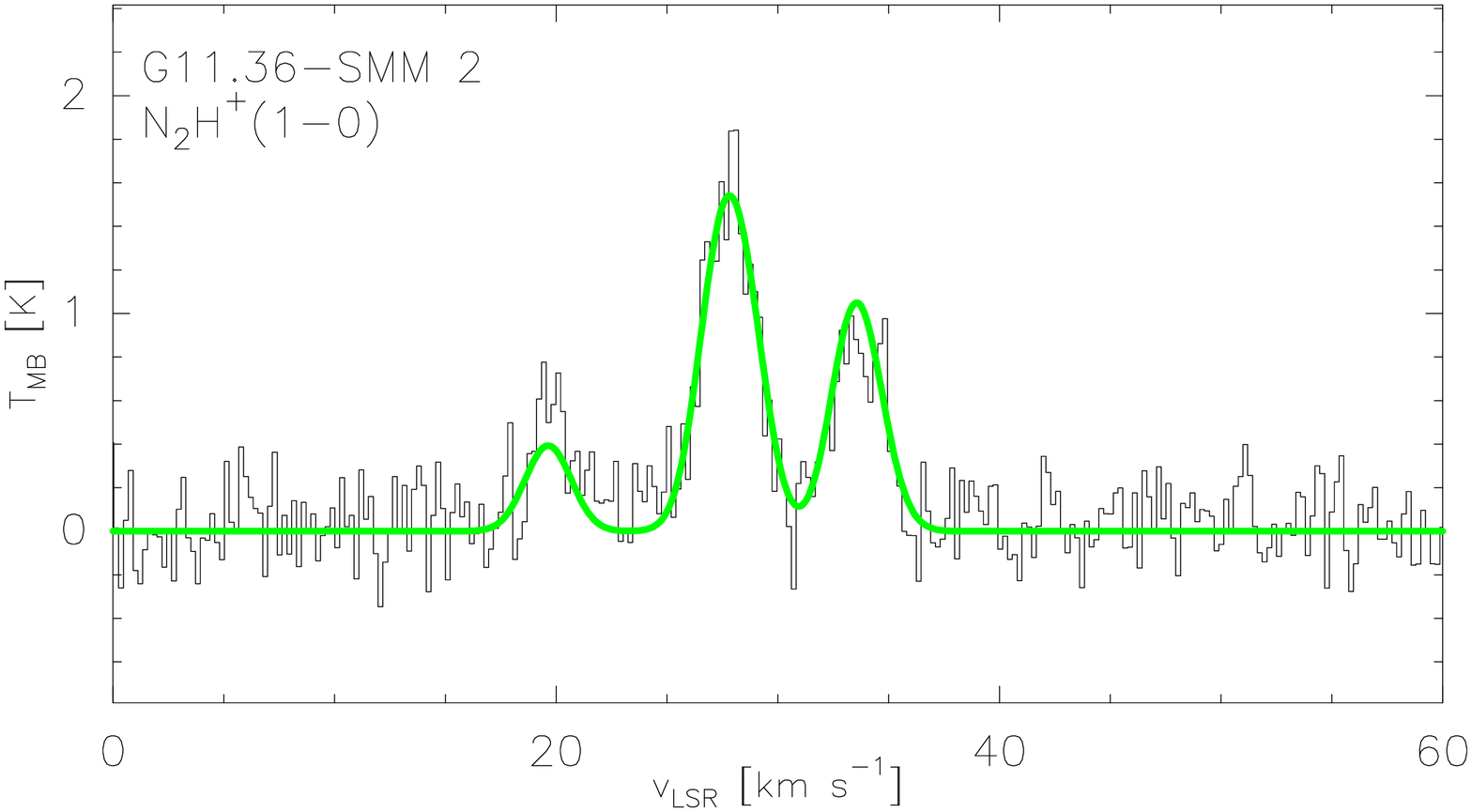}
\caption{Same as Fig.~\ref{figure:G187SMM1_spectra} but towards G11.36--SMM 2.}
\label{figure:G1136SMM2_spectra}
\end{center}
\end{figure*}

\begin{figure*}
\begin{center}
\includegraphics[width=0.245\textwidth]{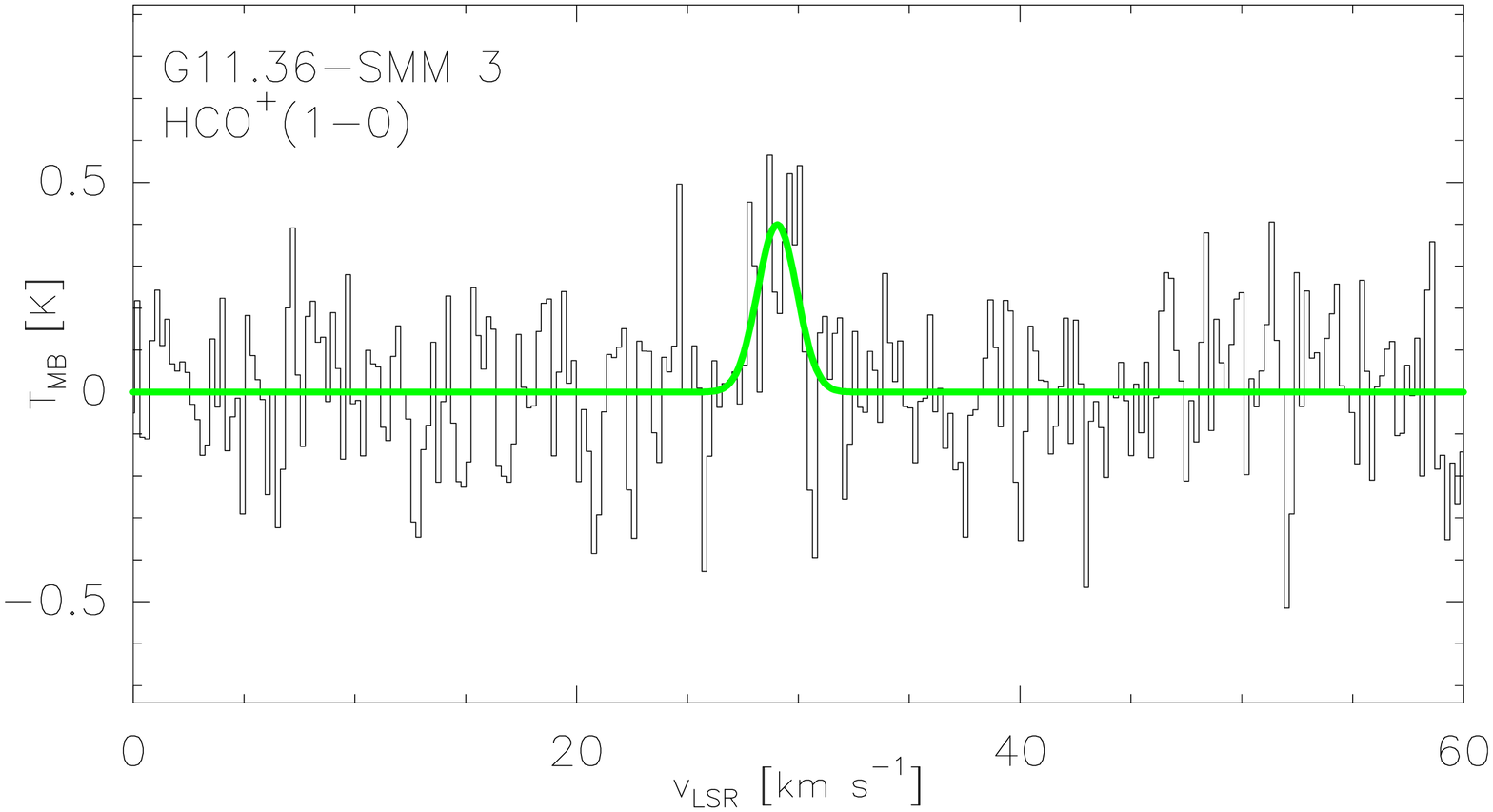}
\includegraphics[width=0.245\textwidth]{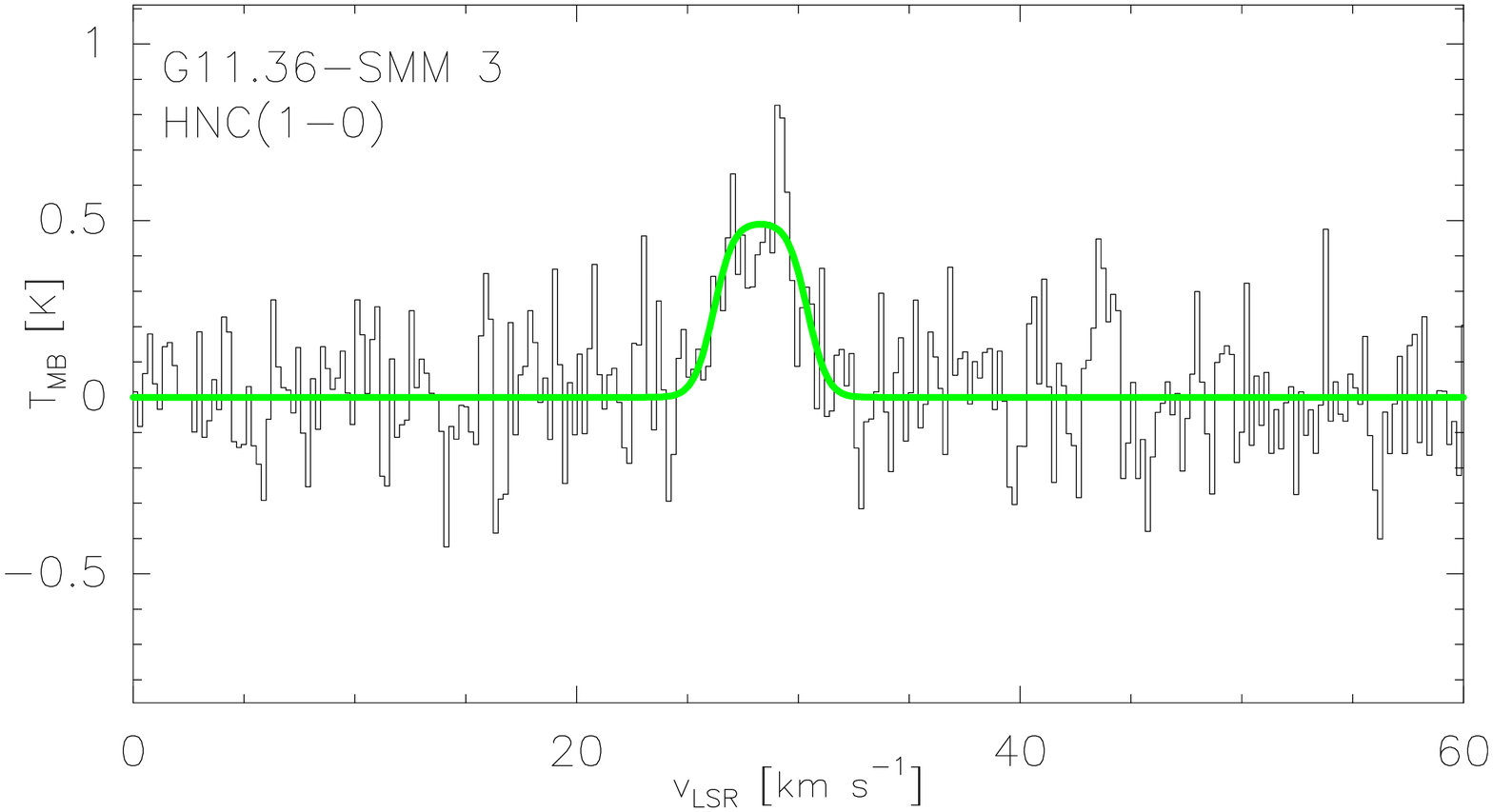}
\includegraphics[width=0.245\textwidth]{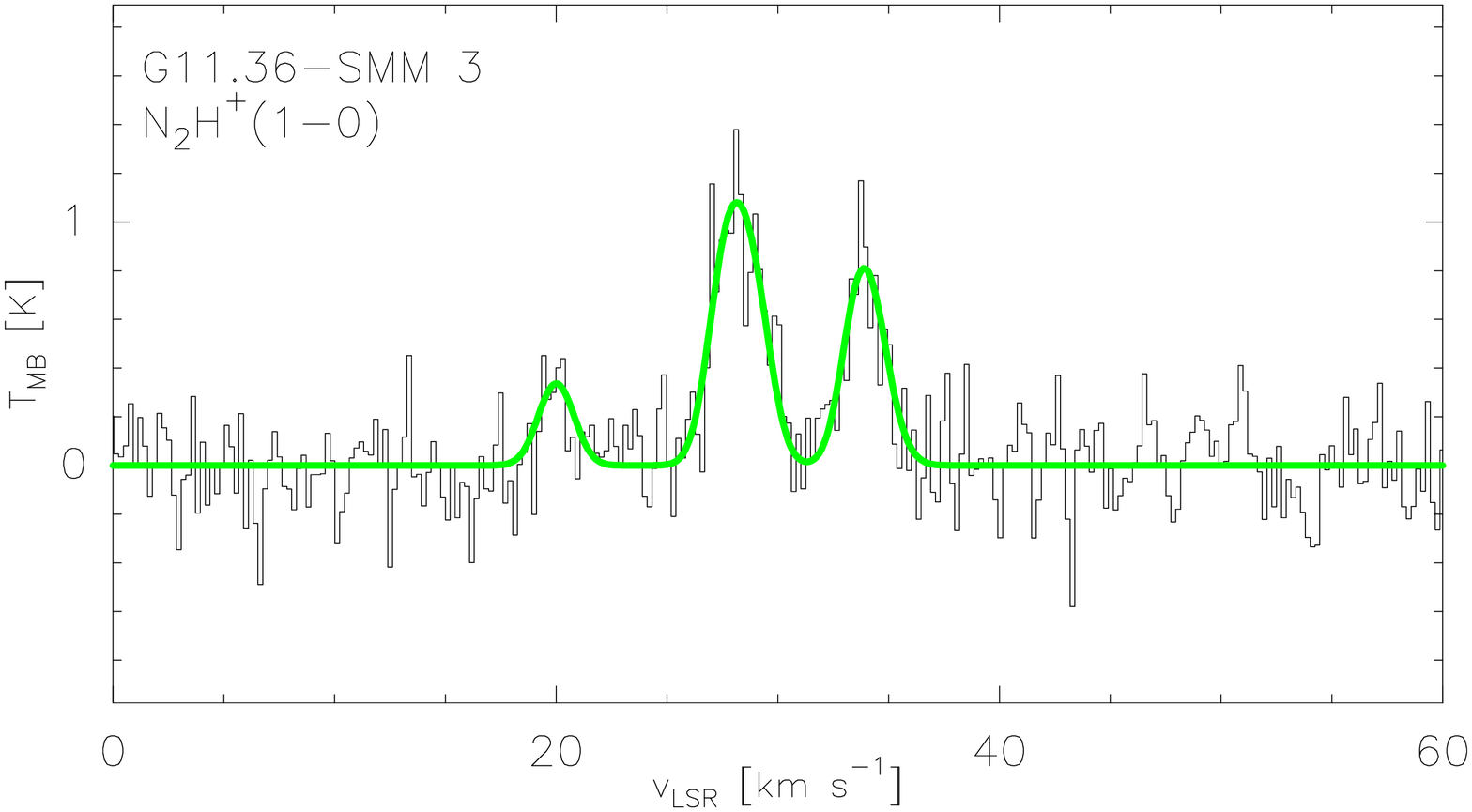}
\caption{Same as Fig.~\ref{figure:G187SMM1_spectra} but towards G11.36--SMM 3.}
\label{figure:G1136SMM3_spectra}
\end{center}
\end{figure*}

\begin{figure*}
\begin{center}
\includegraphics[width=0.245\textwidth]{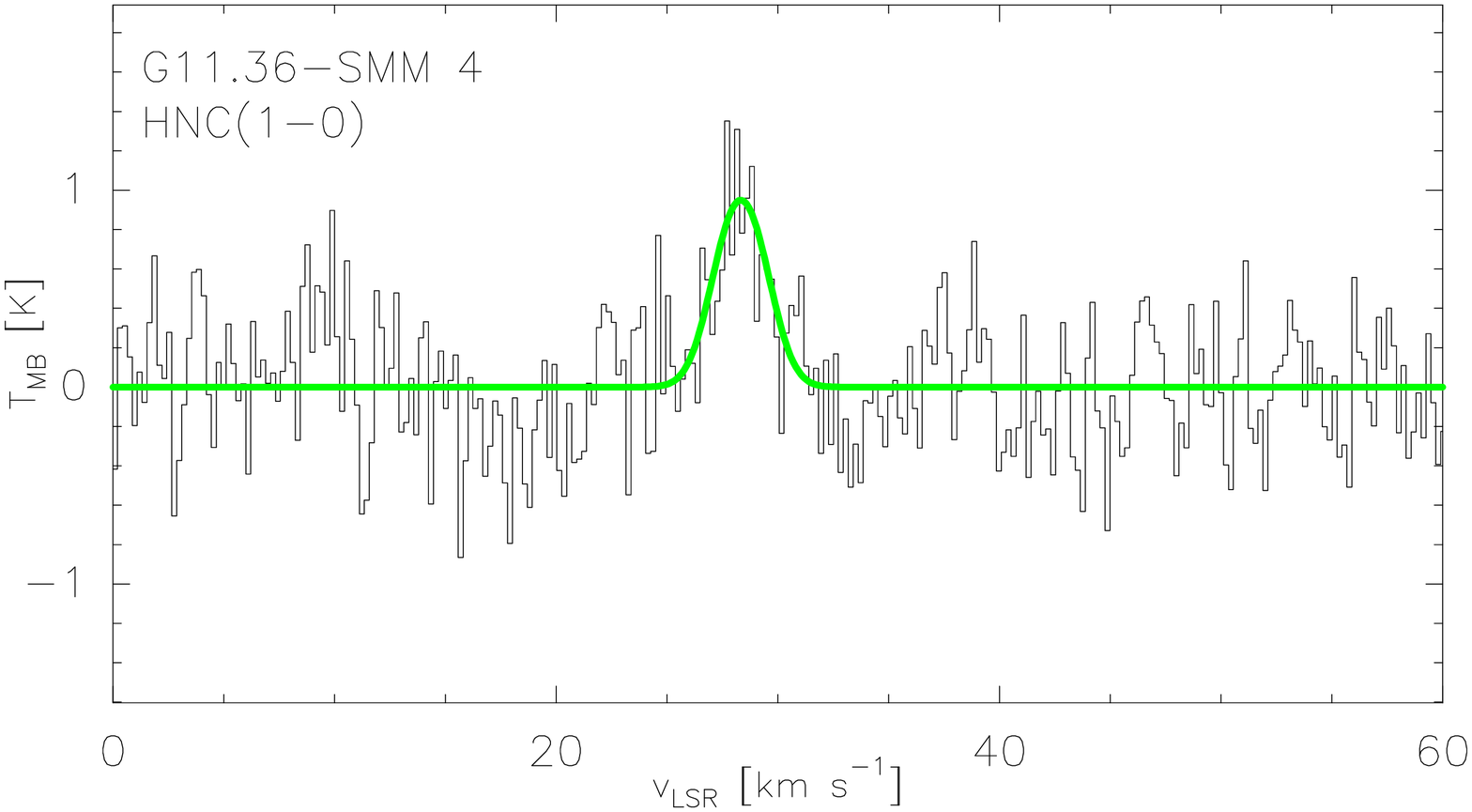}
\includegraphics[width=0.245\textwidth]{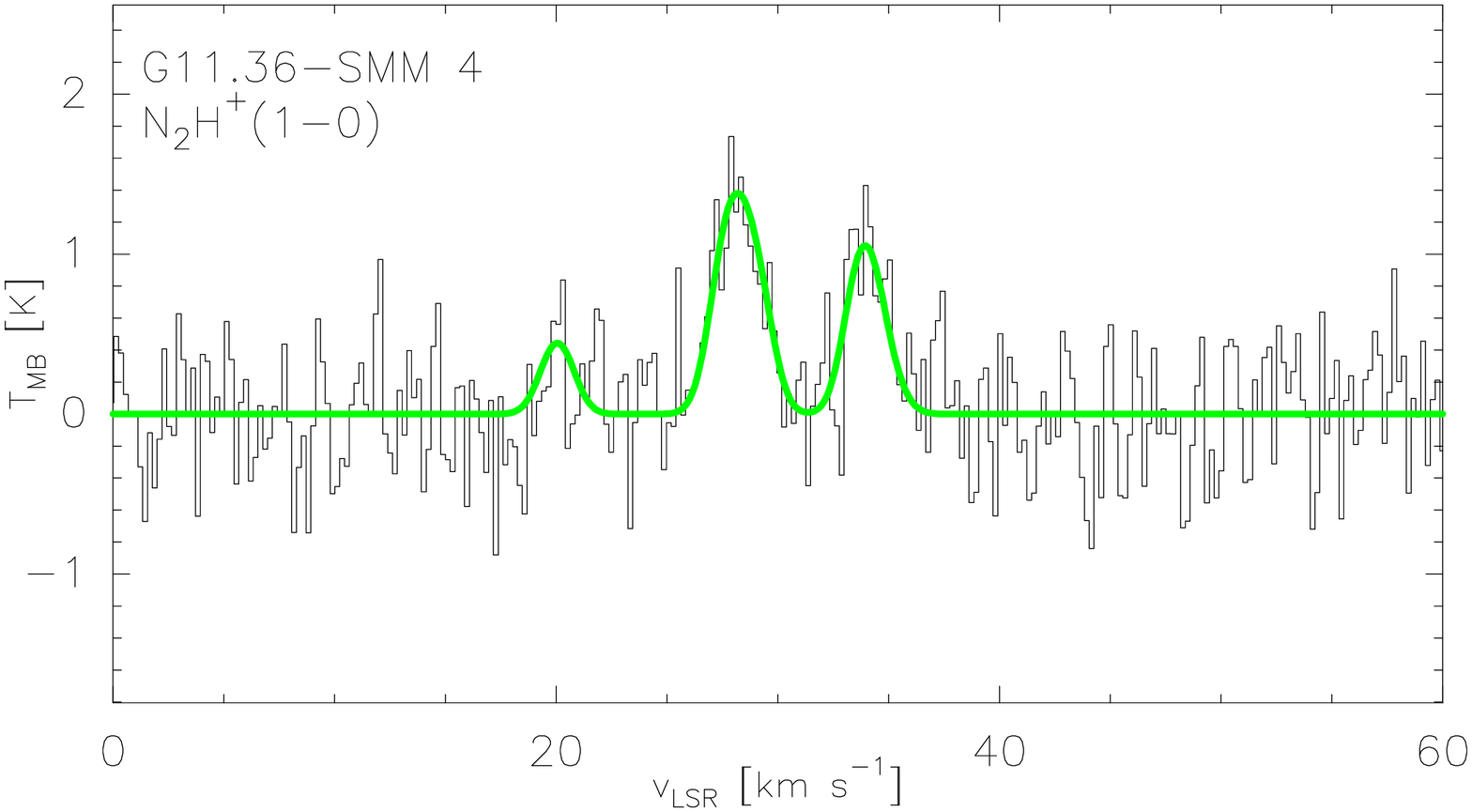}
\caption{Same as Fig.~\ref{figure:G187SMM1_spectra} but towards G11.36--SMM 4.}
\label{figure:G1136SMM4_spectra}
\end{center}
\end{figure*}

\begin{figure*}
\begin{center}
\includegraphics[width=0.245\textwidth]{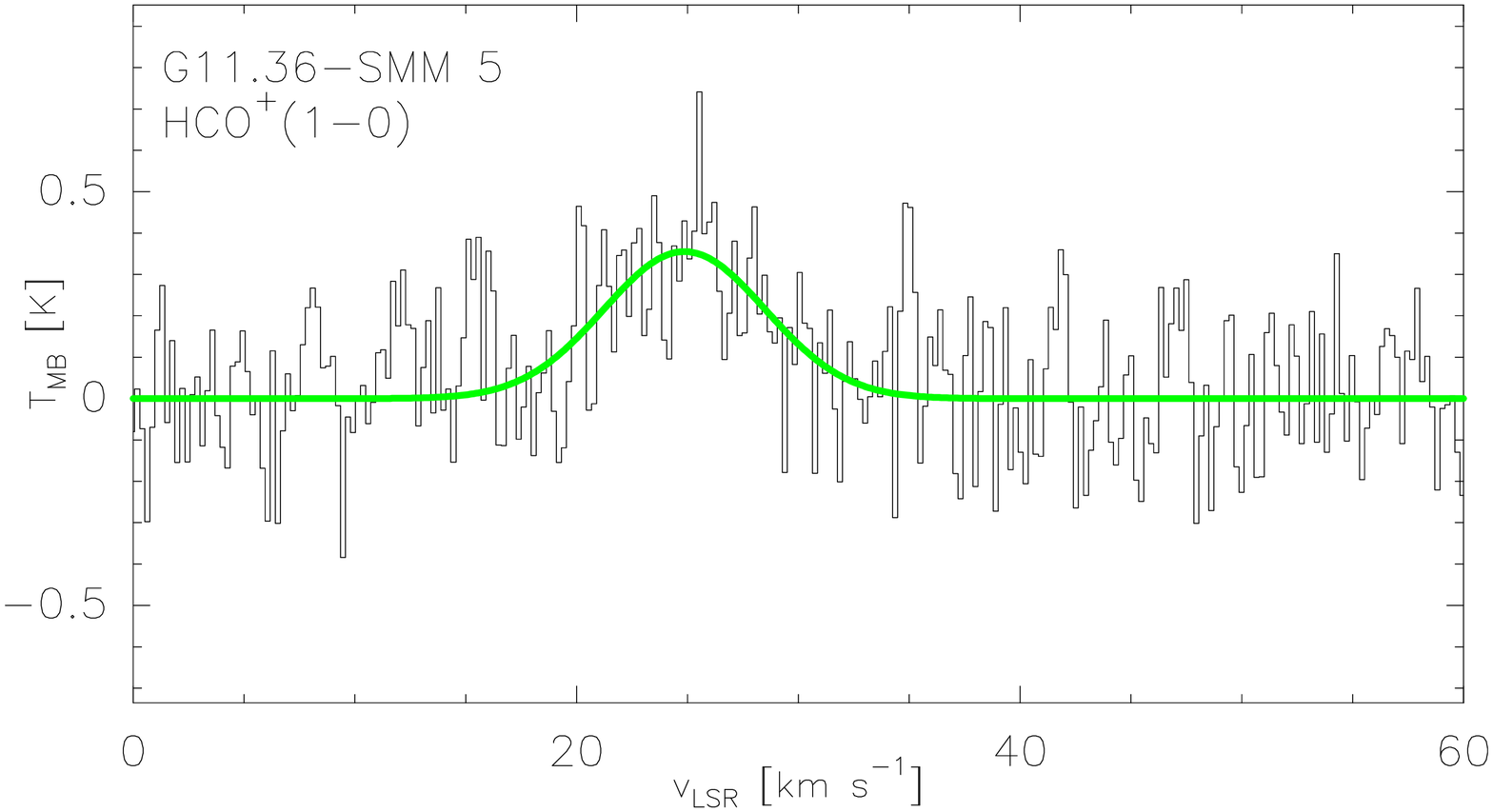}
\includegraphics[width=0.245\textwidth]{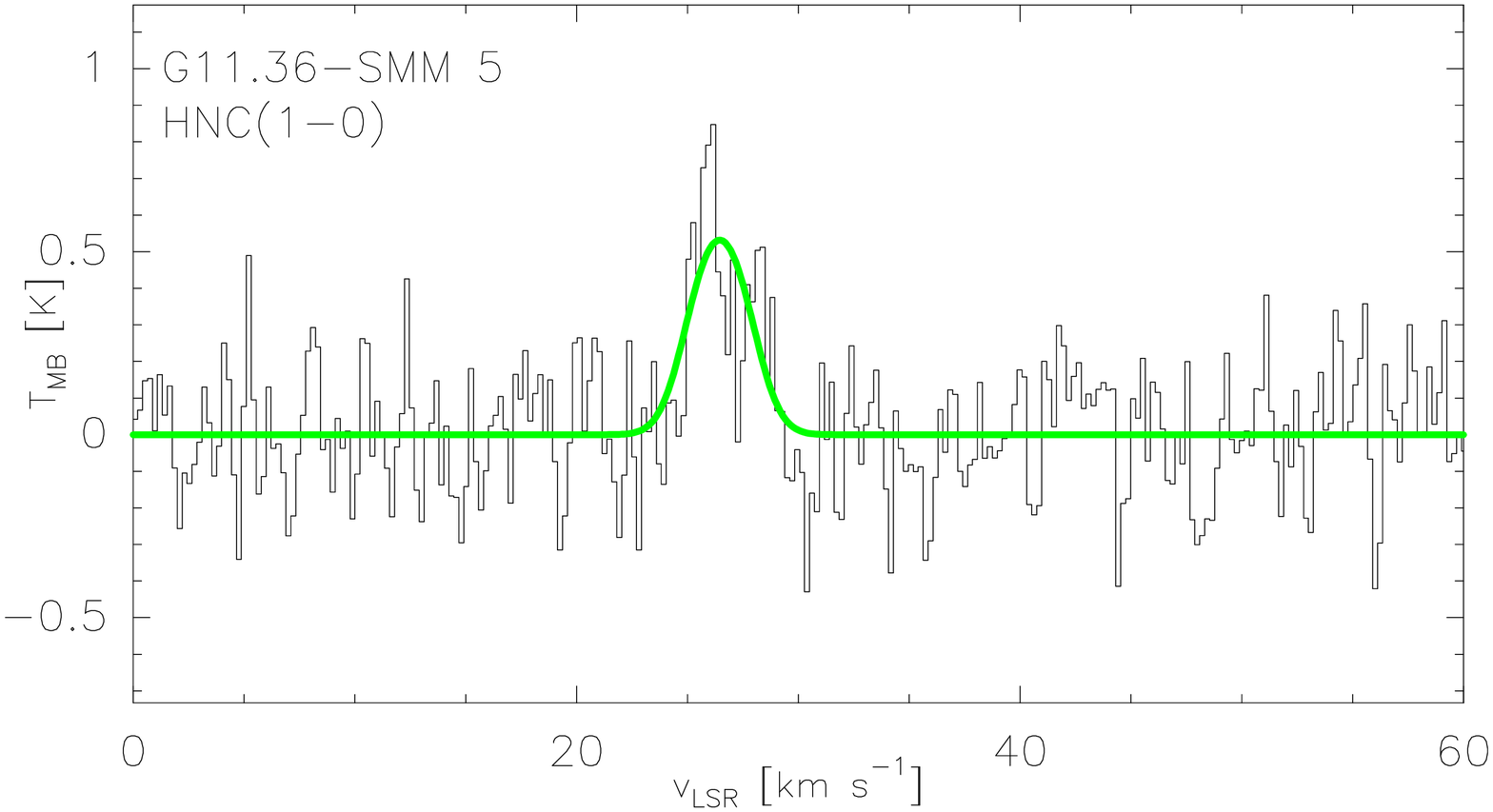}
\includegraphics[width=0.245\textwidth]{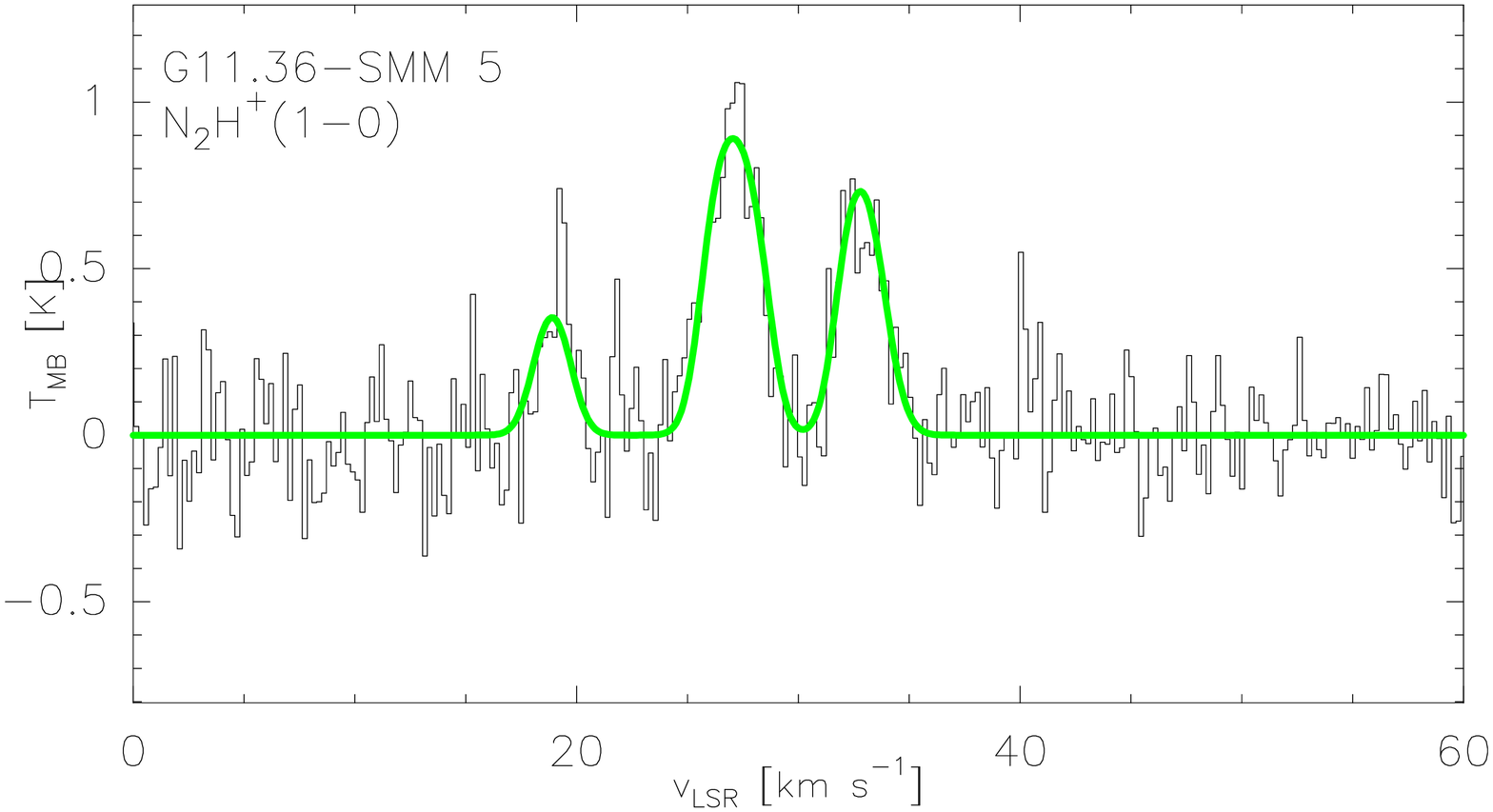}
\caption{Same as Fig.~\ref{figure:G187SMM1_spectra} but towards G11.36--SMM 5.}
\label{figure:G1136SMM5_spectra}
\end{center}
\end{figure*}

\begin{figure*}
\begin{center}
\includegraphics[width=0.245\textwidth]{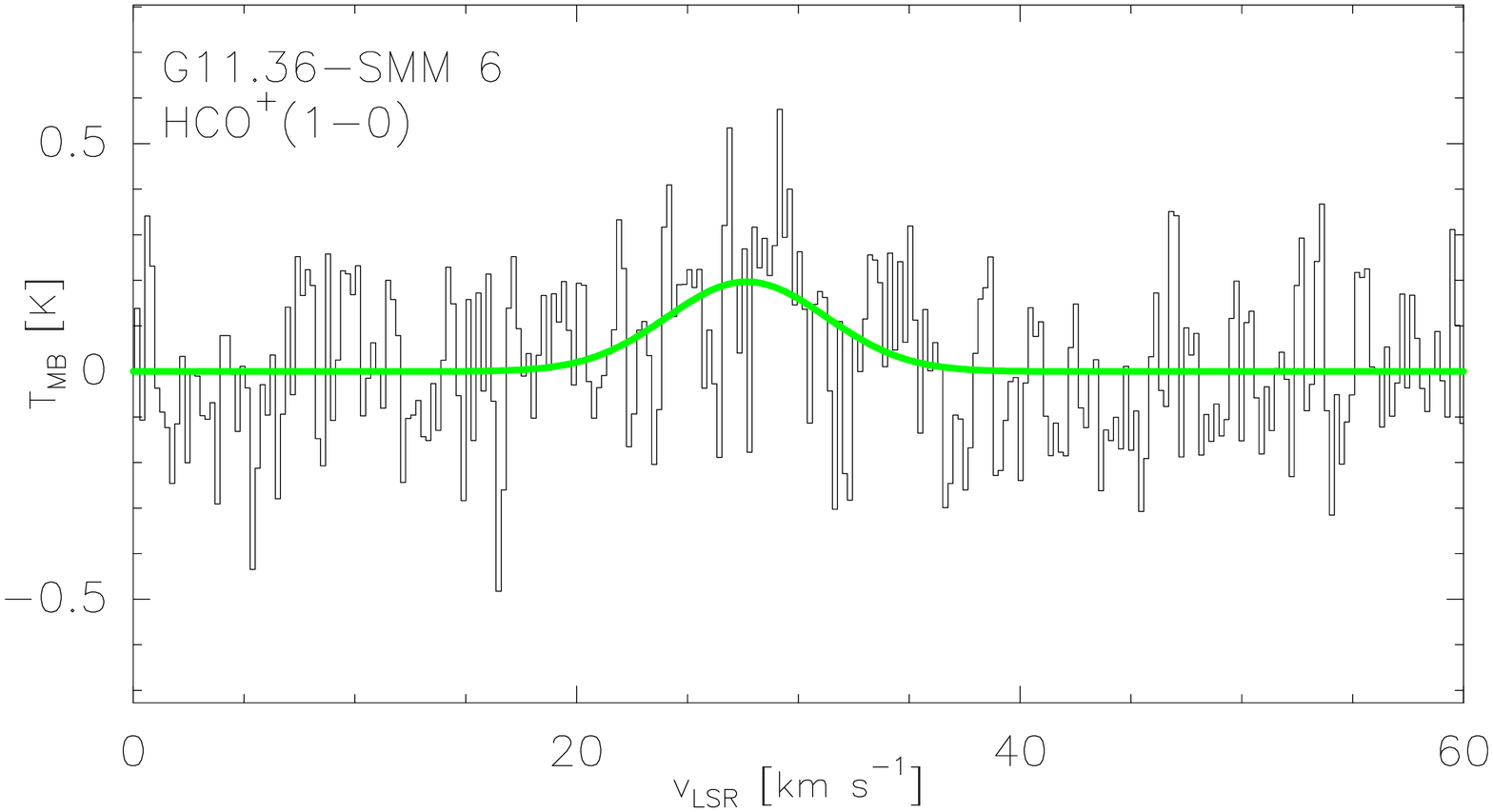}
\includegraphics[width=0.245\textwidth]{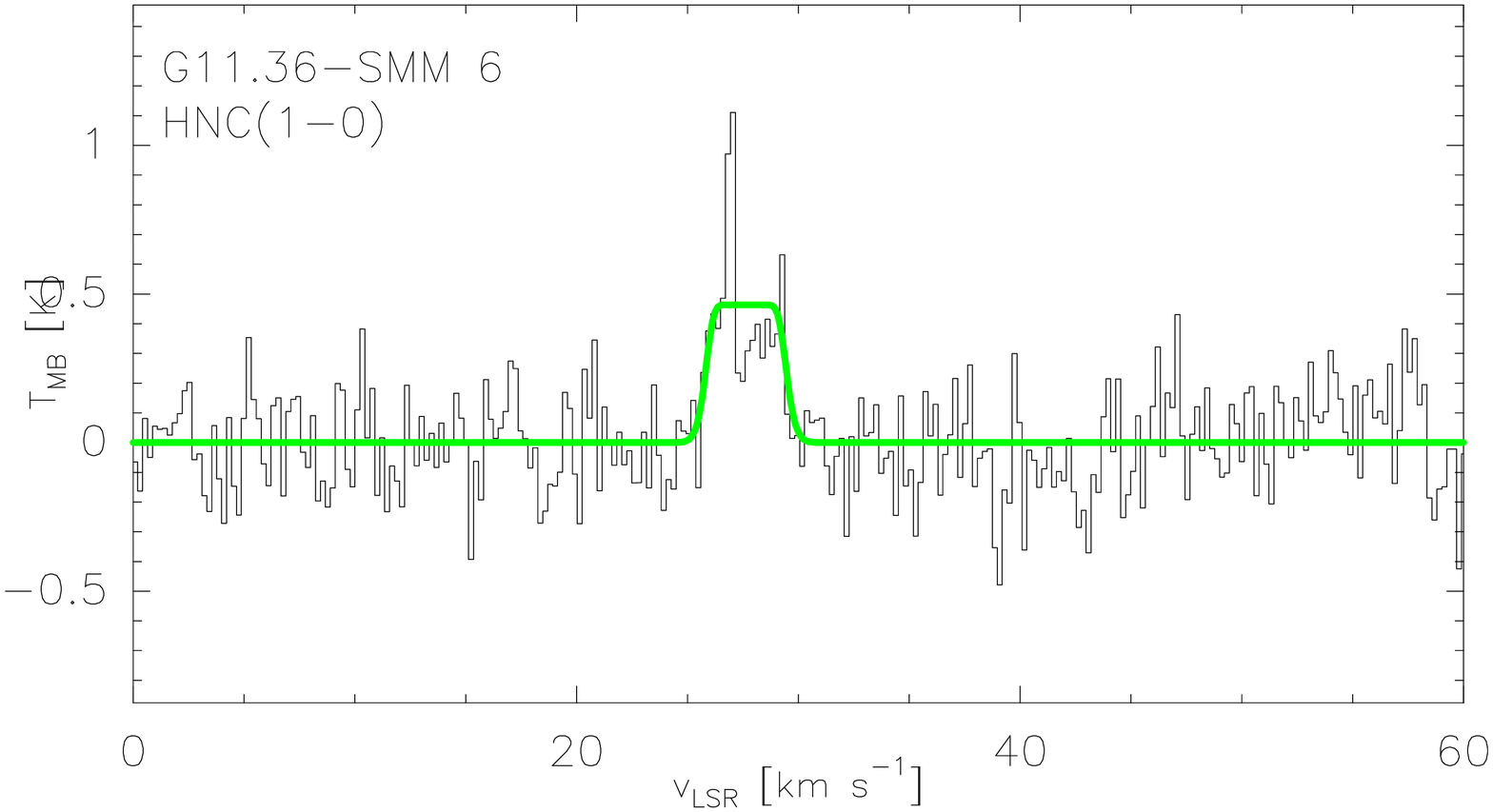}
\includegraphics[width=0.245\textwidth]{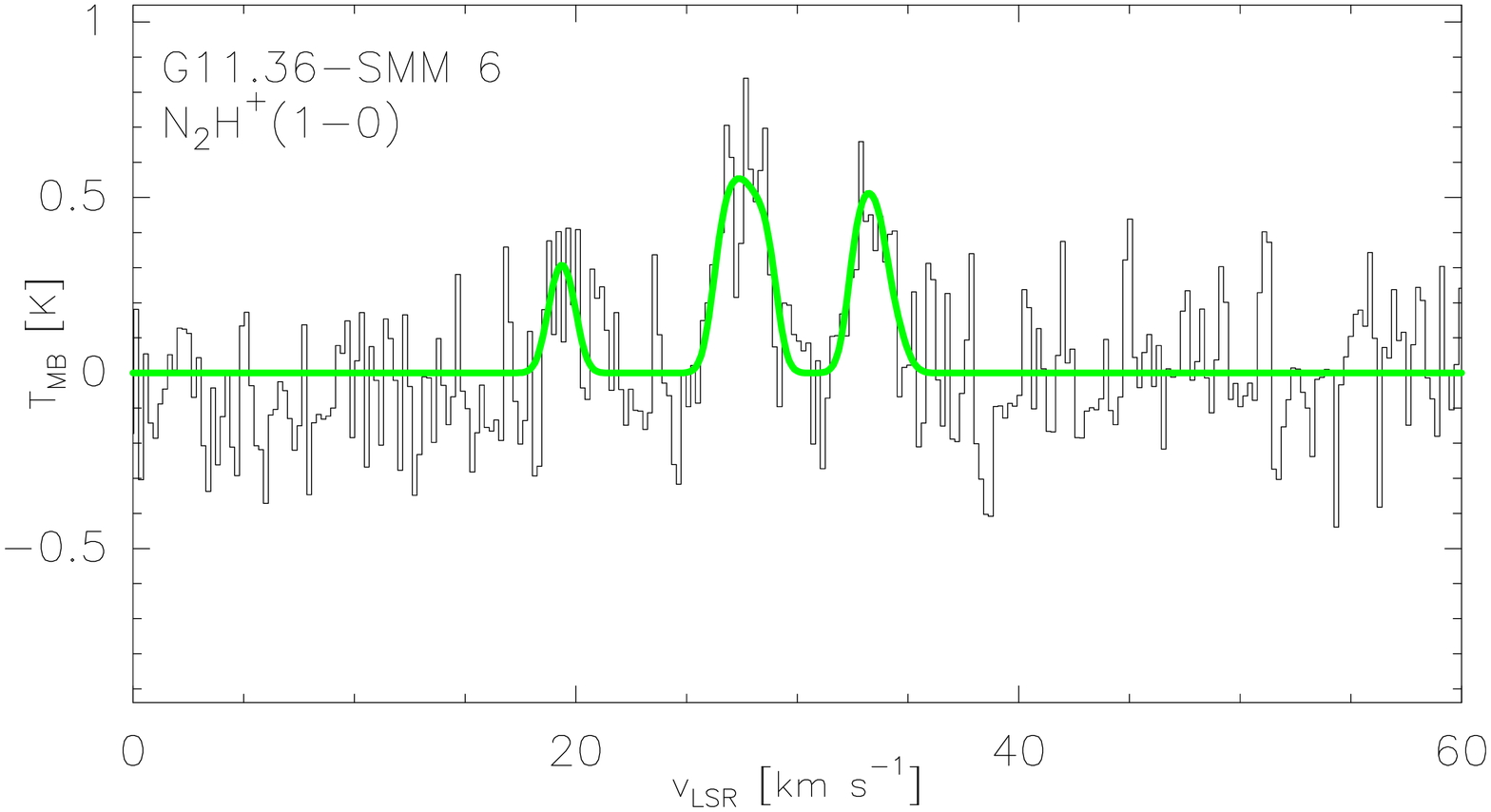}
\caption{Same as Fig.~\ref{figure:G187SMM1_spectra} but towards G11.36--SMM 6.}
\label{figure:G1136SMM6_spectra}
\end{center}
\end{figure*}

\begin{figure*}
\begin{center}
\includegraphics[width=0.245\textwidth]{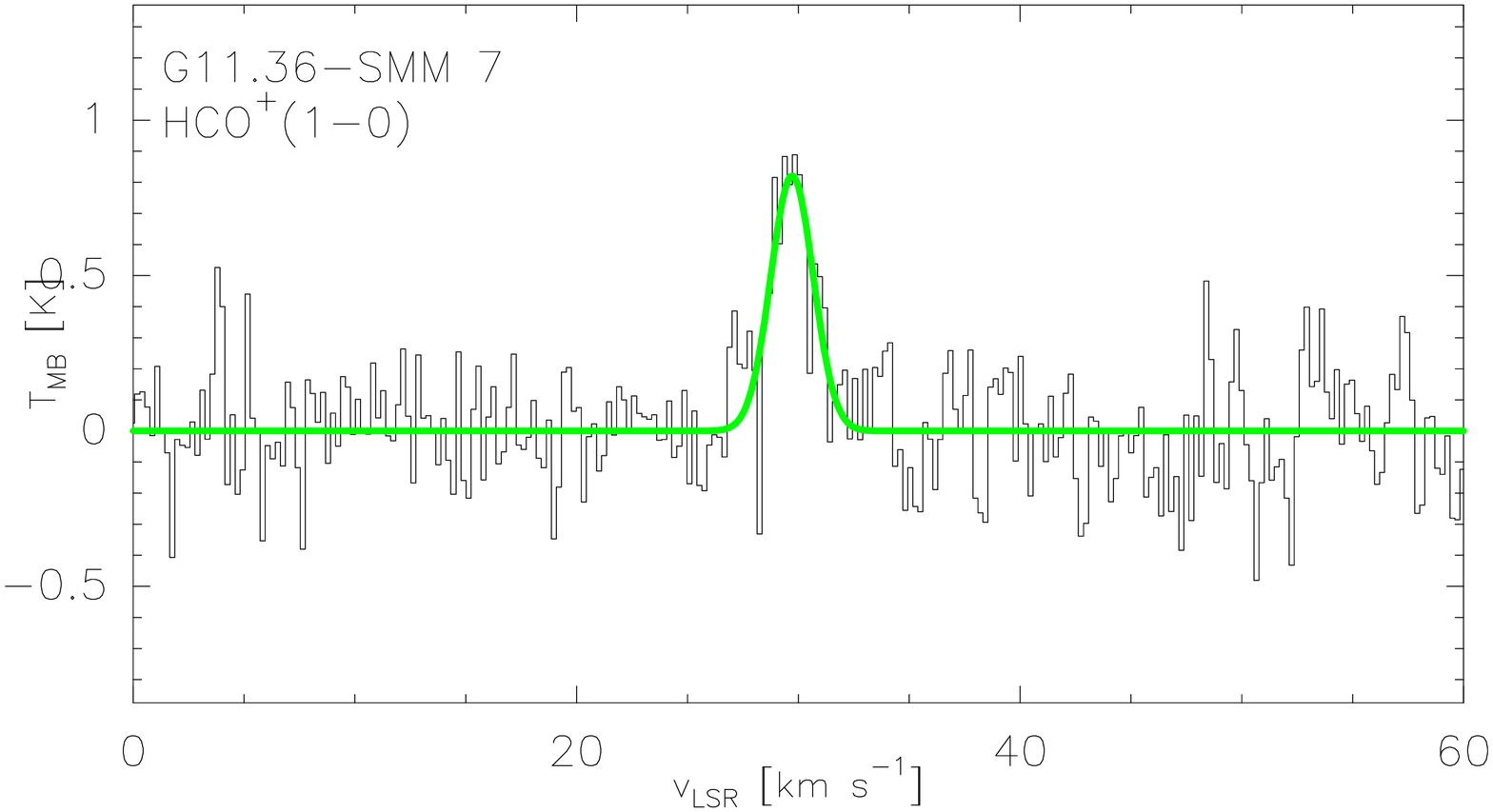}
\includegraphics[width=0.245\textwidth]{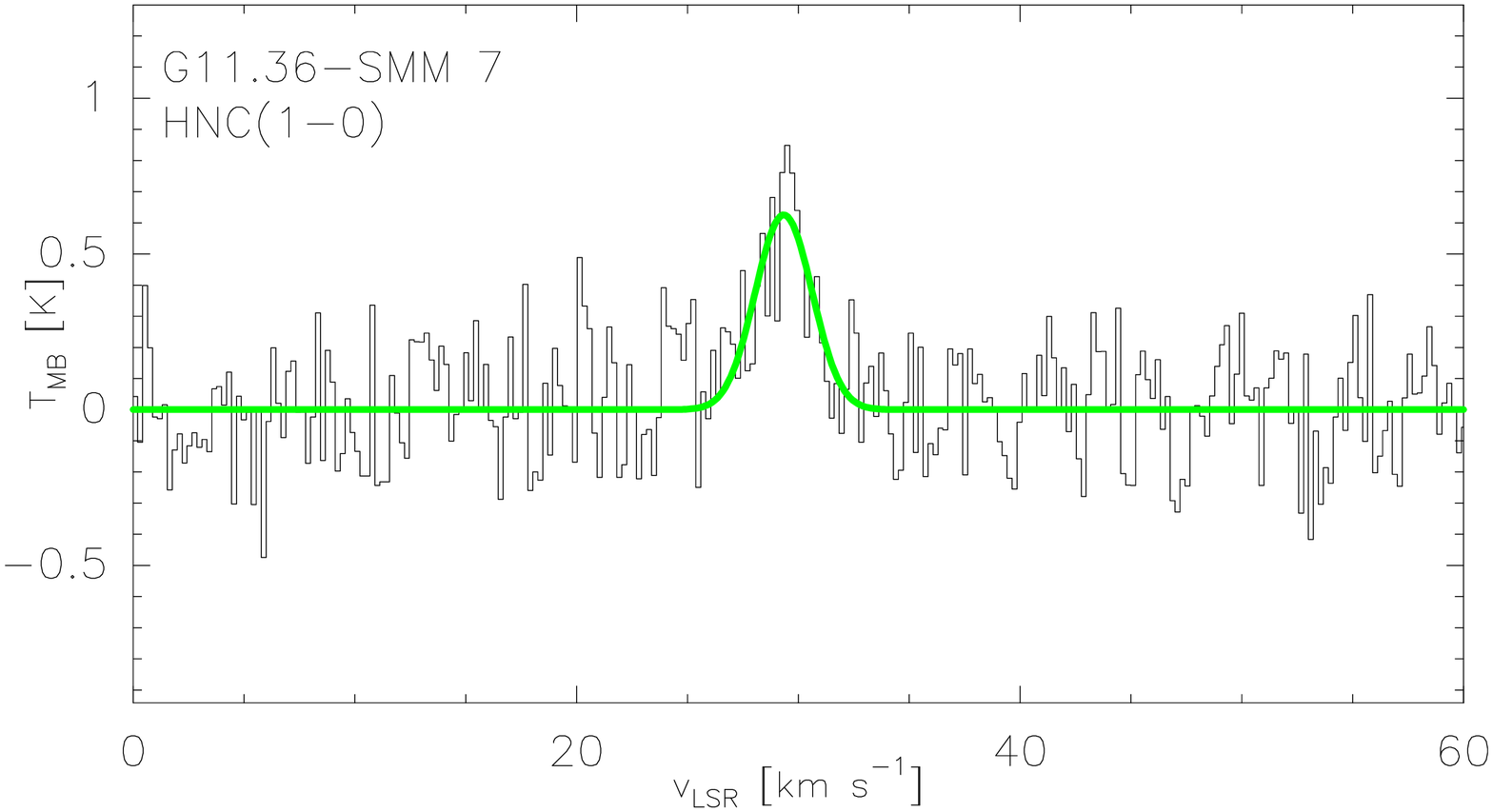}
\includegraphics[width=0.245\textwidth]{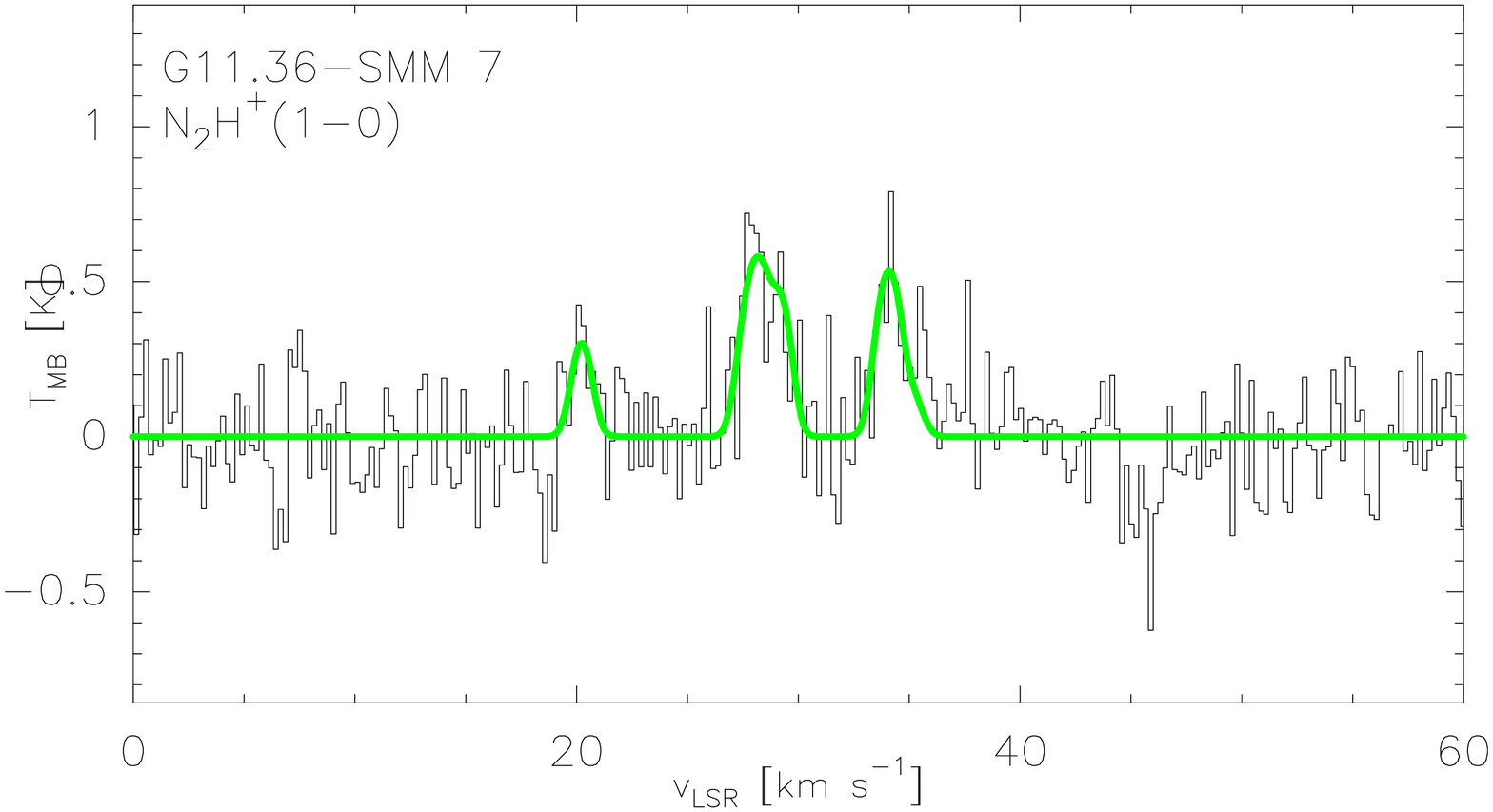}
\caption{Same as Fig.~\ref{figure:G187SMM1_spectra} but towards G11.36--SMM 7.}
\label{figure:G1136SMM7_spectra}
\end{center}
\end{figure*}

\begin{figure*}
\begin{center}
\includegraphics[width=0.245\textwidth]{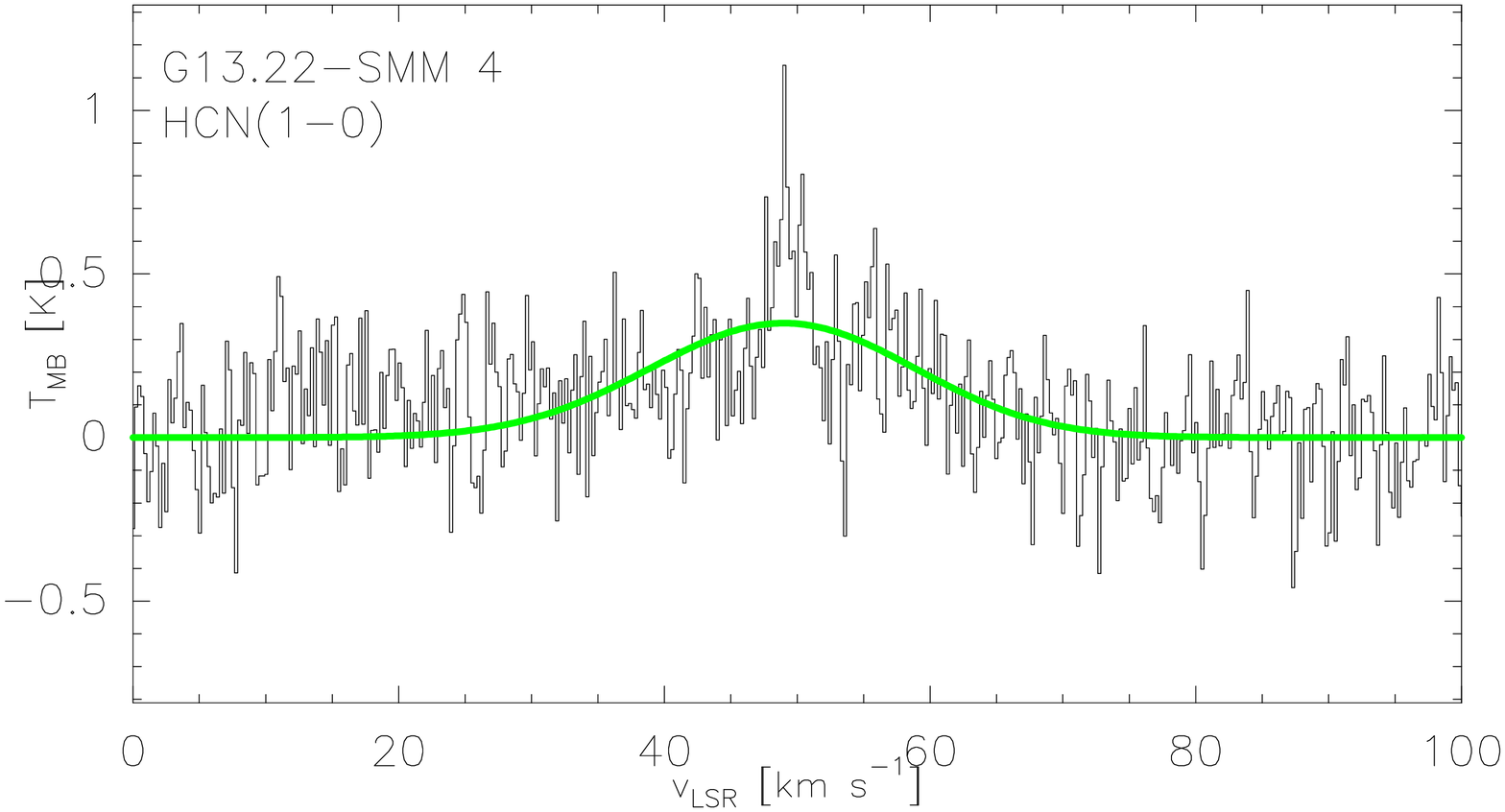}
\includegraphics[width=0.245\textwidth]{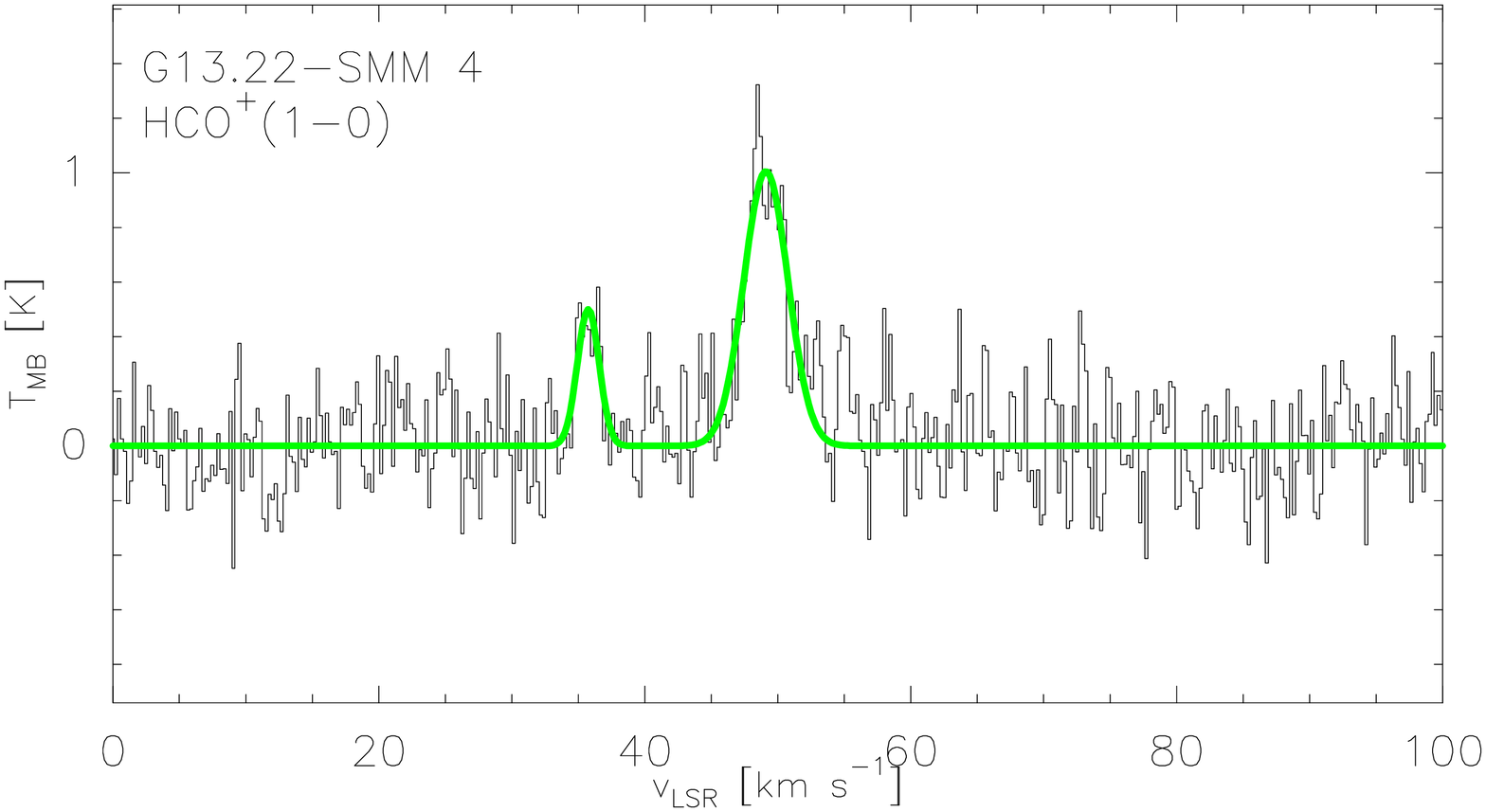}
\includegraphics[width=0.245\textwidth]{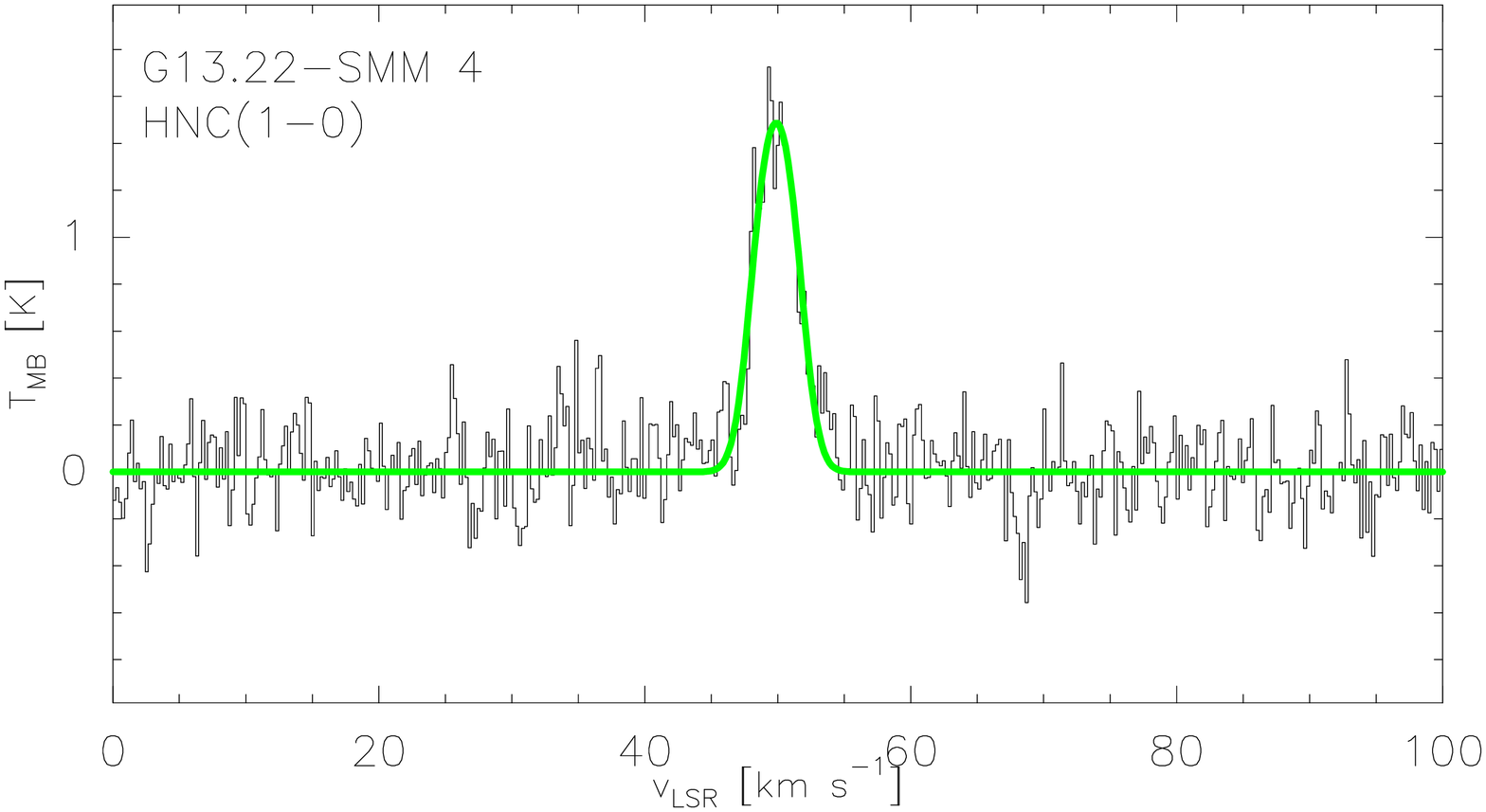}
\includegraphics[width=0.245\textwidth]{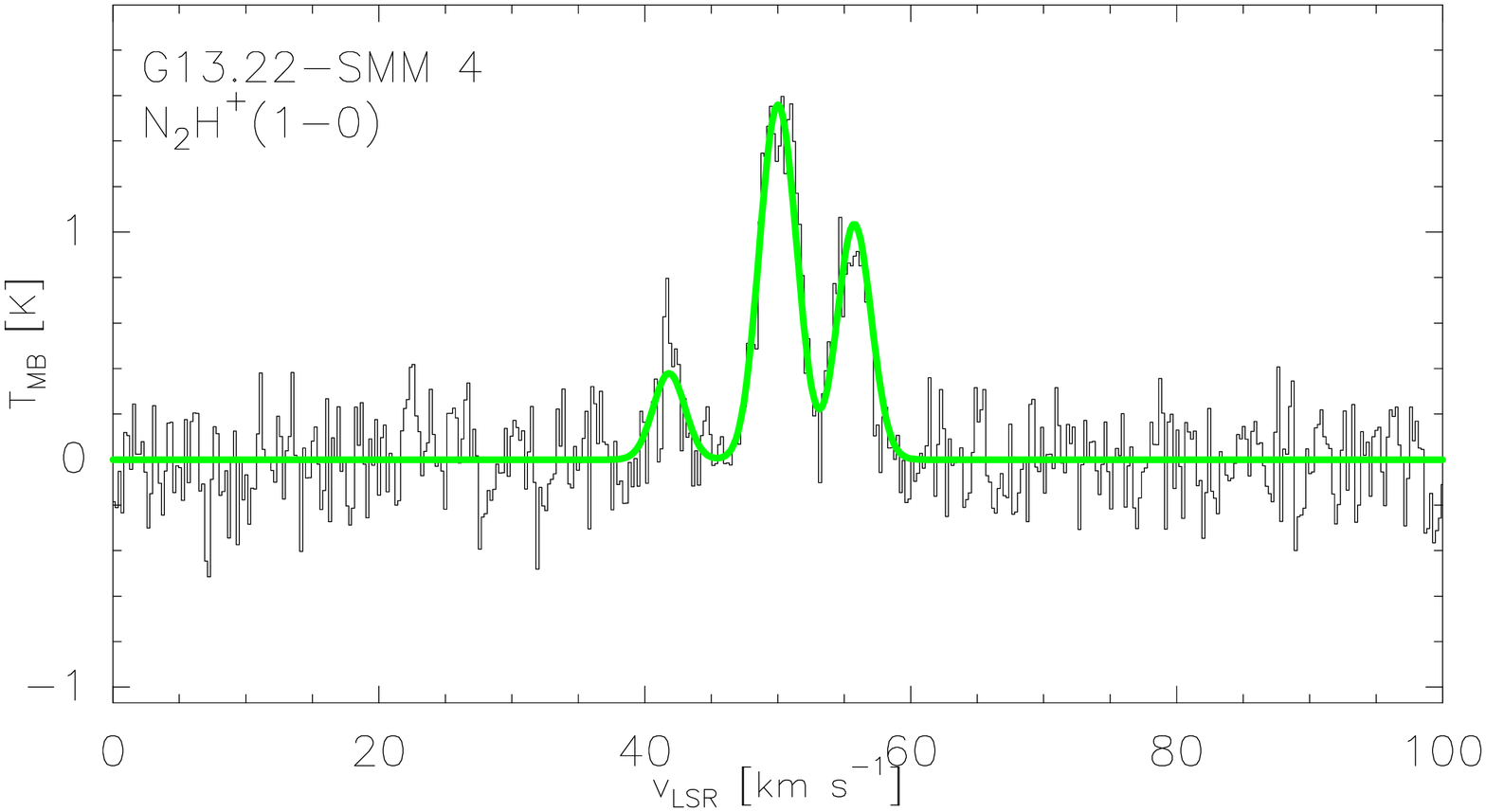}
\caption{Same as Fig.~\ref{figure:G187SMM1_spectra} but towards G13.22--SMM 4. 
There is an additional velocity component in the HCO$^+$ spectrum.}
\label{figure:G1322SMM4_spectra}
\end{center}
\end{figure*}

\begin{figure*}
\begin{center}
\includegraphics[width=0.245\textwidth]{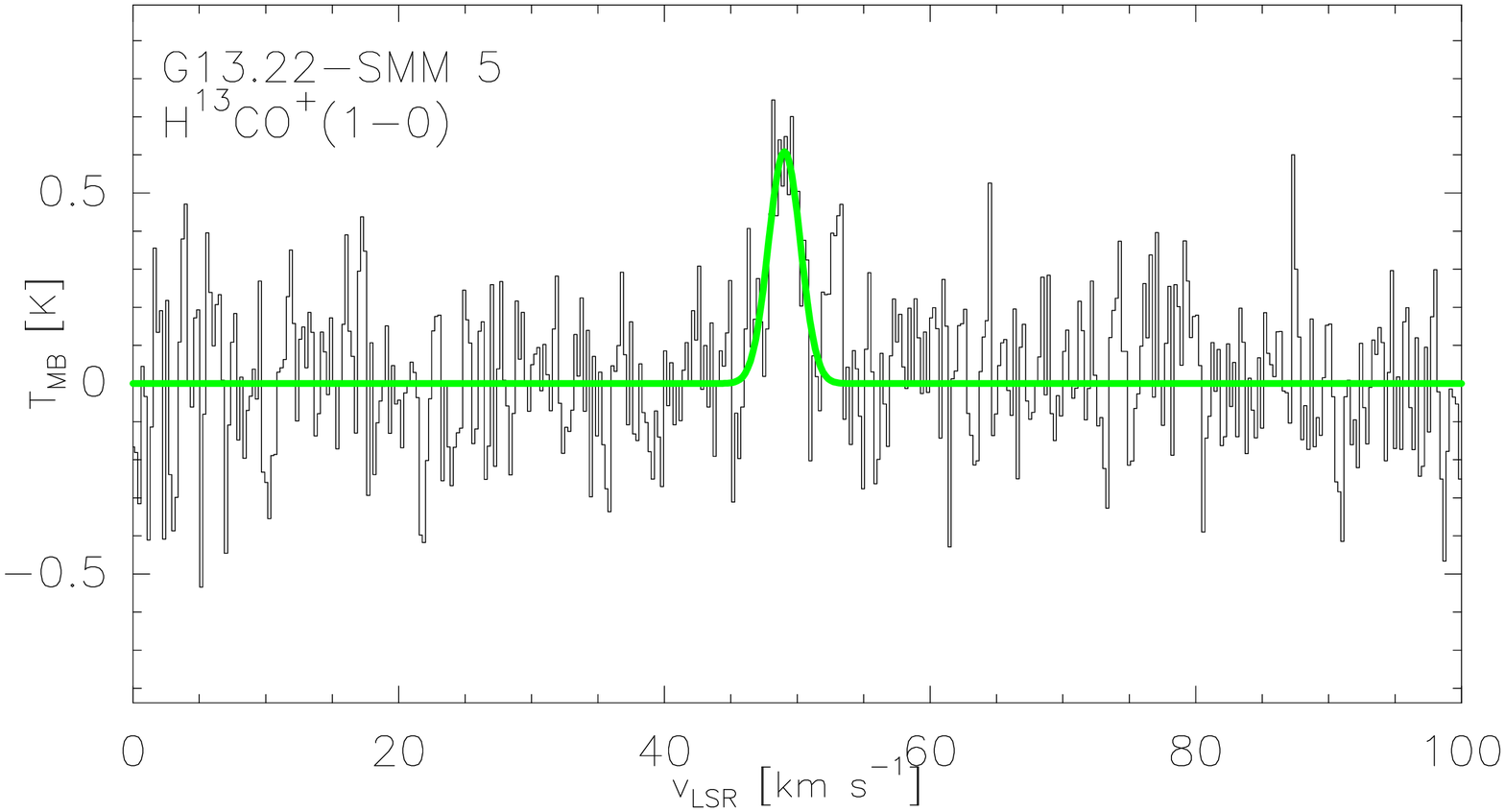}
\includegraphics[width=0.245\textwidth]{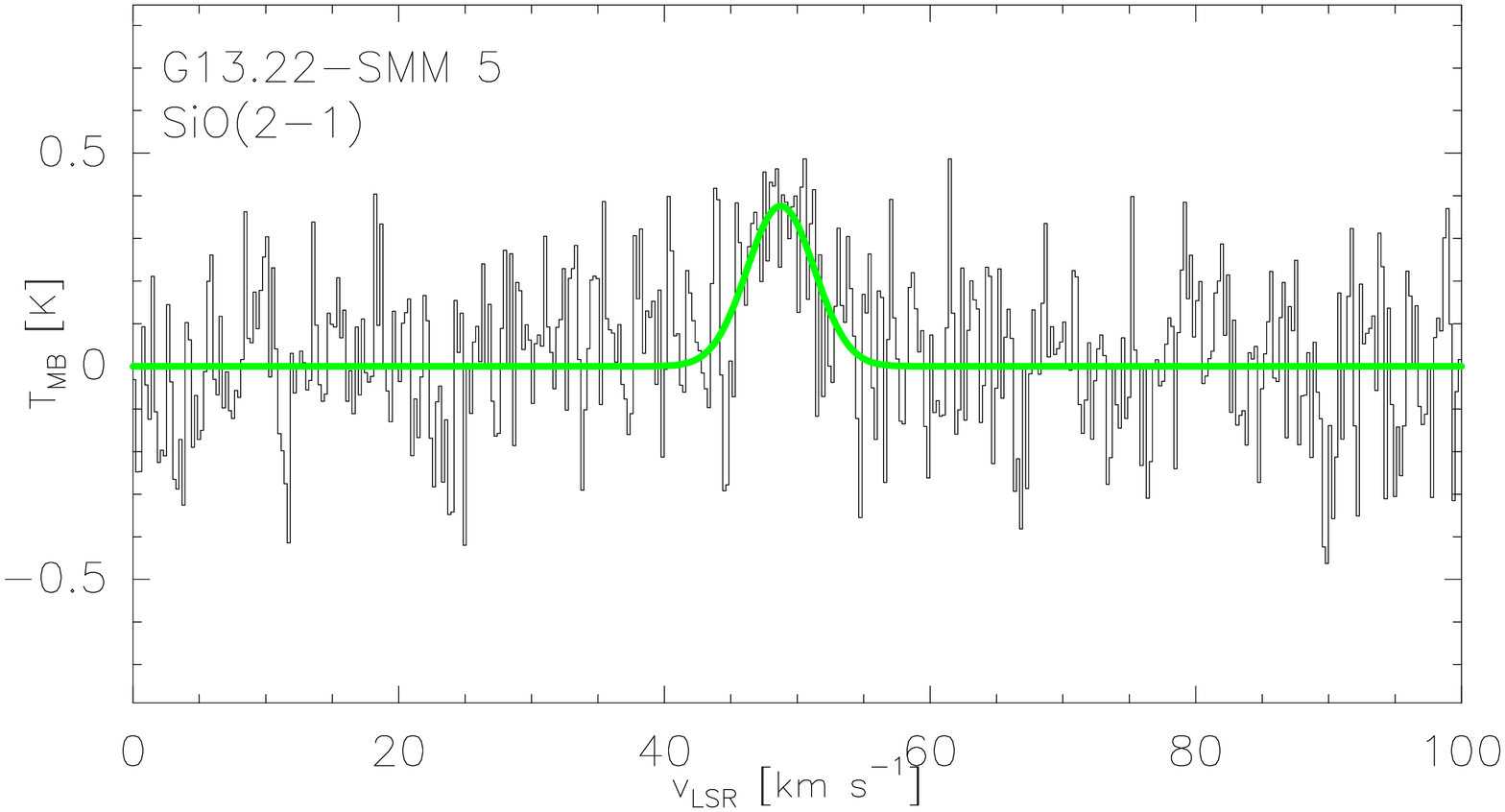}
\includegraphics[width=0.245\textwidth]{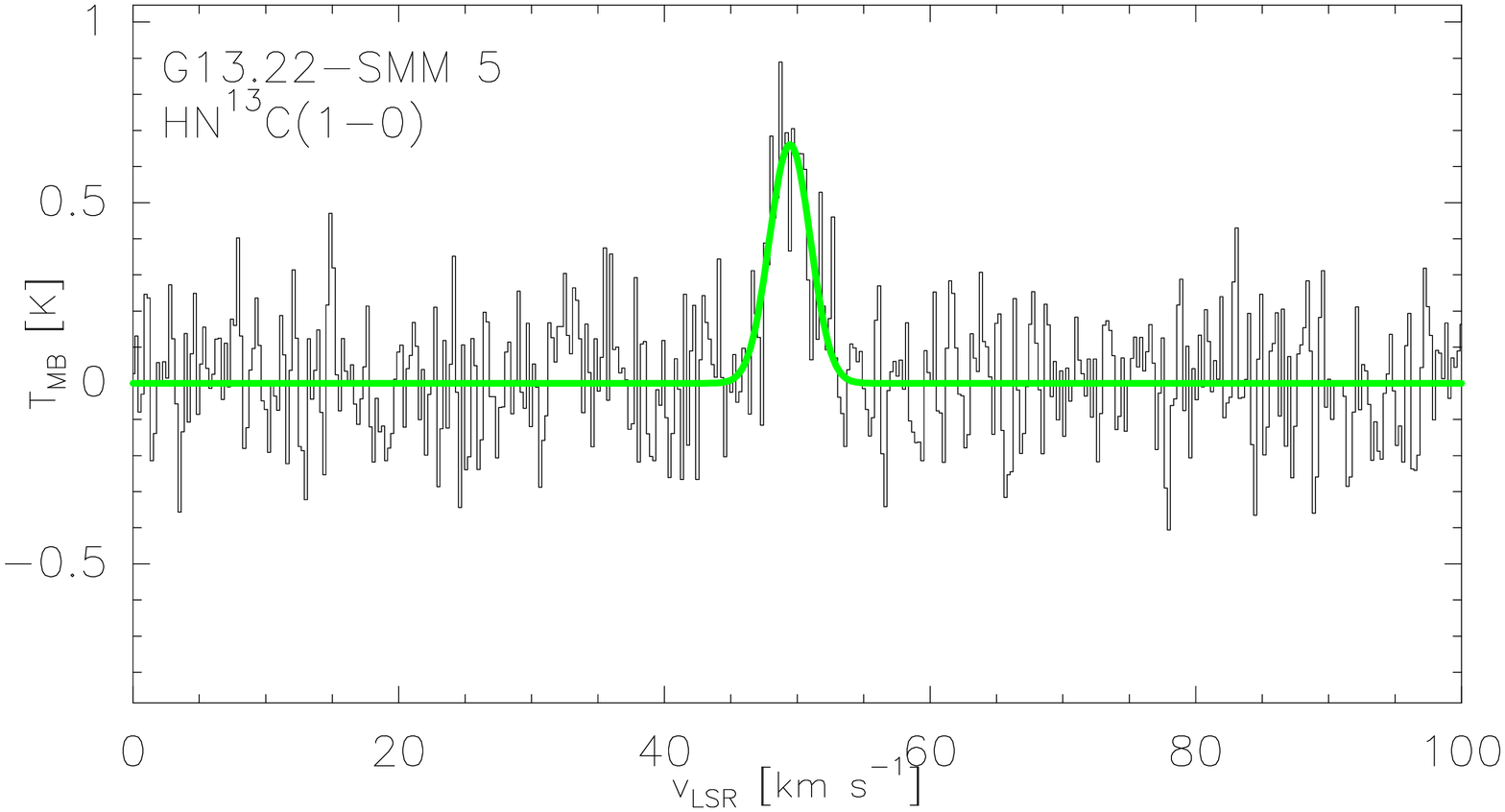}
\includegraphics[width=0.245\textwidth]{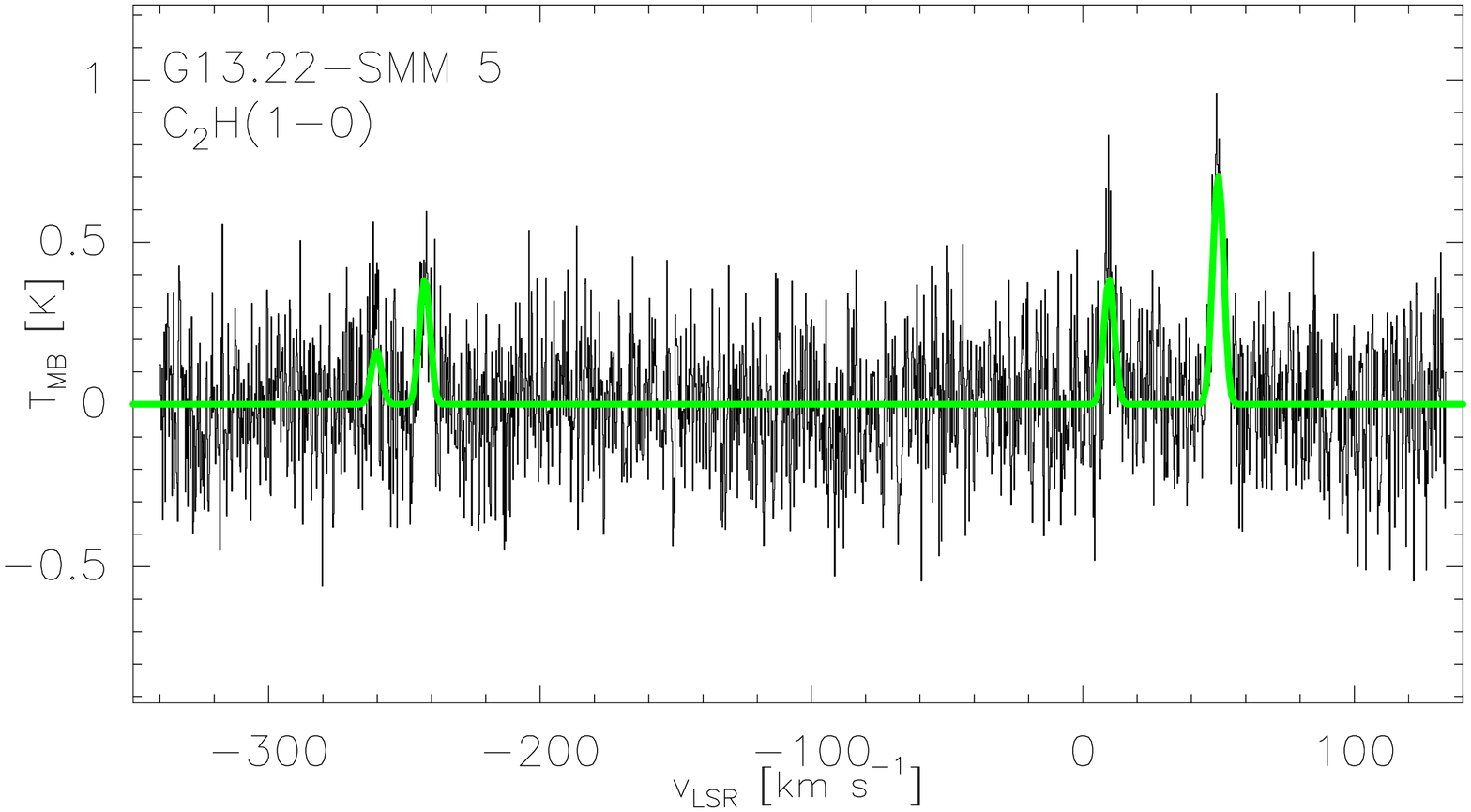}
\includegraphics[width=0.245\textwidth]{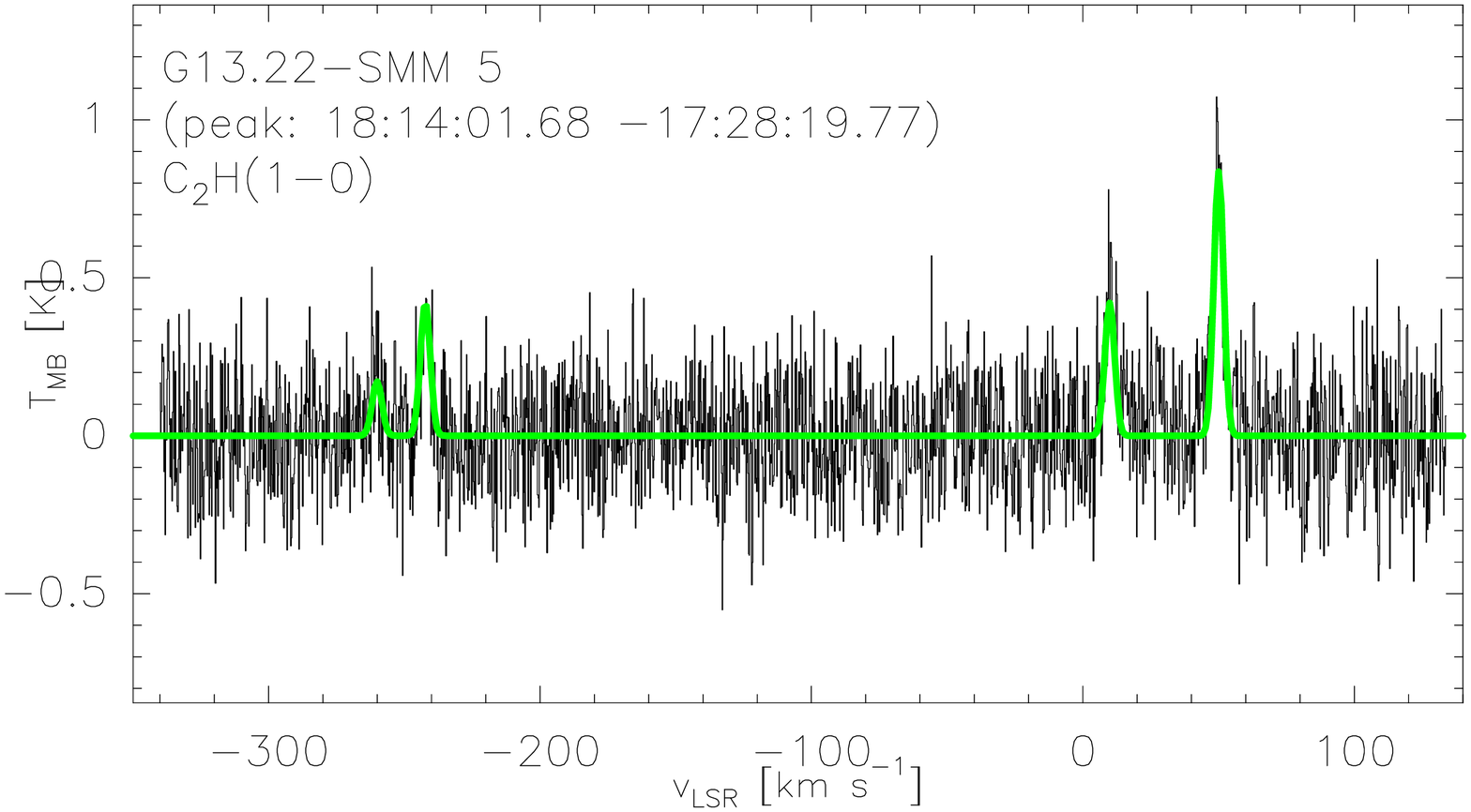}
\includegraphics[width=0.245\textwidth]{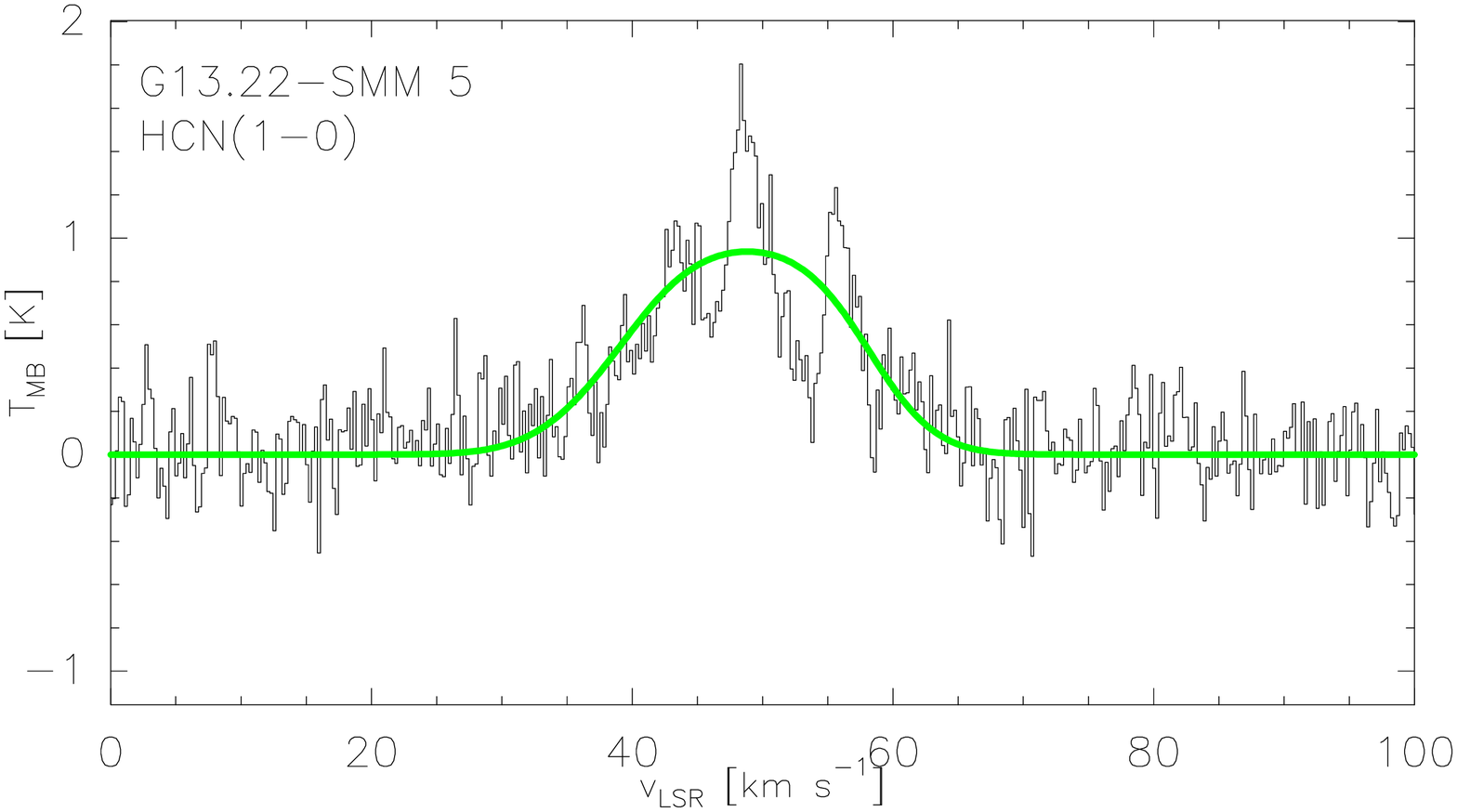}
\includegraphics[width=0.245\textwidth]{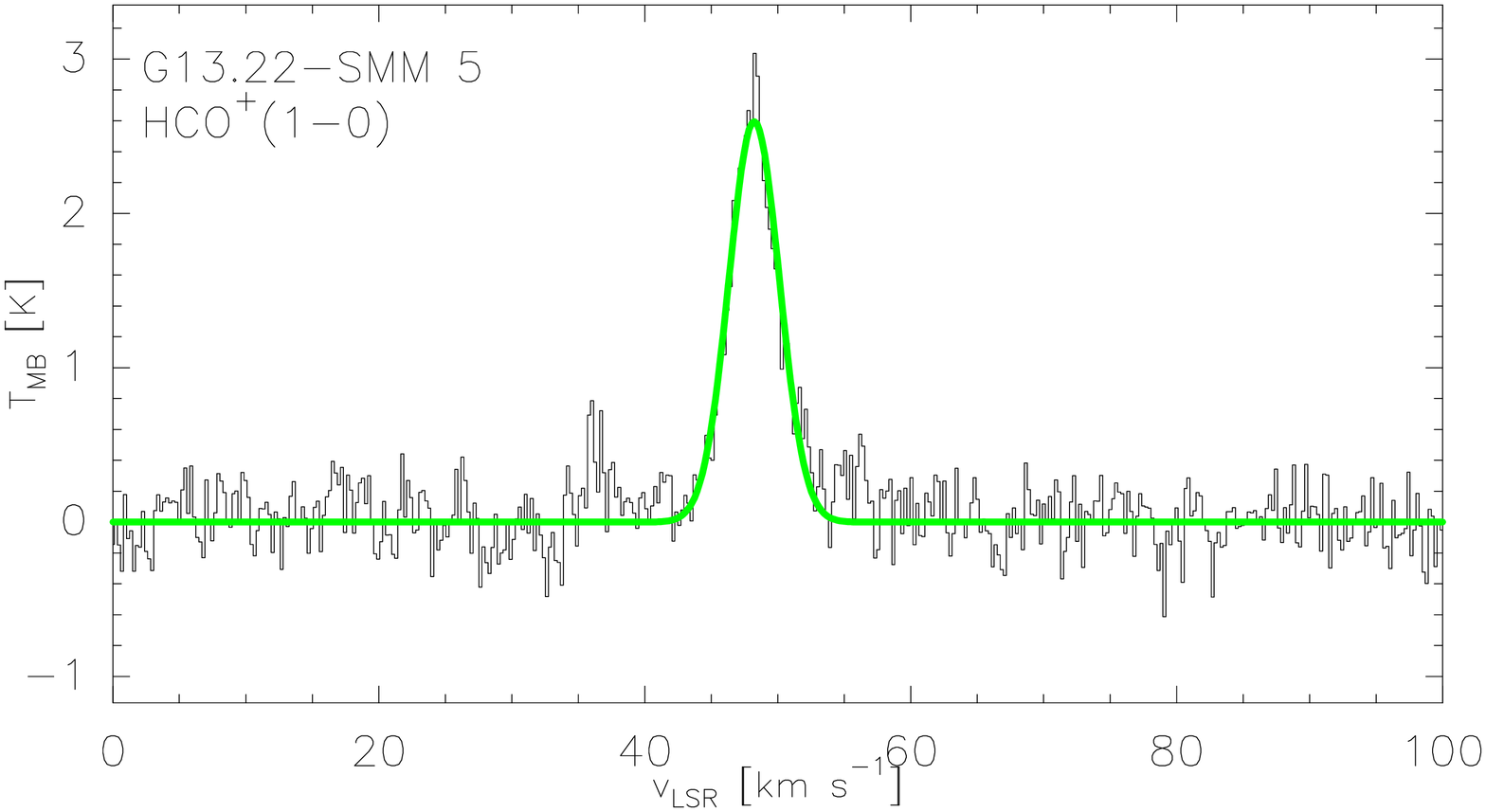}
\includegraphics[width=0.245\textwidth]{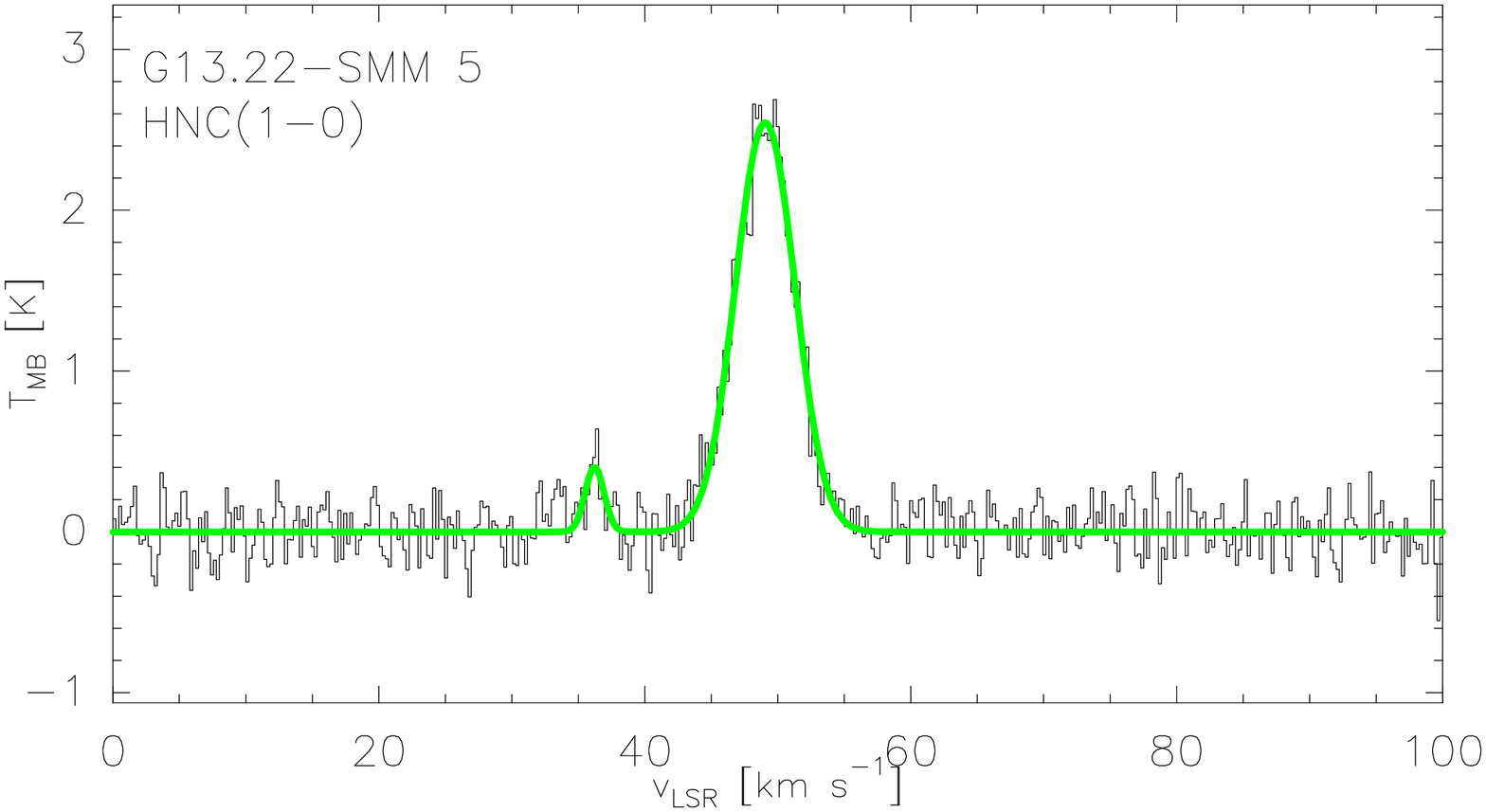}
\includegraphics[width=0.245\textwidth]{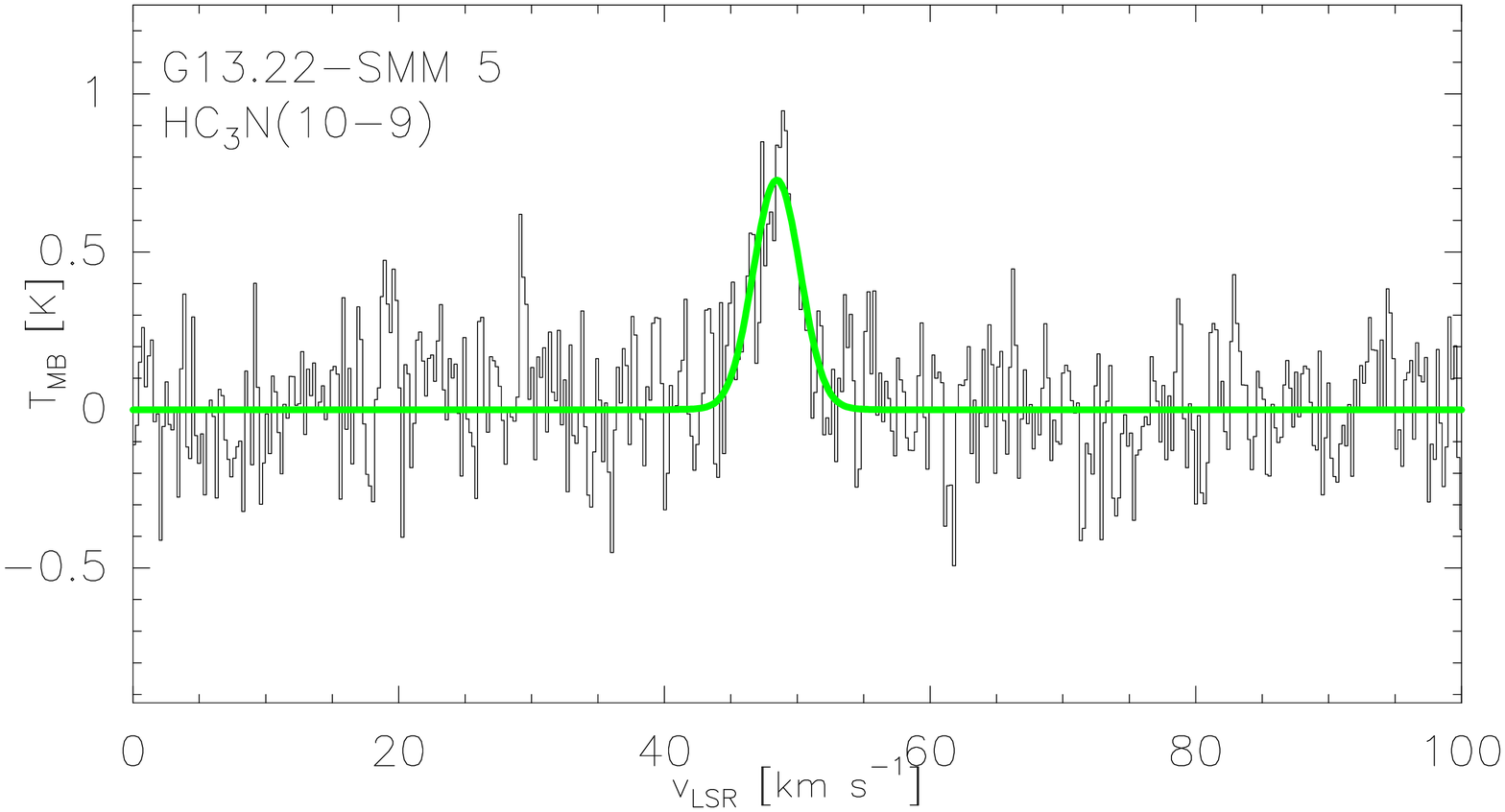}
\includegraphics[width=0.245\textwidth]{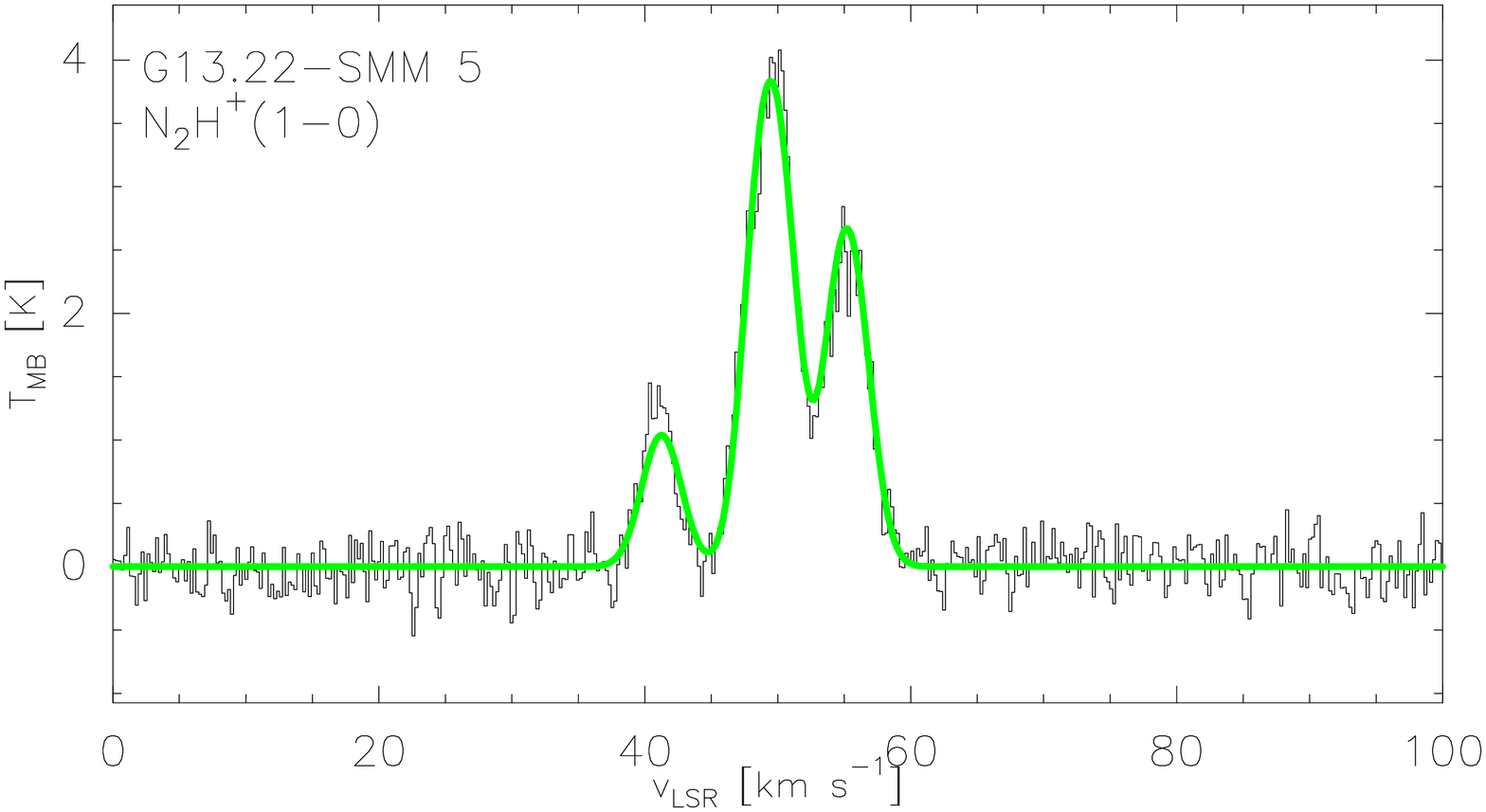}
\caption{Same as Fig.~\ref{figure:G187SMM1_spectra} but towards G13.22--SMM 5. 
The velocity range is wider for the two C$_2$H spectra.}
\label{figure:G1322SMM5_spectra}
\end{center}
\end{figure*}

\begin{figure*}
\begin{center}
\includegraphics[width=0.245\textwidth]{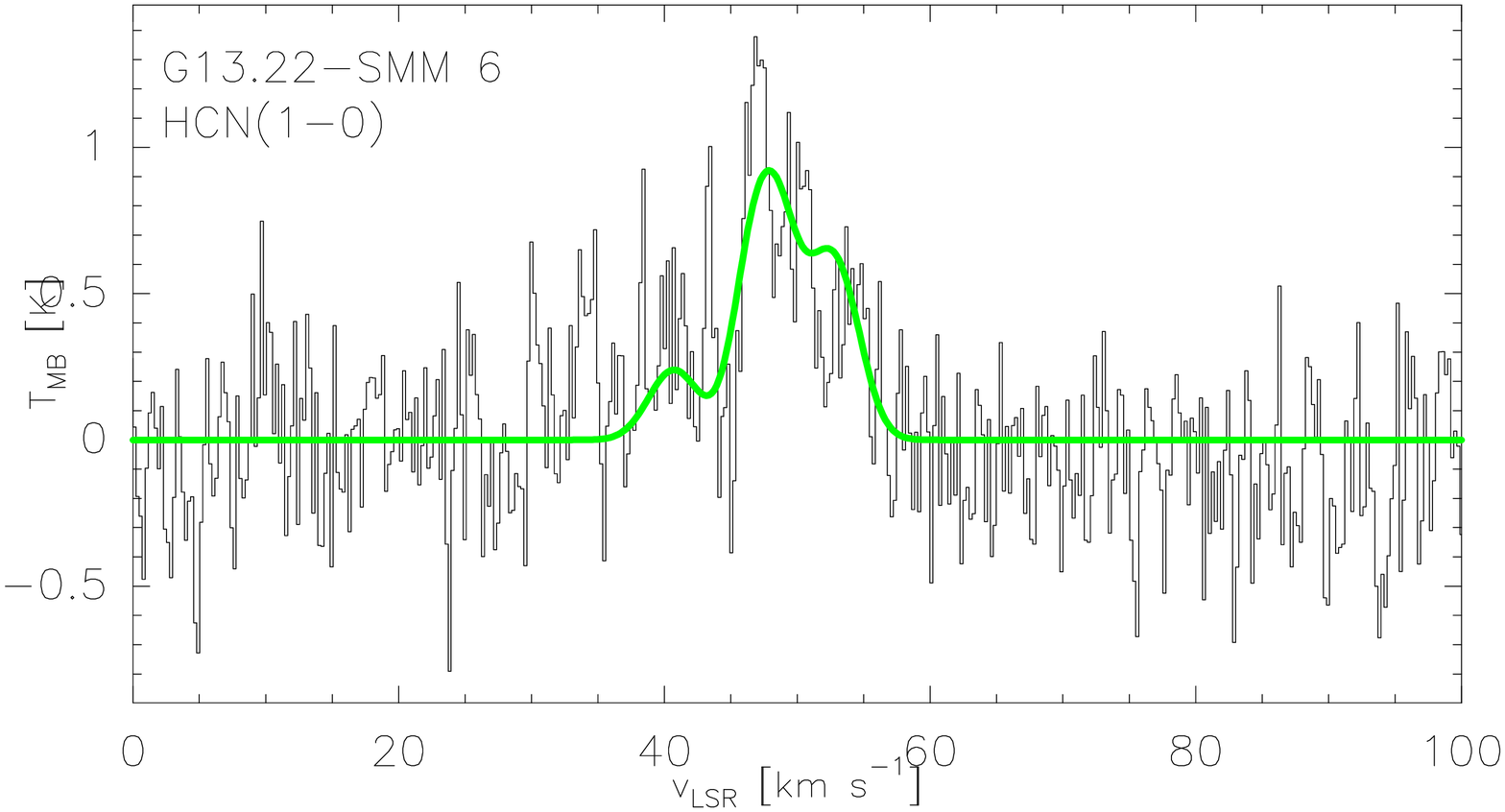}
\includegraphics[width=0.245\textwidth]{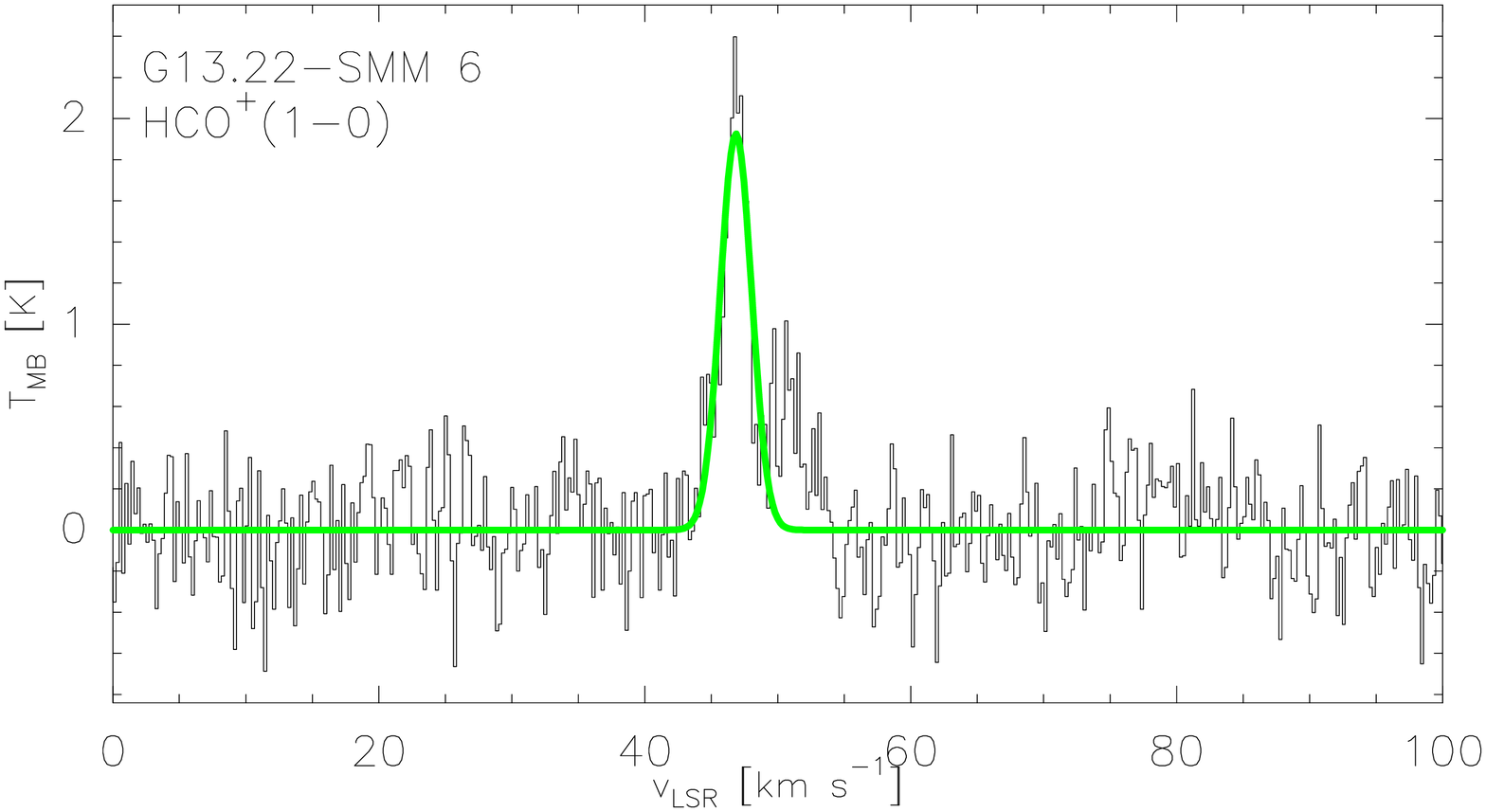}
\includegraphics[width=0.245\textwidth]{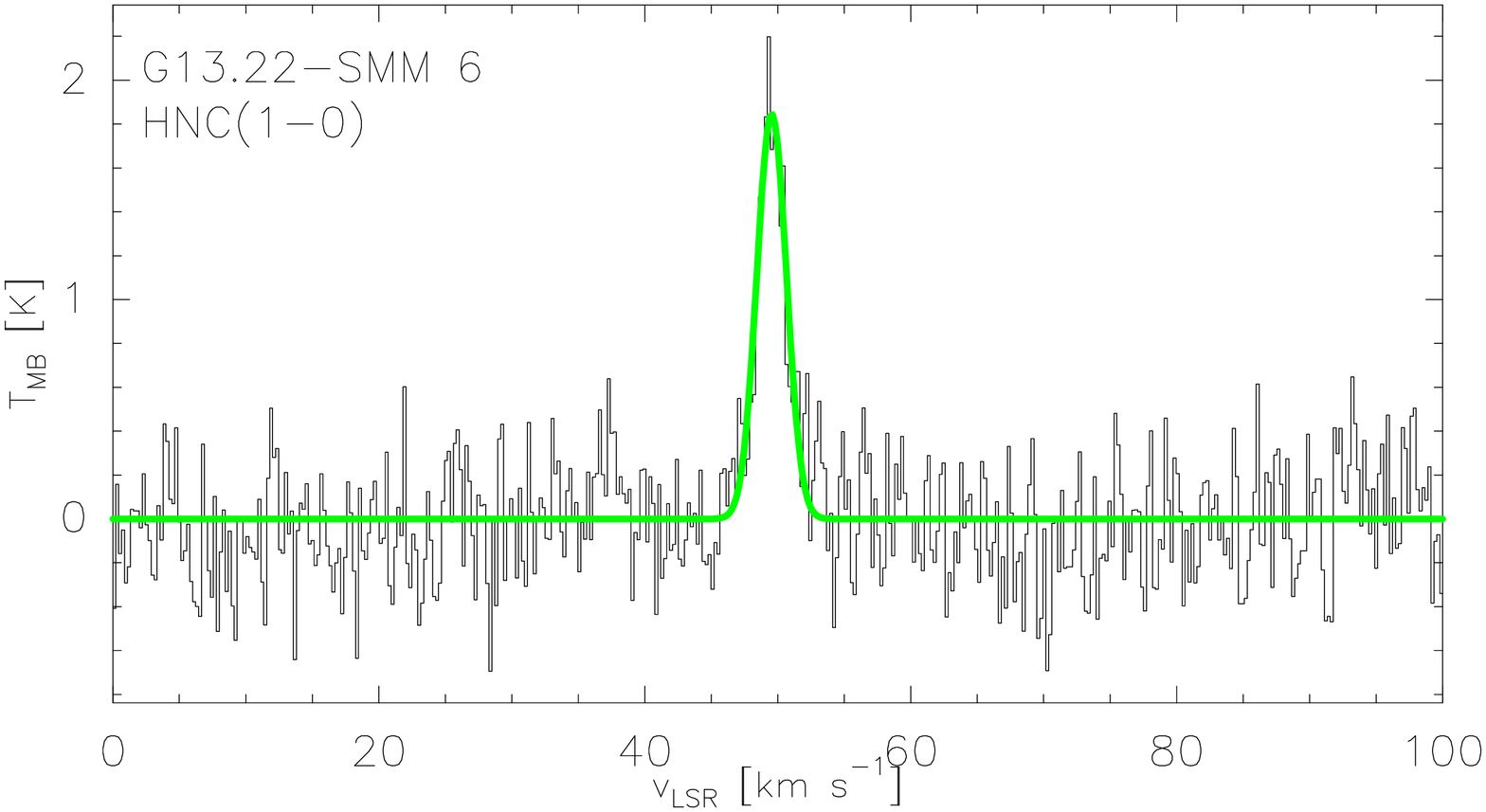}
\includegraphics[width=0.245\textwidth]{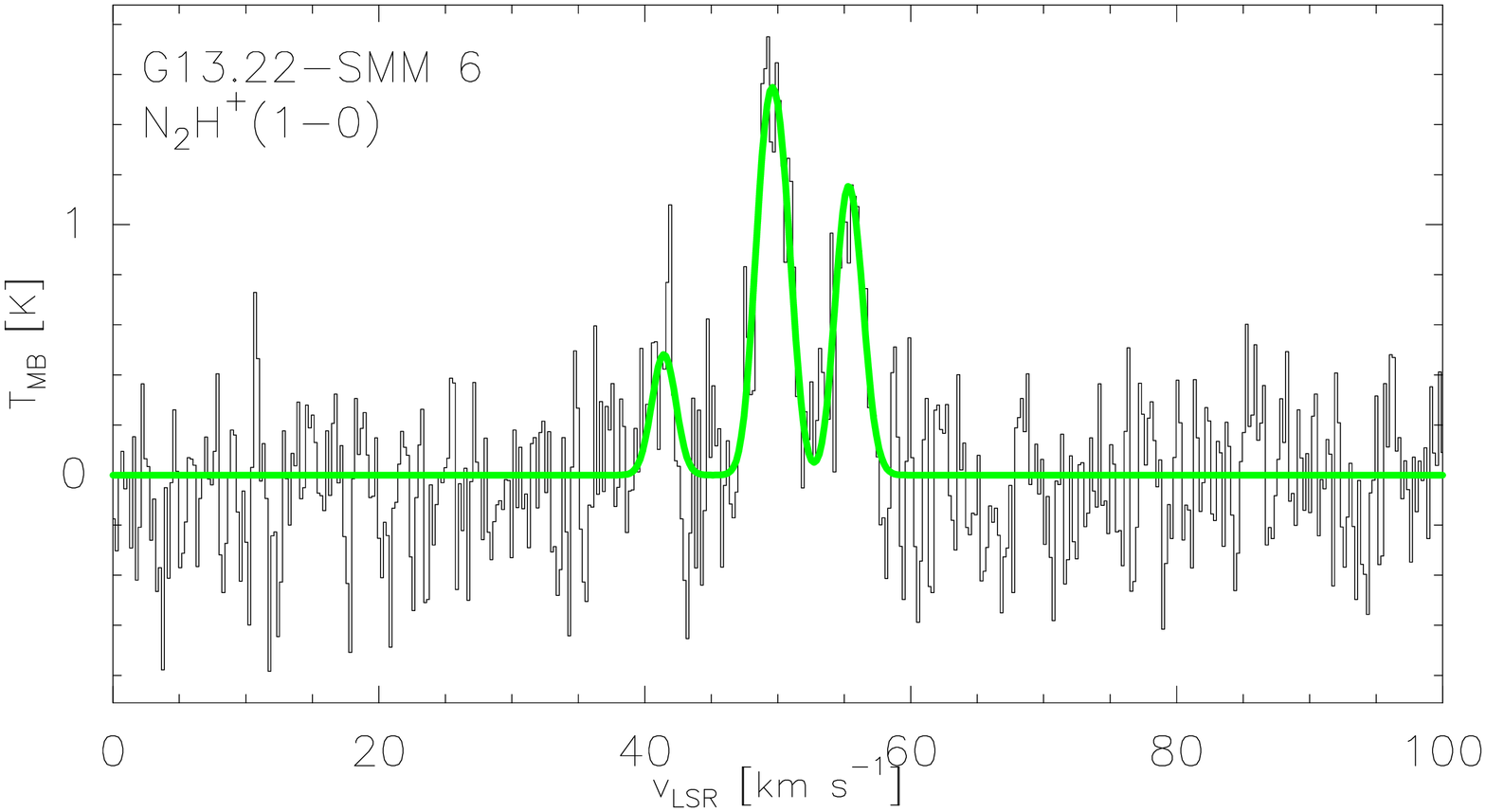}
\caption{Same as Fig.~\ref{figure:G187SMM1_spectra} but towards G13.22--SMM 6.}
\label{figure:G1322SMM6_spectra}
\end{center}
\end{figure*}

\begin{figure*}
\begin{center}
\includegraphics[width=0.245\textwidth]{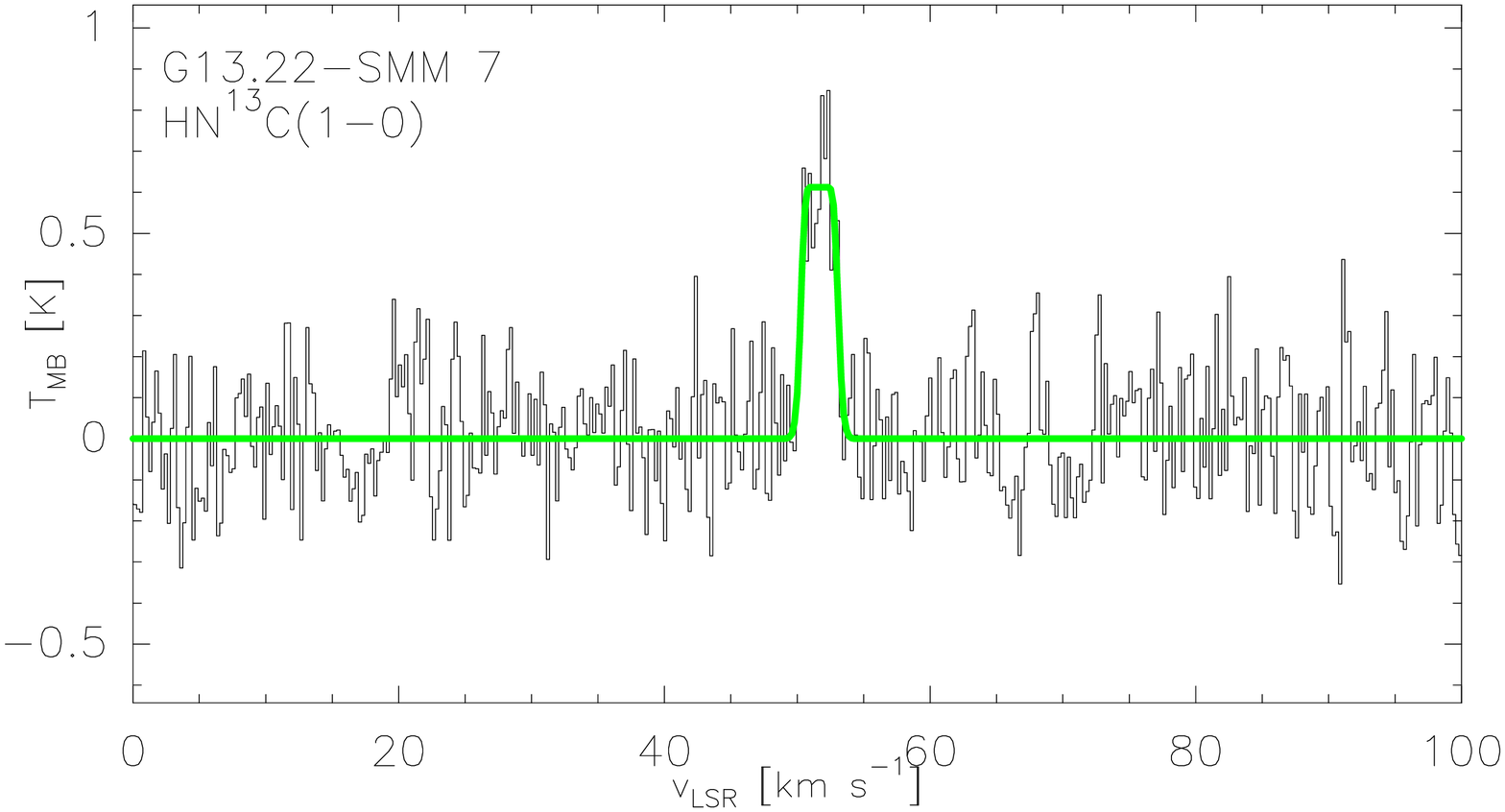}
\includegraphics[width=0.245\textwidth]{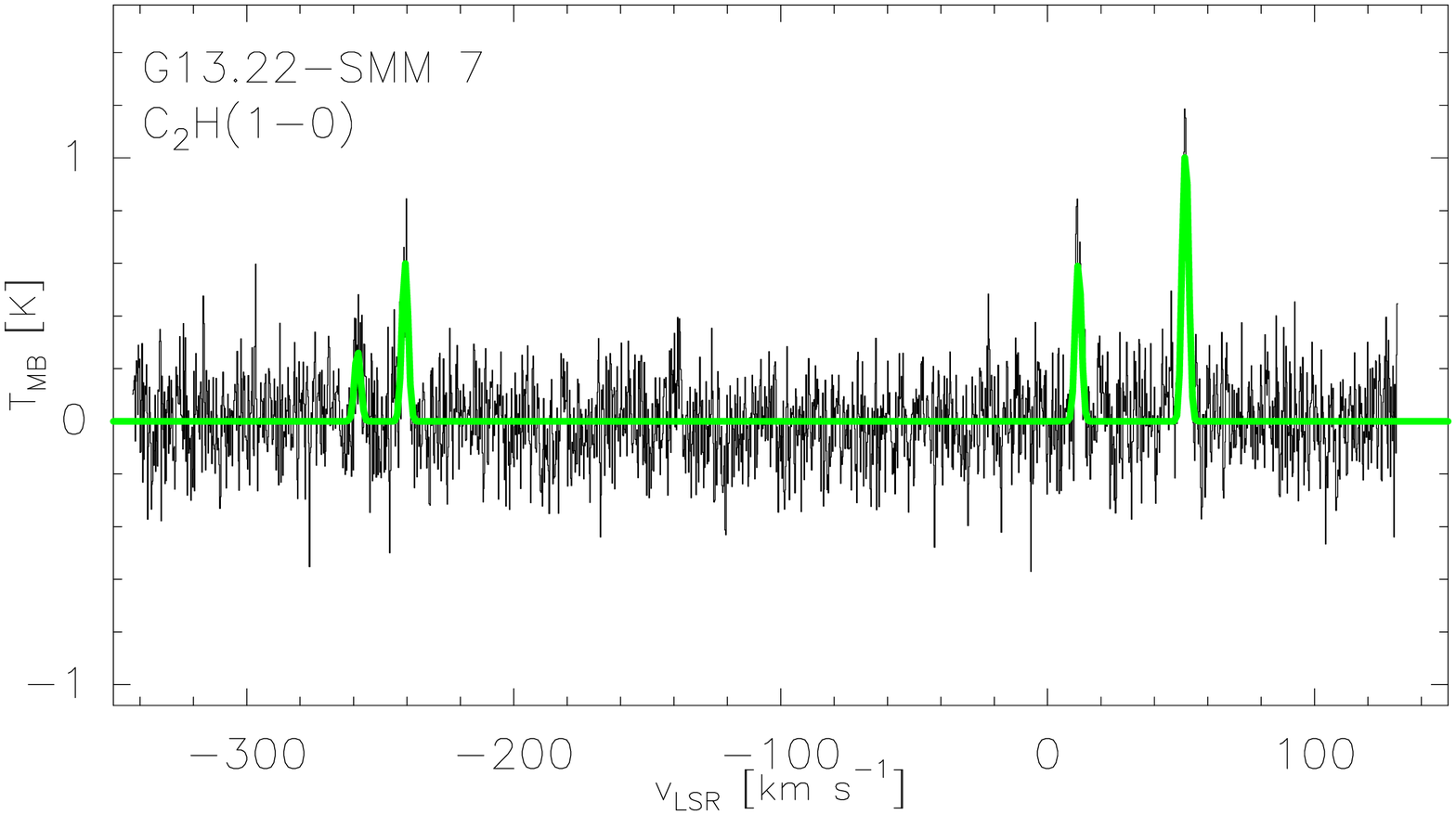}
\includegraphics[width=0.245\textwidth]{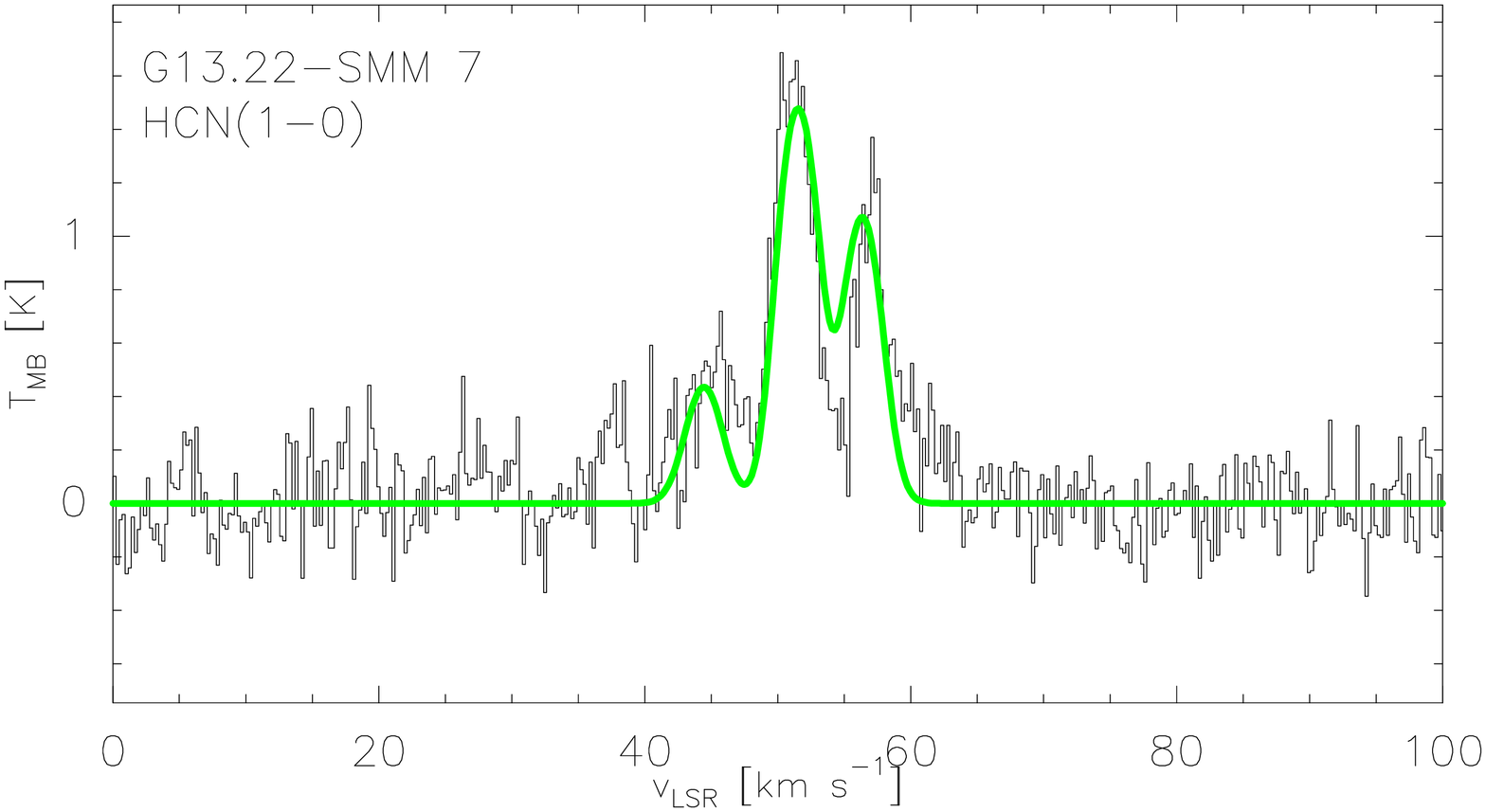}
\includegraphics[width=0.245\textwidth]{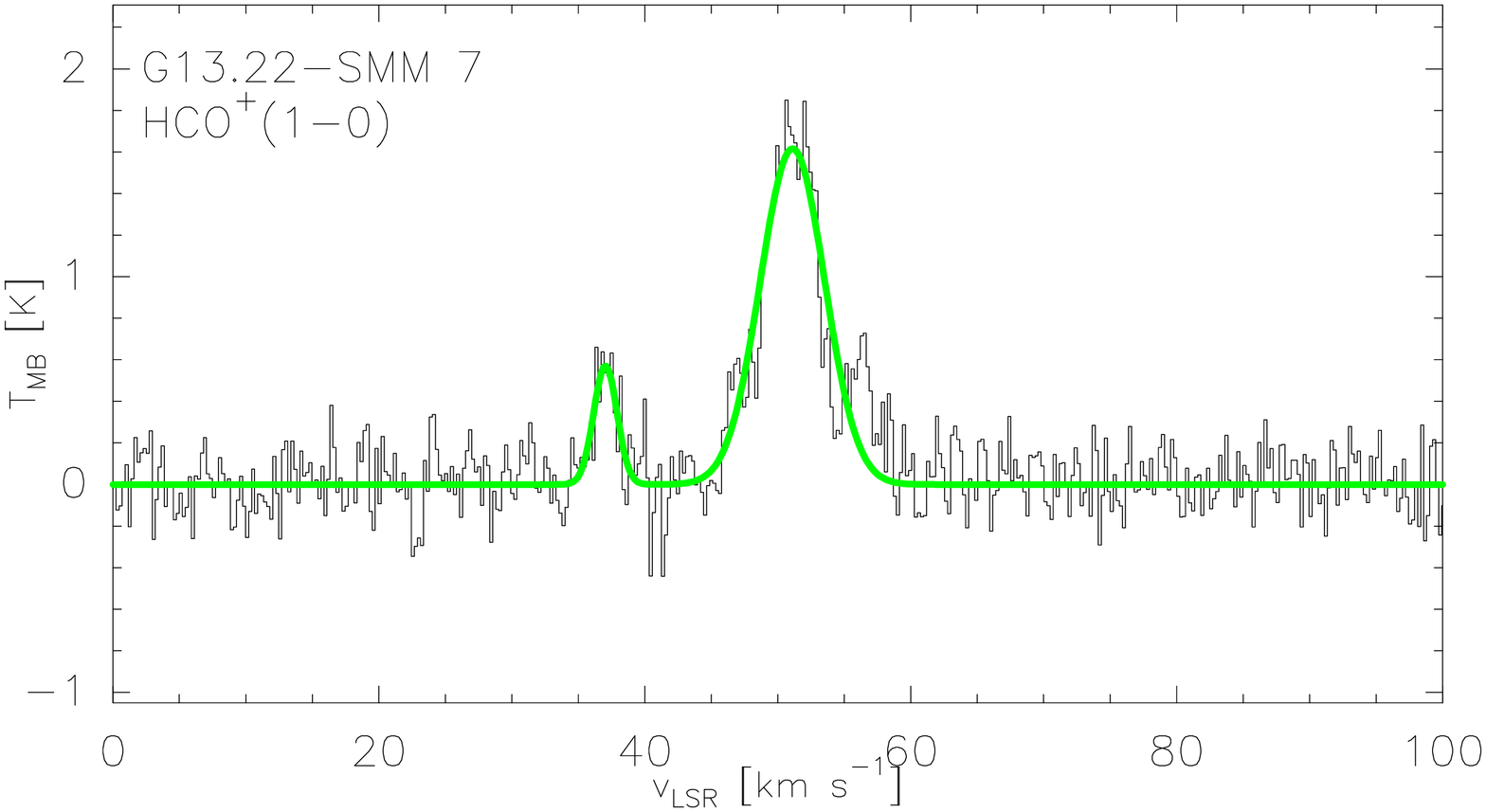}
\includegraphics[width=0.245\textwidth]{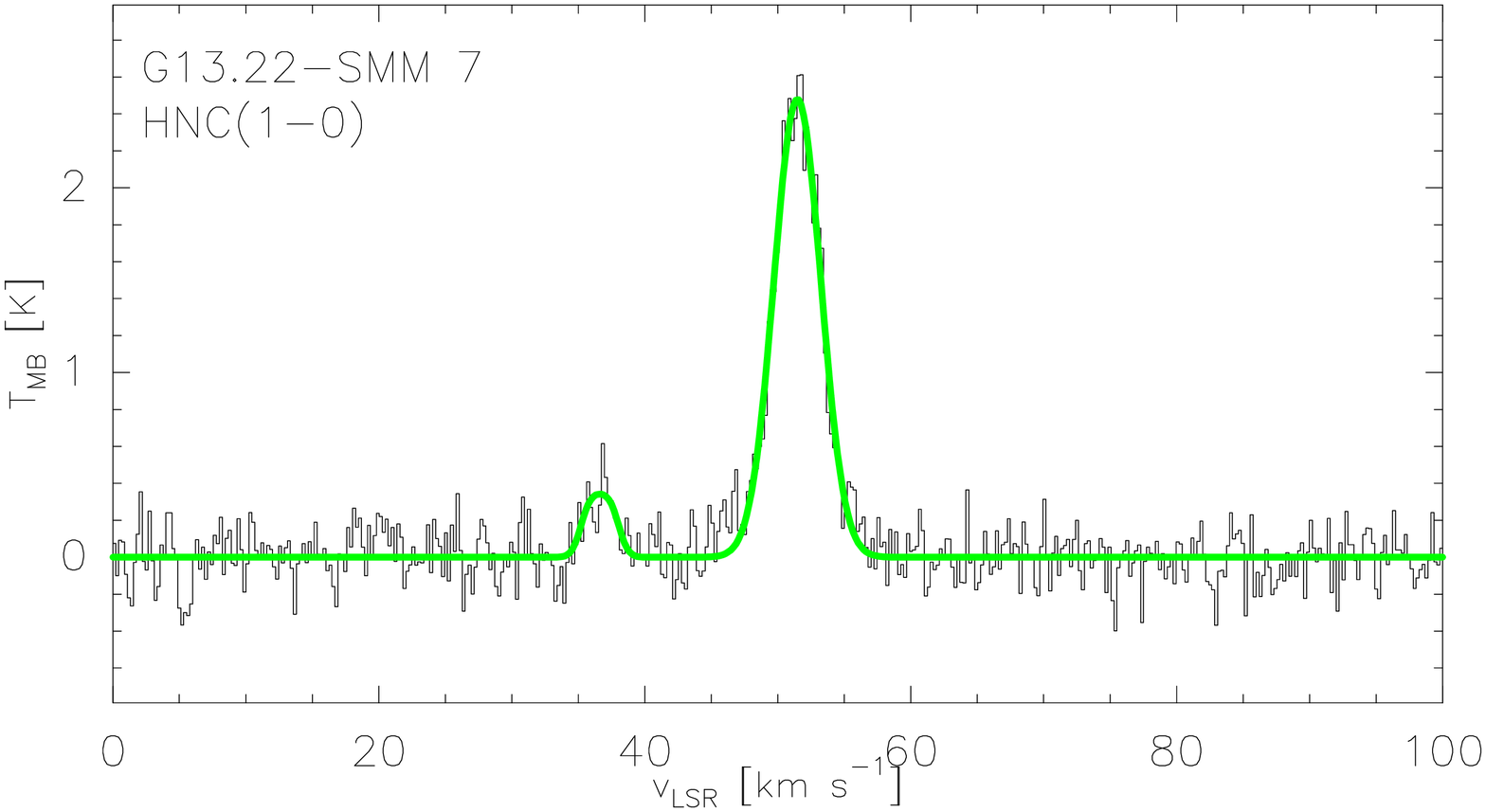}
\includegraphics[width=0.245\textwidth]{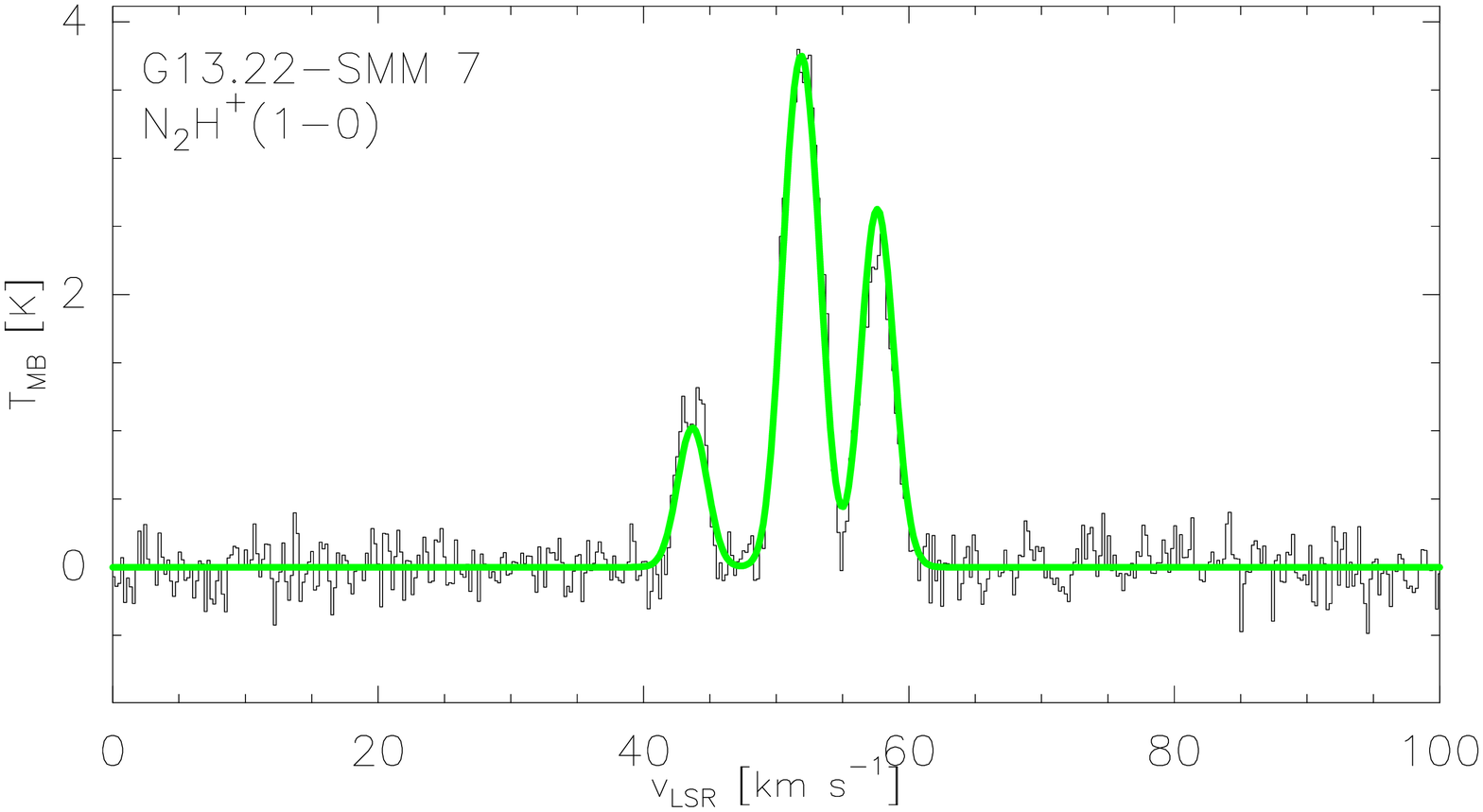}
\caption{Same as Fig.~\ref{figure:G187SMM1_spectra} but towards G13.22--SMM 7. 
The C$_2$H spectrum has a wider velocity range. Two velocity components are 
seen in the HCO$^+$ and HNC spectra.}
\label{figure:G1322SMM7_spectra}
\end{center}
\end{figure*}

\begin{figure*}
\begin{center}
\includegraphics[width=0.245\textwidth]{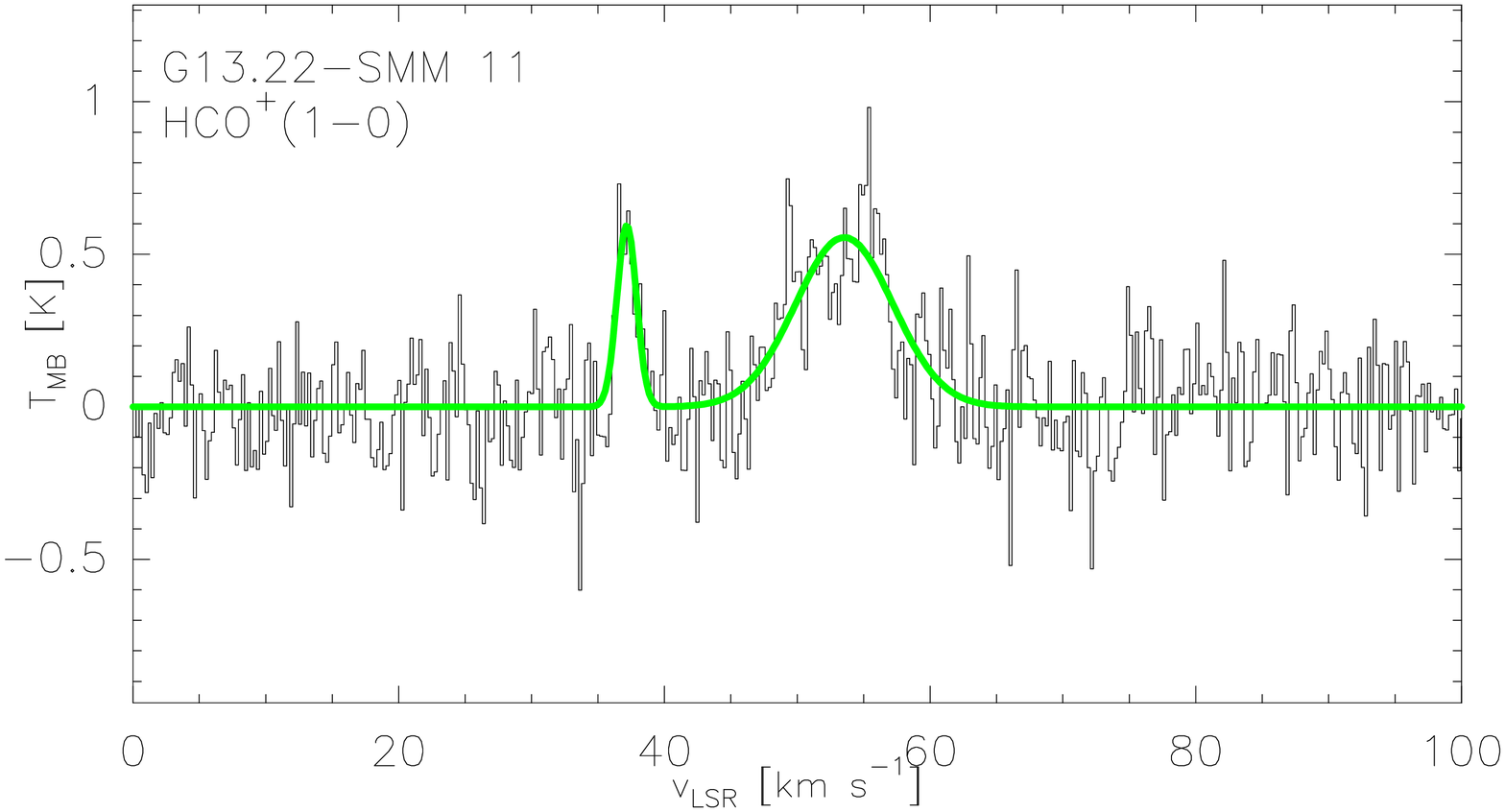}
\includegraphics[width=0.245\textwidth]{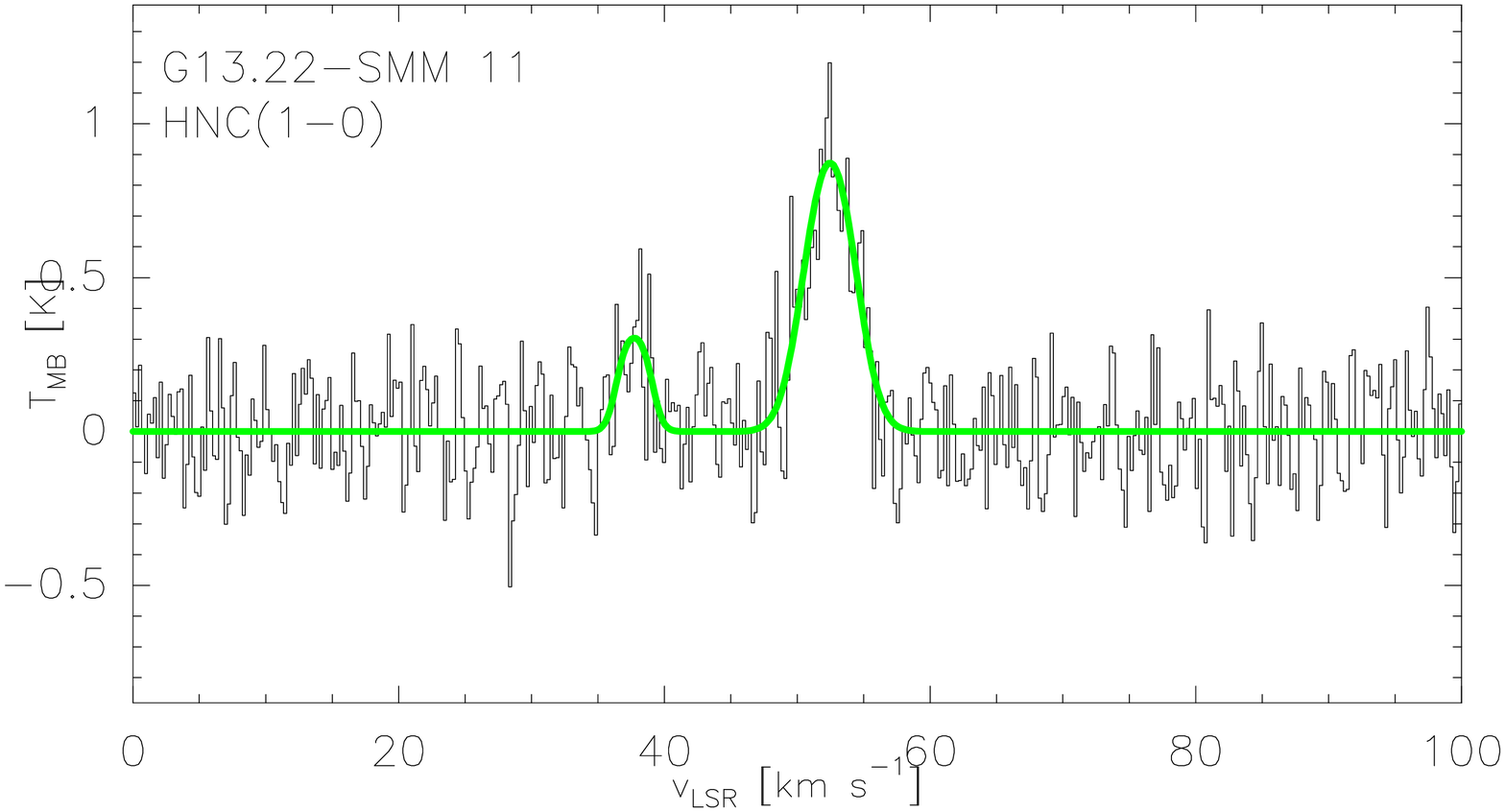}
\includegraphics[width=0.245\textwidth]{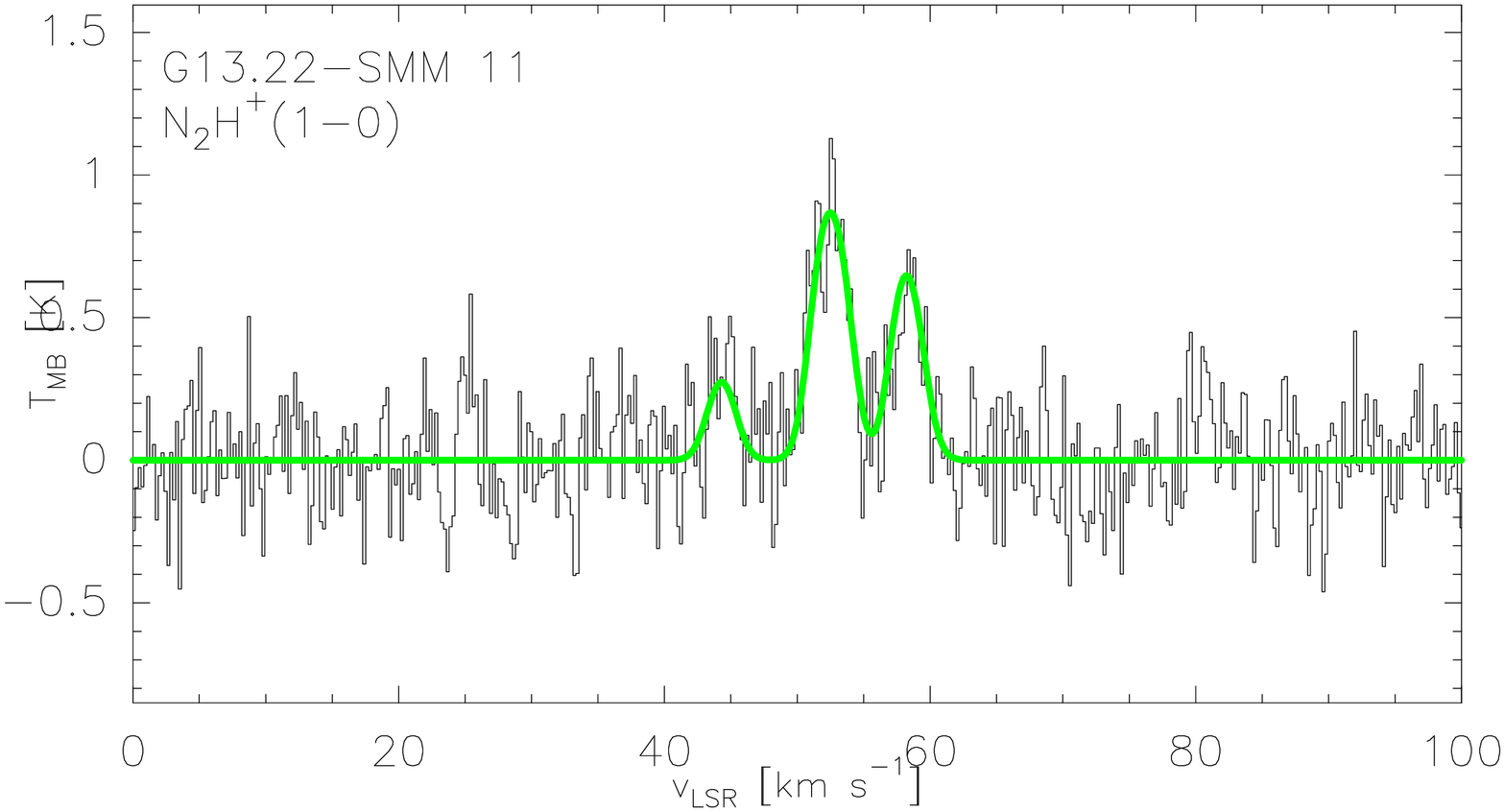}
\caption{Same as Fig.~\ref{figure:G187SMM1_spectra} but towards 
G13.22--SMM 11. Two velocity components are seen in the HCO$^+$ and HNC 
spectra.}
\label{figure:G1322SMM11_spectra}
\end{center}
\end{figure*}

\begin{figure*}
\begin{center}
\includegraphics[width=0.245\textwidth]{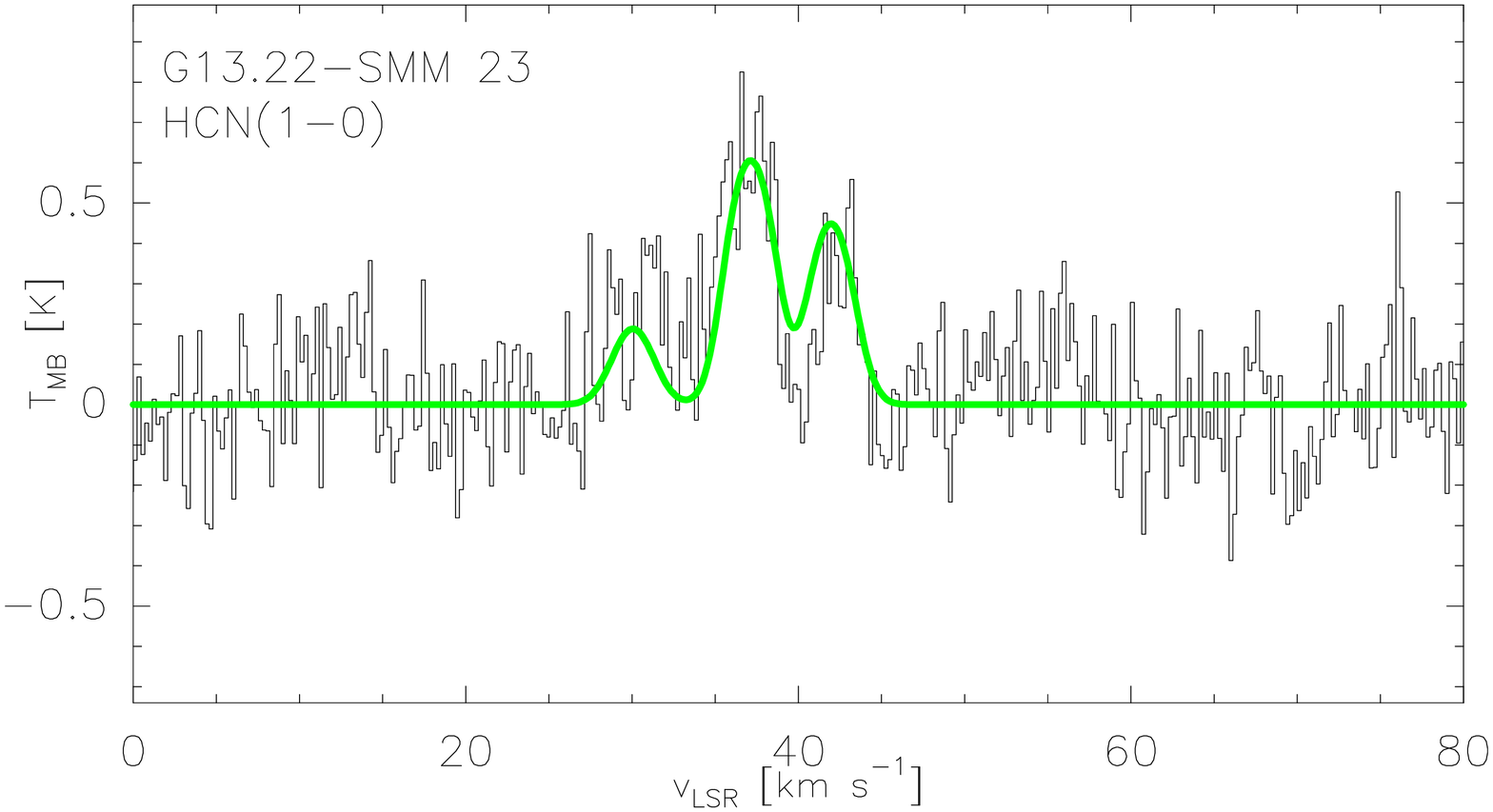}
\includegraphics[width=0.245\textwidth]{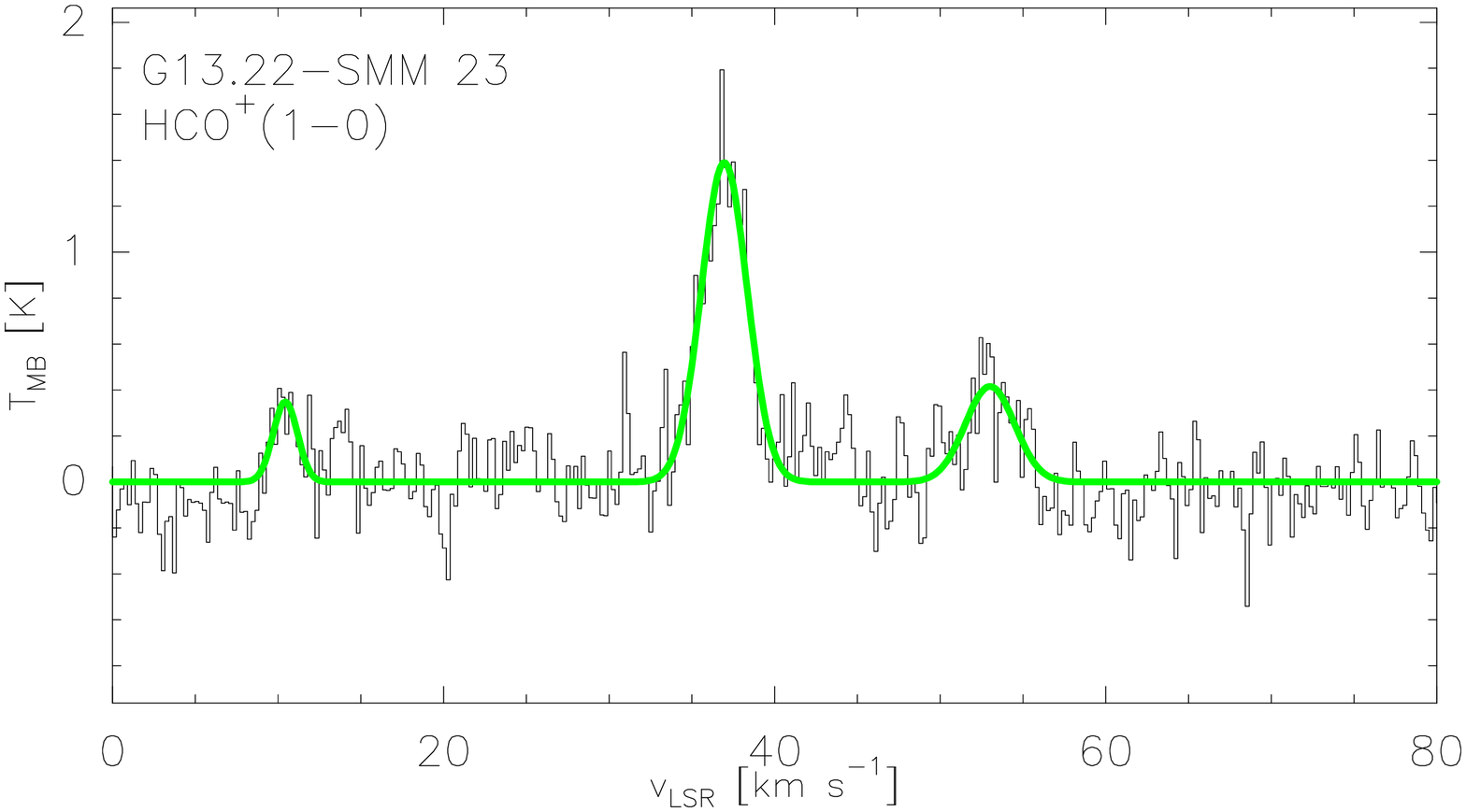}
\includegraphics[width=0.245\textwidth]{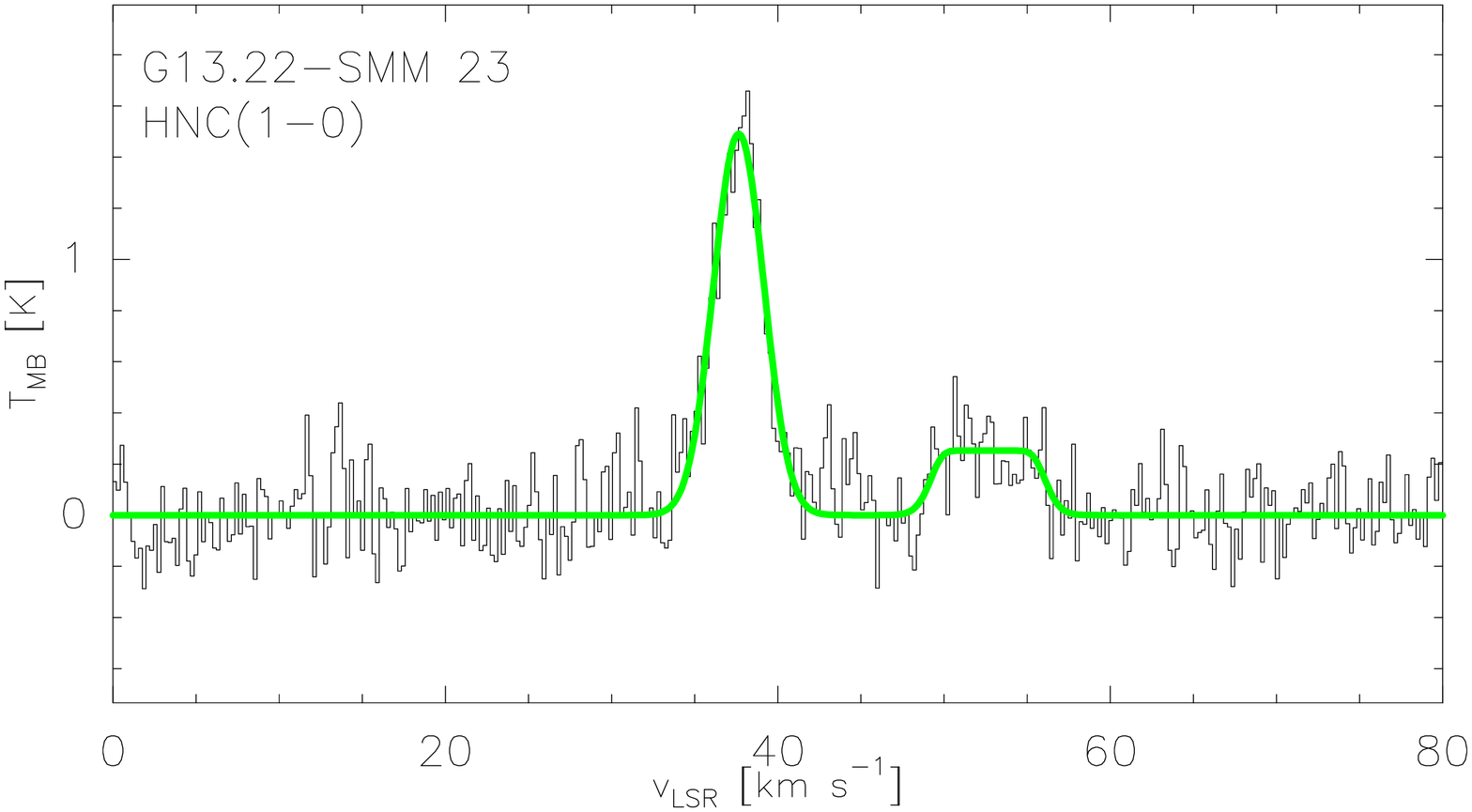}
\includegraphics[width=0.245\textwidth]{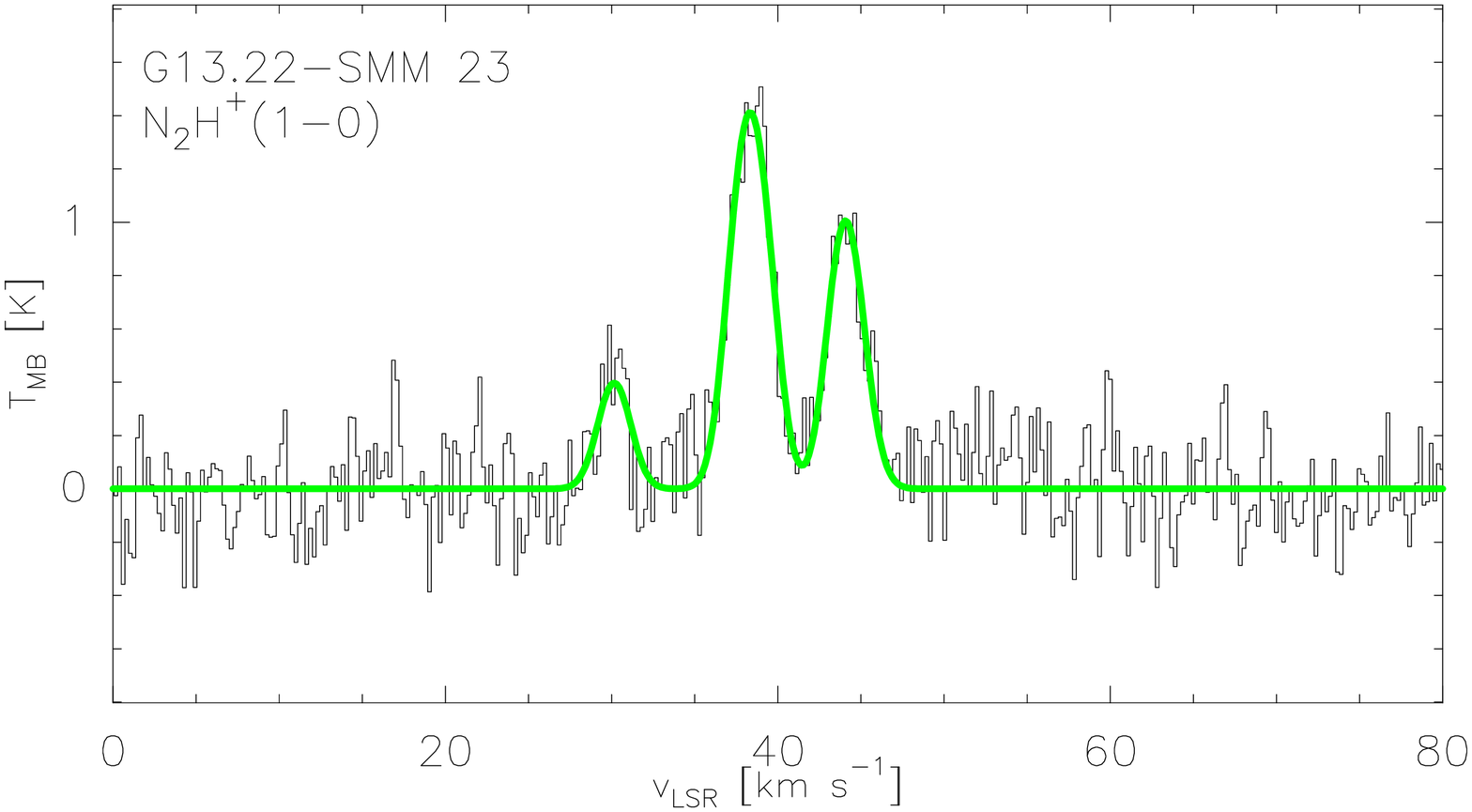}
\caption{Same as Fig.~\ref{figure:G187SMM1_spectra} but towards 
G13.22--SMM 23. Three velocity components are seen in the HCO$^+$ spectrum, 
while two are visible in the HNC spectrum.}
\label{figure:G1322SMM23_spectra}
\end{center}
\end{figure*}

\begin{figure*}
\begin{center}
\includegraphics[width=0.245\textwidth]{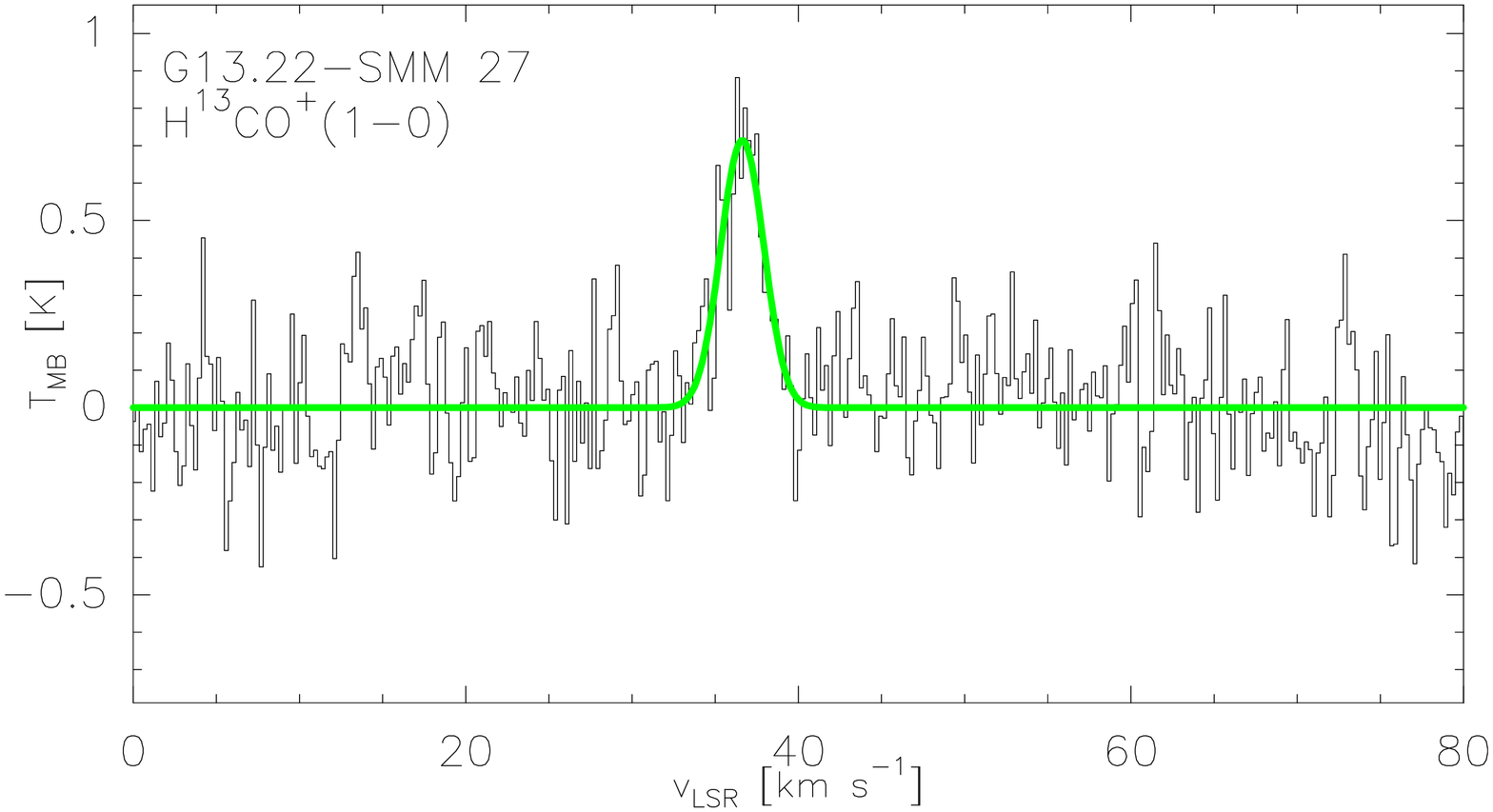}
\includegraphics[width=0.245\textwidth]{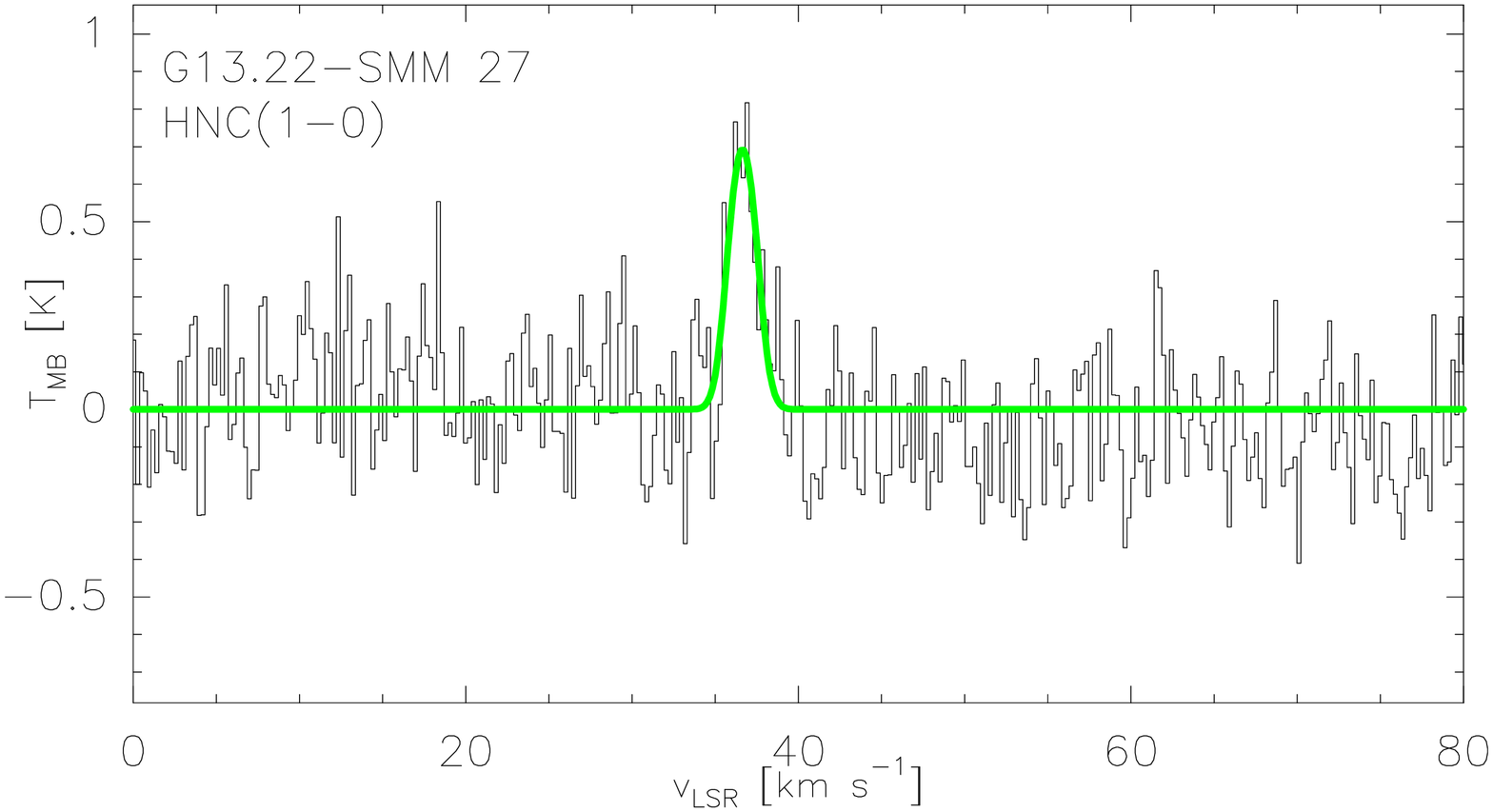}
\includegraphics[width=0.245\textwidth]{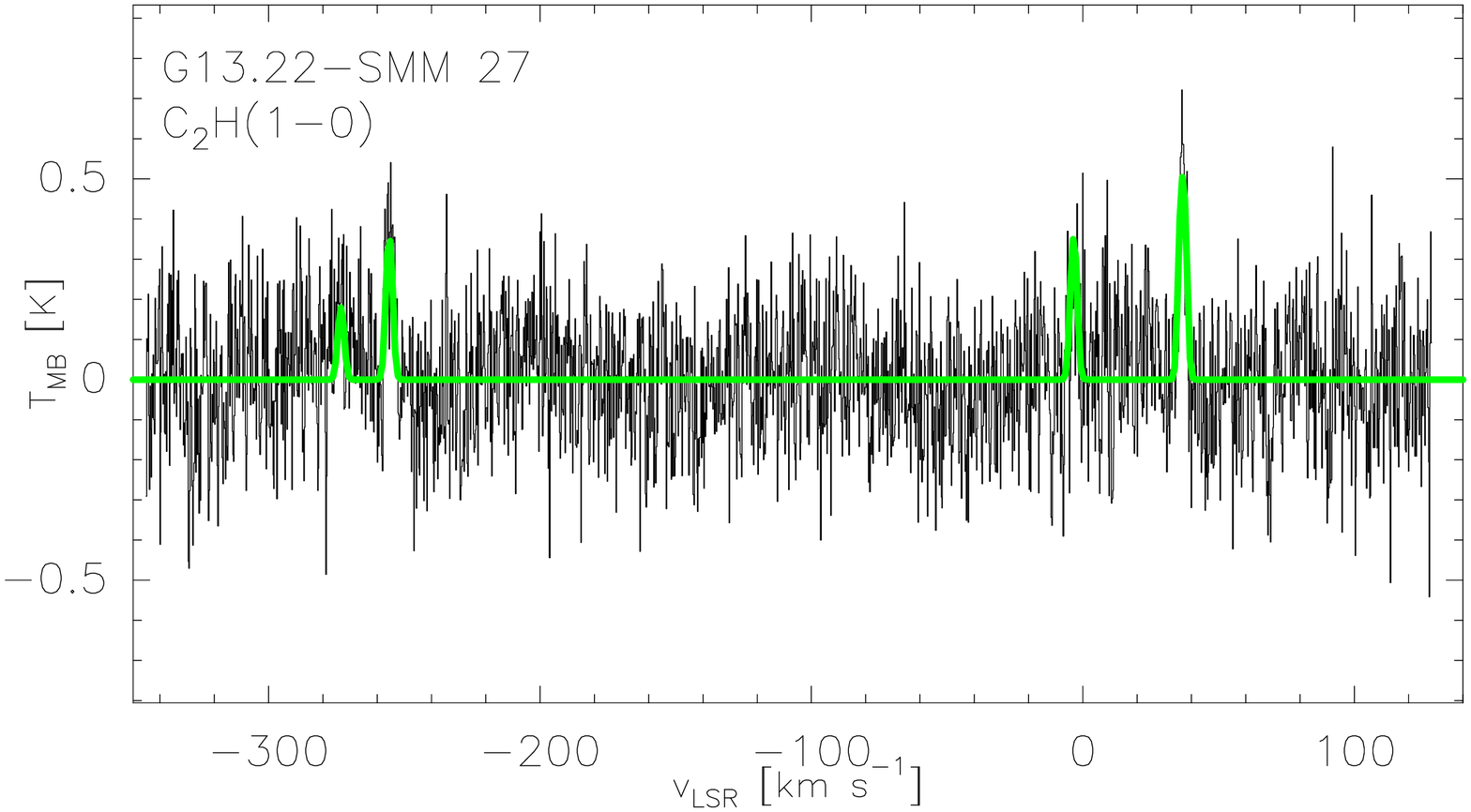}
\includegraphics[width=0.245\textwidth]{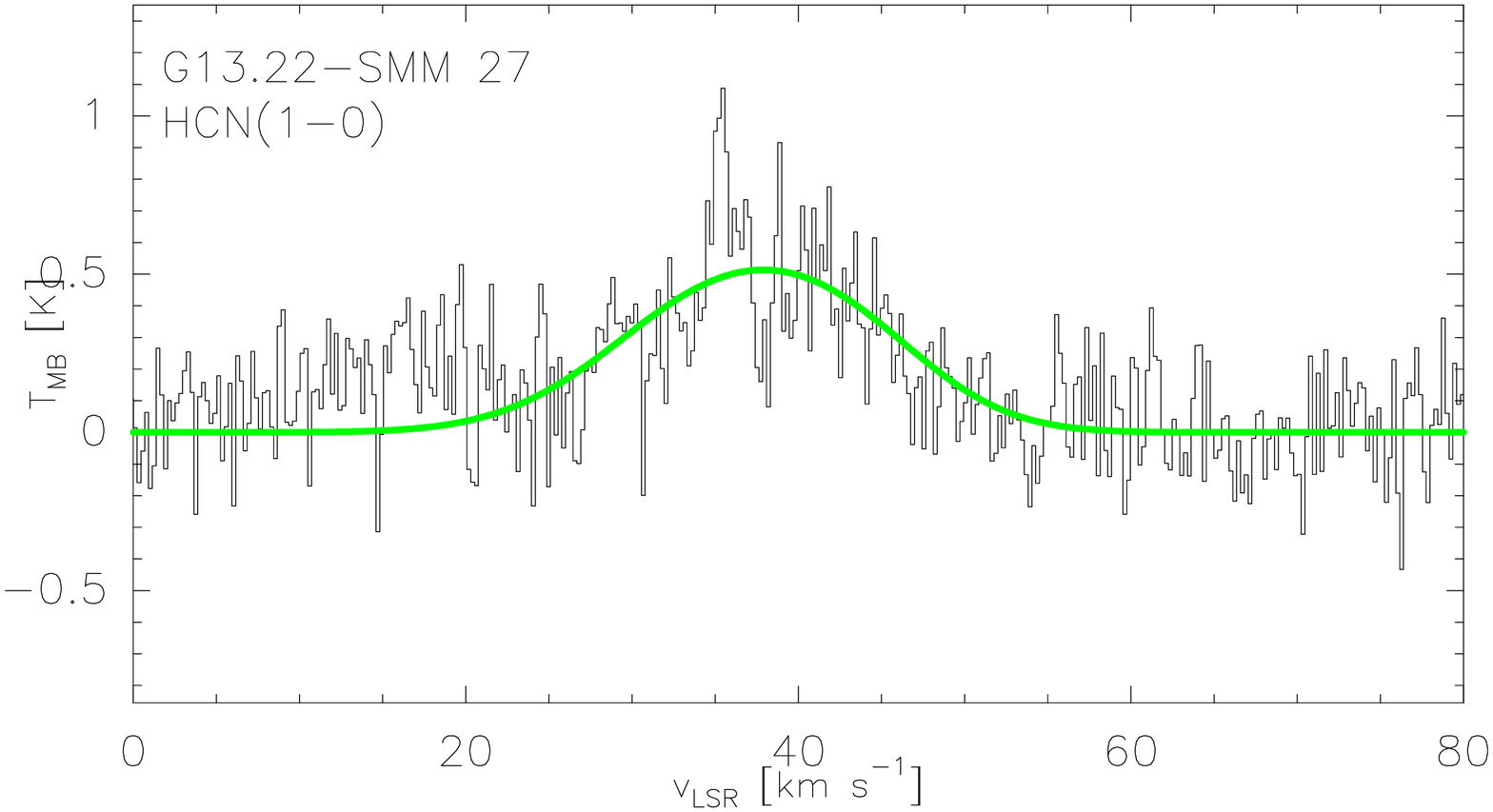}
\includegraphics[width=0.245\textwidth]{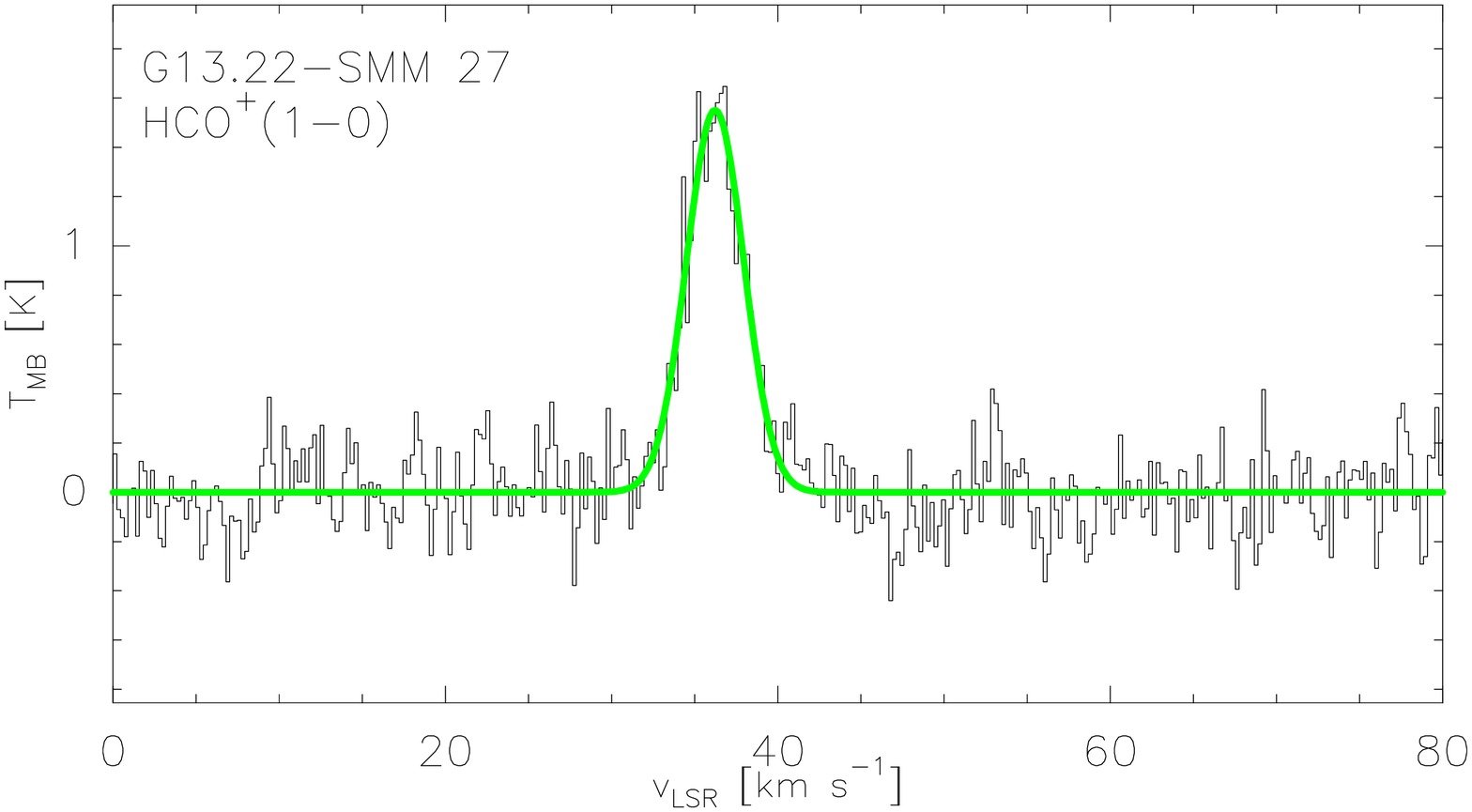}
\includegraphics[width=0.245\textwidth]{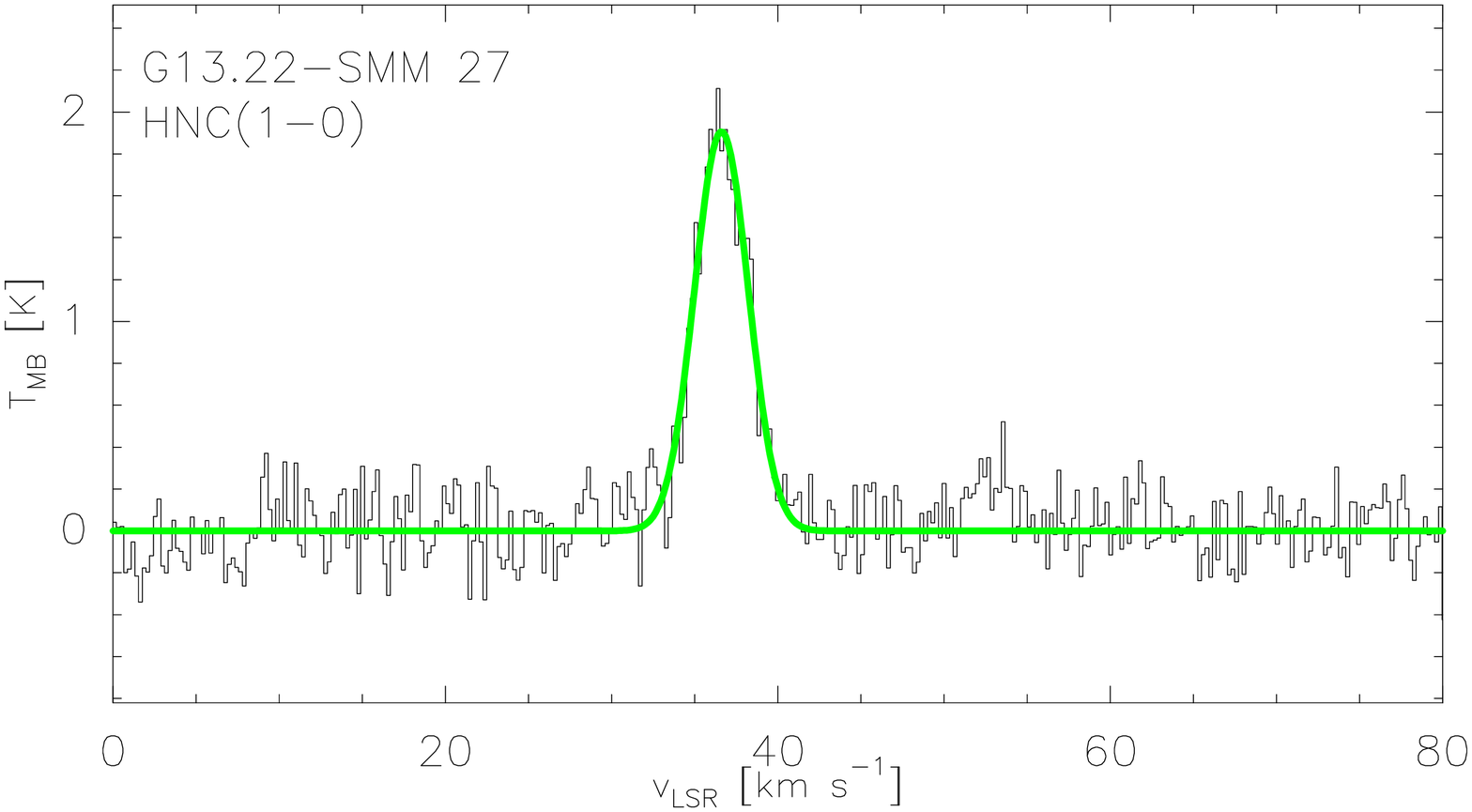}
\includegraphics[width=0.245\textwidth]{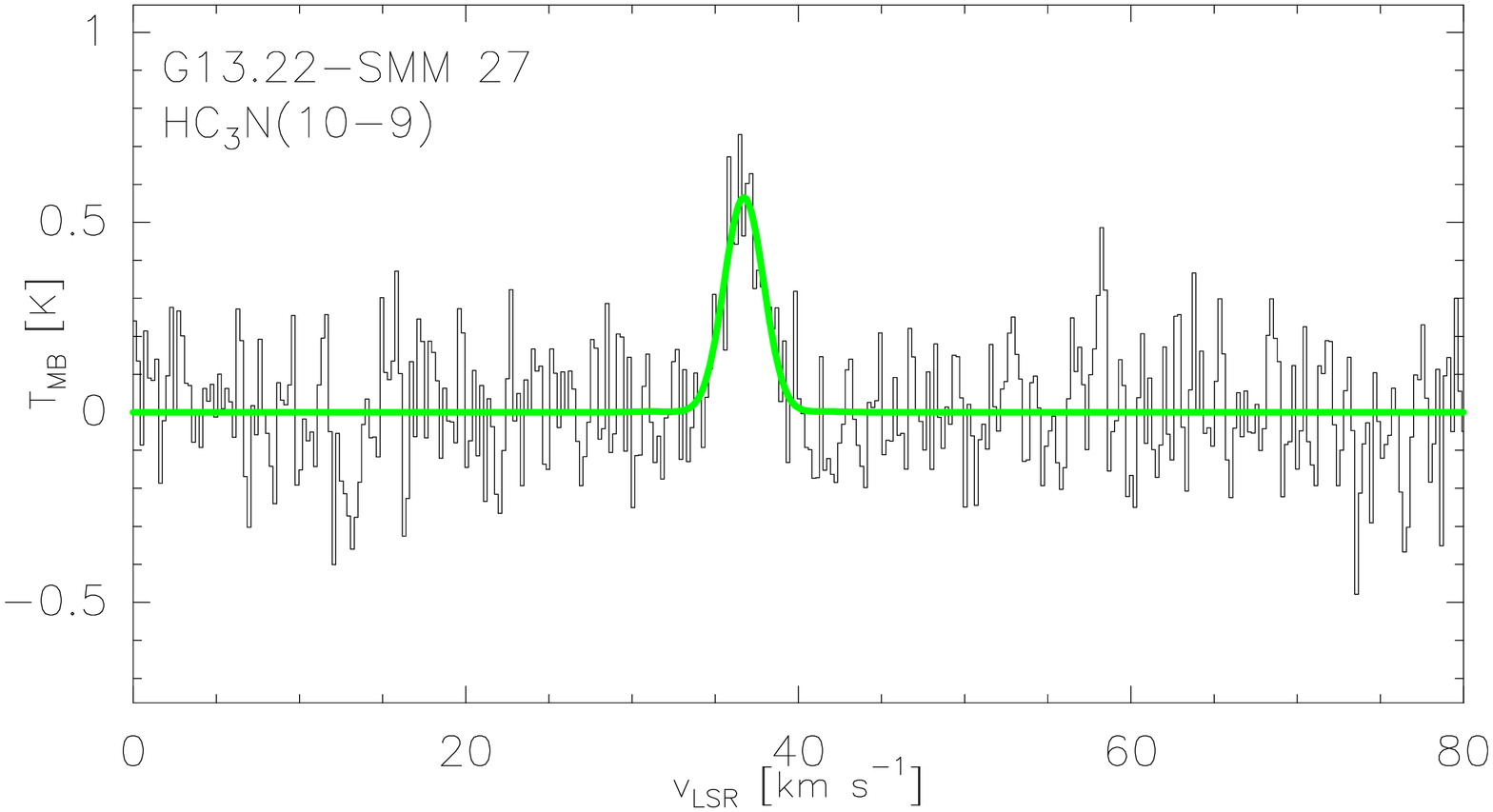}
\includegraphics[width=0.245\textwidth]{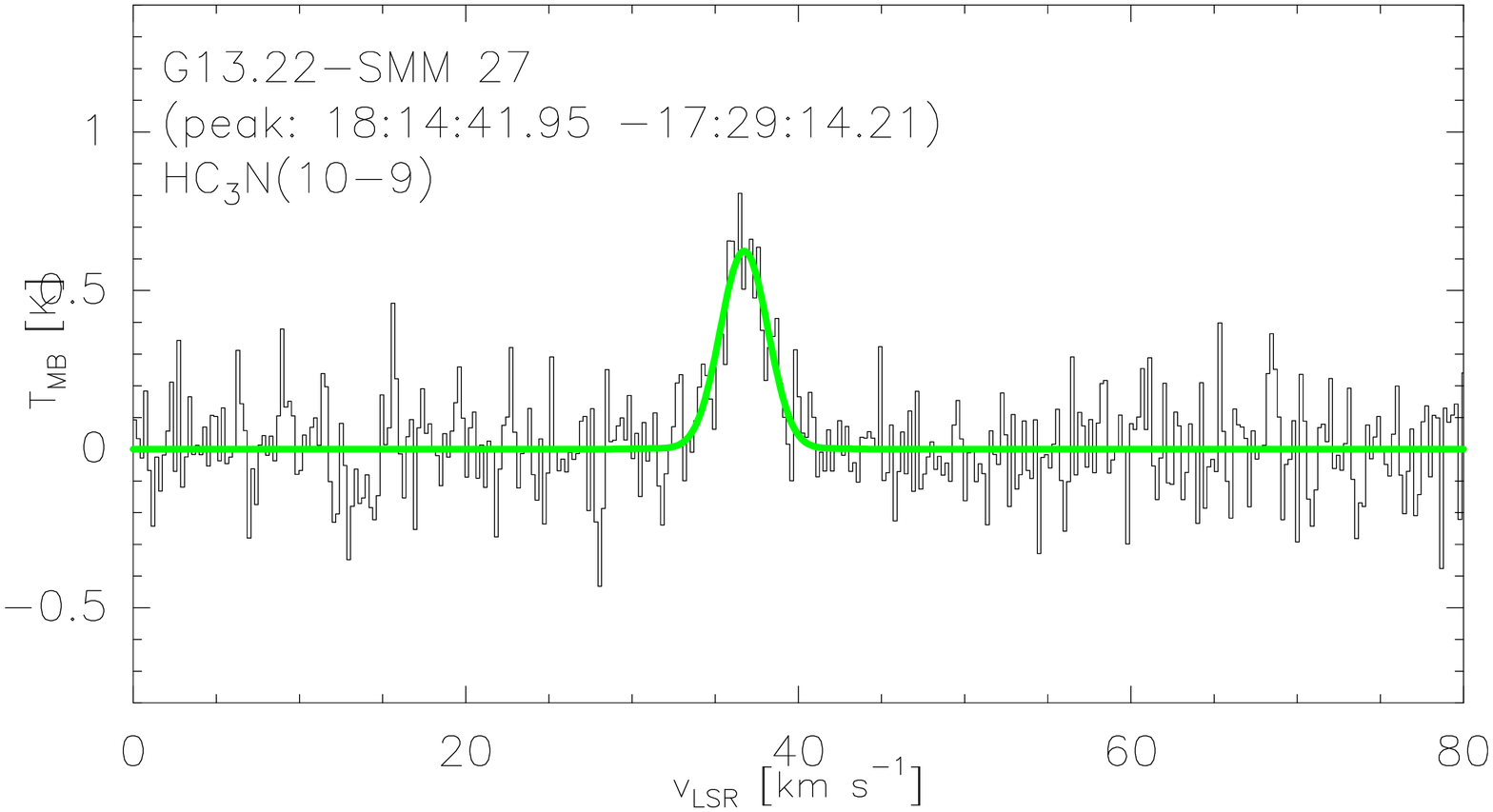}
\includegraphics[width=0.245\textwidth]{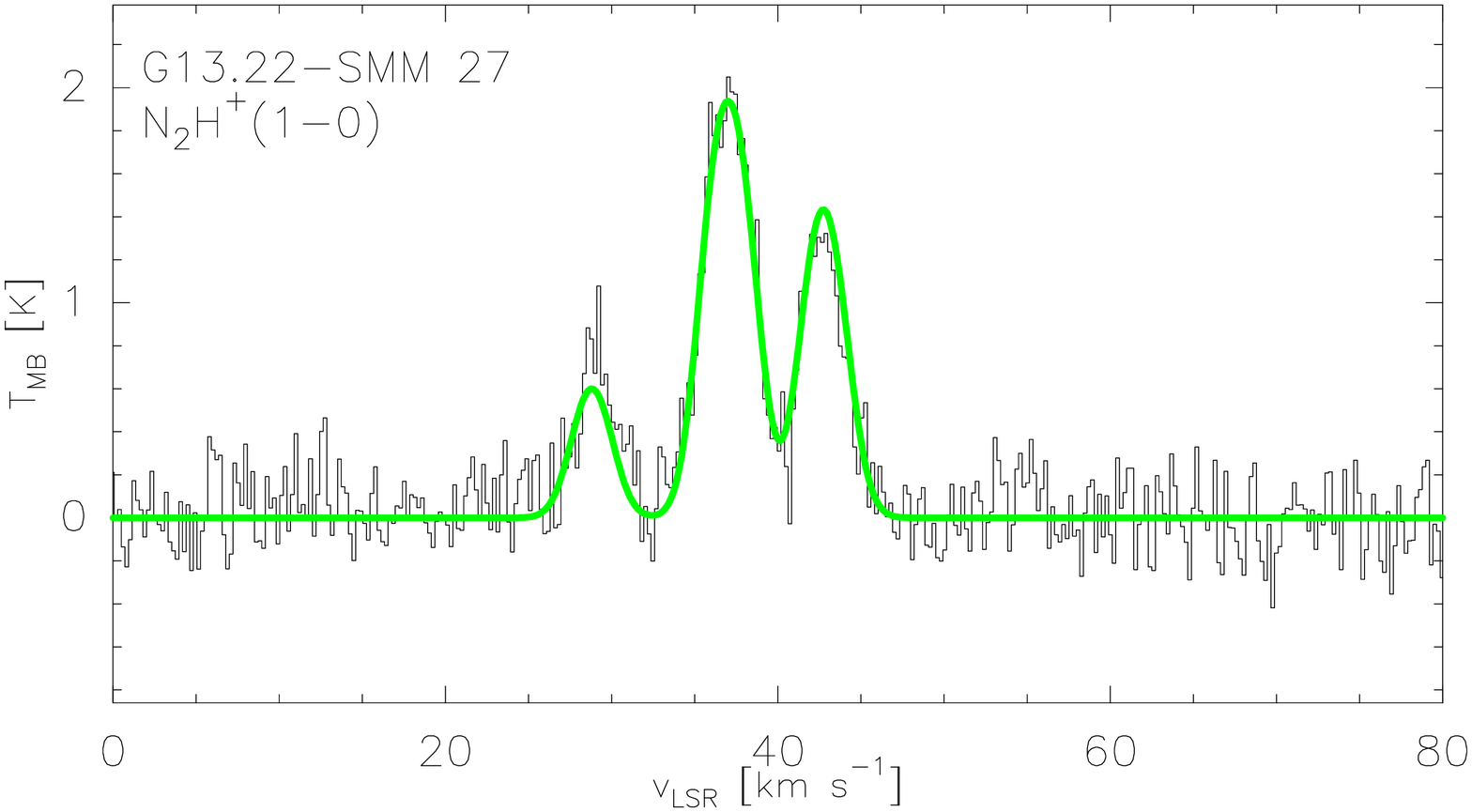}
\includegraphics[width=0.245\textwidth]{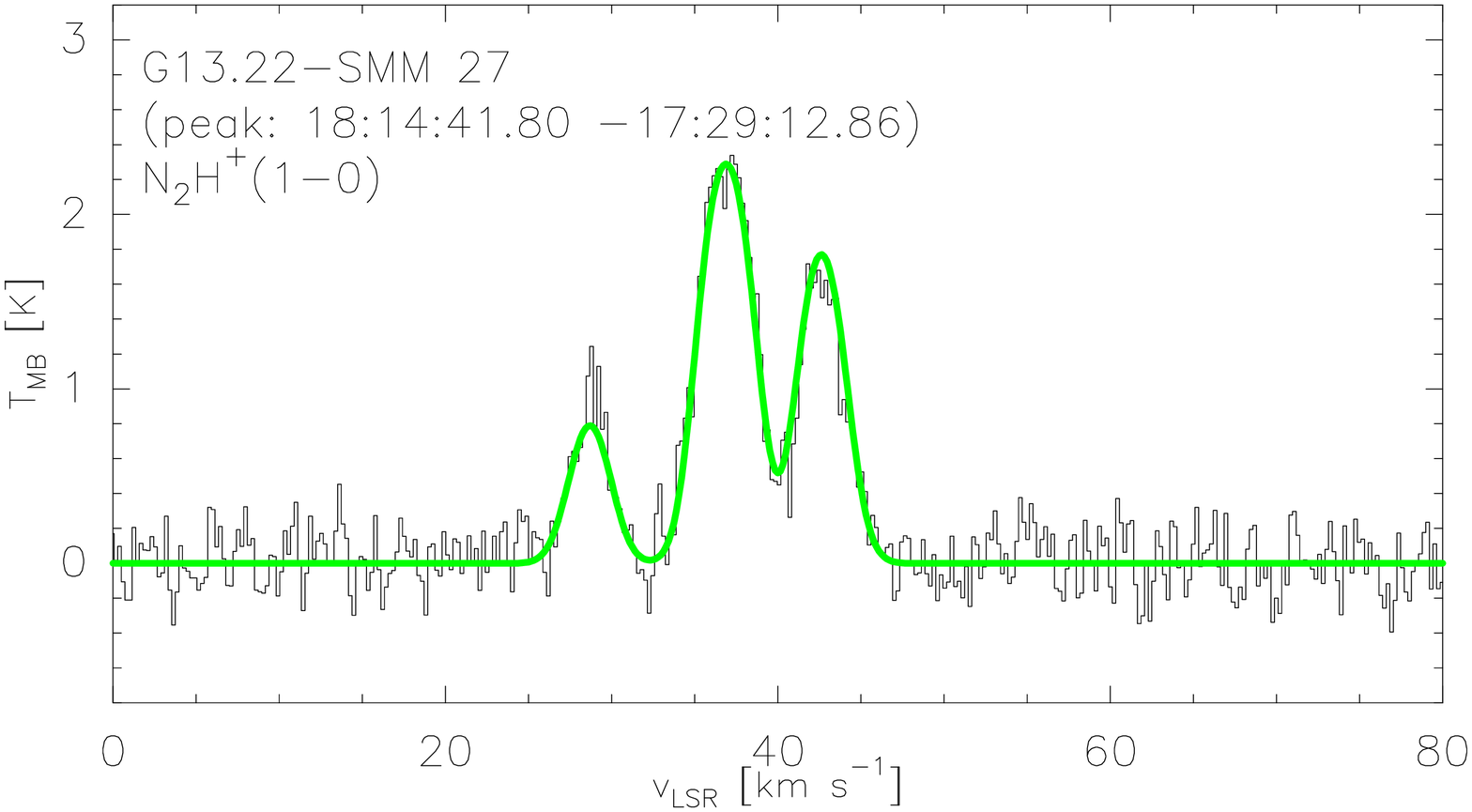}
\caption{Same as Fig.~\ref{figure:G187SMM1_spectra} but towards 
G13.22--SMM 27. The velocity range in the C$_2$H spectrum is wider.}
\label{figure:G1322SMM27_spectra}
\end{center}
\end{figure*}

\begin{figure*}
\begin{center}
\includegraphics[width=0.245\textwidth]{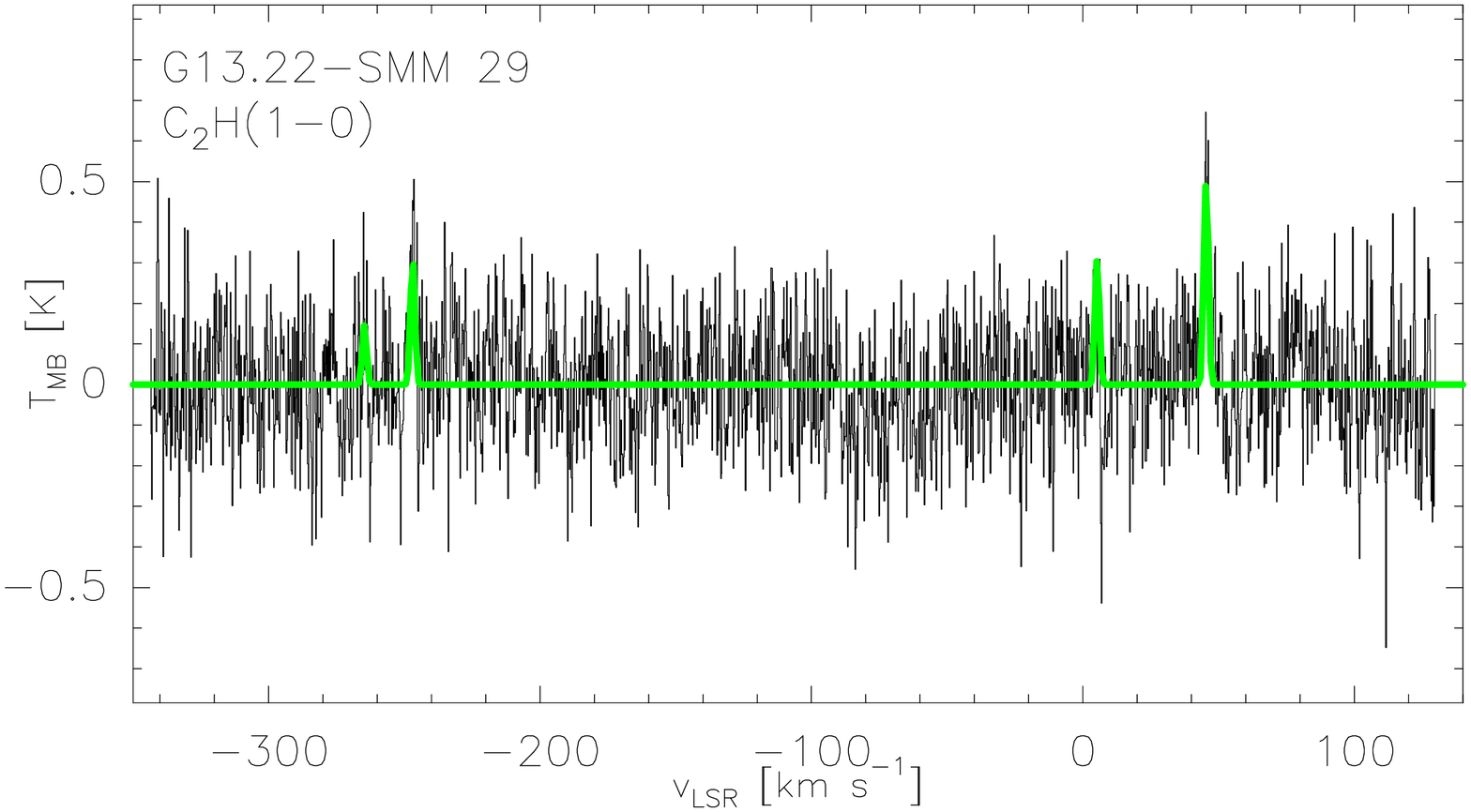}
\includegraphics[width=0.245\textwidth]{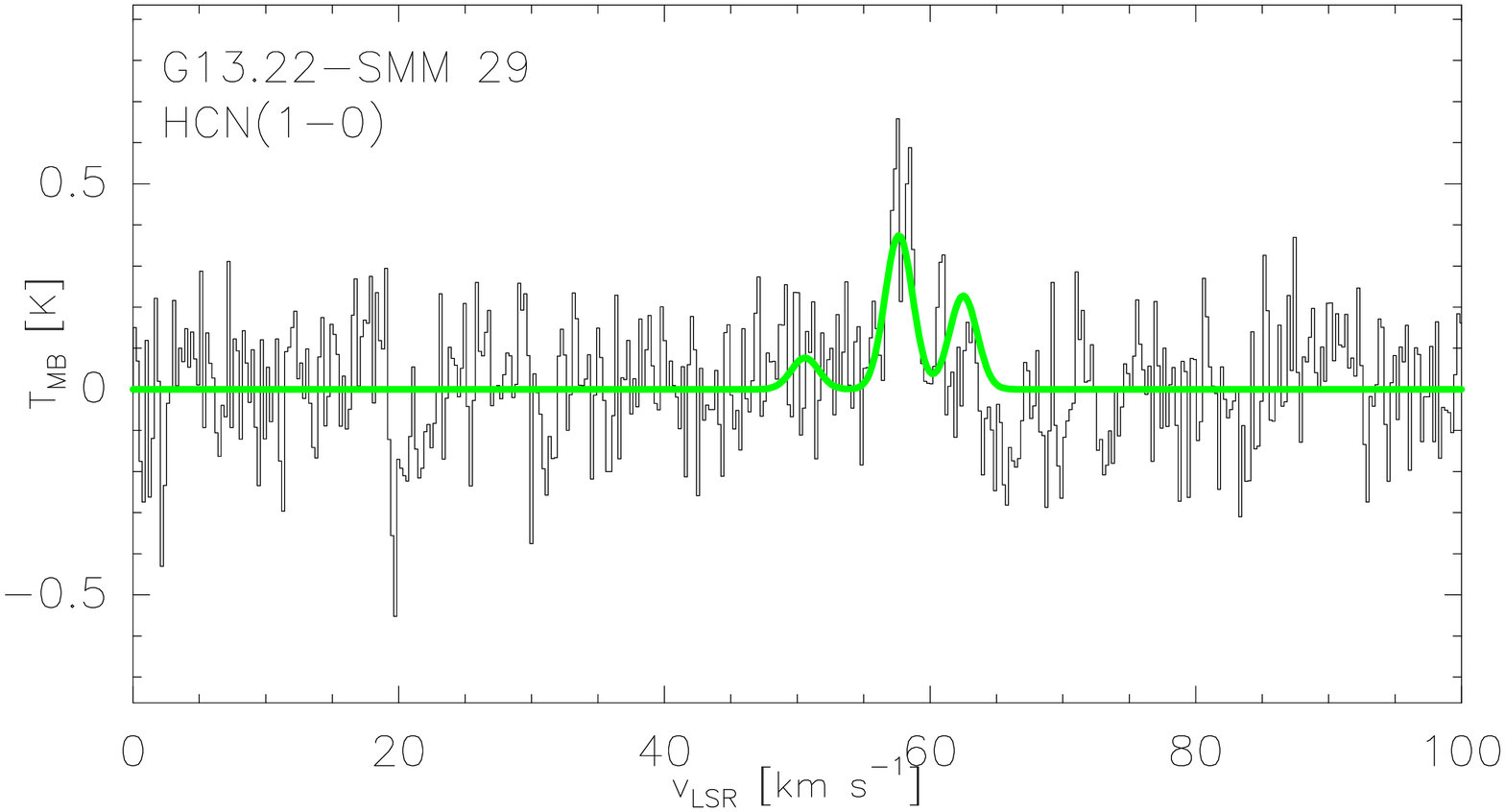}
\includegraphics[width=0.245\textwidth]{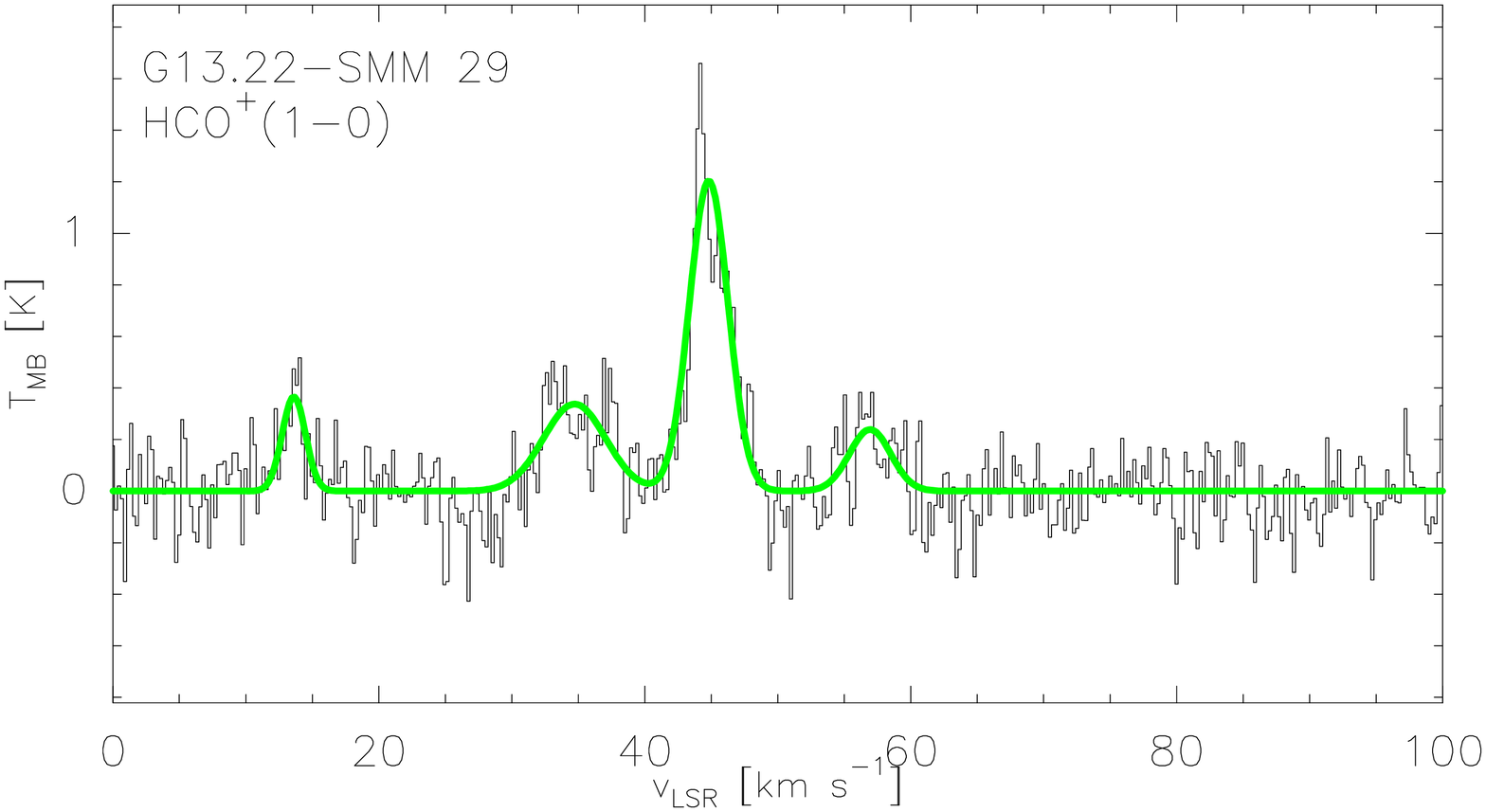}
\includegraphics[width=0.245\textwidth]{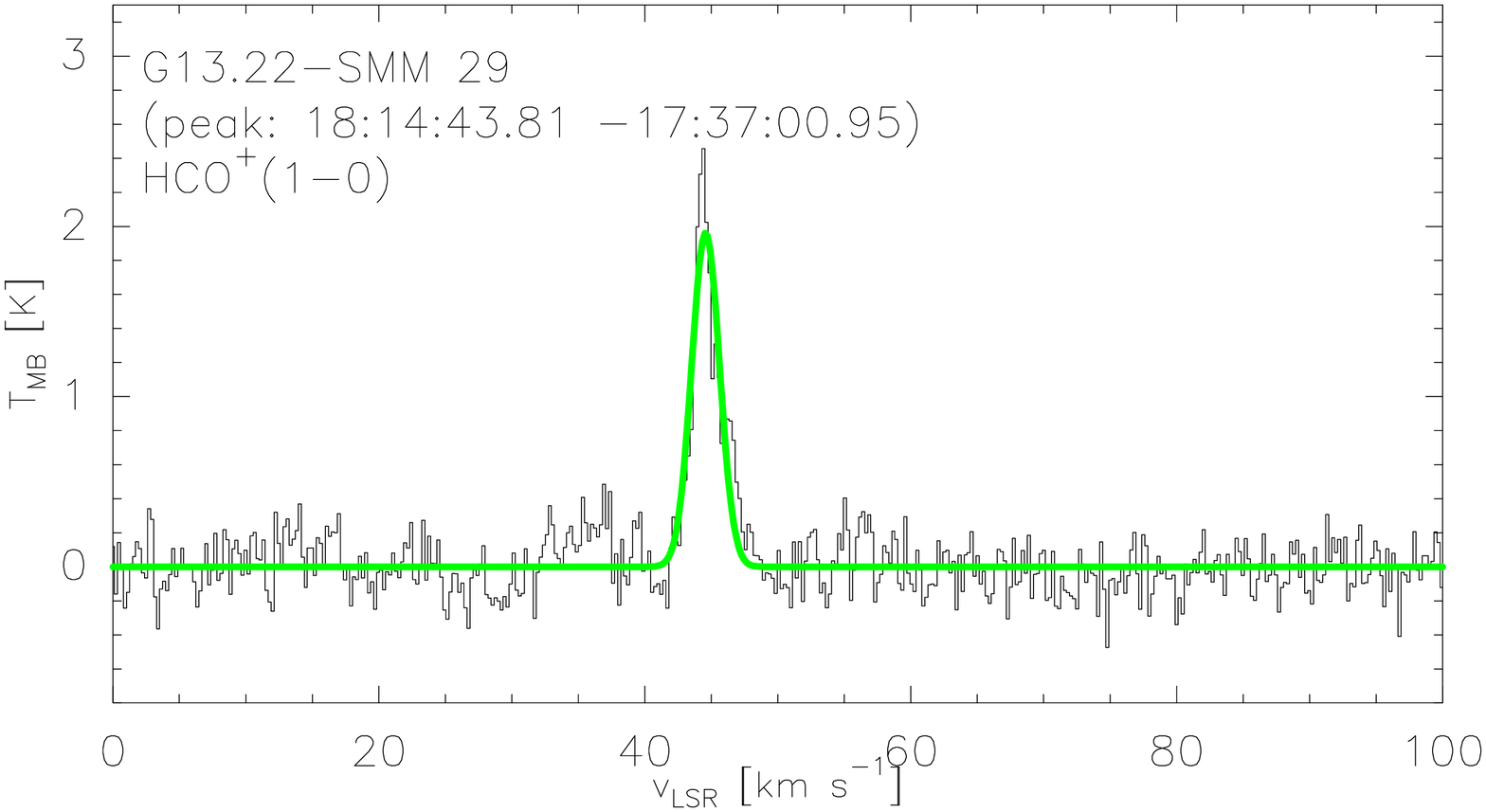}
\includegraphics[width=0.245\textwidth]{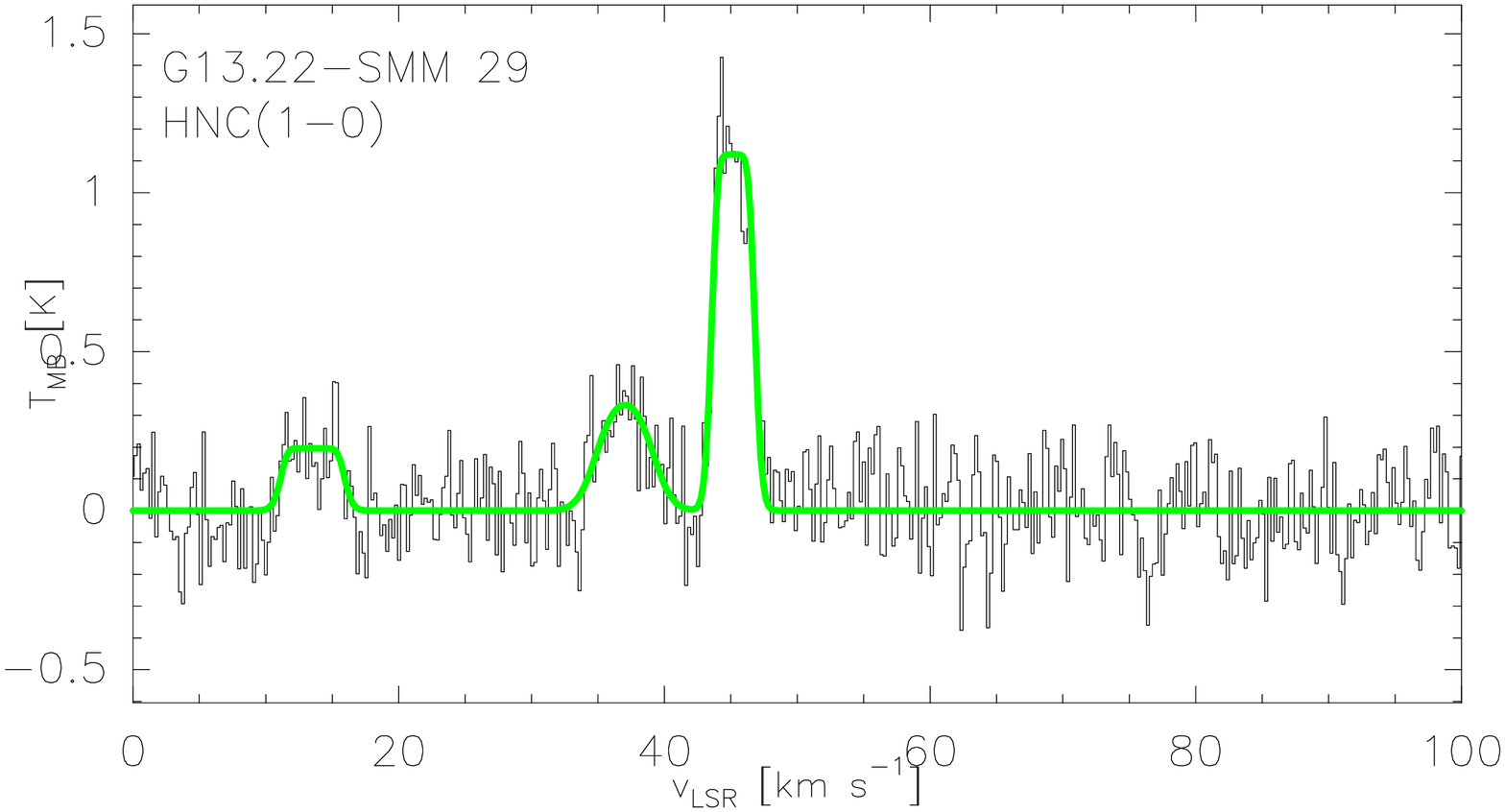}
\includegraphics[width=0.245\textwidth]{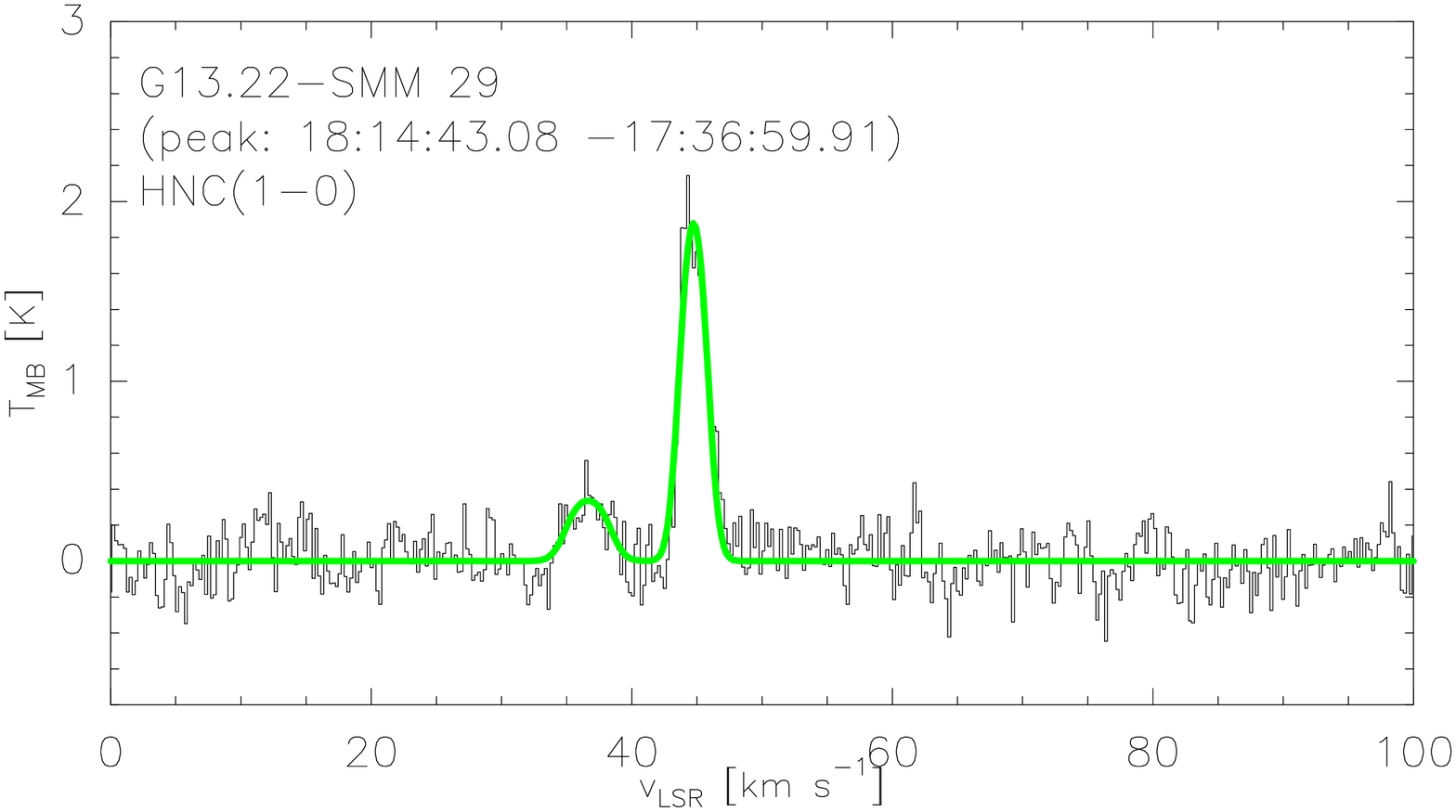}
\includegraphics[width=0.245\textwidth]{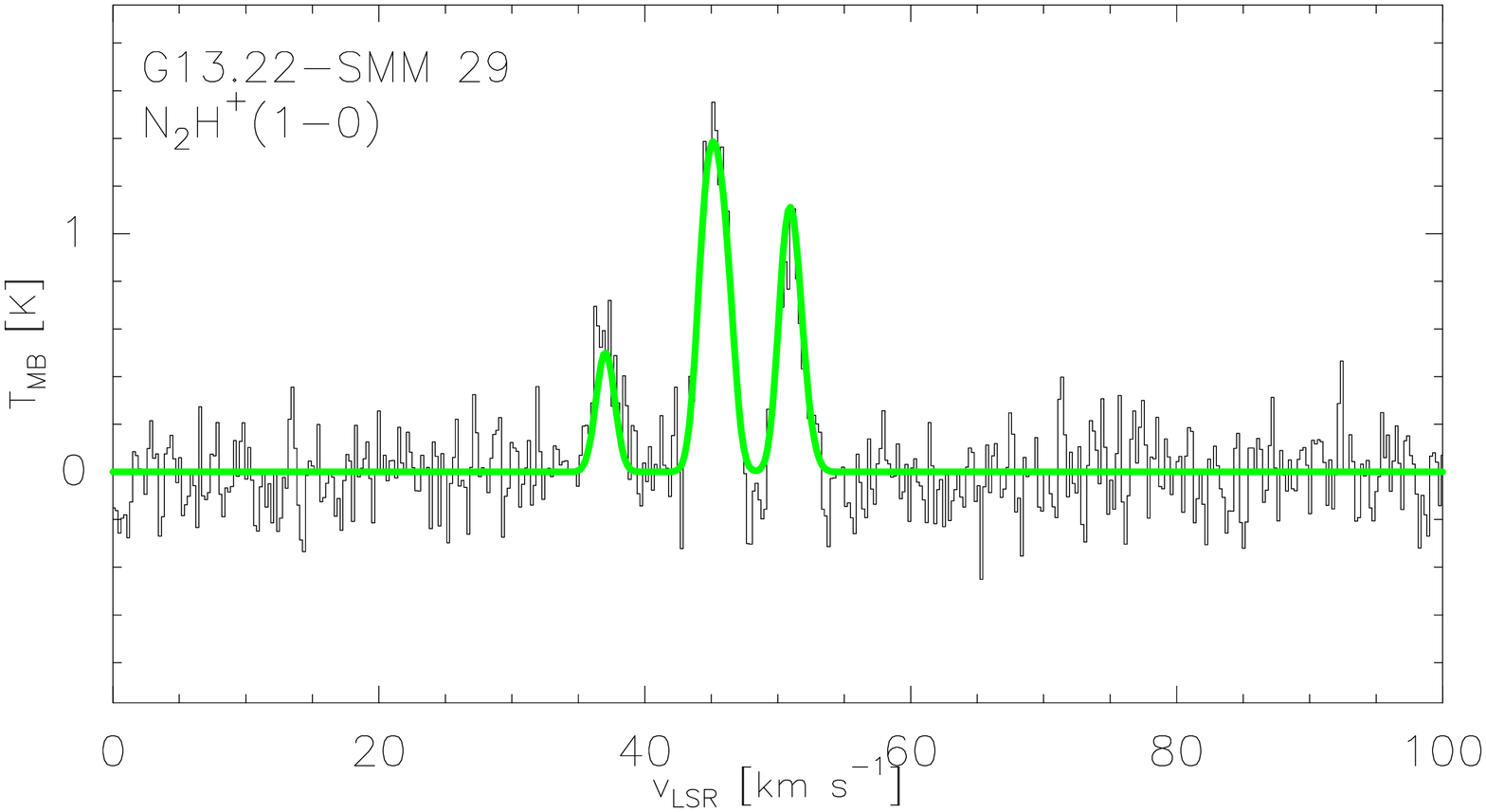}
\includegraphics[width=0.245\textwidth]{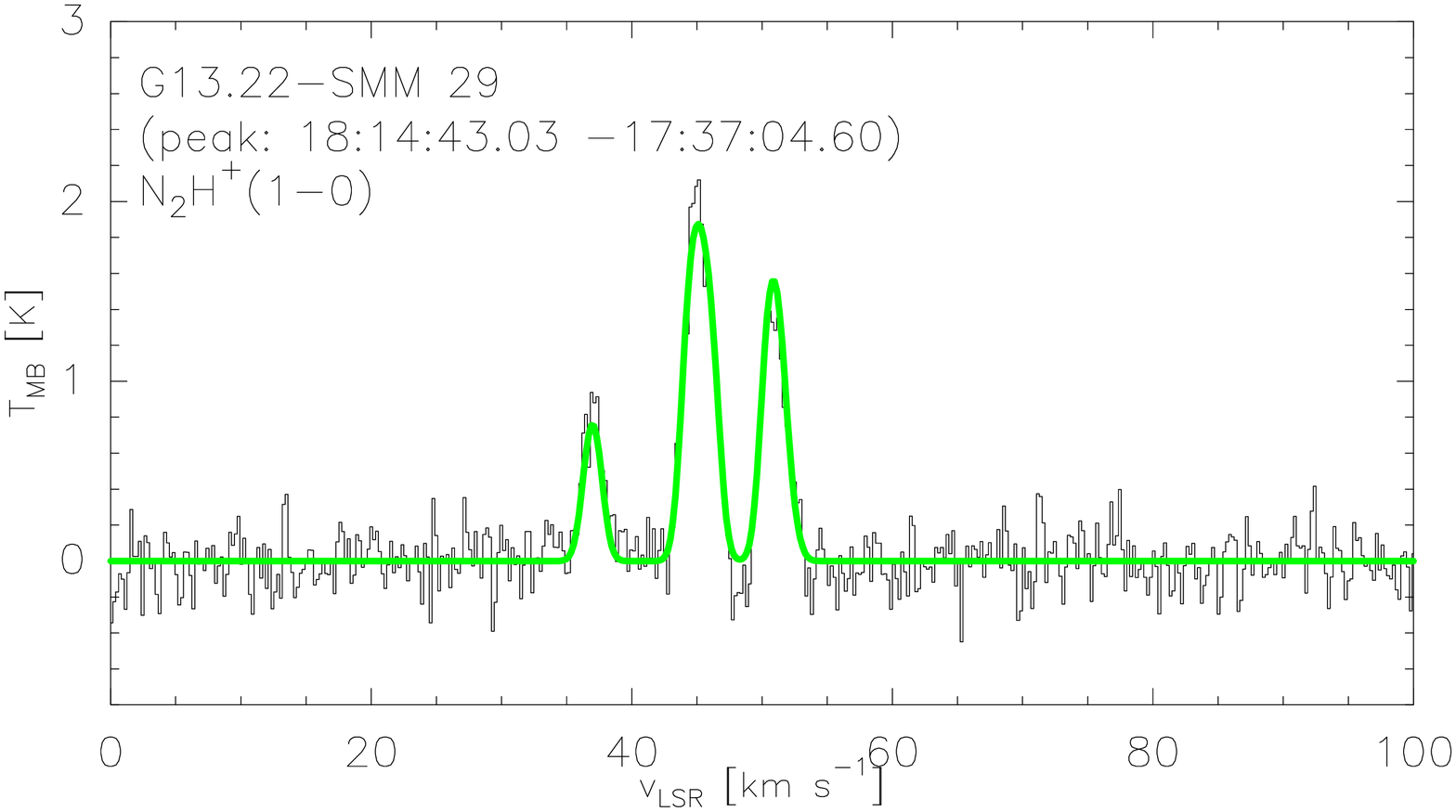}
\caption{Same as Fig.~\ref{figure:G187SMM1_spectra} but towards 
G13.22--SMM 29. The velocity range in the C$_2$H spectrum is wider. Four 
velocity components are seen in the HCO$^+$ spectrum, three in the HNC 
spectrum towards the LABOCA peak, and two in the HNC spectrum towards the line 
emission peak.}
\label{figure:G1322SMM29_spectra}
\end{center}
\end{figure*}

\begin{figure*}
\begin{center}
\includegraphics[width=0.245\textwidth]{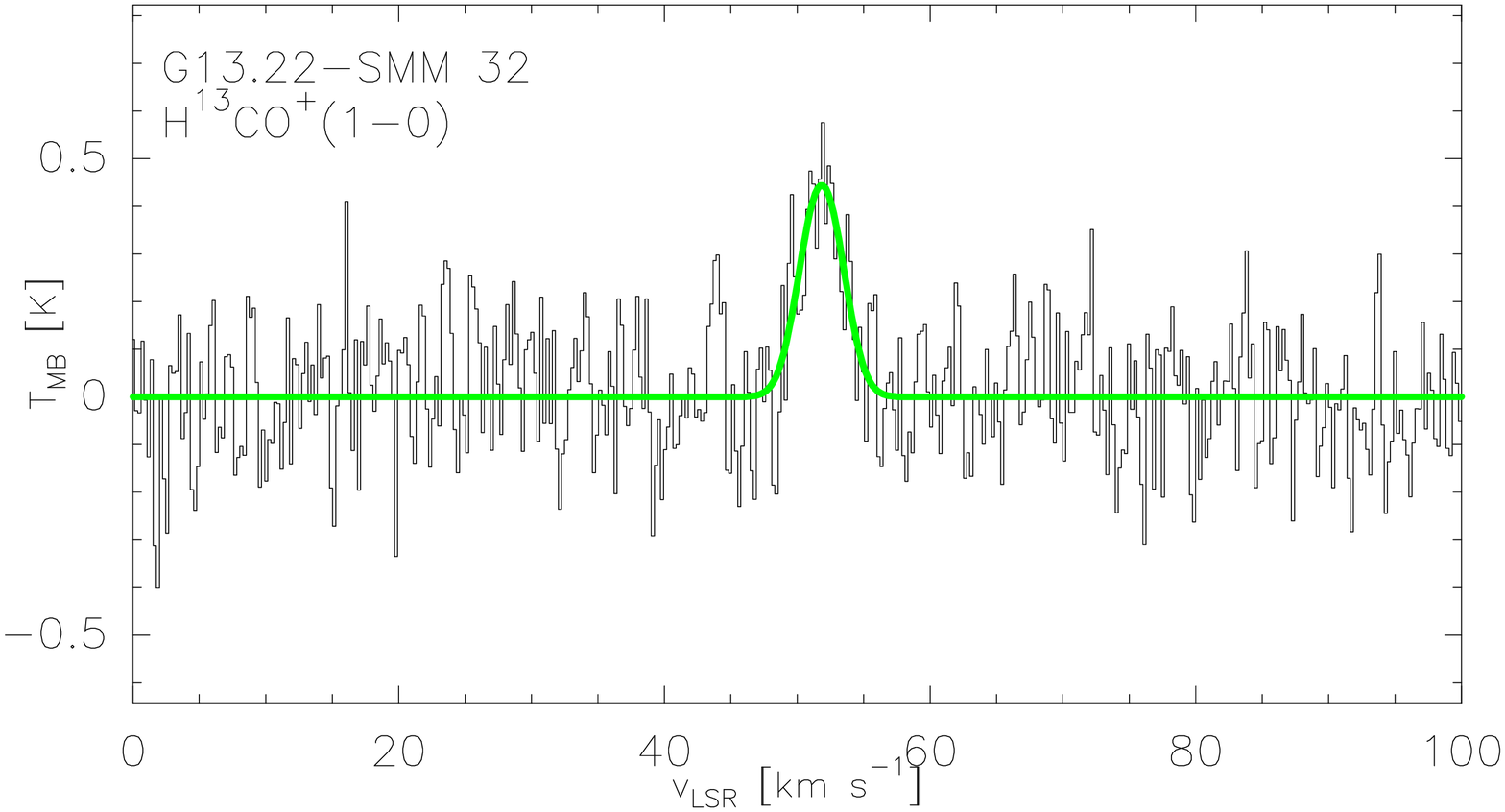}
\includegraphics[width=0.245\textwidth]{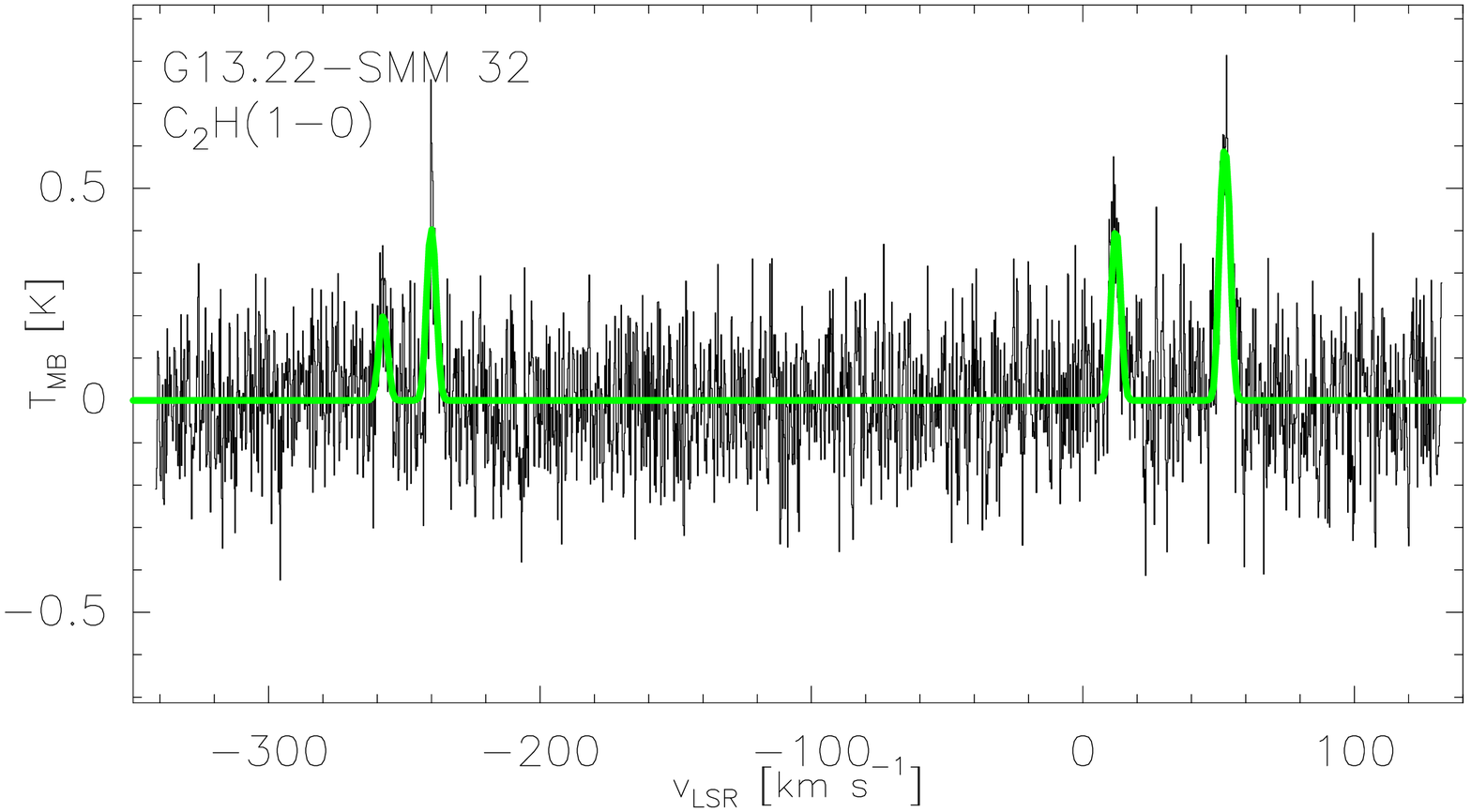}
\includegraphics[width=0.245\textwidth]{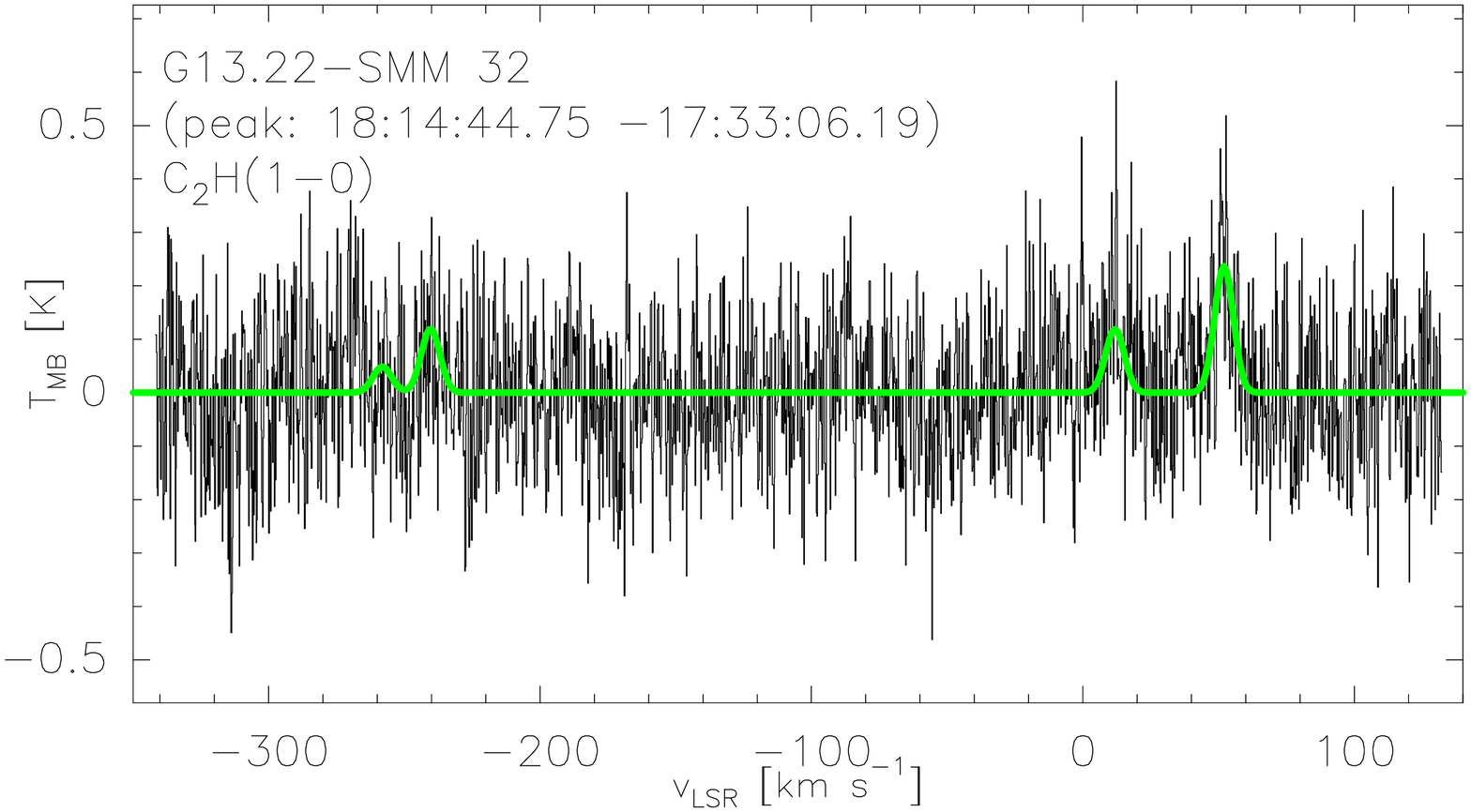}
\includegraphics[width=0.245\textwidth]{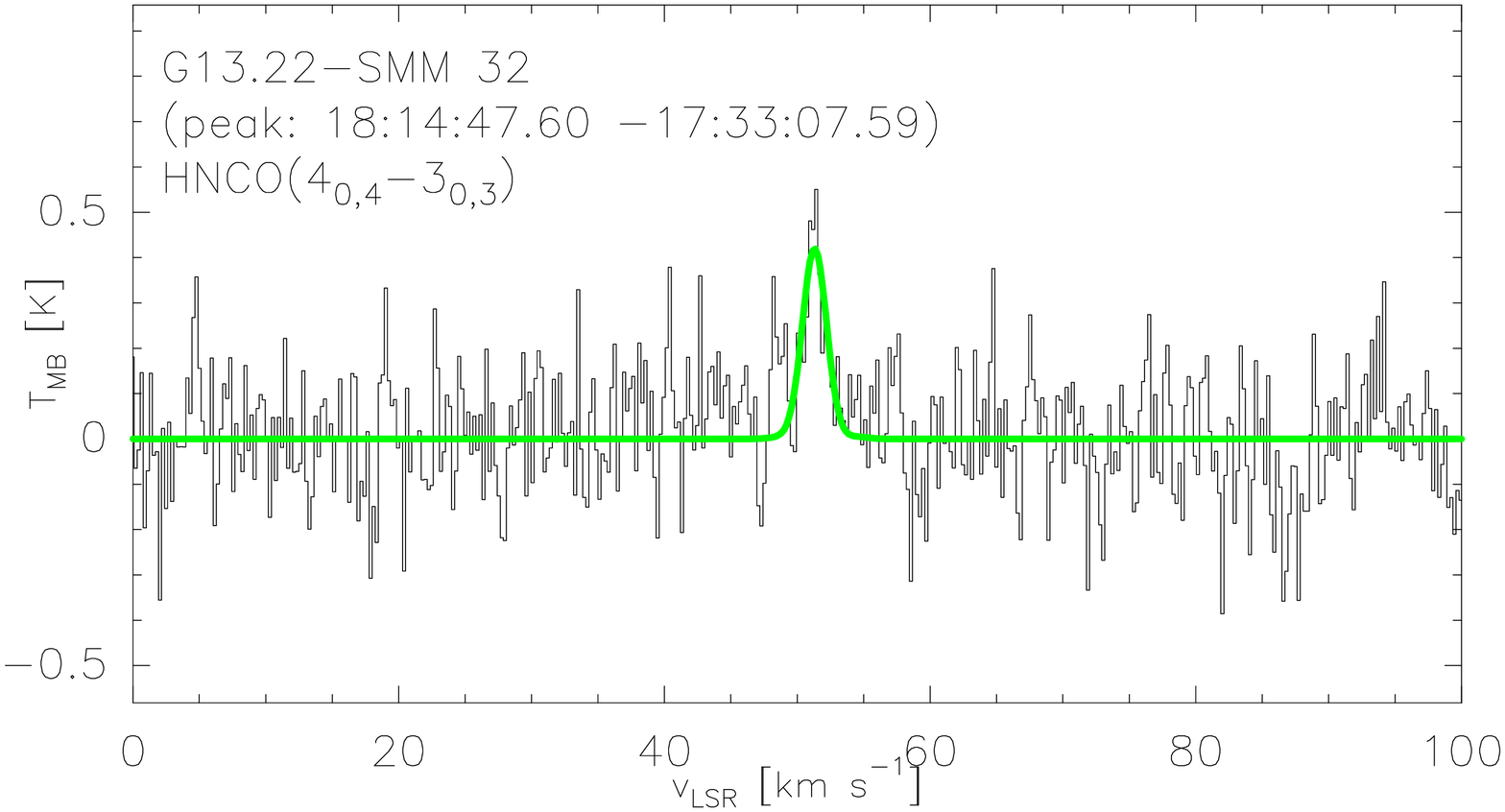}
\includegraphics[width=0.245\textwidth]{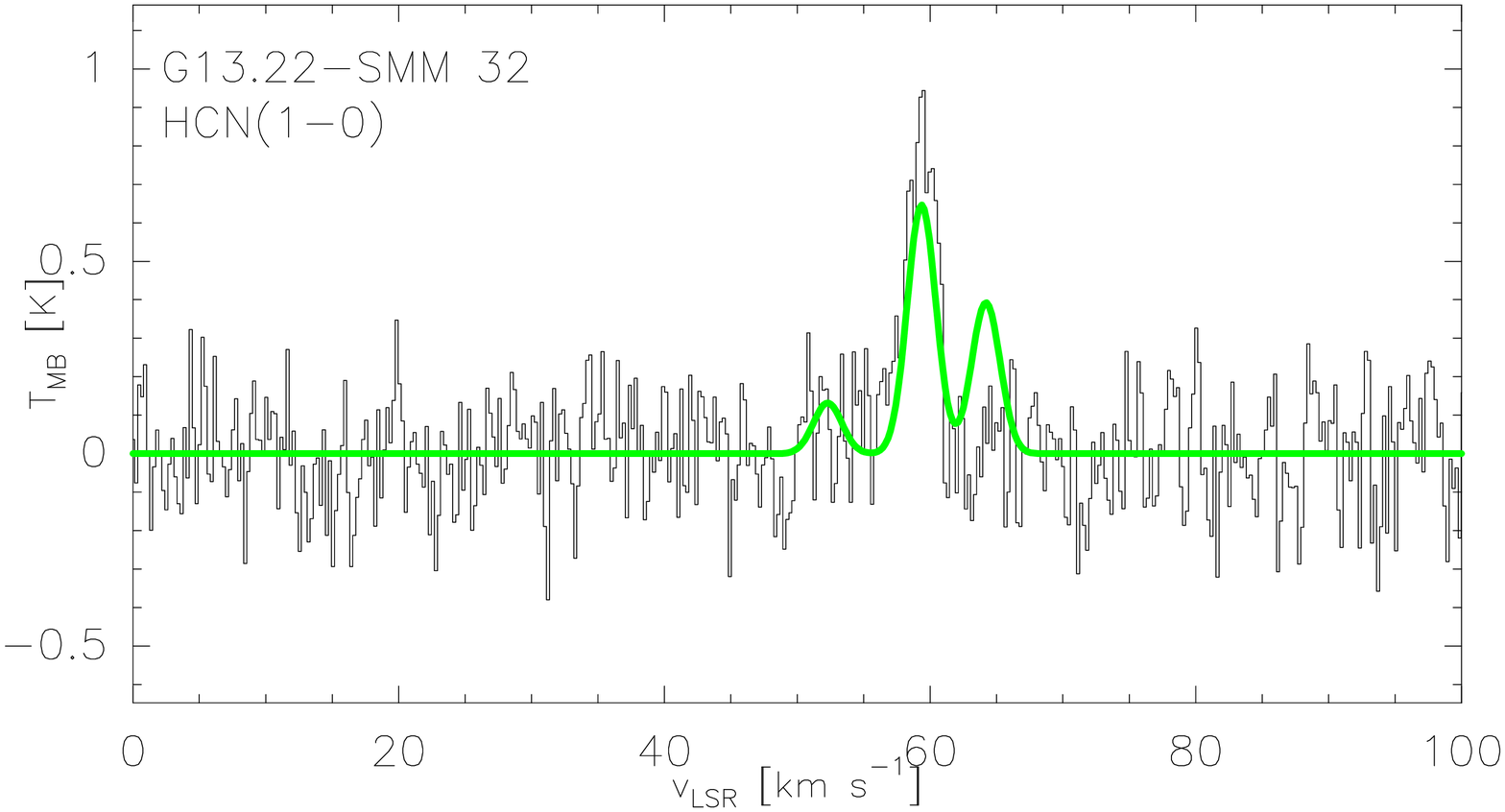}
\includegraphics[width=0.245\textwidth]{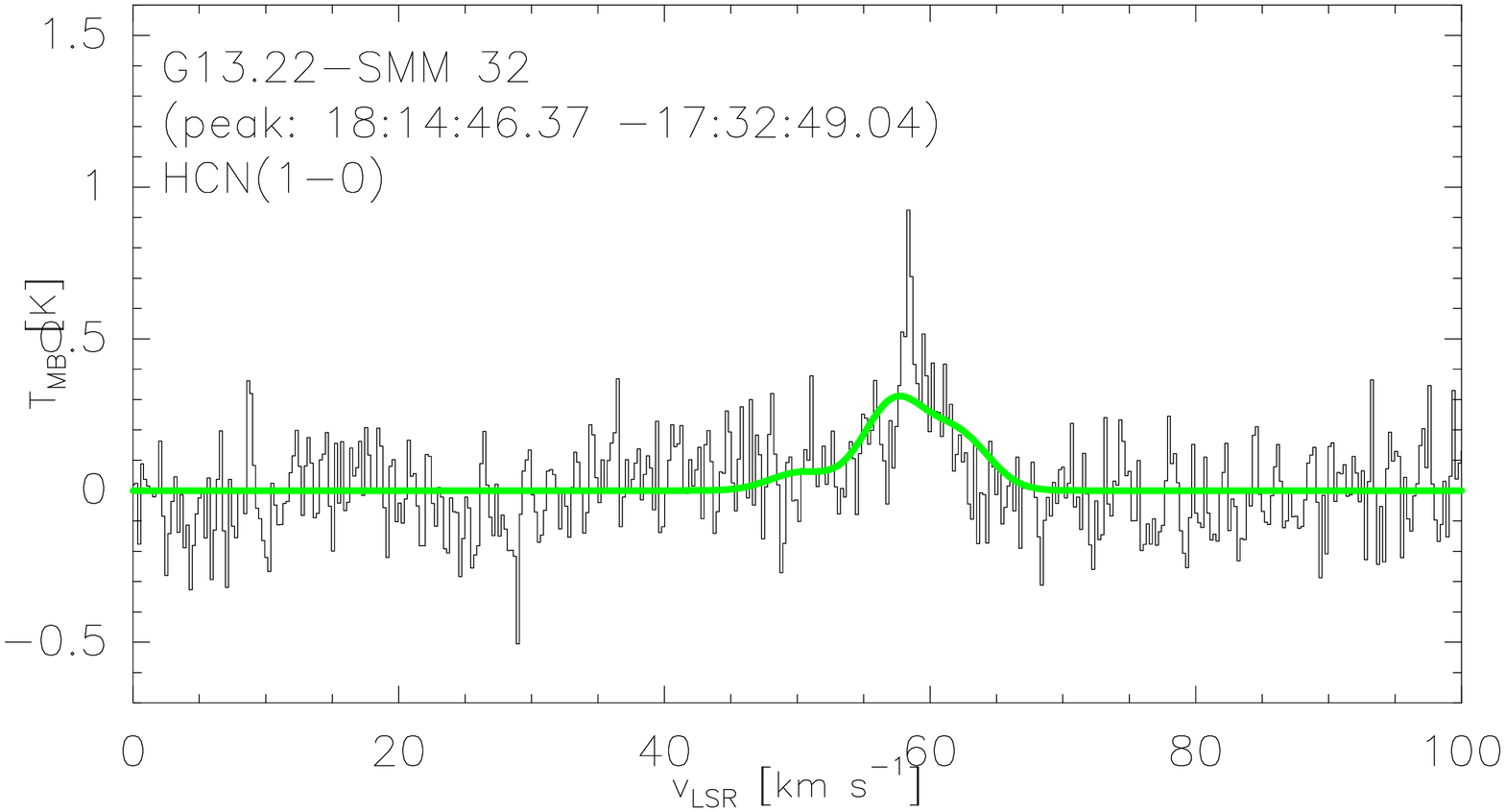}
\includegraphics[width=0.245\textwidth]{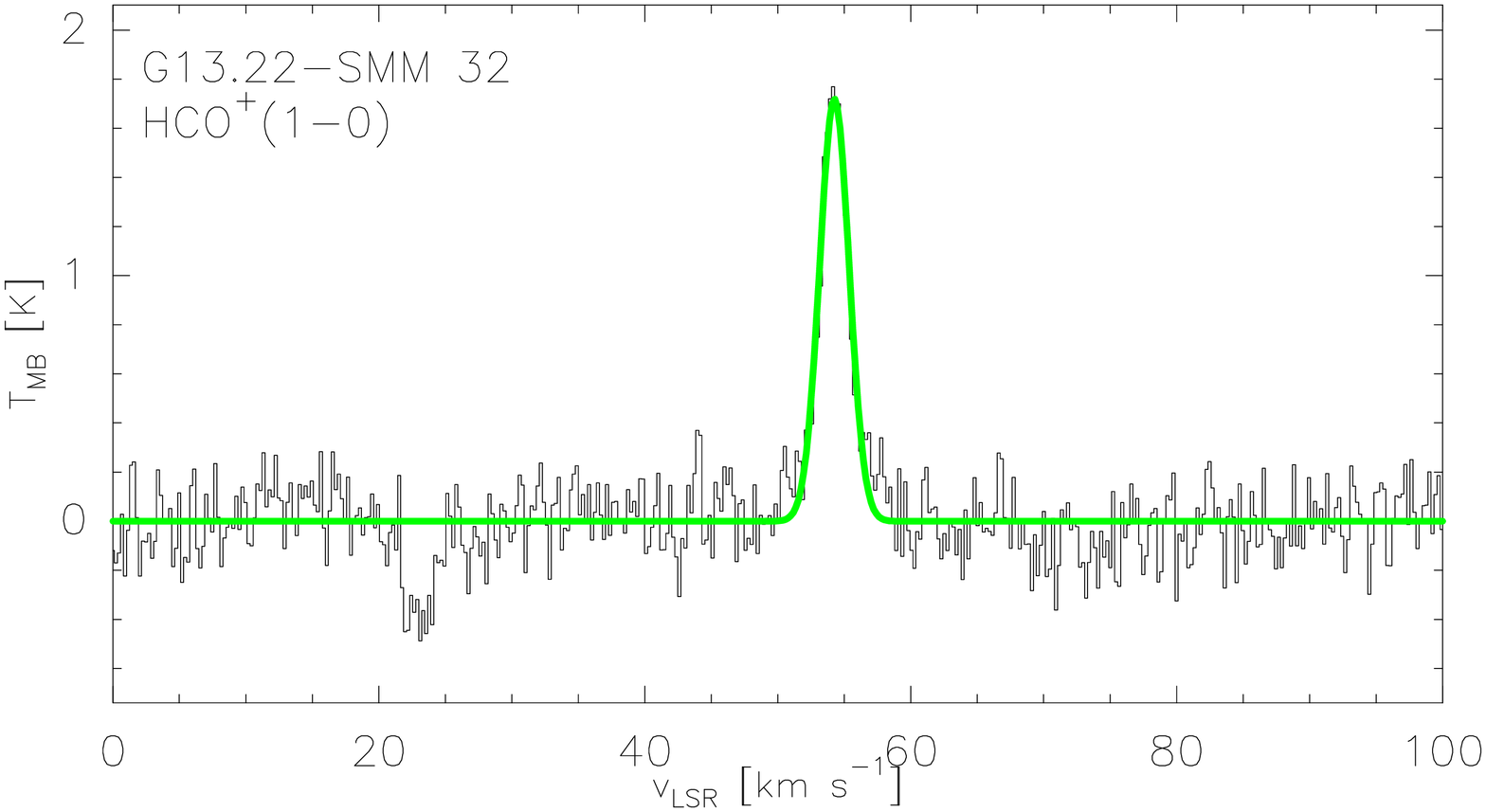}
\includegraphics[width=0.245\textwidth]{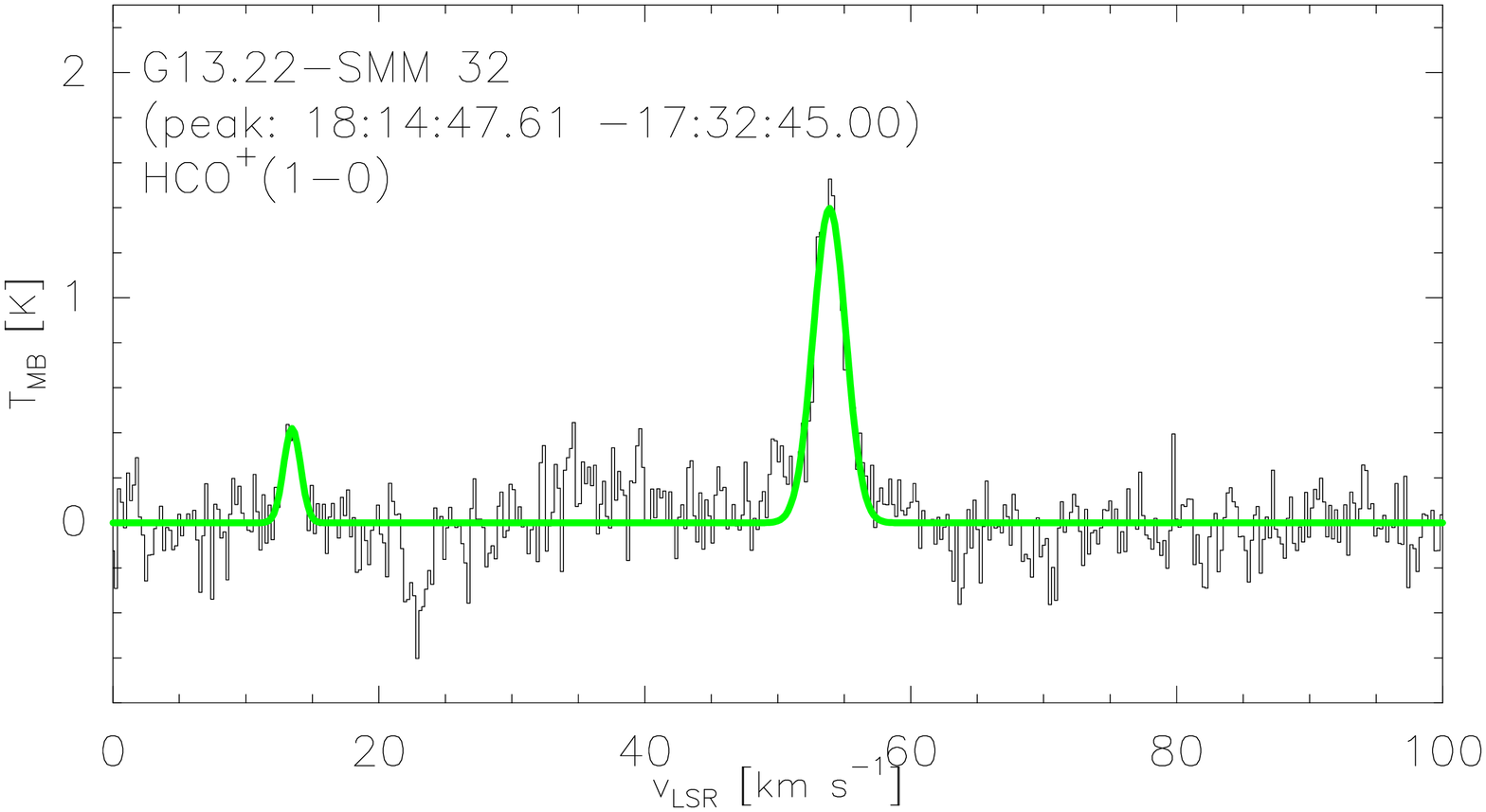}
\includegraphics[width=0.245\textwidth]{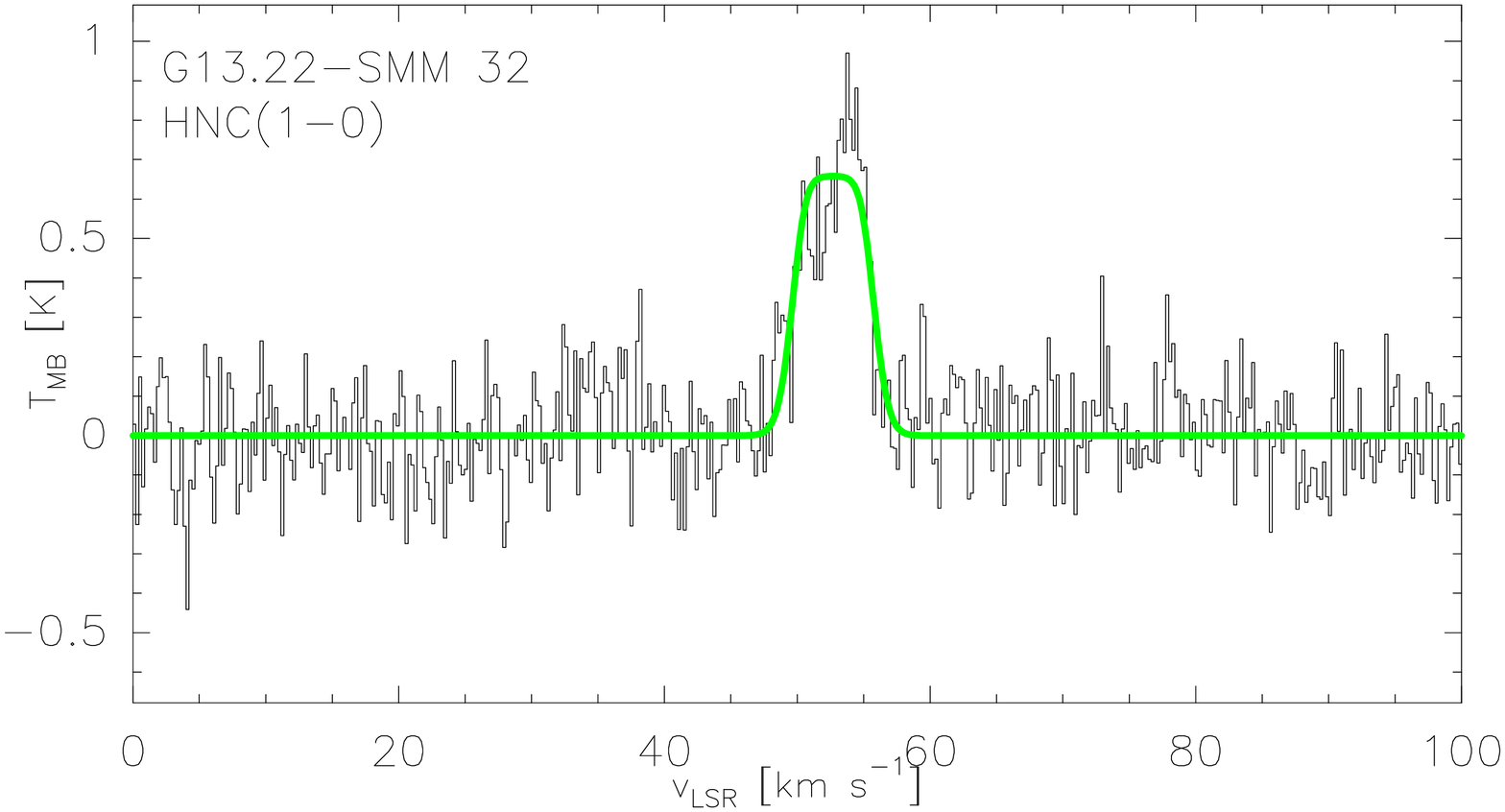}
\includegraphics[width=0.245\textwidth]{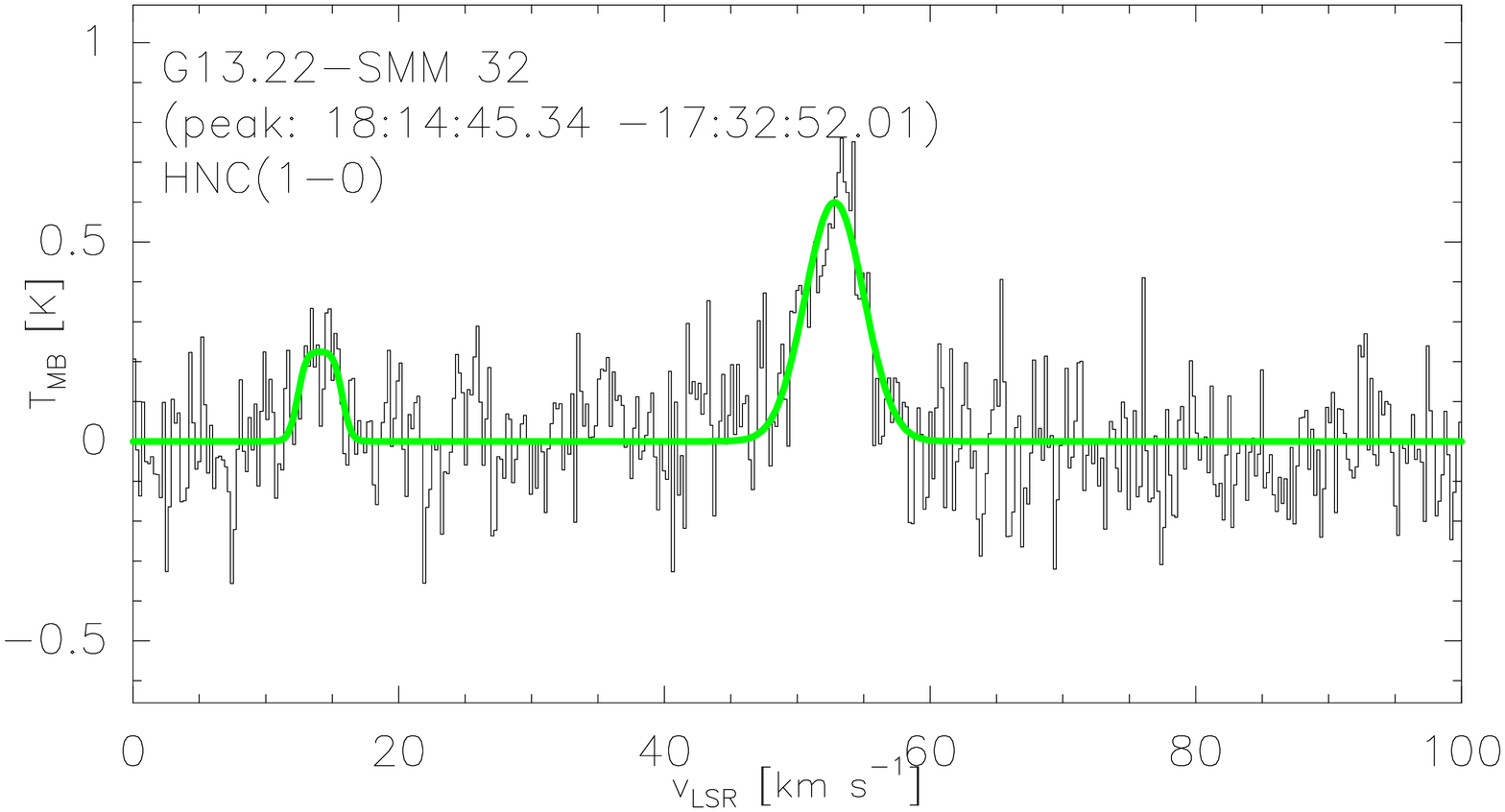}
\includegraphics[width=0.245\textwidth]{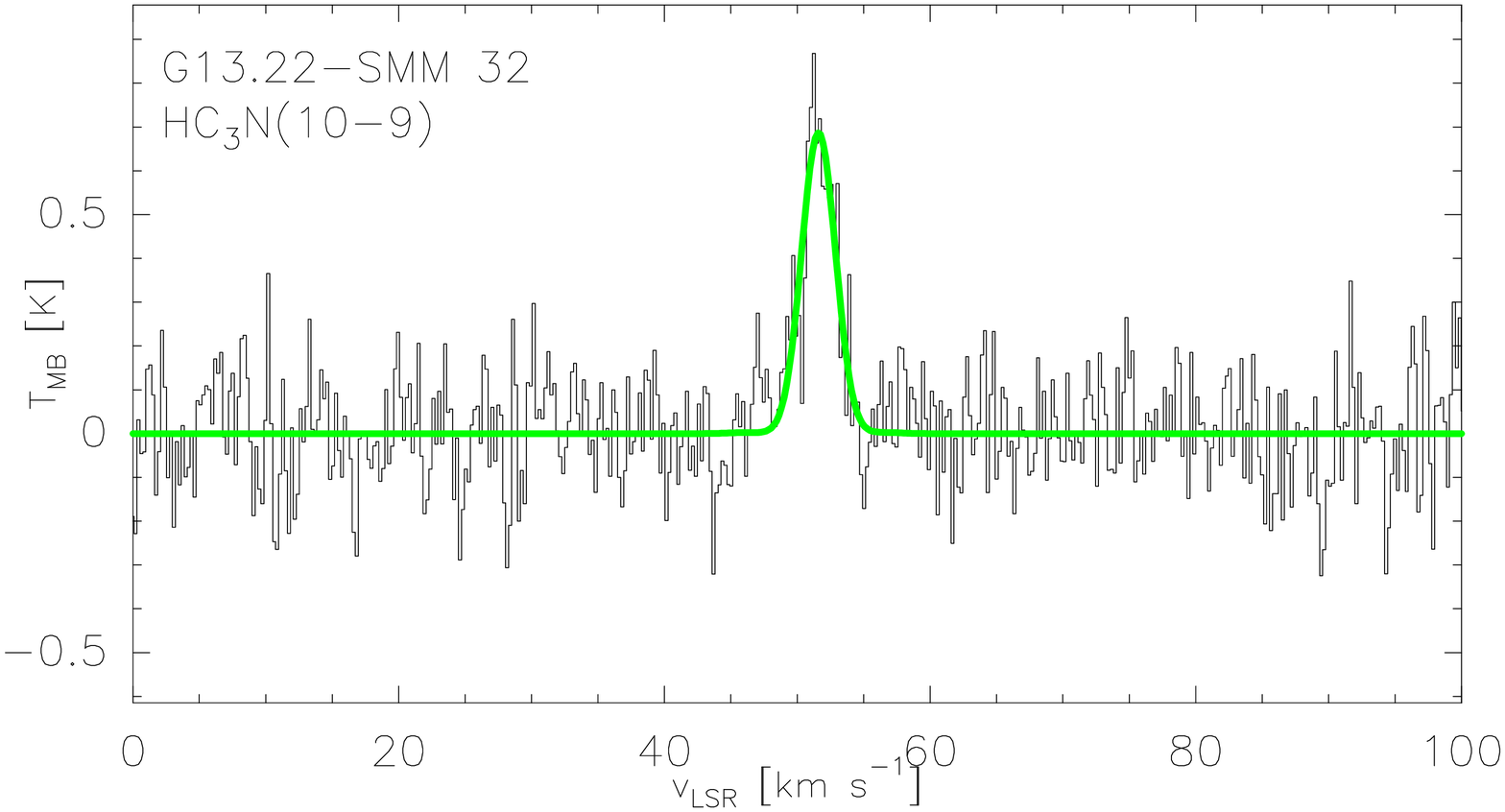}
\includegraphics[width=0.245\textwidth]{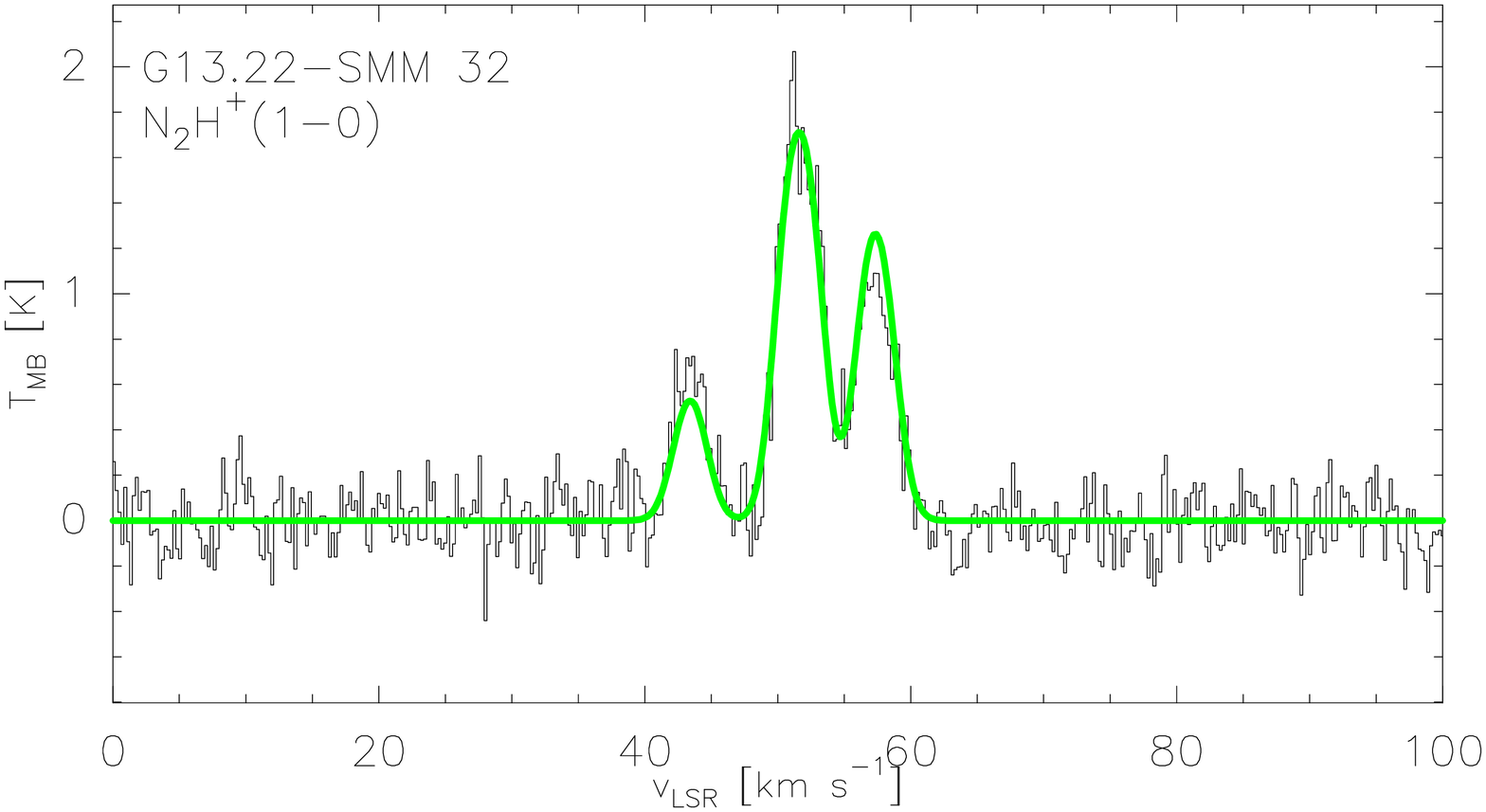}
\includegraphics[width=0.245\textwidth]{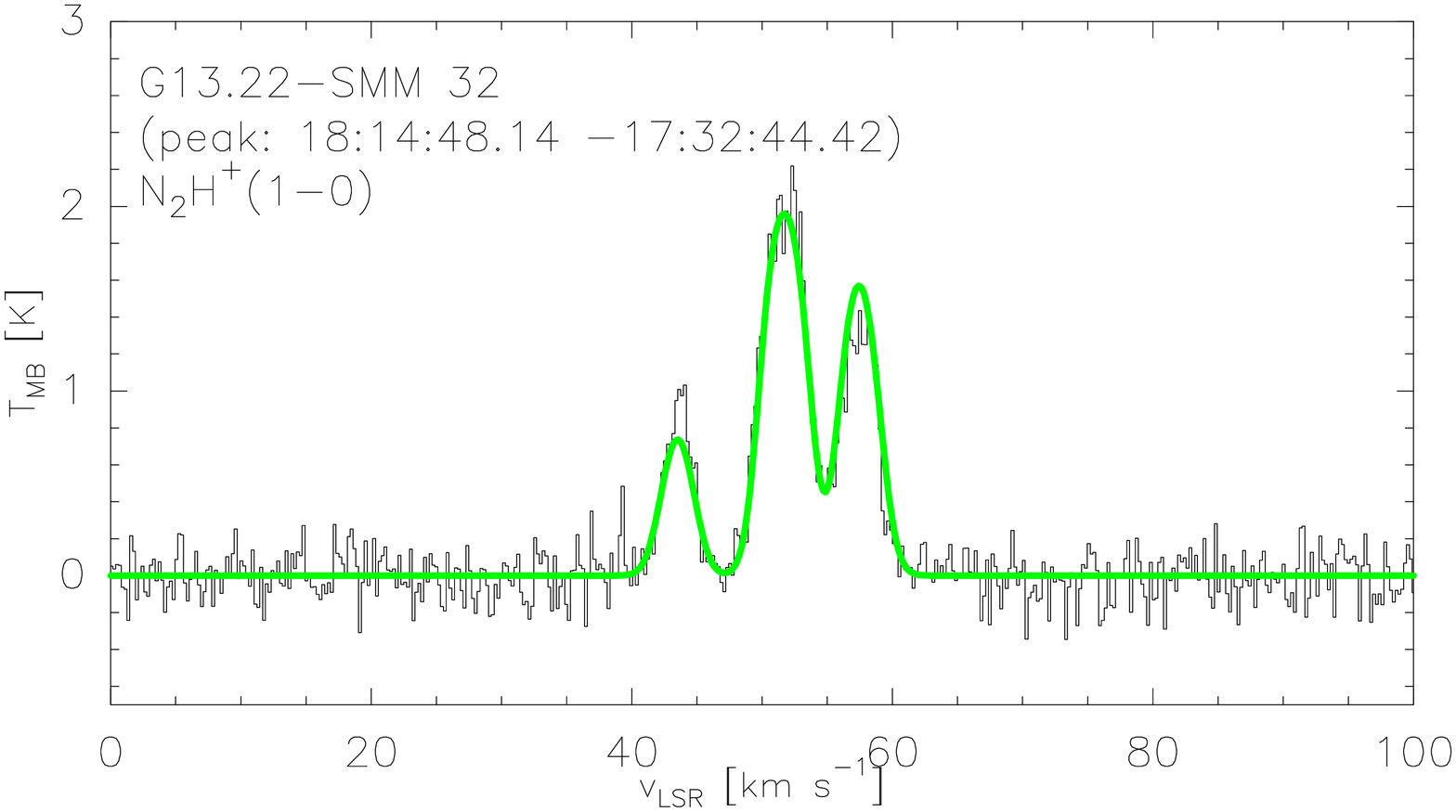}
\caption{Same as Fig.~\ref{figure:G187SMM1_spectra} but towards 
G13.22--SMM 32. The velocity range for the C$_2$H spectra is wider. Two 
velocity components are detected in the HCO$^+$ and HNC spectra towards the 
line emission peaks.}
\label{figure:G1322SMM32_spectra}
\end{center}
\end{figure*}

The Hanning-smoothed spectra of the detected spectral lines are presented in 
Figs.~\ref{figure:G187SMM1_spectra}--\ref{figure:G1322SMM32_spectra}. In each 
panel, the fit to the line is superimposed as a green line.

\end{appendix}

\clearpage

\longtab{3}{
\centering
{\scriptsize
\renewcommand{\footnoterule}{}
\begin{longtable}{c c c c c c c c c c}
\caption{\label{table:lineparameters} Spectral-line parameters, beam-averaged 
molecular column densities, and fractional abundances with respect to H$_2$.}\\
\hline\hline
Source & Transition & ${\rm v}_{\rm LSR}$ & $\Delta {\rm v}$ & $T_{\rm MB}$\tablefootmark{a}  & $\int T_{\rm MB} {\rm dv}$\tablefootmark{a} & $\tau$ & $T_{\rm ex}$\tablefootmark{a} & $N$\tablefootmark{b} & $x$\\
     & & [km~s$^{-1}$] & [km~s$^{-1}$] & [K] & [K~km~s$^{-1}$] & & [K] & [cm$^{-2}$] & [$10^{-10}$]\\
\hline 
\endfirsthead
\caption{continued.}\\
\hline\hline
Source & Transition & ${\rm v}_{\rm LSR}$ & $\Delta {\rm v}$ & $T_{\rm MB}$\tablefootmark{a}  & $\int T_{\rm MB} {\rm dv}$\tablefootmark{a} & $\tau$ & $T_{\rm ex}$\tablefootmark{a} & $N$\tablefootmark{b} & $x$\\
     & & [km~s$^{-1}$] & [km~s$^{-1}$] & [K] & [K~km~s$^{-1}$] & & [K] & [cm$^{-2}$] & [$10^{-10}$]\\
\hline
\endhead
\hline
\endfoot
G1.87-0.14 &\\
SMM 1 & SiO$(2-1)$ & $53.1\pm0.5$ & $19.56\pm1.02$ & $0.52\pm0.24$ & $10.84\pm0.50$ & $0.17\pm0.01$ & $=E_{\rm u}=6.25$ & $2.5\pm0.1(13)$ & $2.6\pm0.3$\\
      & HN$^{13}$C$(1-0)$ & $54.1\pm0.6$ & $9.29\pm2.67$ & $0.26\pm0.21$(G) & $3.83\pm0.40$(G) & $0.29\pm0.03$ & $5.2\pm0.1$(A) & $9.6\pm1.0(12)$ & $1.0\pm0.1$\\
      & C$_2$H$(1-0)$ & $52.0\pm0.3$ & $11.30\pm0.71$ & $0.57\pm0.23$(S) & 
$8.63\pm0.51$(S) & $2.38\pm1.00$ & $3.8\pm0.5$ & $1.2\pm0.5(15)$ & $125.4\pm53.7$ \\
      & HNCO$(4_{0,\,4}-3_{0,\,3})$ & $51.6\pm0.1$ & $11.70\pm0.14$ & $3.04\pm0.25$(G) & $38.68\pm0.42$(G) & $2.86\pm0.39$ & $=2E_{\rm u}/3=7.03$ & $7.9\pm1.1(14)$& $82.5\pm14.1$\\
      & HCN$(1-0)$\tablefootmark{c} & $41.9\pm0.4$(B)/ & $20.10\pm0.89$(B)/ & $0.97\pm0.37$(B)/ & $20.78\pm0.80$(B)/ & \ldots & \ldots & \ldots & \ldots\\
      &                             & $62.3\pm0.1$(R) & $11.55\pm0.25$(R) & $2.77\pm0.37$(R) & $34.09\pm0.72$(R) & \ldots & \ldots & \ldots & \ldots\\
      & HCO$^+(1-0)$\tablefootmark{d} & $56.8\pm0.2$ & $14.76\pm0.43$ & $1.58\pm0.29$ & $24.84\pm0.53$ & $1.67\pm0.74$ & $4.9\pm0.2$(A) & $6.3\pm2.8(13)$ & $6.6\pm3.0$\\
      & HCO$^+(1-0)$\tablefootmark{e, f} & $53.8\pm0.1$ & $16.13\pm0.28$ & $2.24\pm0.31$ & $38.38\pm0.53$ & \ldots\tablefootmark{g} & $4.9\pm0.2$(A) & $3.5\pm0.1(13)$ & $3.7\pm0.4$ \\
      &                                  & $-41.5\pm0.6$ & $3.37\pm1.05$ & $0.52\pm0.22$ & $1.87\pm0.60$ & $0.31\pm0.05$ & $4.9\pm0.2$(A) & $2.3\pm0.7(12)$ & $0.7\pm0.2$\\
      & HNC$(1-0)$\tablefootmark{c, d} & $52.0\pm0.1$ & $10.60\pm0.37$ & \ldots & \ldots & $17.20\pm3.60$ & $4.5\pm0.1$(A) & $5.3\pm1.1(14)$ & $55.4\pm12.7$\\
      &                                & $44.3\pm0.2$(B)/ & $9.47\pm 0.53$(B)/ & $1.21\pm0.24$(B)/ & $12.24\pm0.62$(B)/ & \ldots & \ldots & \ldots & \ldots\\
                  &                     & $57.2\pm0.1$(R) & $9.86\pm0.29$(R) & 
$2.19\pm0.24$(R) & $22.95\pm0.62$(R) & \ldots & \ldots & \ldots & \ldots\\
      & HNC$(1-0)$\tablefootmark{c, e} & $51.8\pm0.1$ & $15.40\pm0.15$ & \ldots & \ldots & $1.43\pm0.01$ & $4.5\pm0.1$(A) & $6.4\pm0.1(13)$ & $14.7\pm1.5$\\
      &                                & $48.6\pm0.4$(B)/ & $17.96\pm0.48$(B)/ & $2.01\pm0.20$(B)/ & $38.37\pm2.03$(B)/ & \ldots  & \ldots & \ldots & \ldots\\ 
      &                                 & $56.5\pm0.1$(R) & $7.93\pm0.46$(R) & $1.72\pm0.20$(R) & $14.53\pm1.77$(R) & \ldots  & \ldots & \ldots & \ldots\\
      & HC$_3$N$(10-9)$ & $51.6\pm0.1$ & $7.74\pm0.25$ & $1.66\pm0.24$(G) & 
$13.92\pm0.35$(G) & $0.23\pm0.01$ & $=E_{\rm u}=24.01$ & $5.8\pm0.1(13)$ & $6.1\pm0.6$\\
      & CH$_3$CN$(5_1-4_1)$ & $48.5\pm0.4$ & $5.85\pm0.69$ & $0.32\pm0.23$(G) & $3.07\pm0.31$(G) & $0.03\pm0.01$ & $=2E_{\rm u}/3=13.59$ & $4.2\pm0.4(11)$ & $0.04\pm0.01$\\
      & N$_2$H$^+(1-0)$\tablefootmark{c} & $53.5\pm0.3$ & $8.85\pm0.53$ & \ldots & \ldots &          $10.70\pm1.84$ & $7.2\pm0.2$(A) & $5.4\pm1.0(14)$ & $56.4\pm11.8$\\
 &                                 & $46.4\pm0.1$(B)/ & $13.98\pm0.11$(B)/ & $0.58\pm0.25$(B)/ & $8.60\pm0.05$(B)/ & \ldots & \ldots & \ldots & \ldots\\
      &                            & $60.5\pm0.1$(R) & $9.71\pm0.11$(R) & $1.14\pm0.25$(R) & $11.80\pm0.05$(R) & \ldots & \ldots & \ldots & \ldots\\
SMM 8 & SiO$(2-1)$ & $53.4\pm1.0$ & $18.54\pm2.48$ & $0.24\pm0.22$ & $4.71\pm0.53$ & $0.08\pm0.01$ & $=E_{\rm u}=6.25$ & $1.1\pm0.1(13)$ & $3.5\pm0.5$\\
      & HNCO$(4_{0,\,4}-3_{0,\,3})$ & $45.5\pm0.5$ & $17.00\pm1.52$ & $0.63\pm0.23$(G) & $14.70\pm0.61$(G) & $0.34\pm0.02$ & $=2E_{\rm u}/3=7.03$ & $6.1\pm0.3(14)$& $193.2\pm21.7$\\
      & HCN$(1-0)$ & $55.0\pm0.1$ & $13.50\pm0.38$ & $1.35\pm0.26$(G) & $25.19\pm0.57$(G) & $0.89\pm0.10$ & $6.5\pm0.8$ & $1.4\pm0.2(14)$ & $44.3\pm7.8$\\ 
      & HCO$^+(1-0)$ & $54.1\pm0.3$ & $18.17\pm0.71$ & $1.06\pm0.26$ & $20.45\pm0.63$ & $0.79\pm0.19$ & $4.9\pm0.2$(A) & $3.6\pm0.9(13)$ & $11.4\pm3.1$\\ 
      & HNC$(1-0)$ & $49.4\pm0.3$ & $17.90\pm0.82$ & $1.05\pm0.23$(G) & $20.18\pm0.79$(G) & $2.00\pm0.55$ & $4.5\pm0.1$(A) & $1.0\pm0.3(14)$ & $31.7\pm10.0$\\
      & N$_2$H$^+(1-0)$ & $46.6\pm0.6$ & $20.10\pm1.69$ & $0.47\pm0.23$(G) & $12.90\pm0.62$(G) & $1.26\pm0.39$ & $4.6\pm1.0$ & $7.2\pm3.9(13)$ & $22.8\pm12.6$\\
SMM 10 & HNCO$(4_{0,\,4}-3_{0,\,3})$ & $44.3\pm0.3$ & $17.00\pm0.85$ & $1.14\pm0.38$(G) & $20.88\pm0.85$ & $0.67\pm0.07$ & $=2E_{\rm u}/3=7.03$ & $2.7\pm0.3(14)$ & $188.5\pm29.3$\\ 
       & HCN$(1-0)$ & $51.7\pm0.3$ & $14.30\pm0.90$ & $1.38\pm0.41$ & $24.74\pm0.91$ & $1.87\pm0.94$ & $5.1\pm0.5$(A) & $2.2\pm1.1(14)$ & $153.6\pm78.6$\\
       & HCO$^+(1-0)$ & $51.8\pm0.6$ & $17.33\pm1.74$ & $0.61\pm0.40$ & $11.35\pm0.89$ & $0.38\pm0.10$ & $4.9\pm0.2$(A) & $1.4\pm0.1(13)$ & $9.8\pm1.3$\\
       & HNC$(1-0)$ & $46.1\pm0.5$ & $18.90\pm1.40$ & $1.09\pm0.39$(G) & $22.21\pm1.21$(G) & $2.15\pm0.98$ & $4.5\pm0.1$(A) & $1.2\pm0.5(14)$ & $83.8\pm36.1$\\
       & N$_2$H$^+(1-0)$ & $45.7\pm0.6$ & $8.33\pm1.45$ & $0.64\pm0.38$(G) & $14.68\pm0.91$(G) & $8.08\pm3.16$ & $3.6\pm0.5$ & $1.4\pm2.6(14)$ & $97.7\pm181.8$ \\
SMM 12 & SiO$(2-1)$ & $34.0\pm1.5$ & $49.27\pm3.32$ & $0.23\pm0.20$ & $11.95\pm0.72$ & $0.07\pm0.01$ & $=E_{\rm u}=6.25$ & $2.8\pm0.2(13)$ & $7.8\pm1.0$\\
       & C$_2$H$(1-0)$ & $44.6\pm0.2$ & $5.67\pm0.61$ & $0.40\pm0.21$(S) & $2.86\pm0.27$(S) & $4.07\pm2.14$ & $3.3\pm0.3$ & $8.8\pm4.7(14)$ & $244.0\pm132.6$\\
       & HNCO$(4_{0,\,4}-3_{0,\,3})$ & $44.2\pm0.1$ & $8.41\pm0.15$ & $2.64\pm0.27$(G) & $24.13\pm0.34$(G) & $2.17\pm0.26$ & $=2E_{\rm u}/3=7.03$ & $4.3\pm0.5(14)$& $119.2\pm18.3$\\
       & HCN$(1-0)$\tablefootmark{c} & $47.4\pm0.4$ & $30.40\pm0.87$ & \ldots & \ldots & \ldots & \ldots & \ldots & \ldots \\
       &                             & $41.5\pm0.9$(B)/ & $37.51\pm1.45$(B)/ & $0.56\pm0.20$(B)/ & $22.48\pm1.03$(B)/ & \ldots & \ldots & \ldots & \ldots\\
       &                             & $54.6\pm0.2$(R) & $9.51\pm0.65$(R) & $0.79\pm0.20$(R) & $8.00\pm0.70$(R) & \ldots & \ldots & \ldots & \ldots\\ 
       & HCO$^+(1-0)$ & $44.9\pm0.7$ & $41.65\pm1.58$ & $0.46\pm0.20$ & $20.44\pm0.65$ & $0.27\pm0.04$ & $4.9\pm0.2$(A) & $2.5\pm0.1(13)$ & $6.9\pm0.7$\\
       & HNC$(1-0)$\tablefootmark{c} & $46.9\pm0.4$ & $22.70\pm1.11$ & \ldots & \ldots & $1.07\pm0.22$ & $4.5\pm0.1$(A) & $7.1\pm1.5(13)$ & $19.7\pm4.6$\\
       & &$41.1\pm0.9$(B)/ & $31.45\pm1.63$(B)/ & $0.41\pm0.13$(B)/ & $13.77\pm0.77$(B)/ & \ldots & \ldots & \ldots & \ldots \\
       & &$50.8\pm0.2$(R) & $6.37\pm0.54$(R) & $0.72\pm0.13$(R) & $4.86\pm0.47$(R) & \ldots & \ldots & \ldots & \ldots \\
       & HC$_3$N$(10-9)$ & $44.5\pm0.1$ & $6.05\pm0.28$ & $1.02\pm0.21$(G) & $6.70\pm0.28$ & $0.14\pm0.01$ & $=E_{\rm u}=24.01$ & $2.8\pm0.1(13)$ & $7.8\pm0.8$\\
       & N$_2$H$^+(1-0)$ & $46.3\pm0.5$ & $8.84\pm1.21$ & $0.47\pm0.20$(G) & $10.03\pm0.50$(G) & $6.54\pm2.19$ & $3.4\pm0.3$ & $1.1\pm2.0(14)$ & $30.5\pm55.5$\\
SMM 14 & SiO$(2-1)$ & $38.7\pm2.0$ & $50.97\pm6.00$ & $0.22\pm0.22$ & $12.07\pm0.95$ & $0.07\pm0.01$ & $=E_{\rm u}=6.25$ & $2.8\pm0.2(13)$ & $5.7\pm0.7$\\
       & C$_2$H$(1-0)$ & $45.7\pm0.5$ & $14.30\pm0.70$ & $0.39\pm0.21$(S) & $6.18\pm0.55$(S) & $0.94\pm0.20$ & $4.1\pm0.1$(A) & $6.5\pm1.3(14)$ & $132.8\pm29.7$\\
       & HNCO$(4_{0,\,4}-3_{0,\,3})$ & $46.1\pm0.1$ & $10.60\pm0.20$ & $2.30\pm0.26$(G) & $26.47\pm0.41$(G) & $1.72\pm0.17$ & $=2E_{\rm u}/3=7.03$ & $4.3\pm0.4(14)$ & $87.9\pm12.0$\\
       & HCN$(1-0)$\tablefootmark{c} & $46.1\pm0.1$ & $41.30\pm0.38$ & \ldots & \ldots & \ldots & \ldots & \ldots & \ldots\\
       &            & $39.3\pm0.2$(B)/ & $45.85\pm1.6$(B)/ & $0.59\pm0.22$(B)/ & $28.99\pm0.98$(B)/ & \ldots & \ldots & \ldots & \ldots\\
       &            & $56.7\pm0.2$(R) & $7.41\pm0.55$(R) & $0.89\pm0.22$(R) & $6.98\pm0.43$(R) & \ldots & \ldots & \ldots & \ldots\\
       & HCO$^+(1-0)$ & $49.1\pm0.8$ & $32.09\pm1.72$ & $0.35\pm0.21$ & $11.94\pm0.58$ & $0.20\pm0.03$ & $4.9\pm0.2$(A) & $1.5\pm0.1(13)$ & $3.1\pm0.4$\\
       & HNC$(1-0)$\tablefootmark{c} & $51.4\pm0.2$ & $10.80\pm0.61$ & \ldots & \ldots & $2.78\pm0.91$ & $4.5\pm0.1$(A) & $8.8\pm2.9(13)$ & $18.0\pm6.2$\\ 
     & & $44.8\pm0.7$(B)/ & $26.50\pm1.32$(B)/ & $0.47\pm0.14$(B)/ & $13.33\pm0.63$(B)/ & \ldots & \ldots & \ldots & \ldots \\
     & & $52.5\pm0.1$(R) & $4.61\pm0.25$(R) & $1.27\pm0.14$(R) & $6.21\pm0.36$(R) & \ldots & \ldots & \ldots & \ldots \\
& HC$_3$N$(10-9)$ & $47.2\pm0.3$ & $10.40\pm0.62$ & $0.56\pm0.19$(G) & $6.28\pm0.35$(G) & $0.07\pm0.01$ & $=E_{\rm u}=24.01$ & $2.6\pm0.1(13)$ & $5.3\pm0.6$\\
       & N$_2$H$^+(1-0)$ & $49.8\pm0.5$ & $10.70\pm2.40$ & $0.56\pm0.22$(G) & $11.41\pm0.51$ & $1.85\pm0.70$ & $4.4\pm0.8$ & $5.3\pm3.6(13)$ & $10.8\pm5.4$\\
SMM 15 & HNCO$(4_{0,\,4}-3_{0,\,3})$ & $43.2\pm0.5$ & $30.40\pm1.39$ & $0.63\pm0.24$(G) & $20.76\pm0.80$(G) & $0.34\pm0.02$ & $=2E_{\rm u}/3=7.03$ & $1.1\pm0.1(14)$ & $48.6\pm6.6$\\
       & HCN$(1-0)$\tablefootmark{c} & $51.5\pm0.5$ & $23.40\pm1.24$ & \ldots &\ldots & \ldots & \ldots & \ldots & \ldots \\
       &                              & $38.5\pm0.1$(B)/ & $37.71\pm0.11$(B)/ & $0.23\pm0.27$(B)/ & $9.41\pm0.06$(B)/ & \ldots & \ldots & \ldots & \ldots\\
       &                               & $54.5\pm0.1$(R) & $18.49\pm0.11$(R) & $0.79\pm0.27$(R) & $15.64\pm0.06$(R) & \ldots & \ldots & \ldots & \ldots\\
      & HCO$^+(1-0)$\tablefootmark{c} & $27.3\pm1.2$(B)/ & $14.04\pm2.38$(B)/ & $0.24\pm0.18$(B)/ & $3.66\pm0.56$(B)/ & \ldots & \ldots & \ldots & \ldots\\
      &                               & $53.7\pm0.4$(R) & $19.26\pm1.03$(R) & $0.75\pm0.18$(R) & $15.43\pm0.66$(R) & \ldots & \ldots & \ldots & \ldots\\
      & HNC$(1-0)$ & $43.4\pm0.8$ & $21.20\pm2.81$ & $0.59\pm0.23$(G) & $18.32\pm1.00$(G) & $3.46\pm2.24$ & $3.5\pm0.3$ & $1.5\pm1.0(14)$ & $66.2\pm44.7$\\
      & N$_2$H$^+(1-0)$ & $39.4\pm0.7$ & $17.00\pm0.70$ & $0.31\pm0.23$(G) & $12.08\pm0.63$(G) & $0.30\pm0.02$ & $7.2\pm0.2$(A) & $1.6\pm0.1(13)$ & $7.1\pm0.8$\\
SMM 16 & C$_2$H$(1-0)$ & $41.5\pm0.9$ & $18.90\pm2.41$ & $0.24\pm0.21$(S) & 
$4.31\pm0.48$(S) & $0.53\pm0.11$ & $4.1\pm0.1$(A) & $1.7\pm0.2(14)$ & $50.8\pm7.9$\\
       & HNCO$(4_{0,\,4}-3_{0,\,3})$ & $41.7\pm0.1$ & $10.80\pm0.30$ & $1.33\pm0.21$(G) & $15.51\pm0.36$(G) & $0.81\pm0.05$ & $=2E_{\rm u}/3=7.03$ & $2.1\pm0.1(14)$ & $62.7\pm7.0$\\
       & HCN$(1-0)$\tablefootmark{c} & $36.6\pm1.1$ & $38.30\pm0.54$ & \ldots & \ldots & $3.66\pm0.33$ & $5.1\pm0.5$(A) & $1.2\pm0.1(15)$ & $358.4\pm46.9$ \\
       &                             & $32.5\pm1.5$(B)/ & $59.84\pm2.58$(B)/ & $0.36\pm0.21$(B)/ & $23.09\pm1.07$(B)/ & \ldots & \ldots & \ldots & \ldots\\
       &                             & $52.8\pm0.3$(R) & $10.48\pm0.98$(R) & $0.54\pm0.21$(R) & $6.03\pm0.59$(R) & \ldots & \ldots & \ldots & \ldots\\
       & HCO$^+(1-0)$\tablefootmark{c} & $13.6\pm3.5$(B)/ & $32.86\pm7.73$(B)/ & $0.10\pm0.14$(B)/ & $3.40\pm0.71$(B)/ & \ldots & \ldots & \ldots & \ldots\\
       &                               & $48.9\pm0.7$(R) & $19.18\pm1.61$(R) & $0.36\pm0.14$(R) & $7.39\pm0.61$(R) & \ldots & \ldots & \ldots & \ldots\\
       & HNC$(1-0)$\tablefootmark{c} & $47.0\pm0.3$ & $11.30\pm1.02$ & \ldots & \ldots & \ldots & \ldots & \ldots & \ldots\\
       &                             & $37.6\pm2.8$(B)/ & $13.79\pm3.82$(B)/ & $0.24\pm0.21$(B)/ & $3.52\pm1.20$(B)/ & \ldots & \ldots & \ldots & \ldots\\
       &                             & $48.3\pm0.5$(R) & $7.42\pm1.05$(R) & $0.68\pm0.21$(R) & $5.34\pm1.24$(R) & \ldots & \ldots & \ldots & \ldots\\
SMM 20 & HNCO$(4_{0,\,4}-3_{0,\,3})$ & $37.4\pm0.2$ & $2.90\pm0.20$ & $0.38\pm0.19$(G) & $3.03\pm0.31$(G) & $0.20\pm0.01$ & $=2E_{\rm u}/3=7.03$ & $1.7\pm0.2(13)$ & $6.1\pm1.0$ \\
       & HNC$(1-0)$ & $39.2\pm0.2$ & $7.16\pm0.54$ & $0.71\pm0.19$(G) & $8.60\pm0.33$(G) & $7.30\pm2.71$ & $3.6\pm0.2$ & $1.1\pm0.4(14)$ & $39.4\pm14.9$\\
       & HC$_3$N$(10-9)$ & $37.4\pm0.6$ & $8.25\pm1.36$ & $0.30\pm0.25$(G) & $3.13\pm0.38$(G) & $1.46\pm0.68$ & $3.6\pm0.7$ & $4.9\pm2.4(14)$ & $175.5\pm87.8$\\
       & N$_2$H$^+(1-0)$ & $38.7\pm0.4$ & $10.00\pm2.47$ & $0.45\pm0.21$(G) & $7.62\pm0.40$(G) & $0.45\pm0.03$ & $7.2\pm0.2$(A) & $9.9\pm0.5(12)$ & $3.5\pm0.4$\\
SMM 21 & HNCO$(4_{0,\,4}-3_{0,\,3})$ & $40.4\pm0.2$ & $11.80\pm0.41$ & $1.34\pm0.26$(G) & $17.12\pm0.48$ & $0.82\pm0.06$ & $=2E_{\rm u}/3=7.03$ & $2.3\pm0.2(14)$ &$116.5\pm15.7$\\
       & HCN$(1-0)$\tablefootmark{c} & $46.9\pm0.6$ & $28.90\pm1.46$ & \ldots & \ldots & \ldots & \ldots & \ldots & \ldots\\
       &                             & $30.1\pm2.2$(B)/ & $19.97\pm4.69$(B)/ & $0.32\pm0.27$(B)/ & $6.79\pm1.71$(B)/ & \ldots & \ldots & \ldots & \ldots\\
       &                             & $51.9\pm0.9$(R) & $18.30\pm1.84$(R) & $0.75\pm0.27$(R) & $14.60\pm1.71$(R) & \ldots & \ldots & \ldots & \ldots\\
       & HCO$^+(1-0)$\tablefootmark{c} & $40.3\pm2.3$(B)/ & $6.95\pm3.76$(B)/ & $0.06\pm0.27$(B)/ & $0.45\pm0.50$(B)/ & \ldots & \ldots & \ldots & \ldots\\
       &                               & $55.0\pm0.8$(R) & $30.89\pm2.76$(R) & $0.45\pm0.27$(R) & $14.72\pm1.16$(R) & \ldots & \ldots & \ldots & \ldots\\
       & HNC$(1-0)$ & $42.6\pm0.5$ & $17.30\pm0.86$ & $0.54\pm0.25$(G) & $11.04\pm0.58$(G) & $0.76\pm0.15$ & $4.5\pm0.1$(A) & $3.8\pm0.8(13)$ & $19.2\pm4.5$\\
       & N$_2$H$^+(1-0)$ & $41.8\pm0.7$ & $16.10\pm2.24$ & $0.39\pm0.26$(G) & $9.70\pm0.63$(G) & $2.24\pm0.71$ & $3.8\pm0.7$ & $7.8\pm1.3(13)$ & $39.5\pm7.7$\\
SMM 23 & SiO$(2-1)$ & $39.2\pm0.5$ & $9.65\pm1.08$ & $0.34\pm0.21$ & $3.50\pm0.33$ & $0.11\pm0.01$ & $=E_{\rm u}=6.25$ & $8.2\pm0.8(12)$ & $1.6\pm0.2$\\
       & HNCO$(4_{0,\,4}-3_{0,\,3})$\tablefootmark{d} & $38.4\pm0.5$ & $9.77\pm2.28$ & $0.46\pm0.21$(G) & $5.42\pm0.38$(G) & $0.24\pm0.013$ & $=2E_{\rm u}/3=7.03$ & $3.0\pm0.2(13)$ & $6.0\pm0.7$\\
       & HNCO$(4_{0,\,4}-3_{0,\,3})$\tablefootmark{e} & $36.6\pm0.1$ & $10.60\pm0.29$ & $1.26\pm0.22$(G) & $14.53\pm0.37$(G) & $0.76\pm0.05$ & $=2E_{\rm u}/3=7.03$& $1.9\pm0.1(14)$ & $43.4\pm5.0$\\
       & HCN$(1-0)$\tablefootmark{c} & $32.1\pm0.7$ & $23.50\pm2.25$ & \ldots & \ldots & $1.02\pm0.48$ & $5.1\pm0.5$(A) & $2.0\pm0.9(14)$ & $40.0\pm18.5$\\
       &            & $22.3\pm1.4$(B)/ & $34.85\pm3.87$(B)/ & $0.40\pm0.22$(B)/ & $14.93\pm1.43$(B)/ & \ldots & \ldots & \ldots & \ldots\\
       &            & $47.8\pm0.6$(R) & $15.18\pm1.40$(R) & $0.44\pm0.22$(R) & $7.16\pm1.13$(R) & \ldots & \ldots & \ldots & \ldots\\
       & HCO$^+(1-0)$ & $34.7\pm0.1$ & $36.23\pm0.92$ & $0.24\pm0.20$ & $9.18\pm0.56$ & $0.13\pm0.02$ & $4.9\pm0.2$(A) & $1.1\pm0.1(13)$ & $2.2\pm0.3$\\
       & HNC$(1-0)$\tablefootmark{d} & $38.0\pm0.2$ & $9.97\pm0.62$ & $1.17\pm0.20$(G) & $14.41\pm0.36$(G) & $1.32\pm0.62$ & $5.2\pm0.9$ & $4.8\pm2.3(13)$ & $9.6\pm4.7$\\
       & HNC$(1-0)$ (N)\tablefootmark{e} & $37.3\pm0.1$ & $10.90\pm0.24$ & $1.64\pm0.21$(G) & $19.22\pm0.35$(G) & \ldots\tablefootmark{g} & $4.5\pm0.1$(A) &  $3.4\pm0.1(13)$ & $6.8\pm0.7$\\ 
       & HNC$(1-0)$ (S)\tablefootmark{e} & $37.7\pm0.1$ & $9.53\pm0.21$ & $1.76\pm0.23$(G) & $18.08\pm0.34$(G) & \ldots\tablefootmark{g} & $4.5\pm0.1$(A) & $3.2\pm0.1(13)$ & $8.0\pm0.9$\\
       & HC$_3$N$(10-9)$\tablefootmark{d} & $38.3\pm0.5$ & $6.53\pm1.19$ & $0.35\pm0.25$(G) & $3.31\pm0.35$(G) & $0.05\pm0.01$ & $=E_{\rm u}=24.01$ & $1.4\pm0.1(13)$ & $2.8\pm0.3$\\
       & HC$_3$N$(10-9)$\tablefootmark{e} & $36.7\pm0.1$ & $7.60\pm0.37$ & $0.97\pm0.26$(G) & $9.22\pm0.392$(G) & $1.12\pm0.10$ & $5.9\pm0.8$ & $7.6\pm0.8(14)$ & $168.5\pm24.4$ \\
       & CH$_3$CN$(5_1-4_1)$\tablefootmark{e} & $34.8\pm0.7$ & $12.20\pm1.10$ & $0.25\pm0.21$(G) & $3.94\pm0.42$(G) & $0.02\pm0.01$ & $=2E_{\rm u}/3=13.59$ & $5.4\pm0.6(11)$ & $0.1\pm0.02$\\
       & N$_2$H$^+(1-0)$\tablefootmark{d} & $37.9\pm0.3$ & $11.60\pm0.93$ & $0.57\pm0.21$(G) & $8.93\pm0.41$(G) & $0.57\pm0.04$ & $7.2\pm0.2$(A) & $3.7\pm0.4(13)$ & $7.4\pm1.1$\\
       & N$_2$H$^+(1-0)$\tablefootmark{e} & $36.0\pm0.2$ & $8.00\pm1.22$ & $1.03\pm0.21$(G) & $15.92\pm0.40$(G) & $1.89\pm1.20$ & $5.7\pm1.5$ & $5.9\pm3.9(13)$ & $14.7\pm9.8$\\  
SMM 24 & HNCO$(4_{0,\,4}-3_{0,\,3})$ & $43.2\pm0.2$ & $11.60\pm0.43$ & $1.03\pm0.25$(G) & $13.01\pm0.46$(G) & $0.60\pm0.04$ & $=2E_{\rm u}/3=7.03$ & $1.6\pm0.1(14)$ & $50.8\pm6.1$\\
       & HCN$(1-0)$ & $43.1\pm2.2$ & $26.60\pm9.90$ & \ldots & \ldots & \ldots &\ldots & \ldots & \ldots\\
       &            & $28.8\pm0.1$(B)/ & $37.53\pm0.11$(B)/ & $0.29\pm0.26$(B)/ & $11.58\pm0.05$(B)/ & \ldots & \ldots & \ldots & \ldots\\
       &            & $52.5\pm0.1$(R) & $16.92\pm0.11$(R) & $0.63\pm0.26$(R) & $11.26\pm0.05$(R) & \ldots & \ldots & \ldots & \ldots\\
       & HCO$^+(1-0)$\tablefootmark{c} & $37.9\pm0.2$(B)/ & $40.07\pm0.23$(B)/ & $0.06\pm0.17$(B)/ & $2.60\pm0.06$(B)/ & \ldots & \ldots & \ldots & \ldots\\
       &                               & $55.3\pm0.2$(R) & $30.66\pm0.23$(R) & $0.34\pm0.17$(R) & $11.20\pm0.06$(R) & \ldots & \ldots & \ldots & \ldots\\
       & HNC$(1-0)$ & $43.5\pm0.6$ & $12.50\pm1.53$ & $0.38\pm0.25$(G) & $6.48\pm0.53$(G) & $3.92\pm1.45$ & $3.2\pm0.3$ & $9.0\pm3.5(13)$ & $28.6\pm11.5$\\
       & N$_2$H$^+(1-0)$ & $44.5\pm0.2$ & $11.80\pm4.08$ & $0.30\pm0.25$(G) & $7.56\pm0.62$(G) & $0.29\pm0.02$ & $7.2\pm0.2$(A) & $2.3\pm1.5(13)$ & $7.3\pm4.8$\\
SMM 27 & SiO$(2-1)$ & $46.6\pm0.6$ & $17.29\pm1.12$ & $0.34\pm0.20$ & $6.28\pm0.41$ & $0.11\pm0.01$ & $=E_{\rm u}=6.25$ & $1.5\pm0.1(13)$ & $4.9\pm0.6$\\
       & C$_2$H$(1-0)$ & $45.0\pm0.6$ & $14.90\pm1.55$ & $0.24\pm0.20$(S) & $4.75\pm0.43$(S) & $0.53\pm0.10$ & $4.1\pm0.1$(A) & $1.7\pm0.4(14)$ & $56.1\pm14.5$\\
       & HNCO$(4_{0,\,4}-3_{0,\,3})$ & $44.5\pm0.2$ & $15.20\pm1.15$ & $0.79\pm0.20$(G) & $17.26\pm0.42$(G) & $0.44\pm0.02$ & $=2E_{\rm u}/3=7.03$ & $9.6\pm0.2(13)$ & $31.7\pm3.4$\\
       & HCN$(1-0)$\tablefootmark{c} & $45.8\pm0.5$ & $28.10\pm2.33$ & \ldots & \ldots & $7.24\pm2.98$ & $5.1\pm0.5$(A) & $1.7\pm0.7(15)$ & $560.7\pm238.3$\\
       &            & $24.6\pm0.5$(B)/ & $15.46\pm1.24$(B)/ & $0.42\pm0.22$(B)/ & $6.84\pm0.58$(B)/ & \ldots & \ldots & \ldots & \ldots\\
       &            & $54.5\pm0.4$(R) & $30.91\pm1.09$(R) & $0.76\pm0.22$(R) & $25.12\pm0.74$(R) & \ldots & \ldots & \ldots & \ldots\\
       & HCO$^+(1-0)$\tablefootmark{c} & $24.0\pm1.8$(B)/ & $16.35\pm2.92$(B)/ & $0.13\pm0.14$(B)/ & $2.23\pm0.45$(B)/ & \ldots & \ldots & \ldots & \ldots\\
       &                               & $60.0\pm0.6$(R) & $32.27\pm1.95$(R) & $0.42\pm0.14$(R) & $14.53\pm0.68$(R) & \ldots & \ldots & \ldots & \ldots\\
       & HNC$(1-0)$ & $44.2\pm0.2$ & $20.20\pm0.45$ & $0.92\pm0.20$(G) & $19.95\pm0.42$(G) & $1.60\pm0.34$ & $4.5\pm0.1$(A) & $9.5\pm2.0(13)$ & $31.3\pm7.4$\\
       & HC$_3$N$(10-9)$ & $46.3\pm0.3$ & $15.50\pm0.78$ & $0.57\pm0.19$(G) & $9.50\pm0.40$(G) & $0.08\pm0.01$ & $=E_{\rm u}=24.01$ & $4.0\pm0.2(13)$ & $13.2\pm1.5$ \\
       & CH$_3$CN$(5_1-4_1)$ & $44.5\pm0.3$ & $11.50\pm4.03$ & $0.20\pm0.19$(G) & $3.08\pm0.37$(G) & $0.02\pm0.01$ & $=2E_{\rm u}/3=13.59$ & $7.0\pm0.8(11)$ & $0.2\pm0.04$\\
       & N$_2$H$^+(1-0)$ & $45.8\pm0.4$ & $18.30\pm3.06$ & $0.52\pm0.20$(G) & $12.03\pm0.46$(G) & $0.52\pm0.04$ & $7.2\pm0.2$(A) & $5.4\pm1.0(13)$ & $17.8\pm3.8$\\
SMM 28 & SiO$(2-1)$ & $37.6\pm0.4$ & $26.50\pm0.83$ & $0.73\pm0.21$ & $20.68\pm  0.58$ & $0.25\pm0.02$ & $=E_{\rm u}=6.25$ & $4.8\pm0.1(13)$ & $17.9\pm1.9$  \\
       & C$_2$H$(1-0)$ & $38.2\pm0.6$ & $19.20\pm1.00$ & $0.28\pm0.20$(S) & $6.16\pm0.52$(S) & $0.63\pm0.12$ & $4.1\pm0.1$(A) & $2.6\pm0.5(14)$ & $97.1\pm21.2$\\
       & HNCO$(4_{0,\,4}-3_{0,\,3})$ & $39.1\pm0.3$ & $21.70\pm0.67$ & $0.77\pm0.21$(G) & $18.06\pm0.53$(G) & $0.43\pm0.02$ & $=2E_{\rm u}/3=7.03$ & $1.0\pm0.1(14)$ & $37.4\pm5.4$\\
       & HCN$(1-0)$\tablefootmark{c} & $44.5\pm0.3$ & $28.90\pm0.29$ & \ldots & \ldots & $23.10\pm0.30$ & $5.1\pm0.5$(A) & $5.5\pm0.1(15)$ & $2055\pm215.3$\\
       &            & $23.4\pm0.2$(B)/ & $16.03\pm0.55$(B)/ & $1.15\pm0.27$(B)/ & $19.57\pm0.78$(B)/ & \ldots & \ldots & \ldots & \ldots \\
       &            & $59.1\pm0.6$(R) & $41.67\pm1.60$(R) & $0.73\pm0.27$(R) & $32.47\pm1.01$(R) & \ldots & \ldots & \ldots & \ldots \\
       & HCO$^+(1-0)$\tablefootmark{c} & $24.2\pm0.6$(B)/ & $11.34\pm1.47$(B)/ & $0.27\pm0.14$(B)/ & $3.25\pm0.41$(B)/ & \ldots & \ldots & \ldots & \ldots \\
       &                               & $64.9\pm0.9$(R) & $40.80\pm2.20$(R) & $0.37\pm0.14$(R) & $15.89\pm0.71$(R) & \ldots & \ldots & \ldots & \ldots \\
       & HNC$(1-0)$ & $37.5\pm0.2$ & $16.30\pm0.81$ & $0.74\pm0.21$(G) & $23.41\pm0.51$(G) & $1.15\pm0.22$ & $4.5\pm0.1$(A) & $5.5\pm1.1(13)$ & $20.6\pm4.6$\\
       & HC$_3$N$(10-9)$ & $38.3\pm0.2$ & $16.00\pm0.96$ & $0.97\pm0.20$(G) & $22.26\pm0.48$(G) & $0.13\pm0.01$ & $=E_{\rm u}=24.01$ & $9.3\pm0.2(13)$ & $34.7\pm3.7$\\
       & N$_2$H$^+(1-0)$ & $37.7\pm0.5$ & $28.70\pm1.31$ & $0.56\pm0.21$(G) & $18.35\pm0.64$(G) & $0.52\pm0.04$ & $7.2\pm0.2$(A) & $8.5\pm0.8(13)$ & $31.8\pm4.4$\\
SMM 30 & SiO$(2-1)$ & $39.0\pm0.4$ & $10.23\pm0.77$ & $0.45\pm0.22$ & $4.91\pm0.34$ & $0.15\pm0.01$ & $=E_{\rm u}=6.25$ & $1.1\pm0.1(13)$ & $3.6\pm0.5$\\
       & HNCO$(4_{0,\,4}-3_{0,\,3})$ & $38.6\pm0.2$ & $10.50\pm0.47$ & $0.81\pm0.25$(G) & $9.20\pm0.37$(G) & $0.45\pm0.03$ & $=2E_{\rm u}/3=7.03$ & $5.1\pm0.2(13)$& $16.8\pm1.9$\\
       & HCN$(1-0)$\tablefootmark{c} & $37.9\pm0.3$ & $12.90\pm0.72$ & \ldots & \ldots & \ldots & \ldots & \ldots & \ldots \\
       &                             & $31.4\pm0.7$(B)/ & $18.81\pm1.64$(B)/ & $0.64\pm0.22$(B)/ & $12.84\pm0.99$(B)/ & \ldots & \ldots & \ldots\\
       &                             & $48.6\pm0.4$(R) & $10.24\pm0.94$(R) & $0.77\pm0.22$(R) & $8.41\pm0.91$(R) & \ldots & \ldots & \ldots \\
       & HCO$^+(1-0)$ & $36.7\pm0.9$ & $29.20\pm1.95$ & $0.29\pm0.21$ & $8.91\pm0.56$ & $0.16\pm0.02$ & $4.9\pm0.2$(A) & $1.1\pm0.1(13)$ & $3.6\pm0.5$ \\     
& HNC$(1-0)$ & $38.5\pm0.2$ & $9.24\pm0.55$ & $0.98\pm0.21$(G) & $15.67\pm0.39$(G) & $1.77\pm0.42$ & $4.5\pm0.1$(A) & $4.8\pm1.2(13)$ & $15.8\pm4.3$\\
       & HC$_3$N$(10-9)$ & $39.0\pm0.4$ & $5.32\pm0.93$ & $0.41\pm0.26$(G) & $3.28\pm0.32$(G) & $0.05\pm0.01$ & $=E_{\rm u}=24.01$ & $1.4\pm0.1(13)$ & $4.6\pm0.6$\\
       & N$_2$H$^+(1-0)$ & $39.6\pm0.4$ & $9.80\pm3.21$ & $0.55\pm0.23$(G) & $9.23\pm0.45$(G) & $0.55\pm0.04$ & $7.2\pm0.2$(A) & $3.1\pm1.0(13)$ & $10.2\pm3.5$\\
SMM 31 & SiO$(2-1)$ & $33.2\pm1.1$ & $21.91\pm2.44$ & $0.28\pm0.26$ & $6.46\pm0.66$ & $0.09\pm0.01$ & $=E_{\rm u}=6.25$ & $1.5\pm0.2(13)$ & $11.6\pm2.0$\\
       & HNCO$(4_{0,\,4}-3_{0,\,3})$ & $33.9\pm0.3$ & $23.60\pm2.03$ & $0.63\pm0.27$(G) & $17.53\pm0.70$(G) & $0.34\pm0.02$ & $=2E_{\rm u}/3=7.03$ & $9.7\pm0.4(13)$ & $75.3\pm8.3$\\
       & HCN$(1-0)$\tablefootmark{c} & $28.7\pm0.6$ & $37.90\pm1.76$ & \ldots & \ldots & \ldots & \ldots & \ldots & \ldots \\ 
       &                             & $26.1\pm0.4$(B)/ & $26.63\pm1.04$(B)/ &  $0.95\pm0.31$(B)/ & $26.97\pm0.96$(B)/ & \ldots & \ldots & \ldots & \ldots  \\
       &                             & $56.8\pm0.9$(R) & $18.71\pm2.36$(R) & $0.35\pm0.31$(R) & $7.06\pm0.67$(R) & \ldots & \ldots & \ldots & \ldots  \\
       & HCO$^+(1-0)$ & $29.3\pm0.6$ & $11.05\pm1.91$ & $0.39\pm0.28$ & $4.60\pm0.59$ & $0.22\pm0.04$ & $4.9\pm0.2$(A) & $5.7\pm0.7(12)$ & $4.4\pm0.7$\\
       & HNC$(1-0)$ & $29.4\pm0.2$ & $12.50\pm0.49$ & $1.19\pm0.25$(G) & $16.11\pm0.51$(G) & $2.57\pm0.91$ & $4.5\pm0.1$(A) & $9.4\pm3.3(13)$ & $73.0\pm26.7$\\
       & HC$_3$N$(10-9)$ & $33.2\pm0.7$ & $22.50\pm1.59$ & $0.41\pm0.27$(G) & $9.90\pm0.63$(G) & $0.05\pm0.01$ & $=E_{\rm u}=24.01$ & $4.1\pm0.3(13)$ & $31.8\pm4.0$\\
       & N$_2$H$^+(1-0)$ & $31.6\pm0.8$ & $19.20\pm2.59$ & $0.40\pm0.29$(G) & $9.08\pm0.70$ & $0.39\pm0.03$ & $7.2\pm0.2$(A) & $1.2\pm0.1(13)$ & $9.3\pm1.2$\\
SMM 38 & SiO$(2-1)$ & $41.7\pm0.6$ & $12.57\pm1.53$ & $0.32\pm0.27$ & $4.32\pm0.45$ & $0.10\pm0.01$ & $=E_{\rm u}=6.25$ & $1.0\pm0.1(13)$ & $3.5\pm0.5$\\
       & HNCO$(4_{0,\,4}-3_{0,\,3})$ & $40.5\pm0.9$ & $25.30\pm2.61$ & $0.35\pm0.25$(G) & $9.71\pm0.75$(G) & $0.18\pm0.01$ & $=2E_{\rm u}/3=7.03$ & $5.4\pm0.4(13)$& $18.8\pm2.4$\\
       & HNC$(1-0)$\tablefootmark{d} & $42.1\pm0.5$ & $15.70\pm1.75$ & $0.62\pm0.24$(G) & $12.36\pm0.53$(G) & $1.65\pm1.18$ & $3.9\pm0.7$ & $6.2\pm4.5(13)$ & $21.9\pm15.8$\\
       & HNC$(1-0)$\tablefootmark{e} & $43.5\pm0.4$ & $10.70\pm1.17$ & $0.63\pm0.25$(G) & $8.85\pm0.44$(G) & $2.28\pm1.40$ & $3.7\pm0.5$ & $5.4\pm3.4(13)$ & $48.1\pm30.8$\\
       & HC$_3$N$(10-9)$ & $40.5\pm0.6$ & $9.72\pm1.18$ & $0.32\pm0.25$(G) & $3.90\pm0.44$(G) & $0.04\pm0.01$ & $=E_{\rm u}=24.01$ & $7.9\pm0.8(14)$ & $274.6\pm39.9$\\
       & N$_2$H$^+(1-0)$\tablefootmark{d} & $40.9\pm0.5$ & $13.40\pm1.30$ & $0.56\pm0.27$(G) & $9.85\pm0.54$(G) & $0.56\pm0.05$ & $7.2\pm0.2$(A) & $4.3\pm0.6(13)$ & $14.9\pm2.6$\\
       & N$_2$H$^+(1-0)$\tablefootmark{e} & $44.7\pm0.4$ & $8.49\pm0.93$ & $0.54\pm0.27$(G) & $6.91\pm0.46$(G) & $0.54\pm0.05$ & $7.2\pm0.2$(A) & $2.6\pm0.4(13)$ & $21.7\pm4.3$\\
G2.11+0.00 & \\
SMM 5 & HCN$(1-0)$ & $59.6\pm0.2$ & $6.83\pm0.86$ & $0.61\pm0.20$(G) & $7.13\pm0.40$(G) & $1.09\pm0.21$ & $5.1\pm0.5$(A) & $6.1\pm1.4(13)$ & $22.4\pm5.6$\\
      & HCO$^+(1-0)$\tablefootmark{f} & $60.2\pm0.1$ & $5.22\pm0.34$ & $0.83\pm0.20$ & $4.59\pm0.25$ & $0.56\pm0.10$ & $4.9\pm0.2$(A) & $7.4\pm1.4(12)$ & $2.7\pm0.6$\\
      &                               & $16.7\pm0.3$ & $2.93\pm0.71$ & $0.32\pm0.20$ & $0.99\pm0.19$ & $0.18\pm0.03$ & $4.9\pm0.2$(A) & $1.2\pm0.2(12)$ & $0.4\pm0.1$\\
      & HNC$(1-0)$ & $60.2\pm0.1$ & $3.46\pm0.22$ & $0.83\pm0.20$(G) & $3.70\pm0.22$(G) & $1.52\pm0.57$ & $4.4\pm0.5$ & $1.5\pm0.6(13)$ & $5.5\pm2.3$\\ 
      & N$_2$H$^+(1-0)$ & $61.5\pm0.1$ & $2.28\pm0.34$ & $0.56\pm0.20$(S) & $1.92\pm0.21$(S) & $2.30\pm2.20$ & $4.2\pm1.1$ & $1.3\pm1.9(13)$ & $4.8\pm7.0$\\
G11.36+0.80 & \\
SMM 1 & HCO$^+(1-0)$ & $28.9\pm0.1$ & $2.03\pm0.36$ & $0.60\pm0.24$ & $1.30\pm0.19$ & $0.37\pm0.07$ & $4.9\pm0.2$(A) & $1.6\pm0.2(12)$ & $0.9\pm0.2$ \\
      & HNC$(1-0)$ & $28.6\pm0.2$ & $3.10\pm0.67$ & $0.69\pm0.33$(G) & $2.29\pm 0.31$(G) & $1.05\pm0.29$ & $4.5\pm0.1$(A) & $9.5\pm3.3(12)$ & $5.6\pm2.0$ \\
      & N$_2$H$^+(1-0)$ & $27.8\pm0.1$ & $2.23\pm0.22$ & $1.30\pm0.28$(S) & $3.98\pm0.24$(S) & $1.46\pm0.14$ & $7.2\pm0.2$(A) & $1.8\pm0.3(13)$ & $10.6\pm2.1$\\  
SMM 2 & HCO$^+(1-0)$ & $28.9\pm0.1$ & $1.84\pm0.32$ & $0.59\pm0.26$ & $1.16\pm0.17$ & $0.36\pm0.07$ & $4.9\pm0.2$(A) & $1.4\pm0.2(12)$ & $0.6\pm0.1$\\
      & HNC$(1-0)$ & $28.2\pm0.2$ & $4.06\pm1.53$ & $0.60\pm0.27$(G) & $2.72\pm0.24$(G) & $0.87\pm0.19$ & $4.5\pm0.1$(A) & $1.0\pm0.4(13)$ & $4.2\pm1.7$\\
      & N$_2$H$^+(1-0)$ & $27.7\pm0.1$ & $2.41\pm0.17$ & $1.61\pm0.28$(S) & $4.90\pm0.26$(S) & $1.91\pm0.20$ & $7.2\pm0.2$(A) & $2.6\pm0.3(13)$& $10.9\pm1.7$\\
SMM 3 & HCO$^+(1-0)$ & $29.1\pm0.2$ & $2.08\pm0.42$ & $0.40\pm0.27$ & $0.89\pm0.17$ & $0.23\pm0.04$ & $4.9\pm0.2$(A) & $1.1\pm0.2(12)$ & $0.7\pm0.1$\\
      & HNC$(1-0)$ & $28.3\pm0.2$ & $2.65\pm0.71$ & $0.53\pm0.23$(G) & $2.18\pm0.24$(G) & $0.74\pm0.14$ & $4.5\pm0.1$(A) & $5.7\pm1.9(12)$ & $3.4\pm1.2$\\
      & N$_2$H$^+(1-0)$ & $28.0\pm0.1$ & $1.83\pm0.22$ & $1.07\pm0.27$(S) & $3.16\pm0.23$(S) & $1.26\pm1.24$ & $6.9\pm3.5$ & $1.2\pm1.2(13)$ & $7.2\pm7.3$\\
SMM 4 & HNC$(1-0)$ & $28.3\pm0.2$ & $2.61\pm0.51$ & $0.97\pm0.45$(G) & $2.77\pm0.42$(G) & $1.74\pm0.80$ & $4.5\pm0.1$(A) & $1.3\pm0.7(13)$ & $10.8\pm5.9$\\       
      & N$_2$H$^+(1-0)$ & $28.1\pm0.1$ & $1.71\pm0.38$ & $1.36\pm0.48$(G) & $3.77\pm0.44$(G) & $1.55\pm0.23$ & $7.2\pm0.2$(A) & $1.5\pm0.4(13)$& $12.4\pm3.5$ \\
SMM 5 & HCO$^+(1-0)$ & $24.9\pm0.4$ & $8.57\pm1.61$ & $0.36\pm0.23$ & $3.24\pm0.43$ & $0.20\pm0.03$ & $4.9\pm0.2$(A) & $4.0\pm0.5(12)$ & $2.8\pm0.5$\\
      & HNC$(1-0)$ & $26.9\pm0.1$ & $1.69\pm0.10$ & $0.56\pm0.28$(G) & $1.79\pm0.20$(G) & $0.80\pm0.18$ & $4.5\pm0.1$(A) & $4.0\pm0.9(12)$ & $2.8\pm0.7$\\
      & N$_2$H$^+(1-0)$ & $26.9\pm0.1$ & $1.83\pm0.21$ & $0.96\pm0.22$(S) & $2.77\pm0.22$(S) & $3.17\pm1.77$ & $4.7\pm0.8$ & $1.7\pm1.1(13)$ & $12.0\pm7.8$\\
SMM 6 & HCO$^+(1-0)$ & $27.7\pm0.8$ & $8.46\pm2.47$ & $0.20\pm0.25$ & $1.77\pm0.38$ & $0.11\pm0.02$ & $4.9\pm0.2$(A) & $2.2\pm0.5(12)$ & $3.4\pm0.9$\\
      & HNC$(1-0)$ & $27.7\pm0.1$ & $1.57\pm0.10$ & $0.56\pm0.31$(G) & $1.76\pm0.21$(G) & $0.80\pm0.19$ & $4.5\pm0.1$(A) & $3.7\pm0.9(12)$ & $5.8\pm1.6$\\ 
& N$_2$H$^+(1-0)$ & $27.4\pm0.1$ & $1.20\pm0.38$ & $0.62\pm0.22$(S) & $1.59\pm0.20$(S) & $6.56\pm5.48$ & $3.6\pm0.4$ & $1.6\pm2.8(13)$ & $25.1\pm43.9$ \\ 
SMM 7 & HCO$^+(1-0)$ & $29.7\pm0.1$ & $2.24\pm0.32$ & $0.83\pm0.31$ & $1.97\pm0.21$ & $0.56\pm0.13$ & $4.9\pm0.2$(A) & $3.2\pm0.9(12)$ & $4.7\pm1.4$\\
      & HNC$(1-0)$ & $29.4\pm0.2$ & $2.89\pm0.53$ & $0.63\pm0.25$(G) & $2.00\pm0.26$(G) & $0.93\pm0.19$ & $4.5\pm0.1$(A) & $7.9\pm2.2(12)$ & $11.7\pm3.5$\\
      & N$_2$H$^+(1-0)$ & $28.2\pm0.1$ & $0.79\pm0.26$ & $0.58\pm0.23$(S) & $1.38\pm0.20$(S) & $3.84\pm3.74$ & $3.8\pm0.7$ & $6.6\pm12.9(12)$ & $9.8\pm19.1$ \\
G13.22-0.06 & \\
SMM 4 & HCN$(1-0)$ & $47.9\pm1.0$ & $21.50\pm3.47$ & $0.35\pm0.27$(G) & $8.75\pm0.84$(G) & $0.54\pm0.12$ & $5.1\pm0.5$(A) & $9.6\pm2.6(13)$ & $28.0\pm8.1$\\
      & HCO$^+(1-0)$\tablefootmark{e} & $49.1\pm0.1$ & $3.89\pm0.38$ & $1.01\pm0.23$ & $4.18\pm0.29$ & $0.73\pm0.16$ & $4.9\pm0.2$(A) & $7.2\pm1.7(12)$ & $2.1\pm0.5$\\
      &                               & $35.8\pm0.2$ & $1.93\pm0.32$ & $0.50\pm0.23$ & $1.03\pm0.17$ & $0.30\pm0.05$ & $4.9\pm0.2$(A) & $1.3\pm0.2(12)$ & $0.4\pm0.1$\\
      & HNC$(1-0)$ & $49.9\pm0.1$ & $3.07\pm0.18$ & $1.52\pm0.25$(G) & $6.17\pm0.24$(G) & $1.53\pm0.12$ & $5.7\pm0.5$ & $2.0\pm0.2(13)$ & $5.8\pm0.8$\\
      & N$_2$H$^+(1-0)$ & $49.8\pm0.1$ & $2.77\pm0.17$ & $1.54\pm0.26$(S) & $6.02\pm0.32$(S) & $1.81\pm0.18$ & $7.2\pm0.2$(A) & $2.8\pm0.3(13)$& $8.2\pm1.2$ \\
SMM 5 & H$^{13}$CO$^+(1-0)$ & $49.0\pm0.2$ & $2.74\pm0.50$ & $0.61\pm0.24$(G) & $1.83\pm0.27$ & $0.65^{+0.31}_{-0.29}$\tablefootmark{h} & $5.6\pm1.5$ & $5.8\pm2.9(12)$ & $0.4\pm0.2$\\      
      & SiO$(2-1)$ & $48.8\pm0.4$ & $5.95\pm1.20$ & $0.38\pm0.24$ & $2.38\pm0.35$ & $0.12\pm0.01$ & $=E_{\rm u}=6.25$ & $5.5\pm0.8(12)$ & $0.4\pm0.1$\\
      & HN$^{13}$C$(1-0)$ & $50.0\pm0.2$ & $3.52\pm0.39$ & $0.66\pm0.25$(G) & $2.54\pm0.24$(G) & $0.69^{+0.40}_{-0.30}$\tablefootmark{h} & $5.4\pm1.4$ & $1.5\pm0.8(13)$ & $1.2\pm0.6$\\
      & C$_2$H$(1-0)$\tablefootmark{d} & $49.9\pm0.2$ & $4.48\pm0.45$ & $0.70\pm0.25$(S) & $3.71\pm0.32$(S) & $2.08\pm0.68$ & $4.1\pm0.1$(A) & $4.5\pm1.3(14)$ & $34.7\pm10.6$\\
      & C$_2$H$(1-0)$\tablefootmark{e} & $50.1\pm0.1$ & $4.09\pm0.32$ & $0.82\pm0.21$(S) & $3.81\pm0.32$(S) & $1.17\pm0.04$ & $5.1\pm0.6$ & $3.4\pm0.3(14)$ & $40.9\pm5.4$\\
      & HCN$(1-0)$ & $47.4\pm0.3$ & $16.20\pm0.79$ & $1.45\pm0.34$(S) & $7.28\pm0.38$(S) & $0.58\pm0.07$\tablefootmark{d} & $5.1\pm0.5$(A) & $7.7\pm1.0(13)$ & $5.9\pm1.0$ \\
      & HCO$^+(1-0)$ & $48.2\pm0.1$ & $4.46\pm0.14$ & $2.60\pm0.25$ & $12.34\pm0.30$ & $10.90^{+5.30}_{-4.80}$\tablefootmark{h} & $5.6\pm0.3$ & $1.5\pm0.7(14)$ & $11.6\pm5.5$\\
      & HNC$(1-0)$\tablefootmark{f} & $49.1\pm0.1$ & $4.65\pm0.37$ & $2.55\pm0.26$(G) & $14.19\pm0.29$(G) & $21.78^{+10.80}_{-9.36}$\tablefootmark{h} & $5.5\pm0.3$ & $2.2\pm1.0(14)$ & $17.0\pm7.9$\\
      &                             & $36.4\pm0.1$ & $0.70\pm0.37$ & $0.51\pm0.26$(G) & $0.57\pm0.15$(G) & $1.22\pm0.10$ & $3.9\pm0.6$ & $2.0\pm1.1(12)$ & $15.4\pm1.5$\\
      & HC$_3$N$(10-9)$ & $48.4\pm0.2$ & $4.07\pm0.43$ & $0.73\pm0.24$(G) & $3.22\pm0.28$ & $0.10\pm0.01$ & $=E_{\rm u}=24.01$ & $1.3\pm0.1(13)$ & $1.0\pm0.1$\\
      & N$_2$H$^+(1-0)$ & $49.3\pm0.1$ & $3.43\pm0.07$ & $3.89\pm0.30$(S) & $18.10\pm0.43$(S) & $1.19\pm0.28$ & $17.9\pm3.2$ & $1.1\pm0.3(14)$ & $8.5\pm2.5$ \\
SMM 6 & HCN$(1-0)$ & $47.9\pm0.3$ & $4.29\pm0.55$ & $0.76\pm0.38$(G) & $11.23\pm0.84$(G) & $0.79\pm0.28$ & $5.1\pm0.5$(A) & $2.8\pm1.1(13)$ & $12.4\pm5.0$ \\ 
      & HCO$^+(1-0)$ & $46.9\pm0.1$ & $2.73\pm0.32$ & $1.94\pm0.38$ & $5.64\pm0.42$ & \ldots\tablefootmark{g} & $4.9\pm0.2$(A) & $6.9\pm0.5(12)$ & $3.1\pm0.4$ \\
      & HNC$(1-0)$ & $49.6\pm0.1$ & $2.60\pm0.23$ & $1.86\pm0.37$(G) & $5.33\pm0.34$(G) & \ldots\tablefootmark{g} & $4.5\pm0.1$(A) & $9.4\pm0.6(12)$ & $4.2\pm0.5$ \\
      & N$_2$H$^+(1-0)$ & $49.4\pm0.1$ & $2.17\pm0.24$ & $1.62\pm0.44$(S) & $5.17\pm0.40$(S) & $1.93\pm0.28$ & $7.2\pm0.2$(A) & $6.9\pm0.4(12)$ & $3.1\pm0.4$ \\ 
SMM 7 & HN$^{13}$C$(1-0)$ & $52.2\pm0.1$ & $1.28\pm0.07$ & $0.72\pm0.20$(G) & $1.87\pm0.16$(G) & $0.79^{+0.30}_{-0.26}$\tablefootmark{h} & $5.5\pm1.1$ & $1.1\pm0.2(13)$ & $1.2\pm0.3$\\
      & C$_2$H$(1-0)$ & $51.7\pm0.1$ & $2.63\pm0.19$ & $1.02\pm0.23$(S) & $3.32\pm0.23$(S) & $1.62\pm0.94$ & $5.0\pm1.0$ & $2.9\pm1.7(14)$ & $32.1\pm19.1$\\
      & HCN$(1-0)$ & $51.5\pm0.1$ & $3.21\pm0.13$ & $1.64\pm0.30$(S) & $6.41\pm 0.12$ & $1.96\pm0.58$ & $5.5\pm0.6$ & $5.8\pm1.7(13)$ & $6.4\pm2.0$\\
      & HCO$^+(1-0)$\tablefootmark{f} & $51.1\pm0.1$ & $5.70\pm0.23$ & $1.62\pm0.28$ & $9.80\pm0.29$ & $1.79\pm0.83$ & $4.9\pm0.2$(A) & $2.6\pm1.2(13)$ & $2.9\pm1.4$\\
      &                               & $37.0\pm0.1$ & $2.05\pm0.38$ & $0.57\pm0.28$ & $1.25\pm0.17$ & $0.35\pm0.07$ & $4.9\pm0.2$(A) & $1.5\pm0.2(12)$ & $0.2\pm0.03$\\
      & HNC$(1-0)$\tablefootmark{f} & $51.5\pm0.1$ & $3.82\pm0.37$ & $2.49\pm0.23$(G) & $10.92\pm0.22$ & $24.84^{+9.36}_{-8.10}$\tablefootmark{h} & $5.5\pm0.2$ & $3.7\pm1.4(14)$ & $41.0\pm16.0$\\
      &                             & $36.8\pm0.1$ & $2.27\pm0.37$ & $0.37\pm0.23$(G) & $1.00\pm0.16$(G) & $1.10\pm0.10$ & $3.7\pm0.6$ & $5.5\pm1.0(12)$ & $0.6\pm0.1$\\
      & N$_2$H$^+(1-0)$ & $51.7\pm0.1$ & $2.59\pm0.06$ & $3.83\pm0.32$(S) & $13.86\pm0.40$(S) & $1.00\pm0.28$ & $20.0\pm4.4$ & $8.9\pm2.5(13)$ & $9.9\pm2.9$\\
SMM 11 & HCO$^+(1-0)$\tablefootmark{f} & $53.5\pm0.3$ & $8.64\pm0.67$ & $0.56\pm0.23$ & $5.10\pm0.36$ & $0.34\pm0.06$ & $4.9\pm0.2$(A) & $6.3\pm0.5(12)$ & $3.3\pm0.4$\\
       &                               & $37.2\pm0.1$ & $1.73\pm0.30$ & $0.60\pm0.23$ & $1.10\pm0.17$ & $0.37\pm0.07$ & $4.9\pm0.2$(A) & $1.4\pm0.2(12)$ & $0.7\pm0.1$\\
       & HNC$(1-0)$\tablefootmark{f} & $52.5\pm0.1$ & $4.26\pm0.32$ & $0.88\pm0.24$(G) & $4.09\pm0.25$(G) & $1.49\pm0.36$ & $4.5\pm0.1$(A) & $1.9\pm0.5(13)$ & $9.9\pm2.8$ \\
       &                             & $37.8\pm0.1$ & $1.26\pm0.16$ & $0.32\pm0.24$(G) & $0.85\pm0.19$(G) & $0.41\pm0.07$ & $4.5\pm0.1$(A) & $1.5\pm0.3(12)$ & $0.8\pm0.2$\\
       & N$_2$H$^+(1-0)$ & $52.3\pm0.1$ & $2.45\pm0.30$ & $0.90\pm0.27$(S) & $3.27\pm0.26$(S) & $1.87\pm1.63$ & $5.3\pm1.8$ & $1.6\pm1.4(13)$ & $8.3\pm7.4$ \\
SMM 23 & HCN$(1-0)$ & $37.1\pm0.1$ & $2.82\pm0.29$ & $0.42\pm0.24$(G) & $4.64\pm0.37$(G) & $2.30\pm1.62$ & $3.4\pm0.4$ & $3.0\pm2.1(13)$ & $18.5\pm13.1$\\
       & HCO$^+(1-0)$\tablefootmark{f} & $37.0\pm0.1$ & $3.25\pm0.18$ & $1.39\pm0.23$ & $4.81\pm0.22$ & $1.26\pm0.37$ & $4.9\pm0.2$(A) & $1.0\pm0.3(13)$ & $6.2\pm2.0$\\
       &                               & $10.4\pm0.2$ & $1.72\pm0.44$ & $0.35\pm0.22$ & $0.64\pm0.15$ & $0.20\pm0.03$ & $4.9\pm0.2$(A) & $7.9\pm1.8(11)$ & $0.5\pm0.1$ \\
       &                               & $53.0\pm0.2$ & $3.50\pm0.55$ & $0.42\pm0.22$ & $1.55\pm0.21$ & $0.24\pm0.03$ & $4.9\pm0.2$(A) & $1.9\pm0.3(12)$ & $1.2\pm0.2$ \\
       & HNC$(1-0)$\tablefootmark{f} & $37.7\pm0.1$ & $3.45\pm0.14$ & $1.50\pm0.20$(G) & $5.62\pm0.20$(G) & $5.74\pm5.54$ & $4.5\pm0.1$(A) & $5.8\pm5.6(13)$ & $35.7\pm34.7$\\
       &                             & $52.7\pm0.2$ & $3.14\pm0.20$ & $0.30\pm0.20$(G) & $1.82\pm0.25$(G) & $0.38\pm0.06$ & $4.5\pm0.1$(A) & $3.2\pm0.4(12)$ & $2.0\pm0.3$\\
       & N$_2$H$^+(1-0)$ & $38.2\pm0.1$ & $2.29\pm0.17$ & $1.43\pm0.18$(S) & $4.46\pm0.24$(S) & $1.02\pm0.77$ & $9.2\pm4.2$ & $2.0\pm1.5(13)$ & $12.3\pm9.3$\\
SMM 27 & H$^{13}$CO$^+(1-0)$ & $36.6\pm0.1$ & $2.97\pm0.31$ & $0.72\pm0.21$(G) & $2.31\pm0.21$(G) & $1.51^{+0.80}_{-0.62}$\tablefootmark{h} & $4.5\pm0.7$ & $1.0\pm0.5(13)$ & $1.4\pm0.7$\\
       & HN$^{13}$C$(1-0)$ & $37.2\pm0.1$ & $1.73\pm0.57$ & $0.71\pm0.22$(G) &  $1.53\pm0.18$ & $1.09^{+0.51}_{-0.42}$\tablefootmark{h} & $4.8\pm0.9$ & $1.0\pm0.5(13)$ & $1.4\pm0.7$\\
       & C$_2$H$(1-0)$ & $36.8\pm0.1$ & $2.64\pm0.28$ & $0.61\pm0.20$(S) & $1.577\pm0.18$(S) & $4.01\pm2.44$ & $3.6\pm0.3$ & $4.6\pm2.8(14)$ & $63.8\pm39.3$\\
       & HCN$(1-0)$ & $36.1\pm0.5$ & $15.90\pm1.92$ & $0.52\pm0.25$(G) & $10.15\pm0.68$(G) & $0.88\pm0.20$ & $5.1\pm0.5$(A) & $1.2\pm0.3(14)$ & $16.6\pm4.5$\\
       & HCO$^+(1-0)$ & $36.2\pm0.1$ & $4.05\pm0.17$ & $1.55\pm0.24$ & $6.70\pm0.23$ & $28.30^{+15.10}_{-11.40}$\tablefootmark{h} & $4.5\pm0.3$ & $2.6\pm1.2(14)$ & $36.1\pm17.0$\\
       & HNC$(1-0)$ & $36.7\pm0.1$ & $3.54\pm0.10$ & $1.91\pm0.23$(G) & $7.72\pm0.23$(G) & $37.80^{+17.82}_{-14.58}$\tablefootmark{h} & $4.9\pm0.2$ & $4.4\pm1.9(14)$ & $61.0\pm27.0$\\
       & HC$_3$N$(10-9)$\tablefootmark{d} & $36.7\pm0.2$ & $2.73\pm0.33$ & $0.57\pm0.21$(G) & $1.67\pm0.19$(G) & $0.08\pm0.01$ & $=E_{\rm u}=24.01$ & $7.0\pm0.8(12)$ & $1.0\pm0.1$\\
       & HC$_3$N$(10-9)$\tablefootmark{e} & $36.7\pm0.1$ & $3.29\pm0.32$ & $0.63\pm0.21$(G) & $2.23\pm0.20$(G) & $0.08\pm0.01$ & $=E_{\rm u}=24.01$ & $9.3\pm0.8(12)$ & $2.0\pm0.2$\\
       & N$_2$H$^+(1-0)$\tablefootmark{d} & $36.8\pm0.1$ & $2.77\pm0.12$ & $2.01\pm0.22$(S) & $7.81\pm0.28$(S) & $1.85\pm0.55$ & $8.4\pm1.4$ & $3.7\pm1.1(13)$ & $5.1\pm1.6$ \\
        & N$_2$H$^+(1-0)$\tablefootmark{e} & $36.7\pm0.1$ & $2.82\pm0.09$ & $2.39\pm0.21$(S) & $9.99\pm0.31$(S) & $2.58\pm0.47$ & $8.0\pm0.8$ & $4.9\pm0.9(13)$ & $9.0\pm1.9$ \\
SMM 29 & C$_2$H$(1-0)$ & $45.4\pm0.1$ & $1.77\pm0.29$ & $0.49\pm0.18$(S) & $1.31\pm0.20$(S) & $1.25\pm0.26$ & $4.1\pm0.1$(A) & $1.1\pm0.3(14)$ & $18.3\pm5.3$\\
       & HCN$(1-0)$ & $57.7\pm0.2$ & $2.41\pm0.46$ & $0.49\pm0.22$(G) & $1.21\pm0.18$(G) & $0.82\pm0.16$ & $5.1\pm0.5$(A) & $1.6\pm0.4(13)$ & $2.7\pm0.7$\\
       & HCO$^+(1-0)$\tablefootmark{d, f} & $44.8\pm0.1$ & $3.60\pm0.11$ & $1.23\pm0.23$ & $4.72\pm0.04$ & $1.00\pm0.25$ & $4.9\pm0.2$(A) & $9.2\pm2.3(12)$ & $1.5\pm0.4$\\
       &                                  & $13.6\pm0.1$ & $1.96\pm0.11$ & $0.37\pm0.23$ & $0.77\pm0.04$ & $0.21\pm0.03$ & $4.9\pm0.2$(A) & $9.5\pm0.5(11)$ & $0.2\pm0.02$\\    
       &                                  & $34.8\pm0.1$ & $5.43\pm0.11$ & $0.34\pm0.23$ & $1.97\pm0.04$ & $0.19\pm0.03$ & $4.9\pm0.2$(A) & $2.4\pm0.1(12)$ & $0.4\pm0.04$\\
         &                                 & $57.0\pm0.1$ & $3.65\pm0.11$ & $0.24\pm0.23$ & $0.93\pm0.04$ & $0.13\pm0.02$ & $4.9\pm0.2$(A) & $1.1\pm0.1(12)$ & $0.2\pm0.02$\\
       & HCO$^+(1-0)$\tablefootmark{e} & $44.6\pm0.1$ & $2.44\pm0.13$ & $1.97\pm0.35$ & $5.13\pm0.20$ & $1.57\pm0.76$\tablefootmark{d} & $4.9\pm0.2$(A) & $9.7\pm4.7(12)$ & $3.2\pm1.6$\\                                
       & HNC$(1-0)$\tablefootmark{d, f} & $45.2\pm0.1$ & $1.57\pm0.24$ & $1.29\pm0.20$(G) & $3.92\pm0.17$(G) & $3.13\pm1.12$ & $4.5\pm0.1$(A) & $1.4\pm0.6(13)$ & $2.3\pm1.0$ \\ 
       &                                & $13.6\pm0.3$ & $2.17\pm0.25$ & $0.22\pm0.20$(G) & $0.92\pm0.24$(G) & $0.27\pm0.04$ & $4.5\pm0.1$(A) & $1.6\pm0.4(12)$ & $0.3\pm0.1$\\
       &                                & $37.1\pm0.3$ & $3.57\pm0.76$ & $0.33\pm0.20$(G) & $1.68\pm0.23$(G) & $0.43\pm0.06$ & $4.5\pm0.1$(A) & $3.0\pm0.4(12)$ & $0.5\pm0.1$\\
       & HNC$(1-0)$\tablefootmark{e, f} & $44.8\pm0.1$ & $1.80\pm0.18$ & $1.93\pm0.30$(G) & $4.74\pm0.16$(G) & $1.87\pm0.91$ & $6.0\pm1.0$ & $1.5\pm0.8(13)$ & $30.6\pm3.1$\\
       &                                & $36.7\pm0.2$ & $2.07\pm0.23$ & $0.36\pm0.30$(G) & $1.21\pm0.35$(G) & $0.47\pm0.10$ & $4.5\pm0.1$(A) & $2.1\pm0.6(12)$ & $0.4\pm0.1$ \\
       & N$_2$H$^+(1-0)$\tablefootmark{d} & $45.1\pm0.1$ & $1.69\pm0.09$ & $1.53\pm0.20$(S) & $3.71\pm0.19$(S) & $0.77\pm0.04$ & $11.6\pm1.2$ & $1.7\pm0.1(13)$ & $2.8\pm0.3$\\
       & N$_2$H$^+(1-0)$\tablefootmark{e} & $45.0\pm0.1$ & $1.59\pm0.08$ & $2.07\pm0.24$(S) & $5.32\pm0.22$(S) & $3.07\pm0.79$ & $6.8\pm0.8$ & $2.5\pm0.7(13)$ & $5.1\pm1.5$\\
SMM 32 & H$^{13}$CO$^+(1-0)$ & $51.8\pm0.2$ & $3.58\pm0.78$ & $0.45\pm0.17$(G) & $1.81\pm0.19$(G) & $0.72^{+0.36}_{-0.31}$\tablefootmark{h} & $4.7\pm1.0$ & $6.4\pm3.3(12)$ & $1.1\pm0.6$\\
       & C$_2$H$(1-0)$\tablefootmark{d} & $52.3\pm0.1$ & $3.69\pm0.32$ & $0.58\pm0.18$(S) & $3.16\pm0.25$(S) & $3.65\pm1.38$ & $3.6\pm0.3$ & $5.8\pm2.2(14)$ & $98.4\pm38.6$\\
       & C$_2$H$(1-0)$\tablefootmark{e} & $52.0\pm0.4$ & $8.07\pm1.33$ & $0.25\pm0.18$(S) & $2.02\pm0.30$(S) & $0.56\pm0.10$ & $4.1\pm0.1$(A) & $7.8\pm1.2(13)$ & $63.4\pm11.6$\\
       & HNCO$(4_{0,\,4}-3_{0,\,3})$\tablefootmark{e} & $51.3\pm0.1$ & $2.01\pm0.57$ & $0.41\pm0.21$(G) & $0.98\pm0.28$(G) & $0.22\pm0.01$ & $=2E_{\rm u}/3=7.03$ &$5.4\pm1.5(12)$ & $1.3\pm0.4$\\
       & HCN$(1-0)$\tablefootmark{d} & $59.4\pm0.1$ & $2.50\pm0.22$ & $0.86\pm0.18$(G) & $2.45\pm0.17$(G) & $1.84\pm0.40$ & $5.1\pm0.5$(A) & $3.8\pm0.9(13)$ & $6.5\pm1.7$\\
       & HCN$(1-0)$\tablefootmark{e} & $57.3\pm0.6$ & $5.43\pm0.84$ & $0.46\pm0.22$(G) & $2.30\pm0.23$(G) & $0.76\pm0.14$ & $5.1\pm0.5$(A) & $3.4\pm0.8(13)$ & $9.3\pm2.4$\\
       & HCO$^+(1-0)$\tablefootmark{d} & $54.3\pm0.1$ & $2.60\pm0.12$ & $1.73\pm0.19$ & $4.77\pm0.17$ & $11.80^{+5.90}_{-5.00}$\tablefootmark{h} & $4.7\pm0.2$ & $7.3\pm3.4(13)$ & $12.4\pm5.9$\\
       & HCO$^+(1-0)$\tablefootmark{e, f} & $53.9\pm0.1$ & $2.87\pm0.17$ & $1.40\pm0.18$ & $4.28\pm0.19$ & $1.27\pm0.33$ & $4.9\pm0.2$(A) & $9.3\pm2.5(12)$ & $1.7\pm0.5$\\
       &                                  & $13.5\pm0.1$ & $1.50\pm0.34$ & $0.42\pm0.18$ & $0.67\pm0.13$ & $0.24\pm0.04$ & $4.9\pm0.2$(A) & $8.2\pm1.6(11)$ & $0.1\pm0.03$\\
       & HNC$(1-0)$\tablefootmark{d} & $53.0\pm0.1$ & $5.30\pm0.16$ & $0.74\pm0.20$(G) & $4.34\pm0.22$(G) & $1.15\pm0.21$ & $4.5\pm0.1$(A) & $1.8\pm0.3(13)$ & $3.1\pm0.6$\\
       & HNC$(1-0)$\tablefootmark{e, f} & $52.9\pm0.2$ & $5.09\pm0.39$ & $0.60\pm0.19$(G) & $3.32\pm0.23$(G) & $0.87\pm0.14$ & $4.5\pm0.1$(A) & $1.3\pm0.2(13)$ & $6.1\pm1.9$\\
       &                                & $14.1\pm0.3$ & $2.10\pm0.98$ & $0.23\pm0.13$(G) & $0.84\pm0.17$(G) & $0.29\pm0.03$ & $4.5\pm0.1$(A) & $1.5\pm0.3(12)$ & $0.7\pm0.2$ \\
       & HC$_3$N$(10-9)$ & $51.6\pm0.1$ & $2.93\pm0.30$ & $0.69\pm0.20$(G) & $2.18\pm0.16$(G) & $0.09\pm0.01$ & $=E_{\rm u}=24.01$ & $9.1\pm0.7(12)$ & $1.5\pm0.2$\\
       & N$_2$H$^+(1-0)$\tablefootmark{d} & $51.4\pm0.1$ & $2.87\pm0.11$ & $1.82\pm0.24$(S) & $7.21\pm0.22$(S) & $1.86\pm0.53$ & $7.9\pm1.3$ & $3.5\pm1.0(13)$ &$5.9\pm1.8$ \\
       & N$_2$H$^+(1-0)$\tablefootmark{e} & $51.5\pm0.1$ & $2.76\pm0.08$ & $2.10\pm0.22$(S) & $8.96\pm0.23$(S) & $3.20\pm0.47$ & $6.8\pm0.5$ & $4.6\pm0.7(13)$ & $7.7\pm1.4$\\
\end{longtable}
\tablefoot{
\tablefoottext{a}{``S'' refers to the value of the strongest hf component, while ``G'' refers to the blended group of hf components. The $T_{\rm ex}$ values marked with ``A'' represent the average values derived for other sources.}\tablefoottext{b}{$a(b)$ stands for $a\times10^b$.}\tablefoottext{c}{The line is double peaked. Hyperfine fit results are given when possible. Gaussian fit parameters are given separately for both peaks (B=blue, R=red).}\tablefoottext{d}{Towards the LABOCA peak position.}\tablefoottext{e}{Towards the line emission peak. ``N'' refers to the northern maximum while ``S'' stands for the southern peak.}\tablefoottext{f}{Two or more velocity components.}\tablefoottext{g}{The optical thickness could not be estimated with the assumptions made. The column density was therefore estimated using the assumption of optically thin emission.}\tablefoottext{h}{Calculated from the isotopologue intensity ratio.}}
}
}

\end{document}